\documentclass[UTF8,oneside,openany]{ctexbook}
\usepackage[a4paper,top=25mm,bottom=25mm,left=28mm,right=22mm]{geometry}
\usepackage{graphicx}
\usepackage{longtable,booktabs,array,tabularx}
\usepackage{calc}
\usepackage{xcolor}
\usepackage{amssymb}
\usepackage{listings}
\usepackage{hyperref}
\usepackage{bookmark}
\usepackage{fvextra}
\usepackage{xurl}
\usepackage{seqsplit}
\ctexset{
  chapter/name={Chapter~},
  chapter/number=\thechapter,
  appendix/name={Appendix~},
  appendix/number=\thechapter
}
\providecommand{\tightlist}{}
\providecommand{\passthrough}[1]{#1}
\providecommand{\pandocbounded}[1]{#1}
\newcommand{\breakablecode}[1]{\texttt{\seqsplit{#1}}}

\AtBeginEnvironment{longtable}{\small\sloppy}
\definecolor{codebackground}{RGB}{247,248,250}
\hypersetup{hidelinks,unicode=true}

\title{OpenAgenet: Architecture, Protocols, Governance, and Conformance for an Internet of Agents Infrastructure}
\author{Jinliang Xu}
\date{November 2026}
\begin{document}
\frontmatter
\begin{titlepage}
\centering
\vspace*{10mm}
\includegraphics[width=0.25\textwidth]{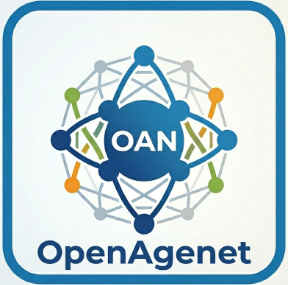}
\vspace{14mm}

{\LARGE\bfseries OpenAgenet\par}
\vspace{5mm}
{\Huge\bfseries Architecture, Protocols, Governance, and Conformance\par}
\vspace{5mm}
{\Large\bfseries for an Internet of Agents Infrastructure\par}
\vspace{14mm}
{\large Technical Monograph and Architecture/Protocol Report\par}
\vspace{3mm}
{\large Version 1.1\par}
\vspace{16mm}
{\large Jinliang Xu\par}
\vspace{3mm}
{\large \href{mailto:xujinliang@caict.ac.cn}{xujinliang@caict.ac.cn}\par}
\vspace{3mm}
{\large \href{mailto:jlxufly@gmail.com}{jlxufly@gmail.com}\par}
\vspace{10mm}
{\large November 2026\par}
\vfill

\href{https://www.openagenet.xyz}{https://www.openagenet.xyz}\par
\vspace{3mm}
\href{https://github.com/OpenAgenet}{https://github.com/OpenAgenet}
\end{titlepage}

\chapter*{Abstract}
\addcontentsline{toc}{chapter}{Abstract}

The Internet of Agents requires infrastructure that allows heterogeneous autonomous agents to identify resources, establish control and authorization, publish verifiable resource descriptions, discover capabilities, and evaluate governance state across independently operated services. This technical monograph presents OpenAgenet (OAN), an open infrastructure architecture for those functions. It defines the principal roles and data objects of the system, including resource decentralized identifiers, DID Documents, verifiable credentials and presentations, signed request envelopes, resource packages, trusted publication records, capability tags, lifecycle states, and governance projections. It then connects these objects through registration, verification, publication, distribution, discovery, invocation, node authorization, and operational workflows.

Particular attention is given to trust boundaries, controller continuity, replay protection, cryptographic profiles, authorization domains, failure semantics, and the distinction between protocol requirements, reference-implementation behavior, and observations from a particular deployment. The monograph also provides a conformance and evidence model that links requirements to fixed fixtures, implementation entry points, configuration profiles, and reproducible checks. Rather than treating successful access to one deployment as proof of protocol correctness, the report records the scope and limits of each form of evidence. The resulting architecture is intended to support interoperable and auditable agent-resource infrastructure while leaving application-specific reasoning and business logic outside the protocol core.

\vspace{1em}
\noindent\textbf{Keywords:} Internet of Agents; autonomous agents; decentralized identifiers; verifiable credentials; resource discovery; trusted publication; governance; conformance testing; distributed systems.

\chapter*{Scholarly Positioning and Contributions}
\addcontentsline{toc}{chapter}{Scholarly Positioning and Contributions}

\section*{Problem and scope}

OAN addresses an infrastructure problem that arises when agent-accessible resources are created and operated by different parties: a resource description alone does not establish control, authorization, publication status, governance state, or fitness for invocation. The report therefore treats identity, registration, trusted publication, distribution, discovery, governance, and caller-side verification as distinct stages with explicit evidence boundaries. It does not prescribe an agent's internal model, planner, or reasoning algorithm, and it does not replace application-specific authorization or business logic.

\section*{Principal contributions}

The report makes five connected technical contributions:

\begin{enumerate}
  \item It defines a role and object model for agent-resource infrastructure, including the responsibilities of resource providers, Registrar nodes, the Root platform, distribution services, Discovery nodes, governance indexers, SDKs, and callers.
  \item It specifies how resource identifiers, DID Documents, credentials, signed envelopes, resource packages, publication records, capability tags, versions, and lifecycle state are bound across registration, publication, discovery, and invocation.
  \item It states security and governance boundaries for controller continuity, replay protection, cryptographic agility, authorization domains, trusted publication proofs, revocation, and degraded operation.
  \item It separates normative protocol requirements, reference-implementation behavior, deployment configuration, experimental observations, and evolution designs so that evidence at one layer is not misrepresented as proof at another.
  \item It provides a conformance and traceability method that connects requirements to fixed inputs, expected outcomes, implementation entry points, configuration profiles, and retainable evidence.
\end{enumerate}

\section*{Relationship to earlier OAN publications}

The OAN White Paper motivates the need for identity, trusted publication, and semantic discovery in an Internet of Agents. The OAN Yellow Paper develops mechanisms, trust boundaries, state transitions, and verification requirements. This monograph consolidates those foundations into a self-contained architecture and protocol report and extends them with detailed object mappings, interface semantics, operational constraints, implementation traceability, and conformance procedures. Where a statement depends on a particular implementation or deployment, the report labels that dependency rather than promoting it to a universal protocol property.

\section*{Evidence and reproducibility}

Design rationale, normative rules, implementation evidence, and runtime evidence are treated as different forms of support. Reproducible claims identify the relevant version, configuration, fixed fixture or workload, expected result, and evidence location. Repository links and source mappings help locate an implementation, but do not by themselves establish conformance or production readiness. Likewise, a successful deployment observation does not establish interoperability across independent implementations.

\section*{Limitations}

The report describes an evolving reference architecture rather than a finalized Internet standard. Some capabilities are supported only by the reference implementation or a named deployment profile. Multi-Root coordination, broad cross-organizational interoperability, and complete live-network conformance evidence remain areas for further implementation and evaluation. Performance observations are conditional on the recorded software revision, configuration, workload, and environment and must not be read as universal service guarantees.

\tableofcontents
\mainmatter
\chapter{Document Scope and Status}\label{chap:00}

This chapter is the entry point for reading this technical manual. It is intended for resource providers, node operators, client and SDK developers, protocol implementers, integration developers, and technical personnel evaluating OpenAgenet (OAN). The manual covers resources such as Agent Services, Skills, MCP Servers, and Tool APIs, as well as supporting mechanisms including subject identities, resource DIDs, DID Documents, registration VCs, Root-verified ResourcePackages, content distribution, discovery indexing, and pre-call verification. Readers can use it as a guide from concepts to implementation: first read Chapters~\ref{chap:01}~through~\ref{chap:04} to understand the goals, roles, architecture, and resource model, and then consult the identity, registration, publication, discovery, operations, and conformance chapters as needed.

This manual classifies its evidence in four layers. Design rationale explains goals and trade-offs. Normative rules establish interoperability, security, and data-integrity baselines. Implementation evidence shows how the reference code realizes those rules. Runtime evidence records the outcome of a particular commit and configuration in a particular environment. The four layers may corroborate one another, but none can substitute for another. A feature statement in a README, for example, cannot replace route and test evidence, and one successful deployment cannot turn a reference configuration into a protocol requirement. When later chapters cite volatile interfaces, configuration values, or experimental results, they also identify the source, applicable version, and conditions under which the claim was verified.

\pandocbounded{\includegraphics[width=\linewidth,height=0.72\textheight,keepaspectratio]{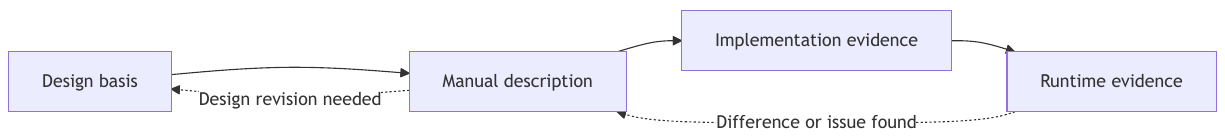}}

\section{Revision History}\label{sec:00.1}

{\def\LTcaptype{none} 
\begin{longtable}[]{@{}>{\raggedright\arraybackslash}p{0.2400\linewidth}>{\raggedright\arraybackslash}p{0.2400\linewidth}>{\raggedright\arraybackslash}p{0.2400\linewidth}>{\raggedright\arraybackslash}p{0.2400\linewidth}@{}}
\toprule\noalign{}
Date & Capability change & Chapters affected & Implementation and verification basis \\
\midrule\noalign{}
\endhead
\bottomrule\noalign{}
\endlastfoot
2026-09-14 & Initial technical-manual release & Entire manual & Initial Chinese Markdown and the White and Yellow Papers available at the time \\
2026-10 & Added Identity Runtime, Identity Gateway, and Identity Sidecar; expanded on-chain bulletin and Trust Indexer coverage & Chapters 3, 7, 13, 16, and 21; Appendices E and F & The three identity-integration repositories, \passthrough{\lstinline!oan-bulletin-contracts/Sui!}, and \passthrough{\lstinline!oan-trust-indexer!} \\
2026-11 & Completed cross-version controller continuity; migrated request nonces to SQLite; optimized the DID verification executor and shared capability tree & Chapters 5, 7, 8, 9, 17, and 18; Appendices A, C, and D & Controller-continuity design and unit/integration gates; Root nonce and verification-path designs \\
2026-11 & Unified typed configuration across four core Rust modules; revised publication, distribution, discovery, and benchmark observability & Chapters 3, 9, 10, 11, 12, 19, 20, 23, and 24 & Configuration-refactoring documents, four pipeline benchmarks, and the White/Yellow Paper experiment audit \\
\end{longtable}
}

\section{Purpose and Scope}\label{sec:00.2}

This manual explains OAN in terms of objects, responsibilities, processes, and evidence. Objects include resource DIDs, DID Documents, registration submissions, Root-verified ResourcePackages, registration VCs, and governance projections. Responsibilities identify who creates, verifies, and stores each object. Processes connect registration, publication, distribution, discovery, and invocation. Evidence enables a reader to check a conclusion against public types, node behavior, or test results. The existence of one object must not be treated as proof that the entire pipeline has completed: a credential returned by a Registrar, for example, does not mean that a Discovery node has already indexed the resource.

The main chapters give priority to mechanisms that should remain stable across versions. The appendices carry types, routes, fixtures, and repository entry points that change more frequently. This does not make implementation detail less important; it prevents a route or field change from leaving the conceptual chapters and the code as two competing sources of truth.

The manual covers the complete path from local preparation of a resource to its discovery and verification: the resource provider prepares the resource and local identity; the client submits the materials to a Registrar; the Registrar performs admission checks; the Root platform verifies the materials and publishes a trusted ResourcePackage; the content-distribution platform hosts retrievable artifacts; a Discovery node synchronizes them and builds indexes; and the caller checks the DID, credentials, version, and governance status before connecting. The corresponding chapters explain the resource model, \passthrough{\lstinline!did:oan!}, DID Documents, ResourcePackages, VCs and VPs, signed messages, node authorization, semantic tags, HTTP APIs, SDKs, the Community Skill, lifecycle, operations, and conformance.

{\def\LTcaptype{none} 
\begin{longtable}[]{@{}>{\raggedright\arraybackslash}p{0.3200\linewidth}>{\raggedright\arraybackslash}p{0.3200\linewidth}>{\raggedright\arraybackslash}p{0.3200\linewidth}@{}}
\toprule\noalign{}
Task for the reader & Suggested reading path & Main result \\
\midrule\noalign{}
\endhead
\bottomrule\noalign{}
\endlastfoot
Understand OAN\textquotesingle s positioning and components & Chapters 1, 2, 3, and 4 & Understand the goals, roles, network topology, and resource types \\
Register or publish a resource & Chapters 5, 6, 7, 8, 9, and 10 & Understand identity, ResourcePackages, registration, trusted publication, and distribution \\
Integrate discovery, identity components, and semantic queries & Chapters 7, 11, 12, 14, 15, 16, and 21 & Understand queries, identity integration, node interoperability, and SDK usage \\
Deploy, troubleshoot, and evaluate & Chapters 17, 19, 20, 23, and 24 plus the appendices & Obtain security, operations, deployment, testing, and version references \\
\end{longtable}
}

The manual does not prescribe which model or reasoning algorithm an agent must use internally, does not define all business logic of a business system, and does not replace external protocols such as A2A or MCP. Each chapter should state its applicable roles, inputs and outputs, responsible party, verifying party, state changes, exception handling, and retainable evidence. When a conclusion applies only to a reference implementation or a current deployment, its version, environment, and boundary should be stated.

\section{Relationship to the OAN White Paper and Yellow Paper}\label{sec:00.3}

The White Paper is the conceptual entry point for the project. It mainly explains why infrastructure for resource identity, trusted publication, and semantic discovery is needed for the Internet of Agents (IoA). The Yellow Paper focuses more on how the mechanisms work, explaining roles, data objects, trust boundaries, publication paths, authorization domains, state machines, security invariants, and reference verification methods. This manual organizes those materials into technical guidance that readers can consult and use, while adding implementation mappings, examples, operational boundaries, and verification paths.

{\def\LTcaptype{none} 
\begin{longtable}[]{@{}>{\raggedright\arraybackslash}p{0.3200\linewidth}>{\raggedright\arraybackslash}p{0.3200\linewidth}>{\raggedright\arraybackslash}p{0.3200\linewidth}@{}}
\toprule\noalign{}
Material & What readers mainly obtain & Boundary of use \\
\midrule\noalign{}
\endhead
\bottomrule\noalign{}
\endlastfoot
White Paper & Background, vision, overall positioning, and design principles & Not directly equivalent to interface fields or deployment commitments \\
Yellow Paper & Roles, mechanisms, trust boundaries, and engineering derivations & Reference algorithms and derivations do not automatically define a fixed format for every implementation \\
This technical manual & Conceptual explanations, processes, implementation mappings, examples, operations, and verification entry points & Normative requirements, reference implementation behavior, and current runtime evidence must still be distinguished \\
\end{longtable}
}

Readers can start with the White Paper to understand ``why,'' use the Yellow Paper to understand ``how the mechanisms are organized,'' and then return to this manual to find ``how to understand, integrate, implement, and verify'' them. When different materials differ only in abstraction level, the manual should explain their relationship. When an actual conflict exists in fields, processes, or security boundaries, it should record the conflict location, impact, applicable version, and revision status rather than conceal it with vague wording.

{\def\LTcaptype{none} 
\begin{longtable}[]{@{}>{\raggedright\arraybackslash}p{0.2400\linewidth}>{\raggedright\arraybackslash}p{0.2400\linewidth}>{\raggedright\arraybackslash}p{0.2400\linewidth}>{\raggedright\arraybackslash}p{0.2400\linewidth}@{}}
\toprule\noalign{}
Material or evidence & Main question answered & What it can directly demonstrate & What it cannot demonstrate alone \\
\midrule\noalign{}
\endhead
\bottomrule\noalign{}
\endlastfoot
White Paper & Why OAN is being built and what its overall vision is & Background, goals, value, and principles & Specific interface fields or a deployed implementation \\
Yellow Paper & How the mechanisms are organized and where the security boundaries are & Roles, processes, states, and mechanism derivations & That every implementation uses the same code \\
Unified technical specification & How different implementations interoperate & Fields, behavior, errors, and conformance requirements & Long-term availability of a particular deployment \\
Code, tests, and configuration & How a reference implementation realizes the rules & Module behavior, test results, and configuration constraints & That the design vision has already been implemented \\
Deployment and runtime records & Whether a particular environment currently works & Build, startup, smoke-test, and runtime status & The protocol\textquotesingle s inherent requirements \\
\end{longtable}
}

\pandocbounded{\includegraphics[width=\linewidth,height=0.72\textheight,keepaspectratio]{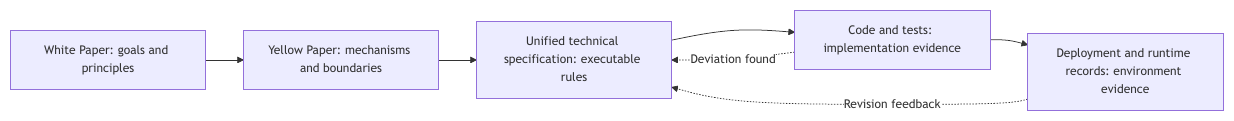}}

During publication or review, record at least the specification version, referenced material versions, implementation commit, configuration version, test time, and applicable environment. When materials differ, first determine whether the difference is caused by abstraction level or is an actual conflict; only the latter should enter the conflict record and revision process.

\section{\texorpdfstring{Relationship to the \texttt{did:oan} Method Specification}{Relationship to the did:oan Method Specification}}\label{sec:00.4}

The \passthrough{\lstinline!did:oan!} method materials define the method-specific DID syntax, method-specific identifiers, DID Document structure, verification methods, public-key representation, service endpoints, resolution behavior, and boundaries for control proofs. The related chapters of this manual do not reinvent those method rules. Instead, they explain how OAN uses them: how clients generate and store local keys, how resource controllers prove control, how Registrar nodes validate DID Documents, how the Root platform associates resource versions with trusted publication records, how Discovery nodes build verifiable indexes, and how callers resolve and verify signatures before connecting.

A minimal status decision can be represented as follows:

\begin{lstlisting}
DID Document valid
    -> Resource control proof valid
    -> Accepted by Registrar node
    -> Published by Root platform
    -> Indexed by Discovery node
    -> Caller completes version, governance, and local permission checks
\end{lstlisting}

The steps above are not different names for one ``usable'' state. A DID Document that is valid in syntax, structure, and signature does not mean that the resource has been accepted by a Registrar node, published by the Root platform, indexed by a Discovery node, or granted permission for use. A Root publication fact cannot replace the caller\textquotesingle s checks of the DID Document, VC, endpoint, version, governance status, and local business permissions. Later chapters should explain how these states connect through fields, processes, and status transitions.

Since the first draft was completed in late September 2026, the OAN implementation baseline has added cross-version controller-continuity checks, unified typed configuration, persistent publication tasks and JetStream notifications, atomic SQLite storage for request nonces, dual structured and semantic indexes on the discovery side, and three integration components: Identity Runtime, Identity Gateway, and Identity Sidecar. This manual describes those capabilities according to the reference implementation and the evidence with which they were actually verified. Multi-Root coordination and live interoperability between independent implementations remain evolution directions rather than current deployment facts.

\section{Relationship to Implementation Repositories and Deployment Materials}\label{sec:00.5}

Implementation mappings in this manual help readers answer ``where is this capability implemented, how can it be verified, and what are its current boundaries''; a repository name is not itself a conformance proof. Shared data structures, serialization, hashing, and cryptographic behavior should first be checked against \passthrough{\lstinline!oan-protocol-commons!}; Root publication, governance state, and publication proofs against \passthrough{\lstinline!oan-root-nodes!}; registration admission and registration credentials against \passthrough{\lstinline!oan-registrar-nodes!}; synchronization, indexing, semantic retrieval, and discovery responses against \passthrough{\lstinline!oan-discovery-nodes!}; projection of on-chain governance events into runtime authorization read models against \passthrough{\lstinline!oan-trust-indexer!}; and client integration against \passthrough{\lstinline!oan-sdk-ts!} and the Community Skill. These node and shared-protocol repositories are the public implementation references. Unit tests, fixtures, and admission results should cite an independent public test suite, public SDK test material, or reproducible evidence from this manual. READMEs, schemas, route handlers, test material, configuration examples, and deployment scripts provide different kinds of evidence, and their purpose should be stated when cited.

Each mapping should record at least the manual section, repository and module or file, corresponding version, implementation status, verification method, and known deviation. Implementation status should distinguish implemented with test evidence, implemented with insufficient evidence, partially implemented, and design-only. Deployment materials should also record environment variables, external services, databases, key boundaries, build artifacts, startup order, and post-deployment smoke tests. Addresses, node counts, and runtime metrics from the official environment are runtime evidence and should not be written as mandatory conditions for every environment.

\section{Normative Language and Conformance Terms}\label{sec:00.6}

This manual uses RFC-style keywords to indicate degrees of obligation: MUST corresponds to ``must,'' MUST NOT to ``must not,'' SHOULD to ``should,'' SHOULD NOT to ``normally should not,'' and MAY to ``may.'' These words are used only for confirmed interoperability, security, data-integrity, or conformance requirements. When describing current code behavior, deployment status, or operational advice, use more precise wording such as ``current implementation,'' ``normally,'' ``recommended,'' or ``may.''

Conformance judgments should target a specific role and configuration profile rather than making a broad judgment about OAN as a whole. The capability boundaries of resource providers, Registrar nodes, the Root platform, the content distribution platform, Discovery nodes, the off-chain trust indexer, resolvers, and SDKs are not identical. Each test or runtime check should state its fixed input, prerequisite configuration, execution steps, expected output, failure impact, and evidence-retention method. Passing tests in a reference implementation does not mean that every third-party implementation is inherently conformant, and normal operation of the official deployment does not mean that the protocol covers every environment.

The manual uses five categories when describing content: current facts supported by code and runtime verification; requirements that form protocol or security baselines; behavior of a reference implementation in a named repository; recommendations for deployment and use; and evolution directions that have not yet been implemented. Experimental results describe observations under a stated commit, configuration, hardware platform, and test script. They do not automatically become performance commitments for every node. When the text says ``current implementation,'' ``default configuration,'' ``experimental result,'' or ``planned support,'' the reader should use the adjacent source or the implementation mapping in Appendix~\ref{chap:appendix-f} to determine its scope.

The Chinese Markdown is the direct maintenance source for this revision. The Chinese and English Markdown, both LaTeX editions, and the website publication must be synchronized against the same revision record. Updating only one representation does not complete a manual release.

\section{Content Status and Version Freeze}\label{sec:00.7}

``Implemented,'' ``passed a fixed fixture,'' ``verified in an isolated environment,'' and ``running in production'' describe four different states. The first two can be supported by source code and automated tests. Isolated-environment verification also requires a record of the binary, configuration, and dependencies. Production status requires deployment gates and runtime observation. When a capability lacks a single evidence chain across these states, this manual retains its design or reference-implementation classification rather than promoting it on the strength of one successful health check.

Four visible labels help qualify easily confused capability statements. \textbf{Current reference implementation} means that a code entry point and targeted verification can be identified. \textbf{Optional deployment profile} means that an implementation exists but activation depends on configuration and runtime dependencies. \textbf{Experimental observation} is valid only for the recorded commit, configuration, dataset, and measurement method. \textbf{Evolution design} states an architectural or specification direction and must not be treated as a current network capability. These labels are convenient presentations of the five content categories above, not a second classification system. Multi-Root coordination and cross-organizational interoperability without live-environment evidence are currently evolution designs.

This Chinese technical-manual revision uses the local repository state of 2026-11-12 as its verification snapshot. The following table records the principal implementation references used to check the main text. Appendix~\ref{chap:appendix-f} maintains repository-level responsibilities and test entry points. If a later release changes any reference, the affected chapters, configuration examples, and figures must be checked again.

{\def\LTcaptype{none} 
\begin{longtable}[]{@{}>{\raggedright\arraybackslash}p{0.3200\linewidth}>{\raggedright\arraybackslash}p{0.3200\linewidth}>{\raggedright\arraybackslash}p{0.3200\linewidth}@{}}
\toprule\noalign{}
Implementation scope & Repository and verification reference & Principal use in this manual \\
\midrule\noalign{}
\endhead
\bottomrule\noalign{}
\endlastfoot
Shared protocols, types, and cryptography & Public release of \passthrough{\lstinline!oan-protocol-commons!} & Public facts for DIDs, ResourcePackages, credentials, signatures, configuration, and publication events \\
Root, Registrar, and Discovery & Public releases of \passthrough{\lstinline!oan-root-nodes!}, \passthrough{\lstinline!oan-registrar-nodes!}, and \passthrough{\lstinline!oan-discovery-nodes!} & Registration admission, authoritative state, publication and distribution, indexing, and query behavior \\
Off-chain governance projection & \passthrough{\lstinline!oan-trust-indexer!} \passthrough{\lstinline!a6c7ce780c85!} & On-chain event reading, cursors, state projection, and queries \\
Client and operations entry points & Public release of \passthrough{\lstinline!oan-sdk-ts!}; public Community Skill and deployment-verification records & Client materials, API calls, deployment gates, and benchmark entry points \\
Identity-integration components & \passthrough{\lstinline!oan-identity-runtime!} \passthrough{\lstinline!1e3f3e4b7c34!}; \passthrough{\lstinline!oan-identity-gateway!} \passthrough{\lstinline!fa788c40539c!}; \passthrough{\lstinline!oan-identity-sidecar!} \passthrough{\lstinline!f5ec07b46dc2!} & Identity and proof integration for A2A, MCP, and HTTP invocations \\
Contracts and verification material & Public release of \passthrough{\lstinline!oan-bulletin-contracts!}; publicly reproducible end-to-end records and conformance evidence & On-chain bulletins, end-to-end flows, and third-party admission verification \\
\end{longtable}
}

A publishable manual snapshot should also freeze the configuration-example version, protocol and schema versions, experiment-results directory, and generation date. The Chinese Markdown is the maintenance source for this revision. The English Markdown and LaTeX outputs must be generated from, or checked item by item against, the same snapshot; an old version number must not remain in use after only one representation has been updated.

\section{References}\label{sec:00.8}

{\def\LTcaptype{none} 
\begin{longtable}[]{@{}>{\raggedright\arraybackslash}p{0.3200\linewidth}>{\raggedright\arraybackslash}p{0.3200\linewidth}>{\raggedright\arraybackslash}p{0.3200\linewidth}@{}}
\toprule\noalign{}
Source & Type & Link \\
\midrule\noalign{}
\endhead
\bottomrule\noalign{}
\endlastfoot
OAN White Paper & arXiv White Paper: project vision and positioning & \url{https://arxiv.org/abs/2606.03161} \\
OAN Yellow Paper & arXiv Yellow Paper: technical architecture and trust model & \url{https://arxiv.org/abs/2606.03163} \\
\passthrough{\lstinline!did:oan!} DID Method Specification & Standard/method specification: \passthrough{\lstinline!did:oan!} identifiers and method rules & \url{https://github.com/OpenAgenet/oan-public-docs/blob/main/did-oan-specs/doc/OAN\%20DID\%20Method\%20Specification.md} \\
OAN Resource Identity and Discovery & IETF Internet-Draft: resource identity and discovery & \url{https://datatracker.ietf.org/doc/draft-xu-oan-resource-identity-discovery/} \\
Agentic Overlay Network Architecture & IETF Internet-Draft: agentic overlay network architecture & \url{https://datatracker.ietf.org/doc/draft-xu-agentic-overlay-network-architecture/} \\
Efficient Agent Discovery Profile & IETF Internet-Draft: efficient agent discovery profile & \url{https://datatracker.ietf.org/doc/draft-xu-efficient-agent-discovery-profile/} \\
\end{longtable}
}

\chapter{System Goals and Design Principles}\label{chap:01}

This chapter explains why OpenAgenet (OAN) needs trusted resource infrastructure for the Internet of Agents (IoA), and how its system goals are realized through identity, registration, trusted publication, content distribution, discovery, governance, and pre-call verification. Rather than listing isolated product features, it explains how these mechanisms jointly address resources that cannot be identified, descriptions whose origins cannot be traced, fragmented discovery across directories, discontinuities in controller updates, and external protocols that lack a common identity context.

\section{OAN and the Internet of Agents}\label{sec:01.1}

OAN provides resource identity, trusted publication, and cross-node discovery infrastructure for the IoA. It is not concerned with the model or conversational capability of one particular agent. Instead, it enables Agent Services, Skills, MCP Servers, and Tool APIs to have stable identities, verifiable control relationships, describable capabilities, traceable publication status, and reusable discovery entry points. OAN treats these objects as discoverable resources, allowing them to participate in cross-organization interconnection through a common \passthrough{\lstinline!did:oan!}, resource metadata, registration protocol, and discovery protocol rather than through a manually maintained directory on a single platform.

The core result of OAN is to connect ``what capability exists'' with ``which evidence supports treating it as a candidate resource.'' The resource provider prepares a DID Document and resource description locally. After admission by a Registrar node, the Root platform verifies node authorization, the resource-control proof, metadata, and version information. A content distribution platform then makes the materials available to authorized Discovery nodes. After receiving search results, the caller still performs pre-call verification using the DID, VC, Root proof, governance status, and endpoint information. OAN does not proxy every business invocation, nor does it continuously prove that a registered description is identical to the runtime content. The resource provider and its controller remain responsible for description accuracy and timely updates.

From the user\textquotesingle s perspective, the goal of OAN can be summarized as four consecutive outcomes: the resource can be identified, its origin can be verified, it can be discovered within an authorized scope, and the caller can obtain sufficient grounds for judgment before connecting. These outcomes are jointly supported by identity materials, registration and publication evidence, discovery indexes, and caller-side verification. None of them can be replaced by a single page, directory, or search score.

{\def\LTcaptype{none} 
\begin{longtable}[]{@{}>{\raggedright\arraybackslash}p{0.3200\linewidth}>{\raggedright\arraybackslash}p{0.3200\linewidth}>{\raggedright\arraybackslash}p{0.3200\linewidth}@{}}
\toprule\noalign{}
Design goal & Main supporting mechanisms & Observable result for readers \\
\midrule\noalign{}
\endhead
\bottomrule\noalign{}
\endlastfoot
Resources are identifiable & \passthrough{\lstinline!did:oan!}, DID Documents, resource metadata & Distinguish the resource subject, type, version, and service references \\
Origins are verifiable & Control proofs, registration VCs, Root-verified ResourcePackages & Check the control relationship, publication origin, and material hashes \\
Resources are discoverable & Authorization domains, capability tags, structured queries, and semantic retrieval & Find candidate resources by DID or task requirements \\
Decisions can be made before invocation & Version, lifecycle, governance status, endpoints, and local policy & Decide whether to proceed to a business invocation \\
\end{longtable}
}

\section{Problems Addressed by OAN}\label{sec:01.2}

{\def\LTcaptype{none} 
\begin{longtable}[]{@{}>{\raggedright\arraybackslash}p{0.3200\linewidth}>{\raggedright\arraybackslash}p{0.3200\linewidth}>{\raggedright\arraybackslash}p{0.3200\linewidth}@{}}
\toprule\noalign{}
Problem & OAN mechanism & Verifiable result \\
\midrule\noalign{}
\endhead
\bottomrule\noalign{}
\endlastfoot
Resources cannot be distinguished stably & \passthrough{\lstinline!did:oan!} and resource type & Subject, type, and version can be referenced unambiguously \\
Description origin and responsible party are unclear & Controller, Registrar verification, registration VC & Submission responsibility and registration source are traceable \\
A new controller could self-sign a takeover & Current-controller authorization and Root atomic state check & Unauthorized successor versions are rejected \\
Cross-directory results lack trusted state & Trusted packages, authorization domains, Discovery validation & Candidates carry version, provenance, and governance evidence \\
External calls lack a unified identity context & Runtime, Gateway, Sidecar, request/response proofs & OAN identity material can enter A2A, MCP, or HTTP paths \\
\end{longtable}
}

When Agent Services, Skills, MCP Servers, and Tool APIs are distributed across different teams, code repositories, or runtime environments, resource providers often have to describe their locations and capabilities through webpages, documentation, configuration files, or manual communication. Such approaches make it difficult to express the controller, version, endpoint, dependencies, applicable protocols, and usage conditions in a stable way. They also make it difficult to determine whether a search result represents the current version, a historical version, or a resource that has already been suspended. As the number of resources grows, directory maintenance, duplicate registration, version updates, and cross-organization collaboration create continuing manual costs.

For callers, name or text matching can answer only ``possibly relevant.'' It cannot by itself answer whether the resource is genuine, whether it was published by an authorized node, whether it remains valid, or whether it can be used within the current authorization domain. If registration, publication, discovery, and invocation do not have clear responsibility boundaries, an intermediate directory may introduce unverified resources into results or continue displaying resources that have been revoked, replaced, or whose origins are unknown. OAN addresses these problems separately through DID control proofs, node authorization, Root publication, ResourcePackage hashes, authorization-domain filtering by Discovery nodes, and pre-call verification. It does not use a single platform label as a substitute for every trust judgment.

The problems can be mapped to OAN\textquotesingle s handling as follows: dispersed resource endpoints require common resource identities and service descriptions; descriptions with unclear origins require controller relationships, registration VCs, and Root trusted-publication proofs; opaque version status requires binding versions, digests, and lifecycle states to resource records; controller transfer risk requires each successor version to be authorized by the currently recorded controller; and difficult cross-organization discovery requires Discovery nodes, authorization domains, structured conditions, and optional semantic queries to work together. OAN therefore does not merely create another resource list. It organizes registration, origin verification, network propagation, and pre-call judgment into a traceable process.

\section{Resource-First Interconnection Model}\label{sec:01.3}

``Resource first'' means that the network first establishes stable identities for capability objects that can be independently identified, registered, published, and discovered. It does not require every object to conform to one agent runtime. An Agent Service may refer to Skills, MCP Servers, and Tool APIs, while each resource retains its own DID, controller, version, endpoint, and evidence chain. Composition can therefore change without simultaneously changing the identities of all component resources.

This model also explains the boundary between OAN and a business runtime. OAN helps a caller locate a resource and verify its origin; actual task orchestration, session state, input and output, and business authorization remain the responsibility of A2A, MCP, HTTP, or application protocols. Network-level trusted publication does not automatically turn a resource declaration into a service-level commitment.

The primary objects of OAN registration, publication, distribution, and discovery are resource versions with a resource type, resource DID, description, capability tags, use cases, endpoints, versions, and states, rather than agent instances that must run on a unified platform. An Agent Service may provide a session- or task-oriented service; a Skill may serve as a reusable capability unit; an MCP Server may expose a collection of tools; and a Tool API may provide a specific interface. They differ in business form, but all enter the network through unified resource identity and metadata.

The unified model standardizes common properties without erasing type differences. The resource type determines required fields, endpoint shape, protocol bindings, invocation constraints, and version handling. The resource DID identifies a specific controllable object; the ResourcePackage binds the representation and proofs for a version; and capability tags and use cases support structured and semantic discovery. Resources may express dependencies and composition through DIDs, versions, or package hashes. However, ``discoverable'' only means that a resource satisfies directory and trust conditions; it does not automatically mean that every business invocation in a composition relationship is authorized.

The resource-first model can be understood through five groups of information: identity fields answer ``which resource is this,'' description fields answer ``what can it do,'' version and digest fields answer ``which version of the material is being viewed,'' endpoint and protocol fields answer ``how can it be integrated,'' and proof and status fields answer ``why may it be treated as a candidate resource.'' Different resource types may share these common fields, but this does not turn a Skill into an Agent Service, an MCP Server into a Tool API, or a discovery result into business invocation authorization. Any OAN resource may serve as the controller of another resource. A dedicated controller commonly carries the control identity of an enterprise or individual, but it is not the only control subject permitted by the protocol. Resolvable-object examples and the differences among the four resource types are centralized in Chapters 4 and 6 and Appendix~\ref{chap:appendix-a}, avoiding a second simplified field model in this chapter.

\section{Trusted Registration, Publication, Discovery, and Invocation}\label{sec:01.4}

The OAN trust path is not a single action completed by one request. It is a continuous process performed jointly by the resource controller, Registrar node, Root platform, content distribution platform, Discovery node, and caller. The resource controller first generates or imports identity materials locally and prepares the resource DID Document, resource metadata, capability tags, endpoint, and control proof. The Registrar node performs admission checks and submits upstream. The Root platform verifies authorization, signatures, hashes, versions, and governance conditions. Downstream nodes only propagate and index trusted facts that have already been formed. The caller independently verifies the relevant evidence before establishing a business connection.

For a resource version, readers can determine ``how far the process has progressed'' in this order:

\begin{enumerate}
\def\labelenumi{\arabic{enumi}.}
\tightlist
\item
  Whether the local materials have been generated and the controller can use the corresponding private key to create the proof;
\item
  Whether the Registrar node has accepted the request and returned the registration result;
\item
  Whether the Root platform has formed the acceptance and trusted-publication facts for the version;
\item
  Whether the content distribution platform and Discovery node have completed validation, synchronization, and authorization-domain filtering;
\item
  Whether the caller has completed checks of the DID, VC, version, status, endpoint, and business policy.
\end{enumerate}

Any failed step should retain a locatable status or error so that readers can determine whether the cause is a material problem, a node-authorization problem, propagation delay, an indexing problem, or a caller-side policy rejection.

\subsection{Resource Registration and Control Proofs}\label{subsec:01.4.1}

Resource registration begins when the resource provider generates a resource version locally. The client should collect the name, description, capability tags, use cases, protocol bindings, endpoint, version, and dependencies according to the resource type, and establish a verifiable relationship among the resource DID, control key, public key, and DID Document. The material submitted to a Registrar node should prove that the requester controls the resource DID. The Registrar node may validate formats, fields, tags, endpoints, and policies, but its acceptance does not mean that the Root platform has accepted the resource.

A control proof answers ``who can represent this resource in a registration request.'' It cannot replace judgments about resource quality, endpoint availability, or invocation permissions. On failure, the service should return locatable field or proof errors while keeping the local draft and key materials outside the service\textquotesingle s control. A second submission for the same DID should distinguish a duplicate request for the same version from an authorized update to a new version, and dispatch them respectively to idempotency handling and version-update rules.

\subsection{Root Verification and Trusted Publication}\label{subsec:01.4.2}

When admitting a resource, the Root platform independently checks node identity, resource control, controller continuity, digests, and versions against current authoritative state. This is not a repetition of the Registrar\textquotesingle s commercial identity check. It is the final writer\textquotesingle s independent enforcement of protocol invariants. Acceptance creates the ResourcePackage and publication task; CDN persistence and Discovery visibility belong to subsequent asynchronous stages.

After receiving a signed upstream request from a Registrar node, the Root platform should first verify request freshness, request-body hash, nonce, and the Registrar node\textquotesingle s valid authorization. It then verifies the resource DID Document, resource control proof, resource type, metadata hash, ResourcePackage, and version-transition conditions. Only when these facts are consistent and satisfy governance and publication policies does the Root platform form an acceptance record and Root-verified ResourcePackage for the resource version. The Root acceptance record should bind the DID, version, representation hash, authorization domain, status, time, or sequence information so that downstream nodes do not have to determine the origin from readable text alone.

After successful Root verification, publication normally proceeds asynchronously through persistent tasks, messages, or synchronization cursors. The resource provider may receive a registration result from the Registrar node, but the resource should not be described as discoverable across the network until content distribution, Root publication, and Discovery-node indexing are complete. Rejection, isolation, retry, and republishing should have observable states; one successful HTTP response must not be used to represent completion of the entire propagation path.

\subsection{Content Distribution and Discovery-Node Indexing}\label{subsec:01.4.3}

The content distribution platform carries ResourcePackages, DID Documents, metadata, and proofs after the Root platform has verified them. A Discovery node obtains candidate versions from the distribution layer. It recalculates digests, verifies the Root proof and governance status, and filters resources by its own authorization domain before writing qualified resources into a structured index. When a semantic engine is enabled, the node then builds a derived semantic index asynchronously. The distribution layer improves accessibility and synchronization efficiency, but it does not acquire responsibility for maintaining the Root platform\textquotesingle s authoritative state.

Distribution and indexing are asynchronous. It is therefore normal for the Root platform to have published a resource while a Discovery node has not yet made it visible, for structured indexing to be complete while semantic indexing is retrying, or for a node to provide only an older snapshot temporarily. Implementations should retain synchronization cursors, version freshness, rejection reasons, and pending tasks so that operators can distinguish not published, not synchronized, validation failure, authorization-domain mismatch, and search-index failure.

\subsection{Pre-Call Verification and Integration}\label{subsec:01.4.4}

After obtaining candidate resources from a Discovery node, the caller must not treat search results directly as invocation permission. At a minimum, it should verify the response origin and freshness, the resource DID Document, the resource control relationship, the registration VC or equivalent registration evidence, the Root-verified ResourcePackage, the ResourcePackage hash, the current version, lifecycle status, the Discovery node\textquotesingle s authorization domain, and the target endpoint\textquotesingle s protocol and authentication declarations. For high-risk operations, the actual request should also include a timestamp, nonce, target DID, request-body hash, and signature to prevent replay or replacement of the result.

Successful verification only means that the candidate satisfies the prerequisites of the OAN trust path and the caller\textquotesingle s local policy. Whether the call can ultimately execute still depends on authentication, authorization, business state, and security policy at the resource endpoint. A client should distinguish DID or signature errors, credential-status errors, Root-proof mismatches, expired versions, invalid governance states, authorization-domain mismatches, and endpoint invocation failures rather than reducing every problem to ``resource not found.''

Registrar response, Root admission, content-distribution persistence, and Discovery-node validation and indexing are four distinct completion points. A caller may use the registration response to confirm the synchronous request result, but a resource becomes visible at a particular Discovery node only after the required publication and discovery states have been reached. This distinction is also the basis for explaining end-to-end latency and fault recovery.

\section{Separation of Resource Control, Node Authorization, and Governance}\label{sec:01.5}

Resource control is expressed through the control key and control proof associated with the resource DID, answering ``who may submit or update the document on behalf of this resource.'' Node authorization is expressed through Root authorization materials and their governance status, answering ``which Registrar node, Discovery node, or other infrastructure node is qualified to perform a particular network action.'' On-chain governance records express governance facts such as changes to node authorization, answering ``when an authorization becomes effective, is suspended, restored, or revoked.'' A Root proof answers ``whether a resource version has passed Root verification and entered the publication path''; it is not a credential of resource ownership.

When OAN operating rules require a dedicated controller to undergo verification of a real enterprise or individual identity at registration time, an authorized Registrar provider performs that verification and issues a registration VC to the controller. Resources registered by that controller record the control relationship in their DID Documents, and subsequent updates remain subject to authorization by the current controller. Registration responsibility can therefore be traced along the control relationship. This traceability establishes who is responsible for the registered materials; it does not automatically prove that an endpoint is currently available or that its runtime content will never change.

These three categories of fact should be checked separately. Possession of a resource private key does not give an unauthorized node registration privileges. Possession of an old node authorization VC does not bypass the current governance status. Root publication of a resource does not replace the caller\textquotesingle s judgment about endpoint and business permissions. This separation allows OAN to manage infrastructure authorization without receiving resource-control private keys, while allowing governance status, resource identity, and publication status to be updated independently.

{\def\LTcaptype{none} 
\begin{longtable}[]{@{}>{\raggedright\arraybackslash}p{0.3200\linewidth}>{\raggedright\arraybackslash}p{0.3200\linewidth}>{\raggedright\arraybackslash}p{0.3200\linewidth}@{}}
\toprule\noalign{}
Judgment question & Main evidence & Judgment it cannot replace \\
\midrule\noalign{}
\endhead
\bottomrule\noalign{}
\endlastfoot
Who may submit or update a resource & Resource DID, control key, and control proof & Whether the node has infrastructure authorization \\
Which node may perform a network action & Node authorization materials and governance status & Whether that node controls a particular resource \\
Which version has entered the trusted publication path & Root proof, version, and hash & Whether the resource endpoint meets the caller\textquotesingle s business requirements \\
Whether an invocation may actually be made & The caller\textquotesingle s combined verification and local business policy & It cannot be decided only by a search result or page status \\
\end{longtable}
}

\section{Federation, Interoperability, and Incremental Adoption}\label{sec:01.6}

OAN is federated because multiple Registrar nodes and Discovery nodes can operate independently under the same Root trust and publication rules. Registrar nodes can provide resource admission for different organizations, teams, or integration scenarios. Discovery nodes can provide indexes based on authorization domain, network location, or service capability. The Root platform unifies verification and publication facts; it does not require every resource to be stored in one business platform. A third-party node may enter the corresponding operating state only after completing identity, authorization, interface, and conformance checks.

Incremental adoption can begin with one organization or an experimental network: first adopt \passthrough{\lstinline!did:oan!} resource identity and the registration-discovery flow, then connect existing Agents, MCP Servers, or Tool APIs, and later add multiple nodes, authorization domains, semantic tags, and governance read models. Interoperability means preserving OAN\textquotesingle s boundaries for identity, proofs, versions, and status while connecting existing business systems through protocol adapters. It does not require a business system to become a unified runtime immediately.

An organization can proceed in the order ``make visible first, make trusted next, federate later.'' In the first phase, it organizes existing resources and establishes resource identities and descriptions. In the second, it integrates a Registrar node and forms control proofs, registration VCs, and Root publication evidence. In the third, it deploys or integrates Discovery nodes and enables structured and semantic queries. In the fourth, it adds nodes from multiple organizations, authorization domains, governance synchronization, and protocol mappings. Existing business endpoints can remain in place at every phase. OAN mainly provides resource identity, registration, publication, discovery, and pre-call judgment; it does not require an immediate replacement of existing business systems.

\begin{lstlisting}
Resource organization and identity establishment
    -> Registration and trusted publication
    -> Discovery-node indexing
    -> Multi-node integration and authorization-domain governance
    -> Mapping to existing Agent, MCP, and Tool/API protocols
\end{lstlisting}

The current implementation permits multiple Registrar nodes and multiple Discovery nodes to join the same network. Federation here does not mean that consensus or global-state replication among multiple Root platforms has already been implemented; multi-Root operation remains a future architectural direction.

\subsection{Relationship Between the Identity Trust Layer and Business Protocols}\label{subsec:01.6.1}

OAN\textquotesingle s identity and trust capabilities span resource registration and the business-invocation integration layer, but they do not define a replacement business protocol for A2A, MCP, or HTTP. Identity Runtime is suitable for using identity materials inside application code or a controlled service; Identity Gateway verifies and forwards proofs at a shared ingress; and Identity Sidecar accompanies a workload when the business process should be changed as little as possible. The three forms may be selected or combined according to key-custody and network boundaries. None acquires additional controller authority or governance power. The business protocol remains responsible for task, tool-invocation, streaming, cancellation, and error semantics.

\section{Control Plane, Data Plane, and Semantic Plane}\label{sec:01.7}

The control plane maintains facts that determine whether a resource can enter the network, propagate, or be treated as valid, including identity, authorization, version, lifecycle, and publication status. The data plane carries DID Documents, ResourcePackages, metadata, proofs, and distribution content for clients. The semantic plane maintains capability-tag trees, tag hierarchies, aliases, descriptions, and query explanations for registration recommendations and semantic matching during discovery. Identity and trust capabilities span the control plane and the invocation-integration layer; this does not require mechanically adding a separate ``identity plane.''

The three planes have a clear dependency order. Semantic matching can help a Discovery node find relevant resources, and data distribution can allow a node to obtain a ResourcePackage. However, a candidate has trusted-discoverable status only after the control plane confirms that the resource version has been accepted and published and is within the Discovery node\textquotesingle s authorization domain. Semantic relevance, endpoint accessibility, or interface presentation cannot grant DID control, node authorization, or governance qualification in reverse.

Control-plane progress is not one large serial loop. The client/Registrar registration-request loop, the Root-to-CDN background publication loop, and the synchronization loop for each Discovery target have separate queues and completion conditions. Backlog in one loop does not invalidate facts already formed by the others, but it does affect when a resource becomes visible at a particular Discovery node. Semantic indexing is a derived task after structured indexing; semantic-service unavailability must not block propagation of the trusted ResourcePackage.

{\def\LTcaptype{none} 
\begin{longtable}[]{@{}>{\raggedright\arraybackslash}p{0.2400\linewidth}>{\raggedright\arraybackslash}p{0.2400\linewidth}>{\raggedright\arraybackslash}p{0.2400\linewidth}>{\raggedright\arraybackslash}p{0.2400\linewidth}@{}}
\toprule\noalign{}
Plane & Main facts maintained & Typical participating components & Constraint imposed on other planes \\
\midrule\noalign{}
\endhead
\bottomrule\noalign{}
\endlastfoot
Control plane & Identity, authorization, version, lifecycle, and publication status & Root platform, governance components, Registrar nodes, off-chain trust indexer & Determines which materials may enter propagation and trusted discovery \\
Data plane & DID Documents, ResourcePackages, metadata, proofs, and distribution content & Content distribution platform, Registrar nodes, Discovery nodes & Must preserve hashes, versions, and origins, and must not rewrite control-plane facts \\
Semantic plane & Capability tags, hierarchies, aliases, descriptions, and query explanations & Root semantic governance, Discovery nodes & Helps understanding and ranking, but cannot replace identity and authorization judgments \\
\end{longtable}
}

\section{Public Visibility, Discoverability, and Invocation Authorization}\label{sec:01.8}

{\def\LTcaptype{none} 
\begin{longtable}[]{@{}>{\raggedright\arraybackslash}p{0.2400\linewidth}>{\raggedright\arraybackslash}p{0.2400\linewidth}>{\raggedright\arraybackslash}p{0.2400\linewidth}>{\raggedright\arraybackslash}p{0.2400\linewidth}@{}}
\toprule\noalign{}
State & Completion condition & Observable basis & Still cannot imply \\
\midrule\noalign{}
\endhead
\bottomrule\noalign{}
\endlastfoot
Public registration & Registrar returns and stores a registration result & Response, registration VC, or record & Root, CDN, or Discovery completed \\
Trusted publication & Root accepts the version and forms trusted material & Current Root version, RootProof, task state & Every Discovery node converged \\
Discoverable & A specified Discovery validates and exposes it & Target cursor, index state, query result, package evidence & Business endpoint permits invocation \\
Callable & Caller and endpoint complete identity, state, and business authorization & Request/response proof, endpoint authentication, local policy & Future continued availability \\
\end{longtable}
}

{\def\LTcaptype{none} 
\begin{longtable}[]{@{}>{\raggedright\arraybackslash}p{0.2400\linewidth}>{\raggedright\arraybackslash}p{0.2400\linewidth}>{\raggedright\arraybackslash}p{0.2400\linewidth}>{\raggedright\arraybackslash}p{0.2400\linewidth}@{}}
\toprule\noalign{}
Stage & Primary responsible party & Fact produced or confirmed & Basis usable by next stage \\
\midrule\noalign{}
\endhead
\bottomrule\noalign{}
\endlastfoot
Preparation & Resource controller/provider & DID, DID Document, metadata, control proof & Submittable registration material \\
Registration & Registrar & Acceptance, field validation, registration result & Registration credential or upstream record \\
Publication & Root & Node authorization, resource version, Root proof & Distributable ResourcePackage \\
Indexing & Discovery & Receipt, validation, domain filtering, index state & Discovery result with provenance state \\
Pre-invocation decision & Caller/verifier & Identity, signature, version, governance, endpoint checks & Local decision whether to connect \\
\end{longtable}
}

Public visibility means that material can be read; discoverability means that it has entered the authorized view of a particular Discovery node; invocation authorization is decided by the resource endpoint according to caller identity, credentials, and business policy. These concepts cannot be collapsed into one Boolean value. A resource may be public but excluded from candidates because it is suspended, or discoverable while still requiring a separate token or business authorization. User interfaces and SDKs should show these states separately so that a search hit is not mistaken for invocation permission.

A successful registration response means that the Registrar has completed the synchronous request cycle and returned a result. Trusted publication means that the Root has accepted the version and created the corresponding authoritative state and publication materials. Discoverable means that a particular authorized Discovery node has validated the ResourcePackage, applied authorization-domain filtering, and written it to its local structured index. Callable is a local judgment made after the caller verifies OAN evidence and then applies endpoint authentication and its own business policy.

These states may advance in sequence, or may temporarily diverge because of asynchronous tasks or faults. An older snapshot may continue to provide a limited set of verified historical results when refresh fails, but it must carry freshness or status information. Resource suspension, revocation, version replacement, and node-authorization changes should affect discovery and pre-call verification after the corresponding state has propagated. Any public portal, statistics page, or cache is only an observation interface; it cannot replace the Root acceptance record, governance status, or signature verification as a trust source.

The following diagram compresses the chapter\textquotesingle s goals into a checkable resource-version flow. Each arrow represents the transfer of a fact or material to the next stage. It does not mean that the previous stage automatically grants authority for the next stage. Asynchronous distribution and indexing may create time differences, and the caller must still verify the result before the final connection.

\pandocbounded{\includegraphics[width=\linewidth,height=0.72\textheight,keepaspectratio]{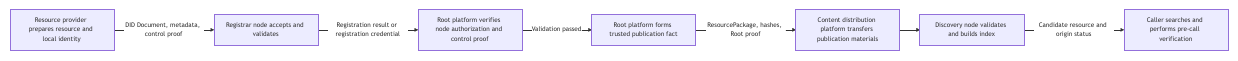}}

\begin{quote}
\textbf{Discoverable does not mean callable.} A Discovery node returns candidate resources, but the caller must still make the final decision based on its permissions, endpoint authentication, resource status, and business risk.
\end{quote}

The design principles in this chapter can be reviewed through three acceptance questions: Does the resource have a verifiable identity? Does publication have traceable evidence? Can the discovery result carry enough origin and status information to support pre-call judgment? If any answer depends only on page text or name matching, the trusted interconnection process is incomplete.

OAN provides resource identity, controller information, registration origin, trusted-publication proof, governance status, and discovery explanations as evidence for pre-call judgment. The resource provider and controller remain responsible for consistency between the registered description and actual runtime content, while the business endpoint remains responsible for invocation authentication and runtime service quality. The four states and three loops described above jointly delimit what this chapter calls ``trusted interconnection''; it cannot be reduced to a resource appearing on a webpage or one HTTP request succeeding.

\section{References}\label{sec:01.9}

{\def\LTcaptype{none} 
\begin{longtable}[]{@{}>{\raggedright\arraybackslash}p{0.3200\linewidth}>{\raggedright\arraybackslash}p{0.3200\linewidth}>{\raggedright\arraybackslash}p{0.3200\linewidth}@{}}
\toprule\noalign{}
Source & Type & Link \\
\midrule\noalign{}
\endhead
\bottomrule\noalign{}
\endlastfoot
OAN White Paper & arXiv White Paper: project vision, ecosystem value, and rationale & \url{https://arxiv.org/abs/2606.03161} \\
OAN Yellow Paper & arXiv Yellow Paper: technical principles, trust boundaries, and system mechanisms & \url{https://arxiv.org/abs/2606.03163} \\
GRAIL Semantic Discovery Paper & arXiv paper: semantic discovery and resource matching & \url{https://arxiv.org/abs/2605.02489} \\
Agentic Overlay Network Architecture & IETF Internet-Draft: agentic network architecture & \url{https://datatracker.ietf.org/doc/draft-xu-agentic-overlay-network-architecture/} \\
\end{longtable}
}

\chapter{Terms, Roles, and Trust Boundaries}\label{chap:02}

This chapter defines the subjects, resources, credentials, nodes, and trust boundaries involved in resource interconnection within OpenAgenet (OAN). The role definitions support the protocols that follow; they are not abstract descriptions of organizational structures or deployment processes. A subject controls identities and resources; a Registrar node accepts and submits registrations; the Root platform validates resource versions under established rules and maintains authoritative publication state; the content distribution platform persists and serves trusted ResourcePackages; a Discovery node synchronizes, filters, indexes, and returns results; and a caller verifies candidate resources and decides whether to connect. Each role\textquotesingle s permissions should match the facts it can produce, verify, or retain.

Three relationships require particular distinction in OAN: whether the resource controller is authorized to represent the resource DID; whether an infrastructure node is recognized by the Root platform and current governance status; and whether a resource version has passed Root publication and been indexed by a Discovery node. DIDs, signatures, VCs, Root proofs, and on-chain governance events carry different facts. No role can replace a verification performed by another role merely because it stores a JSON document, receives a successful HTTP response, or has a search ranking.

When reading this chapter, place each term into four questions: who or what does it describe, what does it prove, who creates or stores it, and can it replace another piece of evidence? For example, a resource DID describes the controlled resource object, a control proof expresses the resource controller\textquotesingle s operational authorization, node authorization expresses an infrastructure node\textquotesingle s qualification to operate, and a Root proof expresses that a particular resource version has passed Root verification. All four may appear in one registration or discovery flow, but they must not be collapsed into a generic ``trusted'' field.

{\def\LTcaptype{none} 
\begin{longtable}[]{@{}>{\raggedright\arraybackslash}p{0.3200\linewidth}>{\raggedright\arraybackslash}p{0.3200\linewidth}>{\raggedright\arraybackslash}p{0.3200\linewidth}@{}}
\toprule\noalign{}
Judgment dimension & Typical question & Main evidence \\
\midrule\noalign{}
\endhead
\bottomrule\noalign{}
\endlastfoot
Object identity & Which subject or resource does the current material describe? & Subject DID, resource DID, DID Document \\
Control relationship & Who can represent this object in a submission or update? & Control public key, control proof, signature \\
Infrastructure qualification & Which node is qualified to perform a network action? & Node authorization, governance status, authorization domain \\
Publication and visibility & Which version has been accepted, distributed, and indexed? & Root proof, synchronization record, discovery response \\
Business invocation & Should the caller connect and invoke the resource? & Local verification result, endpoint authentication, business policy \\
\end{longtable}
}

\section{Core Terms and Abbreviations}\label{sec:02.1}

The names in this section do more than standardize wording: they also delimit the responsibility level of each object. For example, a registration VC is a registration statement made by a Registrar node, whereas a Root trusted-publication proof attests to content that the Root platform has verified and published. Their issuers, covered facts, and propagation locations differ. When a code type with the same name appears later, first determine its conceptual role according to this section, and then consult the implementation mapping in the appendices.

\subsection{Unified Terminology Table}\label{subsec:02.1.1}

{\def\LTcaptype{none} 
\begin{longtable}[]{@{}>{\raggedright\arraybackslash}p{0.4800\linewidth}>{\raggedright\arraybackslash}p{0.4800\linewidth}@{}}
\toprule\noalign{}
Full English term & Abbreviation \\
\midrule\noalign{}
\endhead
\bottomrule\noalign{}
\endlastfoot
OpenAgenet & OAN \\
Internet of Agents & IoA \\
Decentralized Identifier & DID \\
DID Document & \\
\passthrough{\lstinline!did:oan!} Method & \\
Verifiable Credential & VC \\
Verifiable Presentation & VP \\
ResourcePackage & \\
Resource Metadata & \\
\passthrough{\lstinline!oanMetadata!} & \\
Resource Controller & \\
Resource Provider & \\
Agent Service & \\
Skill & \\
MCP Server & \\
Tool/API & \\
Registrar Service Node & \\
Discovery Service Node & \\
Root Platform & \\
Content Delivery Network & CDN \\
Trust Indexer & \\
User Agent & \\
Model Context Protocol & MCP \\
Agent2Agent Protocol & A2A \\
A2A Agent Card & \\
Software Development Kit & SDK \\
Application Programming Interface & API \\
Hypertext Transfer Protocol & HTTP \\
JavaScript Object Notation & JSON \\
Request for Comments & RFC \\
Root-verified ResourcePackage & \\
Root Proof & \\
Registration Verifiable Credential & \\
Resource DID & \\
Local Identity & \\
Node Authorization & \\
Governance Status & \\
Resource Version & \\
Service Endpoint & \\
Capability Tag & \\
Capability Tag Tree & \\
Semantic Discovery & \\
Structured Discovery & \\
Structured Index & \\
Semantic Index & \\
Authorization Domain & \\
Control Plane & \\
Data Plane & \\
Semantic Plane & \\
Governance Authority & \\
Verifiable Control Proof & \\
Public HTTP API & \\
Message Authentication Code & \\
Resource Type & \\
Subject Type & \\
\passthrough{\lstinline!implementationLinks!} & \\
Protocol Binding & \\
Publication Cursor & \\
Publication Sequence & \\
Governance Event Sequence & \\
Synchronization Cursor & \\
Freshness & \\
Governance Event & \\
Publication Fact & \\
Discovery Index & \\
Data Integrity Proof & \\
Public Key Multibase & \\
Crypto Suite & \\
Nonce & \\
Hash Algorithm & \\
Package Claims & \\
Protocol Version & \\
Version Constraint & \\
Previous Version & \\
Challenge & \\
Proof & \\
Holder & \\
Controller & \\
Credential Subject & \\
Request ID & \\
Replay Protection & \\
Idempotency & \\
Rate Limiting & \\
Signature & \\
Discovery Query & \\
Query Projection & \\
Exact Lookup & \\
Lexical Search & \\
Hybrid Search & \\
Search Score & \\
Query Explanation & \\
Pagination Cursor & \\
Publication Manifest & \\
Notification Batch & \\
Publication Job & \\
Watermark & \\
Backoff & \\
Signed Request Envelope & \\
DID Control Challenge & \\
Subject Control Proof Bundle & \\
Registration Submission & \\
Discovery Response & \\
Index Snapshot & \\
Query Fallback & \\
Protocol Envelope & \\
Capability Vocabulary & \\
Governance Projection & \\
Sync Batch & \\
Correlation ID & \\
Checkpoint & \\
Event Cursor & \\
Capability Set & \\
Content Digest & \\
Audience & \\
Cross-Origin Resource Sharing & CORS \\
Entity Tag & ETag \\
Server-Side Request Forgery & SSRF \\
Web Cryptography API & \\
OpenAPI Specification & \\
Identity Runtime & \\
Identity Gateway & \\
Identity Sidecar & \\
Current Controller State & \\
Controller Continuity Proof & \\
Authorization Digest & \\
DID Verification Executor & \\
Node Package Manager & npm \\
Base64url & \\
Base58 & \\
JSON Web Key & JWK \\
Multibase & \\
Ed25519 & \\
SM2 & \\
SHA-256 & \\
SM3 & \\
\end{longtable}
}

DIDs in this manual are decentralized identifiers, and \passthrough{\lstinline!did:oan!} is the method identifier used by OAN. A DID Document is a resolvable document corresponding to a DID. It contains verification methods, service endpoints, and resource-related metadata. A DID control relationship is demonstrated by verifiable keys and signatures. A VC expresses registration, node authorization, or another claim for which the issuer assumes responsibility. A VP organizes credentials for verification. A ResourcePackage is a distributable object that binds a resource version, metadata, digests, and Root publication evidence.

The Registrar node is responsible for resource admission, request validation, control-proof verification, and submission to the Root platform. The Discovery node is responsible for synchronizing Root-published content, building indexes, and returning signed discovery responses. The Root platform performs three hub roles: trust authorization, data distribution, and semantic governance. The content distribution platform carries verified materials and is not a trust authority. MCP Servers, A2A, and Tool/API refer respectively to resource types or services and interfaces in external protocol contexts; similar terms do not imply the same lifecycle or permissions.

Names in the unified terminology table should be used directly in the body, interface descriptions, and examples. For example, registration materials may contain a resource type, resource DID, control proof, and service endpoint at the same time. ``Resource type'' is not ``resource DID,'' and a ``service endpoint'' is not a resource control identity. \passthrough{\lstinline!MCP!} is the name of an external protocol or a protocol identifier in a resource-type context and must not be generalized as a synonym for every tool resource. If an external protocol uses its own object name, explain the correspondence at the first mapping and use the names in Section 2.1.1 thereafter.

{\def\LTcaptype{none} 
\begin{longtable}[]{@{}>{\raggedright\arraybackslash}p{0.4800\linewidth}>{\raggedright\arraybackslash}p{0.4800\linewidth}@{}}
\toprule\noalign{}
Term & Meaning in this manual \\
\midrule\noalign{}
\endhead
\bottomrule\noalign{}
\endlastfoot
Current controller & The controller identified by the Root platform\textquotesingle s current authoritative resource version and therefore able to authorize the next version \\
Controller declared by the new document & The controller that a pending DID Document proposes to use after the new version takes effect; it cannot by itself prove that the transfer is legitimate \\
Publication task & A durable background work item stored after Root admission and awaiting processing by the content distribution platform \\
Publication event & An event that wakes or notifies a background consumer; it does not replace the durable task or authoritative resource state \\
Discovery convergence & A specified Discovery node has fetched, validated, and written the target resource version; it is not equivalent to Root admission \\
Identity Runtime, Identity Gateway, and Identity Sidecar & Three identity-integration forms for in-application, shared-ingress, and sidecar deployment; they are not new network-governance roles \\
Authoritative state & Current resource-version and controller state maintained by Root under protocol rules; ``authoritative'' describes the level of fact and does not mean Root may change control arbitrarily \\
Event notification & A message used to wake background processing; durable tasks and scans recover from allowed duplication or loss, and the notification is not itself a resource fact \\
Derived read model & Rebuildable query state derived from trusted packages or governance events, such as Discovery indexes and Trust Indexer projections \\
Cache snapshot & A time-bounded copy retained for reads or degraded display; it must not overwrite authoritative state or governance facts \\
\end{longtable}
}

\section{Resource Controller and Resource Provider}\label{sec:02.2}

The resource controller holds the control key corresponding to the current controller and is responsible for proving the authority to create, register, or update the resource\textquotesingle s DID Document and resource version. The resource provider supplies the actual capability, resource content, service endpoint, protocol binding, and usage instructions, and assumes business and operational responsibility for those claims. In an individual or small deployment, the two are usually the same subject. In an organizational setting, the controller may be an organization or platform, while the provider may be a team or system operating the service under delegation. Any OAN resource may be the controller of another resource; a dedicated controller is only a common organizational form.

The Registrar node and Root platform verify the controller\textquotesingle s control evidence and the consistency of the submitted materials. They must not infer DID control merely because the provider can access an endpoint, nor infer that a business endpoint is genuinely available merely because the controller holds a private key. Resource metadata should make the relationship between controller and provider explicit. When that relationship changes, it should be expressed through an authorization update, ownership transfer, or new resource-version rule rather than by changing only a display name.

When the controller and provider are separate, the current controller still creates the signature proof, while the provider is responsible for endpoint and capability declarations. Concrete fields cannot be inferred from an organizational relationship diagram; they are defined by the common types in Chapters 5 and 6 and Appendix~\ref{chap:appendix-a}. Operating an endpoint does not automatically give the provider resource control, and a valid control proof does not continuously guarantee the endpoint\textquotesingle s business quality.

\section{Subject Identity and Local Identity}\label{sec:02.3}

Subject identity is the identity context on which a resource control relationship depends. Local Identity is the client-side collection of subject configuration, control keys, resource keys, and necessary credentials. When a Local Identity is created, keys are generated or imported in a client environment controlled by the user. A registration request sends only the DID Document, public keys, signatures, and public credentials required for protocol verification; it does not send resource-control or subject private keys. One Local Identity may manage multiple resources, but every resource retains an independent DID, version, and control-proof boundary. Local Identity is a client-side key and configuration container, not an account on a Registrar node.

\subsection{Subject Identity and Resource Control Identity}\label{subsec:02.3.1}

A subject DID expresses the subject of a control relationship, while a resource DID expresses a resource object that can be registered, published, discovered, and updated independently. Subject identity and resource identity may be associated through the same key system, but they must not be conflated in the protocol. A subject private key proves that the subject can manage related resources; a resource-control signature proves that a particular operation on the resource DID is authorized. A runtime account of the provider, a web-login account, or an account for a Registrar node cannot automatically replace either type of DID control evidence.

\subsection{Local Key Generation, Import, and Backup}\label{subsec:02.3.2}

The Local Identity flow should support creating a subject identity in the client, importing an existing identity backup, storing resource-control materials, and restoring them when needed. A backup file may contain the subject DID, public key, control key, resource configuration, and necessary credentials, but its sensitivity and custody responsibility must be explicit. During import, the structure and correspondence among keys must be checked; a filename or display text must not be treated as proof of identity authenticity. Downloads, imports, and restoration in a browser-based registration flow are client-side operations. The server receives only the public materials required for protocol verification.

The actual backup format, encryption status, and restoration checks are defined by the current SDK and official website implementations, as described in Chapter~\ref{chap:16} and Appendix~\ref{chap:appendix-e}. This chapter specifies only the boundary: private material in a backup must not be uploaded to a Registrar node, the Root platform, or a Discovery node, and a restoration must verify key-pair correspondence and current controller state.

\subsection{Private-Key Custody and Node Visibility}\label{subsec:02.3.3}

Resource-control private keys, subject private keys, and node-runtime private keys are held by their respective controllers. A Registrar node, Root platform, Discovery node, content distribution platform, or off-chain trust indexer must not obtain these private keys from ordinary registration or discovery requests. Nodes may see public keys, signatures, and public credentials in DID Documents, and may retain submission records, Root proofs, and index data, but public evidence cannot be used to derive private keys. Identity Runtime, Identity Gateway, and Identity Sidecar may use controlled keys or remote signing capability only as permitted by their deployment model; this does not give them resource-controller authority.

\section{Root Platform}\label{sec:02.4}

In the current single-Root reference implementation, the Root platform is the network-level validation and publication-state component. It is neither the grantor of resource control nor the execution center for all business capabilities. It receives submissions from infrastructure nodes permitted to operate and, under protocol rules and current authoritative state, verifies identity, control proofs, resource types, versions, digests, and publication conditions. It maintains trusted publication facts for node synchronization and coordinates content distribution. Root stores and issues the proofs and state needed for network verification; it does not replace a resource provider in holding control keys for business endpoints.

\subsection{Trust-Authorization Hub}\label{subsec:02.4.1}

As the trust-authorization hub, the Root platform determines, according to governance status, which Registrar nodes, Discovery nodes, and other infrastructure participants have the required network qualifications, and verifies the available node-authorization materials. The subject of this judgment is node qualification, role, and authorization scope, not a resource controller or the business quality of a resource. Nodes must also check whether authorization remains valid at runtime; they must not continue restricted operations solely because an old VC exists or a previous request succeeded.

The question answered by this hub is ``can this node perform a particular network action,'' not ``is every item of business content returned by the node trustworthy.'' Node authorization materials normally need to be associated with node identity, role, authorization domain, validity period, and governance status. When processing a restricted request, the executing node should check both the authorization material and current status rather than checking only whether the VC signature is valid.

\subsection{Data-Distribution Hub}\label{subsec:02.4.2}

The Root platform creates durable publication facts and pending distribution tasks, but it does not treat each Discovery node\textquotesingle s local index as its own authoritative state. The current asynchronous path comprises the Root publication background loop, persistence confirmation by the content distribution platform, and an independent synchronization loop for each discovery target. A lagging Discovery node does not roll back a version already admitted by Root and cannot block other targets from advancing independently.

As the data-distribution hub, the Root platform organizes verified resource versions, DID Documents, ResourcePackages, digests, Root trusted-publication proofs, and target synchronization state, and provides verifiable materials through durable publication tasks and the content distribution platform. Event notifications wake processing and database tasks support recovery. The content distribution platform may cache and transmit these materials in batches, but it must not modify facts accepted by Root. Successful distribution means only that the material was persisted; it does not mean that any Discovery node has completed validation.

\subsection{Semantic-Governance Hub}\label{subsec:02.4.3}

As the semantic-governance hub, the Root platform maintains capability tags and their hierarchy, identifiers, aliases, and versions for tag selection and recommendation during registration, and for semantic indexes and query explanations at Discovery nodes. Semantic governance determines how tags are organized and evolved, but it does not replace resource control proofs or node authorization. A Discovery node may use local indexing and ranking methods on the basis of Root vocabulary, but it must not extend its authorization domain through local tags.

The capability tag tree answers ``how can capabilities be described and located with shared vocabulary.'' It does not answer ``who controls the resource'' or ``which node can publish the resource.'' A Discovery node may map a natural-language task to one or more capability tags and return tag matches and relevance explanations, but the final result must still be associated with the resource DID, version, origin, governance status, and Root proof.

\subsection{Root Platform Responsibility Boundary}\label{subsec:02.4.4}

The Root platform is responsible for network-level verification, authorization, publication, and coordination. It does not execute specific business invocations for resource providers, guarantee the business quality of resource endpoints, or automatically grant a caller business permission because a resource has been published. Public websites, metrics, maps, and governance-event displays are observation entry points, not replacements for Root facts. When the Root platform is unavailable, nodes may provide limited existing data under explicit freshness and fallback rules, but they must not fabricate new acceptance, authorization, or publication facts.

\section{Registrar Service Node}\label{sec:02.5}

The Registrar node is the admission and submission boundary through which resource providers enter OAN. It may provide forms, SDKs, or Skills, check DID syntax, resource types, required fields, capability tags, endpoints, hashes, and control proofs, and package qualified requests into signed submissions to the Root platform. A registration VC or registration result issued by the Registrar node proves only that it processed the relevant registration fact. It cannot replace a Root proof or allow the Registrar node to expand the resource\textquotesingle s network visibility by itself.

The Registrar node must perform restricted actions using its own valid node authorization and permitted authorization domain, and must retain enough request IDs, verification results, error reasons, and submission status for auditing. It may reject malformed requests or requests with unclear control. When the Root platform rejects a request or asynchronous publication fails, the Registrar node must return the actual status and must not display local admission as network-wide publication completion.

From the evidence-chain perspective, the Registrar node produces at least three distinguishable results: whether the request arrived and passed local validation, whether it was submitted to the Root platform, and whether the Root platform accepted and published it. When troubleshooting registration, a client or operator should query these levels using the request ID, resource DID, version, and submission status rather than relying only on a page saying ``submission successful.''

\section{Discovery Service Node}\label{sec:02.6}

The Discovery node is the authorized read side of Root-published content. It obtains candidate ResourcePackages from the content distribution platform or Root notifications, verifies Root proofs, hashes, versions, and governance status, filters resources according to its authorization domain, and builds structured and semantic indexes. Its responses to users, applications, and agents are locally processed discovery responses. It must therefore retain resource origin, version, freshness, and verification status, and must assume responsibility for response content when signed responses are supported.

A Discovery node may optimize query performance, ranking, and semantic matching, and may retain an older snapshot for limited fallback display. It must not modify Root publication facts, add unauthorized resources to a trusted index, or use local ranking to override revocation, suspension, or version status. Search results are inputs to the caller\textquotesingle s subsequent verification, not automatic invocation permission.

Scores, tag matches, and explanations in a discovery response are derived index- or query-layer results. The resource DID, version, origin, and freshness support subsequent verification. Even a high relevance score cannot bypass DID Document, Root trusted-publication proof, governance-status, or endpoint checks. Executable query examples are centralized in Chapters 11 and 16 so that this chapter does not maintain a response structure that may drift.

\section{Content Distribution Platform}\label{sec:02.7}

The content distribution platform carries ResourcePackages, DID Documents, metadata, proofs, digests, and synchronization materials that have been verified by the Root platform and permitted for propagation. Its goal is to improve the reachability, batch efficiency, and caching capability of node synchronization. It is a data-plane component, not an authority for governance or identity trust. Recipients must reverify distributed content using the Root proof, digests, and status. The distribution layer may be operated by the same organization as the Root platform or by independent infrastructure, but that deployment relationship must not change protocol responsibility boundaries.

\section{Off-Chain Trust Indexer}\label{sec:02.8}

The off-chain trust indexer is the read-side projection component for on-chain governance facts. It reads governance events such as node authorization, authorization-domain changes, suspension, restoration, and revocation, and writes authorization status into a queryable model for the Root platform, Registrar nodes, and Discovery nodes according to sequences or cursors. Projection results may be delayed or affected by faults. They must retain the last processed position, event source, projection time, and error status. Runtime code must not equate ``seen by the indexer'' with the final on-chain fact.

When the Root platform or another executing node uses indexer status, it must define expiration, staleness, and fallback behavior. For actions that must prevent an unauthorized node from continuing, the node should fail closed when status cannot be confirmed or exceeds the allowed freshness. Public statistics or historical displays may show delay and old data, but must not change the actual authorization decision.

A projection is an off-chain read model of on-chain governance events and must retain event identifiers, checkpoints or cursors, projection time, and source. Concrete fields follow the current \passthrough{\lstinline!oan-bulletin!} and Trust Indexer types; a complete event example appears in Chapter~\ref{chap:13} and Appendix~\ref{chap:appendix-a}. If the sequence is interrupted, the status is stale, or the source cannot be verified, the executing node must follow the failure policy for sensitive actions rather than treating the last successful projection as permanent authorization.

\section{Caller, User, Application, and Agent}\label{sec:02.9}

The user is the person expressing an intent to register, discover, or use a business capability. The browser or another client is the tool that carries the Local Identity and requests. The application is software that calls OAN interfaces and executes local policies. The user agent is an agent that can query and initiate calls on behalf of the user. These roles may appear together in one flow, but their sources of authority differ: the user controls local keys, the client protects private materials, the application verifies network evidence, and the user agent selects resources based on the verification result.

During discovery, the caller is a relying party. During an actual business invocation, it must also handle request signing, credential presentation, replay protection, and local authorization policies. It must not trust only search ranking, a website display name, or a resource description forwarded by a third party. Before connecting, it should verify the DID, VC, Root proof, governance status, version, and endpoint declaration, and decide whether to continue according to the concrete business risk.

\section{Governance Authority and Authorization Domain}\label{sec:02.10}

The governance authority decides whether an infrastructure node enters, remains within, or exits OAN\textquotesingle s trusted operating scope. An authorization domain is the capability or resource scope that a node is permitted to process. It may constrain which resources a Registrar node accepts, which resources a Discovery node synchronizes and returns, and which network actions a node may perform. An authorization domain is not a resource control relationship or a capability tag: tags describe and match capabilities, while authorization domains limit permissions and visibility.

Governance status may be formed jointly from on-chain events, Root runtime status, and node authorization materials, but an executing node must state the source and priority of each fact. A historical authorization VC, cached status, or manually maintained display must not override a current revocation or suspension. When authorization-domain changes propagate with delay, the node should follow the prescribed freshness and fail-closed rules for sensitive actions.

An authorization domain can be understood as ``resource scope + action scope + observation scope.'' Resource scope limits which resources or resource types a node can process. Action scope limits whether the node can register, synchronize, index, or perform management actions. Observation scope limits which callers or network areas may receive results. These three scopes jointly determine effective node permissions; a broad \passthrough{\lstinline!active!} status cannot replace them.

A Registrar node has two levels of responsibility. At the protocol-processing level, it verifies the DID Document, control proof, digests, versions, and authorization scope. Under an operator\textquotesingle s real-name admission policy, the operator may additionally verify the enterprise or individual behind a controller and assume responsibility for the registration identity it issues. The latter is a network operation and admission-policy responsibility; it is not a fact automatically proven by a DID signature.

Any OAN resource may act as the controller of another resource. A dedicated controller commonly carries the control identity of an enterprise or individual, but the protocol does not require a controller to belong to a dedicated resource type.

\section{Trust Boundaries and Responsibility Matrix}\label{sec:02.11}

The OAN responsibility matrix should be established along four dimensions: producing facts, verifying facts, storing status, and making external statements. The resource controller produces resource identity and control proof. The Registrar node verifies admission materials and submits registration. The Root platform verifies node authorization and resource versions and forms publication facts. The content distribution platform transfers verified materials. The Discovery node verifies and indexes them within its authorization domain. The off-chain trust indexer projects governance status. The caller verifies candidate resources and executes local invocation policy. Every action should point to a corresponding signature, credential, hash, governance event, interface response, or persistent record.

\subsection{Resource Controller and Resource Provider}\label{subsec:02.11.1}

The resource controller is responsible for the resource DID, control key, and update authorization. The resource provider is responsible for capability descriptions, resource content, endpoints, and business availability. They may be the same subject or may be separated through an explicit delegation or operating relationship. The Registrar node and Root platform should determine resource-operation permission only through a verifiable control proof. They must not infer that the provider controls the resource DID merely because it submitted a service address or successfully logged in.

\subsection{Registrar Node, Root Platform, and Discovery Node}\label{subsec:02.11.2}

The same registration materials have different meanings at the three roles: a Registrar node performs admission, preliminary validation, and credential issuance; the Root platform makes the final admission decision against the current authoritative version and creates publication facts; and a Discovery node consumes only ResourcePackages that passed publication verification and builds a read model. Troubleshooting should therefore distinguish ``request returned,'' ``Root admitted,'' ``content distribution confirmed,'' and ``visible at the target Discovery node,'' rather than inspecting only the final query result.

The HTTP interfaces of these three roles correspond to different trust boundaries. The resource publisher submits a \breakablecode{ResourceRegistrationSubmission} to the Registrar node. The Registrar node submits a \passthrough{\lstinline!ResourceVerifyAndPublishRequest!} containing \passthrough{\lstinline!registrarDid!} and \passthrough{\lstinline!upstreamAuth!} to the Root platform. The Discovery node returns a \passthrough{\lstinline!ResourceDiscoveryResponse!} to the caller. The first interface receives resource-control materials, the second proves that an authorized node is representing the network, and the third provides local-index candidates. Interface reachability, request parseability, and candidate visibility are different facts and cannot replace one another across roles.

The Registrar node accepts, validates, and submits registrations. The Root platform verifies node authorization and resources, performs trusted publication, and coordinates distribution. The Discovery node synchronizes, verifies, filters by authorization domain, and serves queries over published materials. A Registrar result cannot replace a Root publication fact, and a Discovery result cannot rewrite Root status. Interfaces, proofs, and states among the three must remain traceable.

\subsection{Caller and Verifier}\label{subsec:02.11.3}

The caller may be a user, application, or user agent. It submits queries, receives candidate resources, and decides whether to establish a business connection. The verifier may be the caller itself, an SDK, or a verification component trusted by the caller. It checks DIDs, signatures, VCs, Root proofs, governance status, versions, endpoints, and invocation policies. A verification component may provide a verification result, but it must not directly execute a high-risk business action without the caller\textquotesingle s policy authorization.

\subsection{Responsibility for Key Actions}\label{subsec:02.11.4}

Resource creation and updates belong to the resource controller. Registration acceptance and registration credentials belong to the Registrar node. Root acceptance and trusted publication belong to the Root platform. Material transfer belongs to the content distribution platform. Authorization-domain indexing and discovery responses belong to the Discovery node. Governance-event projection belongs to the off-chain trust indexer. Final business invocation authorization belongs to the resource endpoint and the caller\textquotesingle s local policy. A public website, statistics interface, or manually maintained event table is an observation and presentation layer and cannot serve as protocol authority for these key actions.

The boundaries among roles can be checked by asking who produces facts, who verifies facts, who stores status, and who makes external statements. ``Responsible'' in the table below means protocol responsibility; it does not mean responsibility for all business content or runtime environments.

{\def\LTcaptype{none} 
\begin{longtable}[]{@{}>{\raggedright\arraybackslash}p{0.1600\linewidth}>{\raggedright\arraybackslash}p{0.1600\linewidth}>{\raggedright\arraybackslash}p{0.1600\linewidth}>{\raggedright\arraybackslash}p{0.1600\linewidth}>{\raggedright\arraybackslash}p{0.1600\linewidth}>{\raggedright\arraybackslash}p{0.1600\linewidth}@{}}
\toprule\noalign{}
Role & Generates/signs & Verifies & Persists & Derives/displays & Real-world identity responsibility \\
\midrule\noalign{}
\endhead
\bottomrule\noalign{}
\endlastfoot
Resource controller & DID Document, control proof, update authorization & Materials to be signed and local key & Local identity materials & Provides public identity materials & Responsible for its control operations \\
Resource provider & Capability description, endpoint, and usage conditions & Consistency between runtime content and description & Business resource state & Provides the capability externally & Responsible for service content and availability \\
Registrar node & Upstream signature and registration VC & Control proof, fields, node/authorization-domain prerequisites & Registration records and credentials & Registration queries & Verifies the enterprise or individual behind a controller under operating policy \\
Root platform & RootProof, trusted ResourcePackage, and publication task & Node qualification, current controller, version, digest, and rules & Current authoritative state, admission, and publication tasks & Publication state & Does not replace the Registrar\textquotesingle s real-world identity due diligence \\
Content distribution platform & Persistence confirmation & Upstream source and package boundary & Trusted ResourcePackages & Package reads and directory statistics & Does not guarantee resource identity \\
Discovery node & Query response and optional response proof & ResourcePackage, authorization domain, version, and governance state & Structured index and optional semantic index & Candidates, ranking, and explanations & Does not guarantee endpoint business content \\
Off-chain trust indexer & Governance projection & On-chain events, ordering, and finality conditions & Checkpoints, cursors, and projections & Governance queries & Does not create on-chain authorization facts \\
Caller/verification component & Request proof and local decision & DID, VC, RootProof, endpoint, and business policy & Local session/audit state & Decision whether to connect & Responsible for local authorization decisions \\
\end{longtable}
}

{\def\LTcaptype{none} 
\begin{longtable}[]{@{}>{\raggedright\arraybackslash}p{0.2400\linewidth}>{\raggedright\arraybackslash}p{0.2400\linewidth}>{\raggedright\arraybackslash}p{0.2400\linewidth}>{\raggedright\arraybackslash}p{0.2400\linewidth}@{}}
\toprule\noalign{}
Data or state & Primary owner & Fact level & Rebuildable? \\
\midrule\noalign{}
\endhead
\bottomrule\noalign{}
\endlastfoot
Browser/SDK Local Identity and private keys & User or resource controller & Local control material & Cannot be recovered by network nodes without a backup \\
Registrar registration record and registration VC & Registrar node & Registration fact & Depends on Registrar persistent state and issuance material \\
Current resource version and controller state & Root platform & Authoritative resource state & Must not be reconstructed from a Discovery index and written back over it \\
CDN trusted ResourcePackage & Content distribution platform & Copy of Root publication material & Can be redistributed from Root publication state \\
Discovery structured/semantic indexes & Discovery node & Derived read model & Rebuildable from trusted ResourcePackages \\
On-chain governance event & Bulletin contract/governance chain & Governance fact & Off-chain components retain only projections \\
Trust Indexer projection & Off-chain trust indexer & Derived governance read model & Replayable from an on-chain checkpoint \\
Official Network page & Official backend/cache & Display snapshot & Not a protocol authority \\
\end{longtable}
}

\pandocbounded{\includegraphics[width=\linewidth,height=0.72\textheight,keepaspectratio]{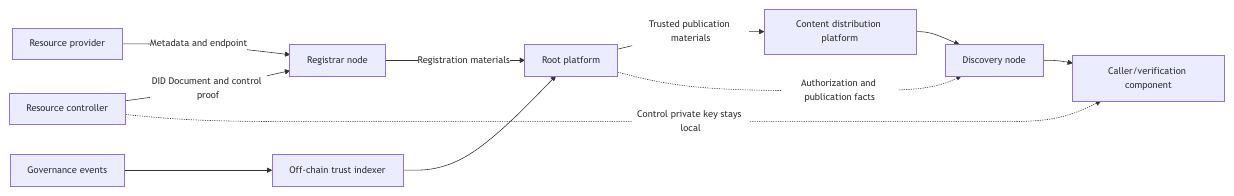}}

When a key action occurs, the corresponding signature, credential, hash, governance event, synchronization cursor, or interface response should be locatable. If a result has only ``a node returned success'' as evidence and its source and verification scope cannot be explained, it can be treated only as an observation of runtime status, not as a cross-role trust proof.

When troubleshooting a resource or node, establish the evidence chain in this order: identify the object and version, confirm the control proof or node authorization, check the publication fact and governance status, and finally inspect the discovery index, endpoint, and caller policy. This order applies to registration failures, stale discovery results, and propagation after node-authorization revocation. If any step lacks a source or cannot be verified, limit the conclusion to the observable scope.

Additional implementation and observation terms are listed below.

{\def\LTcaptype{none} 
\begin{longtable}[]{@{}>{\raggedright\arraybackslash}p{0.4800\linewidth}>{\raggedright\arraybackslash}p{0.4800\linewidth}@{}}
\toprule\noalign{}
Term & Meaning and boundary \\
\midrule\noalign{}
\endhead
\bottomrule\noalign{}
\endlastfoot
DID verification executor & Bounded internal execution resources that isolate blocking DID verification work; they do not change validation rules or external protocols \\
Registration admission complete & Registrar has completed the synchronous request cycle and returned; this does not mean CDN or Discovery convergence \\
CDN persistence complete & The verifiable ResourcePackage for a Root publication task has been confirmed as stored by the content distribution platform \\
Single-target discovery convergence & One Discovery node has validated and applied all applicable publication materials through the target cursor \\
All-target convergence & Every target Discovery node in the observation scope has reached the expected deduplicated count and cursor \\
Real-world responsible party & An enterprise or individual verified under operating policy and responsible for a controller or resource registration; not merely a string field in a DID Document \\
\end{longtable}
}

JetStream messages are event notifications; Discovery indexes and Trust Indexer state are derived read models; and the official Network page is a cache or display snapshot. They may accelerate processing or support observation, but they cannot write back over Root current resource state or on-chain governance facts. Identity Runtime, Identity Gateway, and Identity Sidecar sit on the invocation-integration side and do not become new state authorities.

\section{References}\label{sec:02.12}

{\def\LTcaptype{none} 
\begin{longtable}[]{@{}>{\raggedright\arraybackslash}p{0.3200\linewidth}>{\raggedright\arraybackslash}p{0.3200\linewidth}>{\raggedright\arraybackslash}p{0.3200\linewidth}@{}}
\toprule\noalign{}
Source & Type & Link \\
\midrule\noalign{}
\endhead
\bottomrule\noalign{}
\endlastfoot
\passthrough{\lstinline!oan-protocol-commons!} & Code repository: shared protocol types, identity objects, and trust materials & \url{https://github.com/wolfbrother/oan-protocol-commons} \\
OAN Yellow Paper & arXiv Yellow Paper: roles, trust boundaries, and verification mechanisms & \url{https://arxiv.org/abs/2606.03163} \\
\passthrough{\lstinline!did:oan!} DID Method Specification & Standard/method specification: DID method and identity objects & \url{https://github.com/OpenAgenet/oan-public-docs/blob/main/did-oan-specs/doc/OAN\%20DID\%20Method\%20Specification.md} \\
\end{longtable}
}

\chapter{Overall Architecture and Network Topology}\label{chap:03}

This chapter describes the federated architecture adopted by OpenAgenet (OAN): coordination by a root platform, access through multiple registrar service nodes, propagation through a content distribution platform, indexing by multiple discovery service nodes, and a governance read model provided by an off-chain trust indexer. The OAN network can simultaneously accept registrar service nodes and discovery service nodes operated by the official organization and third parties. Each node participates in the network with an independent DID, role, authorization domain, and runtime status. A resource controller can submit a resource to an appropriate registrar service node, and a caller can query different discovery service nodes. The root platform uniformly verifies node eligibility, resource versions, and publication facts. The content distribution platform propagates verified materials to authorized discovery service nodes, and each discovery service node uses those materials to maintain its own structured and semantic indexes.

\pandocbounded{\includegraphics[width=\linewidth,height=0.72\textheight,keepaspectratio]{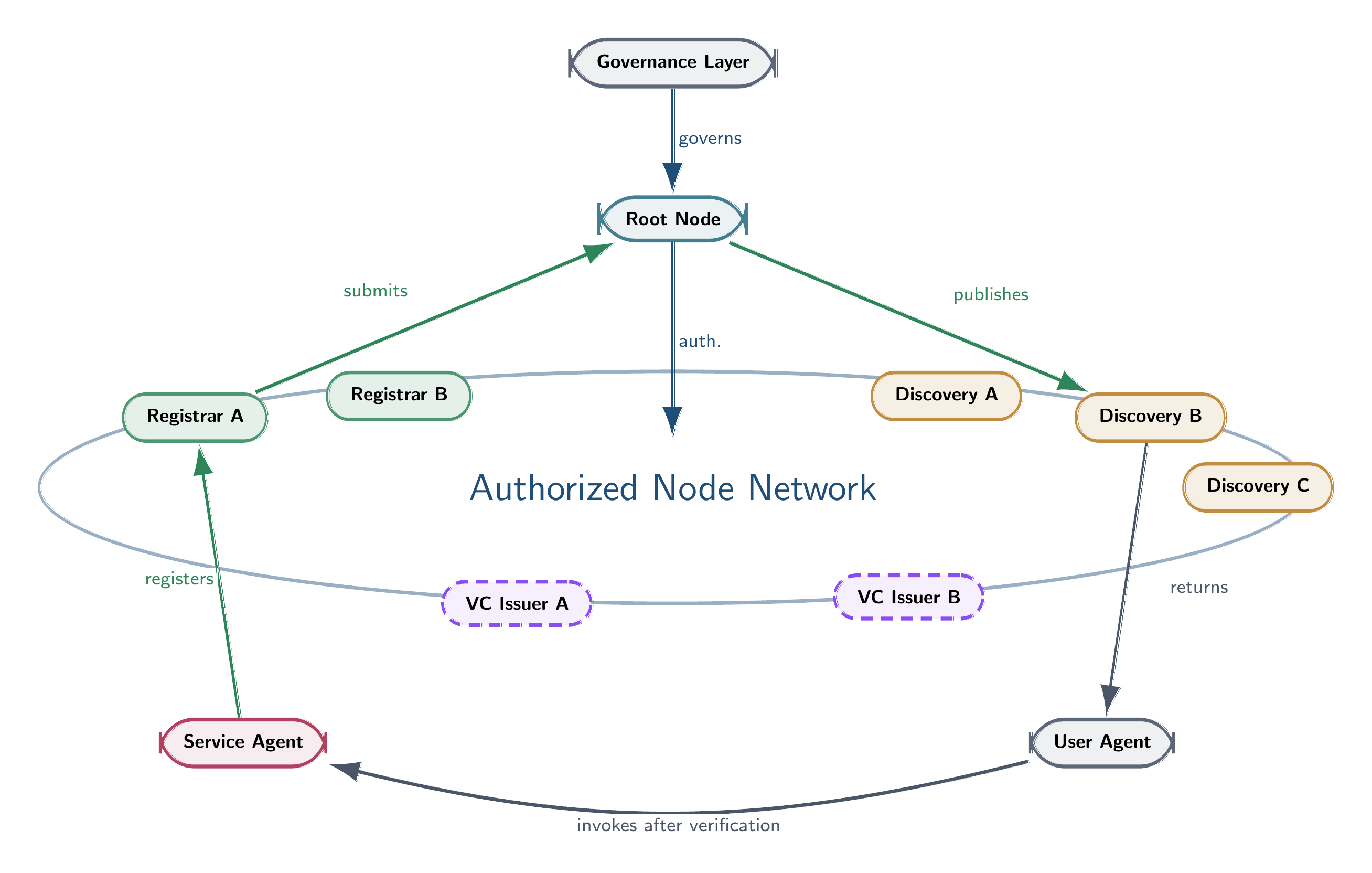}}

Three types of flows are connected within the same resource lifecycle, but they do not replace one another. The data flow carries DID Documents, resource packages, metadata, digests, and proofs; the control flow carries node authorization, request authentication, publication acceptance, lifecycle, and governance status; and the semantic flow carries the capability tag tree, tag aliases, capability descriptions, use cases, and query explanations. An independent deployment may adjust its storage, messaging, and network implementation, but it must not allow the content distribution platform to acquire governance authority, allow a discovery service node to rewrite root-platform publication facts, or allow semantic relevance to replace identity and authorization verification.

\section{Logical Architecture}\label{sec:03.1}

{\def\LTcaptype{none} 
\begin{longtable}[]{@{}>{\raggedright\arraybackslash}p{0.2400\linewidth}>{\raggedright\arraybackslash}p{0.2400\linewidth}>{\raggedright\arraybackslash}p{0.2400\linewidth}>{\raggedright\arraybackslash}p{0.2400\linewidth}@{}}
\toprule\noalign{}
Component & Durable state or material & Primary upstream & Primary downstream \\
\midrule\noalign{}
\endhead
\bottomrule\noalign{}
\endlastfoot
Client/SDK & Local Identity, resource draft, private keys, request state & User or resource controller & Registrar, Discovery, business endpoint \\
Registrar & Registration records, registration VCs, Root submission results & Client & Root \\
Root & Current version/controller, RootProof, publication tasks and target state & Registrar, governance read model & Publisher/CDN, status query \\
CDN Publisher/CDN & Task-claim state, trusted packages and durable confirmations & Root tasks and notifications & Discovery \\
Discovery & Target cursor, validation results, structured and optional semantic indexes & CDN, governance read model & Query clients \\
Trust Indexer & On-chain checkpoint, event cursor, governance projection & Sui bulletin contract & Root and other authorization checks \\
\end{longtable}
}

The client is the primary operational boundary for resource controllers and callers. It stores local identities and private keys, generates or imports resource-control materials, and sends requests to any registrar service node or discovery service node that meets its authorization-domain and service requirements. Each registrar service node independently stores its local registration records, registration credentials, and submission status, and packages validated resource registration requests into requests submitted to the root platform. The root platform stores network-wide authorization status, acceptance records, resource versions, publication proofs, and pending tasks. The content distribution platform provides multiple authorized discovery service nodes with root-platform-verified resource packages, DID Documents, and metadata. Each discovery service node separately stores its synchronization cursor, verified resource packages, rejection records, structured index, and semantic index. The off-chain trust indexer stores a projection of on-chain governance events into node authorization status, providing a unified read model for third-party node onboarding, suspension, and authorization-domain changes.

The logical connections can be summarized as follows: the client-to-registrar-service-node connection is the resource access flow; the registrar-service-node-to-root-platform connection is a protected upstream submission flow; the root-platform-to-content-distribution-platform connection is a publication data flow; the content-distribution-platform-to-discovery-service-node connection is a synchronization flow; the discovery-service-node-to-client connection is a query response flow; and the on-chain governance system-to-off-chain-trust-indexer-to-executing-node connection is an authorization-state flow. The root platform also provides discovery service nodes with publication notifications or summaries, allowing them to obtain the current resource package by cursor instead of relying on registrar service nodes to write directly to discovery indexes.

This topology organizes the network by node roles and authorization relationships rather than limiting its scale to a fixed set of service instances. After completing node identity, authorization-domain, interface, security, and conformance checks, a third-party operator can run a new registrar service node or discovery service node. The new node cooperates with existing nodes through common publication facts and governance status recognized by the root platform, while retaining independent deployment, storage, indexing, and service policies.

\section{Root Platform as the Trust and Authorization Hub}\label{sec:03.2}

The root platform\textquotesingle s trust and authorization hub responsibilities apply to infrastructure participants, not to the business quality of resources controlled by resource controllers. The root platform should maintain verifiable node identities, node roles, authorization domains, and governance status. When processing registration submissions, publication tasks, content distribution, and discovery notifications, it should check whether the relevant nodes remain effectively authorized. The off-chain trust indexer provides the root platform with a read-side projection of on-chain events, but the projection may have synchronization latency. For high-risk writes, the root platform should decide whether to accept, reject, or suspend processing according to explicit freshness, missing-data, and invalidation rules.

Node credentials prove that ``the node has a certain type of network qualification''; resource-control proofs prove that ``the requester can represent a resource DID''; and root-platform publication proofs prove that ``a particular resource version has passed root-platform verification.'' These three types of proof cover different facts, and none can replace another. At a minimum, an authorization check should bind the node DID, role, authorization domain, status, validity period or governance sequence, and record the verification result as an auditable accepted, rejected, or pending status.

The output should at least distinguish \passthrough{\lstinline!allow!}, \passthrough{\lstinline!deny!}, and \passthrough{\lstinline!indeterminate!}. \passthrough{\lstinline!indeterminate!} means that the authorization status, governance projection, or a critical proof cannot currently be confirmed. It must not be silently converted into an allow decision. The specific status codes and error structures are described in the registration, governance, and API chapters.

\section{Root Platform as the Data Distribution Hub}\label{sec:03.3}

After a resource version passes verification, the root platform organizes the DID Document, resource metadata, resource package, DID Document digest, metadata digest, resource package digest, resource type, version, lifecycle status, authorization domain, and root-platform trusted publication proof into a publication artifact that can be carried by the content distribution platform. A publication artifact must have explicit version and digest boundaries. After obtaining the materials, a discovery service node must be able to independently recompute digests, verify the proof statement, and determine the current version without relying on an informal statement or interface status from the content distribution platform.

Consistency among different discovery service nodes does not require them to have identical local databases. It requires them to validate the same root-platform publication fact using the same resource version, digests, proof, and governance status. The root platform establishes a traceable order through the publication cursor, resource DID, and version. The content distribution platform may use caching, batching, or retries to improve transmission efficiency, but it must not present an old version as the current version or modify resource-package content without a new publication fact.

The receiver must be able to bind the resource DID, version, every digest, and the Root trusted-publication proof, and may write the material into its read model only after local verification. Concrete field names follow the shared protocol types rather than a second illustrative object in this chapter.

\section{Root Platform as the Semantic Governance Hub}\label{sec:03.4}

The root platform maintains the capability tag tree and its versions, hierarchical levels, tag identifiers, Chinese or English labels, aliases, and parent-child relationships, and provides this governed vocabulary to both registrar service nodes and discovery service nodes. During registration, a client can select tags for a resource based on its description or receive tag recommendations. During discovery, a discovery service node can convert the user\textquotesingle s natural-language task, tag conditions, and structured filters into local query conditions. The tag tree is used to unify semantic boundaries and support authorization-domain decisions; it does not mean that the root platform evaluates the business quality of a resource on behalf of its provider.

Semantic governance and semantic retrieval have clearly separated responsibilities: the root platform maintains shared vocabulary and governance facts, while discovery service nodes are responsible for local indexing, tokenization, vectors, or other ranking implementations. A discovery service node may retain local semantic aliases and ranking signals, but its returned results should still retain the resource source, version, authorization domain, and verification status. A result with high relevance but without a valid control proof, root-platform publication proof, or governance status must not enter the trusted discovery results.

A discovery node can generate candidate conditions based on aliases and natural-language queries, but returned results must still refer to the resource DID, publication version, and verification status. Changes to the vocabulary version affect query interpretation and ranking; they must not silently change the identity or digest of an already signed resource package.

\section{Publication Path from a Registrar Node to the Root Platform}\label{sec:03.5}

A normal publication consists of local preparation, access validation by a registrar service node, a signed submission from the registrar service node, root-platform authentication and verification, resource-version acceptance, content distribution, and discovery notification. A registrar service node may return a registration result or registration credential, but a resource is eligible to enter the propagation path only after the root platform has completed resource-version verification and generated a publication fact. Publication tasks are usually asynchronous. The registration response, root-platform acceptance, completion of content distribution, and completion of discovery indexing should be recorded and displayed separately.

Failure handling must preserve the stage at which the failure occurred. Client field errors, DID Document structure errors, and control-proof errors are access or verification failures. An unauthorized registrar service node, an expired upstream signature, or a request-body digest mismatch is a node-authentication failure. A resource type, version, digest, or governance condition that does not satisfy root-platform requirements is a publication rejection. Temporary unavailability of content distribution or discovery indexing is a propagation-stage failure. Only temporary failures that can be safely retried should be retried automatically, and retry requests must still satisfy freshness and idempotency constraints.

\begin{lstlisting}
Client -> Registrar service node: Resource materials and control proof
Registrar service node -> Root platform: Node-signed upstream submission
Root platform -> Registrar service node: accepted / rejected / pending
Root platform -> Content distribution platform: Verified publication artifact
Content distribution platform -> Discovery service node: Publication materials and cursor
\end{lstlisting}

Each response in this path describes only the result of the current stage. For example, \passthrough{\lstinline!accepted!} may mean that the root platform accepted the resource version, but it does not mean that a discovery service node has completed indexing. \passthrough{\lstinline!pending!} means that the task is still progressing asynchronously. The caller and operator should continue to determine subsequent results from status interfaces or synchronization records.

\subsection{Access Validation by a Registrar Node}\label{subsec:03.5.1}

A Registrar first checks whether the request conforms to the resource registration model, including the resource DID, DID Document, resource type, subject type, name and description, capability tags, use cases, protocol bindings, service endpoints, authorization domain, version, and digest consistency. It also verifies the relationship among the resource DID, DID Document subject, control keys, and control proof, and confirms that the DID routing code matches the accepting Registrar route. Recommended tags may assist completion but cannot replace required fields or proof of control.

\subsection{Upstream Submission and Identity Authentication}\label{subsec:03.5.2}

The submission sent by a registrar service node to the root platform is not an unprotected form. It is a protected request containing the resource registration content, request identifier, protocol version, target path, timestamp, nonce, request-body digest, sender DID, and signature. The root platform uses this information to verify the registrar service node\textquotesingle s identity, authorization domain, and request freshness, and then verifies the resource registration credential, resource control proof, and resource-package statement contained in the request. The submitter\textquotesingle s private key is used only within the registrar service node\textquotesingle s own node environment; the resource controller continues to retain the resource-control private key.

\subsection{Root-Platform Verification and Publication Confirmation}\label{subsec:03.5.3}

During verification, the root platform simultaneously checks node authorization, the resource DID Document, resource control proof, registration credential, resource-type binding, metadata digest, resource-package digest, version relationship, authorization domain, and lifecycle conditions. After successful verification, the root platform records the resource-version acceptance fact, generates a root-platform trusted publication proof containing the resource DID, version, digests, status, authorization domain, and relevant governance or publication references, and marks subsequent content-distribution and discovery-notification tasks as executable. This confirmation does not mean that a discovery service node has completed indexing.

\subsection{Failure, Rejection, and Retry}\label{subsec:03.5.4}

The root platform should distinguish an idempotent retry of the same resource DID, version, and digest from an update request in which content or version changes under the same DID. The former can return the existing acceptance result or publication status. The latter must re-verify control, version relationship, and publication conditions. An invalid signature, replayed nonce, invalid node authorization, digest mismatch, or inconsistent resource type must not be bypassed through simple retries. Temporary errors such as network timeouts or temporary downstream unavailability can enter a backoff-based task retry, while retaining the association with the original request.

\section{Distribution Path from the Root Platform to Discovery Nodes}\label{sec:03.6}

The root platform provides publication content and synchronization positions to discovery service nodes through the content distribution platform or protected synchronization interfaces. A discovery service node requests incremental materials according to its locally stored cursor and can perform controlled backfill when the cursor is missing or recovery is required. For every candidate resource, the discovery service node must obtain the current resource package, verify the DID Document, metadata, digests, root-platform trusted publication proof, lifecycle status, and authorization domain, and then write the verified version into its local index. Materials that fail verification should enter a traceable rejection record rather than being silently discarded.

The ultimate goal of the synchronization path is for each discovery service node to form a read model that is ``locally queryable, source-traceable, and status-verifiable.'' A discovery index may use relational tables, a full-text index, a semantic index, or a cache, but each index record should retain the resource DID, version, package digest, publication time or cursor, source, and verification status. Ranking and summaries in a query response are derived information and cannot override root-platform publication facts.

\subsection{Distribution Content and Publication Checkpoints}\label{subsec:03.6.1}

Distributable content includes at least the resource DID, resource type, resource version, DID Document, resource metadata, resource-package digest, metadata digest, DID Document digest, authorization domain, lifecycle status, root-platform trusted publication proof, and the necessary publication cursor or source reference. A publication checkpoint should answer whether the resource has been accepted by the root platform, whether the content has been written to the distribution layer, whether a discovery service node has obtained and verified it, and at which stage a failure remains.

A propagation checkpoint must not compress multiple states into one \passthrough{\lstinline!available!} field. When a Discovery node has not yet fetched or verified the material, Root publication may already be complete, but the resource must not be described as discoverable at that node.

\subsection{Discovery-Node Synchronization and Indexing}\label{subsec:03.6.2}

After obtaining materials, a discovery service node first completes source, integrity, and governance verification, then writes the verified resource package to a table or equivalent storage, and subsequently updates the structured and semantic indexes. The structured index can support queries by DID, resource type, version, capability tags, and authorization domain. The semantic index can use capability descriptions, tag aliases, use cases, and protocol bindings for matching. Every index write should be associated with the verification result; the node must not display a result first and perform trust verification later.

\subsection{Sequence Numbers, Freshness, and Convergence}\label{subsec:03.6.3}

A publication cursor expresses the propagation order of root-platform publication materials. A discovery service node\textquotesingle s synchronization cursor expresses the position that the node has processed. The two are not equivalent. A discovery service node should record the last successfully applied cursor, the time at which the material was received, the resource version, and the verification result, and should determine whether an old snapshot can still be used for limited display according to the permitted freshness. Advancement of the cursor means that the synchronization position has advanced; it does not mean that every resource is ready to be invoked.

When the local cursor falls behind, the node can report lag and continue processing verified old snapshots, but it must not disguise \passthrough{\lstinline!localAppliedCursor!} as the latest root-platform publication sequence. After a gap is recovered, the node should re-verify every publication material in the missing range before advancing the local cursor.

\subsection{Distribution Failure and Recovery}\label{subsec:03.6.4}

Content-distribution failure, network interruption, temporary unavailability of a resource package, and discovery-service-node database failure should all support recovery. A recoverable task should retain the resource DID, target version, source cursor, and retry information. When a digest mismatch, proof mismatch, or invalid governance status is encountered, the task should enter a rejected or isolated state and wait for new publication materials or a governance change rather than retrying indefinitely. After recovery, processing should continue from a persisted cursor or controlled backfill point to avoid skipping publication records.

\section{Multiple Registrar Nodes and Discovery Nodes}\label{sec:03.7}

Multiple registrar service nodes can provide access for different organizations, resource types, or authorization domains, while multiple discovery service nodes can provide indexing for different network regions, user groups, or query policies. They share the root platform\textquotesingle s basis for judging identity, authorization, resource versions, and publication proofs, but do not need to share the same database or local ranking. When a resource is submitted repeatedly through multiple registrar service nodes, the root platform should identify an idempotent request or an update request based on the resource DID, version, digest, and control proof. When a discovery service node receives the same publication material repeatedly, its indexing should remain idempotent.

Federation does not mean that arbitrary nodes write to one another. A registrar service node may accept and submit only within its authorization domain. A discovery service node may synchronize and return only the published resources allowed within its authorization domain. Propagation between nodes should primarily pass through root-platform publication facts and the content distribution path. A failure of one discovery service node does not make root-platform facts disappear; another discovery service node that has completed synchronization can continue providing results, while the failed node\textquotesingle s local visibility and freshness should be reported accurately.

\subsection{Access by Multiple Registrar Nodes}\label{subsec:03.7.1}

The root platform maintains an independent node identity, role, and authorization domain for each registrar service node. Different registrar service nodes can submit the same resource, but changing the registrar service node cannot grant new resource control; an update to the same resource still requires a valid control proof from the resource controller and must satisfy the version-update rules. A registrar service node\textquotesingle s local registration record is evidence on the intake side; the root platform\textquotesingle s acceptance record is the basis for determining network publication.

\subsection{Synchronization by Multiple Discovery Nodes}\label{subsec:03.7.2}

Each discovery service node has an independent synchronization cursor and local index state, and can obtain the same root-platform publication facts from the content distribution platform according to its authorization domain. Nodes should not treat local search results as authoritative data for another node. When consistency needs to be compared, the comparison should use the resource DID, version, package digest, root-platform proof, and governance status. Different result counts or rankings over a short period may result from synchronization delay or local policy, but not from arbitrary rewriting of root-platform facts.

\subsection{Data Consistency Between Nodes}\label{subsec:03.7.3}

Consistency between nodes should distinguish fact consistency from read-model consistency. Fact consistency requires the DID Document, metadata, resource-package digest, and root-platform proof of the same publication version to be consistent. Read-model consistency permits differences in indexing time, ranking, caching, and semantic interpretation, but each difference must be explainable by the cursor, freshness, or authorization domain. When a discovery service node detects a version conflict, it should prioritize the root-platform publication sequence and proof, rather than letting local arrival time override the trusted source.

When comparing two discovery service nodes, the following stable fields should be compared before result counts or rankings:

\begin{lstlisting}
(resourceDid, version, packageDigest, rootProof, governanceStatus)
\end{lstlisting}

When these fields are consistent, the two nodes generally have the same core understanding of the publication fact. Differences in indexing time, semantic score, pagination position, and cache time are read-model differences. If the stable fields differ, the root-platform publication records and governance-event sequence should be consulted to locate the cause.

\subsection{Node Failure and Isolation}\label{subsec:03.7.4}

When a node exhibits signature anomalies, expired authorization, persistent synchronization failure, publication-material tampering, or non-conforming interface behavior, it should be isolatable so that restricted writes are no longer accepted from it or new authorization scope is no longer provided to it. Isolation does not mean deleting historical facts. The root platform and other nodes should retain auditable records and decide whether to re-enable the node based on governance status and recovery checks. A failure of an ordinary query node can be handled by another node, but this must not expand the other node\textquotesingle s authorization domain.

\subsection{Identity-Integration Component Topology}\label{subsec:03.7.5}

Identity Runtime, Identity Gateway, and Identity Sidecar are located on the caller, resource-service, or application-ingress side. They are not inserted into the Registrar---Root---CDN---Discovery resource-publication control chain. Runtime is suitable for identity selection, signing, and verification inside application code or a controlled service. Gateway enforces identity boundaries and forwarding at a shared network ingress. Sidecar is deployed alongside one workload to provide request signing, response verification, and verified-subject injection when embedding Runtime is impractical. All three reuse OAN DID and proof rules, while A2A, MCP, or HTTP continues to carry business messages, streaming, and error semantics.

\section{Third-Party Node Onboarding Model}\label{sec:03.8}

Third-party node onboarding starts with node identity materials and an operating role, rather than simply opening an HTTP port. The applicant should submit the node DID Document, public verification methods, service endpoint, proposed node role, authorization domain, and operating contact or operational materials. The governance authority or root platform then performs qualification review and authorization. After obtaining test status, the node must pass tests for identity, signatures, interfaces, resource-package verification, synchronization, error handling, and status updates. A production node then provides services within its formal authorization domain.

Test nodes and production nodes should be isolated in authorization status, visibility, data retention, and network impact. A test node can use test resources and test governance status to validate an implementation, but must not mix test publication facts into a production discovery index. A production node\textquotesingle s authorization revocation, suspension, or scope change must be recognizable at runtime and affect restricted operations.

\subsection{Pre-Onboarding Authorization}\label{subsec:03.8.1}

Pre-onboarding authorization should at least confirm that the node DID can be resolved, verification methods are available, the node role is explicit, endpoint ownership is clear, request signatures can be verified, the authorization domain can be expressed, and the operator can explain key custody, data storage, and failure handling. Authorization materials should bind the node identity and role; a domain name, IP address, or human-readable name alone must not serve as proof of node eligibility.

\subsection{Node Capabilities and Authorization Scope}\label{subsec:03.8.2}

The node capability description should distinguish between ``interfaces the node can provide'' and ``actions the node is authorized to perform.'' A discovery service node may support structured and semantic queries but be authorized to index only certain resource domains. A registrar service node may support access for all four resource types but be authorized to accept only some of them. The authorization scope should map to the node\textquotesingle s actual filtering, submission, and response logic. Requests outside the scope must be rejected or omitted from responses.

\subsection{Protocol Conformance Checks}\label{subsec:03.8.3}

Conformance checks should cover health checks, DID Document parsing, signature and digest verification, node authorization status, resource registration, root-platform publication, content distribution, discovery synchronization, query responses, and error contracts. Test results should record the input, node version, configuration, expected result, actual result, and evidence location. Passing an interface-connectivity test does not mean that the node has passed trust-boundary tests.

The minimum onboarding record retains the actual node DID, role, authorized endpoint, authorization domain, test profile, and result of every check. DID resolution, signature verification, publication synchronization, and query response each record their input, conclusion, and evidence file; a fake DID, unreachable URL, or generic \passthrough{\lstinline!test-passed!} value cannot replace real results. The record describes only the scope and result of the onboarding assessment. Passing a test is not a guarantee of long-term operation; production still checks authorization status, dependencies, and continuous synchronization.

\subsection{Runtime Status Management}\label{subsec:03.8.4}

During production operation, a node should continuously check its own authorization, governance status, dependent endpoints, synchronization cursor, task queue, and database status. After authorization is suspended or revoked, the node should stop the corresponding restricted actions. When the content distribution platform or off-chain trust indexer is temporarily unavailable, the node may use existing state according to freshness rules, but must not fabricate new authorization, acceptance, or publication facts. Recovery should include an observable resynchronization and re-verification process.

\section{Public, Private, and Consortium Network Deployments}\label{sec:03.9}

A public network allows Internet users to access public registration, discovery, or resolution entry points, but public accessibility does not mean that every resource or every node status is visible to every user. A private network can restrict registrar service nodes, discovery service nodes, the content distribution platform, and resource endpoints to an organizational network or identity domain, while using the same resource identity, control proof, and publication-verification boundaries. A consortium network can be operated by multiple organizations that run nodes and exchange resource publication facts under an agreed root platform, governance authority, and authorization domain.

The three deployment forms can share protocol objects, but they should make the network scope, domains and endpoints, certificates, authorization materials, data retention, and cross-domain propagation conditions explicit. Deployment domains, node counts, and database choices are runtime configuration and should not be written into a resource DID or root-platform proof as immutable protocol conditions. What must be interoperable is identity, signatures, resource packages, digests, status, and node authorization.

{\def\LTcaptype{none} 
\begin{longtable}[]{@{}>{\raggedright\arraybackslash}p{0.2400\linewidth}>{\raggedright\arraybackslash}p{0.2400\linewidth}>{\raggedright\arraybackslash}p{0.2400\linewidth}>{\raggedright\arraybackslash}p{0.2400\linewidth}@{}}
\toprule\noalign{}
Deployment form & Primary reachability & Suitable resource propagation boundary & Common facts that must remain consistent \\
\midrule\noalign{}
\endhead
\bottomrule\noalign{}
\endlastfoot
Public network & Internet users or public clients & Resources and discovery results within public authorization domains & DID, control proof, root-platform publication proof, and governance status \\
Private network & Internal users and systems of an organization & Internal authorization or business domains & Resource version, digests, node role, and verification order \\
Consortium network & Authorized members of multiple organizations & Resources and semantic vocabulary propagated under consortium rules & Shared trust root, node authorization, and cross-domain verification boundaries \\
\end{longtable}
}

The deployment form changes reachability and the scope of data propagation. It does not change how resource control is proved, nor does it turn the network entry point itself into a trust authority.

\section{Availability, Isolation, and Failure Boundaries}\label{sec:03.10}

OAN\textquotesingle s degradation strategy must distinguish read-side display from trust-sensitive actions. When the distribution service is temporarily unavailable, a discovery service node can display an old snapshot with a freshness marker, and an operations page can display delayed statistics or historical events. However, resource registration, root-platform publication, node-authorization changes, revocation-state application, and verification before high-risk invocation should be rejected, suspended, or placed in a pending state when critical evidence cannot be confirmed. No degradation may generate a new signed fact or change the resource lifecycle state.

Network interruption, database failure, and index lag should be recorded separately. During recovery, components should continue from persisted tasks, cursors, or event positions and re-verify cross-node materials, avoiding the direct treatment of an old cached version as the latest version. Isolation boundaries include node identity, authorization domain, resource version, and data-storage boundaries. A failure or incorrect data from one node must not automatically spread as a trust fact to other nodes.

{\def\LTcaptype{none} 
\begin{longtable}[]{@{}>{\raggedright\arraybackslash}p{0.2400\linewidth}>{\raggedright\arraybackslash}p{0.2400\linewidth}>{\raggedright\arraybackslash}p{0.2400\linewidth}>{\raggedright\arraybackslash}p{0.2400\linewidth}@{}}
\toprule\noalign{}
Failure symptom & Permitted read-side behavior & Trust-sensitive action & Basis for recovery \\
\midrule\noalign{}
\endhead
\bottomrule\noalign{}
\endlastfoot
Temporary content-distribution unavailability & Display an old snapshot with a freshness marker & Suspend new trusted publication or unverified propagation & Distribution recovery, digest verification, and publication proof \\
Governance projection delay & Display delayed status or historical data & Apply conservative handling to authorization changes & Governance-event sequence and projection cursor \\
Discovery-index failure & Provide results through another node or display unavailability & Do not generate a new trusted indexing conclusion & Database recovery and re-indexing \\
Node identity/signature anomaly & Retain audit records & Reject restricted writes and isolate the node & Authorization recovery and re-onboarding tests \\
\end{longtable}
}

\section{Interaction of the Control, Data, and Semantic Planes}\label{sec:03.11}

The control plane first decides whether a node and a resource version meet the conditions for entering the network, propagation, and validity. The data plane then transfers the DID Document, resource package, metadata, digests, and proofs permitted for publication by the control plane. The semantic plane builds capability-tag, description, and use-case indexes for resources that have already passed identity and publication checks. A query can use structured and semantic conditions at the same time, but the result must contain or be traceable to the resource DID, version, source, root-platform proof, and governance status.

Governance-status changes affect continued propagation through the data and semantic planes. After a node is revoked or a resource version becomes invalid, new synchronization should stop or filter the corresponding materials, and existing indexes should be updated after the status has propagated. A change to the tag tree can affect registration recommendations and query interpretation, but cannot retroactively change the identity or digest of a signed resource package. The dependency order among the three planes is that control facts constrain data propagation, data materials support semantic indexing, and semantic results cannot grant control authority in return.

The boundaries among the three planes also determine the order of failure handling. When control-plane status is unknown, the data plane can retain verified materials for limited reads, but must not expand propagation. When data-plane materials are missing or their digests are inconsistent, the semantic plane must not establish or update a trusted index. When the semantic index is unavailable, the system can still provide exact DID queries or other explicitly supported structured queries. After recovery, it should first restore control status and data integrity, and then rebuild or refresh the semantic index.

The three planes and their dependencies are shown below. In the implementation, the data exchanged between the planes is not undefined free-form JSON. It is a JSON/HTTP payload serialized from shared Rust types: resource registration uses \passthrough{\lstinline!ResourceRegistrationSubmission!}; forwarding from a registrar service node to the root platform wraps the request in \passthrough{\lstinline!ResourceVerifyAndPublishRequest!}; content distribution uses \passthrough{\lstinline!ResourceCdnPublishRequest!} or \passthrough{\lstinline!ResourceCdnBatchPublishRequest!}; and discovery queries use \passthrough{\lstinline!ResourceDiscoveryQuery!} and return \breakablecode{ResourceDiscoveryResponse}. These types define field names and nesting relationships, Axum routes define methods and paths, and signature envelopes and resource control proofs define different trust boundaries. Together, they form the node-cooperation contract.

\pandocbounded{\includegraphics[width=\linewidth,height=0.72\textheight,keepaspectratio]{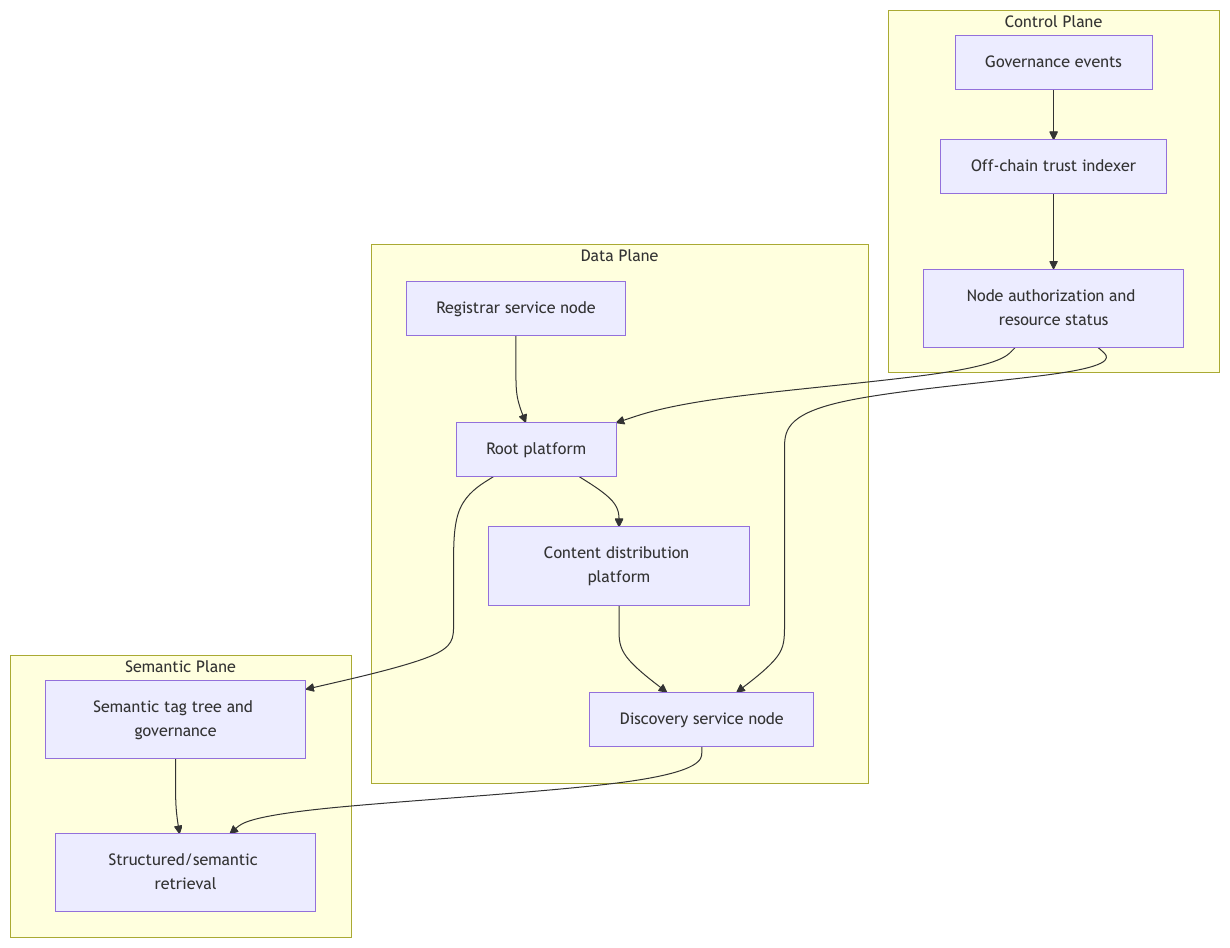}}

A cross-plane query response explains both why a resource was found and which evidence supports using it as a candidate. A Discovery node may filter by capability descriptions, tags, and structured resource types, but every result remains associated with its resource DID, version, ResourcePackage digest, Root trusted-publication proof, and current governance state. Semantic relevance is never interpreted directly as authorization or a security guarantee.

\section{Sources of Facts and Derived Read Models}\label{sec:03.12}

The sources of fact for the resource control relationship are the resource DID Document, control keys, and valid control proof. The source of fact for registration acceptance is the registrar service node\textquotesingle s registration record or registration credential. The sources of fact for root-platform acceptance and publication are the root platform\textquotesingle s persistent records and root-platform trusted publication proof. The source of fact for distribution is the publication artifact and its digests stored by the content distribution platform. The source of fact for discovery visibility is the synchronization, verification, and indexing record of the specific discovery service node. The source of fact for node authorization status is the governance events and their projection in the off-chain trust indexer. Website pages, cached interfaces, and statistics are observation layers and are not the sole authority for any of the facts above.

A derived read model must retain source references, a version or cursor, verification time, freshness, authorization domain, and error status. A read model may combine fields or build a full-text or semantic index for query efficiency, but it must not discard the DID, version, digest, and proof references needed to re-verify the resource. When the read model differs from the authoritative source, it should show a synchronization or verification anomaly rather than allowing the existence of a local database record to override the authoritative status.

{\def\LTcaptype{none} 
\begin{longtable}[]{@{}>{\raggedright\arraybackslash}p{0.2400\linewidth}>{\raggedright\arraybackslash}p{0.2400\linewidth}>{\raggedright\arraybackslash}p{0.2400\linewidth}>{\raggedright\arraybackslash}p{0.2400\linewidth}@{}}
\toprule\noalign{}
Network fact & Authoritative source & References a derived node must retain & Handling when a gap occurs \\
\midrule\noalign{}
\endhead
\bottomrule\noalign{}
\endlastfoot
Resource control relationship & DID Document and control proof & DID, verification method, document digest & Reject or await verification \\
Registration acceptance & Registrar service node record/registration credential & Node identity, resource DID, time, and status & Must not be written as published \\
Root-platform publication & Root-platform record and root-platform trusted publication proof & Resource version, three types of digest, proof & Must not enter trusted distribution \\
Node authorization & Governance events and off-chain read model & Event identifier, sequence, and current status & Fail closed for sensitive actions \\
Discovery index & Discovery service node\textquotesingle s local record & Synchronization cursor, verification time, and freshness & Retain the old snapshot and mark it as lagging \\
\end{longtable}
}

When \passthrough{\lstinline!localCursor!} is behind, source verification fails, or the governance projection has not yet been updated, the read model may still be retained for limited display, but \passthrough{\lstinline!freshness!} should be changed to \passthrough{\lstinline!stale!} or \passthrough{\lstinline!verification\_required!} and the reason for the error should be retained. A row in an index database can only show that the local node once received the material; by itself, it cannot prove that the resource remains valid, propagatable, or invocable.

\section{Network Topology and Sequence Consistency}\label{sec:03.13}

The root-platform publication sequence is produced by the root platform and identifies the position of resource publication materials in the data-distribution path. The on-chain governance-event sequence is produced by the on-chain governance system and expresses the order of node-authorization and status changes. The discovery service node\textquotesingle s synchronization cursor is maintained by the discovery service node and expresses the distribution position that the node has processed. The three values may look numerically similar, but they are not interchangeable. The implementation must store and check them separately, avoiding the use of one global number to infer resource publication, governance finality, or local indexing status.

When a discovery service node encounters a cursor gap, disorder, or rollback, it should suspend application of the affected scope, retrieve missing materials from the root platform or content distribution platform, and continue only after re-verification. Duplicate materials can be processed idempotently by resource DID, version, and package digest. Duplicate governance events should be deduplicated by event identifier or sequence. A rollback in the publication sequence must not overwrite a newer version that has already been confirmed. During cross-node delay, an old snapshot may be retained for limited read-side display, but freshness must be marked, and sensitive actions must stop when revocation, suspension, or authorization status cannot be confirmed.

The three sequences should be stored separately in implementation and operations records. The configuration-style example below expresses their sources and processing boundaries only; it does not mean that they share one numbering format:

\begin{lstlisting}
resource_publication:
  producer: root-platform
  latest: publish-1842
  applies_to: did-document-and-resource-package
  gap_action: pause-affected-publication

governance_events:
  producer: governance-ledger
  latest: event-771
  applies_to: node-authorization-and-resource-status
  gap_action: fail-closed-for-sensitive-actions

discovery_sync:
  producer: discovery-service-node
  local_cursor: publish-1839
  applies_to: local-validation-and-indexing
  gap_action: fetch-missing-publications-and-revalidate
\end{lstlisting}

For example, a discovery service node\textquotesingle s synchronization cursor falling behind the root-platform publication sequence only means that the local index may be incomplete. It does not mean that governance events are behind by the same amount. If a gap appears in governance events, an application that depends on governance status should be suspended even when resource publication materials are complete. A node may mark a resource as a current discoverable result only after the corresponding sequences are continuous, the materials pass verification, and the status projection is complete.

\section{Execution Domains, Completion Conditions, and Evolution Boundaries}\label{sec:03.14}

{\def\LTcaptype{none} 
\begin{longtable}[]{@{}>{\raggedright\arraybackslash}p{0.2400\linewidth}>{\raggedright\arraybackslash}p{0.2400\linewidth}>{\raggedright\arraybackslash}p{0.2400\linewidth}>{\raggedright\arraybackslash}p{0.2400\linewidth}@{}}
\toprule\noalign{}
Execution domain & Starting point & Completion condition & Can fail or accumulate backlog independently \\
\midrule\noalign{}
\endhead
\bottomrule\noalign{}
\endlastfoot
Registration-request loop & Client submits complete signed materials to Registrar & Registrar returns a determinate result & Client, Registrar, and Root admission path \\
CDN-publication background loop & Root creates a durable publication task & CDN confirms that the verifiable ResourcePackage has been persisted & Publisher, notification channel, and CDN \\
Per-discovery-target synchronization loop & Target cursor is behind the publishable high-water mark & That Discovery node validates, indexes, and advances its cursor & Any Discovery target and its database \\
\end{longtable}
}

The Trust Indexer and Sui bulletin contract form a governance side path. The contract produces node-governance facts, the indexer creates an off-chain projection in on-chain order, and core nodes consume that projection for admission decisions. They do not participate in ordinary Root---CDN---Discovery ResourcePackage propagation. The current reference implementation operates with one Root authoritative-state domain. Multi-Root operation is a standards and architectural evolution direction and must not be inferred as implemented merely from a topology containing multiple Registrar and Discovery nodes.

\section{References}\label{sec:03.15}

{\def\LTcaptype{none} 
\begin{longtable}[]{@{}>{\raggedright\arraybackslash}p{0.3200\linewidth}>{\raggedright\arraybackslash}p{0.3200\linewidth}>{\raggedright\arraybackslash}p{0.3200\linewidth}@{}}
\toprule\noalign{}
Source & Type & Link \\
\midrule\noalign{}
\endhead
\bottomrule\noalign{}
\endlastfoot
\passthrough{\lstinline!oan-root-nodes!} & Code repository: root platform, publication coordination, and node directory & \url{https://github.com/OpenAgenet/oan-root-nodes} \\
\passthrough{\lstinline!oan-registrar-nodes!} & Code repository: registrar service node and access path & \url{https://github.com/OpenAgenet/oan-registrar-nodes} \\
\passthrough{\lstinline!oan-discovery-nodes!} & Code repository: discovery service node, synchronization, and indexing & \url{https://github.com/OpenAgenet/oan-discovery-nodes} \\
\passthrough{\lstinline!oan-trust-indexer!} & Code repository: governance-event synchronization and off-chain state projection & \url{https://github.com/OpenAgenet/oan-trust-indexer} \\
OAN Yellow Paper & arXiv yellow paper: network architecture and trust path & \url{https://arxiv.org/abs/2606.03163} \\
Agentic Overlay Network Architecture & IETF draft: agentic overlay network architecture & \url{https://datatracker.ietf.org/doc/draft-xu-agentic-overlay-network-architecture/} \\
\end{longtable}
}

\chapter{OAN Resource Model}\label{chap:04}

This chapter explains how OpenAgenet (OAN) models capability units that can be registered, published, distributed, and discovered as resources. The current core resource types are Agent Service, Skill Package, MCP Service, and Tool API. Infrastructure nodes, organizations, and developers may also appear as identities or subjects in control, publication, and issuance relationships, but they are not among the four discoverable resource types discussed here. Each resource has an independent resource DID and resource type and can form a verifiable representation of one version through its description, capabilities, endpoints, version, and lifecycle information.

The resource model unifies only the common elements required for cross-type interoperability; it does not force the four resource types into the same business object. The resource type determines field interpretation, endpoint shape, protocol binding, and invocation method; the resource DID establishes the control and registration boundary; resource metadata is used for registration and discovery; and the resource package binds the DID document, metadata, hashes, and the root platform\textquotesingle s trusted publication proof into a distributable object. Subsequent chapters specify identity, registration, discovery, publication, and invocation separately, avoiding a duplicate definition of all protocol behavior in this chapter.

\section{Resource Model Overview}\label{sec:04.1}

The common identity fields of the resource model include the resource DID, resource type, and subject type, as well as optional controller DID, publisher DID, and issuer DID; common descriptive fields include the name, description, capability description, capability tags, and use cases; common connectivity fields include protocol bindings, service endpoints, and authentication or credential requirements; and common state fields include the version, lifecycle state, authorization domain, network scope, and update time. \passthrough{\lstinline!oanMetadata!} is the primary metadata object in a DID document for carrying OAN resource semantics and lifecycle information, while the resource metadata in a resource package provides a versioned representation for publication and discovery.

The current common resource enumeration includes \passthrough{\lstinline!agent\_service!}, \passthrough{\lstinline!skill!}, \passthrough{\lstinline!mcp\_server!}, and \passthrough{\lstinline!tool\_api!}. The resource type, subject type, \passthrough{\lstinline!oanMetadata!} in the DID Document, registration submission, and ResourcePackage metadata must be cross-checked. A \passthrough{\lstinline!did:oan!} routing code does not encode resource type. When a field is absent, empty, or present only in the web presentation layer, it must not be used to infer that the resource has the corresponding capability; actual invocation permission is not automatically granted by an endpoint or description field in the resource model.

The four resource types share the common models of resource DID, resource metadata, version, and lifecycle, while using different protocol bindings, endpoints, and package contents according to their capability forms:

\pandocbounded{\includegraphics[width=\linewidth,height=0.72\textheight,keepaspectratio]{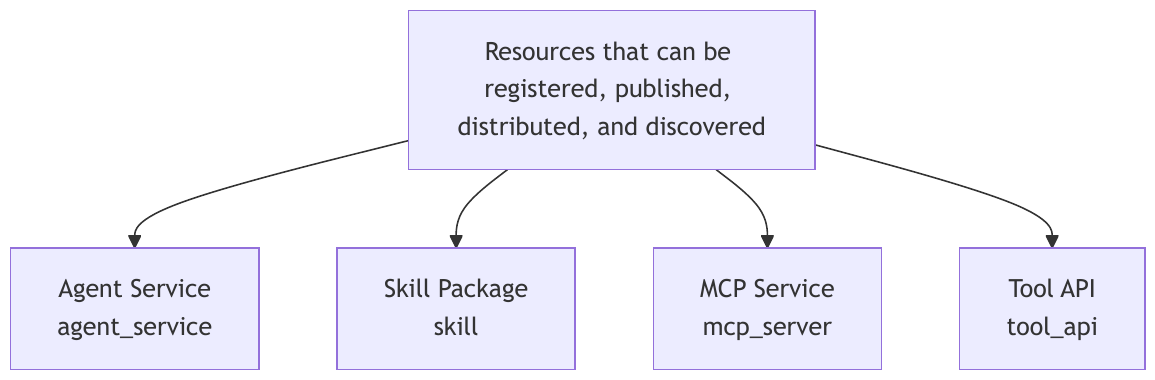}}

The following structural example is adapted from a cross-language fixture in the public common repository. It shows a resource DID accepted by the current \passthrough{\lstinline!did:oan!} parser and the combination of resource-type and description fields in a DID Document. Verification methods and the proof are omitted, so this is a resolvable structural example rather than a registration request that can be submitted directly:

\begin{lstlisting}
{
  "id": "did:oan:K7mQ9:7YpQm9Kx2VnRb6Ts3WfHa4Cd5Ej8LgNz",
  "controller": "did:oan:P9aBc:2LmNo3PqRsTuVwXyZaBcDeFgHiJkLmNo",
  "oanMetadata": {
    "subjectType": "resource",
    "resourceType": "skill",
    "resourceDescription": {
      "name": "Repository Structure Summary",
      "description": "Summarizes the structure of a code repository."
    },
    "capabilityTags": ["code-repository", "project-structure-summary"],
    "lifecycleState": "active"
  }
}
\end{lstlisting}

This object can serve as a minimal reading example of resource metadata, but it cannot independently replace the DID document, control proof, resource package hash, or the root platform\textquotesingle s trusted publication proof. A discovery service node can use its structured fields to build an index, while the caller must still obtain the corresponding evidence and perform subsequent verification according to the resource type.

\section{Agent Services}\label{sec:04.2}

An agent service is a resource type that provides callable agent capabilities, with the resource type value \passthrough{\lstinline!agent\_service!}. It expresses an independent control and publication boundary through its resource DID, describes capabilities, capability tags, use cases, inputs and outputs, and applicable scenarios through its resource description, and expresses how to establish a connection through service endpoints and protocol bindings. An agent service may declare A2A, MCP, or other business protocols, but declaring a protocol only identifies the resource\textquotesingle s interface context; it does not mean that the caller has already obtained business authorization.

The relationship with an A2A Agent Card is an interoperability mapping relationship. OAN capability descriptions, capability tags, use cases, service endpoints, protocol bindings, version, and authentication declarations can provide trusted identity and publication provenance for corresponding A2A Agent Card information; A2A Agent Card fields can also be mapped to OAN resource metadata through an adapter. OAN additionally provides the resource DID, control proof, registration credential, root platform publication proof, governance status, and authorization domain, allowing a capability description to be associated with provenance, version, and network governance facts without changing the interaction semantics of A2A itself.

{\def\LTcaptype{none} 
\begin{longtable}[]{@{}>{\raggedright\arraybackslash}p{0.3200\linewidth}>{\raggedright\arraybackslash}p{0.3200\linewidth}>{\raggedright\arraybackslash}p{0.3200\linewidth}@{}}
\toprule\noalign{}
Interoperability information & Where it is carried in the OAN resource model & Boundary to preserve during mapping \\
\midrule\noalign{}
\endhead
\bottomrule\noalign{}
\endlastfoot
Service name and description & \passthrough{\lstinline!resourceDescription.name!}, \passthrough{\lstinline!resourceDescription.description!} & Explains what the resource is; it is not equivalent to invocation authorization \\
Capabilities and applicable tasks & \passthrough{\lstinline!capabilityDescription!}, \passthrough{\lstinline!capabilityTags!}, \passthrough{\lstinline!useCaseExamples!} & Describes discoverable capabilities; it does not replace runtime capability checks \\
Service entry point & \passthrough{\lstinline!service!} in the DID document and \passthrough{\lstinline!serviceRef!} in the protocol binding & The entry-point address may be public, but the endpoint still requires independent authentication \\
Protocol and version & \passthrough{\lstinline!protocolBindings.protocol!}, \passthrough{\lstinline!version!}, \passthrough{\lstinline!transport!} & OAN records the access context without changing A2A interaction semantics \\
Identity and publication evidence & Resource DID, control proof, registration credential, and the root platform\textquotesingle s trusted publication proof & The content of an A2A Agent Card alone cannot be treated as an OAN publication fact \\
\end{longtable}
}

\section{Skill Packages}\label{sec:04.3}

A skill package is a capability unit that can be published, installed, loaded, or reused independently, with the resource type value \passthrough{\lstinline!skill!}. It should describe its capabilities, inputs and outputs, use cases, dependencies, and usage conditions through its resource description, and explain its acquisition and usage entry points through service endpoints, package information, or implementation links. A skill package may be composed by an agent service or another orchestration system, but its discoverability, version, and provenance should still be based on its own resource DID, resource package, and root platform publication facts.

The ``reusable'' property of a skill package means that a caller can use it repeatedly after satisfying its dependency, version, and authorization conditions; it does not mean that the caller can bypass resource control, installation security, or business permissions. Installation commands, download addresses, scripts, or dependencies in the skill package description are public declarations only. The caller must still verify the resource package hash, publication proof, version state, and local execution policy.

Current integration tests use an \passthrough{\lstinline!https!} manifest URL for a Skill Package and carry the manifest, download URL, and package digest in \passthrough{\lstinline!packageInfo!} to verify that Skill Packages share the same registration---publication---discovery pipeline as other resource types. This is a test-input form, not a mandatory delivery method for every Skill Package. In an actual publication, the manifest, downloadable material, and version remain consistent with the DID Document, metadata, and ResourcePackage digests. When dependencies exist, the target resources and version constraints are verified separately.

\section{MCP Services}\label{sec:04.4}

An MCP service is a service resource that provides tool or resource access capabilities through the Model Context Protocol, with the resource type value \passthrough{\lstinline!mcp\_server!}. Its resource DID identifies the service\textquotesingle s publication and control boundary; its resource metadata describes capability tags, use cases, protocol version, transport method, service endpoint, and authentication requirements; and its MCP capability description can be expressed through protocol bindings or a dedicated capability object. A tool list or resource list is a capability declaration and cannot independently replace endpoint authentication, invocation authorization, or runtime status checks.

The mapping between OAN and MCP metadata should preserve two boundaries: MCP is responsible for tool, resource, and interaction semantics at the protocol layer, while OAN provides the MCP service with resource identity, control proof, trusted publication, governance status, and a discovery entry point. An MCP service endpoint being reachable does not mean that the endpoint has been published by the root platform or is open to every caller; discovery results should not interpret an MCP capability declaration directly as invocation permission.

Current mixed-resource integration tests use \passthrough{\lstinline!mcp!} as the protocol value and an HTTPS URL as the service entry point, verifying that an MCP Service can pass through the common pipeline. The manual does not freeze the test URL or one transport version as a protocol requirement. \passthrough{\lstinline!protocolBindings!} states the integration context, while \passthrough{\lstinline!service!} in the DID Document states the network entry point; both remain associated with the resource DID, publication version, digests, and authentication requirements. A Discovery node cannot mark a resource as directly callable merely because an MCP tool inventory exists.

\section{Tool APIs}\label{sec:04.5}

A tool API is a resource that provides a specific operation capability through a network interface, with the resource type value \passthrough{\lstinline!tool\_api!}. It should describe the interface address, operation or capability name, input parameters, output results, protocol, version, authentication method, and invocation constraints; complex parameters may be expressed through input and output schemas, protocol schema references, or external documentation. A tool API can be referenced by an agent service, skill package, or MCP service, but a reference relationship does not mean that the caller has passed authentication and authorization at the target endpoint.

The resource description of a tool API and the actual service implementation should maintain a verifiable association. The resource controller is responsible for declaring the interface and version; the publication path is responsible for binding the DID document, metadata, hashes, and root platform proof; and the caller is responsible for checking the endpoint, version, evidence, and its own policy before connecting. The existence of an interface, a normal HTTP response, or the presence of a discovery index entry cannot independently prove that the tool\textquotesingle s business result is correct or that invocation is permitted.

Current integration tests express a Tool API through an HTTPS manifest URL and bind the manifest and package digest in \passthrough{\lstinline!packageInfo!}. A real resource may additionally separate ``what it can do'' from ``how to invoke it'' through its description, protocol bindings, and an external schema. The schema constrains input and output shapes, while credential requirements state the authentication material needed before invocation. Actual permission decisions, rate limiting, business validation, and result quality remain the responsibility of the target API and caller policy. The resource model provides only the identity boundary in which these declarations can be registered, published, and discovered.

The four resource types share a verifiable resource identity and discoverable description, but their post-discovery use differs. An Agent Service normally exposes a business-interaction endpoint; a Skill Package describes a reusable capability or deliverable; an MCP Service provides MCP protocol capabilities; and a Tool API provides an interface and invocation constraints. OAN registers and distributes these descriptions and evidence; it does not host or continuously execute their business content.

\section{Resource Relationships and Composition}\label{sec:04.6}

Composition relationships connect resources through DIDs and version constraints rather than copying the referenced object into a parent resource. An Agent Service can therefore declare a dependency on a Skill Package or MCP Service while the caller still verifies each object\textquotesingle s controller, version, Root proof, and lifecycle separately. A relationship containing only a name or URL may be displayed by a Discovery service, but it does not establish that the dependency is a particular registered resource.

Resource relationships express implementation associations or composition semantics. The current common model provides \passthrough{\lstinline!implementationLinks!}, but not every \passthrough{\lstinline!dependency!}, \passthrough{\lstinline!invokes!}, or \passthrough{\lstinline!provided-by!} relationship has a stable, uniform public schema. This manual treats only fields present in common types and validated by nodes as interoperable. Other relationships may be used as extension descriptions, but nodes cannot be assumed to interpret them identically. A ResourcePackage can carry a relationship representation for one version; it cannot extend that representation automatically to future versions.

Before runtime, a composition system should verify the DID, resource type, version or package hash, root platform\textquotesingle s trusted publication proof, lifecycle state, and authorization domain of every referenced resource. If a referenced resource is undiscoverable, does not satisfy the version constraint, lacks proof, has been revoked, or has been suspended, the caller or orchestration system should choose according to risk among rejecting the composition, waiting for synchronization, or displaying it only without invocation eligibility. It must not treat the existence of an index summary as proof of composition integrity.

Relationship verification checks at least the target DID, relationship type, target type, version constraint, and current publication state of the target resource. A target-service reference only helps locate a service; it cannot replace endpoint reachability and authentication. When the target resource version or ResourcePackage digest changes, the relationship is matched and verified again. A relationship field not yet present in the public schema is treated as a namespaced extension, and the publisher must allow for a receiver ignoring it.

\subsection{Dependencies Between Resources}\label{subsec:04.6.1}

A dependency relationship means that normal use of one resource requires another resource or external capability. The dependency declaration should point to the target resource DID and, where necessary, provide the target type, service reference, and version constraint; the business meaning, installation method, and runtime authorization of the dependency can be supplemented through the resource description or protocol binding, but the identity and version boundaries must be expressed structurally. When the dependency target is updated, the caller should recheck the version constraint and publication state.

\subsection{Composition of Agent Services and Skill Packages}\label{subsec:04.6.2}

An agent service may depend on or compose multiple skill packages, using them as reusable capabilities for task execution, information processing, or tool orchestration. The composition record should distinguish the agent service DID, skill package DID, skill package version or package hash, and relationship type; publication of the agent service does not automatically extend the control and publication facts of the skill package. The caller must verify the evidence and compatibility conditions of both the agent service and the skill package.

\subsection{Invocation Relationships Between MCP Services and Tool APIs}\label{subsec:04.6.3}

An MCP service may provide a collection of tools or reference one or more tool APIs through protocol bindings; a tool API may also be called directly by an agent service or skill package. The relationship declaration should specify the target DID, target service or operation, protocol, and version, avoiding a statement that only says ``supports a particular tool.'' Tools discoverable at the MCP protocol layer still require OAN resource identity, publication state, endpoint authentication, and the caller\textquotesingle s authorization decision.

\subsection{Composition Integrity and Version Constraints}\label{subsec:04.6.4}

Composition integrity requires at least that the DIDs, resource types, versions or package hashes, relationship declarations, root platform trusted publication proofs, and current lifecycle states of both sides of a relationship are consistent, and that the target resource is within the authorization domain visible to the caller. When a version constraint is not satisfied, the package hash changes, the proof expires, or the resource is suspended, the old composition declaration must not be used as an unconditional basis for invocation; where necessary, the caller should rediscover, reverify, or fall back to a clearly compatible version.

\section{Resources as Independently Identifiable Objects}\label{sec:04.7}

Each of the four resource types can have an independent resource DID, using that DID as the identity boundary for creation, registration, publication, discovery, and update. A controller DID expresses the control relationship, and a publisher DID expresses registration or operational provenance, but neither replaces the resource\textquotesingle s own DID. Verification methods, service endpoints, and \passthrough{\lstinline!oanMetadata!} in the DID Document jointly support verification of resource identity, control relationships, and capability declarations.

An independent DID allows a skill package, MCP service, or tool API to be discovered and reused independently of a particular agent service, and also allows it to be updated, suspended, revoked, and versioned independently. When resources have composition relationships, each resource must still be verified separately through its own DID and publication evidence; the trusted state of one resource does not automatically transfer to its dependencies.

The current \passthrough{\lstinline!did:oan!} syntax is \passthrough{\lstinline!did:oan:<routing-code>:<suffix-code>!}. The \passthrough{\lstinline!routing-code!} is a five-character, case-sensitive Base58 routing code; the \passthrough{\lstinline!suffix-code!} is a 32-character Base58 suffix; and the complete DID is 46 characters long. The routing code associates the resource with Registrar routing and does not encode resource type. The common protocol also checks that the routing code of a resource DID matches that of the accepting Registrar DID. The following valid structure comes from a cross-language fixture:

{\def\LTcaptype{none} 
\begin{longtable}[]{@{}>{\raggedright\arraybackslash}p{0.3200\linewidth}>{\raggedright\arraybackslash}p{0.3200\linewidth}>{\raggedright\arraybackslash}p{0.3200\linewidth}@{}}
\toprule\noalign{}
Component & Example value & Semantics \\
\midrule\noalign{}
\endhead
\bottomrule\noalign{}
\endlastfoot
Method prefix & \passthrough{\lstinline!did:oan!} & OAN DID method \\
Routing code & \passthrough{\lstinline!K7mQ9!} & Registrar routing identifier, not a resource type \\
Suffix code & \passthrough{\lstinline!7YpQm9Kx2VnRb6Ts3WfHa4Cd5Ej8LgNz!} & Resource identifier suffix under that route \\
Complete DID & \passthrough{\lstinline!did:oan:K7mQ9:7YpQm9Kx2VnRb6Ts3WfHa4Cd5Ej8LgNz!} & Resource identifier accepted by the current parser \\
\end{longtable}
}

Resource type comes from the registration submission, \passthrough{\lstinline!oanMetadata.resourceType!} in the DID Document, and ResourcePackage metadata, and nodes cross-check these values. The current parser rejects old four-character type prefixes such as \passthrough{\lstinline!SKFI!}. The SDK\textquotesingle s DID creation function likewise uses only the supplied routing code and does not derive one from resource type.

\section{Resource Metadata and Capability Descriptions}\label{sec:04.8}

Resource metadata is a structured description used for registration, publication, and discovery in addition to resource identity. Common fields include the resource DID, resource type, subject type, controller or publisher DID, name, description, capability description, capability tags, authorization domain, protocol bindings, service endpoints, credential requirements, package information, network scope, lifecycle state, version, and update time. \passthrough{\lstinline!oanMetadata!} in the DID document and the resource metadata in the resource package should remain consistent in key fields such as resource DID, resource type, version, hash, authorization domain, and state.

The capability description answers what the resource can do; capability tags answer which governed capability categories the resource belongs to; and use cases answer in which tasks those capabilities are used. Names and free text help people and semantic engines understand a resource, but cannot replace structured tags, protocol bindings, endpoints, or control proofs. Metadata is a declaration made by the resource provider. The root platform and discovery service nodes can verify its structure, hash, and publication provenance, but complete fields do not replace an independent assessment of business quality.

Capability fields in \passthrough{\lstinline!oanMetadata.resourceDescription!} can be organized into a minimal object for retrieval:

\begin{lstlisting}
{
  "name": "Repository Structure Summary",
  "description": "A reusable capability for examining code repositories.",
  "capabilityDescription": "Finds repository files and summarizes the project structure.",
  "capabilityTags": ["code-repository", "project-structure-summary"],
  "useCaseExamples": [
    "I need a tool that can search code repositories and summarize the project structure."
  ],
  "version": "1.0.0"
}
\end{lstlisting}

Natural-language fields are suitable for semantic discovery and user understanding, tags are suitable for structured filtering, and the version associates the description with the resource package and publication evidence. Together they describe the resource\textquotesingle s capabilities, but do not mean that a node has verified runtime behavior, nor do they replace identity, publication, and authorization checks before invocation.

\subsection{Basic Identity and Description Fields}\label{subsec:04.8.1}

The basic fields should be sufficient to identify the resource DID, resource type, subject type, name, description, and controller or publisher relationship, and should be consistent with \passthrough{\lstinline!id!}, \passthrough{\lstinline!oanMetadata.resourceType!}, \passthrough{\lstinline!oanMetadata.subjectType!}, and related DID fields in the DID document. Resource type values should use the protocol-defined enumeration; a Chinese display name or alias in the presentation layer must not be the sole basis for machine judgment.

\subsection{Capability Tags, Capability Descriptions, and Use Cases}\label{subsec:04.8.2}

Capability tags should preferably use canonical identifiers from the capability tag tree maintained by the root platform. Extension tags may also be retained when their source is clearly identified. Capability descriptions and use cases should explain executable capabilities, inputs and outputs, and applicable tasks. A registration service node may recommend tags based on a description, and a discovery service node may combine tags, aliases, descriptions, and use cases for structured or semantic search, but relevance results still require identity, publication, and authorization-domain verification.

\subsection{Protocol, Endpoint, and Authentication Information}\label{subsec:04.8.3}

Protocol bindings should specify the protocol name, version, transport method, service reference, or schema reference, and associate these with one or more service endpoints. A service endpoint should specify its address, type, version, protocol, and any necessary port or server information. Authentication information should express the public authentication method, credential requirements, or prerequisites for invocation. Endpoint and authentication declarations are part of the resource capability description; they do not mean that the caller has already obtained access permission.

\subsection{State, Version, and Governance Information}\label{subsec:04.8.4}

Resource metadata should distinguish the resource lifecycle state, resource package version, update time, expiration time, previous version or version constraint, authorization domain, and network scope. The lifecycle state expresses facts such as whether the resource is active, suspended, revoked, or replaced; propagation progress is expressed by root platform publication, content distribution, and discovery synchronization records. These concepts must not be conflated in a single \passthrough{\lstinline!available!} field.

\section{Use Cases, Protocol Bindings, Endpoints, and Authentication Declarations}\label{sec:04.9}

Use cases, protocol bindings, service endpoints, and authentication declarations together form the public explanation of ``how to understand and use this resource.'' A resource may have multiple use cases, protocol bindings, and endpoints. Each binding should, as far as possible, specify the applicable capability, protocol version, transport method, service reference, and schema; authentication declarations should, as far as possible, be associated with a target endpoint, operation, or version. If an endpoint is unavailable, the protocol version is incompatible, or authentication information is incomplete, a discovery service node may return a result with the corresponding state, but should not describe it as directly callable.

\subsection{Use Case Descriptions}\label{subsec:04.9.1}

A use case description should state the task objective, input conditions, expected output, and applicable resource capabilities, and should preferably correspond to the capability description and capability tags. A natural-language use case can support semantic retrieval and user understanding, but cannot independently prove that the resource has the capability; precise constraints should be combined with input and output schemas, protocol bindings, and version information.

\subsection{Protocol Bindings}\label{subsec:04.9.2}

Protocol bindings should record the protocol name, protocol version, transport or interaction method, and references to endpoints or schemas. A2A and MCP may enter the resource description as external protocol contexts. OAN binding fields only express how the resource connects to these protocols; they should not turn OAN resource registration state into a mandatory clause of an external protocol.

\subsection{Service Endpoints}\label{subsec:04.9.3}

A service endpoint is the network entry point through which a resource provides capabilities or obtains materials. It may contain an endpoint identifier, endpoint type, address, version, protocol, server type, and port. When a resource has multiple endpoints, the protocol or version differences between them should be stated. When an endpoint is unreachable, its certificate is invalid, or the service is inactive, the discovery result should preserve that fact or freshness information rather than considering only the static address.

\subsection{Authentication and Invocation Requirements}\label{subsec:04.9.4}

Authentication and invocation requirements explain what identity, credential, signature, permission, or protocol context a caller must provide. A resource may publicly declare the credential type, issuer, scope, and presentation method, but final business authorization is jointly determined by the resource endpoint and the caller\textquotesingle s policy. OAN VCs, the root platform\textquotesingle s trusted publication proof, and node governance status are materials for verification before invocation; they are not automatic authorization for every business operation.

Authentication declarations in the resource model primarily answer ``what must the caller prepare,'' such as credential type, issuer, scope, and presentation method. The endpoint must still check credential validity, request subject, business permissions, and runtime policy after receiving a request. The public declaration of the same resource can be understood in two layers:

{\def\LTcaptype{none} 
\begin{longtable}[]{@{}>{\raggedright\arraybackslash}p{0.3200\linewidth}>{\raggedright\arraybackslash}p{0.3200\linewidth}>{\raggedright\arraybackslash}p{0.3200\linewidth}@{}}
\toprule\noalign{}
Layer & What the resource model can express & Conclusion that cannot be derived from this field alone \\
\midrule\noalign{}
\endhead
\bottomrule\noalign{}
\endlastfoot
Public declaration & Credential type, issuer, scope, presentation method, and applicable endpoint & That the caller definitely possesses the credential \\
Invocation decision & Endpoint authentication result, authorization policy, protocol context, and business conditions & That the resource is definitely available or that the business result is definitely correct \\
\end{longtable}
}

\section{Resource Packages and Publication Artifacts}\label{sec:04.10}

A ResourcePackage is a distributable object bound to one resource version. The current public \passthrough{\lstinline!ResourcePackage!} type contains the package version, resource DID, resource type, DID Document, optional raw DID Document, DID Document digest, metadata digest, package digest, digest algorithm, resource metadata, Root proof, and creation time. The DID Document expresses identity, verification methods, service endpoints, and OAN metadata; resource metadata expresses capabilities, protocols, endpoints, authorization domains, and lifecycle; a control proof expresses the current controller\textquotesingle s authorization of registration or update; a registration VC expresses a registration fact processed by the Registrar; and a Root proof expresses that Root validated and accepted the resource version under the rules.

At node interfaces, the client submits a \passthrough{\lstinline!ResourceRegistrationSubmission!}. In addition to the DID Document and three digests, it contains the package version, registration credential, the resource\textquotesingle s own control proof, and, in an external-controller case, a controller-authorization proof. After verification, Root organizes the corresponding fields into a \passthrough{\lstinline!ResourcePackage!}, which then enters a publication task and CDN distribution. A resource description, registration submission, and distributable ResourcePackage are therefore different objects; a client cannot treat its registration submission as if a Root proof had already been formed.

These objects form a layered validation path rather than parallel attachments. A node first parses the DID Document and checks the resource DID, type, package version, and controller relationship. It then recomputes the DID Document, metadata, and package digests under the prescribed canonicalization and digest algorithm, and subsequently verifies control proofs, the registration credential, Root proof, and current governance state. Registrar admission, Root acceptance, CDN persistence, and Discovery indexing are produced by different components; success at one stage does not replace another stage.

The public types have the following layered relationship. Complete fields and verifiable vectors are in Appendices A and C; this chapter does not duplicate a fictitious ResourcePackage that could drift from the implementation.

{\def\LTcaptype{none} 
\begin{longtable}[]{@{}>{\raggedright\arraybackslash}p{0.1920\linewidth}>{\raggedright\arraybackslash}p{0.1920\linewidth}>{\raggedright\arraybackslash}p{0.1920\linewidth}>{\raggedright\arraybackslash}p{0.1920\linewidth}>{\raggedright\arraybackslash}p{0.1920\linewidth}@{}}
\toprule\noalign{}
Object & Main producer & Key content & What it proves & What it cannot prove alone \\
\midrule\noalign{}
\endhead
\bottomrule\noalign{}
\endlastfoot
DID Document & Resource and its current controller & \passthrough{\lstinline!id!}, controller, verification methods, services, and \passthrough{\lstinline!oanMetadata!} & Resource identity structure and declared control relationship & Registrar admission or Root acceptance \\
\passthrough{\lstinline!ResourceRegistrationSubmission!} & Client/SDK & DID Document, metadata, three digests, registration credential, and control proofs & Complete input to one registration request & Completion of publication or discovery synchronization \\
\passthrough{\lstinline!ResourcePackage!} & Root & Validated content, resource metadata, Root proof, and creation time & A particular resource version was accepted by Root under the rules & Current availability of the business endpoint or invocation authorization \\
CDN persistence object & Publication/CDN path & Downloadable ResourcePackage and publication state & Content entered that CDN\textquotesingle s persistence scope & Every Discovery node completed synchronization \\
Discovery index object & Each Discovery node & Validated structured fields and optional semantic projection & That node processed the corresponding resource version & Another Discovery node has the same progress \\
\end{longtable}
}

A verifier first uses \passthrough{\lstinline!resourceDid!} and \passthrough{\lstinline!resourceType!} to determine object ownership, then recomputes the three digests, checks consistency of key fields between \passthrough{\lstinline!metadata!} and the DID Document, and finally verifies that \passthrough{\lstinline!rootProof.packageClaims!} matches the current package declarations. This order separates content integrity, resource control, registration admission, and Root acceptance instead of interpreting one downloaded file as all network facts.

\subsection{ResourcePackage Contents}\label{subsec:04.10.1}

Root constructs a ResourcePackage after admitting a registration submission. Its core content includes the resource DID, DID Document, resource metadata or package information, version, bound digests, and Root proof. The client and Registrar submit \passthrough{\lstinline!ResourceRegistrationSubmission!}; they must not preconstruct a \passthrough{\lstinline!ResourcePackage!} containing a Root proof. A registration VC is likewise not embedded in the public ResourcePackage. A caller that needs it obtains it from a protected registration-record query or its own local copy.

ResourcePackage content is organized around one resource version. The current type binds \passthrough{\lstinline!resourceDid!}, \passthrough{\lstinline!resourceType!}, \passthrough{\lstinline!packageVersion!}, the DID Document, DID Document digest, metadata digest, package digest, digest algorithm, resource metadata, and Root proof. \passthrough{\lstinline!rootProof.packageClaims!} can associate the resource DID, type, version, digests, lifecycle state, authorization domains, and an optional bulletin reference.

\subsection{Metadata and Resource Content}\label{subsec:04.10.2}

Metadata is a structured description of the resource content, capabilities, and runtime entry points. The resource content may be skill package files, a service description, a tool interface, or an MCP capability declaration. Package information, download addresses, manifest addresses, implementation links, and version fields in the metadata should point to or describe the current resource version; historical metadata must not override the identity and hash of the current resource package.

\subsection{Hashes and Integrity Proofs}\label{subsec:04.10.3}

The resource package should express the DID document hash, metadata hash, and resource package hash separately, and declare the hash algorithm used. The verifier should recompute the corresponding hashes according to the canonical representation, and check whether the internal fields of the resource package are consistent with the declarations in the root platform proof. A matching hash proves content integrity only; it does not independently prove control, node authorization, or trustworthiness of the business endpoint.

\subsection{Publication, Download, and Verification Artifacts}\label{subsec:04.10.4}

A registration service node may return a registration result, registration credential, or downloadable material; the root platform produces an acceptance record and the root platform\textquotesingle s trusted publication proof; the content distribution platform provides downloadable resource packages and related materials; and a discovery service node stores a verified index object. Downloading a resource package does not mean that invocation authorization has been completed. The caller should reverify it according to provenance, version, freshness, governance state, and local policy.

\section{Resource Lifecycle and Visibility State}\label{sec:04.11}

The resource lifecycle describes the state of the resource version itself, while network propagation progress describes whether that version has passed registration acceptance, root platform acceptance, content distribution, and discovery indexing. A resource may be registered but not yet published by the root platform, published but not yet synchronized by a particular discovery service node, or present in a discovery index but no longer visible to a particular caller because of a change in authorization domain or governance state. State judgments should cite both the component that produced the state and the corresponding evidence.

Resource updates should distinguish an idempotent retry of the same version from publication of a new version. A repeated submission with the same resource DID, version, and hashes may reuse the existing result; content changes, version changes, control-key changes, or lifecycle changes must enter a new verification and publication decision. An old snapshot may be used for limited display, but must not override current revocation, suspension, replacement, or authorization-domain restrictions.

Resource version and propagation state can be viewed along two dimensions:

{\def\LTcaptype{none} 
\begin{longtable}[]{@{}>{\raggedright\arraybackslash}p{0.1920\linewidth}>{\raggedright\arraybackslash}p{0.1920\linewidth}>{\raggedright\arraybackslash}p{0.1920\linewidth}>{\raggedright\arraybackslash}p{0.1920\linewidth}>{\raggedright\arraybackslash}p{0.1920\linewidth}@{}}
\toprule\noalign{}
Resource version/lifecycle & Registration service node & Root platform and content distribution platform & Discovery service node & Meaning for the caller \\
\midrule\noalign{}
\endhead
\bottomrule\noalign{}
\endlastfoot
Draft & Not accepted & No publication fact & Should not enter a trusted index & May only be edited or previewed locally \\
Registered & Registration record or registration credential exists & May not yet be published & May be invisible & Does not establish network-wide publication \\
Published & Registration evidence is retained & Root platform trusted publication proof and distribution materials exist & May still be synchronizing & The caller can wait for discovery or directly verify the materials within its authorization domain \\
Indexed & Registration evidence is retained & Publication materials are verifiable & Local validation and indexing are complete & The caller must still check current governance state and invocation authorization \\
Suspended/revoked & The state record must be retained & Further propagation or use is restricted & Query results should be updated or marked stale & An old snapshot must not continue to be treated as current and usable \\
\end{longtable}
}

\subsection{Registration State}\label{subsec:04.11.1}

A local draft means that the resource materials have not been accepted by a registration service node; submitted means that the registration service node has received and processed the request; and registered means that the registration service node has formed a registration record or registration credential. These states do not directly indicate that the root platform has published the resource or that a discovery service node has indexed it. The resource controller should preserve the local DID document, keys, and submission materials so that they can be reverified after a failure or during an update.

\subsection{Publication and Indexing State}\label{subsec:04.11.2}

Publication state distinguishes at least Root admission, persistence confirmation by the content distribution platform, validated receipt by a particular Discovery node, and completion of indexing at that node. Success at one stage provides input to the next but does not replace its completion condition. With multiple discovery targets, state is recorded for each target; resource counts received by different targets must not be added before comparison with the Root total.

Root-platform published means that the root platform has accepted a particular resource version and formed a publication proof; content distributed means that the resource package has entered the serving scope of the content distribution platform; and indexed by a discovery service node means that a particular discovery service node has verified the resource package and written it to its local read model. These three states progress asynchronously and may be temporarily inconsistent, so they should be expressed through cursors, state, and freshness.

\subsection{Suspension, Revocation, and Recovery State}\label{subsec:04.11.3}

Suspension means that a resource or its version may temporarily no longer be propagated or invoked as \passthrough{\lstinline!active!} within a specified scope; revocation means that the corresponding resource, credential, publication fact, or node authorization is no longer valid; and recovery means that a new governance or publication decision has again permitted the corresponding action. A state change should be associated with the version, resource DID, governance fact, or publication record, rather than merely changing web display text. A discovery service node should update its index and query results after the state has propagated.

\subsection{Publicly Visible and Trustfully Discoverable State}\label{subsec:04.11.4}

Publicly visible means that a portal or node can return resource information to a particular query party; trustfully discoverable means that the discovery service node has verified the resource identity, publication proof, hashes, lifecycle, and authorization domain and permits the resource to be returned within its authorized scope. A resource may be displayed publicly without being trustfully discoverable, or may have been published by the root platform but be invisible to a particular authorization domain. The caller must still perform its own verification before connecting.

Lifecycle state, DID/ResourcePackage version, asynchronous propagation state, and current visibility are separate dimensions. For example, a resource may be \passthrough{\lstinline!active!} while its new version is still moving through content distribution or synchronization to one discovery target. Lifecycle alone therefore cannot establish that every Discovery node returns that version.

\section{Resource Ownership, Control, and Delegation Relationships}\label{sec:04.12}

At the protocol layer, OAN directly enforces controller authorization of DID Document updates. ``Ownership,'' ``operational responsibility,'' and ``service provider'' may be recorded as business or verification relationships. Any registered OAN resource may become the controller of another resource; it need not have a dedicated controller \passthrough{\lstinline!resourceType!}. Dedicated controller resources mainly express stable responsible parties such as legal persons, organizations, or individuals. When a relationship changes, the controller in the authoritative DID Document before the update determines who may submit the next version.

Resource ownership is a rights relationship at the business or organizational layer; DID control is a cryptographic relationship in which keys and signatures prove the ability to represent a resource DID and perform prescribed operations; operational management is responsibility for an endpoint, content, or runtime environment; publication delegation is the qualification for a registration service node or another authorized participant to submit or process materials on behalf of someone else; and invocation authorization is a business permission jointly determined by the resource endpoint and the caller\textquotesingle s policy. These roles may be held by the same subject or by different subjects.

Resource metadata may express some relationships through \passthrough{\lstinline!controllerDid!}, \passthrough{\lstinline!publisherDid!}, \passthrough{\lstinline!subjectDid!}, \passthrough{\lstinline!implementationLinks!}, or credential requirements, but the fields themselves are not sufficient proof that the relationships are genuine. The verifier should separately check the resource DID control proof, credentials of the publisher or registration service node, the root platform publication proof, governance state, and business endpoint authentication. It must not infer other permissions merely because a party can upload a file, provide an address, or issue a registration result.

A controller relationship is not a dependency, invocation, containment, or provision relationship between resources. Any resource may control another resource, but only the current authoritative controller may authorize that resource\textquotesingle s next version. The resource provider and controller are responsible for updating the registered description when runtime content changes. OAN verifies update authority and material provenance; it does not continuously prove the truth of content returned by a business endpoint.

\section{Resource Dependencies and Composition Integrity}\label{sec:04.13}

Dependency integrity requires every resource in a composition to be independently identifiable, locatable, verifiable, and usable according to its version. The minimum check set includes the target resource DID, resource type, relationship type, target version or package hash, consistency between the DID document and metadata, the root platform\textquotesingle s trusted publication proof, lifecycle state, authorization domain, and necessary protocol compatibility conditions. When a discovery service node returns only an index summary, the caller should obtain the complete resource package or verifiable proof before establishing a high-risk composition.

When a discovery node is behind, the target resource is temporarily unavailable, or proof is temporarily missing, the composition may be marked as pending synchronization or unverifiable. When the target resource has version drift, is suspended, revoked, outside the authorization domain, or has an inconsistent hash, the composition should not continue to be used unless a clearly constraint-satisfying alternative version exists. Composition integrity is a pre-invocation verification condition, not an incidental result of registration or search ranking.

The four resource types share common information such as resource DID, resource type, capability description, version, endpoint, protocol binding, and lifecycle, but differ in content form and invocation method. The goal of the resource model is to provide a common registration, publication, and discovery framework, not to force the four resource types into the same runtime object.

{\def\LTcaptype{none} 
\begin{longtable}[]{@{}>{\raggedright\arraybackslash}p{0.2400\linewidth}>{\raggedright\arraybackslash}p{0.2400\linewidth}>{\raggedright\arraybackslash}p{0.2400\linewidth}>{\raggedright\arraybackslash}p{0.2400\linewidth}@{}}
\toprule\noalign{}
Resource type & Main content carried & Common discovery basis & Key checks before invocation or use \\
\midrule\noalign{}
\endhead
\bottomrule\noalign{}
\endlastfoot
Agent service & Agent capabilities for tasks or sessions & Capability tags, use cases, interaction protocols, and service endpoints & Service identity, protocol, authentication, and business permissions \\
Skill package & Reusable instructions, workflows, or capability components & Capability description, applicable scenarios, dependencies, and version & Package hash, provenance, compatibility, and execution boundary \\
MCP service & MCP-oriented tool collection or service entry point & MCP protocol binding, tool capabilities, and endpoint & MCP version, server identity, authentication, and tool declaration \\
Tool API & Directly callable interface capability & API capability, input/output schema, and endpoint & API version, schema, authentication, and invocation authorization \\
\end{longtable}
}

\pandocbounded{\includegraphics[width=\linewidth,height=0.72\textheight,keepaspectratio]{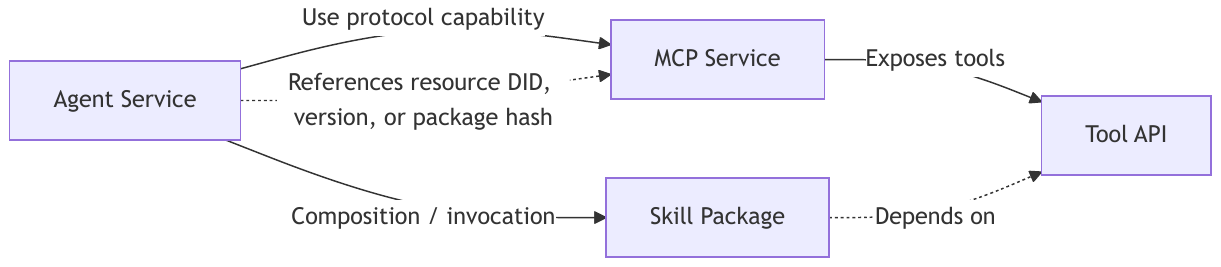}}

Resource relationships must explicitly state the target DID, relationship type, and version or package hash. A name, search relevance score, or reachable endpoint alone cannot prove dependency integrity. When a referenced resource is suspended, revoked, has version drift, or falls outside the authorization domain, the composition party should refuse further use or move it to a pending-verification state.

\begin{quote}
\textbf{Boundary of the resource model:} OAN is responsible for the identity, description, publication, discovery, and verification associations of resources. Model inference, business workflows, tool execution, and final business authorization inside a resource remain the responsibility of the corresponding resource and the caller\textquotesingle s policy.
\end{quote}

The four resource types differ after discovery: an Agent Service normally points to an interactive business endpoint; a Skill can point to a package, instructions, or a loading entry point; an MCP Server represents an MCP protocol service; and a Tool API represents an invocable interface and its constraints. OAN stores and verifies identity, description, provenance, and version materials; it does not host or continuously prove the runtime content. The latest mixed-resource fixtures verify that all four object types share the main registration, publication, and discovery flow, but synthetic benchmark resources do not replace evaluation of real product-description quality.

Stable controller, resource-type, endpoint, tag, use-case, and implementation-link fields can form a resource-product projection. Relationships such as dependency, invokes, and provided-by become interoperable only after they enter the public schema. Any resource can act as a controller, but a control relationship does not create an invocation, containment, dependency, or provision relationship.

\section{References}\label{sec:04.14}

{\def\LTcaptype{none} 
\begin{longtable}[]{@{}>{\raggedright\arraybackslash}p{0.3200\linewidth}>{\raggedright\arraybackslash}p{0.3200\linewidth}>{\raggedright\arraybackslash}p{0.3200\linewidth}@{}}
\toprule\noalign{}
Source & Type & Link \\
\midrule\noalign{}
\endhead
\bottomrule\noalign{}
\endlastfoot
\passthrough{\lstinline!oan-protocol-commons!} & Code repository: resource types, core domain objects, and resource packages & \url{https://github.com/wolfbrother/oan-protocol-commons} \\
\passthrough{\lstinline!oan-sdk-ts!} & Code repository: resource DIDs, resource packages, and client-side material construction & \url{https://github.com/OpenAgenet/oan-sdk-ts} \\
OAN White Paper & arXiv white paper: resource ecosystem and interoperability positioning & \url{https://arxiv.org/abs/2606.03161} \\
OAN Yellow Paper & arXiv yellow paper: resource model and trusted publication mechanism & \url{https://arxiv.org/abs/2606.03163} \\
\passthrough{\lstinline!did:oan!} DID Method Specification & Standard/method specification: resource identity expression & \url{https://github.com/OpenAgenet/oan-public-docs/blob/main/did-oan-specs/doc/OAN\%20DID\%20Method\%20Specification.md} \\
\end{longtable}
}

\chapter{\texorpdfstring{\texttt{did:oan} Method and Resource Identity}{did:oan Method and Resource Identity}}\label{chap:05}

This chapter describes the current rules for \passthrough{\lstinline!did:oan!} as the resource identity method of OpenAgenet (OAN). The method layer answers whether an identifier is valid, how it is associated with a DID Document, and who can provide valid proof for a particular operation. It does not answer whether a resource has been registered, accepted by the Root, or indexed by Discovery. Those states refer to evidence produced by their respective components.

Use of \passthrough{\lstinline!did:oan!} requires a parseable identifier, a structurally valid DID Document, a verifiable control proof, and cross-consistent resource fields. The DID Document carries resource type, capability tags, descriptions, use cases, protocol bindings, and service references through \passthrough{\lstinline!oanMetadata!}, enabling Discovery to build indexes from identity-bound material. A ResourcePackage then fixes those declarations to a particular version through digests and proof. Semantic fields remain provider declarations and are not proof of runtime performance merely because they appear in a DID Document.

\section{Method Purpose and Identifier Scope}\label{sec:05.1}

\passthrough{\lstinline!did:oan!} can identify Agent Services, Skills, MCP Servers, Tool APIs, subjects, organizations, and infrastructure nodes. The current method-specific identifier does not encode object type. Whether an object is a resource, subject, or node is determined from its DID Document, resource type, node credentials, and usage context. Resource, subject, Registrar, Discovery, and Root DIDs serve different purposes and must not be conflated merely because they use the same DID method.

Uniqueness first means string uniqueness within the method namespace. Random generation uses cryptographically secure random values; deterministic derivation uses the routing code, control material, and a nonce. A resource DID does not change when it propagates to different Discovery nodes. Historical versions of the same DID are associated through versions, digests, publication sequences, and historical DID Documents rather than by generating a new DID for each update.

In the OAN network topology, the entry node and the identity object are two different relationships. A client may submit a resource through one Registrar Service Node and query the same resource from multiple Discovery Service Nodes, but the DIDs of those service nodes identify the nodes themselves. The resource DID is still generated or held by the resource controller. A resolver should obtain the corresponding DID Document from the resource DID instead of embedding the submission entry point into the DID.

{\def\LTcaptype{none} 
\begin{longtable}[]{@{}>{\raggedright\arraybackslash}p{0.2400\linewidth}>{\raggedright\arraybackslash}p{0.2400\linewidth}>{\raggedright\arraybackslash}p{0.2400\linewidth}>{\raggedright\arraybackslash}p{0.2400\linewidth}@{}}
\toprule\noalign{}
Object & DID subject & Primary purpose & Object it must not replace \\
\midrule\noalign{}
\endhead
\bottomrule\noalign{}
\endlastfoot
Subject DID & Organization, developer, or controlling subject & Express the control relationship and signing context & Specific resource DID \\
Resource DID & Agent Service, Skill, MCP Server, or Tool API & Express the registration, version, and control boundary of a resource & Subject DID or node DID \\
Registrar Service Node DID & Registration infrastructure node & Express the identity of a registration entry point and node & Resource control \\
Discovery Service Node DID & Discovery infrastructure node & Express the identity of an indexing and query entry point & Resource publication fact \\
Root platform DID & Root platform infrastructure identity & Express the source of Root-related proofs and publication & Resource ownership \\
\end{longtable}
}

\section{DID Syntax and Method-Specific Identifier}\label{sec:05.2}

The current method-specific identifier has the form \passthrough{\lstinline!did:oan:<routing-code>:<suffix-code>!}. The routing code is five case-sensitive Base58 characters and the suffix code is 32 Base58 characters. The alphabet excludes \passthrough{\lstinline!0!}, \passthrough{\lstinline!O!}, \passthrough{\lstinline!I!}, and \passthrough{\lstinline!l!}. The complete DID is 46 characters. A parser validates the method prefix, four-part structure, routing code, suffix code, and complete regular expression separately. The routing code selects a Registrar route; it is not a resource-type or application-domain code.

The public cross-language fixture \passthrough{\lstinline!did:oan:K7mQ9:7YpQm9Kx2VnRb6Ts3WfHa4Cd5Ej8LgNz!} is accepted by the current parser. Changing routing-code case creates a different string; the parser does not uppercase it. Invalid lengths, characters, or segment counts are rejected. The legacy four-character form \passthrough{\lstinline!did:oan:SKFI:...!} returns \passthrough{\lstinline!InvalidRoutingCode!}.

The method library separates a parsed value into the complete value, routing code, and suffix code. The following describes parser behavior; it is not an HTTP response format:

\begin{lstlisting}
{
  "input": "did:oan:K7mQ9:7YpQm9Kx2VnRb6Ts3WfHa4Cd5Ej8LgNz",
  "routingCode": "K7mQ9",
  "suffixCode": "7YpQm9Kx2VnRb6Ts3WfHa4Cd5Ej8LgNz",
  "routingCodeLength": 5,
  "suffixLength": 32,
  "parseResult": "valid"
}
\end{lstlisting}

\passthrough{\lstinline!parseResult!} is explanatory text in this manual. The parser performs no case normalization. Any segment, length, or character error fails directly and cannot be repaired into another DID.

\section{Relationship Between Resource DID and Subject/Controller}\label{sec:05.3}

The routing code serves routing only. It neither identifies an object category nor proves a legal or business relationship among subject, controller, provider, and operator. Resource control is expressed by the DID Document\textquotesingle s top-level \passthrough{\lstinline!controller!}, verification methods, \passthrough{\lstinline!oanMetadata.controllerDid!}, and proof for the current operation. \passthrough{\lstinline!publisherDid!} and \passthrough{\lstinline!issuerDid!} express publication or issuance relationships and do not replace the controller.

A subject may control multiple resource DIDs, and the resource provider may differ from the controller. The DID of a Registrar Service Node represents infrastructure qualification and cannot replace the resource DID. When an update uses the same resource DID, it must include an update proof that can be verified against the current control relationship. The DID string itself does not carry ownership, and public resolution of a DID does not transfer control.

\subsection{Subject DID}\label{subsec:05.3.1}

A subject DID identifies the subject that controls resources or represents an organization or developer. It establishes the context of the control relationship and signing subject. It is not the DID of a specific resource and is not equivalent to a user login account. The DID Document of a subject DID should expose the public keys and service information required for verification. The private key is kept by the subject in the Local Identity.

\subsection{Resource DID}\label{subsec:05.3.2}

A resource DID establishes an identity boundary for a resource that can be registered, published, discovered, updated, and revoked independently. Each of the four product resource types can have its own resource DID. The resource type must agree across the registration submission, DID Document \passthrough{\lstinline!oanMetadata.resourceType!}, and ResourcePackage metadata. The routing code must be syntactically valid and agree with the accepting Registrar\textquotesingle s route; it does not encode resource type.

\subsection{Control and Signature Relationships}\label{subsec:05.3.3}

A control relationship answers ``who may produce the next DID Document,'' while a signature relationship answers ``which key signed this particular material.'' In initial self-controlled registration, both point to the resource itself. Initial registration under an external controller additionally requires authorization by the controller declared in the new document. During an update, verification obtains the old controller from the Root\textquotesingle s current authoritative document rather than trusting only the new document. If A changes the controller to B, A signs that update; B becomes the authorizer only after the update succeeds.

Any resource with a resolvable DID Document and valid control relationship may be declared as another resource\textquotesingle s controller; its resource type need not be \passthrough{\lstinline!controller!}. A dedicated controller resource is convenient for centralized organizational management but gains no Root, Registrar, or Discovery privileges from that role. The arrows below mean only who may authorize the target DID\textquotesingle s next document and related registration operation; they do not mean that a controller holds the business-signing keys of subordinate resources.

\pandocbounded{\includegraphics[width=\linewidth,height=0.72\textheight,keepaspectratio]{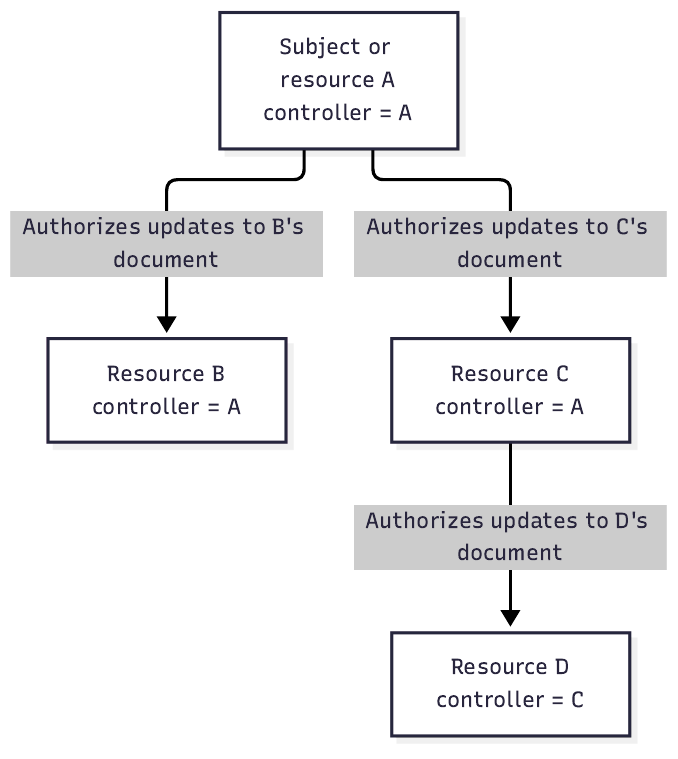}}

Self-control means only that the resource currently authorizes its own updates; external control means that the authorizer is another DID. Regardless of depth, each update is verified only against the direct controller in the target resource\textquotesingle s current authoritative document. Permission is not inferred by traversing the entire relationship tree, and proof is not skipped merely because upper-level subjects share a name.

The control relationship is jointly supported by the verification methods in the DID Document and verifiable control proofs for specific operations. A verifier should check the signer, verification method, target DID, purpose, challenge, time, and DID Document hash in the proof. It must not ignore purpose, target, or freshness checks merely because the proof\textquotesingle s public key is identical to a public key in the document.

\subsection{Separation of Subject, Resource, and Provider}\label{subsec:05.3.4}

The subject controls the identity and resource operations, the resource provider is responsible for capabilities, content, and service endpoints, and the operator is responsible for the runtime environment and availability. These roles may belong to one subject or may be separate. \passthrough{\lstinline!publisherDid!} or \passthrough{\lstinline!subjectDid!} is only a relationship declaration; the truth of the relationship still requires corresponding credentials, delegation materials, and control proofs.

\section{DID Document Structure}\label{sec:05.4}

An OAN DID Document is based on the DID Core data model and must at least contain \passthrough{\lstinline!@context!}, \passthrough{\lstinline!id!}, \passthrough{\lstinline!verificationMethod!}, \passthrough{\lstinline!authentication!}, \passthrough{\lstinline!assertionMethod!}, \passthrough{\lstinline!service!}, and the OAN-resource-oriented \passthrough{\lstinline!oanMetadata!}. \passthrough{\lstinline!id!} must be the normalized \passthrough{\lstinline!did:oan!}. The controller of each verification method, service-endpoint references, and resource identity in the metadata must be mutually verifiable. For Agent Services, Skills, MCP Servers, and Tool APIs, \passthrough{\lstinline!oanMetadata!} is the structured carrier for the resource type, capability semantics, protocol relationships, and publication references.

\passthrough{\lstinline!oanMetadata.resourceDescription!} and \passthrough{\lstinline!oanMetadata.agentDescription!} can express capability descriptions, capability tags, and use cases. A resource description can also carry input and output schemas, examples, audiences, domains, languages, and versions. Top-level \passthrough{\lstinline!authorizedDomains!}, \passthrough{\lstinline!protocolBindings!}, \passthrough{\lstinline!implementationLinks!}, \passthrough{\lstinline!credentialRequirements!}, and \passthrough{\lstinline!packageInfo!} supplement visibility scope, interaction protocols, composition relationships, invocation credential requirements, and publication materials respectively. A Discovery Service Node can use these fields to build tag, lexical, and vector indexes that map natural-language tasks to candidate resources. After obtaining a candidate, a caller can continue checking the DID control relationship, resource version, and publication proof.

The following is a minimal DID Document that illustrates the relationships among fields. The public key, service address, and resource DID in the example are placeholder values. An actual document must use a real public key and proof materials matching the controller:

\begin{lstlisting}
{
  "@context": [
    "https://www.w3.org/ns/did/v1",
    "https://w3id.org/oan/v1"
  ],
  "id": "did:oan:K7mQ9:7YpQm9Kx2VnRb6Ts3WfHa4Cd5Ej8LgNz",
  "verificationMethod": [
    {
      "id": "did:oan:K7mQ9:7YpQm9Kx2VnRb6Ts3WfHa4Cd5Ej8LgNz#key-1",
      "type": "Ed25519VerificationKey2020",
      "controller": "did:oan:K7mQ9:7YpQm9Kx2VnRb6Ts3WfHa4Cd5Ej8LgNz",
      "cryptoSuite": "Ed25519Sha256",
      "publicKeyMultibase": "zExamplePublicKey"
    }
  ],
  "authentication": [
    "did:oan:K7mQ9:7YpQm9Kx2VnRb6Ts3WfHa4Cd5Ej8LgNz#key-1"
  ],
  "assertionMethod": [
    "did:oan:K7mQ9:7YpQm9Kx2VnRb6Ts3WfHa4Cd5Ej8LgNz#key-1"
  ],
  "service": [],
  "oanMetadata": {
    "subjectType": "skill",
    "resourceType": "skill",
    "controllerDid": "did:oan:DVFI:7YpQm9Kx2VnRb6Ts3WfHa4Cd5Ej8LgNz",
    "capabilityTags": ["code-repository", "project-structure-summary"],
    "resourceDescription": {
      "name": "Repository structure analyzer",
      "description": "Searches code repositories and summarizes project structure.",
      "capabilityDescription": "Locate repository content, identify modules and summarize their relationships.",
      "capabilityTags": ["code-repository", "project-structure-summary"],
      "useCaseExamples": [
        "Inspect an unfamiliar repository before development",
        "Summarize modules, entry points and dependencies"
      ]
    },
    "lifecycleState": "active"
  }
}
\end{lstlisting}

A complete document also contains verifiable verification methods and purpose references. A verifier checks the relationships among \passthrough{\lstinline!id!}, the top-level controller, verification-method controllers, authentication and assertion references, and checks resource type against the registration submission and ResourcePackage. Capability tags, descriptions, and use cases refer to the same resource capability. The controller proves a document containing those declarations, and \passthrough{\lstinline!didDocumentHash!} fixes that document version.

\subsection{DID Identifier and Method Metadata}\label{subsec:05.4.1}

\passthrough{\lstinline!id!} is the subject identifier of the DID Document and exactly matches the resource DID in the request or ResourcePackage. \passthrough{\lstinline!@context!} contains at least the DID Core context. The parser obtains the routing code and suffix code from the DID; resource category is expressed by \passthrough{\lstinline!oanMetadata!} and surrounding protocol objects and cannot be inferred from the routing code.

\subsection{verificationMethod}\label{subsec:05.4.2}

\passthrough{\lstinline!verificationMethod!} describes public-key methods that can be used to verify signatures. Each item must contain at least a method ID, type, controller, and one public-key representation. OAN core types support representations such as \passthrough{\lstinline!publicKeyMultibase!} and \passthrough{\lstinline!publicKeyJwk!}, and can use \passthrough{\lstinline!cryptoSuite!} and \passthrough{\lstinline!publicKeyFormat!} to specify the cryptographic suite and encoding. A private key must never appear in a DID Document.

\subsection{authentication and assertionMethod}\label{subsec:05.4.3}

\passthrough{\lstinline!authentication!} identifies verification methods that can be used for identity authentication or request proofs. \passthrough{\lstinline!assertionMethod!} identifies verification methods that can be used to issue or verify assertions. Each reference must resolve to a method ID in the same DID Document. A key used only for assertions must not automatically be treated as an authentication key, and a public key present in \passthrough{\lstinline!verificationMethod!} must not automatically be considered valid for every purpose.

\subsection{service Endpoint}\label{subsec:05.4.4}

Each service item in \passthrough{\lstinline!service!} should contain a unique service ID, service type, and \passthrough{\lstinline!serviceEndpoint!}. It may also specify a version, protocol, server type, and port. An endpoint may point to a resource business service, resolution service, ResourcePackage, or another protocol description. The fact that an endpoint can be resolved proves only that an address exists; it does not prove that the business is available or that the caller is authorized.

\subsection{Resource Metadata and Extension Fields}\label{subsec:05.4.5}

\passthrough{\lstinline!oanMetadata!} carries OAN resource type, subject type, controller, publisher, capabilities, authorization domains, protocol bindings, implementation links, credential requirements, package information, lifecycle, and other semantics. Extension fields must not override the identity and verification meanings of DID Core. If the DID, type, version, authorization domain, or hash conflicts among resource metadata, the ResourcePackage, and the registration submission, the material should be rejected or isolated according to the conflict-handling rules.

The relationship between \passthrough{\lstinline!oanMetadata!} and DID Core fields can be understood as follows:

{\def\LTcaptype{none} 
\begin{longtable}[]{@{}>{\raggedright\arraybackslash}p{0.3200\linewidth}>{\raggedright\arraybackslash}p{0.3200\linewidth}>{\raggedright\arraybackslash}p{0.3200\linewidth}@{}}
\toprule\noalign{}
Field area & Main question answered & Verification focus \\
\midrule\noalign{}
\endhead
\bottomrule\noalign{}
\endlastfoot
\passthrough{\lstinline!id!}, \passthrough{\lstinline!verificationMethod!} & Which DID is this, and which public key verifies it? & DID method syntax, method ID, controller, and public-key format \\
\passthrough{\lstinline!authentication!}, \passthrough{\lstinline!assertionMethod!} & Which methods can be used for authentication or assertion? & Whether references exist and purposes match \\
\passthrough{\lstinline!service!} & Where can the service or protocol entry point be obtained? & Service ID, protocol, version, and endpoint format \\
\passthrough{\lstinline!oanMetadata!} & What is the resource, and what capabilities and status does it have? & Resource type, subject type, authorization domain, version, and package binding \\
\end{longtable}
}

Therefore, a parser may preserve unknown OAN extension fields, but it must not allow extension fields to change the meaning of \passthrough{\lstinline!id!}, verification methods, or purpose references. After successful method resolution, a resource-network verifier should continue checking consistency between metadata and the ResourcePackage.

\section{Verification Methods and Public Keys}\label{sec:05.5}

The \passthrough{\lstinline!id!} of a verification method should belong to the corresponding DID, and its \passthrough{\lstinline!controller!} should match the control relationship. The public-key representation and cryptographic suite must be parsable by the verifier. A control signature should specify \passthrough{\lstinline!proofPurpose!}, \passthrough{\lstinline!verificationMethod!}, the target object, and the signature coverage. The verifier first resolves the public key from the DID Document, and then checks the signature, purpose, hash, and time conditions.

Multiple verification methods may separately serve authentication, assertion, update, or recovery purposes. Key rotation must be authorized by the currently valid control relationship and must preserve the state required for historical verification or old-key invalidation. An old key must not continue issuing valid new updates. Merely writing an attacker\textquotesingle s public key into a DID Document that has not been accepted by the current control relationship does not grant control of the DID.

The acceptability of a verification method is not the single condition that ``a public key exists.'' It is the joint result of multiple fields:

The verifier should first confirm that \passthrough{\lstinline!verificationMethod!} can be resolved from the target DID Document, and then confirm that its controller, purpose, target DID, and signed document hash match. If the public key belongs to another DID, the method is not referenced for the applicable purpose, or the document has been replaced, verification fails. Comparing only whether public-key bytes are identical is insufficient.

\section{Service Endpoints and Discovery References}\label{sec:05.6}

Service endpoints in a DID Document provide location information for resources or protocols. Typical uses include business services, Registrar Services, Discovery Services, DID resolution, and ResourcePackage retrieval. The \passthrough{\lstinline!id!} of a service item should be associable with the resource DID, and \passthrough{\lstinline!serviceEndpoint!} should conform to the declared address or object format. A discovery response may reference an endpoint, but the caller must still verify the Root publication fact, endpoint freshness, and business authentication requirements.

The interfaces of Registrar Service Nodes, Discovery Service Nodes, and resolvers describe network-component behavior. They do not change their authorization boundaries merely because an endpoint is written into a DID Document. If an endpoint is missing, its protocol is incompatible, its address is unreachable, or the referenced object is incompatible, resolution may still successfully return the DID Document. Before discovery or invocation, however, the verifier must clearly mark the corresponding problem.

Resolution results should distinguish ``the document was retrieved successfully'' from ``the endpoint is available.'' A client-oriented result summary may contain:

\passthrough{\lstinline!availability: not\_checked!} means that the resolution phase did not assess business-endpoint availability. A resolver must not rewrite the existence of an HTTP address or successful DID Document retrieval into endpoint health, published-resource status, or invocation authorization.

\section{Resource Metadata Binding}\label{sec:05.7}

The resource DID Document is associated with resource metadata and the ResourcePackage through \passthrough{\lstinline!oanMetadata!} and package information. For the same resource version, the resource DID, resource type, capability tags, capability description, use cases, authorization domains, protocol bindings, version, and hashes should remain verifiably consistent across the DID Document, registration submission, and ResourcePackage. The ResourcePackage\textquotesingle s \passthrough{\lstinline!didDocumentHash!} binds the exact document content; binding only the DID string is insufficient. Semantic discovery can therefore refer simultaneously to a stable DID, governed semantic declarations, and a published version instead of building an index only over free text separated from identity.

After a resolver returns the DID Document and \passthrough{\lstinline!oanMetadata!}, a Discovery Service Node can extract capability tags, capability descriptions, use cases, input and output schemas, and protocol bindings. It can combine the shared capability vocabulary with a local semantic search engine to build an index. During a query, a natural-language task is used for semantic retrieval, capability tags and resource types for structured filtering, and use cases and capability descriptions for explaining the match. Authorization-domain, lifecycle, and publication-status checks are then applied. This chain connects the semantic extension of the DID method with OAN semantic governance and the discovery protocol.

The binding relationship for one resource version can be checked through the following summary. The three hashes serve different purposes: a resource DID only locates an object and cannot replace integrity checks on the document and content:

If the DID Document changes, \passthrough{\lstinline!didDocumentHash!} should change even when the resource DID remains the same, and a new control proof and resource-network verification should be triggered. If the caller queries only the DID string and does not compare the document hash, it cannot determine which document version it has received.

\section{Control Proofs and DID Document Proofs}\label{sec:05.8}

A resource control proof is generated by the controller using the private key corresponding to a verification method in the DID Document. It proves control over a particular DID, document hash, operation purpose, and challenge. The Registrar Service Node verifies this proof, and the Root platform verifies it again before accepting a resource version. A registration credential proves the fact that registration was processed; a Root proof proves the fact of network publication. These three kinds of evidence cannot replace one another.

A DID Document proof should cover the normalized document content or a prescribed document hash, and should include a locatable verification method, creation time, proof purpose, and signature value. Verification must fail if the proof is absent or expired, the target DID does not match, the document hash does not match, the purpose does not match, or the signature cannot be verified. No published state may be produced from such a failure.

The verification input for an OAN control challenge should bind the request context to the signature instead of verifying only a context-free signature. The key checks of a security library can be abstracted as follows:

The Registrar Service Node should compare the subject DID, target Registrar Service Node, purpose, document hash, and validity period in the challenge with the current request item by item. Passing the challenge proves only that the controller authorized this particular operation with its associated context. It does not mean that the Registrar Service Node has formed a registration record or that the Root platform has published the resource.

\section{DID Generation and Local Key Handling}\label{sec:05.9}

The client is the primary generation and custody boundary for DIDs and control keys. A random DID uses cryptographically secure random values, while deterministic derivation uses a routing code, control material, and nonce. The result must pass method parsing, while resource-type consistency is checked across protocol objects. A Registrar receives only the public DID Document, signatures, and proofs needed for verification, never subject or resource private keys.

A Local Identity backup is sensitive material for restoring a control relationship, not a public credential. During import, the DID, keys, public keys, control relationships, and file structure must be checked for consistency. A newly generated control proof after restoration must be rebound to the current operation and document hash; a historical signature cannot be treated as unconditional authorization for every future version.

When handling Local Identity materials, the client can reduce key-leakage and binding errors through the following sequence:

\begin{enumerate}
\def\labelenumi{\arabic{enumi}.}
\tightlist
\item
  Generate or import subject and resource-control keys locally, and check that public and private keys match.
\item
  Generate or resolve \passthrough{\lstinline!did:oan!} with the selected Registrar routing code, and validate the five-character routing code and suffix.
\item
  Construct the DID Document and confirm that verification methods, purpose references, and \passthrough{\lstinline!oanMetadata!} point to the same resource DID.
\item
  Generate a new challenge control proof for the current document and operation instead of reusing a signature bound to an old document.
\item
  Submit only the public DID Document, proof, and necessary resource metadata to the Registrar Service Node; never upload private keys.
\end{enumerate}

A Local Identity backup downloaded by a browser is user-side recovery material and cannot be replaced by storing keys on the server. A page indicating ``created'' or ``download complete'' only means that the client completed the file operation. The user must still verify that the file is complete and keep it according to local security requirements.

\subsection{DID Generation}\label{subsec:05.9.1}

The client can call a method library\textquotesingle s random-generation or deterministic-derivation interface. The routing code in generation parameters must be valid. Deterministic derivation is stable only for identical routing code, control material, nonce, and method version. It is not a mechanism for recovering a private key from a DID.

\subsection{Key Generation and Storage}\label{subsec:05.9.2}

Control keys should be generated or imported in a client environment controlled by the user, with private keys stored through local secure storage or encrypted backup. DID Documents, registration requests, and discovery responses expose only public keys and materials required for verification. Server logs, request bodies, and error responses must not contain private keys or plaintext materials from which private keys can be recovered.

\subsection{Local Identity Backup}\label{subsec:05.9.3}

A Local Identity backup may contain a subject DID, resource DIDs, key materials, resource configuration, and necessary credentials. It should clearly state its sensitivity level, creation time, and recovery purpose. The backup filename, download location, and page text do not provide identity authentication. The user must independently protect the backup and prevent unauthorized copying.

\subsection{Import, Recovery, and Key-Use Boundaries}\label{subsec:05.9.4}

After import or recovery, the client should confirm that public and private keys correspond, that the DID Document subject is consistent, and that the resource type is consistent. It should regenerate the current challenge and hash before signing. A Registrar Service Node, Root platform, or Discovery Service Node may verify public proofs, but should not obtain private keys from the import process. Node-runtime private keys must also be separated from user resource-control private keys.

\section{DID Resolution Requirements}\label{sec:05.10}

A resolution request first validates the DID method and syntax, and then uses a resolution service to obtain the currently available DID Document or an explicitly requested historical version. A successful response should contain the document and its verifiable metadata. ``Not found,'' unsupported method, invalid format, inactive resource, and unavailable historical version should use distinguishable errors or statuses instead of returning an empty document and making the caller guess the cause.

Successful resolution means only that the resolver can return a DID Document. It does not mean that the resource is registered, the Root platform has published it, the credential remains valid, or the business endpoint is available. The return type of the current resource-detail or DID Document HTTP interface is determined by the specific node. The content distribution platform\textquotesingle s \passthrough{\lstinline!GET /cdn/documents/\{did\}!} directly returns \passthrough{\lstinline!DidDocument!}, while some directory interfaces place the DID in a \passthrough{\lstinline!did!} field and the document in a \passthrough{\lstinline!document!} field. Clients must parse according to the actual interface contract and must not assume that every DID read interface uses the same wrapper structure. The caller should continue judging based on resource-network evidence, governance status, version, and local policy.

Resolution errors should allow the caller to distinguish input errors, method capability, and network status. The following result categories can be used; the specific HTTP mapping is determined by the resolution-service interface:

{\def\LTcaptype{none} 
\begin{longtable}[]{@{}>{\raggedright\arraybackslash}p{0.3200\linewidth}>{\raggedright\arraybackslash}p{0.3200\linewidth}>{\raggedright\arraybackslash}p{0.3200\linewidth}@{}}
\toprule\noalign{}
Result category & Method-layer judgment & Caller handling \\
\midrule\noalign{}
\endhead
\bottomrule\noalign{}
\endlastfoot
\passthrough{\lstinline!invalid\_did!} & DID prefix, segments, routing code, or suffix is invalid & Correct the input and do not start a business call \\
\passthrough{\lstinline!method\_not\_supported!} & The current resolver does not support \passthrough{\lstinline!did:oan!} & Switch to a resolver that supports the method \\
\passthrough{\lstinline!not\_found!} & The method is valid but no DID Document can be returned & Distinguish not registered from temporarily unreachable \\
\passthrough{\lstinline!resolved!} & The document has been retrieved and passed basic structural checks & Continue with control, publication, and endpoint verification \\
\passthrough{\lstinline!status\_restricted!} & The document or resource status restricts current use & Retain historical evidence and stop sensitive actions \\
\end{longtable}
}

A resolver must not conceal \passthrough{\lstinline!not\_found!} or \passthrough{\lstinline!method\_not\_supported!} by returning an empty object. It must not treat an old snapshot from a Discovery Service Node as the current result of DID method resolution.

Interoperability with external protocols such as A2A and MCP is a matter of consuming and adapting resolution results. It does not change the identifier, DID Document, or control-relationship rules of \passthrough{\lstinline!did:oan!}. A resolver may preserve \passthrough{\lstinline!oanMetadata!}, service endpoints, and protocol bindings for an adapter to map them into an external protocol description. The caller must still independently verify the DID, ResourcePackage, control proof, and Root publication evidence.

\section{DID Update, Version Management, Suspension, Recovery, and Deactivation}\label{sec:05.11}

An update to the same resource normally keeps the resource DID unchanged and expresses the change through a new DID Document version, ResourcePackage version, hash, and control proof. The updater must sign with the currently valid control relationship. The Root platform and relevant nodes should check version order, reference to the previous version, document hash, and resource lifecycle state. The method library itself provides DID resolution and type validation only; it does not replace the network update protocol.

The minimum decision for an update to the same DID can be summarized as follows:

\begin{lstlisting}
current DID + current control proof + new document hash
    -> verify current control and version relation
    -> accept a new resource version or reject the update
\end{lstlisting}

If the DID in an update request differs from the existing record, it should be treated as a new resource or an invalid replacement. The object to be overwritten must not be determined by receipt time. If the DID is the same but the control proof is invalid, the old document must not be overwritten merely because the resource name, endpoint, or submitter is the same.

Suspension, recovery, and deactivation are resource-network state actions. They must be bound to explicit control proofs, governance facts, or authorization conditions. Historical documents may be retained for auditing and historical verification, but current-state resolution must not treat an old document or old publication proof as the latest state.

\subsection{Update with the Same DID}\label{subsec:05.11.1}

An update using the same DID requires the new material\textquotesingle s \passthrough{\lstinline!id!} to equal the existing resource DID and requires the current control key to sign the new document or prescribed document hash. If the control relationship changes, the key-rotation or recovery rules must be satisfied first. Submitting the same DID with only a new public key cannot overwrite the existing document.

\subsection{Version Selection}\label{subsec:05.11.2}

The current version should be determined jointly by the resource version, document hash, package hash, and publication sequence. A resolver or Discovery Service Node should select the latest version according to trusted publication facts and lifecycle status, not solely according to receipt time or an auto-incrementing local database number. Historical-version queries require an explicit version identifier or publication reference.

\subsection{Suspension and Recovery}\label{subsec:05.11.3}

Suspension should prevent discovery, publication, or pre-call verification within the specified scope from treating the resource as active. Recovery requires new valid status evidence and a renewed check of the version and publication evidence. A cached old document may be used for historical display, but it must not bypass the current suspension state.

\subsection{Deactivation and Historical State}\label{subsec:05.11.4}

{\def\LTcaptype{none} 
\begin{longtable}[]{@{}>{\raggedright\arraybackslash}p{0.2400\linewidth}>{\raggedright\arraybackslash}p{0.2400\linewidth}>{\raggedright\arraybackslash}p{0.2400\linewidth}>{\raggedright\arraybackslash}p{0.2400\linewidth}@{}}
\toprule\noalign{}
Situation & Authorizer for this operation & Controller after acceptance & Key rejection condition \\
\midrule\noalign{}
\endhead
\bottomrule\noalign{}
\endlastfoot
Initial self-controlled registration & Resource\textquotesingle s own verification method & Resource itself & Self-signature does not match candidate document \\
Initial external-controller registration & Controller declared by candidate document & That controller & Controller authorization proof missing or invalid \\
Ordinary update to existing DID & Currently registered controller & Unchanged & Signed only by a key in the new document \\
Transfer A to B & Current controller A & B & A did not authorize, or predecessor changed \\
Next update after B takes control & B & Value declared by new version & Still using A\textquotesingle s authorization \\
\end{longtable}
}

Deactivation means that the DID or resource no longer accepts new valid use within the specified scope. Historical documents and proofs may still be retained for auditing. Deactivation, resource revocation, registration-credential revocation, and endpoint shutdown should be recorded separately. One page-status field cannot replace all of these facts.

The basic control path for same-DID updates, suspension, recovery, and deactivation is shown below. The diagram expresses the verification order between the method layer and the resource-network layer; it does not mean that the method library completes registration or publication by itself:

\pandocbounded{\includegraphics[width=\linewidth,height=0.72\textheight,keepaspectratio]{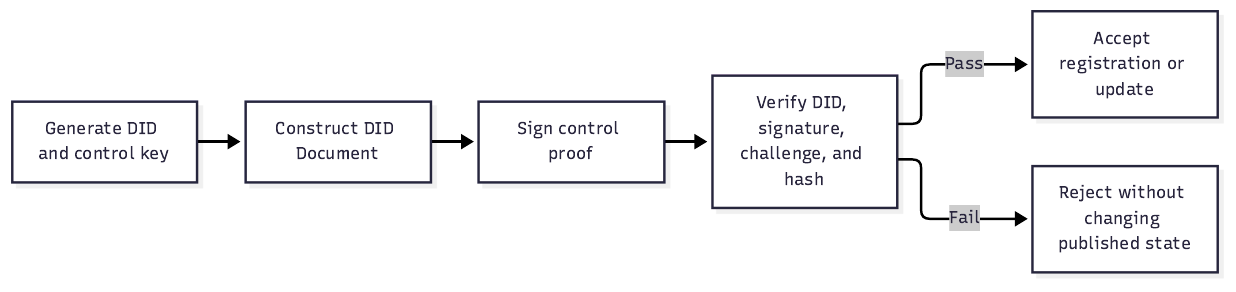}}

\section{DID Method Security and Privacy Considerations}\label{sec:05.12}

Security comes from private-key confidentiality, signature coverage, DID Document public-key binding, challenge, and freshness checks, not from a DID string being hidden or complicated. An attacker can copy a public DID Document and public key, but without a valid control private key, current challenge, and corresponding authorization, the attacker cannot produce an acceptable update proof. A random nonce, request time, document hash, and purpose field should participate in replay protection.

A method implementation should avoid unconditionally exposing subject relationships, resource content, and endpoint privacy. A DID Document should contain only the information required for public verification and discovery. Sensitive credentials, private keys, and internal addresses should be handled through controlled endpoints or external authorization mechanisms. Failed verification must not write a published state, and error logs should retain sufficient audit information without revealing private materials.

\section{DID Method Conformance Requirements and Test Vectors}\label{sec:05.13}

Conformance tests cover valid parsing; rejection of incorrect prefixes, segment counts, routing-code or suffix lengths, and illegal characters; case-sensitive routing; random generation that parses again; and deterministic derivation that is stable for identical inputs and changes with routing code, control material, or nonce. Resource-type consistency is tested across registration, DID Document metadata, and ResourcePackage rather than against characters in the DID.

Signature tests should also cover a verification-method purpose mismatch, target-DID mismatch, document-hash mismatch, expired challenge, reused nonce, wrong public key, and update with an old key. Test evidence should record fixed inputs, method version, expected errors, and actual outputs. Passing method-layer tests does not mean that the registration, publication, or discovery chain has passed.

The method implementation can use the following minimum test vectors as examples of input classification:

{\def\LTcaptype{none} 
\begin{longtable}[]{@{}>{\raggedright\arraybackslash}p{0.3200\linewidth}>{\raggedright\arraybackslash}p{0.3200\linewidth}>{\raggedright\arraybackslash}p{0.3200\linewidth}@{}}
\toprule\noalign{}
Input & Expected result & Basis for rejection or acceptance \\
\midrule\noalign{}
\endhead
\bottomrule\noalign{}
\endlastfoot
\passthrough{\lstinline!did:oan:K7mQ9:7YpQm9Kx2VnRb6Ts3WfHa4Cd5Ej8LgNz!} & Pass & Four-segment structure, five-character Base58 routing code, and 32-character Base58 suffix are valid \\
Same valid DID with routing-code letter case changed & Treat as a different identifier & Routing code is case-sensitive and is not normalized \\
\passthrough{\lstinline!did:oan:K7mQ9:0YpQm9Kx2VnRb6Ts3WfHa4Cd5Ej8LgN!} & Reject & The suffix contains the forbidden Base58 character \passthrough{\lstinline!0!} \\
\passthrough{\lstinline!did:oan:K7mQ:7YpQm9Kx2VnRb6Ts3WfHa4Cd5Ej8LgNz!} & Reject & The routing code does not contain five characters \\
Valid DID + different \passthrough{\lstinline!resourceType!} values in submission and document & Reject registration & Cross-object resource-type validation fails \\
\end{longtable}
}

The DIDs and suffixes in the test vectors are used only to verify the method implementation and must not be used directly as production resource identities. Tests should also record the resolver version, input before and after normalization, error type, and whether any persistent side effect was produced. Returning an error string alone is insufficient to prove that publication state was not modified.

\section{DID Method Interoperability and Resolver Behavior}\label{sec:05.14}

{\def\LTcaptype{none} 
\begin{longtable}[]{@{}>{\raggedright\arraybackslash}p{0.3200\linewidth}>{\raggedright\arraybackslash}p{0.3200\linewidth}>{\raggedright\arraybackslash}p{0.3200\linewidth}@{}}
\toprule\noalign{}
Situation & Authorizer for this operation & Controller after acceptance \\
\midrule\noalign{}
\endhead
\bottomrule\noalign{}
\endlastfoot
Initial self-controlled registration & Resource\textquotesingle s own verification method & Resource itself \\
Initial external-controller registration & Verifiable controller declared by new document & That controller \\
Existing DID update or transfer A to B & Controller in Root current authoritative state & Controller declared by new version \\
\end{longtable}
}

\passthrough{\lstinline!did:oan!} should follow the basic data model of DID Core for DIDs, DID Documents, verification methods, and services, while the method implementation defines its method-specific identifier and resolution behavior. An external resolver should at least recognize the method name and reject identifiers that do not conform to the method syntax. A resolver supporting the OAN method should also preserve \passthrough{\lstinline!oanMetadata!} and its extension fields; it must not lose resource type, capabilities, or publication references during conversion. When consuming public information, an external system must still verify the control relationship, ResourcePackage, and Root publication evidence according to OAN rules.

A \passthrough{\lstinline!did:oan!} identifier consists of the method name and a method-specific identifier. The following example illustrates the structure and field relationships only. The example values do not represent real network objects and must not be treated as control materials that can be published directly.

\begin{lstlisting}
did:oan:<subject-type-or-resource-type-code>:<method-specific-identifier>
\end{lstlisting}

{\def\LTcaptype{none} 
\begin{longtable}[]{@{}>{\raggedright\arraybackslash}p{0.3200\linewidth}>{\raggedright\arraybackslash}p{0.3200\linewidth}>{\raggedright\arraybackslash}p{0.3200\linewidth}@{}}
\toprule\noalign{}
Check object & Question that must be answered & Consequence of failure \\
\midrule\noalign{}
\endhead
\bottomrule\noalign{}
\endlastfoot
DID string & Do the method name, code, characters, and type conform to the rules? & Reject resolution or enter an invalid state \\
DID Document \passthrough{\lstinline!id!} & Does it exactly match the resource DID in the request? & Reject the binding \\
Public key and control relationship & Is the verification method declared by the DID controller and capable of verifying the signature? & Do not accept the control proof or update \\
Signature coverage & Does the signature cover the target DID, document/package hash, challenge, and purpose? & Reject the signature fact \\
Service and metadata & Are the endpoint, protocol, and resource description consistent? & Do not enter trusted discovery or invocation \\
\end{longtable}
}

Method-layer conformance testing should separately record valid resolution, invalid-input rejection, DID-type mapping, document binding, signature purpose, challenge freshness, and update control. Passing method tests proves only that DID-method behavior matches the test vectors. It does not replace business-chain verification by the Registrar Service Node, Root platform, or Discovery Service Node. The specific generation, signing, verification, and rejection paths are described in Sections 5.9, 5.10, 5.11, and 5.13; this section only specifies that cross-resolver consumption must not discard method semantics.

\section{References}\label{sec:05.15}

{\def\LTcaptype{none} 
\begin{longtable}[]{@{}>{\raggedright\arraybackslash}p{0.3200\linewidth}>{\raggedright\arraybackslash}p{0.3200\linewidth}>{\raggedright\arraybackslash}p{0.3200\linewidth}@{}}
\toprule\noalign{}
Source & Type & Link \\
\midrule\noalign{}
\endhead
\bottomrule\noalign{}
\endlastfoot
\passthrough{\lstinline!oan-protocol-commons!} & Code repository: core \passthrough{\lstinline!did:oan!} types and method implementation & \url{https://github.com/wolfbrother/oan-protocol-commons} \\
\passthrough{\lstinline!oan-sdk-ts!} & Code repository: TypeScript DID resolution, generation, and verification & \url{https://github.com/OpenAgenet/oan-sdk-ts} \\
OAN Yellow Paper & arXiv Yellow Paper: identity, control, and resource-publication mechanisms & \url{https://arxiv.org/abs/2606.03163} \\
\passthrough{\lstinline!did:oan!} DID Method Specification & Standard/method specification: \passthrough{\lstinline!did:oan!} identifier format and resolution rules & \url{https://github.com/OpenAgenet/oan-public-docs/blob/main/did-oan-specs/doc/OAN\%20DID\%20Method\%20Specification.md} \\
OAN Resource Identity and Discovery & IETF Internet-Draft: resource identity and discovery & \url{https://datatracker.ietf.org/doc/draft-xu-oan-resource-identity-discovery/} \\
\end{longtable}
}

\chapter{Resource Metadata and Resource Package Types}\label{chap:06}

This chapter defines the data boundaries, field relationships, deterministic hashing, and recipient-side verification methods for resource metadata and \passthrough{\lstinline!ResourcePackage!}. Resource metadata describes what a resource is, what it can do, and how it can be connected. \passthrough{\lstinline!ResourcePackage!} organizes the DID document, metadata, hashes, and root platform trusted publication proof for a specific resource version into a distributable object. A registration submission is the input to the verification process, while a root-platform-verified resource package is the publication product after network-level verification. These three objects are not interchangeable.

This chapter primarily relies on the resource types, serialization implementation, and resource package validation code in \passthrough{\lstinline!oan-protocol-commons!}. The current registration payload schema only provides object-level entry constraints; the semantics of individual fields are jointly determined by the protocol types and validation logic. Where the schema, implementation, and deployment configuration differ, the applicable version and the difference should be stated. A single example must not be treated as mandatory behavior for every implementation.

\section{Resource Metadata Schema}\label{sec:06.1}

In the Rust implementation, resource metadata is represented by \passthrough{\lstinline!ResourceMetadata!}. Its core fields are as follows:

{\def\LTcaptype{none} 
\begin{longtable}[]{@{}>{\raggedright\arraybackslash}p{0.3200\linewidth}>{\raggedright\arraybackslash}p{0.3200\linewidth}>{\raggedright\arraybackslash}p{0.3200\linewidth}@{}}
\toprule\noalign{}
Field & Type & Semantics and validation focus \\
\midrule\noalign{}
\endhead
\bottomrule\noalign{}
\endlastfoot
\passthrough{\lstinline!resourceDid!} & string & The resource DID; it must correspond to the \passthrough{\lstinline!id!} of the \passthrough{\lstinline!ResourcePackage!} and DID document \\
\passthrough{\lstinline!resourceType!}, \passthrough{\lstinline!subjectType!} & enum & The resource and subject types; current resource package validation requires the two values to match \\
\passthrough{\lstinline!publisherDid!}, \passthrough{\lstinline!subjectDid!} & optional string & Declarations of the publisher-subject relationship; they do not independently prove control \\
\passthrough{\lstinline!name!}, \passthrough{\lstinline!description!} & string & The resource name and summary description \\
\passthrough{\lstinline!capabilityTags!} & string array & Capability tags for structured and semantic discovery \\
\passthrough{\lstinline!authorizedDomains!} & string array & Authorized domains in which the resource may be propagated or discovered \\
\passthrough{\lstinline!protocolBindings!} & object array & Protocol, version, transport method, and service or schema references \\
\passthrough{\lstinline!services!} & service endpoint array & The resource\textquotesingle s business entry points or protocol entry points \\
\passthrough{\lstinline!lifecycleState!} & string & The lifecycle state of the current resource version \\
\passthrough{\lstinline!packageVersion!} & string & The current resource package version \\
\passthrough{\lstinline!packageHash!}, \passthrough{\lstinline!metadataHash!} & string & Integrity references for the resource package and metadata \\
\passthrough{\lstinline!hashAlgorithm!} & string & Hash-suite identifier, which must be consistent with the hash references \\
\passthrough{\lstinline!updatedAt!} & UTC time & The update time of the current metadata version \\
\end{longtable}
}

These fields may be exposed for registration, distribution, and discovery, but control private keys, internal access credentials, and operational details that should not be public must not be placed in the metadata. Unknown extension fields may be retained only when they do not change the meaning of core fields, the hash input, or the verification result. A recipient should reject core-field type errors, a missing resource DID, or unsupported enum values.

The following fragment shows the field relationships in a \passthrough{\lstinline!ResourceMetadata!} instance. The hash values are examples and must not be used directly in production materials. During metadata hash computation, \passthrough{\lstinline!packageHash!} and \passthrough{\lstinline!metadataHash!} must first be cleared and then populated with the computed results.

\section{Required and Optional Fields}\label{sec:06.2}

The following fragment is taken from the public cross-language resource fixture. It shows the currently serialized core fields and is not a substitute for the complete \passthrough{\lstinline!ResourceMetadata!} type.

\begin{lstlisting}
{
  "resourceDid": "did:oan:K7mQ9:7YpQm9Kx2VnRb6Ts3WfHa4Cd5Ej8LgNz",
  "resourceType": "skill",
  "version": "1.0.0",
  "lifecycleState": "active",
  "capabilityTags": ["legal.contract.review"],
  "authorizedDomains": ["legal-services"]
}
\end{lstlisting}

The structurally required fields of \passthrough{\lstinline!ResourcePackage!} include \passthrough{\lstinline!packageVersion!}, \passthrough{\lstinline!resourceDid!}, \passthrough{\lstinline!resourceType!}, \passthrough{\lstinline!didDocument!}, \passthrough{\lstinline!didDocumentHash!}, \passthrough{\lstinline!metadataHash!}, \passthrough{\lstinline!packageHash!}, \passthrough{\lstinline!hashAlgorithm!}, \passthrough{\lstinline!metadata!}, \passthrough{\lstinline!rootProof!}, and \passthrough{\lstinline!createdAt!}. In the metadata, the resource DID, resource type, subject type, package version, three key hashes, and hash algorithm are the core fields for forming a verifiable publication object. Capability tags, authorized domains, protocol bindings, and service endpoints may be empty arrays or may be undeclared for a given resource type, but their omission must not be presented as evidence that the resource has the corresponding capability.

The four resource types share common identity, description, version, integrity, and governance fields. An agent service generally needs capability descriptions, use cases, interaction protocols, and service endpoints. A skill package needs package entry points, installation or loading information, and dependencies. An MCP server needs transport methods, tool or resource inventories, and authentication requirements. A tool/API needs operations, parameters, outputs, and invocation constraints. Exact requiredness is determined by the protocol model for the applicable version; web form inputs cannot directly replace the underlying schema.

When a field required by the protocol is empty, the recipient must reject the object. A missing optional field is handled as ``undeclared''. Unknown extension fields may be retained in compatibility mode, but they must not be used for permission decisions that the recipient cannot interpret. The registration service node should perform these checks before registration, and the recipient of a \passthrough{\lstinline!ResourcePackage!} should perform them again.

The four resource types share the resource DID, resource type, subject type, description, capability tags, version, authorized domains, and integrity fields, but the resource type determines which capability descriptions and endpoint information have practical meaning. When fields are missing, the recipient should distinguish ``undeclared'' from ``structurally incomplete''. For example, no optional service endpoint may mean that no invocation entry point has yet been declared, while a missing \passthrough{\lstinline!resourceDid!}, resource type, DID document, or resource package hash prevents the formation of a verifiable resource package.

{\def\LTcaptype{none} 
\begin{longtable}[]{@{}>{\raggedright\arraybackslash}p{0.3200\linewidth}>{\raggedright\arraybackslash}p{0.3200\linewidth}>{\raggedright\arraybackslash}p{0.3200\linewidth}@{}}
\toprule\noalign{}
Field scope & Common requirement for all four resource types & Type-specific supplement \\
\midrule\noalign{}
\endhead
\bottomrule\noalign{}
\endlastfoot
Identity and type & \passthrough{\lstinline!resourceDid!}, \passthrough{\lstinline!resourceType!}, \passthrough{\lstinline!subjectType!} & The DID must be valid; resource and subject types must be mutually consistent \\
Capability description & Name, description, tags, and use cases may be used for discovery & Agent services emphasize tasks and interaction; skill packages emphasize reuse and dependencies \\
Connectivity & Protocol bindings and service endpoints are declared as needed & MCP servers and tool APIs generally need more explicit protocol or interface descriptions \\
Integrity and version & Package version, three classes of hashes, and hash algorithm & A content change must result in a new verification and publication decision \\
\end{longtable}
}

\section{Resource Type and Subject Type Constraints}\label{sec:06.3}

Resource type is independent of the \passthrough{\lstinline!did:oan!} routing code. Current ResourcePackage validation requires \passthrough{\lstinline!resourceType!} and \passthrough{\lstinline!subjectType!} in resource metadata to be consistent with the registration submission and DID Document metadata. Infrastructure nodes, organizations, and developers may also have method-level identities, but a node DID must not be substituted for a product resource DID.

A subject may control multiple resource DIDs, and one resource provider may operate multiple resources. The identity boundaries between the subject DID, resource DID, and node DID are not merged by relationship fields. A verifier must reject a resource package when its resource type, subject type, DID document \passthrough{\lstinline!id!}, or package metadata is inconsistent.

\section{Name, Description, Tags, and Use Cases}\label{sec:06.4}

The name and description are public claims by the resource provider about the resource\textquotesingle s capabilities. Capability tags and use cases are used for registration recommendations, semantic governance, and discovery queries. Tags should preferably reference identifiers in the root platform capability tag tree; an extended tag should state its source or namespace. Natural language may use Chinese, English, or another declared language, but field types and array structures must remain stable. HTML, scripts, or unparseable objects must not be mixed into these fields. A detailed description does not mean that the capability has been verified, and relevance ranking is not proof of control or business quality.

The name and description are public claims by the resource provider about the resource\textquotesingle s capabilities. Capability tags and use cases are used for registration recommendations, semantic governance, and discovery queries. Tags should preferably reference identifiers in the root platform capability tag tree; an extended tag should state its source or namespace. Natural language may use Chinese, English, or another declared language, but field types and array structures must remain stable. HTML, scripts, or unparseable objects must not be placed in text fields.

A more detailed description or use case does not mean that the capability has been verified. Relevance ranking by a discovery service node is also not proof of control, endpoint availability, or business quality. Length, quantity, and content filtering should be specified by deployment or conformity configuration and enforced before registration and indexing.

When capability tags and use cases enter the discovery process, they may serve both as structured conditions and as semantic retrieval text, while the original values and their sources should be retained in the result. For example, a query such as ``I need a tool that can inspect a code repository and summarize its project structure'' may match the relevant capability description and use case. Whether the resource can be invoked still requires checking the resource DID, publication proof, endpoint, and credential requirements. A tag match is evidence of relevance, not proof of resource control or business effectiveness.

\section{Semantic Tags and Protocol Bindings}\label{sec:06.5}

\passthrough{\lstinline!capabilityTags!} expresses the resource capability classification, while \passthrough{\lstinline!authorizedDomains!} expresses the scope of propagation and discovery permissions. They cannot substitute for one another. A binding object in \passthrough{\lstinline!protocolBindings!} must at least state the protocol name and may include a version, transport method, \passthrough{\lstinline!serviceRef!}, and \passthrough{\lstinline!schemaRef!}. A protocol binding describes how to connect; it does not automatically grant invocation permission. A registration service node may use these fields to recommend tags, and a discovery service node may use them for structured or semantic retrieval, but returned results must still retain their source and verification status.

\passthrough{\lstinline!capabilityTags!} expresses the resource capability classification, while \passthrough{\lstinline!authorizedDomains!} expresses the scope of propagation and discovery permissions. They cannot substitute for one another. A binding object in \passthrough{\lstinline!protocolBindings!} must at least state the protocol name and may include a version, transport method, \passthrough{\lstinline!serviceRef!}, and \passthrough{\lstinline!schemaRef!}. A protocol binding describes how to connect; it does not automatically grant invocation permission. The tag-tree version, tag identifier, and aliases should retain their source so that different nodes can interpret the same tag.

A registration service node may provide recommendations based on capability descriptions and the tag tree. A discovery service node may use them for structured or semantic retrieval, but externally returned results must still retain the original tags, protocol bindings, source, and verification status. A tag match without a root platform publication proof must not enter the trusted discovery result set.

The minimum expression of a protocol binding may be written as the following object. \passthrough{\lstinline!serviceRef!} and \passthrough{\lstinline!schemaRef!} are internal resource references or external schema references; they must not be interpreted as new identities:

\section{Service Endpoints and Invocation Requirements}\label{sec:06.6}

Resource endpoints are expressed as a service endpoint array. The core structure includes an endpoint ID, service type, and \passthrough{\lstinline!serviceEndpoint!}, and may include a version, protocol, server type, and port. An endpoint may be an agent interaction entry point, an MCP service entry point, a tool API address, a resource package address, or a protocol description address. The endpoint type and protocol should explain its purpose, and version or authentication differences between multiple endpoints should be explicit. An authentication declaration may express the credential type, issuer, scope, presentation method, or invocation prerequisite, but it is only a public statement of invocation requirements. The existence of an endpoint or HTTP reachability does not mean that the caller has been granted access. Sensitive tokens, private keys, and internal network credentials must not be published as public metadata.

\section{Resource Versions and Compatibility Fields}\label{sec:06.7}

\passthrough{\lstinline!packageVersion!} identifies the resource package version. A \passthrough{\lstinline!version!} in the resource description or a version in a protocol binding may express a resource capability version or protocol version; they must not be substituted for one another without explanation. \passthrough{\lstinline!previousVersion!}, version constraints, or release notes may establish an update chain and compatibility relationship, but a version string alone does not prove the source of the content. Every version must be bound to the resource DID, DID document hash, metadata hash, resource package hash, and hash algorithm. A new version under the same resource DID should enter the verification process through a new resource package and control proof.

Version comparison should at least keep the resource package version, protocol version, and resource lifecycle separate:

{\def\LTcaptype{none} 
\begin{longtable}[]{@{}>{\raggedright\arraybackslash}p{0.3200\linewidth}>{\raggedright\arraybackslash}p{0.3200\linewidth}>{\raggedright\arraybackslash}p{0.3200\linewidth}@{}}
\toprule\noalign{}
Information & Example & Object being judged \\
\midrule\noalign{}
\endhead
\bottomrule\noalign{}
\endlastfoot
Resource package version & \passthrough{\lstinline!1.2.0!} & Whether the current resource materials and their hashes are a newer version \\
Protocol version & \passthrough{\lstinline!MCP 2025-06-18!} & Whether the caller and endpoint can interact under the same protocol \\
Version constraint & \passthrough{\lstinline!>=1.0.0!} & Which target versions are permitted by the composition relationship \\
Lifecycle state & \passthrough{\lstinline!active!} & Whether the current version remains available for continued propagation \\
\end{longtable}
}

The same version string does not mean that the content is identical. A recipient should still check the resource DID, three classes of hashes, root platform proof, and governance state. If content changes while an old hash is retained, the recipient should treat it as an integrity failure.

\section{Package Hash and Metadata Hash}\label{sec:06.8}

The current implementation separately computes the DID document hash, resource metadata hash, and resource package hash. A hash value is expressed with an algorithm prefix, such as \passthrough{\lstinline!sha256:<digest>!}. \passthrough{\lstinline!hashAlgorithm!} must be consistent with the algorithm prefix in the hash references. Before computing the metadata hash, the metadata\textquotesingle s own \passthrough{\lstinline!metadataHash!} and \passthrough{\lstinline!packageHash!} are cleared to avoid circular dependencies. The resource package hash is computed only from fixed declarations such as the package version, resource DID, resource type, DID document hash, metadata hash, and hash algorithm. A verifier should recompute all three hashes using the same cryptographic suite and serialization method. A matching hash alone does not prove control, node authorization, or endpoint availability.

The input to the resource package hash is not a recursive serialization of the entire resource package. It is the set of package declaration fields explicitly selected by the current implementation:

This input illustrates the field boundary of \passthrough{\lstinline!verify\_package\_hash!}. The actual digest must be computed by the implementation using the unified cryptographic suite; the example digest must not be written into a production package. The metadata hash is computed over \passthrough{\lstinline!ResourceMetadata!} after its two own hash fields have been cleared. The two input sets must not be mixed.

\section{Canonical Serialization and Deterministic Hashing}\label{sec:06.9}

Deterministic hashing requires implementations to use the same field names, enum serialization, array order, time format, null handling, and character encoding. The current Rust implementation serializes the object with serde and then passes it to a unified hash function. Cross-language implementations must not change field names, add arbitrary default values, or sort arrays before hashing unless the specification version explicitly requires it. The metadata hash and package hash must not include their own output fields in the input again. If canonicalization fails, an object cannot be serialized, or the algorithm is unsupported, construction should stop and return an error.

When reproducing the process across languages, the object before hashing should be recorded so that ``different fields'' and ``different serialization'' can be distinguished:

\begin{lstlisting}
ResourceMetadata for hashing
  -> clear metadataHash and packageHash
  -> serialize with protocol field names and stable enum values
  -> hash with the selected crypto suite
  -> write metadataHash back to the package

Package hash input
  -> packageVersion, resourceDid, resourceType
  -> didDocumentHash, metadataHash, hashAlgorithm
  -> serialize and hash
\end{lstlisting}

If only the description, tag order, authorized domains, or time field in the metadata changes, the canonicalized metadata input changes and the recipient should obtain a different metadata hash. If only the computed result is written back without changing any other input, the existing hash must not change.

\section{ResourcePackage Construction Rules}\label{sec:06.10}

The construction order for \passthrough{\lstinline!ResourcePackage!} is: determine the resource DID, resource type, DID document, and metadata; compute the DID document hash and metadata hash; compute the resource package hash; organize the version, hashes, lifecycle state, and authorized-domain declarations; have the resource controller submit the control proof; and let the registration service node and root platform perform their respective registration and publication duties. A root platform trusted publication proof can be produced only after the root platform has completed verification. A client must not construct one in advance as if publication had already occurred.

The responsibility boundaries among resource metadata, the DID document, the resource package, the root platform trusted publication proof, and the discovery index record are as follows:

{\def\LTcaptype{none} 
\begin{longtable}[]{@{}>{\raggedright\arraybackslash}p{0.2400\linewidth}>{\raggedright\arraybackslash}p{0.2400\linewidth}>{\raggedright\arraybackslash}p{0.2400\linewidth}>{\raggedright\arraybackslash}p{0.2400\linewidth}@{}}
\toprule\noalign{}
Object & Main content & Producer & Core checks by the recipient \\
\midrule\noalign{}
\endhead
\bottomrule\noalign{}
\endlastfoot
Resource metadata & Type, description, tags, use cases, endpoints, version, and authorized domains & Resource provider/controller & Fields, types, semantics, and metadata hash \\
DID document & DID, verification methods, public keys, and services & Resource controller & DID resolution, control relationship, and document hash \\
Resource package & DID document, metadata, version, hashes, and state & Constructor & Structure, three classes of hashes, and field consistency \\
Root platform trusted publication proof & The root platform\textquotesingle s verification and publication statement for a specific version & Root platform & Issuer, scope, signature, and state \\
Discovery index record & Local verification, authorized domain, freshness, and synchronization position & Discovery service node & Source reference, governance state, and index freshness \\
Content distribution material & Verified resource package and its propagation references & Content distribution platform & Source, hash, version, and publication authorization \\
\end{longtable}
}

The relationship between construction and publication can be represented as follows:

\pandocbounded{\includegraphics[width=\linewidth,height=0.72\textheight,keepaspectratio]{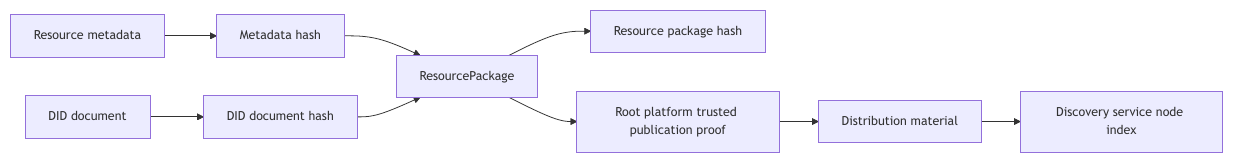}}

The construction process can be represented as a reviewable dependency chain:

\pandocbounded{\includegraphics[width=\linewidth,height=0.72\textheight,keepaspectratio]{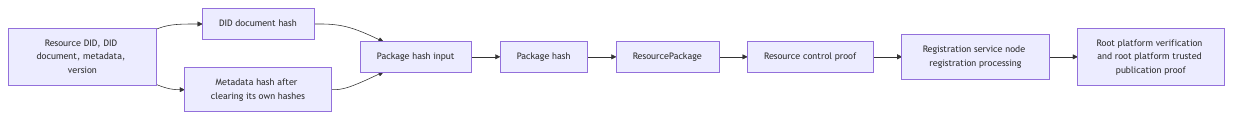}}

The registration processing and root platform verification shown in the diagrams belong to the resource network stage and must not be written into an ordinary resource package constructed by the client in advance. Each step should retain the input version and output hash. If any input changes, all subsequent dependent steps must be performed again.

\subsection{Input Materials}\label{subsec:06.10.1}

Input materials include the resource DID, resource type, DID Document, resource metadata, package version, hash algorithm, resource control proof, and necessary credentials or provenance references. The DID Document \passthrough{\lstinline!id!}, metadata \passthrough{\lstinline!resourceDid!}, package \passthrough{\lstinline!resourceDid!}, and DID targeted by the control challenge must agree. Resource and subject types must agree across the submitted materials; they are not inferred from the routing code.

\subsection{Field Canonicalization}\label{subsec:06.10.2}

The constructor should use protocol field names and deterministic serialization rules, represent time as a parseable UTC value, and keep array order and enum values stable. Before computing the metadata hash, clear \passthrough{\lstinline!metadataHash!} and \passthrough{\lstinline!packageHash!} in the metadata. When computing the package hash, include only the specified package declaration fields.

These fields are mapped to HTTP JSON through \passthrough{\lstinline!serde!}; a separate interface schema is not automatically generated at compile time. \passthrough{\lstinline!ResourceRegistrationSubmission.didDocument!} uses the DID document type, while \passthrough{\lstinline!metadata!} uses dynamic JSON. \passthrough{\lstinline!ResourcePackage.metadata!} uses the resource metadata type, and \passthrough{\lstinline!rootProof!} uses the root platform proof type. The recipient therefore needs both Rust-type-based deserialization and business validation of resource types, hashes, and proofs. Successful deserialization does not mean that the object can be published.

\subsection{Hash Computation}\label{subsec:06.10.3}

Compute the DID document hash, metadata hash, and resource package hash in sequence, and write the algorithm prefix into the corresponding fields. After hashes are populated, none of the other fields participating in the hash may be changed. A change in any recomputed result indicates that the content of the current version has changed.

\subsection{Signature and Proof Generation}\label{subsec:06.10.4}

The resource controller signs the control challenge and the prescribed resource-control materials. A registration service node may issue a registration credential for the registration fact. After checking node authorization, the control proof, hashes, and publication conditions, the root platform generates a root platform trusted publication proof. The issuer, verification method, target DID, purpose, and covered hashes of each proof must be verifiable.

\subsection{Post-construction Validation}\label{subsec:06.10.5}

After construction, check resource type consistency, metadata consistency, the DID document hash, metadata hash, resource package hash, and root platform statement binding in sequence. A failure in any check must prevent the object from being marked distributable or trusted-published. A failed object may be retained as local diagnostic material, but it should not enter a public index.

The post-construction check result can be recorded as a machine-readable diagnostic summary:

\begin{lstlisting}
{
  "resourceDid": "did:oan:K7mQ9:7YpQm9Kx2VnRb6Ts3WfHa4Cd5Ej8LgNz",
  "packageVersion": "1.0.0",
  "checks": [
    { "name": "resource_type_consistency", "status": "passed" },
    { "name": "metadata_consistency", "status": "passed" },
    { "name": "did_document_hash", "status": "passed" },
    { "name": "metadata_hash", "status": "passed" },
    { "name": "package_hash", "status": "passed" },
    { "name": "root_claim_binding", "status": "passed" }
  ],
  "publishable": true
}
\end{lstlisting}

\passthrough{\lstinline!publishable!} means only that the local object passed the current package checks. It does not replace acceptance by a registration service node or publication by the root platform. If any check fails, the diagnostic result should include the corresponding error category, and the package must not enter trusted distribution or a discovery index.

\section{Structure of a Root-Platform-Verified ResourcePackage}\label{sec:06.11}

The resource package after root platform verification is still based on the DID document, resource metadata, and hashes provided by the resource controller, but it adds or binds the root platform\textquotesingle s verification result for that specific resource version. The root platform DID, \passthrough{\lstinline!packageClaims!}, signature or data-integrity proof in \passthrough{\lstinline!rootProof!}, together with any necessary announcement-event reference, expresses which facts the root platform has checked and accepted. The publication sequence, publication time, and synchronization cursor belong to the publication or propagation layer and must not be written back as original claims by the resource provider.

\subsection{Resource Identity Section}\label{subsec:06.11.1}

The resource identity section includes \passthrough{\lstinline!resourceDid!}, \passthrough{\lstinline!resourceType!}, the DID document in the package, and \passthrough{\lstinline!didDocumentHash!}. The recipient should check that the package fields are consistent with the DID document \passthrough{\lstinline!id!}, resolve \passthrough{\lstinline!did:oan!}, and verify the resource-type mapping. A valid identity section only means that the package is bound to a resolvable resource DID; it does not mean that the resource has been published by the root platform.

\subsection{Resource Metadata Section}\label{subsec:06.11.2}

The resource metadata section includes the name, description, capability tags, use cases, authorized domains, protocol bindings, service endpoints, lifecycle, and version information. The recipient should check that the metadata hash, resource type, resource DID, package version, authorized domains, and lifecycle are consistent with the declarations outside the package, and should retain the metadata source and update time.

\subsection{Root Platform Verification Proof}\label{subsec:06.11.3}

The root platform verification proof must at least associate the root platform DID, resource DID, resource type, package version, DID document hash, metadata hash, package hash, hash algorithm, lifecycle state, and authorized domains. If \passthrough{\lstinline!packageClaims!} is inconsistent with the package contents, the proof signature cannot be verified, or the proof source is outside the trusted scope, the package must not be treated as a root-platform-verified resource package.

\subsection{Fields Required for Distribution and Indexing}\label{subsec:06.11.4}

Distribution and indexing require at least the resource DID, resource type, package version, three classes of hashes, root platform proof, lifecycle state, authorized domains, source reference, and publication or synchronization cursor. A discovery service node should also record the local receipt time, verification time, current synchronization position, and rejection reason, so that an old snapshot, indexing lag, or cross-node inconsistency can be explained.

\section{ResourcePackage Integrity and Provenance Verification}\label{sec:06.12}

The recipient should first verify the structure and resource identity, then verify the DID document hash, metadata hash, resource package hash, field consistency, and root platform proof in that order. Finally, it should combine lifecycle, authorized domain, version, and governance state to decide whether the object may be written to a distribution cache or discovery index. A hash or proof failure must prevent the object from entering the trusted read model. Temporary network unavailability may place it in a pending-retry state, but an object with incomplete verification must not be marked as published.

The recipient may classify verification results into three categories: ``passed'', ``rejected'', and ``pending retry'':

{\def\LTcaptype{none} 
\begin{longtable}[]{@{}>{\raggedright\arraybackslash}p{0.2400\linewidth}>{\raggedright\arraybackslash}p{0.2400\linewidth}>{\raggedright\arraybackslash}p{0.2400\linewidth}>{\raggedright\arraybackslash}p{0.2400\linewidth}@{}}
\toprule\noalign{}
Result & Typical reason & Enters the trusted read model? & Follow-up action \\
\midrule\noalign{}
\endhead
\bottomrule\noalign{}
\endlastfoot
Passed & Structure, hashes, source, and state are all verifiable & Yes & Write to distribution or index according to version and authorized domain \\
Rejected & Hash mismatch, invalid proof, type conflict, or untrusted source & No & Save the error and evidence; request correction or resubmission \\
Pending retry & An upstream service is temporarily unreachable or the governance projection has not synchronized & Must not be marked as published & Retain a pending task and retry after recovery \\
\end{longtable}
}

``Pending retry'' is not a verified state. Even when the JSON structure of a package is correct, if the root platform proof or a necessary source cannot be confirmed, the recipient may retain only an incomplete state.

\section{ResourcePackage Provenance Chain and Attribution}\label{sec:06.13}

The resource provenance chain contains at least five classes of facts: the resource controller produces the DID document, metadata, and control proof; the registration service node produces the registration acceptance and registration credential; the root platform produces the resource-version acceptance, root platform trusted publication proof, and publication reference; the content distribution platform forwards or caches materials whose propagation is permitted by the root platform; and the discovery service node produces the local verification and indexing record. Each fact should identify the source subject, target resource DID, version, time or cursor, and preserve its signature or hash association.

The provenance chain can be read in five stages: ``claim, registration, publication, transport, and indexing'':

\begin{lstlisting}
Resource controller
  -> DID document, resource metadata, control proof
Registration service node
  -> registration record, registration credential
Root platform
  -> acceptance record, root platform trusted publication proof, publication reference
Content distribution platform
  -> transport or cache record for verified materials
Discovery service node
  -> local validation, synchronization position, and discovery index record
\end{lstlisting}

Each later stage may add facts about its own processing, but it cannot issue a proof that belongs to an earlier stage. For example, an index record from a discovery service node may show that the node verified and stored materials for a version; it cannot replace the resource controller\textquotesingle s control proof or the root platform\textquotesingle s publication proof.

\subsection{Resource Controller Provenance}\label{subsec:06.13.1}

Resource controller provenance is determined by the DID document\textquotesingle s control relationship, the resource control proof, and its signature verification method. \passthrough{\lstinline!publisherDid!} or \passthrough{\lstinline!subjectDid!} may supplement a business relationship, but cannot replace the control proof. The recipient should check the resource party\textquotesingle s claim against the current DID document and proof together.

\subsection{Registration Service Node Provenance}\label{subsec:06.13.2}

Registration service node provenance is expressed by the receipt record for the registration request, the node DID, node authorization materials, the registration credential, and the processing state. The scope of a registration credential is the registration fact; it does not automatically prove that the root platform has accepted the resource or that a discovery service node has indexed it.

\subsection{Root Platform Publication Provenance}\label{subsec:06.13.3}

Root platform publication provenance is expressed by the root platform DID, root platform trusted publication proof, resource package declarations, and publication or announcement references. The root platform proof must be bound to a specific resource version and its hashes. A proof for another version under the same resource DID cannot be reused directly.

\subsection{Provenance Chain Verification}\label{subsec:06.13.4}

Provenance-chain verification requires checking the control proof, registration credential, root platform trusted publication proof, and distribution record layer by layer. A content distribution platform proves only transport or caching behavior, while a discovery service node proves only local retrieval, validation, and indexing behavior. A missing layer must be reflected in the response; the existence of a later layer must not conceal the failure of an earlier one.

\section{Package Size, Content Limits, and Validation Strategy}\label{sec:06.14}

Size, length, nesting depth, number of endpoints, number of tags, and number of protocol bindings should be specified by deployment or conformity configuration and enforced by the registration service node and \passthrough{\lstinline!ResourcePackage!} recipient using the same limits or limits no lower than the network requirements. When the current schema does not define a unified numerical business limit, this chapter must not invent a fixed number. The implementation must return an identifiable error when a limit is exceeded, preventing oversized text, deeply nested structures, or large numbers of endpoints from exhausting parsing, indexing, and hashing resources. Limit checks must occur before hash computation, so that truncation, omission, or automatic compression cannot produce a different hash.

Limit checks should be applied to the original input, and the failure reason should be associated with a field path rather than silently truncating the content:

\begin{lstlisting}
{
  "validation": "rejected",
  "errors": [
    {
      "path": "metadata.capabilityTags",
      "code": "limit_exceeded",
      "message": "capability tag count exceeds the active profile"
    }
  ],
  "hashingStarted": false
}
\end{lstlisting}

The limit values in the example are determined by the actual deployment configuration. Once input has been truncated, a field has been discarded, or an array has been automatically reordered, the object being hashed has changed and the original signature or hash cannot continue to be used.

\section{Schema Evolution and Backward Compatibility}\label{sec:06.15}

Schema evolution should distinguish the addition of an optional field, a change to existing semantics, field removal, and a change to the hash input or signature coverage. If a new field does not affect parsing by older implementations or existing hash rules, it may be introduced through a versioned extension. If it changes resource identity, resource type, version, hash input, root platform statement, or the verification process, the schema or protocol version must be advanced and the behavior of older nodes must be stated. An older node may ignore an unknown optional field, but it must not silently discard a field covered by a signed object when recomputing it. An unknown required field, an unrecognized hash algorithm, or an unverifiable proof should cause rejection or entry into a compatibility-pending state.

This section specifies only the effects of schema changes on parsing and integrity validation. The responsibility boundaries of resource metadata, the DID document, the resource package, the root platform trusted publication proof, and the discovery index record are described in Sections 6.1, 6.4, 6.10 through 6.13, and Chapters 9 and 10 respectively. A schema-version change does not automatically change resource control, publication state, or discovery visibility. The recipient must still recompute the relevant hashes according to the established validation order and retain the reason for any failure.

\section{References}\label{sec:06.16}

{\def\LTcaptype{none} 
\begin{longtable}[]{@{}>{\raggedright\arraybackslash}p{0.3200\linewidth}>{\raggedright\arraybackslash}p{0.3200\linewidth}>{\raggedright\arraybackslash}p{0.3200\linewidth}@{}}
\toprule\noalign{}
Source & Type & Link \\
\midrule\noalign{}
\endhead
\bottomrule\noalign{}
\endlastfoot
\passthrough{\lstinline!oan-protocol-commons!} & Code repository: resource metadata, resource package, and schema & \url{https://github.com/wolfbrother/oan-protocol-commons} \\
\passthrough{\lstinline!oan-sdk-ts!} & Code repository: resource package construction, hashing, and shape validation & \url{https://github.com/OpenAgenet/oan-sdk-ts} \\
OAN Yellow Paper & arXiv yellow paper: resource packages, hashing, and trusted publication & \url{https://arxiv.org/abs/2606.03163} \\
OAN Resource Identity and Discovery & IETF draft: resource identity and discovery materials & \url{https://datatracker.ietf.org/doc/draft-xu-oan-resource-identity-discovery/} \\
\end{longtable}
}

\chapter{Identity, Keys, Signatures, VCs, and VPs}\label{chap:07}

{\def\LTcaptype{none} 
\begin{longtable}[]{@{}>{\raggedright\arraybackslash}p{0.2400\linewidth}>{\raggedright\arraybackslash}p{0.2400\linewidth}>{\raggedright\arraybackslash}p{0.2400\linewidth}>{\raggedright\arraybackslash}p{0.2400\linewidth}@{}}
\toprule\noalign{}
Object & Primary role & Contains a control private key & Cannot replace \\
\midrule\noalign{}
\endhead
\bottomrule\noalign{}
\endlastfoot
Local Identity & Locally holds identity and signing capability & Yes, under protected storage & DID Document or network publication fact \\
DID Document & Publishes verification methods and service declarations & No & Registration VC or Root proof \\
Registration VC & Registrar\textquotesingle s registration assertion & No & Resource-control authorization \\
Node authorization VC & Infrastructure role and scope & No & Control of a resource \\
Root proof & Root acceptance of a resource version & No & CDN persistence or Discovery indexing \\
\end{longtable}
}

This chapter provides a unified description of the semantic boundaries of OpenAgenet (OAN) identity materials and cryptographic proofs. Identity materials answer who an object is and who can represent the object in making a statement; a signature answers whether specified content was produced by the corresponding private key; a verifiable credential (VC) answers which facts an issuer has made a verifiable statement about; and a verifiable presentation (VP) answers which credentials a holder presents for a specific verification purpose. These four concepts are related, but they cannot substitute for one another.

Resource registration commonly involves a subject DID, a resource DID, a resource control key, a Registrar node key, a node authorization VC, a resource control proof, a registration VC, and a Root proof. A client can generate and hold private keys, a node can verify public keys and proofs, and the Root platform can verify node authorization and publication conditions; these verification capabilities do not mean that a node or the Root platform obtains the resource control private key, nor do they mean that the resource has obtained permission for business invocation.

\section{Subject Identity and Resource Identity}\label{sec:07.1}

A subject is an identity object capable of managing one or more resource control relationships. A resource identity is an object boundary that can be registered, published, discovered, updated, and revoked independently. The subject DID, resource DID, Registrar node DID, Discovery node DID, and Root platform DID may all use the \passthrough{\lstinline!did:oan!} method, but their object categories, key purposes, and trust sources differ. They must not be mixed merely because they use the same method.

{\def\LTcaptype{none} 
\begin{longtable}[]{@{}>{\raggedright\arraybackslash}p{0.2400\linewidth}>{\raggedright\arraybackslash}p{0.2400\linewidth}>{\raggedright\arraybackslash}p{0.2400\linewidth}>{\raggedright\arraybackslash}p{0.2400\linewidth}@{}}
\toprule\noalign{}
Object & Primary question proved & Typical materials & Facts not directly proved \\
\midrule\noalign{}
\endhead
\bottomrule\noalign{}
\endlastfoot
Subject & Who manages the resource control relationship & Subject DID Document, subject control key & That the resource endpoint is necessarily available \\
Resource & Which resource and version have been registered & Resource DID Document, ResourcePackage, version, and hash & That the caller has been permitted to invoke it \\
Registrar node & Whether the node is qualified to accept registration & Node DID Document, node authorization VC & Resource business quality \\
Discovery node & Whether the resource has been synchronized and indexed, and whether it is returned according to the current node state & Node DID Document, index record, discovery response, and freshness information & Resource control, Root platform publication, or business quality \\
Root platform & Whether a network-level publication fact has been formed & Root proof, governance state & The legal relationship of the resource provider \\
\end{longtable}
}

At registration time, the minimum public materials required for verification should be disclosed, such as the resource DID Document, public key, resource metadata, hash, and control proof. The subject\textquotesingle s name, contact information, local identity backup, and private key are not public fields required for resource discovery. Registration submission must not by default place them in node logs, indexes, or public responses.

Identity materials in the same registration request can be divided into three layers according to their purposes:

{\def\LTcaptype{none} 
\begin{longtable}[]{@{}>{\raggedright\arraybackslash}p{0.3200\linewidth}>{\raggedright\arraybackslash}p{0.3200\linewidth}>{\raggedright\arraybackslash}p{0.3200\linewidth}@{}}
\toprule\noalign{}
Layer & Typical materials & Description \\
\midrule\noalign{}
\endhead
\bottomrule\noalign{}
\endlastfoot
Identity layer & Subject DID, resource DID, DID Document, and public key & Identifies the object and indicates where a verifier can obtain the public key \\
Control layer & Control challenge, control proof, and document hash & Shows whether the current controller authorized this specific operation \\
Network-fact layer & Registration VC, Root proof, and governance state & Shows which node processed or published which fact \\
\end{longtable}
}

A verifier should start from the identity layer and check the control layer and network-fact layer in sequence. Taking the resource DID text, a registration VC, or a signature from a particular node alone cannot reconstruct the complete control and publication relationship.

\section{Key Generation, Custody, Backup, and Recovery}\label{sec:07.2}

Keys should be generated, imported, and used in a client or node secure execution environment managed by the controller. The current \passthrough{\lstinline!oan-crypto!} provides Ed25519 and SM2/SM3-related key processing and supports multiple public-key encodings and JWK public-key representations. The cryptographic suite, public-key representation, and proof type must be identifiable by the recipient. When materials are unknown, truncated, or incompatible, the operation should fail explicitly rather than silently switching to another algorithm suite.

{\def\LTcaptype{none} 
\begin{longtable}[]{@{}>{\raggedright\arraybackslash}p{0.2400\linewidth}>{\raggedright\arraybackslash}p{0.2400\linewidth}>{\raggedright\arraybackslash}p{0.2400\linewidth}>{\raggedright\arraybackslash}p{0.2400\linewidth}@{}}
\toprule\noalign{}
Key material & Holder & Typical purpose & External materials \\
\midrule\noalign{}
\endhead
\bottomrule\noalign{}
\endlastfoot
Subject control private key & Subject or authorized controller & Managing the subject and related resources & Subject DID, public key, and control proof \\
Resource control private key & Resource controller & Creating or updating the resource DID Document and versions & Resource DID, public key, and operation proof \\
Registrar node private key & Registrar node & Node authentication, signing upstream requests, and issuing registration VCs & Node DID, public key, and authorization VC \\
Root platform key & Controlled Root platform execution boundary & Node authorization, Root proof, and service signatures & Root platform public key and verifiable proofs \\
\end{longtable}
}

A local identity backup may contain subject configurations, resource configurations, public keys, private keys, and credentials at the same time, and is therefore sensitive material. During import or recovery, the client should check the file structure, the correspondence between DIDs, public keys, and private keys, the resource type, and the control relationship. Successful recovery only means that the client has regained local control materials; it does not mean that the resource has been registered, published, or re-indexed.

The boundary between a local identity backup and a public DID Document can be represented by the following minimal structure:

This structure is for illustrating the boundary only; the actual backup fields are determined by the client implementation. Public identity materials can be sent to a verifier, while \passthrough{\lstinline!localSecret!} must remain within the local secure boundary managed by the controller. Obtaining a public key from a server does not allow a private key to be derived.

\section{Resource Control Signatures}\label{sec:07.3}

A resource control signature must be bound to a specific operation rather than being an isolated signature over arbitrary text. \passthrough{\lstinline!DidControlChallenge!} combines a challenge identifier, draft identifier, target resource DID, DID Document hash, Registrar node DID, operation purpose, verification method, nonce, and validity period. \passthrough{\lstinline!SubjectControlProofBundle!} then associates the control proof, verification time, actual verification method, and proof hash. Both the Registrar node and the Root platform should check these bindings.

At minimum, the following conditions should be checked:

\begin{enumerate}
\def\labelenumi{\arabic{enumi}.}
\tightlist
\item
  \passthrough{\lstinline!subjectDid!} is consistent with the submitted resource DID.
\item
  \passthrough{\lstinline!didDocumentHash!} is consistent with the submitted DID Document.
\item
  \passthrough{\lstinline!verificationMethod!} belongs to the target DID Document and its purpose permits the current operation.
\item
  \passthrough{\lstinline!purpose!} is consistent with the current registration, update, or publication operation.
\item
  \passthrough{\lstinline!nonce!}, issuance time, and expiration time satisfy freshness and replay-prevention requirements.
\end{enumerate}

Successful proof verification only shows that the corresponding private key signed the specified payload. Whether the resource provider is authorized to represent an organization, whether the endpoint is secure, and whether the caller has obtained business authorization must still be determined separately through delegation materials, governance state, and business policies.

The core of a control proof is not the signature value itself, but the context covered by the signature. A verifier can check a registration operation according to the following relationship:

The DID, target Registrar node, document hash, purpose, and nonce are all binding relationships to be checked. The example values do not represent a real request. If any binding field differs from the current request, registration or update must not be accepted merely because the signature algorithm verification succeeds.

\section{Infrastructure Node Identity Keys and Node Signing Keys}\label{sec:07.4}

The node DID Document expresses the node identity. A communication signing key proves that a request comes from the corresponding node execution boundary. A node authorization VC proves the node\textquotesingle s role and authorization domain. Runtime credentials support service startup, key loading, or access to controlled infrastructure. A deployment may host multiple categories of materials within the same security boundary, but protocol semantics should still separate them by purpose.

When sending a resource verification request to the Root platform, \passthrough{\lstinline!SignedRequestEnvelope!} binds \passthrough{\lstinline!requestId!}, protocol version, purpose, HTTP method, path, audience, request time, request nonce, request-body hash, and proof. The recipient should check the audience, path, time window, nonce replay, node DID Document, public key, and node authorization before processing the business content.

The verification sequence for \passthrough{\lstinline!SignedRequestEnvelope!} can be summarized as follows:

\begin{enumerate}
\def\labelenumi{\arabic{enumi}.}
\tightlist
\item
  Read the node DID, audience, HTTP method, path, and request-body hash from the request.
\item
  Parse the node DID Document and locate the verification method and public key used by the request.
\item
  Verify the purpose, creation time, request ID, nonce, and signature payload of the proof.
\item
  Check the role, authorization domain, and current status of the node authorization VC.
\item
  Process resource verification, registration, or publication business only after the checks pass.
\end{enumerate}

This separates the question of whether ``the request comes from a particular node'' from the question of whether ``the node is authorized to perform the current action.''

\section{Root Platform Authorization VC for Infrastructure Nodes}\label{sec:07.5}

The core relationships in a node authorization VC are as follows: \passthrough{\lstinline!issuer!} is the Root platform or a governance-approved issuer, \passthrough{\lstinline!subject!} is the authorized node DID, \passthrough{\lstinline!role!} indicates the node role, \passthrough{\lstinline!claims!} carries the authorization domain and capability scope, and \passthrough{\lstinline!status!}, \passthrough{\lstinline!issuedAt!}, and \passthrough{\lstinline!expiresAt!} describe status and the validity window. The proof of \passthrough{\lstinline!NodeAuthorizationCredential!} must be issued by the issuer over the credential content without a proof. The verifier should obtain the public key from the issuer DID Document and verify the proof.

\begin{lstlisting}
Node DID Document -> obtain node verification public key
      -> verify authorization VC proof
      -> check issuer, subject, role, status, issuedAt, expiresAt
      -> check authorization domain and capability scope
      -> decide whether the current node action is authorized
\end{lstlisting}

When the proof is absent, the issuer cannot be resolved, the subject does not match, the role is not permitted, the authorization has expired, the status is not active, or the authorization domain does not cover the requested scope, the restricted action must be rejected. A node authorization VC proves infrastructure eligibility; it does not prove resource control.

The relationships among node authorization VC fields can be checked using the following example:

The role, authorization domain, and capability values in this example are used to illustrate the validation relationships. The verifier must confirm that the issuer, subject, proof, and governance state match one another, and then compare the authorization scope with the current request. The validity of a node authorization VC does not change the controller of a resource DID.

\section{Registration VC Issued to the Resource Controller}\label{sec:07.6}

{\def\LTcaptype{none} 
\begin{longtable}[]{@{}>{\raggedright\arraybackslash}p{0.3200\linewidth}>{\raggedright\arraybackslash}p{0.3200\linewidth}>{\raggedright\arraybackslash}p{0.3200\linewidth}@{}}
\toprule\noalign{}
VC type & Fact expressed & Usage boundary \\
\midrule\noalign{}
\endhead
\bottomrule\noalign{}
\endlastfoot
\passthrough{\lstinline!OANInfrastructureAuthorizationCredential!} & Infrastructure role, scope, and authorization state & Still check current governance state \\
\passthrough{\lstinline!OANResourceRegistrationCredential!} & Registrar assertion for specific registration materials & Not Root publication or business callability \\
\passthrough{\lstinline!OANBusinessFactCredential!} & Business fact about a subject or resource & Depends on issuer competence for that fact \\
\passthrough{\lstinline!OANQualificationCredential!} & Qualification & Does not replace resource control or publication evidence \\
\passthrough{\lstinline!OANAuditResultCredential!} & Audit result for a scope and time & Not a permanent quality guarantee \\
\passthrough{\lstinline!OANSelfClaimedCapabilityCredential!} & Capability claim by controller or issuer & Self-claim alone does not become official authorization \\
\end{longtable}
}

In the current Registrar node flow, the Registrar node first calls the Root platform resource verification and publication interface. After the upstream operation succeeds, the Registrar node issues a registration VC and writes the registration record to storage. The credential contains the OAN credential context, credential type, issuer, resource credential subject, resource DID, resource type, DID Document hash, metadata hash, resource package hash, ResourcePackage version, hash algorithm, authorization domain, lifecycle state, and credential status.

The following is an illustrative structure; field order has no additional semantic meaning:

A registration VC proves that the Registrar node processed the specified registration materials and the registration facts explicitly listed in the credential. It does not independently prove that the Root platform has completed network-wide publication, that a Discovery node has completed indexing, or that the caller has obtained permission for business invocation.

The most important binding relationships in a registration VC are \passthrough{\lstinline!issuer!}, \passthrough{\lstinline!subject!}, the resource version, and the three classes of hashes. A verifier should compare the credential claims again with the received DID Document, resource metadata, and ResourcePackage:

The example values do not represent a real credential. If \passthrough{\lstinline!subject!}, \passthrough{\lstinline!resourceDid!}, the version, or any hash differs from the current materials, the registration VC cannot be used to prove that the current materials have been registered. Even when the credential signature itself is valid, it only shows that the issuer signed another set of claims.

\section{Governance and Authorization Evidence}\label{sec:07.7}

{\def\LTcaptype{none} 
\begin{longtable}[]{@{}>{\raggedright\arraybackslash}p{0.2400\linewidth}>{\raggedright\arraybackslash}p{0.2400\linewidth}>{\raggedright\arraybackslash}p{0.2400\linewidth}>{\raggedright\arraybackslash}p{0.2400\linewidth}@{}}
\toprule\noalign{}
Evidence & Producer & Verifier & Fact proved \\
\midrule\noalign{}
\endhead
\bottomrule\noalign{}
\endlastfoot
Resource control signature & Resource controller & Registrar node, Root platform & The controller authorized a specific DID, Document, version, and operation \\
Node authorization VC & Root platform or governance issuer & Node, Root platform, and trusted components & Node role, authorization domain, and valid status \\
Registration VC & Registrar node & Controller, auditor, and client & Registration processing and the material relationships listed in the credential \\
Root proof & Root platform & Content distribution platform, Discovery node, and caller & The specified version passed Root platform verification and entered the publication flow \\
Governance event & Governance system or on-chain record & Off-chain trust indexer, node, and auditor & A time-sequenced fact about authorization or governance state \\
\end{longtable}
}

A verifier must combine evidence according to the type of fact being established. A valid registration VC only means that registration processing was completed. Without a Root proof, the registration result must not be described as globally discoverable. When the resource control proof does not match, a valid node authorization VC is not sufficient to accept a resource update.

\section{VC and VP Structural Requirements}\label{sec:07.8}

A VC is a verifiable expression of facts asserted by an issuer. A VP is a presentation organized by a holder for a verification purpose. The current core code directly models node authorization VCs and resource registration VCs. When a VP is used, it should add a holder, presentation purpose, challenge, validity period, and presentation proof. An arbitrary collection of VC JSON objects must not be called a VP directly.

\subsection{Basic VC Structure}\label{subsec:07.8.1}

A VC should be able to identify its context, type, issuer, credentialSubject, issuance time, optional expiration time, status, and proof at minimum. OAN\textquotesingle s Rust credential structures use \passthrough{\lstinline!type!}, \passthrough{\lstinline!issuer!}, \passthrough{\lstinline!subject!}, \passthrough{\lstinline!status!}, \passthrough{\lstinline!issuedAt!}, \passthrough{\lstinline!expiresAt!}, \passthrough{\lstinline!claims!}, and \passthrough{\lstinline!proof!}. An HTTP JSON structure may use interoperability names such as \passthrough{\lstinline!credentialSubject!} and \passthrough{\lstinline!issuanceDate!}, but the serialization mapping must be stable and must not change the meaning of the claims.

\subsection{Basic VP Structure}\label{subsec:07.8.2}

A VP should contain a presentation type, holder, one or more \passthrough{\lstinline!verifiableCredential!} objects, a presentation purpose, and a proof for the current verification. When replay prevention is required, it should also bind a challenge or nonce provided by the verifier. A collection of credentials without holder proof, purpose binding, or the current verification context should not be accepted as a fresh VP.

The minimum VP example can be represented as follows:

This example distinguishes the issuer proof of a VC from the holder presentation proof of a VP. The verifier must still validate \passthrough{\lstinline!issuer!}, \passthrough{\lstinline!subject!}, the VC proof, \passthrough{\lstinline!holder!}, the VP proof, \passthrough{\lstinline!challenge!}, and status according to the actual credential structure. The credential object in the example must not be used directly as an available credential.

\subsection{Issuer, Holder, and Subject}\label{subsec:07.8.3}

The issuer is responsible for the facts asserted by the VC, \passthrough{\lstinline!credentialSubject!} is the object described by the assertion, the holder is the party currently controlling the presentation materials, and the controller is the party controlling verification methods in a DID Document. These four roles may be the same or different. A Registrar node may be the issuer of a registration VC, the resource DID may be its credentialSubject, and the resource controller may be the holder; the node operator therefore does not automatically become the resource controller.

\subsection{Credential Status and Verification}\label{subsec:07.8.4}

Verification must not stop at proof verification. It should also check the credential type, issuer and DID Document, public key, subject, purpose, validity period, status, and required claims. When the status source is unavailable, the verifier should apply an explicit failure or degradation policy according to the credential purpose. ``Temporarily unable to query'' must not be interpreted as active.

For different materials, the verification result should preserve two levels: ``signature valid'' and ``business fact valid''.

{\def\LTcaptype{none} 
\begin{longtable}[]{@{}>{\raggedright\arraybackslash}p{0.3200\linewidth}>{\raggedright\arraybackslash}p{0.3200\linewidth}>{\raggedright\arraybackslash}p{0.3200\linewidth}@{}}
\toprule\noalign{}
Check result & Fact that can be confirmed & Content that still requires checking \\
\midrule\noalign{}
\endhead
\bottomrule\noalign{}
\endlastfoot
Proof verification succeeds & The specified key signed the specified payload & Issuer identity, purpose, validity period, and status \\
VC status is active & The credential is not currently marked invalid by that status source & Whether the claims match the current resource materials \\
Node authorization is valid & The node has infrastructure eligibility within the declared scope & The role, authorization domain, and resource control proof for the current request \\
Resource control proof is valid & The controller authorized a specific resource and operation & Registration acceptance, Root platform publication, and business invocation permission \\
\end{longtable}
}

\section{Relationship among Issuer, Holder, Controller, and Credential Subject}\label{sec:07.9}

These fields express different relationships. The complete ownership and authorization chain cannot be inferred merely from the presence of a field. Verification should check the field relationships, DID Document, public key, proof purpose, governance policy, and business policy together.

\subsection{Issuer}\label{subsec:07.9.1}

The issuer signs the credential content without a proof using the issuance key and bears traceable responsibility for the claims. The verifier should obtain the verification method from the issuer DID Document and check the consistency of the proof creator, verification method, and issuer.

\subsection{Holder}\label{subsec:07.9.2}

The holder is the party currently controlling the credential presentation materials. It may be a resource controller, client, or delegated user assistant agent. Holding a public VC does not mean that the holder has obtained control of the credentialSubject; the relationship of representation still requires control proof, delegation, or business authorization.

\subsection{Resource Controller}\label{subsec:07.9.3}

The resource controller controls registration, update, or other protected operations for the resource identified by the resource DID. The controller may differ from the resource provider, registration VC holder, or endpoint operator. Its permission decision depends on the target DID, verification method, operation-bound proof, and current version rules.

\subsection{Credential Subject}\label{subsec:07.9.4}

\passthrough{\lstinline!credentialSubject!} is the object described by the VC. The subject of a node authorization VC is normally an infrastructure node DID, while the subject of a registration VC is normally a resource DID. If the subject is replaced, the DID Document does not match, or the claims conflict with the subject type, the credential must not be used for the corresponding decision.

\section{Signature Verification and Status Verification}\label{sec:07.10}

After a verification service receives JSON, it should first parse it according to the type or field structure declared by the interface, and then separately verify the signature coverage and the status source. The \passthrough{\lstinline!upstreamAuth!} of an inter-node request proves that the request comes from a node with the corresponding authorization. \passthrough{\lstinline!subjectControlProof!} in resource registration proves that the resource controller authorized the current registration or update. A registration VC describes the registration fact produced by the Registrar node. A Root proof describes that a specific resource version passed Root platform verification. These materials may appear consecutively in the same HTTP business flow, but they must not be conflated into one credential object.

\begin{lstlisting}
Request can be parsed
  -> request signature and audience are valid
  -> resource control proof is valid
  -> registration VC issuer and subject match
  -> Root publication proof matches the version and hashes
  -> governance state and freshness are acceptable
\end{lstlisting}

\pandocbounded{\includegraphics[width=\linewidth,height=0.72\textheight,keepaspectratio]{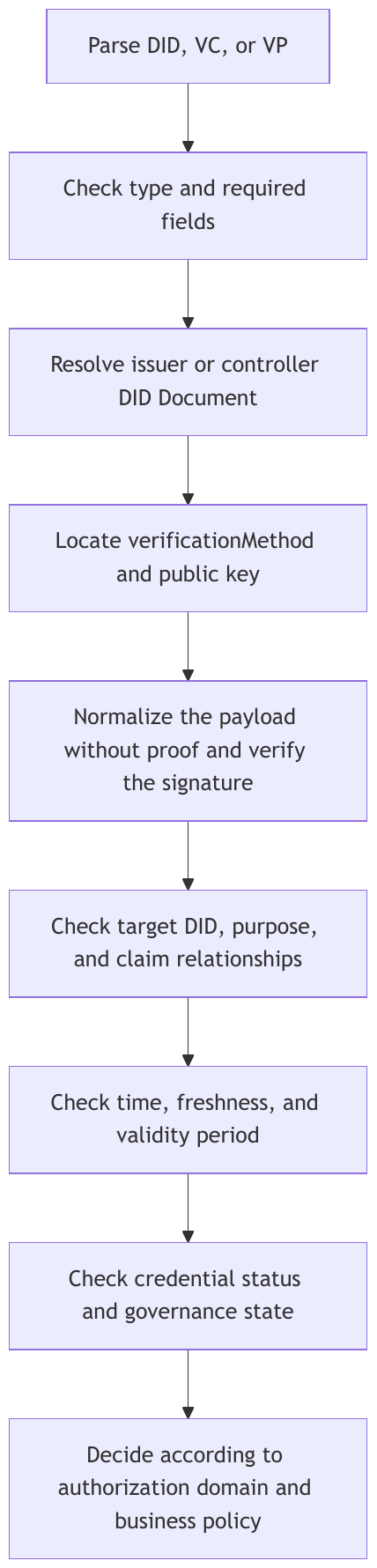}}

Any failure must terminate the flow before a side effect such as publication, authorization, or invocation occurs, and should return a locatable failure reason. Verification records may retain the verification object, proof type, verification method, verification time, status source, and error code, but must not retain private keys or materials from which private keys can be recovered.

\section{Credential Privacy and Disclosure Boundaries}\label{sec:07.11}

The resource DID, public key, capability tags, resource description, endpoint, version, hash, and Root publication evidence required for resource discovery may enter a public ResourcePackage or discovery response. Subject personal information, private keys, local identity backups, undisclosed delegation materials, and unnecessary internal configuration should remain within a controlled boundary.

{\def\LTcaptype{none} 
\begin{longtable}[]{@{}>{\raggedright\arraybackslash}p{0.3200\linewidth}>{\raggedright\arraybackslash}p{0.3200\linewidth}>{\raggedright\arraybackslash}p{0.3200\linewidth}@{}}
\toprule\noalign{}
Material & Default scope & Processing requirement \\
\midrule\noalign{}
\endhead
\bottomrule\noalign{}
\endlastfoot
Resource DID Document & Publicly resolvable & Must not contain private keys; verify core and extension fields \\
Registration VC & As required for registration and audit & Check issuer, subject, status, and hash relationships \\
Node authorization VC & For nodes and verification components & Expose publicly verifiable parts and protect runtime credentials \\
Root proof & For distribution and discovery verification & Bind the resource version and ResourcePackage hash \\
VP & Only to the designated verifier & Bind holder, purpose, challenge, and validity period \\
Local identity backup & Retained only by the controller & Highly sensitive material; prohibit entry into logs and public indexes \\
\end{longtable}
}

Minimal disclosure must not remove binding fields required for verification. If hiding a field prevents the verifier from confirming the issuer, subject, version, hash, or proof, controlled presentation or a verifiable digest should be used instead of unverifiable display text.

\section{Key Rotation and Compromise Response}\label{sec:07.12}

Key rotation should first establish a verifiable control relationship between the new public key and the existing DID, and only then make the old public key unavailable for new operations. A resource DID update should describe the new verification method, its purpose, and its effective version in the new DID Document. Node key rotation must also synchronize the node authorization VC, node configuration, and upstream trust materials. Root platform key rotation must take downstream caches, historical proofs, and verification windows into account.

When a key is compromised, the affected key should be suspended, the old public key and status required to verify historical materials should be retained, and the revocation, replacement, or recovery fact should be published through a new control relationship. If the control private key is completely lost and no recovery key or transfer mechanism exists, the system cannot restore control merely from the DID string; it must enter a governance intervention or re-registration process.

\section{Registration VC and Resource Publication Evidence}\label{sec:07.13}

Two types of proof form a continuous but independent fact layer. After the Registrar node receives a successful response from the Root platform interface, it generates a registration VC. The Root platform produces a Root proof after verifying the resource control proof, metadata, ResourcePackage, version, hashes, and publication conditions. An HTTP success code or \passthrough{\lstinline!rootResponse!} is only a transport result and cannot replace verification of proof content and status.

\begin{lstlisting}
Resource control proof -> Registrar node accepts the registration
       -> Root platform verifies resource, version, hashes, and publication conditions
       -> Root proof
       -> Content distribution platform distributes the material
       -> Discovery node verifies and indexes it
       -> Caller verifies it again and invokes it according to policy
\end{lstlisting}

Registration acceptance, Root platform acceptance, material distribution, Discovery node indexing, and caller authorization must be expressed separately. No single VC or publication proof may compress these states into ``the resource is already trusted and available.''

\section{Node Authorization VC and Infrastructure Permissions}\label{sec:07.14}

Node actions are jointly constrained by role, authorization domain, capability scope, and governance state. A Registrar node may process only resources covered by its authorization domain, and a Discovery node may only synchronize and return resources within its permitted scope. A node\textquotesingle s ability to verify a material does not mean that the node is authorized to perform every network action. The authorization domain should be checked for format, ordering, duplicates, hierarchical coverage, and consistency with the node VC.

\begin{lstlisting}
allow(action) = valid(node authorization VC)
             && active(governance status)
             && role permits(action)
             && granted domains cover(requested domains)
             && freshness is acceptable
\end{lstlisting}

This expression represents a decision relationship rather than fixed implementation code. Errors should distinguish node authorization failure, resource control failure, resource content failure, and upstream service failure so that auditing, client handling, and operations diagnosis remain possible.

\section{Credential Revocation, Suspension, and Status Propagation}\label{sec:07.15}

A credential with a valid signature may still be unusable for a new authorization decision because of a governance event, a status record, or \passthrough{\lstinline!expiresAt!}. Status propagation may pass through on-chain governance events, Root platform status, the off-chain trust indexer, and node caches. Therefore, the verification result should carry the status source, observation time, and freshness information.

\subsection{Credential Status Changes}\label{subsec:07.15.1}

Every status change should be associated with the credential identifier, subject, old status, new status, reason, event sequence, and time, and should be produced by an authorized party. Restoring a suspended credential and issuing a new credential are different operations. Whether a new credential may be issued after revocation should be determined by governance and issuance policies.

\subsection{Node Authorization VC Status}\label{subsec:07.15.2}

After a node authorization VC is suspended or revoked, a Registrar node or Discovery node must not continue to perform new restricted actions. Whether existing materials may continue to be provided can only be determined according to cache freshness and risk policy. A historical VC must not be used to infer that the node still has permission to submit or index new content.

\subsection{Resource Registration VC Status}\label{subsec:07.15.3}

A change in registration VC status affects the current validity of the registration proof, but does not necessarily mean that the resource DID must be deleted. Historical Documents and audit materials may be retained while the current version is marked suspended, revoked, or undiscoverable. The verifier should check the VC status, resource lifecycle, and Root publication status separately.

\subsection{Status Propagation and Cache Updates}\label{subsec:07.15.4}

When a node receives a newer governance event or status cursor, it should verify sequence continuity, event completeness, and proof before updating its local cache. When the cursor is behind, events are missing, the status source is unavailable, or the cache exceeds the allowed freshness, sensitive actions should be suspended or return an indeterminate status rather than continuing to use the old active judgment.

\section{Signature Formats, Normalization, and Verification Test Vectors}\label{sec:07.16}

Cross-language implementations must unify the signed source text, field names, field ordering, null handling, time format, hash algorithm, public-key encoding, and proof fields. \passthrough{\lstinline!signature\_input!}, \passthrough{\lstinline!build\_data\_integrity\_proof!}, and \passthrough{\lstinline!verify\_payload\_with\_proof!} in the reference implementation show the boundaries for payload construction, proof generation, and verification. Other language implementations should produce an equivalent byte sequence rather than compare only in-memory objects.

{\def\LTcaptype{none} 
\begin{longtable}[]{@{}>{\raggedright\arraybackslash}p{0.2400\linewidth}>{\raggedright\arraybackslash}p{0.2400\linewidth}>{\raggedright\arraybackslash}p{0.2400\linewidth}>{\raggedright\arraybackslash}p{0.2400\linewidth}@{}}
\toprule\noalign{}
Vector category & Normal example & Failure example & Acceptance criterion \\
\midrule\noalign{}
\endhead
\bottomrule\noalign{}
\endlastfoot
Public-key parsing & Multiple encodings and JWK restore the same public key & Invalid prefix, length, or coordinate & Explicit public-key encoding error \\
Payload signature & Verification succeeds over the same normalized source & Any field or \passthrough{\lstinline!proofValue!} is modified & Verification fails \\
Control proof & DID, hash, purpose, and challenge match & Any binding field differs & Reject before publication \\
VC status & Active and not expired & Revoked, suspended, or expired & Authorization decision fails \\
Node request & Audience, path, nonce, and time are valid & Replayed nonce or wrong audience & Request rejected with no publication side effect \\
\end{longtable}
}

Test vectors should be versioned and should cover the Rust reference implementation, TypeScript SDK, and node HTTP layer. Test private keys must use dedicated materials; production private keys must not enter test files or logs. Logs may retain non-sensitive evidence such as public keys, hashes, request IDs, and error codes.

\section{Reference Sources}\label{sec:07.17}

{\def\LTcaptype{none} 
\begin{longtable}[]{@{}>{\raggedright\arraybackslash}p{0.3200\linewidth}>{\raggedright\arraybackslash}p{0.3200\linewidth}>{\raggedright\arraybackslash}p{0.3200\linewidth}@{}}
\toprule\noalign{}
Source & Type & Link \\
\midrule\noalign{}
\endhead
\bottomrule\noalign{}
\endlastfoot
\passthrough{\lstinline!oan-protocol-commons!} & Code repository: cryptography, credentials, signatures, and identity structures & \url{https://github.com/wolfbrother/oan-protocol-commons} \\
\passthrough{\lstinline!oan-sdk-ts!} & Code repository: identity bundles, key, and credential client processing & \url{https://github.com/OpenAgenet/oan-sdk-ts} \\
OAN Yellow Paper & arXiv yellow paper: identity, VCs, proofs, and verification boundaries & \url{https://arxiv.org/abs/2606.03163} \\
\passthrough{\lstinline!did:oan!} DID Method Specification & Standard/method specification: DID Documents and verification methods & \url{https://github.com/OpenAgenet/oan-public-docs/blob/main/did-oan-specs/doc/OAN\%20DID\%20Method\%20Specification.md} \\
\end{longtable}
}

\chapter{Registration Protocol}\label{chap:08}

This chapter specifies the process by which resources enter OpenAgenet (OAN) from a client. After the client creates the resource DID, DID Document, resource metadata, ResourcePackage, and resource control proof, it submits them to a Registrar node. The Registrar node performs entry validation and authorization-domain validation, and then calls the Root platform\textquotesingle s resource verification and publication interface through a signed upstream request. The current reference implementation uses \passthrough{\lstinline!/resources/register!} and the compatible \passthrough{\lstinline!/resources/submit!} entry points. The successful path includes a successful Root platform request, registration VC issuance, registration-record persistence, and response delivery.

When the Registrar node returns \passthrough{\lstinline!status: submitted!}, this only means that the registration process has obtained successful results from that node and its upstream call. It does not mean that the resource has completed network-wide publication, content distribution, or discovery indexing. The client should save and display the registration result, Root proof, and the indexing status of Discovery service nodes separately.

The registration path can be summarized as ``the client prepares the materials, the Registrar node accepts and validates them, the Root platform verifies and publishes them, and the Registrar node returns the registration result.'' A client must not infer discovery visibility solely from a success message on a page. It should determine the three stages separately from the resource registration record, the content returned by the Root platform, and the query result from a Discovery service node. The public interface of a Registrar node is the resource submission entry point; upstream calls between nodes belong to a protected service-to-service protocol. Their request formats and signing subjects must not be mixed.

\pandocbounded{\includegraphics[width=\linewidth,height=0.72\textheight,keepaspectratio]{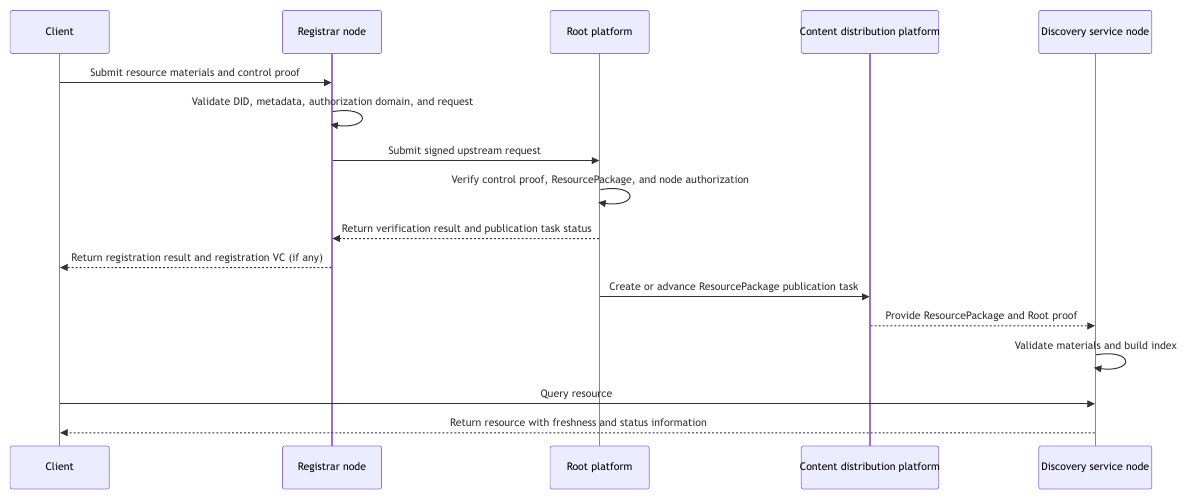}}

The Root platform verification, content distribution, and discovery indexing shown above are consecutive stages that may complete asynchronously. \passthrough{\lstinline!resource-verified-and-queued!} or \passthrough{\lstinline!status: submitted!} returned by a Registrar node only represents the corresponding registration or queueing fact. It cannot be used to infer that the content distribution platform has completed hosting, or that a Discovery service node has built a queryable index.

{\def\LTcaptype{none} 
\begin{longtable}[]{@{}>{\raggedright\arraybackslash}p{0.2400\linewidth}>{\raggedright\arraybackslash}p{0.2400\linewidth}>{\raggedright\arraybackslash}p{0.2400\linewidth}>{\raggedright\arraybackslash}p{0.2400\linewidth}@{}}
\toprule\noalign{}
Registration scenario & Controller in candidate document & Control material required now & Permission boundary after registration \\
\midrule\noalign{}
\endhead
\bottomrule\noalign{}
\endlastfoot
Self-controlled resource & Resource DID itself & Resource Local Identity signs document and challenge & Resource authorizes later updates itself \\
Existing controller & Another registered DID & Resource material plus that controller\textquotesingle s authorization & Controller authorizes document updates; it does not automatically hold business keys \\
New dedicated controller & New controller resource itself & New controller Local Identity; register it as an ordinary resource first & May be declared by other resources; receives no infrastructure privilege \\
\end{longtable}
}

\begin{lstlisting}
Initial registration + controller == resourceDid
  -> verify the resource's self-control proof

Initial registration + controller != resourceDid
  -> verify authorization by the controller declared in the submitted document

Existing DID update
  -> read the current controller from Root authoritative state
  -> verify that current controller's authorization of the new document
\end{lstlisting}

\section{Registration Targets and States}\label{sec:08.1}

Registration state should distinguish the client\textquotesingle s local state, the Registrar node\textquotesingle s processing state, and the Root platform\textquotesingle s publication state. The following logical states are recommended:

{\def\LTcaptype{none} 
\begin{longtable}[]{@{}>{\raggedright\arraybackslash}p{0.2400\linewidth}>{\raggedright\arraybackslash}p{0.2400\linewidth}>{\raggedright\arraybackslash}p{0.2400\linewidth}>{\raggedright\arraybackslash}p{0.2400\linewidth}@{}}
\toprule\noalign{}
State & Entry condition & Permitted operations & Termination or transition condition \\
\midrule\noalign{}
\endhead
\bottomrule\noalign{}
\endlastfoot
Draft & Form or SDK resource materials have not been submitted & Edit, preview, regenerate & Materials pass local validation or are modified \\
Locally validated & Local Identity, fields, hashes, and control proof pass validation & Submit or continue editing & Returns to Draft when any bound field changes \\
Submitting & The client has sent a registration request & Wait for response or query status & A success, failure, or timeout result is received \\
Registered & The Registrar node has obtained the required upstream success result & Save VC and query publication status & Subsequent Root publication and discovery proceed independently \\
Credential pending & A registration result exists but the VC is missing or empty & Query or retry according to the interface rules & VC is obtained and verified, or exception handling begins \\
Duplicate & DID, version, and hashes are identical to existing materials & Return the existing result & No new business version should be created \\
Conflict or failed & Validation, control, authorization, version, or network conditions are not satisfied & Correct materials or handle according to retryability & Must not be displayed as network-wide publication success \\
\end{longtable}
}

\pandocbounded{\includegraphics[width=\linewidth,height=0.72\textheight,keepaspectratio]{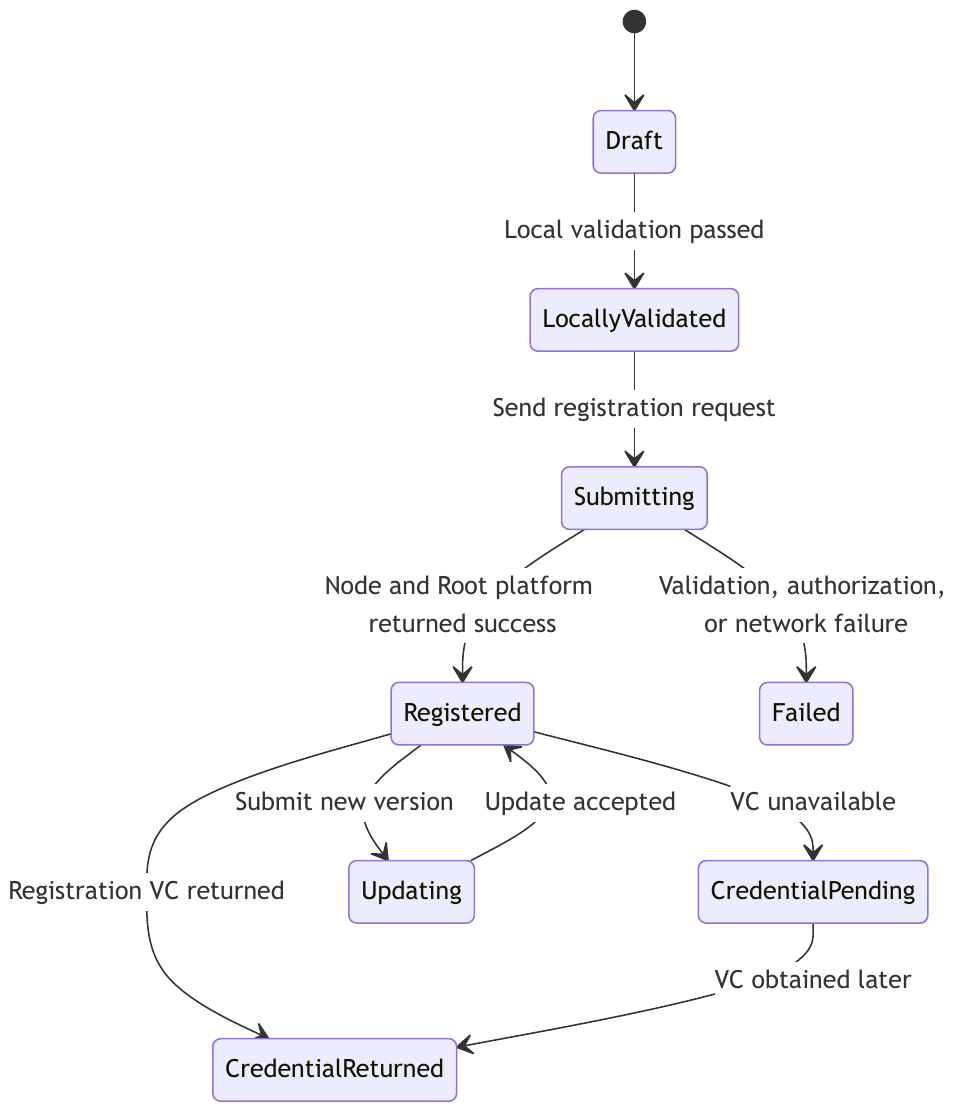}}

Passing browser preview or local validation does not create a registration record; a network timeout must not directly generate a new DID or new version. A retry with the same business materials should reuse the original resource DID, version, and hashes, and the server or query interface should be used to confirm the actual result.

\section{Registration Request Structure}\label{sec:08.2}

The core payload of the registration entry point is \passthrough{\lstinline!ResourceRegistrationSubmission!}. Its main fields and purposes are as follows:

{\def\LTcaptype{none} 
\begin{longtable}[]{@{}>{\raggedright\arraybackslash}p{0.3200\linewidth}>{\raggedright\arraybackslash}p{0.3200\linewidth}>{\raggedright\arraybackslash}p{0.3200\linewidth}@{}}
\toprule\noalign{}
Field & Source & Validation and purpose \\
\midrule\noalign{}
\endhead
\bottomrule\noalign{}
\endlastfoot
\passthrough{\lstinline!resourceDid!} & Resource identity generated by the client & Serves as the resource identity and primary key of the registration record \\
\passthrough{\lstinline!resourceType!} & Selected by the client & Must be consistent with DID Document metadata and package metadata; it is not encoded by the routing code \\
\passthrough{\lstinline!didDocument!} & Client draft & Provides public keys, verification methods, service endpoints, and OAN metadata \\
\passthrough{\lstinline!didDocumentHash!} & Computed by the client according to the protocol & Binds the control challenge to the publication materials \\
\passthrough{\lstinline!metadata!} & Client form or resource description & Validates the resource declaration and authorization domain \\
\passthrough{\lstinline!packageVersion!} & Client resource version & Distinguishes different versions under the same DID \\
\passthrough{\lstinline!packageHash!}, \passthrough{\lstinline!metadataHash!} & Computed by the client & Binds the integrity of the ResourcePackage and metadata \\
\passthrough{\lstinline!hashAlgorithm!} & Protocol and client configuration & Explains the algorithm prefix used by hash fields \\
\passthrough{\lstinline!registrationCredential!} & Protocol compatibility field & Cleared by the Registrar node when submitting to the Root platform \\
\passthrough{\lstinline!subjectControlProof!} & Signature from the resource controller & Proves that the current registration or update has control authorization \\
\end{longtable}
}

These fields can be divided into four groups according to validation order:

{\def\LTcaptype{none} 
\begin{longtable}[]{@{}>{\raggedright\arraybackslash}p{0.3200\linewidth}>{\raggedright\arraybackslash}p{0.3200\linewidth}>{\raggedright\arraybackslash}p{0.3200\linewidth}@{}}
\toprule\noalign{}
Field group & Fields & Main determination \\
\midrule\noalign{}
\endhead
\bottomrule\noalign{}
\endlastfoot
Identity and type & \passthrough{\lstinline!resourceDid!}, \passthrough{\lstinline!resourceType!}, \passthrough{\lstinline!didDocument!} & Whether the resource identity, resource type, and DID Document are mutually consistent \\
Integrity binding & \passthrough{\lstinline!didDocumentHash!}, \passthrough{\lstinline!metadataHash!}, \passthrough{\lstinline!packageHash!}, \passthrough{\lstinline!packageVersion!}, \passthrough{\lstinline!hashAlgorithm!} & Whether the submitted materials correspond to the same version and the same content \\
Business declarations & \passthrough{\lstinline!metadata!} & Whether capability, endpoint, authorization-domain, and lifecycle descriptions are complete \\
Authorization proof & \passthrough{\lstinline!subjectControlProof!}, \passthrough{\lstinline!registrationCredential!} & Whether the resource controller authorized the operation; the Registrar node does not forward an existing registration VC to the Root platform \\
\end{longtable}
}

The registration request is the complete \passthrough{\lstinline!ResourceRegistrationSubmission!} defined by the common protocol types; this chapter does not present an abbreviated object with ellipses as a usable request template.

Before calling the Root platform, the Registrar node should reject requests whose resource DID, DID Document identifier, challenge target, or hashes are inconsistent. Whether unknown extension fields are retained is determined by the protocol version. Unknown core fields, invalid types, or unparseable structures must not be treated as valid requests.

A valid request must at least satisfy all of the following relationships: \passthrough{\lstinline!didDocument.id == resourceDid!}; the resource type declared by the DID Document must be consistent with \passthrough{\lstinline!resourceType!}; the three hash fields must be interpretable according to \passthrough{\lstinline!hashAlgorithm!}; and the subject and document hash in the control challenge must correspond to the resource DID and \passthrough{\lstinline!didDocumentHash!}, respectively. For example, a request that changes \passthrough{\lstinline!resourceType!} from \passthrough{\lstinline!skill!} to \passthrough{\lstinline!agent\_service!} without simultaneously updating the DID Document and metadata should be rejected during local validation by the Registrar node. \passthrough{\lstinline!registrationCredential!} remains in the shared request model for compatibility and record-keeping, but the Registrar node clears it when constructing the upstream request to the Root platform, preventing a credential previously issued by the Registrar node from being used as the Root platform\textquotesingle s verification basis.

The key interaction for forwarding a request from the Registrar node to the Root platform can be summarized as follows:

\pandocbounded{\includegraphics[width=\linewidth,height=0.72\textheight,keepaspectratio]{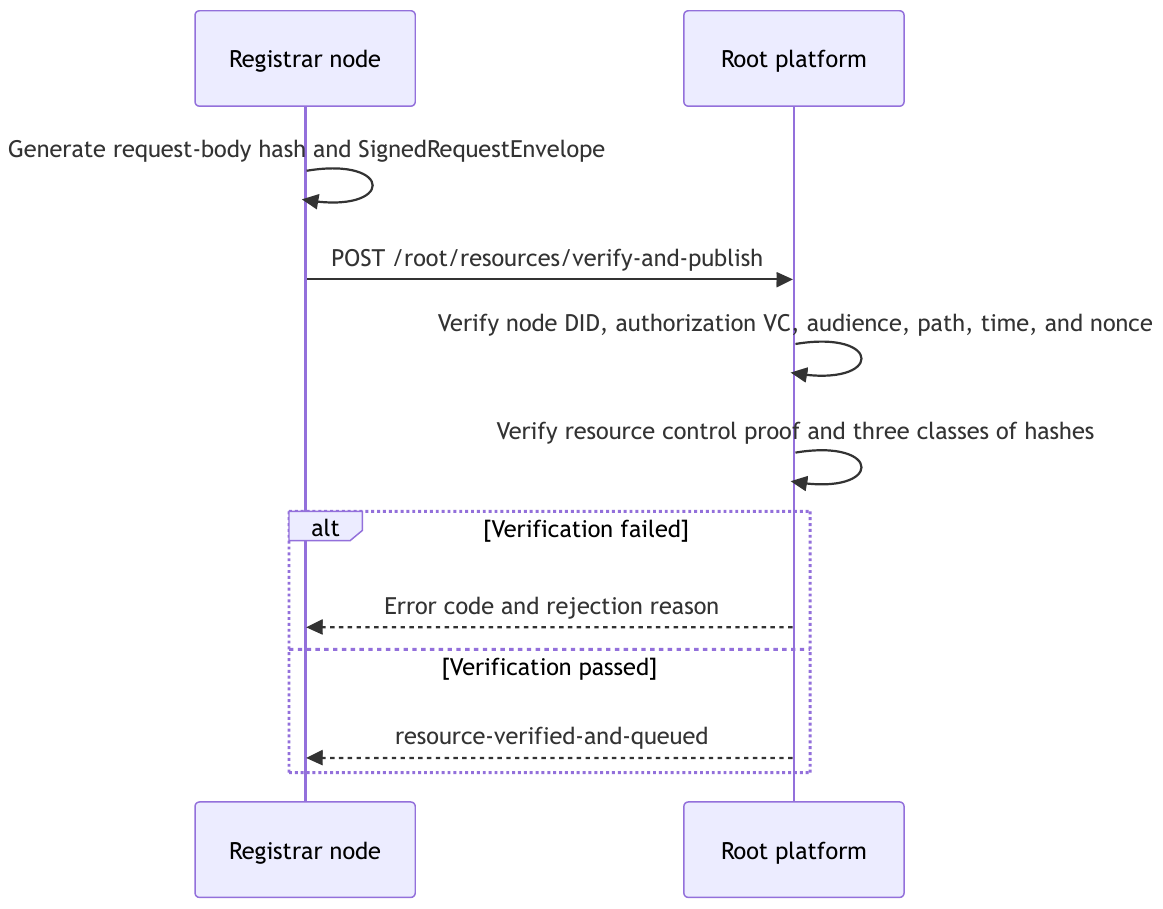}}

This interaction describes only the protected upstream call between the Registrar node and the Root platform. The client submission entry point, registration VC issuance, and subsequent content distribution are handled separately in the other sections of this chapter. A successful Root platform response does not change the fact that the Discovery service node may not yet have completed indexing.

\section{Local Identity Initialization and Import}\label{sec:08.3}

The Local Identity is a collection of materials in which the client stores subject configuration, resource control keys, and necessary credentials. It is not a user account on a Registrar node. It is used to create or restore control relationships and to issue control proofs for specific registration operations. A Registrar node should receive only the DID Document, public keys, proofs, and registration materials required for verification. It must not receive subject or resource control private keys.

The resource control relationship can be represented as follows: ``the Local Identity holds the control key, the resource DID Document publishes the verification method, and the control proof binds one specific operation to that resource and document version.'' Therefore, a Local Identity and a resource are not the same object. One subject configuration may control multiple resources, but each registration request must identify one \passthrough{\lstinline!resourceDid!} and use a proof matching that resource\textquotesingle s DID Document. The Registrar node verifies public materials and signature results, rather than the client\textquotesingle s assertion that ``this is my resource.''

\subsection{Create a New Local Identity}\label{subsec:08.3.1}

The client generates a subject control key and subject DID, and saves the public verification material and local configuration. It can then create an independent resource DID and resource control material for each resource. Creating a Local Identity does not itself create a Registrar node record and does not mean that a resource has obtained a network publication state.

\subsection{Import a Local Identity}\label{subsec:08.3.2}

The client parses a Local Identity backup and validates the correspondence among the subject DID, public key, private key, resource configuration, and credentials. After a successful import, the client can continue controlling existing resources, but a new registration, update, or retry must still create a new proof for the target DID, version, hash, and operation purpose.

\subsection{Private Keys and Subject Configuration}\label{subsec:08.3.3}

Private keys are used only on the client to issue control proofs or other protected signatures. Subject configuration describes the relationship between the subject DID and resources. Server requests and logs may contain public DIDs, public keys, signatures, hashes, and credentials, but must not contain private keys or plaintext materials from which private keys can be recovered.

\subsection{Identity Backup and Recovery}\label{subsec:08.3.4}

An identity backup should contain the structured information required to restore control relationships, and should indicate its sensitivity and version. After recovery, the client should check the key correspondence and DID Document consistency again. The existence of a backup file does not mean that the resource has been registered, and a recovery failure must not overwrite an already valid identity on the client.

\section{DID Document Draft Generation}\label{sec:08.4}

The client generates a DID Document from the resource type, name and description, capability tags, authorization domains, endpoints, version, Local Identity, and control key. \passthrough{\lstinline!id!} must equal \passthrough{\lstinline!resourceDid!}; \passthrough{\lstinline!oanMetadata.resourceType!} and \passthrough{\lstinline!subjectType!} must match the request; verification methods must provide resolvable public keys; and service endpoints and ResourcePackage information must be bound to the resource metadata.

The preview button may generate or reuse a local DID Document and display validation results, but expanding the preview must not issue a registration network request. If the form, Local Identity, advanced access settings, or ResourcePackage information has not changed, the client may reuse the generated document. When an input affecting the DID, document hash, version, or control proof changes, it must regenerate the document and sign it again.

To prevent the client from repeatedly producing different resource identities when the user has not changed the materials, the client should form an internal snapshot from the form values, the Local Identity subject DID, resource control configuration, endpoints, authorization domains, version, and ResourcePackage digest used to generate the DID Document. If the snapshot is identical and a preview document already exists, clicking the button only toggles the expanded state of the preview area. If the snapshot changes, the client regenerates the document and related hashes. This cache is only a client-experience optimization and cannot replace the Registrar node\textquotesingle s revalidation of the request materials.

\section{Pre-submission Local Validation}\label{sec:08.5}

Before submission, the client should at least check the following conditions: a Local Identity has been created or imported; a resource type has been selected; required fields such as name, description, endpoint, authorization domain, capability tags, and version are complete; the DID syntax matches the resource type; the DID Document, metadata, ResourcePackage, and hashes are consistent; the target, purpose, document hash, and verification method of the control challenge are correct; and the control proof can be verified with a public key in the DID Document. Error messages should identify the missing condition, such as ``Local Identity is required,'' ``Resource endpoint is required,'' ``No authorization domain selected,'' or ``DID Document hash is inconsistent.''

The preview button performs only local generation and validation. Before sending a request, the submit button reuses the validation result. Neither button should submit automatically when the page is opened. A failed validation must not write a registration record to the Registrar node, a publication record to the Root platform, or a successful registration VC.

Local validation results should preferably be returned by field so that the UI and SDK can take deterministic actions:

\begin{lstlisting}
{
  "valid": false,
  "errors": [
    {"field": "localIdentity", "code": "required", "message": "Local Identity is required"},
    {"field": "serviceEndpoint", "code": "required", "message": "Resource endpoint is required"}
  ]
}
\end{lstlisting}

When \passthrough{\lstinline!valid: false!}, the client must not enter the submission request. Network unavailability is a submission-stage error and must not be disguised as a local-field error. A successful preview proves only that the current browser can generate and validate the draft; it does not prove that the Registrar node or Root platform has accepted the materials.

\section{Resource Control Proof Verification}\label{sec:08.6}

The Registrar node should verify the proof payload, signature, and business context together. The target resource DID, DID Document hash, verification method, operation purpose, challenge Registrar node, time window, and nonce must match the current request; the public key and signature algorithm must be parseable. A missing, expired, wrong-purpose, wrong-target, hash-inconsistent, or reused proof should be rejected before the Root platform is called.

A failed control proof must not produce a successful registration state or registration VC. The node may save the challenge ID, resource DID, document hash, verification method, verification time, and error code for auditing, but must not save the signing private key.

The validation order should first check structure and context, then verify the signature, and only then begin the publication call:

\begin{lstlisting}
Request is parseable
  -> challenge.subjectDid == resourceDid
  -> challenge.didDocumentHash == request.didDocumentHash
  -> challenge.registrarDid == current Registrar node DID
  -> purpose, issuedAt, expiresAt, and nonce are valid
  -> verificationMethod is in the DID Document's assertionMethod
  -> Resolve public key and verify signature
  -> Continue with the registration flow
\end{lstlisting}

Any failure must stop the subsequent Root platform call. A successful signature check proves only that the signer possesses the corresponding private key and that the signed content has not been modified. Node authorization scope, resource type, version, and Root platform publication conditions must still be verified separately.

\section{Registrar Node Authorization and Scope Validation}\label{sec:08.7}

The current Registrar node reads the authorization domains in its own DID Document, checks consistency between the authorization domains in the resource DID Document and those in the request metadata, and then determines whether the node\textquotesingle s authorization domains cover the resource domains. Authorization domains must be non-empty, correctly formatted, deterministically ordered, and free of privilege escalation. When they are inconsistent or exceed the node\textquotesingle s scope, the node should return an error before calling the Root platform.

The node signing key proves only that the request came from the node\textquotesingle s operating boundary. The node authorization VC proves that the node has the corresponding role and scope. The resource control proof is a separate condition.

The request sent by a Registrar node to the Root platform is bound by \passthrough{\lstinline!SignedRequestEnvelope!} to a request ID, protocol version, purpose, HTTP method, path, audience, timestamp, nonce, request-body hash, and node proof. The Root platform can use these values to check which Registrar node submitted which materials, for what purpose, and at what time. The node-to-node signature does not replace the resource control proof, and does not automatically make the Registrar node the resource controller.

\section{Resource Type, Metadata, Endpoint, and ResourcePackage Validation}\label{sec:08.8}

The resource type in the submission, DID Document metadata, resource metadata, and ResourcePackage forms one consistency chain; the DID routing code is checked separately against the Registrar route. The resource DID, type, version, DID Document digest, metadata digest, and ResourcePackage digest must refer to the same material. Endpoints must conform to their declared protocol and format. Capability tags follow the governed tree or permitted extension rules, and free-form text cannot bypass authorization domains.

The current \passthrough{\lstinline!validate\_shape()!} covers DID syntax, resource-type relationships, DID Document identifiers, resource structure, ResourcePackage version, hash algorithm, three classes of hash references, challenge target DID, document hash, and operation purpose. Whether an endpoint is reachable, whether the business is safe, and whether the resource meets the caller\textquotesingle s quality requirements are judgments for later publication or pre-invocation stages and cannot be replaced by shape validation.

\section{Duplicate Registration and Idempotency}\label{sec:08.9}

Business idempotency should be determined by the resource DID, resource type, version, DID Document hash, metadata hash, and ResourcePackage hash, rather than by submission time or VC issuance time. The request ID and nonce provide request correlation, freshness, and replay protection; they do not alone determine whether two requests represent the same resource version.

\subsection{Repeated Registration of the Same DID}\label{subsec:08.9.1}

When the same DID is submitted again, the server should verify control first, and then compare the version and hashes. Identical materials should return the existing registration result or an explicit idempotent result. The same DID with a different version or any different hash should enter the same-DID update decision. Invalid control proof must not overwrite the existing record.

\subsection{Basis for Idempotency Decisions}\label{subsec:08.9.2}

An idempotency decision depends at least on the resource DID, resource type, version, DID Document hash, metadata hash, and ResourcePackage hash. The system should also save the request ID, processing time, and result to distinguish a business retry from a new version. The current Registrar node writes records using an upsert by resource DID to SQLite, PostgreSQL, or file storage. This can preserve the current record, but it is not by itself evidence of a complete version-level idempotency history.

\subsection{VC Reissuance}\label{subsec:08.9.3}

A retry for the same business version may return the saved registration VC, or may issue a content-equivalent VC with a different issuance time according to an explicit policy. The client should identify the credential by the resource DID, version, hashes, and state, and must not treat a new issuance time as a resource version update.

\subsection{Duplicate Publication Control at the Root Platform}\label{subsec:08.9.4}

The Root platform should make repeated publication of the same DID, version, and hashes produce one business result. A different hash or version should enter an update, conflict, or rejection path. A successful upstream HTTP response received by a Registrar node is not sufficient to prove that the Root platform completed duplicate-publication deduplication; the publication proof and sequence record are the evidence for that determination.

\section{Resource Updates and Same-DID Version Updates}\label{sec:08.10}

An update under the same DID preserves the resource identity, but must produce a new version, document hash, or ResourcePackage hash and express the predecessor relationship through fields such as \passthrough{\lstinline!previousVersion!}. The new version must use the current control relationship to generate a new control proof. An old registration VC or old ResourcePackage cannot automatically authorize a future update. Version increments, concurrent conflicts, history retention, and rollback are jointly enforced by the Registrar node and Root platform. The specific publication replacement rules are described in Chapter~\ref{chap:09}.

\subsection{Update Authorization Verification}\label{subsec:08.10.1}

The verifier must confirm that the submitter is still authorized by the current DID control relationship, and must check that the control proof is bound to the new document hash, version, purpose, and challenge. An update supported only by an old credential without a new control proof for the new version must be rejected.

\subsection{DID Document Updates}\label{subsec:08.10.2}

An updated DID Document continues to use the original resource DID, and its document hash must be included in the registration submission and ResourcePackage. Adding, replacing, or deactivating verification methods requires a verifiable control chain. If the update cannot be proven to have been authorized by the current controller, the current document must not be replaced.

\subsection{Resource Metadata and Endpoint Updates}\label{subsec:08.10.3}

Changes to the resource description, capability tags, authorization domains, protocol bindings, or endpoints should produce a new metadata hash and ResourcePackage hash. Expanding the authorization domain cannot be achieved by changing text fields alone. Endpoint reachability also cannot replace business security and invocation-authorization checks.

\subsection{New Version Publication and Old-Version Handling}\label{subsec:08.10.4}

After a new version is accepted, the Root platform should clearly identify the current version, historical versions, and old-version states. Discovery service nodes synchronize according to the publication sequence. An old version may be retained for auditing or rollback, but must not be confused with the latest version in query results.

\section{Registration VC Issuance and Retrieval}\label{sec:08.11}

In the current reference implementation, the registration credential generation logic is called only after the Root platform\textquotesingle s resource verification and publication interface returns success, and the registration VC is written to the registration record and successful response. The issuer of the registration VC is the Registrar node, and its credential subject is the resource DID. Its claims include registration facts such as resource type, version, hashes, authorization domains, and lifecycle.

\subsection{Conditions for VC Issuance}\label{subsec:08.11.1}

Before issuance, entry-shape validation, resource authorization-domain validation, control-proof processing, and the Root platform request must be completed. Passing local preview, clicking Submit, or the Registrar node receiving a request is not by itself a condition for issuance.

\subsection{VC Delivery}\label{subsec:08.11.2}

The current successful response contains \passthrough{\lstinline!status!}, \passthrough{\lstinline!resourceDid!}, \passthrough{\lstinline!resourceType!}, \passthrough{\lstinline!registrationCredential!}, and \passthrough{\lstinline!rootResponse!}. The client should save the original VC and associate it with the DID Document, version, and hashes. \passthrough{\lstinline!rootResponse!} must still be parsed according to the Root proof and status rules.

\subsection{Handling a Missing or Empty VC}\label{subsec:08.11.3}

If the registration result is successful but the VC is missing or empty, the client should record the registration state and credential state separately, and clearly indicate that the credential has not yet been obtained. It must not fabricate an empty VC or treat an empty value as a valid credential. Queries or retries should retain the original DID, version, and hashes.

\subsection{Local VC Storage}\label{subsec:08.11.4}

The registration VC should be associated with the resource DID Document and resource version in local files. When the client packages and downloads the VC, DID Document, and Local Identity backup, the private key remains sensitive material that may be kept only by the controller. It must not be sent to a node or public storage.

\section{Registration Responses and Error Handling}\label{sec:08.12}

The current Registrar node handler receives \passthrough{\lstinline!Json<ResourceRegistrationSubmission>!} and returns dynamic JSON on success. The stable core fields of a successful response are \passthrough{\lstinline!status!}, \passthrough{\lstinline!resourceDid!}, \passthrough{\lstinline!resourceType!}, \passthrough{\lstinline!registrationCredential!}, and \passthrough{\lstinline!rootResponse!}. \passthrough{\lstinline!rootResponse!} preserves dynamic content returned by the Root platform and must not be assumed by the client to be a fixed VC or an indication that network-wide publication is complete.

{\def\LTcaptype{none} 
\begin{longtable}[]{@{}>{\raggedright\arraybackslash}p{0.3200\linewidth}>{\raggedright\arraybackslash}p{0.3200\linewidth}>{\raggedright\arraybackslash}p{0.3200\linewidth}@{}}
\toprule\noalign{}
Stage & Typical result & Client interpretation \\
\midrule\noalign{}
\endhead
\bottomrule\noalign{}
\endlastfoot
Local validation by Registrar node & \passthrough{\lstinline!4xx!} or structured error & The request did not enter the Root publication flow; correct fields, hashes, or control proof first \\
Root verification and queueing & \passthrough{\lstinline!status: resource-verified-and-queued!} & The Root platform has verified the request and created a subsequent task; content distribution and discovery indexing are not necessarily complete \\
Successful registration by Registrar node & \passthrough{\lstinline!status: submitted!} & The Registrar node has recorded the registration result; the response may contain a registration VC and Root response \\
Temporary network or upstream failure & \passthrough{\lstinline!5xx!}, connection error, or timeout & Query the status using the resource DID, version, hashes, and request identifier before deciding whether to retry \\
\end{longtable}
}

The following handler relationship describes the actual input and output types. It is a concise representation of the interface contract, not a complete implementation that can be substituted directly:

\begin{lstlisting}
async fn register_resource(
    State(state): State<AppState>,
    Json(submission): Json<ResourceRegistrationSubmission>,
) -> ApiResult<serde_json::Value>;
\end{lstlisting}

The \passthrough{\lstinline!ResourceVerifyAndPublishRequest!} exchanged between the Registrar node and Root platform also contains \passthrough{\lstinline!registrarDid!}, the original registration submission, and \passthrough{\lstinline!upstreamAuth!}. A client must not construct this internal forwarding object itself, and must not treat the Registrar node\textquotesingle s \passthrough{\lstinline!registrationCredential!} as the Root platform\textquotesingle s Root proof. The \passthrough{\lstinline!error!}, status code, and request-correlation identifier in an error response should be used together to decide whether to correct, query, retry, or stop.

{\def\LTcaptype{none} 
\begin{longtable}[]{@{}>{\raggedright\arraybackslash}p{0.3200\linewidth}>{\raggedright\arraybackslash}p{0.3200\linewidth}>{\raggedright\arraybackslash}p{0.3200\linewidth}@{}}
\toprule\noalign{}
Result & Meaning & Client handling \\
\midrule\noalign{}
\endhead
\bottomrule\noalign{}
\endlastfoot
Success with VC & The Registrar node completed the current flow and returned the registration credential & Save the VC and separately confirm Root publication and discovery status \\
Success but VC unavailable & The registration result exists but the credential is incomplete & Preserve the DID, version, and hashes; query or retry \\
Validation failure & The request structure, type, hash, or proof is invalid & Correct the specific field; do not retry blindly \\
Authorization failure & The node scope or resource control authority is insufficient & Check the authorization VC, authorization domains, and control relationship \\
Upstream failure & The Root platform rejected the request or is unavailable & Handle according to retryability; do not declare publication success \\
Network timeout & The client did not receive a definitive response & Preserve the original materials, query, or retry with the same business materials \\
Duplicate or conflict & The same version already exists or different content is present & Return the existing result, or enter update and conflict handling \\
\end{longtable}
}

In the current implementation, errors are returned mainly through the HTTP status and \passthrough{\lstinline!error!} text. A stable future protocol should add a request ID, error code, registration state, upstream state, and retryable flag so that clients do not have to infer network state from free text.

Response handling is recommended to follow these deterministic rules:

When an HTTP request times out without a response, the client can record only ``result unknown,'' retain the original DID, version, and hashes, and query or retry with the same materials. It must not generate a new DID on that basis. When a non-success HTTP status is received, the result must not be saved as a successful registration. When a success status is received but the VC is empty, the registration state and credential state should be displayed separately; only a credential that was actually obtained and verified may be downloaded as a VC file.

\section{Registration Audit Records}\label{sec:08.13}

A registration record should at least associate the resource DID, resource type, version, DID Document hash, metadata hash, ResourcePackage hash, authorization domains, Registrar node, registration VC, Root response summary, submission time, and processing result. The current \passthrough{\lstinline!write\_resource\_record()!} writes by resource DID to SQLite, PostgreSQL, or file storage. A complete audit implementation should therefore also retain the operation type, request ID, previous and new versions, and failure reason, so that an upsert does not erase the historical context.

Audit records must not store resource control private keys, Local Identity backups, or recoverable secrets. DID, public key, hash, challenge ID, request ID, state, and error code may be used as trace evidence. Audit-read permissions must be separated from the public discovery interface.

\section{Resource Update Authorization and Same-DID Replacement}\label{sec:08.14}

After receiving new materials for the same DID, the server should confirm the control proof, version relationship, and hash relationship in sequence. The same version with the same hashes is an idempotent retry; the same version with different hashes is a conflict; a new version with a valid predecessor relationship may enter the update flow; version rollback, a missing predecessor, an unapproved authorization change, or an invalid control proof should be rejected or sent for manual handling. The Registrar node may save the acceptance record, but the Root platform determines the network publication fact.

The following table can be used to check replacement decisions:

{\def\LTcaptype{none} 
\begin{longtable}[]{@{}>{\raggedright\arraybackslash}p{0.1920\linewidth}>{\raggedright\arraybackslash}p{0.1920\linewidth}>{\raggedright\arraybackslash}p{0.1920\linewidth}>{\raggedright\arraybackslash}p{0.1920\linewidth}>{\raggedright\arraybackslash}p{0.1920\linewidth}@{}}
\toprule\noalign{}
DID & Version & Content hash & Control proof & Handling \\
\midrule\noalign{}
\endhead
\bottomrule\noalign{}
\endlastfoot
Same & Same & Same & Valid & Handle idempotently as the same business version \\
Same & Same & Different & Valid & Reject as a conflict; do not overwrite the current version \\
Same & Updated & Different & Valid and bound to new materials & Process through the update flow \\
Same & Updated or same & Any & Invalid & Reject without replacement side effects \\
Different & Any & Any & Valid & Process as a new resource registration \\
\end{longtable}
}

\section{Submission Retries, Idempotency Keys, and Duplicate Handling}\label{sec:08.15}

The upstream request from a node to the Root platform uses \passthrough{\lstinline!SignedRequestEnvelope!} to bind the request ID, protocol version, purpose, method, path, audience, request time, request nonce, request-body hash, and proof. The security layer uses nonce storage and a time window for replay protection. Business idempotency depends on the resource DID, version, and content hashes. After a client timeout, it should preserve the original materials, first query the existing result or retry with the same business materials, and must not immediately generate a new DID to conceal an unknown result.

\section{Registration Rate Limiting and Abuse Control}\label{sec:08.16}

The public registration entry point should limit the request rate per time period, request-body and ResourcePackage size, failure count per source, same-DID conflict count, and upstream-forwarding concurrency. Control-proof, node-authorization, and authorization-domain checks should be performed early to reduce the storage, network, and computation pressure caused by forged materials. Rate limiting must not misclassify a legitimate same-DID idempotent query as a new version, and throughput must not be gained by weakening control-proof checks.

The current reference implementation explicitly includes shape validation, authorization-domain coverage checks, upstream request signing, and nonce replay protection. If a reverse proxy, WAF, or service-level rate limiter is enabled in deployment, the actual policy, response codes, and recovery method should be documented in the deployment documentation.

\textbf{Registration Flow Acceptance Checklist}

\begin{itemize}
\tightlist
\item[$\square$]
  For an empty form, missing Local Identity, missing endpoint, or missing authorization domain, the client identifies the specific condition.
\item[$\square$]
  Before calling the Root platform, the Registrar node completes request-shape, resource-authorization-domain, and control-proof validation.
\item[$\square$]
  When the Root platform fails or the network times out, the client does not display network-wide publication success and does not lose the original DID and version materials.
\item[$\square$]
  The registration VC in a successful response can be associated with the resource DID, version, and three classes of hashes.
\item[$\square$]
  Retries with the same DID, version, and hashes receive different handling from updates with a different hash or version.
\item[$\square$]
  Registration, Root publication, content distribution, and discovery indexing states can be observed and verified separately.
\item[$\square$]
  Audit records can trace requests without containing private keys, Local Identity backups, or other sensitive secrets.
\end{itemize}

\section{References}\label{sec:08.17}

{\def\LTcaptype{none} 
\begin{longtable}[]{@{}>{\raggedright\arraybackslash}p{0.3200\linewidth}>{\raggedright\arraybackslash}p{0.3200\linewidth}>{\raggedright\arraybackslash}p{0.3200\linewidth}@{}}
\toprule\noalign{}
Source & Type & Link \\
\midrule\noalign{}
\endhead
\bottomrule\noalign{}
\endlastfoot
\passthrough{\lstinline!oan-registrar-nodes!} & Code repository: registration entry point, validation, registration records, and registration VC & \url{https://github.com/OpenAgenet/oan-registrar-nodes} \\
\passthrough{\lstinline!oan-root-nodes!} & Code repository: Root platform resource verification and publication interfaces & \url{https://github.com/OpenAgenet/oan-root-nodes} \\
\passthrough{\lstinline!oan-protocol-commons!} & Code repository: registration payloads, error codes, and cross-node protocol types & \url{https://github.com/wolfbrother/oan-protocol-commons} \\
\passthrough{\lstinline!oan-sdk-ts!} & Code repository: client registration materials and registration calls & \url{https://github.com/OpenAgenet/oan-sdk-ts} \\
\passthrough{\lstinline!oan-community-skill!} & Code repository: registration-flow scripts and developer invocation examples & \url{https://github.com/OpenAgenet/oan-community-skill} \\
OAN Yellow Paper & arXiv yellow paper: registration, publication, and verification flow & \url{https://arxiv.org/abs/2606.03163} \\
OAN Resource Identity and Discovery & IETF draft: relationship between resource identity, registration, and discovery & \url{https://datatracker.ietf.org/doc/draft-xu-oan-resource-identity-discovery/} \\
\end{longtable}
}

\chapter{Root Platform Verification and Trusted Publication}\label{chap:09}

This chapter explains how the Root platform converts resource materials submitted by a Registrar Service Node into verifiable network-level publication facts. It covers upstream request authentication, resource DID and control-proof verification, node authorization, ResourcePackage integrity, publication policy, issuance of the Root Platform Trusted Publication Proof, version selection, distribution tasks, and failure recovery. It continuously distinguishes among ``verified and queued,'' ``publication fact formed,'' ``content distributed,'' and ``indexed by a Discovery Service Node.''

\section{Root Platform Publication Responsibilities}\label{sec:09.1}

\pandocbounded{\includegraphics[width=\linewidth,height=0.72\textheight,keepaspectratio]{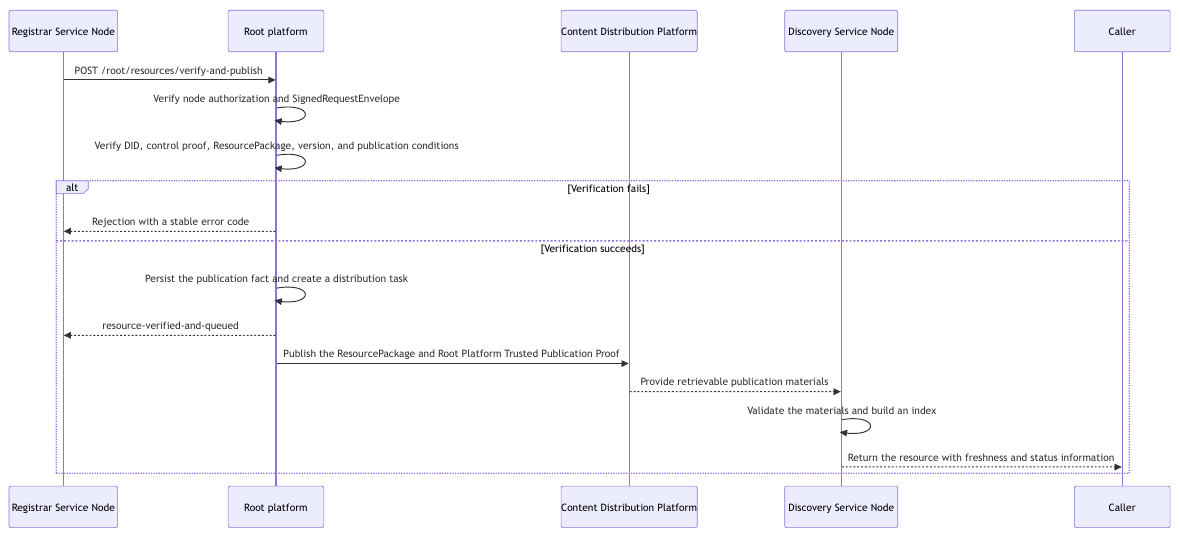}}

``Verified and queued,'' content distribution, and discovery indexing in this flow correspond to different completion conditions. A successful response from the Root platform means only that the resource has passed Root-platform processing and entered the distribution task chain. The Content Distribution Platform and Discovery Service Node must still complete their own validation, publication, or indexing, and the caller should independently verify the DID, registration VC, Root Platform Trusted Publication Proof, and status information.

The Root platform is an independent verification boundary before a resource enters the network-level publication path. It receives registration materials submitted by an authorized Registrar Service Node, re-verifies the resource DID, DID Document, resource control proof, ResourcePackage, node authorization, version, and publication conditions, and writes the verified resource version to the publication record. Pre-validation by a Registrar Service Node is an access-layer check and cannot replace the Root platform's independent confirmation of the publication fact.

In operational terms, Root-platform ``acceptance'' contains two consecutive but distinguishable facts: first, the resource materials pass Root-platform verification and are persisted; second, a publication task for the Content Distribution Platform is created. The current successful response is \passthrough{\lstinline!resource-verified-and-queued!}. Therefore, a Registrar Service Node or client should interpret it as ``verified and queued,'' not directly display it as ``network-wide publication completed.''

\section{Upstream Submission by a Registrar Service Node}\label{sec:09.2}

A Registrar Service Node submits a resource registration request with upstream authentication, rather than uploading an unproven ResourcePackage. The request should be associated with the Registrar Service Node DID, request identifier, protocol version, target path, time, nonce, request-body hash, and signature, and should carry the resource registration submission and its control proof. The Root platform should first verify the submission source and request integrity, and only then process the resource content. A request that is duplicated, expired, or addressed to the wrong audience must not enter the publication state.

The current entry point is \passthrough{\lstinline!POST /root/resources/verify-and-publish!}. After receiving the request, the Root platform loads authorization materials according to the Registrar Service Node DID, verifies the \passthrough{\lstinline!SignedRequestEnvelope!} signature, method, path, audience, time, nonce, and request-body hash, and only then performs structural and content validation of the resource submission. The upstream signature proves ``which Registrar Service Node submitted these materials.'' It is neither a resource control proof nor a Root Platform Trusted Publication Proof.

The request type for this entry point is \passthrough{\lstinline!ResourceVerifyAndPublishRequest!}. Its minimum relationship can be represented as follows:

The current successful response is dynamic JSON. Its core result is \passthrough{\lstinline!status: resource-verified-and-queued!}, together with the resource DID, resource type, package version, three classes of hashes, lifecycle state, and \passthrough{\lstinline!cdnDispatchStatus: queued!}. This means that the Root platform has completed the current verification and created a Content Distribution Platform task. Receipt by the Content Distribution Platform and indexing by a Discovery Service Node must be confirmed through their respective status or synchronization records.

\section{Resource DID and Control-Proof Verification}\label{sec:09.3}

The Root first parses the resource DID and verifies that its routing code matches the Registrar route, then checks the DID Document \passthrough{\lstinline!id!}, resource type, verification methods, package information, and service declarations. Initial self-controlled registration verifies the resource\textquotesingle s own proof; initial registration under an external controller verifies that controller\textquotesingle s authorization; an update to an existing DID obtains the authorizer from the Root\textquotesingle s current accepted state rather than trusting only the new document. A failure must not write the current version, controller binding, or publication task.

A reproducible negative test is to prepare a submission that passes validation, record the Root platform's current resource version and publication-task count, and then modify only one of \passthrough{\lstinline!resourceDid!}, \passthrough{\lstinline!didDocument.id!}, \passthrough{\lstinline!didDocumentHash!}, or the control challenge's \passthrough{\lstinline!subjectDid!}. The Root platform should return the corresponding error, and the failed request should not add or overwrite an acceptance record or Content Distribution Platform task. This verifies that the Root platform does not mistake ``the request was reachable'' for ``resource control was proved.''

\section{Registrar Service Node Authorization Verification}\label{sec:09.4}

The object of Registrar Service Node authorization is infrastructure-node qualification, not resource control. The Root platform should determine whether a node may submit the applicable resource by using the node DID, target role, authorization VC, authorization domain, stable DID Document hash, issuer, and current governance state. If authorization is expired or revoked, the role does not match, or the authorization domain does not cover the target resource, the submission should be rejected or isolated. The check must not be relaxed merely because the request came from a known address.

The Root platform's decision contains at least two steps: whether the node remains in an available governance state, and whether the node's authorization scope covers the resource's authorization domain. The first handles node revocation, suspension, or authorization expiry. The second handles cases in which a Registrar Service Node may serve only part of the resource authorization domain. Even if a Registrar Service Node has successfully issued a registration VC, the Root platform must still re-evaluate the current authorization state before accepting publication. A registration VC cannot replace the Root platform's node-authorization check.

\section{ResourcePackage and Hash Verification}\label{sec:09.5}

The Root platform should recompute the DID Document hash, metadata hash, and ResourcePackage hash according to the unified canonicalization rules, and check consistency among the hash-algorithm prefix, resource DID, resource type, package version, and nested fields. Any hash mismatch, inconsistent treatment of circular fields, or inconsistency between package content and its declarations is an integrity failure. Failed materials may be retained as diagnostic records, but must not enter the distributable-resource set.

After successful recomputation, the Root platform organizes the resource DID, resource type, version, three hashes, lifecycle, and authorization domain into \passthrough{\lstinline!packageClaims!}, and writes the Root proof into the ResourcePackage's \passthrough{\lstinline!rootProof!}. A verifier should use consistency between these claims and the actual ResourcePackage fields as evidence. Merely seeing a client-submitted \passthrough{\lstinline!packageHash!}, or checking only its string format, is insufficient to prove ResourcePackage integrity.

\section{Metadata and Endpoint Policy Checks}\label{sec:09.6}

Policy checks determine whether materials that have passed identity and integrity checks are suitable for the network publication path. The Root platform should check consistency among resource type, capability tags, authorization domain, protocol bindings, service endpoints, lifecycle, and version declarations, and prevent private keys, access tokens, and internal credentials that should not be public from entering public resource materials. Passing a policy check does not guarantee endpoint quality; endpoint invocation is still authorized jointly by the caller and the resource service.

Policy-check results should be recorded separately from subsequent propagation stages. Materials that pass the checks may enter the Root platform acceptance record and Content Distribution Platform task. Whether an endpoint is online, whether the Content Distribution Platform has completed hosting, and whether a Discovery Service Node has completed indexing must be confirmed by the respective later services. If a resource contains a private key, access token, or an internal address that should not be public, it must be rejected or isolated before public distribution materials are formed; the system must not rely on a Discovery Service Node to remedy this during indexing.

\section{Issuance of the Root Platform Trusted Publication Proof}\label{sec:09.7}

\subsection{Issuance Preconditions}\label{subsec:09.7.1}

The Root Platform Trusted Publication Proof may be issued only after the submission source, Registrar Service Node authorization, resource control proof, DID Document, three classes of hashes, resource version, and publication policy have all passed verification. Before issuing the proof, the Root platform should also confirm that the current governance state does not prevent the node or resource version from entering the publication path. Pending verification, pending review, or distribution failure must not be converted directly into published.

\subsection{Proof Contents}\label{subsec:09.7.2}

The proof should bind at least the Root platform DID, resource DID, resource type, ResourcePackage version, DID Document hash, metadata hash, ResourcePackage hash, hash algorithm, lifecycle state, and issuance time. If the proof contains an authorization domain, publication cursor, or announcement reference, it must clearly state whether those fields cover a publication fact or a propagation fact. They must not include business conclusions that were not declared by the resource controller.

\subsection{Proof Signature}\label{subsec:09.7.3}

The Root platform should sign the prescribed proof contents with its protected signing key and specify the cryptographic suite, signature coverage, and verification-method reference. A verifier should first confirm that the issuer belongs to the trusted Root platform, and then check that the signature input exactly matches the current ResourcePackage. Being able to parse JSON or see a Root platform DID is insufficient to prove that the signature is valid.

\subsection{Proof Retrieval and Verification}\label{subsec:09.7.4}

The resource provider, Content Distribution Platform, and Discovery Service Node may obtain the proof through the Root platform publication interface, publication materials, or an associated address. A verifier should check the proof status, target resource DID, version, hashes, issuance time, and Root-platform authorization state, and retain the verification result. Successfully obtaining the proof does not mean that a Discovery Service Node has completed indexing.

The Root Platform Trusted Publication Proof should be used with one determined resource version. A verifier should at least associate the Root platform DID, resource DID, package version, DID Document hash, metadata hash, ResourcePackage hash, proof, and verification time. The verifier must recheck the hashes and proof status when retrieving the same resource again. A registration VC issued by a Registrar Service Node, the Root Platform Trusted Publication Proof, and a Discovery Service Node index record prove different stages and cannot replace one another.

\section{Publication Sequence and Version Selection}\label{sec:09.8}

\subsection{Publication Sequence}\label{subsec:09.8.1}

The publication sequence identifies the order of resource publication facts in the distribution path. It should advance monotonically and be bound to the resource DID, resource version, and package hash. Re-reading or retrying the same publication task must not create a new logical publication fact, and a Discovery Service Node synchronization cursor must not be treated as the Root platform publication sequence.

\subsection{Version Selection for the Same DID}\label{subsec:09.8.2}

Version selection for the same resource DID should consider the ResourcePackage version, predecessor-version relationship, package hash, control proof, and Root platform publication sequence. Identical materials can be confirmed idempotently. Materials with different content but no valid control proof or version relationship should be rejected or isolated, rather than allowing the arrival time to overwrite an existing published version.

\subsection{Duplicate Submission Handling}\label{subsec:09.8.3}

When the Root platform receives a submission whose resource DID, package version, and package hash are exactly the same as an existing resource, it may return the existing publication result or a duplicate-submission status while keeping the original publication cursor and proof unchanged. The duplicate-submission record should still be associated with the new request identifier so that network retries can be distinguished from actual version updates during audit.

\subsection{Conflicting-Version Handling}\label{subsec:09.8.4}

If the same DID has the same version with a different hash, the same publication sequence with different content, an inconsistent predecessor version, or a proof with inconsistent coverage, the submission should enter conflict or isolation handling. The system must not automatically choose the first-arriving version. The Root platform should retain a conflict summary and rejection reason. Only new materials from the controller that satisfy the version and control conditions may be verified again.

When investigating a version problem, inspect the resource DID, package version, three hashes, predecessor relationship, and publication cursor together. Do not select a version only by request arrival time. The publication cursor is ordering evidence in the Root platform publication record and distribution path; it is neither resource control nor a Discovery Service Node synchronization cursor. Even after a Discovery Service Node sees the cursor advance, it must still validate the specific ResourcePackage and Root proof.

\section{Publication Status and Rejection Reasons}\label{sec:09.9}

The current successful response from the Root platform is \passthrough{\lstinline!resource-verified-and-queued!}. It means that the resource has passed Root-platform processing and entered the Content Distribution Platform task chain; it does not mean that the Content Distribution Platform has completed publication. External processing should distinguish at least the following states:

{\def\LTcaptype{none} 
\begin{longtable}[]{@{}llrl@{}}
\toprule\noalign{}
Status & Meaning & Allowed into distribution & Failure or follow-up action \\
\midrule\noalign{}
\endhead
\bottomrule\noalign{}
\endlastfoot
Pending verification & Received, but necessary checks are not complete & No & Query the processing result \\
Verified and queued & Resource verification is complete and a publication task has been created & Yes & Wait for the Content Distribution Platform \\
Published & The publication fact is persisted and the proof is verifiable & Yes & Continue checking propagation and indexing \\
Duplicate submission & The same DID, version, and hash arrived again & Reuse the existing fact & Return the existing result \\
Rejected & Control, authorization, hash, or policy checks failed & No & Correct the materials and resubmit \\
Pending retry & The external publication path temporarily failed & Use the existing task & Retain the task and retry according to policy \\
\end{longtable}
}

Rejection reasons should preferentially use stable codes such as \passthrough{\lstinline!registrar\_not\_authorized!}, \passthrough{\lstinline!did\_document\_hash\_mismatch!}, \passthrough{\lstinline!metadata\_hash\_mismatch!}, and \passthrough{\lstinline!package\_hash\_mismatch!}. An invalid control proof, unauthorized node, and inconsistent hash are verification failures that block publication. Event-delivery failure or temporary Content Distribution Platform unavailability is a recoverable subsequent failure. These two categories must not both be displayed only as ``publication failed.''

\section{Content Distribution Preparation}\label{sec:09.10}

After accepting a resource, the Root platform persists the ResourcePackage, publication task, and event-publication intent, and hands the task to the Content Distribution Platform through a content-distribution event. The event may carry only replayable references such as the resource DID, package version, publication cursor, Root platform DID, and hashes; the Content Distribution Platform then retrieves the ResourcePackage from the Root platform. The current design uses a transactional outbox to keep the resource acceptance record, publication task, and event intent consistent. If event sending fails, the outbox relay retries it.

Acceptance by the message system does not mean that content publication is complete. The Content Distribution Platform should confirm the task after retrieving the ResourcePackage, verifying the Root proof and hashes, and completing its own publication. A Discovery Service Node should also independently validate the ResourcePackage and proof after receiving propagation materials.

\section{Root Platform Publication Audit Trail}\label{sec:09.11}

The audit chain should trace a complete publication decision from a request identifier or task key: what the Registrar Service Node submitted, which fields the Root platform verified, what the ResourcePackage and hashes were, which publication cursor was generated, whether a Content Distribution Platform task was created, whether the event was delivered, and how the task was ultimately processed. At least the following field groups are recommended:

{\def\LTcaptype{none} 
\begin{longtable}[]{@{}>{\raggedright\arraybackslash}p{0.3200\linewidth}>{\raggedright\arraybackslash}p{0.3200\linewidth}>{\raggedright\arraybackslash}p{0.3200\linewidth}@{}}
\toprule\noalign{}
Field group & Typical fields & Traceability purpose \\
\midrule\noalign{}
\endhead
\bottomrule\noalign{}
\endlastfoot
Request association & \passthrough{\lstinline!requestId!}, Registrar Service Node DID, receipt time & Locate one upstream submission \\
Resource version & Resource DID, \passthrough{\lstinline!packageVersion!}, three hashes & Fix the materials being processed \\
Verification result & Control proof, node authorization, policy-check result & Explain acceptance or rejection \\
Publication evidence & Root platform DID, proof, \passthrough{\lstinline!publicationCursor!} & Prove the Root platform publication fact \\
Distribution status & \passthrough{\lstinline!jobKey!}, event status, retry count, last error & Track subsequent propagation \\
\end{longtable}
}

Audit records may retain summaries and references. They must not contain subject private keys, resource-control private keys, identity backups, or access tokens. Public resource fields and internal diagnostic fields should have separate read permissions so that an audit interface does not become a side channel for sensitive materials.

\section{Republish, Rollback, and Recovery}\label{sec:09.12}

When distribution fails, first retry the publication task with the same resource DID, package version, hashes, and task key. Do not generate a new resource identity merely to resend a transmission, and do not interpret a transport retry as a new version. After data recovery, restore the Root platform acceptance record, publication task, and outbox state first, and then rebuild incomplete event delivery. When rolling back to a historical version, the historical version's DID, version, hashes, and Root Platform Trusted Publication Proof still correspond to the original fact. If the resource content changes, a new control proof must be generated, hashes recomputed, and Root-platform verification performed again.

\begin{lstlisting}
Distribution failure
  -> Retain the original jobKey, resource DID, version, and hashes
  -> Check the task lease, retry count, and last error
  -> Retrieve the same ResourcePackage and verify the Root proof
  -> Confirm the task after the Content Distribution Platform succeeds
  -> Let Discovery Service Nodes continue synchronization by publication cursor
\end{lstlisting}

\section{Same-DID Update Authorization and Version Acceptance}\label{sec:09.13}

An update under the same DID must not be decided by comparing version strings alone. The Root platform should confirm that the new materials remain authorized by the current control relationship, that the DID Document, resource metadata, and ResourcePackage hashes correspond to the new materials, and that the new version can establish a predecessor relationship with the existing version. The same version with the same hash is handled as an idempotent retry. The same version with a different hash is a conflict. A new version whose control proof remains bound to the old DID Document hash must be rejected before the acceptance record is written.

A reproducible acceptance test can fix an accepted \passthrough{\lstinline!v1!} and submit, separately, an identical \passthrough{\lstinline!v1!}, a \passthrough{\lstinline!v1!} with a different hash, a \passthrough{\lstinline!v2!} with a valid control proof, and a \passthrough{\lstinline!v2!} without a predecessor relationship. The acceptance criteria are: the first case produces no second logical publication fact; the second does not overwrite \passthrough{\lstinline!v1!}; the third enters new-version processing; the fourth is rejected or isolated; and every result can be explained from the request, version, and hash fields.

\section{Publication Rejection, Isolation, and Reprocessing}\label{sec:09.14}

A clear verification failure should be rejected directly with its reason. Isolation is appropriate only when the materials cannot yet be determined, depend on external governance state, or require review. An isolation record should retain the original submission summary, resource DID, version, hashes, Registrar Service Node, failure stage, and reprocessing conditions, but must not appear in a Content Distribution Platform manifest or discovery index. Reprocessing must check current authorization, control-proof validity, hashes, and governance state again; it must not merely change the old status to ``published.''

\section{Root Platform Publication State Machine}\label{sec:09.15}

\subsection{Pending Verification}\label{subsec:09.15.1}

Pending verification means that the Root platform has received a submission or publication task but has not completed all necessary checks. This state may allow processing-status and diagnostic queries, but must not enter the trusted distribution manifest or be presented externally as registration success, Root-platform publication, or discovery visibility.

\subsection{Verification Passed}\label{subsec:09.15.2}

Verification passed means that the resource materials, control proof, node authorization, and publication conditions have passed Root-platform checks. If the Root proof has not been issued or the publication record has not been persisted, however, this state is still not equivalent to published. The transition should have a clear submission identifier, verification time, and evidence reference.

\subsection{Published}\label{subsec:09.15.3}

Published means that the Root platform has formed a publication fact and Root Platform Trusted Publication Proof bound to a specific resource version, and permits the related materials to enter the Content Distribution Platform path. It does not mean that the Content Distribution Platform has completed hosting, that a Discovery Service Node has completed indexing, or that the caller has obtained permission to invoke the business service.

\subsection{Rejected, Isolated, and Recovered}\label{subsec:09.15.4}

Rejected means that the current submission does not satisfy a necessary condition. Isolated means that the materials require review, completion, or a change in external state. Recovery must recheck current authorization, signatures, hashes, and governance state. A distribution failure may trigger pending retry; it must not alter the existing resource version or forge a new Root Platform Trusted Publication Proof.

During recovery, use the original submission's resource DID, version, and material summary as the association basis, and run the current authorization, control-proof, hash, and governance-state checks again. Only after passing verification may isolated materials return to the publication flow. Network retries must not change the resource version or generate a new Root Platform Trusted Publication Proof. The Root platform's ``published'' state still means only that a publication fact has been formed; it does not mean that every Discovery Service Node has indexed the resource.

\section{References}\label{sec:09.16}

{\def\LTcaptype{none} 
\begin{longtable}[]{@{}>{\raggedright\arraybackslash}p{0.3200\linewidth}>{\raggedright\arraybackslash}p{0.3200\linewidth}>{\raggedright\arraybackslash}p{0.3200\linewidth}@{}}
\toprule\noalign{}
Source & Type & Link \\
\midrule\noalign{}
\endhead
\bottomrule\noalign{}
\endlastfoot
\passthrough{\lstinline!oan-root-nodes!} & Code repository: Root-platform resource verification, publication proofs, and distribution tasks & \url{https://github.com/OpenAgenet/oan-root-nodes} \\
\passthrough{\lstinline!oan-registrar-nodes!} & Code repository: Registrar Service Node upstream submission and registration result & \url{https://github.com/OpenAgenet/oan-registrar-nodes} \\
\passthrough{\lstinline!oan-protocol-commons!} & Code repository: shared ResourcePackage, proof, and publication-event types & \url{https://github.com/wolfbrother/oan-protocol-commons} \\
OAN Yellow Paper & arXiv Yellow Paper: Root platform, trusted publication, and security boundaries & \url{https://arxiv.org/abs/2606.03163} \\
OAN Resource Identity and Discovery & IETF draft: resource identity, publication, and discovery relationships & \url{https://datatracker.ietf.org/doc/draft-xu-oan-resource-identity-discovery/} \\
\end{longtable}
}

\chapter{Content Distribution and Network Propagation}\label{chap:10}

This chapter explains how a ResourcePackage, including its DID Document, metadata, digests, and Root proof, propagates through the CDN to authorized Discovery nodes after Root accepts a resource version. It distinguishes publication tasks, CDN persistence, synchronization cursors, receipt validation, and local-index states, and explains how batch transfer, retry, partial failure, idempotency, and recovery keep propagation traceable without treating content availability as proof that every Discovery node has applied the resource.

\section{Distribution Model and Content Categories}\label{sec:10.1}

Content distribution follows the model ``form the Root acceptance fact first, transmit propagation materials afterward, and reverify at Discovery.'' The main payload is a \passthrough{\lstinline!ResourcePackage!}, which already contains the DID Document, resource metadata, three digests, Root proof, lifecycle state, and authorization domains. A registration VC is issued by a Registrar to the resource controller; it is not a public field of the \passthrough{\lstinline!ResourcePackage!} and does not replace Discovery indexing evidence.

There are three types of objects that are easily confused in the propagation path: the Root platform creates publication tasks and publication cursors, the content distribution platform stores retrievable resource materials, and Discovery nodes convert validated materials into local indexes. They respectively represent ``what the Root platform decides to propagate,'' ``where the content can be retrieved,'' and ``what a node has already applied.'' The same resource DID may run through all three, but the three states must not be merged into one network-wide state.

\section{Content Distribution Service Responsibilities}\label{sec:10.2}

The content distribution platform is responsible for hosting, retrieving, batch-transferring, caching, and reporting the transfer status of publication materials. A content distribution publisher can obtain publication tasks from the Root platform, organize resource packages according to the publication cursor, write them to the content distribution platform, and report successful or failed results back to the Root platform. This service does not create resource control rights, node authorization, or Root-verified ResourcePackage; Discovery nodes must still decide whether to apply materials based on their own validation results.

The current reference implementation\textquotesingle s batch publication response contains published items, failed items, and the corresponding publication cursor. A publisher confirms the result of a batch only when the response is parseable and the number of items is consistent with the request. If a request succeeds but the response lacks a cursor or contains incomplete item statuses, it must be handled as an exception rather than being treated as confirmation that the entire batch has been published.

Root-to-CDN publication and synchronization for one Discovery target advance in two independent background loops. They share resource DIDs and publication positions, but not batch boundaries:

\pandocbounded{\includegraphics[width=\linewidth,height=0.72\textheight,keepaspectratio]{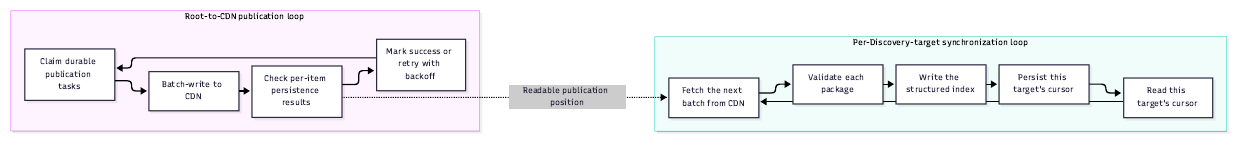}}

The publisher does not wait for one Discovery node to finish a batch before beginning its next iteration. Each Discovery node advances independently according to its authorization domain, batch size, validation results, and cursor. A single target must keep cursor semantics explainable, while the implementation may use bounded concurrency within that constraint.

\section{Distribution Manifest and Publication Checkpoints}\label{sec:10.3}

A publication manifest answers ``which Root-admitted ResourcePackages are in this batch,'' while a checkpoint answers ``how far has the processor advanced reliably.'' A manifest entry associates at least the resource DID, version, package digest, Root proof, and publication sequence. A checkpoint advances after durable confirmation, not merely when an HTTP request is sent or a message notification arrives. A scan of the manifest can therefore recover from a lost notification, and sequence and digest make a duplicate notification idempotent.

The publication manifest is the entry point through which a Discovery node determines the synchronization scope and validation objects. Publication checkpoints divide propagation state into sent, received, validated, and applied. The resource DID, package version, package hash, and publication cursor in the manifest associate a resource version; the target nodes and sequence range in a notification batch associate one propagation operation. A Discovery node may report that an item has been applied only after it has completed validation and written the item to its index.

\section{Distribution of Resource Packages, Metadata, Proofs, and Index Data}\label{sec:10.4}

The propagation chain distributes only facts required for verification and discovery. It does not publish a Discovery node\textquotesingle s local vectors or ranking results as Root content. The ResourcePackage and Root proof are verifiable inputs, and metadata supplies structured fields for indexing. Tokenization, embeddings, semantic views, and query statistics are generated locally by each Discovery node and can be rebuilt from ResourcePackages after a model upgrade.

\subsection{Resource Package Distribution}\label{subsec:10.4.1}

The resource package is the primary complete payload for Discovery nodes. It must at least be associated with the publication manifest through the resource DID, resource package version, and package hash. The content distribution platform may provide a complete package or incremental materials according to the manifest, but the receiver must recompute the hash and validate the Root-verified ResourcePackage before writing the material to a trusted discovery index.

The current content distribution platform provides \passthrough{\lstinline!POST /cdn/resources!} as the single-item publication endpoint and \passthrough{\lstinline!POST /cdn/resources/batch!} as the batch publication endpoint. A single-item request is carried by \passthrough{\lstinline!ResourceCdnPublishRequest!} and contains \passthrough{\lstinline!ResourcePackage!} and \passthrough{\lstinline!upstreamAuth!}; a batch request is carried by \passthrough{\lstinline!ResourceCdnBatchPublishRequest!} and contains entries with \passthrough{\lstinline!publicationCursor!} and shared upstream authentication. A successful single-item response contains \passthrough{\lstinline!status: published!}, the resource DID, resource type, package version, and \passthrough{\lstinline!publicationCursor!}. A batch response additionally distinguishes \passthrough{\lstinline!acceptedCount!}, \passthrough{\lstinline!failedCount!}, successful items, and failed items. Batch HTTP success cannot be used as a substitute for every item having been applied.

Discovery nodes can use \passthrough{\lstinline!POST /cdn/resources/batch-get!} to read resource packages in batches by resource DID, or use the resource index endpoint to read incrementally by \passthrough{\lstinline!afterCursor!}. A batch read returns the requested count, hit count, and resource package items; an incremental index response returns \passthrough{\lstinline!items!}, \passthrough{\lstinline!count!}, \passthrough{\lstinline!afterCursor!}, \passthrough{\lstinline!nextCursor!}, and \passthrough{\lstinline!hasMore!}. These cursors serve batch reads, list pagination, and propagation synchronization respectively and cannot replace one another.

\subsection{DID Document and Metadata Distribution}\label{subsec:10.4.2}

The DID Document and metadata are bound by digests within one resource version. A receiver verifies the DID Document\textquotesingle s \passthrough{\lstinline!id!}, controller, and verification methods, and also checks that the resource type, endpoints, authorization domains, and version in metadata match the package declarations. If only one part is downloaded, the digest reference to the other is retained; an old cache must not be combined into an apparently complete new version.

DID documents and resource metadata may be transferred together with the resource package, or retrieved separately by resource DID, version, and hash. When they are sent separately, a Discovery node must not apply one part while losing the association between the parts. If the DID document, metadata, and package-external summary are inconsistent, the node should reject the combination and record the missing or conflicting fields.

\subsection{VC and Root Proof Distribution}\label{subsec:10.4.3}

A registration VC describes a Registrar\textquotesingle s registration fact and may be retained by the resource controller, an audit interface, or the corresponding registration record. A Root proof enters distribution and Discovery validation with the \passthrough{\lstinline!ResourcePackage!}. The two can be associated by resource DID, version, and digests, but have different origins and propagation paths. Discovery cannot skip the Root proof because it has obtained a registration VC, nor infer the current state of a registration VC from a valid Root proof.

\subsection{Index Data Distribution}\label{subsec:10.4.4}

Index data is a derived read model formed by a Discovery node from published materials. It may contain resource summaries, capability tags, endpoint references, lifecycle status, and freshness information. Index data must not be returned as a new Root platform fact. A Discovery node must be able to identify the publication cursor, validation time, and authorization domain corresponding to an index record.

\section{Distribution-Boundary Integrity Validation}\label{sec:10.5}

After receiving materials, a Discovery node should first validate the notification target, manifest scope, and payload structure, and then validate the resource DID, resource type, version, the three categories of hashes, Root-verified ResourcePackage, lifecycle, and authorization domain. Objects that fail validation must not be written to the trusted discovery index. A rejection record may be written for diagnosis and retry. Successful transfer with corrupted content is an integrity failure; missing material is a recoverable retrieval failure; a revoked governance status is a state failure that blocks application.

Negative acceptance tests at the receiver can independently change only the package content, the resource DID bound to the proof, or the authorization domain of the current Discovery node. In the first two cases, the receiver should reject the material respectively at the hash or proof-binding stage; in the third case, it should block index application. A failure may leave a record containing the resource DID, cursor, failure stage, and error code, but the affected synchronization cursor must not advance.

The Root-to-CDN publication batch and the batch used by a Discovery node to retrieve resources are two independent implementation layers. They must not be assumed to correspond one-to-one merely because both are called batches. The former advances through durable publication tasks and CDN confirmations; the latter advances according to each Discovery target\textquotesingle s own cursor, batch size, and validation results.

JetStream publication events notify consumers that new tasks are available, while database task records preserve recoverable state. On duplicate delivery, a consumer remains idempotent by task identifier and resource version. If an event is temporarily lost, periodic scanning or recovery can still rediscover unfinished tasks.

\section{Version Freshness and Synchronization Cursors}\label{sec:10.6}

Freshness describes how far the current view is from its authoritative source; a cursor is only one piece of evidence used to calculate that distance. A target cursor may stop because the upstream has no new content, the target loop is not running, one resource failed validation, or an indexing transaction has not completed. A status interface therefore combines the cursor with the latest successful time, failed items, and the upstream latest cursor rather than displaying one number directly as healthy or faulty.

The publication cursor describes the position of Root platform publication facts in the propagation path; the synchronization cursor describes the position already processed by a Discovery node; the resource version describes the version of the resource content itself; freshness describes the time relationship between publication or receipt and the current observation moment. They serve ordering, recovery, version selection, and latency explanation respectively, and one field cannot replace the others.

\subsection{Synchronization Cursor}\label{subsec:10.6.1}

A synchronization cursor is maintained by a Discovery node to represent the content distribution position that the node has processed. Advancing the cursor must be conditional on the corresponding materials completing structural, hash, proof, and status validation. Advancing the cursor to a position does not mean that all resources after that position have been applied successfully.

\subsection{Version Freshness}\label{subsec:10.6.2}

Version freshness should consider the Root platform publication cursor, resource package version, publication or update time, local receipt time, and index application time together. Time differences are used to explain propagation latency and must not alone determine version priority. Content selection for the same DID must still be based on a verifiable version, hash, and Root platform publication fact.

\subsection{Expired Data Determination}\label{subsec:10.6.3}

A Discovery node should determine whether data is expired according to the permitted delay, governance status, and resource lifecycle. Ordinary information display may retain an old snapshot marked with freshness information. Before processing revocation, suspension, node authorization changes, or high-risk calls, if the latest status cannot be confirmed, the node should pause the sensitive action instead of continuing to use an unconfirmed old status.

\subsection{Gap Compensation}\label{subsec:10.6.4}

When a cursor gap, discontinuous batch sequence, or missing resource material is detected, the node should pause cursor advancement for the affected scope, retrieve the missing content according to the manifest or update endpoint, and continue after revalidation. Before the gap is filled, the node may report synchronization lag but must not mark subsequent items as applied without evidence.

For example, if a node has applied cursor 100 and receives cursor 102 while 101 is missing, it may save the material for 102 as pending and report the gap, but its local ``applied position'' should remain 100. After retrieving and validating 101, it should process 102 in order. If 101 is explicitly rejected, it should retain a traceable rejection record and determine whether the cursor may cross that item according to the implementation. Receiving a larger cursor value alone is not sufficient to advance the cursor.

\section{Initial Synchronization and Incremental Synchronization}\label{sec:10.7}

Initial synchronization establishes a complete baseline for a target node, while incremental synchronization processes only publication sequences after that baseline. Both use the same ResourcePackage validation and authorization-domain filtering; they differ in pagination starting point and recovery strategy. If incremental notifications arrive before the full baseline completes, the node merges them under one sequence or replays them after the baseline checkpoint so that a newer version seen first is not later overwritten by an older full-sync version.

\subsection{Initial Full Synchronization}\label{subsec:10.7.1}

Initial synchronization should first obtain a verifiable publication manifest, starting cursor, and target authorization domain, and then retrieve resource packages and their proofs according to the manifest. The node should validate, persist, and apply each item to the index. After the entire scope is completed or every rejection has been explicitly recorded, it may report the synchronization result for that scope.

\subsection{Incremental Synchronization}\label{subsec:10.7.2}

Incremental synchronization continues from the persisted synchronization cursor or notification-batch sequence, retrieves publication materials after that position, and processes them in publication order. Duplicate batches and duplicate resource packages should be processed idempotently. A new cursor may advance only after the corresponding items reach an applied or explicitly rejected state.

\subsection{Synchronization Confirmation}\label{subsec:10.7.3}

Synchronization confirmation should distinguish at least received, validated, indexed, and externally queryable. Confirmation responses, logs, and database records should associate the target Discovery node, sequence range, successful or rejected items, current cursor, and last error. An HTTP request succeeding cannot substitute for confirmation that content has converged.

\subsection{Recovery After Interruption}\label{subsec:10.7.4}

After a node restarts or the network is interrupted, it should recover from the last persisted cursor or incomplete batch, replay necessary materials, and process them according to idempotency rules. If the cursor is invalid, the manifest has changed, or governance status cannot be confirmed, the node should retrieve the manifest again or enter a review-pending state instead of skipping the missing interval.

\section{Retry, Backoff, and Interruption Recovery}\label{sec:10.8}

Retries apply only to recoverable transfer, timeout, service-unavailable, or temporary-lock failures. The publisher and Discovery node should retain the task or batch identifier, target cursor, processed items, retry count, and last error, and use backoff to avoid continuously impacting the upstream service. Invalid signatures, hash mismatches, resource-type errors, and authorization-domain mismatches are material or permission failures; they should be recorded and repeated infinite retries of the unchanged payload should be stopped.

A retry record should at least answer three questions: which resource version was retried, whether the failure occurred during retrieval, validation, writing, or result reporting, and what condition the next retry depends on. For batch processing, results should be updated separately for successful and failed items. A successful overall batch response must not cause already successful items to be applied again, and one failed item must not erase the processing evidence for other items. Repeated delivery of the same task should use the resource DID, version, hash, and publication cursor for idempotency checks.

\section{Cache Behavior and Expired-Data Handling}\label{sec:10.9}

Caches reduce query and transfer cost but do not change authoritative relationships. Browser caches are suitable for presentation data; Discovery caches are suitable for validated ResourcePackages or query results. A decision involving a controller, revocation, suspension, or node authorization carries source time and freshness. An expired cache may be displayed as ``last known,'' but must not be presented as currently trusted or currently callable.

\subsection{Browser or Client Cache}\label{subsec:10.9.1}

A browser or client cache stores public materials, network snapshots, or resource summaries already retrieved by the page or client. It is not the authoritative publication record of the Root platform. Cache keys should include the interface scope, resource DID, or page-data version and other necessary context. Clearing browser data may remove the cache, but it does not affect the server-side publication fact. When reading cached data, the client should include the retrieval time, source, and freshness marker so that a local hit is not displayed as if it had just been confirmed by the network.

\subsection{Node-Side Cache}\label{subsec:10.9.2}

The content distribution platform and Discovery nodes may cache resource packages or indexed read models that have passed the corresponding validation. The content distribution platform caches materials available for propagation, whereas a Discovery node caches query data after local application. Node-side caches should retain the resource version, package hash, publication cursor, and validation time, and replace entries in a verifiable order when updating. A cache must not bypass Root-verified ResourcePackage, resource-package hash, or current governance-status checks.

\subsection{Preferential Display of Old Snapshots}\label{subsec:10.9.3}

When new page or query data has not yet been retrieved, the client may first display the most recently saved old snapshot whose source can be identified, while marking the snapshot time, data source, and possible synchronization delay. An old snapshot can improve the waiting experience for ordinary reads, but it is not suitable as the real-time basis for revocation, suspension, authorization changes, or high-risk calls. Snapshot display should be separated from background refresh status; the display time of old data must not be updated to the current confirmation time.

\subsection{Background Refresh and Failure Degradation}\label{subsec:10.9.4}

After a background refresh succeeds, the client or node should complete structural, hash, proof, and status checks on the new materials and then atomically replace the old snapshot. When refresh fails, it should retain the old snapshot and record the failure reason, last successful time, and freshness. If the failure concerns revocation or the authorization status cannot be confirmed, the node should reduce the availability state or pause sensitive actions rather than unconditionally continuing to use the old snapshot. Refresh logic must not block basic page rendering or clear valid existing data on failure and thereby create unnecessary blank space.

Discovery targets advance their synchronization cursors independently. If one target falls behind because of a network, validation, or indexing failure, already completed targets must not be rolled back, and another target\textquotesingle s success must not be counted as its success. ``All Discovery nodes completed'' can be established only by aggregating the state of every target.

\section{Propagation Observability and Audit Records}\label{sec:10.10}

Propagation audits should associate the resource DID, package version, package digest, publication task, publication cursor, notification batch, target Discovery node, send/receive/validation/application times, retry count, and error reason. The Root and Publisher should be able to state which items completed durable CDN persistence; a Discovery node should be able to state which items were rejected or remain pending synchronization. These records trace the propagation process and do not change the original proof of the resource.

For batch propagation, both the overall batch result and each resource-item result should be recorded. An overall successful response does not mean that every item has been applied, and partial success must not erase the evidence of completed items. During operational troubleshooting, \passthrough{\lstinline!jobKey!}, request ID, and publication cursor can be used to associate Root platform tasks, content distribution platform history, and Discovery node synchronization records. These fields trace propagation and are not equivalent to a resource controller signature or governance authorization.

\section{Publication Ordering and Conflict Resolution}\label{sec:10.11}

Conflict resolution first locates the same evolution chain by resource DID, then compares predecessor state, version, and content digest. An identical duplicate can be accepted idempotently; an older version at a higher sequence cannot replace the current version; the same version with a different digest must be quarantined with its source retained; and a missing predecessor may be deferred or trigger a compensating fetch. Directly overwriting by arrival time would allow network jitter to change the final result and is therefore not a consistency rule.

Propagation order is based on the Root platform publication cursor and notification sequence. Resource-version selection is based on the resource package version, predecessor-version relationship, package hash, and Root-verified ResourcePackage. Duplicate batches containing the same materials should be processed idempotently. If different hashes appear at the same position, an old version attempts to overwrite a new version, or predecessor materials are missing, application of the affected scope should be paused and the conflict recorded. Materials must not be overwritten arbitrarily according to arrival time.

\section{Distribution Consistency and Freshness Guarantees}\label{sec:10.12}

Consistency metrics should separately state the scope that the Root platform has published, the scope available from the content distribution platform, and the scope received, validated, and applied by Discovery nodes. Freshness metrics should state the publication or receipt time from which they are calculated and the observation or application time to which they are calculated. Metrics must include the sample scope, failed items, and authorization-domain definition; success by one node or a cache hit cannot represent network-wide convergence.

Suspension, revocation, or replacement of a version must also propagate through content distribution and Discovery synchronization. An old cache may support explicitly marked degraded-freshness display, but a revoked resource must not remain a currently trusted candidate. Recovery must compensate from durable tasks and target cursors rather than depending only on in-memory notifications.

\section{Partial Failure and Recovery Semantics}\label{sec:10.13}

Partial failures are isolated by resource item and target node. Items that have completed validation and application may advance their own state. Failed items should retain independent error, retry, and missing-material information, and must not be skipped because the overall batch response succeeded. During recovery, the node should first fill missing materials, then complete item-level validation and persistence, and finally advance the corresponding cursor or record an explicit rejection. Applied items are not rolled back merely because another item in the same batch failed, and failed items must not be hidden because an old snapshot remains available.

Final convergence requires every item in the target sequence range to reach ``applied'' or have a clear and traceable rejection record. If hashes, proofs, lifecycle, or authorization domains change during recovery, the corresponding validation process must be entered again. An old snapshot may be used only for limited read-side results marked with freshness information; it must not conceal revocation, suspension, or authorization changes.

Discovery synchronization is isolated per target. After the Publisher receives durable confirmation for a CDN package, each Discovery target independently pulls, validates, and applies it according to its own authorization domains and cursor. A database failure or unavailable semantic service at one target must not block structured convergence at another target. For one target, progress for the current batch may be committed only after package validation, structured-index writing, and cursor persistence all succeed.

Bounded concurrency may reduce idle time between batches, but it must not skip a failed batch and leave an inexplicable cursor gap. Recovery distinguishes a missing notification, an unconfirmed CDN write, a package-digest error, a target authorization-domain mismatch, and an index-write failure instead of classifying every case as a network retry.

\section{References}\label{sec:10.14}

{\def\LTcaptype{none} 
\begin{longtable}[]{@{}>{\raggedright\arraybackslash}p{0.3200\linewidth}>{\raggedright\arraybackslash}p{0.3200\linewidth}>{\raggedright\arraybackslash}p{0.3200\linewidth}@{}}
\toprule\noalign{}
Source & Type & Link \\
\midrule\noalign{}
\endhead
\bottomrule\noalign{}
\endlastfoot
\passthrough{\lstinline!oan-root-nodes!} & Code repository: publication tasks, content distribution coordination, and publication status & \url{https://github.com/OpenAgenet/oan-root-nodes} \\
\passthrough{\lstinline!oan-discovery-nodes!} & Code repository: propagation-material receipt, validation, and indexing & \url{https://github.com/OpenAgenet/oan-discovery-nodes} \\
\passthrough{\lstinline!oan-protocol-commons!} & Code repository: resource-package, publication-event, and hash structures & \url{https://github.com/wolfbrother/oan-protocol-commons} \\
OAN Yellow Paper & arXiv yellow paper: trusted publication, propagation, and discovery path & \url{https://arxiv.org/abs/2606.03163} \\
Agentic Overlay Network Architecture & IETF draft: node coordination and overlay-network propagation & \url{https://datatracker.ietf.org/doc/draft-xu-agentic-overlay-network-architecture/} \\
\end{longtable}
}

\chapter{Discovery Protocol and Query Semantics}\label{chap:11}

This chapter explains how a discovery service node processes exact DID queries, structured queries, tag queries, text queries, and semantic queries from a local index that has been verified and filtered by the applicable authorization scope. It covers query input, version selection, provenance and freshness, result deduplication, pagination, explanations, errors, and degradation boundaries. A discovery response is explicitly a candidate result produced from a node\textquotesingle s index; the caller must still perform pre-invocation verification based on the DID, credentials, root platform trusted publication proof, governance status, and endpoint information.

\section{Discovery Objectives and Query Types}\label{sec:11.1}

The query type determines how candidates are generated, but it does not change the common filters applied before candidates enter the result. Exact DID queries directly locate an index key; structured queries use resource type, protocol, tag, and version conditions; text and semantic queries generate relevant candidates. Authorization scope, lifecycle, version selection, and evidence-validity checks are then applied to all of them. This prevents a semantically high-scoring resource from bypassing structured trust boundaries.

A discovery service node provides five query paths for resources that have entered its local index: exact lookup by a complete resource DID; structured queries by resource type, version, protocol, and capability tags; tag queries by capability tags; text or natural-language queries based on resource names, descriptions, tags, use cases, and protocol materials; and combined queries that join natural-language and structured conditions. Query results are derived responses from the discovery service node rather than the root platform\textquotesingle s publication records. The response should retain the resource DID, resource type, version, lifecycle state, provenance, and necessary verification references.

The current reference implementation accepts a unified query object through \passthrough{\lstinline!POST /discovery/resources/query!}. When the query text is a complete, whitespace-free \passthrough{\lstinline!did:oan:!} identifier, it enters the exact DID branch; other text and explicit fields enter structured, lexical, or semantic paths. The query node constructs candidates only from locally indexed material. It must not directly forward the query to a registration service node or the root platform to manufacture a temporary result.

Before retrieval begins, query input should pass boundary validation and then be converted into an internal query object. \passthrough{\lstinline!resourceType!}, \passthrough{\lstinline!capabilityTags!}, \passthrough{\lstinline!protocol!}, and \passthrough{\lstinline!version!} form explicit conditions; \passthrough{\lstinline!versionMode!} specifies the version selection method; \passthrough{\lstinline!limit!} restricts the response size; and \passthrough{\lstinline!query!} carries a DID or a text/natural-language query. When text and structured fields are supplied in the same request, text performs recall and ranking while structured fields narrow the candidate set. The two kinds of conditions must not override each other.

\begin{lstlisting}
{
  "query": "risk list for a commercial contract",
  "capabilityTags": ["legal.contract.review"],
  "versionMode": "latest",
  "limit": 10
}
\end{lstlisting}

This request comes from an actual Discovery resource-regression fixture and is accepted by the current query type. \passthrough{\lstinline!capabilityTags!} are hard conditions, while the query text participates in candidate retrieval and ranking. A candidate\textquotesingle s \passthrough{\lstinline!score!} is relevance under the current index and model, not a resource trust level.

\subsection{Exact DID Discovery}\label{subsec:11.1.1}

Exact discovery matches the complete DID; it performs no prefix completion, case normalization, or similar-string fallback. A miss should produce a stable empty result or an explicit not-found result. If the index contains several versions, current version-selection rules apply, while historical-version retrieval remains distinct from determining whether the resource is currently callable.

A complete resource DID must be interpreted as an exact lookup condition. It must not degrade into a name, prefix, or semantic-similarity search. The node should read the local index object corresponding to the DID and continue checking explicit conditions, lifecycle state, and the node\textquotesingle s authorization scope.

\subsection{Structured-Condition Discovery}\label{subsec:11.1.2}

Fields such as resource type, version, protocol, and capability tags should be converted into reviewable filter conditions. Multiple explicit conditions use intersection semantics by default. An invalid enumeration, an unresolvable version mode, or an unsupported protocol value should produce a clear request error.

\subsection{Capability-Tag Discovery}\label{subsec:11.1.3}

Capability-tag queries use \passthrough{\lstinline!capabilityTags!} in the resource metadata. Parent-child relationships, aliases, and extension tags in the tag tree may participate in the query projection, but the original tags and their sources must be retained. A tag match does not represent control or invocation authorization.

\subsection{Text and Semantic Discovery}\label{subsec:11.1.4}

Text and natural-language queries may use lexical retrieval, vector retrieval, or a hybrid of both. When the semantic index is unavailable, the node may fall back to lexical and structured paths, but the query explanation or diagnostic information should state which backend was actually used.

\subsection{Combined-Condition Discovery}\label{subsec:11.1.5}

A combined query should first apply authorization-scope, lifecycle, and explicit structured-condition filters, and then rank the remaining candidates lexically or semantically. Relevance must not remove hard filters or allow unpublished or unauthorized resources to enter a trusted result.

\section{Exact DID Queries}\label{sec:11.2}

An exact query first validates the DID method syntax and resource-type semantics, and then reads the resource package from the local index. A hit must use the DID document, metadata, endpoint, hash, and root platform proof reference for the same resource version. A miss must not be interpreted simply as proof that the resource does not exist on the entire network, because it may not yet have propagated to this node or may have been filtered by the authorization scope. The node should distinguish the following results:

{\def\LTcaptype{none} 
\begin{longtable}[]{@{}>{\raggedright\arraybackslash}p{0.3200\linewidth}>{\raggedright\arraybackslash}p{0.3200\linewidth}>{\raggedright\arraybackslash}p{0.3200\linewidth}@{}}
\toprule\noalign{}
Result & Explanation & Caller handling \\
\midrule\noalign{}
\endhead
\bottomrule\noalign{}
\endlastfoot
Candidate returned & This node has a visible indexed object that satisfies the conditions & Continue verifying the DID, proof, and endpoint \\
Empty candidates & The resource does not exist here, has not been indexed, or is outside this node\textquotesingle s authorization scope & Do not infer that it does not exist network-wide \\
Suspended/revoked & The indexed object exists, but its current state restricts use & Do not invoke it as a currently valid resource \\
Service error & The database, index, or governance read model is temporarily unavailable & Retry or degrade according to temporary-failure handling \\
\end{longtable}
}

The current exact-query implementation identifies a DID through the \passthrough{\lstinline!query!} field and passes the remaining explicit conditions to resource-type, version, capability-tag, and protocol matching. Therefore, a query containing a DID may still return empty candidates when its version or authorization-scope conditions are not satisfied. An exact hit only means that the discovery service node has a visible index record. The caller must still obtain the complete DID document and proof materials for pre-invocation verification.

The result boundary of an exact DID query can be checked using the following table. Here, ``not found'' means the result under the current index and authorization scope of this node; it is not a statement about the state of the Internet of Agents.

{\def\LTcaptype{none} 
\begin{longtable}[]{@{}>{\raggedright\arraybackslash}p{0.3200\linewidth}>{\raggedright\arraybackslash}p{0.3200\linewidth}>{\raggedright\arraybackslash}p{0.3200\linewidth}@{}}
\toprule\noalign{}
Query stage & Object to check & Meaning of the result \\
\midrule\noalign{}
\endhead
\bottomrule\noalign{}
\endlastfoot
Input identification & Whether \passthrough{\lstinline!query!} is a complete \passthrough{\lstinline!did:oan:!} string & Determines whether the exact path or the ordinary retrieval path is used \\
Index read & Whether this node has an indexed resource-package record & Reflects only whether this node has stored the record \\
Condition matching & Explicit conditions such as version, resource type, protocol, and capability tags & Any unsatisfied hard condition can produce empty candidates \\
Visibility check & This node\textquotesingle s authorization scope and the resource lifecycle & An invisible, suspended, or revoked resource must not be returned as an ordinary candidate \\
Pre-invocation decision & DID document, registration credential, root platform trusted publication proof, and endpoint & Performed by the caller; not replaced by \passthrough{\lstinline!score!} or an index hit \\
\end{longtable}
}

\section{Structured Resource Queries}\label{sec:11.3}

A structured query maps fields in the request to database or in-memory filter conditions. The current reference implementation filters by resource lifecycle, resource type, version, capability tags, and protocol. For an ordinary query without an explicit resource type, it uses a default scope oriented toward resources intended for user consumption. Explicit capability tags require the resource to contain all requested tags. A protocol condition requires the resource service endpoint or protocol binding to declare or explain that protocol. An empty query with structured conditions may be executed, while a query with no conditions at all should be constrained by result count and resource scope limits.

After filtering, results still pass the node\textquotesingle s authorization-scope check and latest-valid-version selection. Display names, descriptions, and retrieval scores are index fields; they are not evidence of identity, publication, or governance.

Explicit filters should be applied before candidates enter relevance ranking. The current implementation performs resource-level matching for \passthrough{\lstinline!resourceType!}, \passthrough{\lstinline!version!}, \passthrough{\lstinline!capabilityTags!}, and \passthrough{\lstinline!protocol!}. Every capability tag supplied by the request must appear in the candidate; a protocol may match through a protocol binding or a service endpoint declaration. A missing field is not automatically a wildcard, and an unresolvable request value must not be silently converted into another condition.

Structured filtering must be reviewable: request fields, resource fields, and the decision should correspond item by item. For example, if a request specifies both \passthrough{\lstinline!resourceType = "mcp\_server"!} and \passthrough{\lstinline!capabilityTags = ["file.search", "desktop.automation"]!}, a candidate must first be of the requested resource type and contain both tags before it participates in text-relevance ranking. A candidate that is semantically close to ``file search'' but has the wrong resource type must not enter the result through ranking. If a field is absent from the resource package, that condition should be treated as unsatisfied rather than as an arbitrary match.

For unknown enumerations, blank fields, duplicate tags, and conflicting version conditions, the node should define deterministic handling during request parsing. Duplicate tags may first be normalized into one condition; an unknown resource type or an unresolvable version mode should be a request error; and a conflict between version and lifecycle should narrow the result under version-selection and state rules rather than be relaxed by the retrieval score.

\section{Capability-Tag Queries}\label{sec:11.4}

Capability tags are declarations by a resource provider that classify the resource\textquotesingle s capabilities. The root platform capability-tag tree provides governance context for tag identifiers, hierarchy, aliases, and versions. A discovery node may project aliases in a query to canonical tags or maintain local retrieval aliases, but the response should retain the resource\textquotesingle s original tags and the tags that actually matched. Whether a parent tag covers a child tag and whether an extension tag enters the trusted index must be defined by the tag-tree version and node configuration, not implicitly determined by model relevance.

A tag query should distinguish three kinds of information: the resource\textquotesingle s original \passthrough{\lstinline!capabilityTags!}; the retrieval terms produced by expanding aliases in query text or explicit parameters; and the tags ultimately used to explain the match. Alias expansion can improve recall, but it must not modify the resource\textquotesingle s original declaration. If an extension tag is not supported by the current node\textquotesingle s tag tree or configuration, it should be treated as an ordinary miss rather than inferred to be a parent tag or synonym.

\section{Text and Semantic Queries}\label{sec:11.5}

Semantic queries may retrieve public information such as resource names, descriptions, capability tags, use cases, protocol bindings, and resource types. The reference implementation may organize resource material into tag, context, and intent views, and combine vector, lexical, and structured scores to rank candidates. \passthrough{\lstinline!score!} expresses relevance to the current query and index snapshot; it does not express a trust level, service quality, or security level.

When the semantic index, embedding service, or pgvector is unavailable, the node may fall back to lexical and structured queries. Degradation must not relax authorization-scope, lifecycle, or explicit conditions. Empty text should not force a semantic backend call. Input that is too long, cannot be encoded, or uses an unsupported language should produce a clear error or an explainable degraded result.

The current semantic implementation projects resource type, protocol, capability tags, and query text into retrieval terms, and constructs search documents from public material such as resource names, descriptions, tags, context, examples, and use cases. When the semantic backend is available, results may combine BM25-like lexical scores, tag scores, context scores, intent scores, and structured scores. These internal scores are used for ranking and explanation and must not be treated by external callers as a governed quality assessment.

Text used to build a semantic index must come from resource description material that has already been made public. Index construction should retain the association among the resource DID, resource-package version, and semantic source content so that the corresponding index can be rebuilt after a resource update and old descriptions do not continue to produce highly relevant results. Multilingual queries may improve recall through normalization, aliases, or embedding models, but a model\textquotesingle s inferred capability must not be written back into resource metadata, and model similarity must not replace resource control, a registration credential, or the root platform trusted publication proof.

A discovery node should be able to distinguish the semantic backend, lexical path, and structured path through runtime logs or a diagnostic interface. For the same hard conditions, a semantic-service failure may change candidate recall and ranking quality only. If the index database, authorization-scope read model, or resource-package read fails, the failure must be handled as unavailable data or a service error rather than returning an apparently normal but incomplete empty result.

\section{Combined Semantic and Structured-Condition Queries}\label{sec:11.6}

A combined query can be abstracted into the following processing order:

\begin{lstlisting}
visible = authorizedDomain AND active AND explicitFilters
ranked  = rank(queryText, visible)
result  = stableSort(ranked)[0:limit]
\end{lstlisting}

Here, \passthrough{\lstinline!explicitFilters!} includes resource type, version, protocol, and explicit capability tags; \passthrough{\lstinline!rank!} may use lexical, vector, or hybrid retrieval; and \passthrough{\lstinline!stableSort!} must retain a stable secondary key. If no resource satisfies the hard conditions, the node should return empty candidates rather than a ``similar'' resource that does not satisfy them.

An important boundary of combined queries is that semantic results must still pass resource-package validity, lifecycle, and authorization-scope checks at the node. A record in the semantic index does not mean that it is visible at the time of the query. For implicit natural-language requirements such as ``latest'', ``trusted'', or ``callable'', the node may convert the portion that maps to protocol fields into structured conditions. A portion that cannot be mapped to fields may be used only as ranking or explanation information; it must not automatically produce an authorization conclusion.

A minimal reproducible example of combined-query processing is:

\begin{lstlisting}
{
  "query": "public HTTPS API for weather forecast current temperature wind speed JSON",
  "resourceType": "tool_api",
  "capabilityTags": [],
  "versionMode": "latest",
  "limit": 10
}
\end{lstlisting}

This combined request also comes from the current regression fixture. A valid result must first satisfy the \passthrough{\lstinline!tool\_api!} type condition; natural language only contributes to relevance over fields such as names, descriptions, and use cases. If the query asks for a ``trusted'' or ``callable'' resource, the node may make that determination only when corresponding structured fields and verifiable facts exist. Otherwise, it preserves the phrase as retrieval intent and leaves validation to the caller\textquotesingle s subsequent checks.

A Discovery node\textquotesingle s query processing can be summarized as input validation, hard-condition filtering, candidate generation, ranking, deduplication, explanation, and response assembly. Resource type, authorization scope, lifecycle, version validity, and provenance validation are hard conditions. Text or vector similarity participates only in candidate generation and ranking and cannot reintroduce a resource excluded by a hard condition.

\section{Authorization-Scope Filtering}\label{sec:11.7}

A discovery node obtains its own authorization scope from its DID document or governance read model and compares it with \passthrough{\lstinline!authorizedDomains!} in resource metadata. The resource domain must be covered by a domain for which the node is authorized. A domain mismatch must keep the resource out of the query result. This filter determines discovery visibility provided by this node to the caller; it does not change the resource DID control relationship or mean that the resource is invisible at other discovery nodes. When governance status or node authorization cannot be confirmed, ordinary display should mark the result as not fresh, while sensitive queries and pre-invocation verification should apply restrictive handling.

\section{Latest Valid Version Selection}\label{sec:11.8}

By default, a query selects, for each resource DID, the latest version that satisfies the node\textquotesingle s authorization scope, has an active lifecycle state, has completed the necessary validation, and has a confirmable synchronization position. Version selection for the same DID should consider the resource-package version, root-platform publication cursor, package hash, predecessor relationship, and governance state. It must not rely only on update time or query arrival order. If a discovery node is behind, it may return a verified older version for limited display, but it must include freshness and synchronization position and must not claim that version to be the network latest.

Version selection must verify at least three conditions: the candidate belongs to the same resource DID; its version satisfies the requested \passthrough{\lstinline!versionMode!} and \passthrough{\lstinline!version!} conditions; and its lifecycle and governance state still allow discovery. For a \passthrough{\lstinline!latest!} query, the selected item is the latest currently verifiable valid version, not the record most recently written to the database. For a specified-version query, a missing version record, inconsistent hash, or unconfirmable state should produce an empty result or a clear data-unavailable error; it must not silently substitute another version.

A reproducible check can use two records with the same DID: record A has an earlier publication sequence but an active state, while record B has a later update time but is suspended. A conforming implementation should select record A or return a restricted result; it must not treat B as the latest valid version merely because its timestamp is newer. The check should also verify that repeated queries over the same index snapshot return the same resource DID, version, and state, while retaining the publication sequence, package hash, and state source as decision evidence.

\section{Search Ranking and Deterministic Result Ordering}\label{sec:11.9}

Ranking occurs only after candidates pass security and governance filters. Relevance scores answer which resources better match the user\textquotesingle s need, while versions, freshness, and stable tie-breakers make the result reproducible. Authorization scope, suspension or revocation, and Root proof are not weighted terms that relevance can offset. The same input, index snapshot, and model version should produce a stable order; a changed model or view should be identifiable in the response explanation or run record.

\subsection{Matching Conditions}\label{subsec:11.9.1}

Matching conditions are the first basis for candidate ranking. Explicit conditions such as resource type, version, protocol, and capability tags are hard filters. Text matches in names, descriptions, use cases, and protocol materials are relevance signals. A candidate must satisfy explicit conditions before it participates in relevance ranking. A resource that is textually similar but has the wrong resource type or version must not enter the result.

\subsection{Trust and Governance State}\label{subsec:11.9.2}

Only candidates with completed root-platform trusted publication, valid governance state, and confirmable node authorization may enter a trusted discoverable result. Trust and governance state are not ordinary features that merely increase a retrieval score; they determine whether the candidate is eligible to be returned. A candidate with unknown state, missing proof, or suspended or revoked status should be filtered out or returned with an explicit restricted state. High relevance must not conceal these conditions.

\subsection{Version and Freshness}\label{subsec:11.9.3}

By default, only the current valid version of a resource DID is retained. When several candidates have similar relevance, the node should prefer the newer valid version whose publication sequence and governance state can both be confirmed. Freshness describes how far the index is from publication and governance facts; it must not be equated with retrieval relevance. An old snapshot may be used for limited display, but it must carry \passthrough{\lstinline!stale!} or an equivalent state.

\subsection{Ranking Stability}\label{subsec:11.9.4}

Under the same index snapshot, authorization scope, and query conditions, ranking should produce repeatable results. Ranking first uses the actually enabled relevance combination and then uses a deterministic resource DID, version, publication sequence, or other stable key as a secondary key. The node should not use database natural order, concurrent completion order, or random perturbation as an implicit ranking rule; otherwise pagination and client-side deduplication cannot work reliably.

\subsection{Duplicate-Result Handling}\label{subsec:11.9.5}

Resource-identity deduplication must occur either before or after ranking; duplicate synchronization records for the same DID must not occupy multiple result positions. Deduplication retains the current valid version, the newer publication sequence, and the record with more complete verifiable evidence, while retaining the source node and duplicate diagnostics. If duplicate records have conflicting hashes or governance states, they must enter exception handling rather than being arbitrarily selected as a trusted result.

\section{Search Explanations and Evidence Display}\label{sec:11.10}

An explanation should identify the path that produced a candidate, the conditions it matched, the filters applied, and the resource version used, rather than merely giving a natural-language conclusion. For a semantic query, it reports text, tag, or vector match sources and raw scores. For a trust decision, it reports authorization scope, lifecycle, and proof state. These two categories are presented separately so that a match reason is not mistaken for a trust reason.

\subsection{Match Reasons}\label{subsec:11.10.1}

Match reasons should point, as far as possible, to observable fields such as capability-tag intersection, resource-type match, protocol match, text-field match, or natural-language intent match. An explanation may be a summary and need not expose the internal model or index implementation, but it must not invent a capability that the resource did not declare. When a query uses a lexical, vector, or hybrid path, the response should identify the actual path or at least state the determinacy boundary of the explanation.

\subsection{Resource Provenance}\label{subsec:11.10.2}

The result should identify the discovery service node, index update time, resource source node or source reference, and available publication sequence and resource-package hash. Provenance fields trace where a fact was synchronized from; they do not mean that the source node guarantees the resource\textquotesingle s business quality. When the resource is obtained through a content distribution platform, the response should retain the reference relationship to the root platform publication material rather than recording only a cache address.

\subsection{DID, VC, and Root Platform Proof}\label{subsec:11.10.3}

The resource DID identifies the resource; the DID document provides public keys, service endpoints, and resource metadata; the registration credential expresses the registration service node\textquotesingle s processing of a registration fact; and the root platform trusted publication proof expresses the root platform\textquotesingle s verifiable validation and publication fact for a particular resource version and its materials. A discovery response may return summaries or references to these pieces of evidence, but before connecting, the caller must obtain and verify the complete materials. No reference field alone constitutes completed verification.

Response fields should be understood in two layers: ``index display'' and ``verifiable evidence''. \passthrough{\lstinline!resourceDid!}, \passthrough{\lstinline!resourceType!}, \passthrough{\lstinline!score!}, version, tags, and endpoints are candidate-display information. The DID document, registration credential, root platform trusted publication proof, and their signature materials are the basis for the caller\textquotesingle s verification of identity, control, and publication facts. A discovery node returning a candidate does not mean that it has performed endpoint connectivity testing, business compatibility assessment, or authorization on behalf of the caller.

\subsection{Governance State and Version Information}\label{subsec:11.10.4}

The explanation area should display the resource lifecycle, resource version, publication sequence, governance state, and index freshness together, allowing the caller to distinguish ``highly relevant'' from ``currently verifiable''. Governance state comes from a governance-event projection or another explicit fact source, while version information comes from the resource publication chain. When they conflict, the response should mark the conflict and apply restrictive handling; a newer display time must not override a revocation or suspension fact.

Search explanations should keep verifiable facts and retrieval inferences separate. \passthrough{\lstinline!matchReason!}, tag intersection, resource-type match, and protocol match are query explanations. The DID document, public key, registration credential, root platform trusted publication proof, publication sequence, and governance state are verification materials. The former answer ``why did this appear in the result?''; the latter answer ``which facts can the caller use for the next decision?'' They must not be merged into one recommendation score.

\section{Discovery Response Model}\label{sec:11.11}

A discovery response should allow the caller to identify at least four things: which discovery service node processed the query, which resource DID corresponds to each candidate, which index state produced the candidate, and why the candidate was returned. A reference implementation response may contain top-level fields such as \passthrough{\lstinline!discoveryDid!}, \passthrough{\lstinline!candidates!}, \passthrough{\lstinline!createdAt!}, and \passthrough{\lstinline!proof!}. A candidate should include the resource DID, resource type, retrieval score, version, lifecycle state, capability tags, authorization domains, service endpoints, protocol bindings, resource-package information, and root platform proof reference. A deployment may add fields, but display fields must not be mistaken for proof fields.

This section does not duplicate a response JSON containing abbreviated DIDs, fictitious endpoints, and non-current fields. A parseable sample should come directly from the cross-language \passthrough{\lstinline!ResourceDiscoveryResponse!} fixture or Discovery service test output and preserve its \passthrough{\lstinline!discoveryDid!}, \passthrough{\lstinline!candidates!}, \passthrough{\lstinline!createdAt!}, and optional \passthrough{\lstinline!proof!}. The response proof and a candidate\textquotesingle s \passthrough{\lstinline!rootProof!} identify the Discovery response source and resource publication source respectively; the caller still validates them under Chapters 7 and 9. An optional absent field is omitted or accompanied by an explicit status, not filled with an unverified default. Compatible clients may ignore unknown extensions, but core identity, state, and provenance fields must retain their meanings.

The minimum response skeleton corresponding to the current protocol structure is \passthrough{\lstinline!discoveryDid!}, \passthrough{\lstinline!candidates!}, \passthrough{\lstinline!createdAt!}, and optional \passthrough{\lstinline!proof!}. In a candidate, \passthrough{\lstinline!resourceDid!}, \passthrough{\lstinline!resourceType!}, \passthrough{\lstinline!score!}, \passthrough{\lstinline!version!}, \passthrough{\lstinline!lifecycleState!}, \passthrough{\lstinline!capabilityTags!}, \passthrough{\lstinline!authorizedDomains!}, \passthrough{\lstinline!services!}, \passthrough{\lstinline!protocolBindings!}, \passthrough{\lstinline!packageInfo!}, and \passthrough{\lstinline!rootProof!} respectively provide display responsibilities for identity, retrieval, state, capability, scope, service, and proof references. The current interface may not return a separate \passthrough{\lstinline!matchReason!} or a complete freshness object. A client must not fill absent fields with facts that have supposedly been verified.

The current discovery-query entry point is \passthrough{\lstinline!POST /discovery/resources/query!}, and its request body corresponds to \passthrough{\lstinline!ResourceDiscoveryQuery!}. This type may contain \passthrough{\lstinline!query!}, \passthrough{\lstinline!resourceType!}, \passthrough{\lstinline!capabilityTags!}, \passthrough{\lstinline!protocol!}, \passthrough{\lstinline!version!}, \passthrough{\lstinline!versionMode!}, and \passthrough{\lstinline!limit!}; \passthrough{\lstinline!limit!} defaults to 10 and \passthrough{\lstinline!versionMode!} defaults to \passthrough{\lstinline!latest!}. The response corresponds to \passthrough{\lstinline!ResourceDiscoveryResponse!}, whose candidates use \passthrough{\lstinline!ResourceDiscoveryCandidate!}. The returned \passthrough{\lstinline!score!} is a relevance score, not an authorization or trust level.

\passthrough{\lstinline!POST /discovery/query/explain!} also accepts \passthrough{\lstinline!ResourceDiscoveryQuery!}, but returns dynamic explanation JSON, including matched resource DIDs, resource types, match flags, scores, tag overlap, resource-type and protocol match status, and semantic-search status. The explanation response is for diagnosing the retrieval process. It is not an ordinary discovery result and cannot replace verification that the resource control relationship or publication proof is valid.

Field names in the current public protocol should be treated as the compatibility baseline for clients. When a resource candidate uses fields such as \passthrough{\lstinline!lifecycleState!}, \passthrough{\lstinline!services!}, \passthrough{\lstinline!packageInfo!}, and \passthrough{\lstinline!rootProof!}, clients should treat optional information according to field presence. Fields such as \passthrough{\lstinline!lifecycle!}, \passthrough{\lstinline!serviceEndpoints!}, \passthrough{\lstinline!rootProofRef!}, or \passthrough{\lstinline!freshness!} in examples are illustrative extensions; they must not be used to claim that all current deployments return those names. New response fields should preserve the ability of older clients to ignore unknown fields and should describe the source and verification method of each fact.

\section{Discovery-Result Freshness and State}\label{sec:11.12}

The freshness of a discovery result is described jointly by the index update time, the processed publication cursor, the governance-event sequence, and the local error state, rather than by the HTTP response time alone. A response should provide, where possible, \passthrough{\lstinline!createdAt!}, index update time, publication sequence or synchronization cursor, governance-state read position, and source-node DID. The caller uses these values to determine whether the result is a current snapshot, an acceptable old snapshot, or a partially available result.

{\def\LTcaptype{none} 
\begin{longtable}[]{@{}>{\raggedright\arraybackslash}p{0.3200\linewidth}>{\raggedright\arraybackslash}p{0.3200\linewidth}>{\raggedright\arraybackslash}p{0.3200\linewidth}@{}}
\toprule\noalign{}
State & Meaning & Permitted use \\
\midrule\noalign{}
\endhead
\bottomrule\noalign{}
\endlastfoot
current & Necessary publication and governance-state synchronization is complete, with no known update gap & May proceed to pre-invocation verification \\
stale & A verifiable local snapshot remains, but publication or governance synchronization is behind the permitted range & Read-only reference with an explicit freshness mark; state must be reconfirmed before invocation \\
partial & Some resources or index shards are available, and the completeness scope has been reduced & Use only within the scope stated by the response; do not claim a full result set \\
unknown & Synchronization position, governance state, or proof freshness cannot be confirmed & Must not be used as a trusted basis for sensitive actions \\
\end{longtable}
}

When version state and governance state conflict, restrictive states such as revocation, suspension, or loss of node authorization take precedence over relevance and time ordering. A discovery service node may retain historical snapshots to improve access during network failures, but it must mark the snapshot state explicitly and must not describe a cache hit as the network latest state.

State judgment for a discovery result should be based on local index material and its synchronization records. \passthrough{\lstinline!createdAt!} represents only the time when the response was generated; it is not the resource publication or index-application time. The publication cursor represents propagation position, the index update time represents processing time at this node, and the governance-event sequence represents the position of the governance read model. The caller must compare these sources and positions to determine whether the result may be behind.

An error response should allow the caller to distinguish a request error, a resource miss, a temporarily unavailable index, and a temporary backend failure. An invalid DID format, an unknown resource type, an unresolvable version mode, or an excessive query limit is a request-side error. A legal query with no visible candidate is an empty result. A database, semantic-index, governance-read-model, or upstream-synchronization failure is a service- or data-side error. The error should contain a machine-readable category, a request identifier, and necessary retry guidance, but should not disclose detailed information about restricted resources.

Public errors currently follow the status code and error object of the concrete handler; this manual does not present an \passthrough{\lstinline!INDEX\_STALE!} envelope not yet fixed by a public type as an established protocol. A client must at least distinguish invalid requests, no visible result, semantic degradation, unavailable base indexes, and synchronization lag from actual responses or status interfaces. If a unified error type is introduced later, Appendix~\ref{chap:appendix-b} should maintain executable examples and field versions.

When the semantic backend is unavailable, the node may fall back to lexical and structured paths. The fallback changes ranking capability only; it must not relax resource type, lifecycle, authorization scope, version, or explicit tag conditions. If the base index itself is unavailable, the node must not use an empty array to pretend that there are ``no resources''. It should return an identifiable service error or a partial result with an explicit scope. The client should retry based on \passthrough{\lstinline!retryable!}, backoff guidance, and its own request budget; the node must not amplify requests through automatic retries.

Error handling should preserve the boundary between ``invalid request'' and ``data temporarily unavailable''. A format error, unknown enumeration, or excessive parameter must not cause the same request to be retried. A temporary semantic-backend failure may use lexical degradation under the same hard conditions. If the base index or governance state cannot be read, a partial result must state its scope and state explicitly. A discovery node may record query statistics and backend names for operations, but its public error must not disclose the existence of resources outside the authorization scope.

Error handling for the discovery interface should keep the response structure stable and allow clients to decide whether to correct the request, wait and retry, or stop using the result. The following boundary may be used:

{\def\LTcaptype{none} 
\begin{longtable}[]{@{}>{\raggedright\arraybackslash}p{0.3200\linewidth}>{\raggedright\arraybackslash}p{0.3200\linewidth}>{\raggedright\arraybackslash}p{0.3200\linewidth}@{}}
\toprule\noalign{}
Error category & Typical cause & Client action \\
\midrule\noalign{}
\endhead
\bottomrule\noalign{}
\endlastfoot
Invalid request & DID format, resource type, version mode, or \passthrough{\lstinline!limit!} is invalid & Correct the request and issue it again; do not repeat the original request \\
No visible result & The conditions are valid, but this node has no candidate that both satisfies them and is visible & Adjust the query conditions if appropriate, but do not infer that the resource does not exist network-wide \\
Semantic degradation & The semantic index or embedding service is temporarily unavailable while the base index remains readable & Accept lexical/structured results and lower expectations for relevance \\
Index unavailable & The resource index, database, or necessary governance read model cannot be read & Handle as a service error, wait for recovery, or switch to a known node \\
Synchronization lag & The local publication cursor is behind or the snapshot is incomplete & Use only within the scope marked by the response and retry later if necessary \\
\end{longtable}
}

Semantic degradation and base-index failure must be distinguished in implementation and operational observation. The former can usually still return candidates satisfying hard conditions, with reduced relevance-ranking capability. The latter cannot establish that the candidate set is complete or that visibility was correctly determined and must not be disguised as an ordinary empty result. Error logs may record the backend, latency, candidate count, and failure reason, while the public response should contain only the minimum information needed for client handling.

\section{Discovery Errors and Degraded Responses}\label{sec:11.13}

The fact that a resource is reachable through an HTTP endpoint means only that the endpoint is reachable for the current request. Entering a discovery index means that a discovery service node has stored a corresponding index record. Having a root platform trusted publication proof means that the root platform has formed a verifiable publication fact for a particular version and its materials. Passing pre-invocation verification additionally requires the caller to recheck the DID control relationship, proof, governance state, endpoint, and its own business policy. These are progressively stronger facts; no level automatically replaces another.

{\def\LTcaptype{none} 
\begin{longtable}[]{@{}>{\raggedright\arraybackslash}p{0.3200\linewidth}>{\raggedright\arraybackslash}p{0.3200\linewidth}>{\raggedright\arraybackslash}p{0.3200\linewidth}@{}}
\toprule\noalign{}
Fact & Primary source & Conclusion that cannot be drawn \\
\midrule\noalign{}
\endhead
\bottomrule\noalign{}
\endlastfoot
Endpoint is reachable & Resource service endpoint & The resource is registered or trustworthy \\
Entered a discovery index & Discovery service node index & The resource is network-latest or permitted for invocation \\
Has a root platform trusted publication proof & Root platform publication material & The business endpoint is necessarily available or suitable for a particular scenario \\
Passed pre-invocation verification & Caller and its verification components & The caller has obtained all business permissions \\
\end{longtable}
}

A public discovery response may display the necessary DID, resource summary, endpoint, and proof reference. For a restricted resource, it should return only the minimum summary within the authorized scope. Before establishing a connection, the caller should obtain or verify the complete DID document, registration credential, root platform trusted publication proof, latest governance state, and endpoint declaration according to the risk of the resource.

Public visibility is jointly determined by the current authorization scope and index state of the discovery service node. A resource absent from public results may not yet be synchronized, may have been filtered, may be suspended, or may be affected by an unavailable index. An empty result alone cannot establish that the resource does not exist. When the caller needs a deterministic conclusion, it should use an exact DID query, read state information, and perform complete pre-invocation verification instead of inferring a hidden inventory from tag enumeration or paginated results.

Visibility involves at least three independent questions: whether the resource has been registered, whether it has propagated to this node, and whether the current caller can see it. A discovery service node can answer only from locally verifiable index material and authorization scope. Therefore, ``candidate returned'' means that the resource has a discovery record within the visible scope of the current node, while ``not returned'' means only that the current query did not produce a visible candidate. For sensitive resources, success, empty results, and permission filtering must not reveal additional information through error codes, field counts, or response latency.

Trusted discoverability also requires the candidate to have a DID document, registration credential, or root platform trusted publication proof reference corresponding to the current version, with no known suspension, revocation, or authorization-loss state. The discovery node provides an index and an evidence entry point. Whether the caller accepts the resource also depends on its own risk policy and its verification of control, proof validity, endpoint security, and business authorization.

\section{Public Visibility and Trusted Discoverability}\label{sec:11.14}

{\def\LTcaptype{none} 
\begin{longtable}[]{@{}>{\raggedright\arraybackslash}p{0.3200\linewidth}>{\raggedright\arraybackslash}p{0.3200\linewidth}>{\raggedright\arraybackslash}p{0.3200\linewidth}@{}}
\toprule\noalign{}
Record relationship & Discovery-result handling & Basis that must be retained \\
\midrule\noalign{}
\endhead
\bottomrule\noalign{}
\endlastfoot
Same DID, version, and package digest & Merge into one candidate & Source node and synchronization batch \\
Same DID, different versions & Select by \passthrough{\lstinline!versionMode!} or return separately & Version, publication sequence, predecessor relation \\
Same DID and version, different digest & Do not merge automatically as trusted & Conflicting digests, sources, diagnostics \\
Different DIDs, identical content & Return as different resources & Each DID, control relationship, and proof \\
\end{longtable}
}

\subsection{DID-Level Deduplication}\label{subsec:11.14.1}

Within one query response, the same resource DID should normally have only one representative record. The representative should be a current valid version with a newer publication sequence, while retaining a traceable source node, root platform proof reference, and synchronization position. If governance states conflict between records, the response should retain conflict diagnostics and use a restrictive state; arbitrary ordering must not hide the conflict.

\subsection{Version-Level Deduplication}\label{subsec:11.14.2}

Different valid versions of the same DID should be returned together only when the query explicitly requests historical versions or specifies a version constraint. A default query returns the current valid version. A historical query should include the version identifier, predecessor version, and publication sequence so that the client cannot mistake a historical record for the current version. The same version identifier with different package hashes is a consistency anomaly; the records should not be merged and should enter diagnostic or quarantine handling.

\subsection{Duplicate Records from Registration Nodes}\label{subsec:11.14.3}

Repeated submission of the same resource to one or more registration service nodes must not create multiple resource identities. A discovery node should deduplicate according to the root platform publication fact, resource DID, version, and package hash. A registration service node\textquotesingle s local acceptance record may be retained for source auditing, but it must not by itself create a public candidate. Whether a repeated submission causes a new registration credential to be issued is a registration-protocol behavior and does not change discovery-side resource-identity rules.

\subsection{Handling Identical Content with Different DIDs}\label{subsec:11.14.4}

Records with the same content, name, or endpoint but different DIDs remain different resource identities and must not be merged solely because their content hashes match. A node may indicate content similarity or a shared package hash in an explanation field, but it must display each control relationship, registration state, version, and root platform proof separately. Whether the same content may be represented by multiple DIDs is determined by the resource controller and governance rules.

The primary deduplication key is the complete resource DID. Records from multiple registration service nodes, content distribution platforms, or synchronization batches are normally the same resource identity when they point to the same resource DID. If records also contain versions, the resource DID and version identify the version. Package hashes, root-platform publication sequences, and source nodes are used to confirm content consistency and retain evidence; records must not be incorrectly merged merely because they have the same name or similar descriptions.

A deduplication decision should retain the ``basis for merging'' and the source of each discarded record so that the reason a candidate appears only once can be explained. A default query may use the following priority when selecting a representative record: first exclude records with restricted lifecycle states, mismatched authorization scopes, or unverifiable proofs; then compare the requested version mode and publication sequence; finally use package hash, source node, and deterministic DID order to resolve ties. This priority only constructs the discovery result. It does not change the original acceptance records of a registration service node or delete historical audit material.

Deduplication must occur before pagination. Otherwise, duplicate records for the same resource consume multiple result positions and cause omissions when the client merges pages. Historical-version queries are an exception: they may return multiple versions of the same DID, but every version must carry an explicit version and state so that the client cannot mistake a historical candidate for the current valid version.

\section{Result Deduplication and Resource-Identity Matching}\label{sec:11.15}

A node should limit query-text length, the number of structured conditions, the number of results returned per request, and semantic-retrieval resource consumption. \passthrough{\lstinline!limit!} must be validated against a server-side maximum; a client must not bypass the limit through repeated requests or abnormal parameters. An invalid, expired, or query-mismatched cursor should produce a clear error rather than silently returning the first page, so that the client does not mistake an incomplete pagination result for a complete one.

Within one index snapshot, pagination must use the same filtering, deduplication, and ranking rules. Ranking should include at least the retrieval score and a deterministic resource DID or other stable key. It must not use a random value per request, undefined database natural order, or unstable concurrent completion order. If the snapshot changes during pagination, the response should communicate this through a snapshot identifier, cursor invalidation, or freshness field so that cross-snapshot duplicates or omissions are not mistaken for a protocol error.

{\def\LTcaptype{none} 
\begin{longtable}[]{@{}>{\raggedright\arraybackslash}p{0.4800\linewidth}>{\raggedright\arraybackslash}p{0.4800\linewidth}@{}}
\toprule\noalign{}
Situation & Recommended handling \\
\midrule\noalign{}
\endhead
\bottomrule\noalign{}
\endlastfoot
Results are fewer than \passthrough{\lstinline!limit!} & Return all visible candidates that satisfy the conditions and indicate that no more results are available \\
Results reach the limit & Return a stable pagination cursor and current snapshot information \\
Query exceeds length or resource budget & Return a non-retryable request error \\
Cursor cannot be parsed or does not match the query & Return a cursor error and require the client to start the query again \\
Index updates during pagination & Continue using the original snapshot or explicitly invalidate the cursor \\
\end{longtable}
}

Pagination solves only transport and display scale. It does not change authorization scope or trusted-discoverability decisions. When merging pages, a client should use resource DID and version as deduplication keys and retain the source and snapshot information for each page.

\section{Query Limits, Pagination, and Result Stability}\label{sec:11.16}

Authorization-scope filtering must be completed before results are returned, rather than returning candidates first and relying on the client to hide them. For resources outside the requester\textquotesingle s or node\textquotesingle s authorization scope, the node must not disclose their existence through names, tags, endpoints, proof references, or error differences. A management interface that needs to distinguish insufficient permission from a missing resource may provide more detailed reasons to an authorized administrator. A public query should by default return only the minimum necessary fields of resources that are published, discoverable, and within this node\textquotesingle s service domain.

Index material should be classified by field sensitivity. Resource DID, resource type, and a public capability summary may enter public results. Private endpoints, internal tags, sensitive configuration other than the control public key, the submitter\textquotesingle s private identity information, and unpublished proof material must not be exposed through semantic indexing. Vector indexes, lexical indexes, and query logs must follow the same authorization boundary; a retrieval backend\textquotesingle s need for text does not justify copying content that should not be public.

Query logs should minimize storage of original natural-language input, caller identifiers, and filter conditions, and should use the deployment\textquotesingle s privacy policy to define retention, access control, and redaction. A requester must not infer the complete inventory of restricted resources by enumerating DIDs, tag combinations, pagination cursors, or error responses. During propagation of authorization-scope changes, resource revocation, or node suspension, the discovery node should restrict visibility first and retain the state source and processing time for audit.

\pandocbounded{\includegraphics[width=\linewidth,height=0.72\textheight,keepaspectratio]{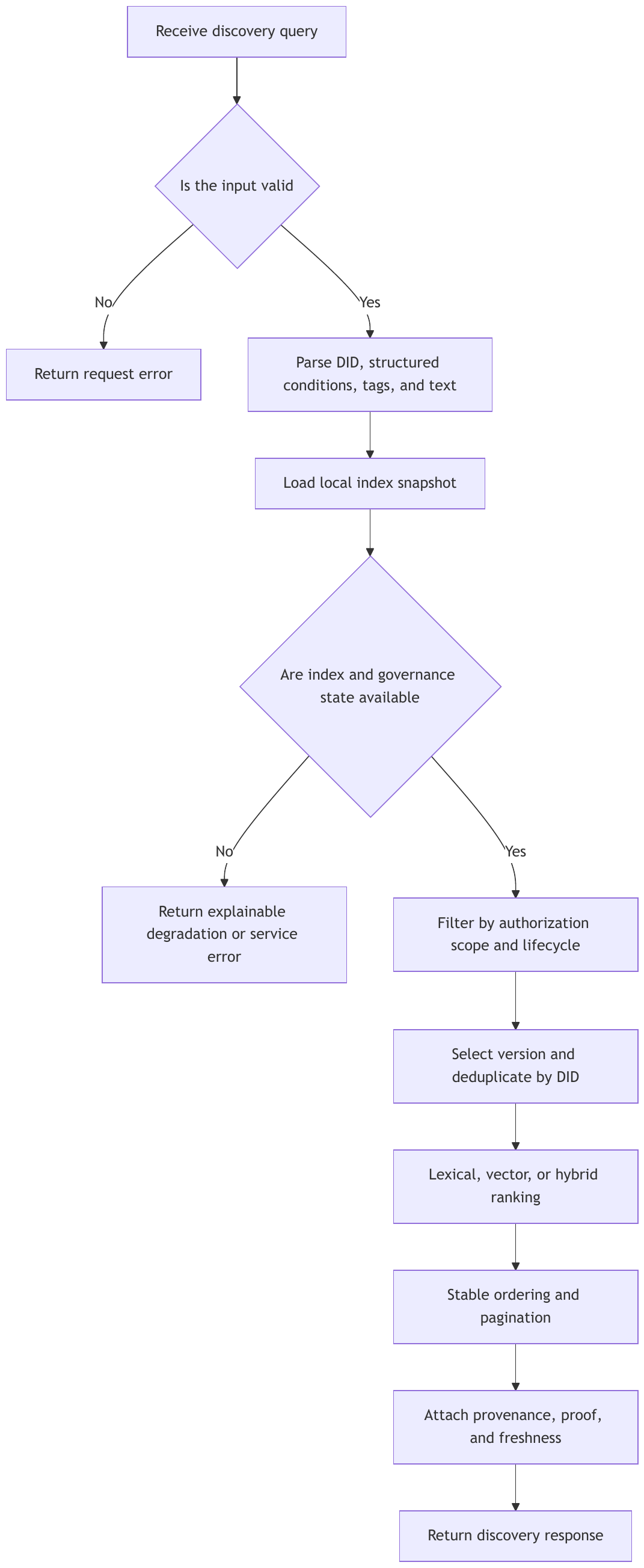}}

This flow expresses the order of processing responsibilities: input validation and hard filtering precede relevance ranking; version selection and deduplication precede pagination; and provenance, proof, and freshness are returned with the response. Use of a semantic backend provides ranking capability and must not change the preceding identity, authorization, or lifecycle decisions. Every degradation path must state which validations it completed and which validations it did not complete.

{\def\LTcaptype{none} 
\begin{longtable}[]{@{}>{\raggedright\arraybackslash}p{0.4800\linewidth}>{\raggedright\arraybackslash}p{0.4800\linewidth}@{}}
\toprule\noalign{}
Check item & Conformance criterion \\
\midrule\noalign{}
\endhead
\bottomrule\noalign{}
\endlastfoot
DID query & A complete resource DID uses the exact path and does not degrade into fuzzy search \\
Structured filtering & Resource type, version, protocol, capability tags, and lifecycle conditions use explicit intersection semantics \\
Authorization scope & Unauthorized resources do not appear in public results; an unknown scope state is handled restrictively \\
Relevance explanation & The retrieval score expresses relevance only and is not called a trust level or invocation permission \\
Version selection & The default returns the current valid version; historical-version queries do not confuse historical state with current state \\
Deduplication & The same DID does not create duplicate candidates through multi-node synchronization or repeated registration \\
Response provenance & The result can point to the discovery service node, index state, publication sequence, and proof reference \\
Freshness & Old snapshots, partial results, and unknown governance state are explicitly marked \\
Error degradation & Semantic-backend failure does not relax hard conditions, and base-index failure is not disguised as an empty result \\
Pagination stability & Ordering is stable within one snapshot, and the cursor is bound to the query and snapshot \\
Privacy boundary & Queries, indexes, and error responses do not disclose sensitive information about resources outside the authorization scope \\
Pre-invocation verification & The discovery result is only candidate input; the caller re-verifies identity, proof, and business permissions \\
\end{longtable}
}

\section{Discovery Privacy and Authorization-Scope Visibility}\label{sec:11.17}

Authorization scope is the boundary of discovery visibility, not a substitute for resource control or invocation permission. A discovery service node should first determine whether a candidate is visible by comparing the node\textquotesingle s authorized service domains with \passthrough{\lstinline!authorizedDomains!} in the resource package, and then perform structured and semantic processing only over visible candidates. If the resource and node domains do not intersect, the candidate must be filtered before the public response. A caller must not infer detailed information about a filtered resource from its name, tags, endpoint, result count, or error differences.

Indexing and display should follow minimum disclosure. A public candidate may provide the resource DID, resource type, public capability tags, necessary service endpoints, version, and proof references. Control private keys, the registrant\textquotesingle s local identity backup, internal governance materials, unauthorized endpoints, and configuration outside the public resource package must not enter the index or query response. Text, tag aliases, and vector representations used by semantic indexing inherit the authorization boundary of the original fields; entering a retrieval backend must not expand visibility.

While authorization scope, lifecycle, or governance state is synchronizing and the local node cannot confirm the latest state, public visibility should be restricted. For management and audit purposes, a controlled interface may record the filtering reason, source node, governance-event sequence, and processing time. For ordinary discovery requests, it should return only a uniform result that does not disclose the existence of a restricted resource. Query logs should also be retained according to minimum-necessary principles, with limits on access to and retention of the original natural-language input, caller information, and query conditions.

{\def\LTcaptype{none} 
\begin{longtable}[]{@{}>{\raggedright\arraybackslash}p{0.3200\linewidth}>{\raggedright\arraybackslash}p{0.3200\linewidth}>{\raggedright\arraybackslash}p{0.3200\linewidth}@{}}
\toprule\noalign{}
Information category & May enter a public discovery result & Handling requirement \\
\midrule\noalign{}
\endhead
\bottomrule\noalign{}
\endlastfoot
Resource identity & Resource DID and resource type & Bind to the currently visible version \\
Capability description & Public capability tags, name, and summary & Must not expand into a capability the resource did not declare \\
Service information & Service endpoints and protocol bindings within the authorized scope & The endpoint must still be verified before invocation \\
Trust material & Necessary reference to the root platform trusted publication proof & Do not treat a reference as completed verification \\
Private material & Private keys, local identity backup, internal configuration, and unpublished proof & Must not enter public indexes, responses, or error details \\
\end{longtable}
}

\section{References}\label{sec:11.18}

{\def\LTcaptype{none} 
\begin{longtable}[]{@{}>{\raggedright\arraybackslash}p{0.3200\linewidth}>{\raggedright\arraybackslash}p{0.3200\linewidth}>{\raggedright\arraybackslash}p{0.3200\linewidth}@{}}
\toprule\noalign{}
Source & Type & Link \\
\midrule\noalign{}
\endhead
\bottomrule\noalign{}
\endlastfoot
\passthrough{\lstinline!oan-discovery-nodes!} & Code repository: synchronization, indexing, structured queries, and semantic queries & \url{https://github.com/OpenAgenet/oan-discovery-nodes} \\
\passthrough{\lstinline!oan-protocol-commons!} & Code repository: discovery request, response, and query types & \url{https://github.com/wolfbrother/oan-protocol-commons} \\
\passthrough{\lstinline!oan-sdk-ts!} & Code repository: discovery client and query-result handling & \url{https://github.com/OpenAgenet/oan-sdk-ts} \\
GRAIL Semantic Discovery Paper & arXiv paper: hybrid semantic retrieval and discovery explanations & \url{https://arxiv.org/abs/2605.02489} \\
Efficient Agent Discovery Profile & IETF draft: agent-discovery metadata and query configuration & \url{https://datatracker.ietf.org/doc/draft-xu-efficient-agent-discovery-profile/} \\
\end{longtable}
}

\chapter{Semantic Governance and Semantic Discovery}\label{chap:12}

This chapter explains how OpenAgenet (OAN) governs capability tags, aliases, authorization domains, and semantic configuration, and how it uses semantic information such as resource names, descriptions, use cases, and protocol bindings for registration recommendations, structured filtering, text retrieval, semantic ranking, and result explanation. It also draws a clear boundary between semantic relevance and identity, control, publication proof, lifecycle, and governance status, ensuring that semantic discovery improves resource matching without changing trusted publication or authorization decisions.

\section{Semantic Model and Capability Vocabulary}\label{sec:12.1}

In OpenAgenet (OAN), the semantic model separates what a resource ``can do'' into several types of information that can be reused by registration and discovery: resource names and descriptions provide natural-language context, capability tags provide relatively stable classification keys, use cases describe applicable tasks, protocol bindings describe interaction methods, and resource types describe object categories. These elements jointly participate in recommendation and retrieval, but they cover different kinds of facts. The capability tag tree answers ``which capability category does this term belong to,'' the resource package answers ``which capabilities does a specific resource declare,'' and the DID Document and credential materials answer ``who controls the resource, whether it has completed registration, and whether the publication fact can be verified.''

Therefore, semantic matching results should be understood as candidate discovery clues rather than trusted conclusions. A Registrar node may generate tag recommendations based on names, descriptions, use cases, and protocols, and the provider may confirm them before they are written into Resource Metadata. A Discovery node may reuse these fields for filtering, lexical retrieval, or semantic ranking. No matter how high the recommendation score is, it cannot replace control verification, registration credentials, Root proof, lifecycle status, or authorization-domain checks.

\section{Root Platform-Managed Capability Tag Tree}\label{sec:12.2}

The capability tag tree is published by the Root platform together with its version and snapshot for use as shared semantic material by registration recommenders and Discovery nodes. The tag tree object in the current public implementation contains \passthrough{\lstinline!version!}, an optional \passthrough{\lstinline!snapshotHash!}, and \passthrough{\lstinline!tags!}; each tag node contains at least a stable \passthrough{\lstinline!id!} and a display \passthrough{\lstinline!label!}, and may also contain \passthrough{\lstinline!aliases!} and recursive \passthrough{\lstinline!children!}. The following is a minimal example organized according to this data structure; all values are illustrative placeholders:

\begin{lstlisting}
{
  "version": 1,
  "snapshotHash": "sha256:example",
  "tags": [
    {
      "id": "code.repository.search",
      "label": "Code Repository Search",
      "aliases": ["Search Codebases"],
      "children": []
    }
  ]
}
\end{lstlisting}

The structure of the tag tree organizes semantics and makes queries reproducible; it is not a resource authorization list. After the Root platform adds or adjusts tags, registration recommendations may use the new version, and Discovery nodes may rebuild their indexes accordingly. The original descriptions and original tags submitted by existing resources must not be silently rewritten because the tag tree changed. If a node uses an older snapshot, it should record the version used so that differences in recommendations or discovery results can be explained.

\subsection{Tag Tree Organization}\label{subsec:12.2.1}

The capability tag tree uses the organization of ``stable identifier + parent-child relationship + display name + vocabulary version.'' A tree node expresses a capability category; it does not directly express whether a resource has been registered, published, or trusted. A resource may be associated with multiple leaf nodes or only with a higher-level node, but the Registrar node should store the tags actually selected by the resource rather than only the recommended path.

\subsection{Tag Identifiers and Hierarchical Relationships}\label{subsec:12.2.2}

The tag ID is a cross-node association key. Parent-child relationships organize the retrieval scope, but capability inheritance must not be inferred without vocabulary rules. Query results should retain the matched tag ID, matching path, and vocabulary version so that cross-node results can be reproduced rather than returning only localized names.

{\def\LTcaptype{none} 
\begin{longtable}[]{@{}>{\raggedright\arraybackslash}p{0.3200\linewidth}>{\raggedright\arraybackslash}p{0.3200\linewidth}>{\raggedright\arraybackslash}p{0.3200\linewidth}@{}}
\toprule\noalign{}
Information & Purpose & Can it serve as trusted evidence \\
\midrule\noalign{}
\endhead
\bottomrule\noalign{}
\endlastfoot
Tag ID & Cross-node association and stable queries & No \\
Parent-child path & Classification and filtering & No \\
Tag alias & Query expansion & No \\
Resource control proof and publication proof & Control and publication facts & Yes, subject to independent verification \\
\end{longtable}
}

\subsection{Tag Recommendation and Discovery Use}\label{subsec:12.2.3}

During registration, recommendation results enter Resource Metadata only after confirmation by the resource provider. During discovery, the same tag ID, aliases, and hierarchical paths may be used for filtering or ranking. Recommendation hits, tag hits, and trusted-resource verification should be expressed separately in the response.

\begin{lstlisting}
Resource description -> Recommended candidates -> Provider confirmation -> Written to ResourcePackage
                                                                         -> Discovery index
Query text -> Intent and alias parsing -> Tag conditions -> Candidate ranking
\end{lstlisting}

\subsection{Tag Tree Version Management}\label{subsec:12.2.4}

When a tag tree is published, it should have a comparable version identifier and effective time. Tags from an older version may retain parsing mappings during a migration period, but the original resource declarations must not be silently rewritten. Until index rebuilding is complete, query results should identify the vocabulary version used.

\section{Tag Hierarchies, Identifiers, Aliases, and Localization}\label{sec:12.3}

A tag object should at least distinguish its normative identifier, display name, aliases, language, and version:

{\def\LTcaptype{none} 
\begin{longtable}[]{@{}>{\raggedright\arraybackslash}p{0.3200\linewidth}>{\raggedright\arraybackslash}p{0.3200\linewidth}>{\raggedright\arraybackslash}p{0.3200\linewidth}@{}}
\toprule\noalign{}
Field & Example & Processing rule \\
\midrule\noalign{}
\endhead
\bottomrule\noalign{}
\endlastfoot
\passthrough{\lstinline!id!} & \passthrough{\lstinline!code.repository.search!} & Stable; does not change with translation \\
\passthrough{\lstinline!label!} & \passthrough{\lstinline!Code Repository Search!} & Used for display \\
\passthrough{\lstinline!aliases!} & \passthrough{\lstinline!Search Codebases!} & Used for query expansion \\
\passthrough{\lstinline!language!} & \passthrough{\lstinline!en-US!} & Identifies the language of the text \\
\passthrough{\lstinline!vocabularyVersion!} & \passthrough{\lstinline!v1!} & Used for reproduction and migration \\
\end{longtable}
}

An unknown tag may be retained as ungoverned provider text, but it must not be presented as an official Root platform tag.

The responsibilities of tag IDs, display names, aliases, and hierarchical paths should remain separate. A query may first use a localized name or alias to find the normative ID and then use the normative ID for structured filtering; the returned result should retain both the actual resource declaration and the matched normative tag. This supports Chinese, English, and synonym queries without treating a display translation as a new capability declaration.

\section{Tag Selection and Recommendation During Registration}\label{sec:12.4}

\subsection{Resource Provider Selection}\label{subsec:12.4.1}

The resource provider may accept, cancel, or modify recommended tags. The final submission should retain both the provider\textquotesingle s original description and the selected normative tags so that it remains possible to explain ``what the resource declared'' and ``what the system recommended.''

\subsection{Root Platform Recommendation}\label{subsec:12.4.2}

The Root platform or its governance materials provide the tag tree, aliases, and version. A recommender generates candidates according to the resource type, name, description, use case, and protocol. The recommender does not issue a resource control proof and does not replace registration validation.

\subsection{Custom Tag Supplement}\label{subsec:12.4.3}

Custom tags express business terms that are not yet covered by the official vocabulary. A custom tag must carry a source or namespace and must be stored separately from official tags. It may participate in text retrieval, but it cannot directly change the resource type, authorization domain, or publication status.

\subsection{Tag Validation and Persistence}\label{subsec:12.4.4}

Before saving, the system should at least validate the tag ID, vocabulary version, parent-node relationship, language, duplicates, and compatibility with the resource type. The validation result should be associated with the resource DID, resource version, and ResourcePackage digest. A recommendation failure or an expired tag must not cause a silent rewrite.

\section{Custom Tags and Provider Descriptions}\label{sec:12.5}

Custom tags are suitable for specific business terms, organization-specific names, or capabilities that have not yet entered the public vocabulary. They should be stored separately from official Root platform tags and should carry provider namespace or source information. Names, descriptions, use cases, and other free text should also retain the original text and language marker. An official tag ID must come from the currently available tag tree; a custom vocabulary must not be promoted to an official tag merely because its format or semantics appear similar.

A Discovery node may use custom tags and provider descriptions for lexical or semantic retrieval, but the result explanation must distinguish among ``provider declaration,'' ``official tag hit,'' and ``retrieval-model inference.'' Duplicate tags, excessively long text, invalid characters, tags that clearly conflict with the resource type, and keywords intended to manipulate ranking should be rejected, truncated, or degraded to ordinary text according to the node\textquotesingle s input validation and indexing strategy. The processing result should retain a traceable reason and must not overwrite the original submission.

\section{Semantic Governance Change Process}\label{sec:12.6}

Semantic governance changes should be expressed through new versioned materials rather than by directly rewriting old tags or historical resources. A change record should at least identify the change type, affected tag IDs, old and new versions, alias or hierarchy mappings, effective time, and the effects on registration recommendations and discovery indexes. Additions and renamings can usually be made compatible through new mappings; merges and splits require an explicit explanation of how old tags are queried; deprecated tags should stop being recommended for new resources while the original values and parseable relationships of historical resources are retained.

After a change is published, the registration recommender may load the new vocabulary, and the Discovery node may gradually rebuild related indexes by resource DID and resource version. Until rebuilding is complete, query results should be able to state the vocabulary or index version used. If an old tag cannot be mapped, the node may retain its original text for text retrieval, but must not present it as a definite hit of a new tag. Tag governance changes semantic interpretation and retrieval materials only; they do not change the resource DID, control relationship, registration credential, or Root platform publication fact.

\section{Semantic Index Construction by Discovery Nodes}\label{sec:12.7}

\subsection{Index Inputs}\label{subsec:12.7.1}

The index inputs of a Discovery node fall into three categories:

\begin{enumerate}
\def\labelenumi{\arabic{enumi}.}
\tightlist
\item
  \textbf{Resource facts}: resource DID, version, public DID Document, and Resource Metadata.
\item
  \textbf{Semantic materials}: normative tags, aliases, names, descriptions, use cases, protocols, and languages.
\item
  \textbf{Governance materials}: lifecycle, Root proof reference, authorization domain, tag tree version, and source node.
\end{enumerate}

Private keys, unpublished drafts, and internal configuration must not enter the index.

\subsection{Tag Index}\label{subsec:12.7.2}

The tag index should at least support filtering by normative tag ID, hierarchical path, language, and vocabulary version, and should retain the tag source. An index hit only means that the resource declaration or index record matches; the node must still check whether the resource is in a discoverable state.

\subsection{Text and Semantic Indexes}\label{subsec:12.7.3}

The text index may cover names, descriptions, and use cases; the semantic index may use a recommender or an optional vector implementation. Both should return the matched fields, index version, and matching basis, and neither may add capabilities that the resource did not declare.

\subsection{Index Update and Deletion}\label{subsec:12.7.4}

When the resource version, lifecycle, Root proof, or tag tree version changes, the node should perform an incremental update by resource DID and version. After a revocation, suspension, or deletion event arrives, the index may retain an audit record, but the resource must not continue to be returned as a normal discoverable candidate.

\section{Matching Tags, Text, Protocols, and Resource Types}\label{sec:12.8}

The four types of conditions have different roles in a query: resource type and an explicitly specified version are generally used to limit candidate categories; capability tags and protocols may serve as explicit filters; text in names, descriptions, and use cases is mainly used for relevance ranking. Multiple explicit conditions are generally processed as an intersection. A tag query requires a candidate to satisfy the requested tag set, while a protocol may match a protocol binding or service declaration. Semantic similarity may improve ranking, but it must not bring a candidate that fails a hard condition back into the result set.

\begin{lstlisting}
Candidate eligibility = Visibility and lifecycle + Explicit structured conditions
Relevance ranking = Tag/protocol hits + Text relevance + Semantic similarity
Trust decision = DID Document + Credentials/Root proof + Caller policy
\end{lstlisting}

The first two lines define the discovery retrieval scope, while the last line is an independent pre-invocation check for identity, publication, and calling policy. When the semantic service is unavailable, a Discovery node may fall back to lexical or structured retrieval, but it must not relax the authorization domain, lifecycle, or explicit conditions.

\section{Query Explanation and Matching Details}\label{sec:12.9}

\subsection{Query Intent Parsing}\label{subsec:12.9.1}

Query parsing should output auditable intermediate conditions, such as resource type, capability tags, protocol, language, and status filters. An uncertain term may be retained as a text condition and must not be forcibly converted into an official tag.

\subsection{Tag and Text Matching}\label{subsec:12.9.2}

Tag hits, alias hits, original-text hits, and semantic-inference hits should be returned in separate categories. When a hard filter is not satisfied, a relevance score must not cause the candidate to re-enter the result set.

\subsection{Structured Condition Matching}\label{subsec:12.9.3}

Resource type, protocol, authorization domain, lifecycle, and publication status are structured conditions. The result for a structured condition should be explicitly expressed as satisfied, unsatisfied, or unknown; unknown must not be treated as satisfied.

\subsection{Result Explanation}\label{subsec:12.9.4}

The query explanation interface of the current Discovery node accepts the same \passthrough{\lstinline!ResourceDiscoveryQuery!} as an ordinary discovery query, but returns dynamic JSON rather than \passthrough{\lstinline!ResourceDiscoveryResponse!}. The explanation may contain \passthrough{\lstinline!items!}, \passthrough{\lstinline!candidateCount!}, \passthrough{\lstinline!matched!}, various matching scores, tag overlap, structured-condition hits, and semantic-search status. These fields are observable retrieval results, not new resource proofs. A client should use the explanation to answer ``why was this returned,'' while relying on the candidate resource\textquotesingle s DID Document, ResourcePackage, Root proof, and governance status to answer ``can this be trusted or invoked.''

An explanation should at least contain the resource DID, matched fields or tags, vocabulary/index version, and the source of trusted status. The distinction between the two types of information can be expressed as follows:

\begin{quote}
``Relevant'' means that the query matches the resource description; ``trusted'' additionally requires independent verification of the DID, control proof, Root proof, authorization status, and lifecycle.
\end{quote}

Query explanations should retain the observable path from natural language to structured conditions. For example, when the user enters ``I need a tool that can search code repositories and summarize project structures,'' the system may identify ``tool'' as a candidate resource type, and ``search code repositories'' and ``summarize project structures'' as capability or text-retrieval clues. If the current vocabulary has no reliable mapping to an official tag, the system should retain the original sentence and ordinary text hit rather than create a new tag based on model inference. The explanation may display matched tags, fields, aliases, and index version, but should not expose unnecessary internal model details.

\section{Multilingual Semantic Discovery}\label{sec:12.10}

\subsection{Chinese and English Resource Descriptions}\label{subsec:12.10.1}

Chinese and English descriptions may be stored together, and the original text should carry a language marker. Translated or machine-generated text may serve as a retrieval projection only and must not replace the provider\textquotesingle s original description.

\subsection{Multilingual Tags}\label{subsec:12.10.2}

Tag names in different languages should map to the same normative tag ID. If no reliable mapping exists, the name should be retained as ordinary text rather than used to create an ungoverned new synonym relationship.

\subsection{Query Language Identification}\label{subsec:12.10.3}

Language identification selects only the lexical, alias, and tokenization paths. It must not change resource type, status, or trust-filter conditions. A mixed-language query should retain the parsing basis for each segment.

\subsection{Cross-Language Matching}\label{subsec:12.10.4}

Cross-language hits should return the normative tag ID, the language and alias actually matched, and an indication of whether the hit came from tag mapping or text translation. A cross-language hit does not mean that the resource has an additional capability.

\section{Extensibility for Vector and Hybrid Retrieval}\label{sec:12.11}

The semantic extension of the current Discovery node may use PostgreSQL/pgvector and an HTTP embedding service to build resource-level semantic indexes; a local semantic recommender provides auxiliary recommendations for capability tags, resource types, and protocols. The choice of backend is a deployment and reference-implementation choice and must not change the processing rules for resource DID, ResourcePackage, authorization domain, lifecycle, or Root proof. A semantic index should also record the model name, model version, vector dimension, index version, and semantic-source hash so that results before and after rebuilding can be explained.

{\def\LTcaptype{none} 
\begin{longtable}[]{@{}>{\raggedright\arraybackslash}p{0.3200\linewidth}>{\raggedright\arraybackslash}p{0.3200\linewidth}>{\raggedright\arraybackslash}p{0.3200\linewidth}@{}}
\toprule\noalign{}
Stage & Information to retain & Boundary on failure \\
\midrule\noalign{}
\endhead
\bottomrule\noalign{}
\endlastfoot
Resource projection & DID, version, public description, tags, use cases, and protocols & Do not write private materials to the semantic index \\
Embedding generation & Model identifier, model version, dimension, and source-content association & A failure for one resource must not fabricate a vector \\
Query retrieval & Query text, hard conditions, index version, and backend & Semantic failure may degrade, but must not relax hard conditions \\
Result explanation & Relevance, matched fields, and evidence references & Relevance is not trust or invocation permission \\
\end{longtable}
}

When the model, dimension, or semantic vocabulary changes incompatibly, the corresponding index should be rebuilt first, and the new results should then be served as a single index version to callers. An old index may be retained during migration, but scores produced by different models must not be compared directly.

\section{Semantic Governance Compatibility and Migration}\label{sec:12.12}

During migration, three types of information should be retained together: the original tags and descriptions submitted by the resource provider, the vocabulary version used at that time, and the parseable mapping in the new version. A compatibility mapping may allow old queries to continue finding a resource, but it must not rewrite the original fields in historical ResourcePackages. After a Discovery node completes index migration, it should update the index version. If migration is incomplete or a tag cannot be mapped, a result may continue to serve as limited text-based reference, but its semantic-governance status should be stated.

\section{Semantic Vocabulary Versioning and Change Impact}\label{sec:12.13}

Vocabulary versions and semantic index versions should be recorded separately: the former identifies the version of the tag tree and alias materials, while the latter identifies the inputs actually used by a particular Discovery node to construct its retrieval data. The effects of the two types of changes can be understood as follows:

{\def\LTcaptype{none} 
\begin{longtable}[]{@{}>{\raggedright\arraybackslash}p{0.2400\linewidth}>{\raggedright\arraybackslash}p{0.2400\linewidth}>{\raggedright\arraybackslash}p{0.2400\linewidth}>{\raggedright\arraybackslash}p{0.2400\linewidth}@{}}
\toprule\noalign{}
Change type & Registration recommendation & Original value of existing resources & Discovery index \\
\midrule\noalign{}
\endhead
\bottomrule\noalign{}
\endlastfoot
Add a tag or alias & Available for new recommendations & Not affected & Incremental update is possible \\
Rename or localize & Retain the normative ID & Retain the original text & Update display and retrieval mappings \\
Merge tags & Establish an old-to-new query mapping & Do not rewrite automatically & Rebuild projections for related resources \\
Split tags & Require additional information from the provider & Do not infer automatically & Retain and mark the old value when uncertain \\
Deprecate a tag & Stop new recommendations & Retain a historical parseable value & Degrade or remove from normal recommendations \\
\end{longtable}
}

When rolling back, the system should restore the previous verifiable vocabulary and index versions rather than deleting resource history. Query results and operational diagnostics should at least identify the vocabulary version, index version, and update time used, so that it is possible to explain why the same query produces different rankings on different nodes.

\section{Semantic Governance Auditability}\label{sec:12.14}

The audit records for semantic governance need to answer ``who changed what in which vocabulary version and at what time, and which recommendations and indexes were affected.'' At minimum, the records should associate the operator or governance subject, operation time, change type, old and new values, vocabulary version, reason, impact scope, index rebuilding status, and rollback basis. Audit records explain changes in semantic materials; a tag-change record must not be mistaken for a resource registration credential or resource control proof.

Registration recommendations and discovery queries should also retain necessary references. For example, a recommendation may be associated with an input digest, vocabulary version, and recommendation candidates; a discovery explanation may be associated with matched tag IDs, aliases or text fields, index version, and resource DID. Natural-language originals, caller information, and internal model outputs should follow the deployment operator\textquotesingle s access-control and minimum-retention policies. A public interface should expose only the minimum content needed for explanation.

\section{Mapping Natural-Language Queries to Structured Conditions}\label{sec:12.15}

Natural-language mapping should distinguish between conditions that can be assigned to structured fields and expressions that can only serve as retrieval clues. For example, ``find a tool that can search code repositories and summarize project structures'' may produce a candidate resource type, candidate capability tags, and text-retrieval terms. Words such as ``latest,'' ``trusted,'' and ``invocable'' may participate in deterministic filtering only when the protocol and governance materials provide corresponding fields or facts; otherwise, they can serve only as ranking or explanation hints. An uncertain term must retain its original text and must not be forcibly mapped to an official tag or authorization condition.

The object above is an illustrative intermediate result for explanation and is not the fixed response format of the current discovery interface. A formal request must still use the discovery-query fields specified in Chapter~\ref{chap:11}. The value \passthrough{\lstinline!suggested!} indicates a semantic suggestion; the caller or resource provider must still confirm the tag, resource type, and protocol conditions. If mapping fails, the vocabulary version is incompatible, or the language does not match, the system may fall back to original-text retrieval or return an explicit result indicating that no usable mapping is available, but it must not conceal the failure as an incorrect formal condition.

\section{References}\label{sec:12.16}

{\def\LTcaptype{none} 
\begin{longtable}[]{@{}>{\raggedright\arraybackslash}p{0.3200\linewidth}>{\raggedright\arraybackslash}p{0.3200\linewidth}>{\raggedright\arraybackslash}p{0.3200\linewidth}@{}}
\toprule\noalign{}
Source & Type & Link \\
\midrule\noalign{}
\endhead
\bottomrule\noalign{}
\endlastfoot
\passthrough{\lstinline!oan-discovery-nodes!} & Code repository: semantic aliases, indexes, and semantic queries & \url{https://github.com/OpenAgenet/oan-discovery-nodes} \\
\passthrough{\lstinline!oan-protocol-commons!} & Code repository: capability tags, authorization domains, and semantic support materials & \url{https://github.com/wolfbrother/oan-protocol-commons} \\
GRAIL Semantic Discovery Paper & arXiv paper: semantic discovery, hybrid retrieval, and ranking explanations & \url{https://arxiv.org/abs/2605.02489} \\
Efficient Agent Discovery Profile & IETF draft: discovery metadata and query profile & \url{https://datatracker.ietf.org/doc/draft-xu-efficient-agent-discovery-profile/} \\
\end{longtable}
}

\chapter{Node Authorization and On-Chain Governance}\label{chap:13}

This chapter explains how OpenAgenet (OAN) converts on-chain governance facts into authorization states that infrastructure nodes can use. Governance contracts record node authorization, authorization domains, capability scopes, and status changes; the off-chain trust indexer reads and stores these events and produces queryable subject states; the Root platform, Registrar nodes, and Discovery nodes use these states to decide whether to accept requests, propagate resources, or use governance materials produced by a node. On-chain authorization, node authorization credentials, resource registration, and resource publication prove different facts and cannot substitute for one another.

In this chapter, ``current state'' means the judgment made by a component based on its synchronized read model. The read model may temporarily lag behind on-chain events. Therefore, when handling new authorization, suspension, resumption, revocation, or high-risk actions, the event sequence, synchronization position, state validity period, and evidence source should all be considered.

\section{Governance Objectives and the On-Chain/Off-Chain Boundary}\label{sec:13.1}

The on-chain governance layer and the off-chain business layer are connected through ``read, verify, project, and execute''; they do not share one state table. On-chain governance contracts produce reviewable raw events, the indexer stores events and subject states, and the Root platform or node services read this projection and make runtime decisions together with the node DID, authorization credentials, and request content. Resource control proofs, resource packages, and publication status belong to the resource business path. Their existence does not allow a service to skip node governance-state checks.

\subsection{On-Chain Governance Facts}\label{subsec:13.1.1}

On-chain governance facts include node authorization, suspension, resumption, and revocation, together with their times, sequences, and sources. On-chain facts are the authoritative source of governance state, but an off-chain service may retain an old read model before synchronization is complete. Responses should distinguish the on-chain position, indexing time, and state currently adopted by the service.

In the current governance contract, governance events are expressed by \passthrough{\lstinline!GovernanceExecutionEvent!}. The event contains fields such as the action type, event sequence, target node DID, node type, status, authorization domain, effective time, expiration time, policy hash, metadata hash, and the previous-event and current-event digests. The raw payload, transaction, event sequence, and on-chain position such as the checkpoint should be stored together. In this way, the read side can obtain both a node's current state and the governance execution that produced that state.

The relationship between on-chain events and off-chain services can be summarized as follows: the chain records that ``a governance action has occurred, together with its order and parameters''; the indexer constructs a queryable state from these events; and the Root platform and other nodes use that state according to their own policies. The indexer cannot modify on-chain events, reissue governance permissions, or fill in missing events on its own. When the raw event and the read model disagree, the original position and event digest should be the starting point for verification.

{\def\LTcaptype{none} 
\begin{longtable}[]{@{}>{\raggedright\arraybackslash}p{0.4800\linewidth}>{\raggedright\arraybackslash}p{0.4800\linewidth}@{}}
\toprule\noalign{}
Field & Purpose \\
\midrule\noalign{}
\endhead
\bottomrule\noalign{}
\endlastfoot
Node DID & Identifies the governed node \\
Event type & Indicates authorization, suspension, resumption, or revocation \\
Authorization domain and capabilities & Limits the operations that may be performed \\
On-chain position & Supports ordering and verification \\
Effective time & Distinguishes the on-chain occurrence time from the service observation time \\
\end{longtable}
}

\subsection{Root Platform Runtime State}\label{subsec:13.1.2}

Root platform runtime state indicates whether the service has read and applied the current governance result, including whether it accepts node submissions, permits publication, and permits downstream distribution. It is a runtime projection of on-chain governance facts and must not replace those facts.

At runtime, the Root platform must distinguish at least between ``the governance state is known'' and ``this request has been authorized.'' The former depends on the state, sequence, and freshness provided by the off-chain trust indexer or another governance read side. The latter also requires matching the requesting node DID, requested action, authorization domain, and capability scope. Even if a node is in \passthrough{\lstinline!active!} status, this only indicates that it may have the corresponding infrastructure responsibility. Resource control, resource content, version, and publication materials still require separate verification.

\subsection{Registration and Publication Status}\label{subsec:13.1.3}

Registration acceptance, Root platform publication, content distribution, and Discovery visibility are different states. Valid node authorization is a qualification for executing infrastructure actions; it does not mean that a resource has passed control, content, and version verification.

These states are produced by different components: the Registrar node records the resource registration request and its validation result; the Root platform decides whether to form a Root-verified publication fact; the content distribution platform propagates approved materials; and the Discovery node adds visible resources to its local index. On-chain node authorization only determines whether an infrastructure node is qualified to perform a certain type of action. It cannot replace the resource provider's signature, resource package integrity checks, publication process, or Discovery index update.

\begin{lstlisting}
Node authorized -> may submit a request
Request validated -> may enter the publication process
Root platform publication -> may enter the distribution process
Discovery node indexing -> resource may become visible to queries
\end{lstlisting}

\subsection{Discovery Node Read Model}\label{subsec:13.1.4}

The Discovery node uses the off-chain trust indexer or governance synchronization result to filter resources. The read model must record its source and synchronization progress. When a governance event has not yet been synchronized, the node should use a conservative state and must not interpret an old cache directly as a current authorization conclusion.

The Discovery node consumes a governance read model in normal operation, rather than scanning the entire on-chain history for every query. The read model must at least associate the node DID, node type, current status, authorization domain, last governance sequence, event digest, and update time. The source status in a query result should also remain separate from the resource's own lifecycle and publication proof. If the indexer is behind, a Discovery node may retain a read-only snapshot to improve access continuity, but it must not label that snapshot as the current governance fact.

\section{Infrastructure Node Authorization Model}\label{sec:13.2}

Infrastructure nodes use the node DID as the governance subject and distinguish Registrar nodes, Discovery nodes, and VC issuers by node type. The current indexer's \passthrough{\lstinline!SubjectState!} stores the node type, node DID, status, authorization domain, policy hash, metadata hash, effective time, expiration time, state version, last update time, last governance sequence, and last event digest. A node authorization credential can be used as off-chain evidence, but it and the on-chain state are two evidence chains that must be checked together: the credential states the issuer's claim about the node's qualification, while the on-chain read model states the current governance status and authorization scope.

A minimal state judgment can be represented as follows:

The DID, sequence, and authorization domain in the example are illustrative values. At runtime, the state can be used as an input for ``governance-level authorization'' only when the node type matches, the status is valid, the effective time has been reached, and the expiration time has not been exceeded. The requested action must then still be checked against the authorization domain and capability scope.

\section{Authorization Domains and Capability Scopes}\label{sec:13.3}

The authorization domain answers ``which business or resource scopes may the node process,'' while the capability scope answers ``which infrastructure actions may the node perform.'' For example, a Discovery node may be authorized to serve a particular capability domain, but this does not mean that it may represent the resource controller to modify a resource DID Document, nor that it can issue a registration credential. Authorization domains should use normalized and comparable values. The current contract also constrains empty domains, quantity, length, canonical order, duplicate entries, and the mixed use of wildcards.

Multiple authorization domains normally represent the union of the node's service scopes. Whether a specific resource may be processed still requires a combined judgment over node status, the resource authorization domain, and governance state. A change to the authorization domain should produce a new governance event and state version. After the scope is reduced, the node must not continue accepting new requests under the old scope. Historical publication facts that have already been formed should retain their source and time and must not simply be deleted.

\section{Governance Event Model}\label{sec:13.4}

The governance contract handles proposal creation, voting, and final execution separately; the indexer focuses on the emitted initialization events and governance execution events. After parsing, the indexer stores both raw event records and structured governance events. Raw records retain the transaction digest, event sequence, checkpoint, package identifier, event type, time, and original JSON. Structured records support queries by node DID, governance sequence, status, and authorization domain. Event names and descriptions shown on a page are projection fields and cannot replace the raw event payload.

The key fields of a governance execution event can be understood as follows:

{\def\LTcaptype{none} 
\begin{longtable}[]{@{}>{\raggedright\arraybackslash}p{0.3200\linewidth}>{\raggedright\arraybackslash}p{0.3200\linewidth}>{\raggedright\arraybackslash}p{0.3200\linewidth}@{}}
\toprule\noalign{}
Field category & Typical fields & Purpose \\
\midrule\noalign{}
\endhead
\bottomrule\noalign{}
\endlastfoot
Identity relationship & \passthrough{\lstinline!subjectDid!}, \passthrough{\lstinline!subjectType!}, \passthrough{\lstinline!targetAddress!} & Associates the governed node with its on-chain target \\
Action and status & \passthrough{\lstinline!eventType!}, \passthrough{\lstinline!eventLabel!}, \passthrough{\lstinline!subjectStatus!} & Indicates authorization, suspension, resumption, revocation, or domain change \\
Scope and time & \passthrough{\lstinline!authorizedDomains!}, \passthrough{\lstinline!effectiveFromMs!}, \passthrough{\lstinline!expiresAtMs!} & Constrains the applicable scope and validity window \\
Ordering evidence & \passthrough{\lstinline!sequence!}, \passthrough{\lstinline!previousEventDigest!}, \passthrough{\lstinline!eventDigest!} & Supports event-chain and state-replay verification \\
Change basis & \passthrough{\lstinline!proposalId!}, \passthrough{\lstinline!policyHash!}, \passthrough{\lstinline!metadataHash!} & Associates the event with governance proposals and policy materials \\
On-chain source & Transaction digest, event sequence, checkpoint & Points to the original on-chain record and read position \\
\end{longtable}
}

The official Network page exposes the event list in the current read model through the public endpoint \href{https://www.openagenet.xyz/api/public/network/governance-events}{\passthrough{\lstinline!/api/public/network/governance-events!}}. At the time of this review, the endpoint's \passthrough{\lstinline!generatedAt!} was \passthrough{\lstinline!2026-09-12T02:01:05.463553057+00:00!}. The following table records the time, node, and page description in the order currently returned by the endpoint. It gives readers a direct view of publicly observed governance actions, but does not replace the raw event payload, transaction, or checkpoint in the governance contract. The list changes as on-chain governance and indexer synchronization change:

{\def\LTcaptype{none} 
\begin{longtable}[]{@{}>{\raggedright\arraybackslash}p{0.3200\linewidth}>{\raggedright\arraybackslash}p{0.3200\linewidth}>{\raggedright\arraybackslash}p{0.3200\linewidth}@{}}
\toprule\noalign{}
Time (UTC) & Node & Event description \\
\midrule\noalign{}
\endhead
\bottomrule\noalign{}
\endlastfoot
2026-08-25 03:11:30 & Test Discovery & Updated the authorization domain of the test Discovery node to the full domain. \\
2026-06-29 02:16:44 & Test Discovery & Updated the authorization domain of the test Discovery node to a set of 28 industry/domain categories. \\
2026-06-29 02:15:47 & Registrar 1 & Updated the authorization domain of the Registrar node to a set of 28 industry/domain categories. \\
2026-06-29 01:26:06 & Test Discovery & Updated the authorization domain of the test Discovery node to a set of 16 industry/domain categories. \\
2026-06-29 01:25:10 & Registrar 1 & Updated the authorization domain of the Registrar node to a set of 16 industry/domain categories. \\
2026-06-29 01:12:12 & Test Discovery & Updated the authorization domain of the test Discovery node to \passthrough{\lstinline!technology.software\_engineering!}. \\
2026-06-29 01:11:05 & Registrar 1 & Updated the authorization domain of the Registrar node to \passthrough{\lstinline!technology.software\_engineering!}. \\
2026-06-28 22:48:58 & Official Discovery & Updated the authorization domain of the official Discovery node to the full domain. \\
2026-06-28 22:47:41 & Test Discovery & Updated the authorization domain of the test Discovery node to the full domain. \\
2026-06-28 22:46:14 & Test Registrar & Updated the authorization domain of the test Registrar node to the full domain. \\
2026-06-28 22:45:59 & Official Registrar & Updated the authorization domain of the official Registrar node to the full domain. \\
2026-06-28 22:45:35 & Registrar 1 & Updated the authorization domain of the Registrar node to the full domain. \\
2026-06-04 09:16:16 & Official Discovery & Resumed the official Discovery node with \passthrough{\lstinline!active!} status. \\
2026-06-04 09:15:47 & Official Discovery & Suspended the authorization of the official Discovery node. \\
2026-06-04 09:15:17 & Test Discovery & Resumed the test Discovery node with \passthrough{\lstinline!active!} status. \\
2026-06-04 09:14:48 & Test Discovery & Suspended the authorization of the test Discovery node. \\
2026-06-04 09:12:10 & Test Registrar & Resumed the test Registrar with \passthrough{\lstinline!active!} status. \\
2026-06-04 09:11:41 & Test Registrar & Suspended the authorization of the test Registrar. \\
2026-06-04 09:11:12 & Official Registrar & Resumed the official Registrar with \passthrough{\lstinline!active!} status. \\
2026-06-04 09:10:42 & Official Registrar & Suspended the authorization of the official Registrar. \\
2026-06-04 09:10:14 & Registrar 1 & Resumed the Registrar node with \passthrough{\lstinline!active!} status. \\
2026-06-04 09:09:46 & Registrar 1 & Suspended the authorization of the Registrar node. \\
2026-06-04 07:21:26 & Legacy Discovery 2 & Revoked the Discovery node authorization. \\
2026-06-04 07:19:09 & Legacy Discovery 1 & Revoked the Discovery node authorization. \\
2026-06-04 07:13:42 & Legacy Registrar 3 & Revoked the Registrar node authorization. \\
2026-06-04 07:12:51 & Legacy Registrar 2 & Revoked the Registrar node authorization. \\
2026-06-04 07:11:21 & Legacy Registrar 1 & Revoked the Registrar node authorization. \\
2026-06-04 07:06:52 & Official Discovery & Authorized the official Discovery node with \passthrough{\lstinline!active!} status; authorization domains: \passthrough{\lstinline!genesis.openagenet.local!}, \passthrough{\lstinline!openagenet.local!}. \\
2026-06-04 07:05:45 & Test Discovery & Authorized the test Discovery node with \passthrough{\lstinline!active!} status; authorization domains: \passthrough{\lstinline!genesis.openagenet.local!}, \passthrough{\lstinline!openagenet.local!}. \\
2026-06-04 07:04:36 & Test Registrar & Authorized the test Registrar with \passthrough{\lstinline!active!} status; no explicit authorization domain was set. \\
2026-06-04 07:03:36 & Official Registrar & Authorized the official Registrar with \passthrough{\lstinline!active!} status; no explicit authorization domain was set. \\
2026-06-04 07:02:43 & Registrar 1 & Authorized the Registrar node with \passthrough{\lstinline!active!} status; no explicit authorization domain was set. \\
2026-06-04 06:24:16 & Legacy Discovery 2 & Authorized the Discovery node with \passthrough{\lstinline!active!} status; authorization domains: \passthrough{\lstinline!genesis.openagenet.local!}, \passthrough{\lstinline!openagenet.local!}. \\
2026-06-04 06:23:35 & Legacy Discovery 1 & Authorized the Discovery node with \passthrough{\lstinline!active!} status; authorization domains: \passthrough{\lstinline!genesis.openagenet.local!}, \passthrough{\lstinline!openagenet.local!}. \\
2026-06-04 06:21:59 & Legacy Registrar 3 & Authorized the Registrar node with \passthrough{\lstinline!active!} status; no explicit authorization domain was set. \\
2026-06-04 06:21:40 & Legacy Registrar 2 & Authorized the Registrar node with \passthrough{\lstinline!active!} status; no explicit authorization domain was set. \\
2026-06-04 06:19:58 & Legacy Registrar 1 & Authorized the Registrar node with \passthrough{\lstinline!active!} status; no explicit authorization domain was set. \\
\end{longtable}
}

\subsection{Node Authorization Events}\label{subsec:13.4.1}

An authorization event places a node in the candidate set of usable nodes. The recipient must still check finality, authorization expiration, capability scope, and the binding of the node DID. It must not allow a request solely because of the event name.

In the current action encoding, authorization of a Registrar node is \passthrough{\lstinline!21!}, authorization of a Discovery node is \passthrough{\lstinline!31!}, and authorization of a VC issuer is \passthrough{\lstinline!41!}. These codes come from the action mapping in the governance contract and indexer. The actual effect of an authorization event is still determined jointly by status, the time window, the authorization domain, and whether the read model has been synchronized. If the target DID and target address in the authorization event do not match the node DID Document and runtime connection identity, automatic acceptance should stop and the case should enter verification.

\subsection{Node Suspension and Resumption Events}\label{subsec:13.4.2}

Suspension events take precedence over ordinary business requests. A resumption event becomes effective only after governance conditions are met and the indexer observes it in the correct order. The service should retain the raw suspension and resumption events rather than only replacing a current Boolean state.

Suspension and resumption do not delete or overwrite historical events. The indexer should store them as a continuous governance sequence and project the current state from the last valid sequence. After resumption, the system only allows subsequent actions that meet the conditions again. Whether a request rejected during suspension should be resubmitted is decided by the caller based on the request identifier and business result; state recovery must not resubmit it automatically.

\subsection{Node Revocation Events}\label{subsec:13.4.3}

A revocation event indicates that a node should no longer perform new operations within its authorization scope. Completed resource publications and historical queries should not be rewritten unconditionally, but downstream services must stop adopting new governance conclusions produced by the revoked node.

Revocation affects the node's subsequent governance qualification; it does not erase all data previously produced by that node from history. After processing a revocation event, downstream services should stop accepting new registrations, publications, distributions, or governance-state updates from that node, and should recheck materials from the node that have not completed verification. Historical resource records independently confirmed by other facts may be retained, but must preserve their source node, event position, and state at that time.

\subsection{Event Time, Sequence, and Source}\label{subsec:13.4.4}

Event ordering should prioritize the on-chain checkpoint, transaction sequence, event sequence, or an explicitly defined composite key. The event occurrence time, indexing time, and page display time must be recorded separately.

After paginated reads, the current event source orders events by checkpoint, transaction digest, and event sequence, and saves the read cursor together with the last processed position in the indexer state. \passthrough{\lstinline!emittedAtMs!} represents the on-chain time in the event payload. The indexer write time represents read-side observation and persistence, while the page generation time represents only the display moment. These values cannot replace one another. If an event has no usable time, ordering and diagnosis should rely on the on-chain position and indexing time.

\section{Node Authorization, Suspension, Resumption, and Revocation}\label{sec:13.5}

Node status is the current projection of a governance sequence onto a subject state. Authorization, suspension, resumption, and revocation all change subject availability through governance events, but their concrete effect is also constrained by the authorization domain, node capability, effective time, expiration time, event order, and index freshness. Before executing an action, a service node should combine the status judgment with the request source, request type, and resource facts. A successful historical request must not be treated as proof that current authorization remains valid.

\subsection{Authorization}\label{subsec:13.5.1}

After authorization, a node obtains an explicit authorization domain and capability scope. When receiving a node request, the Root platform should validate the current status and scope. It must not interpret an authorization event as proof that resource content has already passed verification.

\subsection{Suspension}\label{subsec:13.5.2}

While suspended, a service may retain read-only status queries and audit capabilities, but must not continue accepting new submissions or propagation actions that require the node's authorization, unless governance policy explicitly permits them.

\subsection{Resumption}\label{subsec:13.5.3}

Resumption should re-confirm the node identity, authorization scope, and governance event order. Resumption does not automatically resubmit requests rejected during suspension. The client must re-confirm them according to the request ID and current status.

\subsection{Revocation}\label{subsec:13.5.4}

After revocation, new actions within the authorization scope must stop. The revocation reason, time, operator, and on-chain position should be retained. The Discovery node and distribution service need to refresh their governance read models and converge progressively.

\subsection{Effects of Status on Registration, Distribution, and Discovery}\label{subsec:13.5.5}

{\def\LTcaptype{none} 
\begin{longtable}[]{@{}>{\raggedright\arraybackslash}p{0.2400\linewidth}>{\raggedright\arraybackslash}p{0.2400\linewidth}>{\raggedright\arraybackslash}p{0.2400\linewidth}>{\raggedright\arraybackslash}p{0.2400\linewidth}@{}}
\toprule\noalign{}
Node status & Registration submission & Content distribution & Discovery indexing \\
\midrule\noalign{}
\endhead
\bottomrule\noalign{}
\endlastfoot
active & Process according to the authorization domain & Process approved materials & Provide results that satisfy the filtering conditions \\
suspended & Reject or pause new submissions & Pause related propagation & Do not adopt new governance conclusions from the node \\
revoked & Reject new submissions & Stop its authorized propagation & Revoke or isolate records sourced from the node \\
unknown & Conservatively reject high-risk actions & Do not advance an uncertain state & Read-only results must carry an uncertainty marker \\
\end{longtable}
}

State-transition decisions should not rely on a single Boolean field. When projecting state, the indexer must also check whether the event sequence is continuous, whether the previous-event digest matches, whether events are duplicated, and whether effective and expiration times are valid. Runtime authorization checks must compare the current time with \passthrough{\lstinline!effectiveFromMs!} and \passthrough{\lstinline!expiresAtMs!}: an action cannot be allowed before the effective time, and a nonzero expiration time makes the state invalid once reached. When the state is unknown, the sequence has a gap, or the digest chain is inconsistent, high-risk new actions should be blocked rather than guessing the next state.

\section{Governance Notifications and Authorization State}\label{sec:13.6}

Governance notifications are suitable for communicating that a state has changed and providing an entry point for verification to service nodes and operations interfaces, but a notification itself is not a new authorization credential. A notification or snapshot should contain at least the node DID, node type, current status, authorization domain, last governance sequence, event digest, on-chain position, and indexer observation time. After receiving a notification, the recipient should read the status interface or original on-chain evidence according to its security requirements and verify it. It must not execute a high-risk action solely from the notification text.

When notification delivery fails, the read model lags, or a state snapshot disagrees with the raw event, a service may temporarily retain the known state for low-risk read-only display. It should stop advancing new high-risk authorization actions and record the synchronization position that needs recovery. After notification delivery resumes, the service should recheck according to the governance sequence rather than processing only the last notification.

\section{The Off-Chain Trust Indexer as a Governance Read-Side Projection}\label{sec:13.7}

The off-chain trust indexer is a component for reading and projecting governance information. It reads target governance events from the configured chain endpoint, stores raw events, structured events, and current subject states in local storage, and provides query endpoints for the Root platform and node services over HTTP. Through chain adapters and a unified state projection, the indexer decouples the blockchain implementation from OAN nodes and business logic, allowing the same node-authorization, governance-reading, and state-judgment mechanisms to support consortium chains such as Xinghuo Chain Network and well-known public chains such as Ethereum. The event model, finality, query interface, and deployment adaptation of each chain still require separate implementation and verification; the current reference implementation mainly covers Sui. The value of the indexer is to convert on-chain events into an operational read model. Its boundary is that it does not hold governance keys, submit governance transactions, or create or revoke node authorization based on its own judgment.

\subsection{Reading On-Chain Events}\label{subsec:13.7.1}

Event reading should support initial synchronization, cursor-based incremental reading, and resume from a checkpoint. When reading fails, the last confirmed position must be retained. An empty result must not overwrite an existing valid read model.

The indexer's event source supports paginated queries by event type and combines the GraphQL cursor, transaction digest, and event sequence associated with the event type into a resumable cursor. The configured event-type package, accepted emitting packages, and synchronization mode must remain consistent with the actual on-chain upgrade history. Otherwise, the service may appear healthy while being unable to read target events. Initial synchronization establishes the complete history, incremental synchronization continues from the saved cursor, and checkpoint recovery must reread from a safe position and use idempotent writes to eliminate duplicates.

\subsection{State Projection}\label{subsec:13.7.2}

State projection updates node status, authorization domains, and capability scopes according to event order and retains the necessary event history. Duplicate events should be handled idempotently, and an old event must not overwrite a confirmed newer state.

The current storage layer uses the governance sequence as the primary key for governance events and uses ``node type + node DID'' as the primary key for subject state. A repeated event with the same sequence and event digest can be treated as an idempotent replay. If the same sequence arrives with a different digest, the service should report a governance-event conflict and stop silent overwriting. Only an event with a sequence greater than the subject state's last sequence may update the current projection.

\subsection{Synchronization Cursor and Lag}\label{subsec:13.7.3}

The cursor is used for recovery and observation; it is not equivalent to on-chain finality. The interface should distinguish the currently read position, confirmed position, and position adopted by the service, and should make synchronization lag computable.

\passthrough{\lstinline!/v1/status!} may provide the package identifier, governance-object identifier, last processed checkpoint, event cursor, event source, latest governance sequence, latest event digest, sequence-gap flag, digest-gap flag, and last synchronization time. \passthrough{\lstinline!latestSequence!} only indicates the highest governance sequence written by the indexer. It does not by itself prove that the indexer has caught up with the latest on-chain checkpoint. Lag judgment must also consider the latest position returned by the event source and the synchronization result.

\subsection{Query Interfaces}\label{subsec:13.7.4}

Query interfaces should return the state source, on-chain position, indexing time, and reason for the state. Callers may use these values for filtering, but high-risk authorization must still be verified against original facts or verifiable proofs.

The public subject query, authorization-state query, and governance-event query provided by the indexer serve different purposes: subject queries are suitable for listing node states, authorization-state queries are suitable for runtime decisions, and event queries are suitable for auditing and fault tracing. Query responses come from the indexer read model. Callers should judge freshness by combining the last governance sequence, event digest, and synchronization status. An \passthrough{\lstinline!authorized!} result must not be interpreted as meaning that the indexer has verified all node keys, endpoints, or resource behavior on behalf of the caller.

The current off-chain trust indexer provides read interfaces including \passthrough{\lstinline!GET /v1/status!}, \passthrough{\lstinline!GET /v1/summary!}, \passthrough{\lstinline!GET /v1/subjects!}, \passthrough{\lstinline!GET /v1/registrars!}, \passthrough{\lstinline!GET /v1/discoveries!}, \passthrough{\lstinline!GET /v1/vc-issuers!}, and \passthrough{\lstinline!GET /v1/events!}. Subject interfaces usually return a \passthrough{\lstinline!subjects!} array, governance-event interfaces return an \passthrough{\lstinline!events!} array, and the status interface returns an index-status structure. These responses use dynamic JSON because index state and on-chain event fields are serialized from the indexer read model. \passthrough{\lstinline!POST /v1/admin/sync-once!} is a synchronization management operation and should not be mixed with public query interfaces.

The indexer's chain-access implementation uses official Sui gRPC/protobuf types, such as checkpoint and service-information requests. These types belong to the external Sui RPC protocol and are not OAN's own node-coordination \passthrough{\lstinline!.proto!}. OAN nodes therefore continue to use JSON/HTTP protocols serialized from shared Rust types as business interfaces, while chain endpoints and protobuf remain at the boundary where governance facts are read.

\subsection{Synchronization Failure and Recovery}\label{subsec:13.7.5}

{\def\LTcaptype{none} 
\begin{longtable}[]{@{}>{\raggedright\arraybackslash}p{0.2400\linewidth}>{\raggedright\arraybackslash}p{0.2400\linewidth}>{\raggedright\arraybackslash}p{0.2400\linewidth}>{\raggedright\arraybackslash}p{0.2400\linewidth}@{}}
\toprule\noalign{}
Stage & Authority or evidence & Observable location & Conclusion not yet justified \\
\midrule\noalign{}
\endhead
\bottomrule\noalign{}
\endlastfoot
On-chain confirmation & Transaction, event, checkpoint, event-digest chain & Sui event source & Indexer has read it \\
Indexer read & Persisted cursor and raw event & \passthrough{\lstinline!/v1/status!}, event store & Subject projection committed \\
Projection write & \passthrough{\lstinline!SubjectState!} and latest governance sequence & Subject/authorization query API & Node has refreshed and adopted it \\
Node adoption & Local governance cache and request decision & Node status, logs, gate tests & Resource control and content have passed \\
\end{longtable}
}

When synchronization fails, the service should record the error, retry count, and last successful cursor. After recovery, it should replay from a safe position and perform consistency checks. A temporary indexer outage must not cause all nodes to be marked as revoked.

Recovery should at least include identifying the reason for the read failure, retaining the last successful cursor, rereading affected pages, checking the event sequence and digest chain, confirming that a subject state has not been overwritten by an old event, and recording the new cursor and processed count after recovery. When the indexer is unavailable, dependents should mark the state as unknown or temporarily unavailable rather than writing \passthrough{\lstinline!revoked!} in bulk. Revocation must come from an actual governance event or a verifiable governance state.

\section{Enforcing Governance State at Root Platform Runtime}\label{sec:13.8}

When processing a request, the Root platform should perform the following steps in order:

\begin{enumerate}
\def\labelenumi{\arabic{enumi}.}
\tightlist
\item
  Identify the requesting node DID and verify the communication signature.
\item
  Query the current governance state and its synchronization freshness.
\item
  Check whether the requested action belongs to the authorization domain and capability scope.
\item
  Then verify resource control, version, hash, and content.
\item
  Write the acceptance, rejection, or delay reason to traceable logs.
\end{enumerate}

Valid authorization is only a prerequisite for entering the resource-verification process; it is not a resource-publication conclusion.

The first step confirms which node sent the communication request. The second confirms that node's state and freshness on the governance read side. The third confirms that the action, node type, authorization domain, and capability scope match. Only the fourth step enters verification of the resource DID Document, control signature, resource package, version, and content. Insufficient evidence at any step cannot be compensated for merely by success in a later step. For example, a valid resource-package signature may show that the resource control proof is potentially valid, but it does not prove that the Registrar node submitting it still has current authorization.

\section{Governance Filtering on the Discovery Side}\label{sec:13.9}

Discovery filtering must distinguish at least the source node status, resource lifecycle, Root platform publication status, and index freshness. Semantic relevance or tag matches may participate in ranking only after these hard conditions pass.

The Discovery node may use node authorization status to filter sources, but it should not write that authorization status directly into the resource lifecycle field. When a source node is suspended, revoked, or unknown, newly synchronized materials should be rejected, isolated, or marked as restricted. Existing resource records confirmed by other evidence may retain their historical presence, but they must not continue to be treated as normal new candidates without current governance confirmation. When the authorization domain does not match, the resource is not visible to that Discovery node; this does not mean that the resource DID does not exist across the entire network.

\section{Governance Event History and Public Observability}\label{sec:13.10}

Public event queries should use stable pagination and explicit ordering, and should display the on-chain position and indexing time together:

{\def\LTcaptype{none} 
\begin{longtable}[]{@{}>{\raggedright\arraybackslash}p{0.4800\linewidth}>{\raggedright\arraybackslash}p{0.4800\linewidth}@{}}
\toprule\noalign{}
Field & Description \\
\midrule\noalign{}
\endhead
\bottomrule\noalign{}
\endlastfoot
Event type & Authorization, suspension, resumption, or revocation \\
Node DID & Affected node \\
Authorization domain/capability & Scope involved in the event \\
On-chain position & Checkpoint, transaction, or event sequence \\
\passthrough{\lstinline!observedAt!} & Indexer observation time \\
State reason & Explanation for operations and auditing \\
\end{longtable}
}

Public display may use \passthrough{\lstinline!eventLabel!} as a readable action name, but should retain the raw \passthrough{\lstinline!eventType!}, governance sequence, and source position. ``Recent'' in an event list should be interpreted by governance sequence or on-chain position, not by the response time of the web interface. Pagination boundaries should also be tied to stable ordering so that a client does not display the same event twice or miss an event when new events arrive.

\section{Chain Endpoints, Packages, Objects, and Deployment Configuration}\label{sec:13.11}

Indexer configuration must record the network, current operation package, event-type definition package, allowed emitting packages, governance bulletin object, event-source mode, and synchronization parameters together. OAN currently uses Sui Mainnet. The governance bulletin object is \passthrough{\lstinline!0xb900fb0dc1ebd58fbc4d0c697d644c861578ed948e26638836e338de0d1a3b46!}; the initial package \passthrough{\lstinline!0x1b602d575ddf1ced075f9d2986bfe53cd7d4e59967ec26a0d949009c7b2c9d6c!} defines the governance event types; and the current operation package is \passthrough{\lstinline!0x232f984a25ceb8358290a8bae594078deca2d670a80c8d57d291d64a491d7fd5!}. During upgrades, \passthrough{\lstinline!0x2fd18248c0cd42f371409eb1003176dc0d5744cccfdf2e49e5224b4249a5ffb4!} also emitted governance events. Therefore, the indexer must accept events produced by all three packages to reconstruct the governance sequence continuously across the upgrade.

\begin{lstlisting}
[chain]
network = "sui-mainnet"
grpc_endpoint = "https://fullnode.mainnet.sui.io"
event_query_endpoint = "https://graphql.mainnet.sui.io/graphql"
package_id = "0x232f984a25ceb8358290a8bae594078deca2d670a80c8d57d291d64a491d7fd5"
event_type_package_id = "0x1b602d575ddf1ced075f9d2986bfe53cd7d4e59967ec26a0d949009c7b2c9d6c"
emitting_package_ids = [
  "0x1b602d575ddf1ced075f9d2986bfe53cd7d4e59967ec26a0d949009c7b2c9d6c",
  "0x2fd18248c0cd42f371409eb1003176dc0d5744cccfdf2e49e5224b4249a5ffb4",
  "0x232f984a25ceb8358290a8bae594078deca2d670a80c8d57d291d64a491d7fd5",
]
bulletin_object_id = "0xb900fb0dc1ebd58fbc4d0c697d644c861578ed948e26638836e338de0d1a3b46"

[sync]
event_source = "event_query"
poll_enabled = true
sync_on_start = true
poll_interval_ms = 10000
event_page_size = 50
max_pages_per_sync = 2
\end{lstlisting}

The values above correspond to the current Mainnet event-synchronization configuration. \passthrough{\lstinline!package\_id!} points to the package that executes current governance operations. \passthrough{\lstinline!event\_type\_package\_id!} remains directed to the package that originally defined the event structure. \passthrough{\lstinline!emitting\_package\_ids!} covers the packages that actually emitted governance events during the upgrade history. \passthrough{\lstinline!bulletin\_object\_id!} points to the governance bulletin object that evolves continuously. These four values have different roles: updating only \passthrough{\lstinline!package\_id!} to the latest package while omitting the old event-type package or an intermediate emitting package creates gaps in event queries; configuring only the bulletin object without checking the event source may produce a service that starts normally while its governance sequence has stopped growing.

\section{Governance Upgrade and Migration Rules}\label{sec:13.12}

Before an upgrade, the compatibility relationship among the contract package, event structures, indexer, and node services should be declared, followed by migration and replay. Old events must remain parseable or have an explicit conversion rule. If conversion fails, the raw data must be retained and an uncertain state must be prevented from overwriting a valid state.

An upgrade should also distinguish a ``package-address change'' from an ``event-type-definition change.'' An upgradeable Sui package's current operation package may differ from its event-type definition package. Therefore, event-query configuration, checkpoint-audit configuration, and the set of emitting packages accepted by the indexer cannot be inferred from a package name. After migration, a set of known governance events should be used to verify the raw event count, structured event count, latest sequence, digest chain, and subject states, confirming that the upgrade introduced neither duplicate projections nor skipped events.

\section{Governance Event Ordering and Finality}\label{sec:13.13}

Event processing can be understood through the following states:

\begin{lstlisting}
observed -> ordered -> confirmed -> projected -> exposed
\end{lstlisting}

An event that arrives later but has an earlier on-chain position should be inserted into the history, and affected projections should be recalculated. A confirmed higher-version state must not be directly overwritten by an old event.

\passthrough{\lstinline!observed!} means that the event has been read from the event source. \passthrough{\lstinline!ordered!} means that the indexer has established an order by on-chain position or governance sequence. \passthrough{\lstinline!confirmed!} means that the event has passed the implementation-defined integrity and position checks. \passthrough{\lstinline!projected!} means that it has been written to subject state. \passthrough{\lstinline!exposed!} means that the query interface can read the projection. The fact that an event has entered the public read model does not mean that its on-chain occurrence time equals the interface observation time. When a sequence gap or digest gap exists, public state should carry uncertainty so that incomplete history is not treated as a continuous governance chain.

\section{Consistency Among On-Chain Facts, Off-Chain Publication State, and Read Models}\label{sec:13.14}

Consistency does not require on-chain events, the indexer, the Root platform, and the Discovery node to update at exactly the same moment. It requires every layer to explain the fact position and delay it is using. On-chain events are the source of governance facts; the indexer provides a read model with cursors and times; the Root platform uses it to make publication decisions; and the Discovery node uses it to filter sources and resource visibility. When any layer lags or conflicts, confirmed facts should be retained, differences and their sources should be marked, and restrictive handling should be applied to high-risk actions.

\subsection{Final On-Chain Facts}\label{subsec:13.14.1}

Final on-chain facts are based on a verifiable event position and contract state and should retain at least the object, action, operator, time, and reference to the original event. An indexer interface can provide an access path for reading, but cannot change the original fact.

\passthrough{\lstinline!SubjectState!} in the governance contract stores the node's current status, authorization domain, policy and metadata hashes, effective and expiration times, state version, and update time. \passthrough{\lstinline!GovernanceExecutionEvent!} stores the event context that caused the state change. Together, they can answer ``what is the current state'' and ``which governance execution formed it,'' but the result should still be verified against the original on-chain position of the transaction, event, and governance object.

\subsection{Root Platform Publication State}\label{subsec:13.14.2}

Root platform publication state represents the result of the off-chain verification and publication process. It may lag behind on-chain authorization, or may not be produced because resource verification failed. Therefore, resource publication cannot be inferred from node authorization.

The Root platform may read node governance state before accepting a submission from a Registrar node, but resource publication must still pass checks for the resource DID Document, control signature, resource package integrity, version relationship, content policy, and publication process. Publication state should retain the resource DID, resource version, publication-proof reference, processing time, and source node. It is stored separately from on-chain node authorization state so that downstream propagation eligibility can be reconsidered when the node's governance state changes.

\subsection{Discovery Node Index State}\label{subsec:13.14.3}

Discovery index state indicates whether a Discovery node has received and indexed a propagatable resource. It should associate the resource DID, version, publication-proof reference, indexing time, and governance filtering result.

Discovery index state is a local read-model state, not an on-chain resource state. An index record should identify the publication or distribution source, the resource version used, the time written to the local index, the node governance state adopted at that time, and whether the resource was filtered by authorization domain or lifecycle. Semantic-index hits, capability-tag hits, and governance-authorization hits should be explained separately. A retrieval score must not conceal that the source node has been revoked or that its governance state is unknown.

\subsection{State Inconsistency and Convergence}\label{subsec:13.14.4}

When states are inconsistent, the system should act conservatively: pause high-risk actions, retain confirmed facts, mark the source of the difference and synchronization lag, and update the external state after the governance read model, Root platform, and Discovery index converge.

Common inconsistencies include an on-chain governance sequence that the indexer has not read, an event read by the indexer that the node service has not refreshed, a Root platform publication that the Discovery node has not indexed, and an old local resource record whose source-node status has changed. Each inconsistency should record the compared sequences, times, sources, and handling action. During recovery, the system should recompute from the fact source rather than overwriting the difference with the web access time or the last successful request.

\section{Governance Authority Key Management and Emergency Procedures}\label{sec:13.15}

Governance-key operations must be strictly separated from the local control private keys of resource providers. Emergency procedures should at least include pausing high-risk governance operations, rotating or isolating suspected leaked keys, confirming on-chain events, resynchronizing the indexer, notifying nodes, and verifying state convergence. An operations marker that has not been confirmed through on-chain or authorization procedures must not be used directly as the final governance fact.

Governance keys are used to initiate or approve governance actions. Resource control private keys prove the control relationship for a specific resource DID or resource version. The indexer and ordinary operations processes should not hold governance keys. After emergency handling, the system should separately verify the permissions of the governance account or object, the on-chain event sequence, the indexer state, node runtime decisions, and downstream Discovery results. If any one of these remains unknown or lagging, the external state should retain a restrictive marker; service-process recovery alone is not enough to declare governance recovery complete.

\section{References}\label{sec:13.16}

{\def\LTcaptype{none} 
\begin{longtable}[]{@{}>{\raggedright\arraybackslash}p{0.3200\linewidth}>{\raggedright\arraybackslash}p{0.3200\linewidth}>{\raggedright\arraybackslash}p{0.3200\linewidth}@{}}
\toprule\noalign{}
Source & Type & Link \\
\midrule\noalign{}
\endhead
\bottomrule\noalign{}
\endlastfoot
\passthrough{\lstinline!oan-trust-indexer!} & Code repository: on-chain governance event synchronization and off-chain trust-state projection & \url{https://github.com/OpenAgenet/oan-trust-indexer} \\
\passthrough{\lstinline!oan-root-nodes!} & Code repository: node authorization, governance state, and Root platform decisions & \url{https://github.com/OpenAgenet/oan-root-nodes} \\
\passthrough{\lstinline!oan-protocol-commons!} & Code repository: shared authorization, governance, and state types & \url{https://github.com/wolfbrother/oan-protocol-commons} \\
OAN Yellow Paper & arXiv yellow paper: governance, authorization, and trust boundaries & \url{https://arxiv.org/abs/2606.03163} \\
Agentic Overlay Network Architecture & IETF draft: network-node and governance coordination architecture & \url{https://datatracker.ietf.org/doc/draft-xu-agentic-overlay-network-architecture/} \\
\end{longtable}
}

\chapter{Network Node Protocols and Interoperability}\label{chap:14}

This chapter explains how inter-node protocols deliver an interaction to the correct node while preserving verifiable context; it does not extend "request delivered" into "resource published," nor "node authenticated" into "resource control verified."

\begin{lstlisting}
Envelope -> node authentication -> request signing -> business validation -> status handling -> response recording
\end{lstlisting}

In the OpenAgenet (OAN) implementation, inter-node operations are usually provided with structured request and response types by a shared protocol crate, while the concrete node service is responsible for HTTP routing, storage, and background tasks. Therefore, when reading a request path, verify all three layers together: the HTTP method and path, the JSON envelope fields, and the node\textquotesingle s internal acceptance or publication state. \passthrough{\lstinline!oan-protocol-commons/schemas/agent-contract/openapi.yaml!} is currently the contract entry point and a version placeholder, not a complete automatically generated description of every node interface; this file alone cannot establish that an interface is enabled in every deployment.

\section{Common Protocol Envelope and Request Headers}\label{sec:14.1}

{\def\LTcaptype{none} 
\begin{longtable}[]{@{}>{\raggedright\arraybackslash}p{0.4800\linewidth}>{\raggedright\arraybackslash}p{0.4800\linewidth}@{}}
\toprule\noalign{}
Field & Purpose \\
\midrule\noalign{}
\endhead
\bottomrule\noalign{}
\endlastfoot
\passthrough{\lstinline!requestId!} & Correlate one logical request, response, and cross-service logs \\
\passthrough{\lstinline!protocolVersion!}, \passthrough{\lstinline!purpose!} & Fix field interpretation and business purpose \\
\passthrough{\lstinline!method!}, \passthrough{\lstinline!path!}, \passthrough{\lstinline!aud!} & Bind signature to method, target path, and audience \\
\passthrough{\lstinline!requestTimestamp!}, \passthrough{\lstinline!requestNonce!} & Limit validity and prevent replay \\
\passthrough{\lstinline!bodyHash!} & Bind the envelope to the exact business body \\
\passthrough{\lstinline!proof!} & Record signer, verification method, suite, and signature \\
\end{longtable}
}

The common envelope may use the following field semantics, while the business body remains defined by the corresponding interface:

The example corresponds to the fields of \passthrough{\lstinline!SignedRequestEnvelope!}, rather than adding another set of \passthrough{\lstinline!sourceDid!}, \passthrough{\lstinline!targetDid!}, \passthrough{\lstinline!timestamp!}, or \passthrough{\lstinline!sequence!} fields. \passthrough{\lstinline!aud!} in the envelope denotes the target audience, and \passthrough{\lstinline!proof!} carries the request signature; business payloads such as a ResourcePackage, registration submission, or batch item are carried outside the envelope by the specific request structure. The sender generates \passthrough{\lstinline!requestId!}, \passthrough{\lstinline!requestTimestamp!}, \passthrough{\lstinline!requestNonce!}, and \passthrough{\lstinline!bodyHash!}; the receiver validates the audience, time, nonce, digest, signing subject, and signature coverage. Whether unknown fields are accepted is determined by the protocol version and the deserialization rules of the specific structure. Missing required fields or an inconsistent digest must be rejected.

\section{Inter-Node Authentication}\label{sec:14.2}

{\def\LTcaptype{none} 
\begin{longtable}[]{@{}>{\raggedright\arraybackslash}p{0.3200\linewidth}>{\raggedright\arraybackslash}p{0.3200\linewidth}>{\raggedright\arraybackslash}p{0.3200\linewidth}@{}}
\toprule\noalign{}
Check & What it proves & Failure handling \\
\midrule\noalign{}
\endhead
\bottomrule\noalign{}
\endlastfoot
Node DID & Who the communicating party is & Reject authentication \\
Node authorization VC/governance status & Whether it has the current communication or business qualification & Reject or degrade to read-only \\
Validity period and status & Whether the credential is still valid & Do not use an expired or revoked credential \\
Authorization domain and capabilities & Whether the requested action is within scope & Reject an unauthorized action \\
Resource control proof & Whether the resource is controlled by the submitter & Enter business-validation failure \\
\end{longtable}
}

Authentication decisions should be completed in order: first confirm the connection target and node DID, then resolve the verification method in the node DID Document, next verify the node authorization VC or governance status, and finally check that the requested action falls within the authorization domain. Node authorization and resource control are two different paths: the former indicates that a node may communicate as a particular network role, while the latter is established by the control proof in the resource submission, showing that the submitter can control the resource DID. Even after node authentication succeeds, the receiver must still perform business validation on the ResourcePackage, DID Document, and control proof.

\section{Signing Upstream Requests}\label{sec:14.3}

The signature must cover at least the canonicalized method, path, target node, time, nonce, request ID, and \passthrough{\lstinline!bodyHash!}. The verifier should resolve the public key according to \passthrough{\lstinline!proof.verificationMethod!} or a compatible verification-method identifier, and confirm that \passthrough{\lstinline!proof.creator!}, the verification-method controller, and the trusted upstream node identity are consistent.

The current shared protocol structure uses \passthrough{\lstinline!proof!}, rather than defining separate \passthrough{\lstinline!kid!} and \passthrough{\lstinline!signature!} fields in \passthrough{\lstinline!SignedRequestEnvelope!}. \passthrough{\lstinline!proof.verificationMethod!} points to a verification method in the DID Document, while \passthrough{\lstinline!proof.creator!} identifies the signature creator; \passthrough{\lstinline!oan-crypto!} canonicalizes the signable value according to the cryptographic suite before signing or verifying. For \passthrough{\lstinline!Ed25519Sha256!} and \passthrough{\lstinline!Sm2Sm3!}, the signature input is the bytes of canonicalized JSON; legacy compatibility suites may first hash the canonicalized content, so the protocol version and cryptographic suite must jointly define the signature input. The HTTP header example applies only when a specific deployment maps envelope fields to request headers; it cannot replace the JSON envelope itself.

Signature verification failure, a mismatched target \passthrough{\lstinline!aud!}, a mismatched \passthrough{\lstinline!bodyHash!}, a verification method that does not belong to the signing subject, or an expired request should all be rejected before entering the resource-write and publication queues, with the \passthrough{\lstinline!requestId!} and an error category useful for operations recorded.

\section{Node Authorization VC Exchange}\label{sec:14.4}

Node authorization VC exchange should follow the order "carry or reference -\textgreater{} resolve issuer DID -\textgreater{} verify signature -\textgreater{} query status -\textgreater{} check authorization domain and capabilities." Caching may reduce repeated resolution, but it cannot bypass validity periods, revocation status, or governance-event refreshes.

The issuer of a node authorization VC is a trusted institution with governance responsibility or the Root platform. The receiver should verify the issuer DID Document and signature, and check the subject node DID, role, authorization domain, capabilities, validity period, and revocation or suspension status in the VC. The VC may be carried as a complete credential in the request or obtained through a reference agreed by both parties; in the latter case, the referenced content must be traceable and its digest verifiable, and a cache hit must not cause the status check to be skipped. A resource registration VC is the result of resource registration; it cannot in turn prove that a Registrar node is qualified for inter-node communication.

\section{Registrar node to Root platform Protocol}\label{sec:14.5}

\pandocbounded{\includegraphics[width=\linewidth,height=0.72\textheight,keepaspectratio]{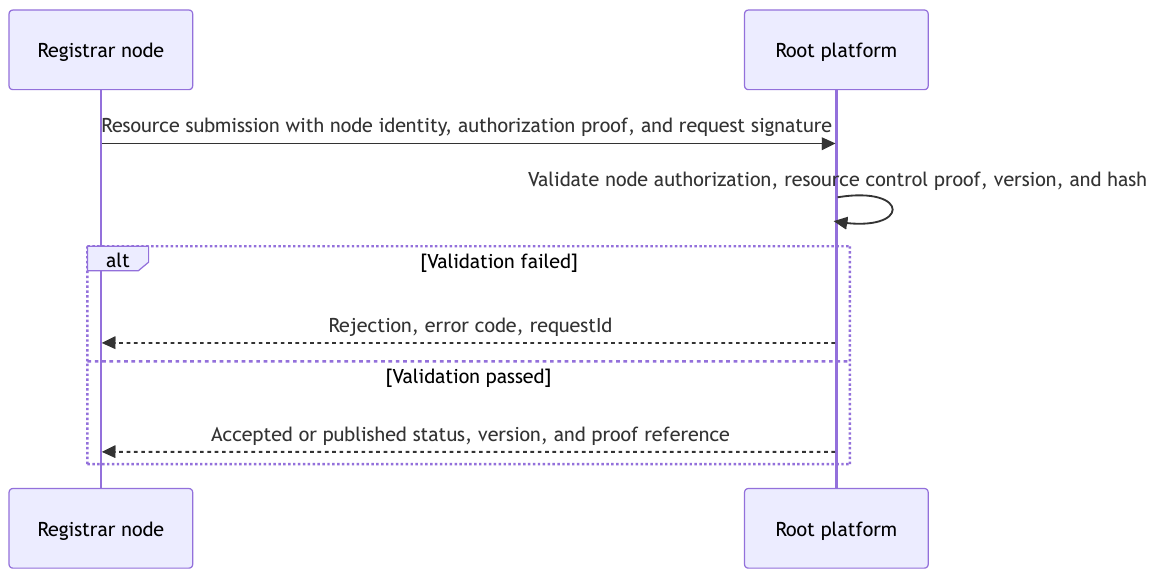}}

A Root platform response of "accepted" only means that the request has entered the processing flow; the registration VC, Root proof, content distribution, and Discovery node index status should be expressed separately.

The business payload submitted by a Registrar node to the Root platform is represented by \passthrough{\lstinline!ResourceVerifyAndPublishRequest!}, which includes the Registrar node DID, the resource registration submission, and \passthrough{\lstinline!upstreamAuth!}. The registration submission includes the resource DID, resource type, DID Document, metadata, version, individual digests, registration credential, and subject control proof. The Root platform first verifies the upstream envelope and node qualification, then validates the binding among the resource control proof, DID Document, and digests; only after that does it enter the Root platform\textquotesingle s persistence and subsequent publication flow. On timeout or connection interruption, the Registrar node should query the acceptance status using the original \passthrough{\lstinline!requestId!}; it must not determine whether the resource has been published solely from receiving a response again or from the number of local retries.

\section{Root platform to content distribution platform Protocol}\label{sec:14.6}

The distribution request should be associated with the resource DID, version, ResourcePackage digest, Root proof reference, and distribution sequence. After receiving it, the content distribution platform recomputes the digest, confirms the source and version, and then returns a received, completed, or failed status; successful receipt does not mean that a Discovery node has indexed the resource.

In the shared protocol, a single content distribution request uses \passthrough{\lstinline!ResourceCdnPublishRequest!}, while a batch request uses \breakablecode{ResourceCdnBatchPublishRequest}; batch items are associated with the Root platform\textquotesingle s publication order through \passthrough{\lstinline!publicationCursor!}, and the common envelope proves the upstream call context through \passthrough{\lstinline!upstreamAuth!}. The corresponding content distribution path constants in the code are \passthrough{\lstinline!/cdn/resources!} and \passthrough{\lstinline!/cdn/resources/batch!}, but the actual accessible address still depends on the service route prefix and reverse-proxy configuration. After validation, the content distribution platform forms its own readable resource or manifest, and the Root platform\textquotesingle s publication-completed status and the Discovery node\textquotesingle s indexing-completed status should be recorded separately.

\section{content distribution platform to Discovery node Protocol}\label{sec:14.7}

Discovery node synchronization should use a cursor or batch sequence, and should revalidate the digest and Root proof reference for each resource version. Duplicate batches should be processed idempotently, and the cursor must advance only after the batch content has been successfully validated.

For incremental data read from the content distribution platform, the shared protocol uses \passthrough{\lstinline!ResourceCdnIndexResponse!} to express \passthrough{\lstinline!items!}, \passthrough{\lstinline!count!}, \passthrough{\lstinline!afterCursor!}, \passthrough{\lstinline!nextCursor!}, and \passthrough{\lstinline!hasMore!}; each item carries the current cursor and ResourcePackage. Notifications sent by the Root platform to a Discovery node instead use information such as the notification batch, start and end sequences, and content distribution address; the two forms must not directly mix field names. A Discovery node should write to its local index and advance the next cursor only after the ResourcePackage, DID Document, digests, and Root proof association have all passed validation; a failed batch retains the original cursor and is retried after remediation.

\pandocbounded{\includegraphics[width=\linewidth,height=0.72\textheight,keepaspectratio]{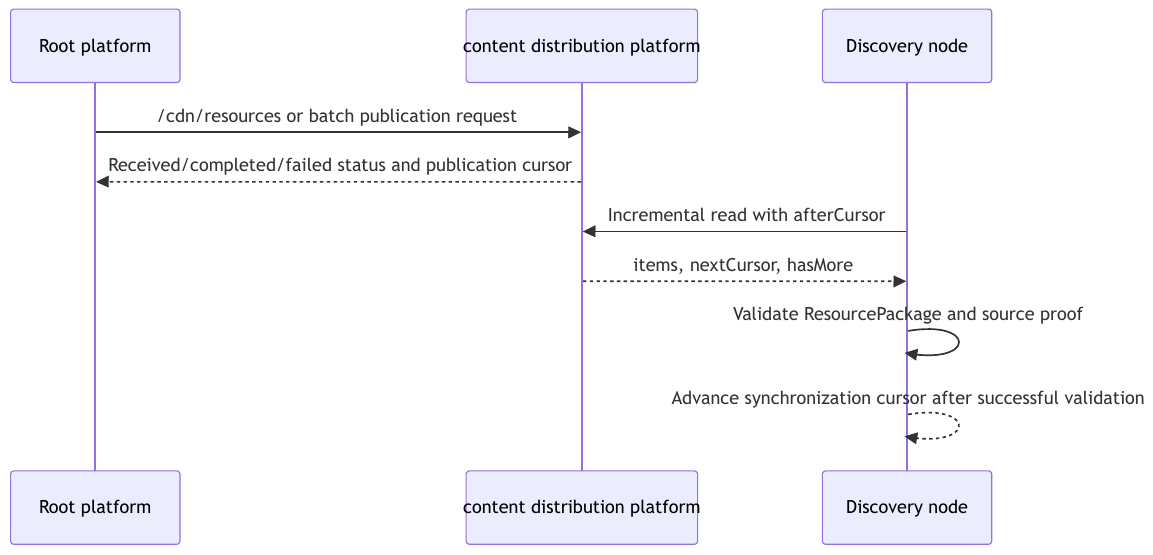}}

\section{Discovery Response Signatures and Source Proofs}\label{sec:14.8}

A Discovery response should be able to distinguish at least the following information:

{\def\LTcaptype{none} 
\begin{longtable}[]{@{}>{\raggedright\arraybackslash}p{0.4800\linewidth}>{\raggedright\arraybackslash}p{0.4800\linewidth}@{}}
\toprule\noalign{}
Information & Meaning \\
\midrule\noalign{}
\endhead
\bottomrule\noalign{}
\endlastfoot
Query node DID & Who generated this response \\
Data source & Which distribution or index source supplied the data \\
Index version/time & Freshness of the read model used for the query \\
Match reason & Which fields or tags matched \\
Root proof reference & External evidence entry point for the publication fact \\
Response signature & Prevents the response from being replaced \\
\end{longtable}
}

A response signature proves that the node is responsible for the response content; it does not replace the resource control proof or Root proof.

The current \passthrough{\lstinline!ResourceDiscoveryResponse!} structure contains \passthrough{\lstinline!discoveryDid!}, \passthrough{\lstinline!candidates!}, \passthrough{\lstinline!createdAt!}, and an optional \passthrough{\lstinline!proof!}; the Discovery node query implementation may return an empty \passthrough{\lstinline!proof!}, so response signatures are a reserved protocol capability or a capability of deployments after enablement and should not be described as the default behavior of all current instances. If enabled, the signature should cover the query result, creation time, query node DID, and the necessary index or source digest; the caller must still independently check the DID Document, resource control proof, and Root proof of the candidate resource. The \passthrough{\lstinline!score!} in a candidate only indicates matching relevance; it is not a trust level or authorization conclusion.

\section{Sequence Numbers, Request Identifiers, and Replay Protection}\label{sec:14.9}

{\def\LTcaptype{none} 
\begin{longtable}[]{@{}>{\raggedright\arraybackslash}p{0.3200\linewidth}>{\raggedright\arraybackslash}p{0.3200\linewidth}>{\raggedright\arraybackslash}p{0.3200\linewidth}@{}}
\toprule\noalign{}
Field & Primary use & Reuse on retry \\
\midrule\noalign{}
\endhead
\bottomrule\noalign{}
\endlastfoot
\passthrough{\lstinline!requestId!} & Associates one business request with its logs & Reuse for the same logical request \\
Idempotency key & Prevents repeated execution of submission side effects & Reuse for the same operation \\
\passthrough{\lstinline!nonce!} & Prevents replay of signed messages & Usually generate again \\
\passthrough{\lstinline!sequence!}/cursor & Ensures batch or event order & Use according to the synchronization protocol \\
Time window & Limits request freshness & Update when resigning \\
\end{longtable}
}

\passthrough{\lstinline!requestId!} connects one logical operation across the Registrar node, Root platform, content distribution platform, and Discovery node; \passthrough{\lstinline!requestNonce!} prevents the same signed envelope from being accepted again; \passthrough{\lstinline!publicationCursor!}, \passthrough{\lstinline!sequenceFrom!}, \passthrough{\lstinline!sequenceTo!}, \passthrough{\lstinline!afterCursor!}, and \passthrough{\lstinline!nextCursor!} propagate order or support incremental reads. Their numeric spaces and advancement rules differ: a request ID must not be treated as an event sequence, and pagination parameters from a Discovery query must not replace a publication synchronization cursor. When retrying the same logical request, \passthrough{\lstinline!requestId!} may be retained, but a new signed message should normally use a new nonce, and the receiver should determine whether it has already been processed according to idempotency rules.

\section{Pagination, Filtering, and Compatibility Rules}\label{sec:14.10}

Client page pagination, Discovery query pagination, and node synchronization cursors must not be mixed: page pagination serves presentation, Discovery pagination serves candidate results, and synchronization cursors serve incremental propagation. When a cursor is invalid, return an identifiable error or require resynchronization; do not silently continue from an arbitrary position.

The current shared query structure has a default \passthrough{\lstinline!limit!} of 10. A Discovery query may also carry conditions such as resource type, capability tags, protocol, version, and version mode; this is only the default behavior of the Discovery query structure and does not represent the default page size of every management list interface. \passthrough{\lstinline!hasMore!} and \passthrough{\lstinline!nextCursor!} in a synchronization response serve incremental propagation, while \passthrough{\lstinline!score!}, version mode, and candidate ordering in query results serve user retrieval. During protocol evolution, new filter fields may be handled according to compatibility rules, but changes to cursor meaning, ordering stability, or version mode should be explicitly communicated to callers through version documentation.

\section{Timeouts, Retries, and Idempotent Operations}\label{sec:14.11}

{\def\LTcaptype{none} 
\begin{longtable}[]{@{}>{\raggedright\arraybackslash}p{0.3200\linewidth}>{\raggedright\arraybackslash}p{0.3200\linewidth}>{\raggedright\arraybackslash}p{0.3200\linewidth}@{}}
\toprule\noalign{}
Operation & Default action after a network timeout & Reason \\
\midrule\noalign{}
\endhead
\bottomrule\noalign{}
\endlastfoot
Read-only query & Retry after limited backoff & Usually has no business side effect \\
Status query & Use the original request ID to query for confirmation & Distinguishes incomplete from not started \\
Resource submission & Do not blindly replay; query idempotent status first & May already have caused registration or publication side effects \\
Synchronization batch & Retry with the original cursor and batch & Preserves incremental order and idempotency \\
\end{longtable}
}

Retries must distinguish "no response received" from "explicit failure received": a network timeout only means that the result is unknown, so resource submission, publication, and state-change operations should first query with the original request identifier; explicit parameter, authorization, or signature errors should not be retried automatically. A background publisher may resume processing at the same publication cursor, and the receiver should identify duplicate messages using the combination of resource DID, version, digest, and cursor to avoid duplicate writes or incorrect progress. Specific timeouts, backoff, and concurrency are determined by deployment configuration; this chapter does not fix the values of a particular environment as protocol constants.

\section{Protocol Version Negotiation}\label{sec:14.12}

Version negotiation should declare at least the protocol version, supported capabilities, and incompatible constraints. When there is no common version, fail explicitly; do not infer from fields or silently downgrade to an insecure path.

The example expresses negotiation information; it does not mean that all current nodes provide an independent capability-negotiation endpoint. The current shared protocol uses \passthrough{\lstinline!oan-resource-2026!} as the resource protocol version constant and uses \passthrough{\lstinline!protocolVersion!} in the envelope; \passthrough{\lstinline!info.version!} in \passthrough{\lstinline!openapi.yaml!} is the contract-file version, and the two must not be treated as equivalent. The negotiation result must at least fix field interpretations, signature coverage, cryptographic suite, cursor semantics, and error handling; if the other party supports only an insecure or uninterpretable downgrade path, refuse to establish that business link.

\section{Message Size, Compression, and Transport Limits}\label{sec:14.13}

Limits should account for the message before compression, the compressed transfer body, and the decompressed content at the same time; the receiver should apply size and format protections before resource writing or expensive validation. Compression changes only the transport method, not the logical content covered by the signature and digest.

The current shared protocol types define ResourcePackage, batch items, and envelope fields, but the common structure file does not uniformly declare message sizes, compression algorithms, or HTTP body limits for all deployments. Therefore, these limits should be jointly determined by the concrete service configuration, reverse proxy, and runtime environment, with verifiable values disclosed in inter-node capability or deployment documentation. Batch synchronization in particular should limit item size, batch item count, and cumulative decompressed size at the same time; when a limit is exceeded, return a clear error before persistence and retain the request ID, without truncating and continuing signature verification or indexing.

\section{Correlation, Traceability, and Audit Identifiers}\label{sec:14.14}

A resource propagation path should retain at least the following correlation fields:

\begin{lstlisting}
requestId -> resourceDid -> resourceVersion -> publishSequence
          -> syncCursor -> discoveryIndexVersion -> errorCode
\end{lstlisting}

Nodes may add local trace identifiers, but must not lose the upstream \passthrough{\lstinline!requestId!}, resource DID, or version; logs must not record private keys or complete sensitive credentials.

For the publication path, the following searchable correlations should be retained at minimum:

{\def\LTcaptype{none} 
\begin{longtable}[]{@{}>{\raggedright\arraybackslash}p{0.3200\linewidth}>{\raggedright\arraybackslash}p{0.3200\linewidth}>{\raggedright\arraybackslash}p{0.3200\linewidth}@{}}
\toprule\noalign{}
Path stage & Key correlation fields & Purpose \\
\midrule\noalign{}
\endhead
\bottomrule\noalign{}
\endlastfoot
Registration acceptance & \passthrough{\lstinline!requestId!}, resource DID, resource version & Locate one registration request \\
Root platform publication & ResourcePackage digest, Root proof reference, \passthrough{\lstinline!publicationCursor!} & Determine the publication fact and order \\
Incremental synchronization & \passthrough{\lstinline!afterCursor!}, \passthrough{\lstinline!nextCursor!}, batch sequence number & Determine whether synchronization was missed or duplicated \\
Discovery index & Discovery node DID, index time, error category & Explain the freshness of query results \\
\end{longtable}
}

Audit records should retain validation results and state transitions, rather than recording only HTTP 200. For sensitive VCs, record their type, digest or reference, and verification conclusion, and avoid writing complete credential contents or any private key to ordinary business logs.

\section{Protocol Compatibility and Deprecation Rules}\label{sec:14.15}

Field or version deprecation should pass through the stages of "announcement -\textgreater{} compatibility period -\textgreater{} migration -\textgreater{} stop accepting," while retaining the error code and request ID. Changes involving signature coverage, authentication semantics, cursor meaning, or state semantics cannot be solved merely by adding optional fields; the protocol version should be upgraded or an explicit migration rule provided.

Compatibility tests should cover at least: whether an old-version envelope can be explicitly rejected or parsed according to compatibility rules, whether unknown optional fields leave signature semantics unchanged, whether an old cursor can never be mistaken for a new cursor, and whether the signing subject and \passthrough{\lstinline!aud!} are still verified under the same rules. During version migration, the sender should prefer the common version confirmed by both parties, and the receiver should return an associable request ID and the reason for incompatibility in an error response; before stopping acceptance of the old version, node configuration, background publication tasks, and historical cursors should first be migrated. Protocol files, shared structures, and service implementations must be updated together; changing only the version string in the documentation is insufficient.

\section{References}\label{sec:14.16}

{\def\LTcaptype{none} 
\begin{longtable}[]{@{}>{\raggedright\arraybackslash}p{0.3200\linewidth}>{\raggedright\arraybackslash}p{0.3200\linewidth}>{\raggedright\arraybackslash}p{0.3200\linewidth}@{}}
\toprule\noalign{}
Source & Type & Link \\
\midrule\noalign{}
\endhead
\bottomrule\noalign{}
\endlastfoot
\passthrough{\lstinline!oan-protocol-commons!} & Code repository: inter-node request, response, error, and event types & \url{https://github.com/wolfbrother/oan-protocol-commons} \\
\passthrough{\lstinline!oan-root-nodes!} & Code repository: Root platform inter-node interfaces & \url{https://github.com/OpenAgenet/oan-root-nodes} \\
\passthrough{\lstinline!oan-registrar-nodes!} & Code repository: Registrar node interfaces and upstream calls & \url{https://github.com/OpenAgenet/oan-registrar-nodes} \\
\passthrough{\lstinline!oan-discovery-nodes!} & Code repository: Discovery synchronization and query interfaces & \url{https://github.com/OpenAgenet/oan-discovery-nodes} \\
\passthrough{\lstinline!oan-trust-indexer!} & Code repository: off-chain trust indexer governance-status query interfaces & \url{https://github.com/OpenAgenet/oan-trust-indexer} \\
Agentic Overlay Network Architecture & IETF draft: node interoperability architecture & \url{https://datatracker.ietf.org/doc/draft-xu-agentic-overlay-network-architecture/} \\
OAN Resource Identity and Discovery & IETF draft: resource identity and discovery interoperability & \url{https://datatracker.ietf.org/doc/draft-xu-oan-resource-identity-discovery/} \\
\end{longtable}
}

\chapter{Public HTTP API}\label{chap:15}

This chapter organizes the public HTTP APIs of OpenAgenet (OAN) into implementable and testable interface contracts. It covers the public and node-to-node calls of the official website backend, Registrar nodes, the Root platform, the content distribution platform, Discovery nodes, and the off-chain trust indexer. The paths, fields, and error semantics used by the browser, TypeScript SDK, and community skill package should remain consistent. Public APIs return only the materials required by public services; they do not return private keys, local identity backups, database administration capabilities, or access-statistics back-office data. An HTTP success response does not automatically mean that a resource has completed Root platform publication, network-wide distribution, or Discovery indexing.

\section{API Conventions and Base URLs}\label{sec:15.1}

{\def\LTcaptype{none} 
\begin{longtable}[]{@{}>{\raggedright\arraybackslash}p{0.3200\linewidth}>{\raggedright\arraybackslash}p{0.3200\linewidth}>{\raggedright\arraybackslash}p{0.3200\linewidth}@{}}
\toprule\noalign{}
Contract item & Unified requirement & Failure behavior \\
\midrule\noalign{}
\endhead
\bottomrule\noalign{}
\endlastfoot
Method and path & Follow the actual route declaration of the service & Return an identifiable route or method error \\
Media type & Use an explicit Content-Type for JSON requests and responses & Reject a body that cannot be parsed \\
Request ID & The client may provide one; the server generates one when absent & Keep the value consistent in the response and logs \\
Time & RFC 3339 strings & Do not silently convert an incorrectly formatted value \\
Caching & Set it according to data freshness and status & Expired data must carry a degradation indicator \\
\end{longtable}
}

OAN interfaces transmit JSON using UTF-8. Time fields use RFC 3339 strings. DID, version, hash, and cursor values are handled as strings and should not be converted by clients into lossy numeric types. The base URLs for production environments are determined by deployment configuration. The protocol constrains paths and semantics, but does not make the current website domain a fixed condition for every deployment. Requests should carry a request ID that can run through one complete call chain. Responses should express results through stable machine-readable fields; presentation-layer text must not be used as the basis for program decisions.

\subsection{Service Base Addresses}\label{subsec:15.1.1}

The current public backend address of the official website and the addresses of individual nodes are jointly determined by the frontend API configuration, node configuration, and deployment environment. The SDK provides independent endpoint configuration for Registrar nodes, Discovery nodes, the Root platform, and the content distribution platform. Callers must not assume that all services are deployed on the same host or under the same domain. Cross-node requests should also record the target node DID, so that the communication peer is not identified solely by a network address.

\subsection{Path and Version Conventions}\label{subsec:15.1.2}

The version in a path should be expressed through an explicit path, protocol field, or capability negotiation. It must not be inferred from a page title or service name. When fields are added compatibly, clients should ignore unknown non-required fields. When field meaning, signature coverage, or status semantics change, the protocol version must be increased or a migration period must be provided. Interface paths, methods, and media types are part of the contract; clients must not guess based on differences in case or trailing slashes.

\subsection{Request Authentication}\label{subsec:15.1.3}

Public queries and health interfaces may be accessed anonymously according to the deployment policy. Resource registration, node-to-node submission, and governance operations require, respectively, a resource control proof, node authorization materials, or operations authentication. Browser CORS permission only determines whether a page can read a response; it does not grant resource control. Authentication failures should return stable errors and should not hide permission problems behind an ambiguous 404 or an empty successful result.

\subsection{Response Format}\label{subsec:15.1.4}

Successful responses should use JSON objects to express data, status, version, freshness, or source. Error responses should contain at least a machine-readable error code, a request ID, and necessary field locations. A service must not return an HTML error page where JSON is expected. For asynchronous publication and indexing, the response should clearly distinguish accepted, processing, completed, and failed states. An HTTP 200 response alone must not represent completion of the full chain.

\section{Health, Readiness, and Service Status Endpoints}\label{sec:15.2}

Health checks should distinguish process liveness, service readiness, availability of critical dependencies, and successful completion of business smoke tests. The website backend\textquotesingle s \passthrough{\lstinline!/health!} only indicates that the backend process can respond; it does not prove that the Registrar node, Discovery node, Root platform, or database is operating normally. After an operations deployment, node health, database connectivity, indexer synchronization status, and the minimum registration and Discovery business paths should be checked separately. Public status interfaces should not disclose internal connection strings, tokens, or stack traces.

\section{Registrar Node Registration Endpoints}\label{sec:15.3}

{\def\LTcaptype{none} 
\begin{longtable}[]{@{}>{\raggedright\arraybackslash}p{0.4800\linewidth}>{\raggedright\arraybackslash}p{0.4800\linewidth}@{}}
\toprule\noalign{}
Field & Current meaning \\
\midrule\noalign{}
\endhead
\bottomrule\noalign{}
\endlastfoot
\passthrough{\lstinline!status!} & \passthrough{\lstinline!submitted!} on current successful implementation path \\
\passthrough{\lstinline!resourceDid!}, \passthrough{\lstinline!resourceType!} & Resource identity and type registered in this request \\
\passthrough{\lstinline!registrationCredential!} & Registration VC issued after Root acceptance \\
\passthrough{\lstinline!rootResponse!} & Raw business response from Root acceptance \\
\end{longtable}
}

{\def\LTcaptype{none} 
\begin{longtable}[]{@{}>{\raggedright\arraybackslash}p{0.3200\linewidth}>{\raggedright\arraybackslash}p{0.3200\linewidth}>{\raggedright\arraybackslash}p{0.3200\linewidth}@{}}
\toprule\noalign{}
Request stage & Must be verified & A success response does not mean \\
\midrule\noalign{}
\endhead
\bottomrule\noalign{}
\endlastfoot
Draft validation & Local identity, required fields, DID Document, and metadata & Submitted or published \\
Resource submission & Control proof, node authorization, and resource package digest & Root platform publication completed \\
Status query & Registration, publication, distribution, and indexing stages & All nodes have converged \\
VC retrieval & Issuer, subject, version, digest, and status & A guarantee of resource business quality \\
Same-DID update & Original control relationship, preceding version, and new digest & Automatic acquisition of update authority \\
\end{longtable}
}

The reference implementation accepts both \passthrough{\lstinline!POST /resources/register!} and the compatibility entry point \passthrough{\lstinline!POST /resources/submit!}; both enter the same resource registration handler. The browser or SDK generates the resource DID, DID Document, metadata, version, and digest locally, and submits them as a \passthrough{\lstinline!ResourceRegistrationSubmission!}. Private keys and local identity backups are not part of the request body. The Registrar node first checks the structure, resource type, authorization domain, and subject control proof, and then calls the Root platform\textquotesingle s verification and publication endpoint with a signed upstream request.

The core fields of a successful response are as follows. The example values are provided only to explain the structure:

\passthrough{\lstinline!status: submitted!} means that the Registrar node has completed its processing of this request and obtained a Root platform response. It does not mean that content distribution and indexing by all Discovery nodes have completed. The reference implementation issues and returns \passthrough{\lstinline!registrationCredential!}; the client should still allow the credential field to be empty or unparseable while preserving the submission result and original response, so that ``registration failed'' can be distinguished from ``credential missing.'' \passthrough{\lstinline!GET /resources!}, \passthrough{\lstinline!GET /resources/\{did\}!}, \passthrough{\lstinline!GET /registrar/status!}, and \passthrough{\lstinline!GET /registrar/stats!} are used to read registration records and node status, but the public scope is still determined by the deployed reverse proxy and access policy.

\section{Root Platform Publication and Status Endpoints}\label{sec:15.4}

The Root platform separates public read routes from management write routes. The resource publication entry point is \passthrough{\lstinline!POST /root/resources/verify-and-publish!}, called by a Registrar node with \passthrough{\lstinline!upstreamAuth!}. The Root platform re-verifies the source node identity, node authorization, resource DID, DID Document, digest, and control proof. Resource queries use \passthrough{\lstinline!GET /root/resources/\{did\}!}. The historical version list and a specified version use \passthrough{\lstinline!/root/resources/\{did\}/versions!} and \passthrough{\lstinline!/root/resources/\{did\}/versions/\{version\}!}, respectively.

{\def\LTcaptype{none} 
\begin{longtable}[]{@{}>{\raggedright\arraybackslash}p{0.3200\linewidth}>{\raggedright\arraybackslash}p{0.3200\linewidth}>{\raggedright\arraybackslash}p{0.3200\linewidth}@{}}
\toprule\noalign{}
Interface category & Typical path & Main boundary \\
\midrule\noalign{}
\endhead
\bottomrule\noalign{}
\endlastfoot
Root platform identity and status & \passthrough{\lstinline!/root/did!}, \passthrough{\lstinline!/root/status!} & Provides public identity or runtime summaries; does not expose management capabilities \\
Resource reads & \passthrough{\lstinline!/root/resources/\{did\}!} and version paths & Returns saved resource versions and publication materials \\
Publication writes & \passthrough{\lstinline!/root/resources/verify-and-publish!} & Requires a trusted upstream envelope from a Registrar node \\
Publication queues & \passthrough{\lstinline!/root/queues/cdn-publish!}, \passthrough{\lstinline!/root/queues/discovery-notify!} & Used to observe asynchronous propagation; whether they are public depends on the deployment policy \\
Governance management & Authorization, domain-update, and revocation paths & Management routes that require separate administrator authentication \\
\end{longtable}
}

A successful publication response means that the Root platform accepted and saved a compliant resource version. The content distribution publisher and Discovery notification tasks then continue processing it. Callers should confirm propagation progress through the resource version, publication queue, or subsequent status, and must not interpret one 2xx response from the Root platform as network-wide visibility.

\section{Content Distribution Endpoints}\label{sec:15.5}

The content distribution platform provides single-resource reads, batch reads, incremental indexing, and split-material reads. \passthrough{\lstinline!GET /cdn/resources/\{did\}!} returns a resource package. \passthrough{\lstinline!POST /cdn/resources/batch-get!} reads multiple resources by DID. \passthrough{\lstinline!GET /cdn/documents/\{did\}!} and \passthrough{\lstinline!GET /cdn/metadata/\{did\}!} return the DID Document and metadata, respectively. Catalog interfaces are located under \passthrough{\lstinline!/cdn/catalog/...!}. The Root platform publisher writes through the protected \passthrough{\lstinline!POST /cdn/resources!} or \passthrough{\lstinline!POST /cdn/resources/batch!}. \passthrough{\lstinline!POST /cdn/purge!} is a management operation and should not be exposed as a public write interface.

Incremental synchronization uses \passthrough{\lstinline!GET /cdn/resources/index?afterCursor=<n>\&limit=<n>!}. In the current reference implementation, omitting parameters may return a compatibility format; supplying cursor or quantity parameters returns a structured page:

\begin{lstlisting}
{
  "items": [],
  "count": 0,
  "afterCursor": 42,
  "nextCursor": 42,
  "hasMore": false
}
\end{lstlisting}

A Discovery node should save \passthrough{\lstinline!nextCursor!} only after fully validating the resource packages in the batch; it must not advance the cursor early merely because the HTTP request succeeded. The current service imposes an implementation-level body limit on batch publication requests. This is a deployment safeguard of the reference implementation, not a protocol value permanently fixed for every OAN node.

\section{Discovery Query and Resource Detail Endpoints}\label{sec:15.6}

{\def\LTcaptype{none} 
\begin{longtable}[]{@{}>{\raggedright\arraybackslash}p{0.3200\linewidth}>{\raggedright\arraybackslash}p{0.3200\linewidth}>{\raggedright\arraybackslash}p{0.3200\linewidth}@{}}
\toprule\noalign{}
Query result field & Purpose & Freshness or trust boundary \\
\midrule\noalign{}
\endhead
\bottomrule\noalign{}
\endlastfoot
resourceDid & Precisely identifies the resource & Still requires resolution and control verification \\
resourceType/version & Filtering and version selection & Does not replace publication proof \\
sourceNode & Tracks the result source & Node authorization requires a separate check \\
indexedAt/cursor & Determines index freshness & Does not equal on-chain finality \\
matchExplanation & Explains a semantic match & Does not equal invocation authorization \\
\end{longtable}
}

The Discovery query entry point is \passthrough{\lstinline!POST /discovery/resources/query!}. The request may contain a natural-language description, resource type, capability tags, protocol, version, version mode, and quantity limit. If the query text can be recognized as a complete resource DID, the reference implementation prioritizes an exact-DID query. Otherwise, it prioritizes an available semantic index and falls back to a keyword-candidate flow when the semantic backend is unavailable or produces no result. \passthrough{\lstinline!discoveryDid!} in the response identifies the Discovery node that returned the results. \passthrough{\lstinline!createdAt!} identifies when the response was generated, and a candidate\textquotesingle s \passthrough{\lstinline!score!} expresses only retrieval relevance.

\begin{lstlisting}
{
  "query": "I need a tool that can search code repositories and summarize the project structure.",
  "resourceType": "skill",
  "capabilityTags": [],
  "versionMode": "latest",
  "limit": 10
}
\end{lstlisting}

Indexed-resource reads and diagnostic interfaces include \passthrough{\lstinline!/discovery/index/resources!}, \passthrough{\lstinline!/discovery/index/resources/\{did\}!}, \passthrough{\lstinline!/discovery/index/stats!}, and \passthrough{\lstinline!/discovery/query/stats!}. Whether these are exposed to the Internet should be determined by the deployment exposure policy. An empty candidate array is a valid query result and should not be converted into a network error. Semantic-backend fallback, index lag, or snapshot degradation should be reflected in observable information; duplicate candidates must not be manufactured to hide differences between data sources.

\section{DID Resolution Endpoints}\label{sec:15.7}

Infrastructure nodes return their own DID Documents through \passthrough{\lstinline!GET /registrar/did!}, \passthrough{\lstinline!GET /root/did!}, and \passthrough{\lstinline!GET /discovery/did!}, respectively. A resource DID can be resolved through the Root platform resource detail, the content distribution platform\textquotesingle s \passthrough{\lstinline!/cdn/documents/\{did\}!}, or the detail of a resource already indexed by a Discovery node. The appropriate choice depends on whether the caller needs the authoritative saved version, a distributed copy, or a Discovery read model.

The \passthrough{\lstinline!id!} in a resolution response should exactly match the requested DID and preserve public verification methods, verification relationships, service endpoints, and OAN metadata. Clients need to distinguish three types of result: an invalid DID format is a request problem, a nonexistent resource is an absent record, and temporary unavailability of a dependency or node is a recoverable failure. No resolution path should return a browser local identity backup, private key, or server key file.

\section{VC and Proof Retrieval Endpoints}\label{sec:15.8}

Registration VCs, node authorization VCs, and Root proofs of trusted publication cover different facts. A Registrar node stores \passthrough{\lstinline!registrationCredential!} in the successful registration response and resource record. The Root platform may return the resource package and its Root proof through resource detail and version interfaces. The \passthrough{\lstinline!/.../root-authorization!} status interfaces of Registrar nodes and Discovery nodes expose the authorization materials or verification status observed by the respective node. The Root platform\textquotesingle s infrastructure-authorization VC issuance entry point is a protected operation and should not be treated as an anonymous self-service retrieval interface.

{\def\LTcaptype{none} 
\begin{longtable}[]{@{}>{\raggedright\arraybackslash}p{0.3200\linewidth}>{\raggedright\arraybackslash}p{0.3200\linewidth}>{\raggedright\arraybackslash}p{0.3200\linewidth}@{}}
\toprule\noalign{}
Evidence & What it proves & What must still be checked after retrieval \\
\midrule\noalign{}
\endhead
\bottomrule\noalign{}
\endlastfoot
Registration VC & A particular Registrar node processed a particular resource registration & Issuer, subject, resource digest, signature, and status \\
Node authorization VC & An infrastructure node was granted a role and authorization domain & Issuer, node DID, role, validity period, and revocation status \\
Root proof of trusted publication & A particular resource version and digest were accepted by the Root platform & Resource DID, version, digest, signature, and governance status \\
\end{longtable}
}

Returning a credential object from an interface does not mean that the credential has been verified. The caller should resolve the issuer\textquotesingle s public key according to \passthrough{\lstinline!proof.verificationMethod!} and check credential time and governance status. When status cannot be obtained, the caller should explicitly return ``indeterminate'' or tighten high-risk operations rather than assuming validity.

\section{Governance Status and Event Endpoints}\label{sec:15.9}

The Root platform provides infrastructure node directories through \passthrough{\lstinline!/root/registrars!}, \passthrough{\lstinline!/root/discovery-nodes!}, and their DID detail interfaces. It provides a governance-event read model through \passthrough{\lstinline!/root/bulletin/events!} and \passthrough{\lstinline!/root/bulletin/events/\{sequence\}!}. Registrar nodes and Discovery nodes expose the authorization status observed by each node through \passthrough{\lstinline!/registrar/root-authorization!} and \passthrough{\lstinline!/discovery/root-authorization!}, respectively. A Discovery node also expresses its current authorization domain through \passthrough{\lstinline!/discovery/authorized-domains!}.

The off-chain trust indexer\textquotesingle s data is an off-chain projection of on-chain governance facts. Status responses should carry the latest sequence, observation time, or synchronization position so that callers can assess freshness. The Root platform\textquotesingle s authorization, authorization-domain update, and revocation interfaces are management routes and require administrator authentication. If on-chain reads fail, the index sequence does not increase for an extended period, or local state cannot be confirmed, the Root platform and Discovery nodes should mark results as stale or indeterminate and tighten publication, synchronization, or Discovery-visibility decisions.

\section{Metrics and Observability Endpoints}\label{sec:15.10}

The website backend organizes multiple nodes and data sources through public read interfaces such as \passthrough{\lstinline!/api/public/home/summary!}, \passthrough{\lstinline!/api/public/network/summary!}, \passthrough{\lstinline!/api/public/network/governance-events!}, node directories, and node statistics. Node-side interfaces also include \passthrough{\lstinline!/registrar/stats!}, \passthrough{\lstinline!/discovery/index/stats!}, \passthrough{\lstinline!/discovery/query/stats!}, and \passthrough{\lstinline!/cdn/catalog/resources/stats!}. The website aggregation is a display read model and should not be treated as a replacement for the authoritative raw status of each node.

Metric fields should provide the generation time, statistical window, source, and missing-data status together. Windows such as \passthrough{\lstinline!TODAY!}, \passthrough{\lstinline!THIS WEEK!}, and \passthrough{\lstinline!THIS MONTH!} require a common time zone and explicit start and end boundaries. The cumulative resource count must not be confused with the number of additions within a window. Public interfaces return only necessary aggregate information. Access-statistics details, complete logs, database paths, management tokens, and internal error stacks remain operations data. The internal reporting entry point used by nodes to report statistics to the website should have authentication and exposure boundaries independent from public read interfaces.

\section{Request and Response Models}\label{sec:15.11}

Requests and responses should be able to express at least four results: success, processing, failure, and indeterminate:

\begin{lstlisting}
status = accepted | processing | completed | failed | unknown
requestId = used to trace one call across nodes
\end{lstlisting}

Clients must determine asynchronous stages from machine-readable status and must not infer that publication, distribution, and Discovery indexing have completed from HTTP 200 or page text alone.

Request and response models are stable boundaries for interface fields, statuses, and errors. Field presence, types, enumerations, hash algorithms, time formats, DID syntax, and nesting relationships should be consistent with \passthrough{\lstinline!oan-protocol-commons!} and the actual node types. Handling of unknown fields should be determined by the protocol version. Missing required fields, duplicate fields, empty values, and incompatible versions should produce errors that can be located.

The actual type relationships of the core node interfaces are as follows. \passthrough{\lstinline!Value!} in the table means that the current handler returns dynamic JSON; it does not mean that every field forms a stable contract. Specific fields still need to be checked against the corresponding handler, shared structures, and error tests.

{\def\LTcaptype{none} 
\begin{longtable}[]{@{}>{\raggedright\arraybackslash}p{0.2400\linewidth}>{\raggedright\arraybackslash}p{0.2400\linewidth}>{\raggedright\arraybackslash}p{0.2400\linewidth}>{\raggedright\arraybackslash}p{0.2400\linewidth}@{}}
\toprule\noalign{}
Service & Method and path & Input type & Output type or stable fields \\
\midrule\noalign{}
\endhead
\bottomrule\noalign{}
\endlastfoot
Registrar node & \passthrough{\lstinline!POST /resources/register!}, \passthrough{\lstinline!POST /resources/submit!} & \passthrough{\lstinline!ResourceRegistrationSubmission!} & \passthrough{\lstinline!Value!}: \passthrough{\lstinline!status!}, \passthrough{\lstinline!resourceDid!}, \passthrough{\lstinline!resourceType!}, \passthrough{\lstinline!registrationCredential!}, \passthrough{\lstinline!rootResponse!} \\
Root platform & \passthrough{\lstinline!POST /root/resources/verify-and-publish!} & \passthrough{\lstinline!ResourceVerifyAndPublishRequest!} & \passthrough{\lstinline!Value!}: verification queue status, resource version, three hashes, lifecycle, and CDN queue status \\
Content distribution platform & \passthrough{\lstinline!POST /cdn/resources!}, \passthrough{\lstinline!POST /cdn/resources/batch!} & \passthrough{\lstinline!ResourceCdnPublishRequest!}, \passthrough{\lstinline!ResourceCdnBatchPublishRequest!} & \passthrough{\lstinline!Value!}: publication status, cursor, and batch accepted/failed items \\
Content distribution platform & \passthrough{\lstinline!GET /cdn/resources/\{did\}!} & Path parameter \passthrough{\lstinline!did!} & \passthrough{\lstinline!ResourcePackage!} \\
Discovery node & \passthrough{\lstinline!POST /discovery/resources/query!} & \passthrough{\lstinline!ResourceDiscoveryQuery!} & \passthrough{\lstinline!ResourceDiscoveryResponse!} \\
Off-chain trust indexer & \passthrough{\lstinline!GET /v1/status!}, \passthrough{\lstinline!GET /v1/events!} & Optional \passthrough{\lstinline!limit!} query parameter & Dynamic status JSON; the event response contains an \passthrough{\lstinline!events!} array \\
\end{longtable}
}

The business stage of an interface cannot be inferred only from the HTTP status code. When a Registrar node returns \passthrough{\lstinline!submitted!}, the registration record has been written. When the Root platform returns \passthrough{\lstinline!resource-verified-and-queued!}, the resource has passed the Root platform\textquotesingle s current verification and entered the distribution queue. When the content distribution platform returns \passthrough{\lstinline!published!}, the resource package has been written to the distribution platform. When a Discovery response returns candidates, it only means that the current Discovery node\textquotesingle s index can see them. Callers still need to make subsequent judgments using the resource DID, version, hashes, Root proof, governance status, and freshness.

\subsection{Registration Requests and Responses}\label{subsec:15.11.1}

A registration request must associate the resource DID, DID Document, resource type, metadata, resource package, version, hashes, Registrar node, and control proof. Before signature verification, the server should confirm the request path, target, challenge, nonce, and content digest; only after verification should it perform duplicate checks and writes. The response should separately express registration acceptance, the registration VC, the Root platform submission result, and retryable errors. It must not make the client infer completion of all stages from a single ``success'' message.

The reference implementation\textquotesingle s public registration entry point directly accepts \passthrough{\lstinline!ResourceRegistrationSubmission!}. When a Registrar node forwards the request to the Root platform, it constructs a \passthrough{\lstinline!ResourceVerifyAndPublishRequest!} and binds the method, path, target, time, nonce, and business-body digest in \passthrough{\lstinline!upstreamAuth!}. The browser submits only resource materials and a control proof; it should not construct or hold the upstream signing identity of an infrastructure node.

\subsection{Publication Requests and Responses}\label{subsec:15.11.2}

A publication request must associate the Registrar node source, resource version, resource package digest, DID Document hash, metadata hash, authorization materials, and request ID. The publication response should carry a Root proof of trusted publication or clearly state that it has not yet been produced, and should return a publication status, batch reference, or cursor reference. Hash mismatches, an unauthorized node, an invalid control proof, and temporary dependency unavailability should be distinguished as non-retryable and retryable failures.

When the Root platform submits to the content distribution platform, single-item and batch requests use \passthrough{\lstinline!ResourceCdnPublishRequest!} and \passthrough{\lstinline!ResourceCdnBatchPublishRequest!}, respectively. Batch items are associated with the Root platform publication order through \passthrough{\lstinline!publicationCursor!}. The accepted count, failed items, or completion cursor in the response is used to resume a publication task; it does not mean that a Discovery node has completed indexing.

\subsection{Discovery Requests and Responses}\label{subsec:15.11.3}

A Discovery request may use a resource DID, resource type, capability tags, protocol, authorization domain, lifecycle, keywords, or a natural-language task description. The response should return candidate DIDs, versions, resource types, match explanations, the source Discovery node, index version, governance status, and proof references. A pagination cursor must be bound to the query conditions and the ordering snapshot. An empty result is not an interface failure; service unavailability, index lag, and snapshot degradation should be clearly marked.

The current \passthrough{\lstinline!ResourceDiscoveryResponse!} contains \passthrough{\lstinline!discoveryDid!}, \passthrough{\lstinline!candidates!}, \passthrough{\lstinline!createdAt!}, and an optional \passthrough{\lstinline!proof!}. A candidate contains the resource DID, resource type, score, version, lifecycle, capability tags, authorization domain, service endpoints, protocol bindings, resource-package information, and Root proof, among other fields. When an optional field is absent, the client must not use presentation-layer default text to fabricate a trusted fact.

\subsection{Status and Error Responses}\label{subsec:15.11.4}

Status and error responses use HTTP status codes for coarse-grained results, JSON error codes for business reasons, and \passthrough{\lstinline!requestId!} for trace correlation. An error may contain a field path, retryability, recommended wait time, and current processing status, but must not echo private keys, complete control proofs, or internal exception stacks. The client should retain the original error for auditing while giving the user an explanation appropriate to its sensitivity.

The current error bodies of multiple Rust nodes use the concise structure \passthrough{\lstinline!\{"error":"machine\_readable\_reason"\}!}, with the HTTP status code distinguishing request errors, permission errors, or service errors. Not every existing response contains \passthrough{\lstinline!requestId!}, a field path, and retry guidance. These fields are therefore directions for interface consistency, and callers must not assume that every node currently returns a complete error envelope.

\section{HTTP Status Codes and Common Errors}\label{sec:15.12}

{\def\LTcaptype{none} 
\begin{longtable}[]{@{}>{\raggedright\arraybackslash}p{0.3200\linewidth}>{\raggedright\arraybackslash}p{0.3200\linewidth}>{\raggedright\arraybackslash}p{0.3200\linewidth}@{}}
\toprule\noalign{}
Status & Typical meaning & Client action \\
\midrule\noalign{}
\endhead
\bottomrule\noalign{}
\endlastfoot
400 & Invalid request format or field & Correct the input and retry \\
401/403 & Identity or authorization requirements are not met & Do not retry blindly; check the credential and authorization domain \\
404 & Resource or route does not exist & Distinguish an unknown resource from a path error \\
409 & DID, version, or idempotency conflict & Query the current status before deciding whether to update \\
429 & Rate or concurrency limit exceeded & Retry according to \passthrough{\lstinline!Retry-After!} or a backoff policy \\
5xx/timeout & Service or dependency failure & Retry a limited number of times while preserving the request ID \\
\end{longtable}
}

It is recommended that 2xx indicate that the request was processed according to the interface semantics, 4xx indicate that the request, authentication, authorization, or business preconditions were not satisfied, and 5xx indicate a server-side or dependency failure. \passthrough{\lstinline!202 Accepted!} can indicate only that an asynchronous task has been accepted; it cannot indicate that publication is complete. \passthrough{\lstinline!409 Conflict!} can be used for a same-DID version conflict or a duplicate state that requires client confirmation. \passthrough{\lstinline!429 Too Many Requests!} should be accompanied by rate-limit information. Services should avoid returning arbitrary different status codes for the same business reason.

The client should first inspect the HTTP status and then parse \passthrough{\lstinline!error!} or the business status in the JSON. A connection failure, TLS failure, or browser CORS rejection has no usable HTTP response and must not uniformly be displayed as ``the server returned 500.'' Sending POST to an endpoint that supports only GET normally produces \passthrough{\lstinline!405 Method Not Allowed!} at the routing layer; the method and reverse-proxy forwarding rules should be checked first.

\section{Rate Limits, Caching, and Security Headers}\label{sec:15.13}

{\def\LTcaptype{none} 
\begin{longtable}[]{@{}>{\raggedright\arraybackslash}p{0.3200\linewidth}>{\raggedright\arraybackslash}p{0.3200\linewidth}>{\raggedright\arraybackslash}p{0.3200\linewidth}@{}}
\toprule\noalign{}
Control & Public queries & Registration or node-to-node requests \\
\midrule\noalign{}
\endhead
\bottomrule\noalign{}
\endlastfoot
Rate & May be limited by source and interface & Should consider subject, node, and idempotency key \\
Request size & Limit query descriptions and page sizes & Limit resource packages and batch sizes \\
Caching & Allow read-only caches with freshness information & Do not cache results that change authorization decisions \\
Security headers & TLS, CORS, and necessary security headers & Add authentication and source verification \\
\end{longtable}
}

Rate limiting should consider at least the source IP, authenticated subject, node DID, resource DID, and interface category, and should use different quotas for registration, Discovery, synchronization, and health requests. Read-only public queries may use short-lived caches, but DID Documents, VCs, Root proofs, governance status, and Discovery results must carry a version or freshness boundary. A conditional request\textquotesingle s ETag must correspond to the response content version. Security headers, content-type checks, and response-body size limits should be enforced at the gateway or service layer. Cache testing should at least verify the following: the first request obtains data and freshness information; a conditional request does not return an incorrect version; after governance revocation, an old cache does not continue authorizing writes; and when the backend is unavailable, an old snapshot is clearly marked rather than disguised as a real-time result. Write interfaces and signed requests should not be stored by a shared cache.

Request-body limits, query-count boundaries, and CORS configuration in the current node code provide the first layer of protection. Production deployment also requires the reverse proxy to handle connection counts, rates, timeouts, and request-size limits.

\section{CORS, Browser Clients, and Cross-Origin Policy}\label{sec:15.14}

When a browser calls a node API, the server should allow only the necessary official website origins, methods, and request headers. A preflight request must not broaden business authorization. \passthrough{\lstinline!Access-Control-Allow-Origin: *!} must not be used together with credentials to transmit sensitive materials. Public cross-origin reads by the official website and node-to-node authentication are two separate controls: the former serves the browser security model, while the latter verifies the request subject and permissions. CORS changes should be a deployment check item to prevent the frontend from producing an opaque \passthrough{\lstinline!Failed to fetch!} error.

Registrar nodes, Discovery nodes, and the content distribution platform all construct their CORS layer from configuration. The website backend preferentially uses the configured frontend origin and falls back to a more permissive policy only when the configuration value cannot be parsed. Production environments should avoid triggering this fallback branch and should perform registration, Discovery, and preflight smoke tests from both Chinese and English pages after deployment. When a browser displays \passthrough{\lstinline!Failed to fetch!}, check the request URL, HTTPS mixed content, certificate, DNS, reverse proxy, preflight response, and allowed origins in sequence; do not inspect only the backend business logs.

\section{Boundary Between Public Endpoints and Operations-Only Endpoints}\label{sec:15.15}

The public boundary is divided into ``public reads, resource-controller signatures, node-to-node authorization, and operations-only'' categories. Public interfaces may provide health information, public catalogs, published resources, and limited metrics. Resource-controller interfaces receive a control proof but not a private key. Node-to-node interfaces require node identity, authorization VCs, and request signatures. Operations interfaces are exposed only within a controlled network or authentication boundary. Access-statistics data may be read only by back-office or operations-authorized subjects; ordinary visitors must not query details through the website or public APIs.

\subsection{Public Query Interfaces}\label{subsec:15.15.1}

Public query interfaces include public catalogs, Discovery queries, published resource details, public DID Documents, public proofs, and health information. Returned data should be filtered by a field allowlist, hiding local file paths, private endpoints, database-management fields, and unpublished credentials. Public queries must still observe lifecycle, governance filtering, and authorization-domain visibility; anonymous access must not bypass these decisions.

The website\textquotesingle s \passthrough{\lstinline!/api/public/...!} routes mainly provide aggregated display data and do not mean that all node routes are automatically public. Deployment should establish an allowlist of public paths and verify that anonymous access can read only filtered data. \passthrough{\lstinline!POST /api/public/site-analytics/visit!}, which records page visits, is a one-way collection entry point and does not provide an interface for reading statistical details.

\subsection{Node-to-Node Interfaces}\label{subsec:15.15.2}

Node-to-node interfaces support cooperation among Registrar nodes, the Root platform, the content distribution platform, Discovery nodes, and the off-chain trust indexer. Requests should carry the source and target identities, authorization materials, protocol version, request ID, time window, nonce, content digest, and signature. Node-to-node interfaces should not be called directly by the website browser in place of a node; browser resource submissions should pass through the Registrar node\textquotesingle s public entry point and server-side validation.

Typical node-to-node writes include a Registrar node calling \passthrough{\lstinline!/root/resources/verify-and-publish!}, the Root platform publisher calling \passthrough{\lstinline!/cdn/resources/batch!}, and the Root platform sending authorization-scoped synchronization notifications to Discovery nodes. Even when a path is reachable from the Internet, the receiver must still verify the trusted upstream envelope and node authorization; network reachability itself is not an authorization basis.

\subsection{Operations Interfaces}\label{subsec:15.15.3}

Operations interfaces include deployment status, internal diagnostics, access-statistics details, database administration, key rotation, and recovery operations. They must not appear in the public API catalog or browser-side code, and should use separate authentication, least privilege, audit logs, and network isolation. Errors returned by operations interfaces must not be fully passed through a public proxy to ordinary users.

The Root platform\textquotesingle s authorization, revocation, and publication-management routes, as well as the website\textquotesingle s internal node-statistics reporting entry point, should be separated from public read routes at the reverse-proxy layer. Combining routes in the same process in service code does not mean that they should share the same Internet exposure policy. Operations personnel should check application authentication and proxy path allowlists together.

\subsection{Access Control and Exposure Boundaries}\label{subsec:15.15.4}

Access control should check the route, method, authenticated subject, node role, authorization domain, and resource scope together. Exposure boundaries should be enforced through ports, the reverse proxy, and CORS configuration. Before publication, four identities should be used to test allowed and denied paths separately: an anonymous browser, a resource controller, an unauthorized node, and an operations subject. The test should preserve status codes, response fields, and log-correlation evidence.

{\def\LTcaptype{none} 
\begin{longtable}[]{@{}>{\raggedright\arraybackslash}p{0.3200\linewidth}>{\raggedright\arraybackslash}p{0.3200\linewidth}>{\raggedright\arraybackslash}p{0.3200\linewidth}@{}}
\toprule\noalign{}
Test identity & Typical operations that should be allowed & Typical operations that should be denied \\
\midrule\noalign{}
\endhead
\bottomrule\noalign{}
\endlastfoot
Anonymous visitor & Health, public catalogs, Discovery, and public resource reads & Authorization, revocation, database and statistical-detail reads \\
Resource controller & Submit a resource with a valid control proof & Submit a same-DID update not controlled by the subject \\
Authorized node & Perform node-to-node calls within its role and authorization domain & Access management operations belonging to another role \\
Operations subject & Perform management and diagnostics under separate authentication & Bypass auditing or obtain a business subject\textquotesingle s private key \\
\end{longtable}
}

\section{API Examples and Machine-Readable Interface Descriptions}\label{sec:15.16}

\begin{lstlisting}
openapi: 3.1.0
info:
  title: OAN Agent Contract
  version: 0.1.0
paths: {}
\end{lstlisting}

The example above is the actual skeleton of \passthrough{\lstinline!oan-protocol-commons/schemas/agent-contract/openapi.yaml!}. It identifies the OpenAPI version and contract-file version, but \passthrough{\lstinline!paths!} is still empty. Therefore, it can serve only as a machine-readable contract entry point and must not be treated as a complete description of all public and node-to-node APIs. Interface verification should use the \passthrough{\lstinline!Router!} definitions of each service, shared protocol structures, configuration, and deployment proxy rules together.

The minimum executable example for a user should begin with a health or public query that requires no signature. The hostname in the following command is a placeholder:

\begin{lstlisting}[language=bash]
curl --fail-with-body \
  --header "Accept: application/json" \
  "https://registrar.example/health"
\end{lstlisting}

Examples that write to the registration service, Root platform, or content distribution platform require a real DID, digest, control proof, or node-to-node signed envelope. A static fake signature must not be presented as an executable request. When the OpenAPI description is further developed, methods, request bodies, responses, authentication, errors, and examples should be completed route by route, and automated tests should check whether the description remains synchronized with the actual routes.

\section{References}\label{sec:15.17}

{\def\LTcaptype{none} 
\begin{longtable}[]{@{}>{\raggedright\arraybackslash}p{0.3200\linewidth}>{\raggedright\arraybackslash}p{0.3200\linewidth}>{\raggedright\arraybackslash}p{0.3200\linewidth}@{}}
\toprule\noalign{}
Source & Type & Link \\
\midrule\noalign{}
\endhead
\bottomrule\noalign{}
\endlastfoot
\passthrough{\lstinline!oan-protocol-commons!} & Code repository: HTTP payloads, error codes, and OpenAPI contract entry point & \url{https://github.com/wolfbrother/oan-protocol-commons} \\
\passthrough{\lstinline!oan-root-nodes!} & Code repository: Root platform public and internal APIs & \url{https://github.com/OpenAgenet/oan-root-nodes} \\
\passthrough{\lstinline!oan-registrar-nodes!} & Code repository: Registrar service APIs & \url{https://github.com/OpenAgenet/oan-registrar-nodes} \\
\passthrough{\lstinline!oan-discovery-nodes!} & Code repository: Discovery service APIs & \url{https://github.com/OpenAgenet/oan-discovery-nodes} \\
\passthrough{\lstinline!oan-sdk-ts!} & Code repository: Client HTTP calls and type wrappers & \url{https://github.com/OpenAgenet/oan-sdk-ts} \\
OAN Resource Identity and Discovery & IETF draft: resource identity and Discovery interface direction & \url{https://datatracker.ietf.org/doc/draft-xu-oan-resource-identity-discovery/} \\
Efficient Agent Discovery Profile & IETF draft: Discovery query profile & \url{https://datatracker.ietf.org/doc/draft-xu-efficient-agent-discovery-profile/} \\
\end{longtable}
}

\chapter{Client SDK, Community Skill, and Developer Experience}\label{chap:16}

This chapter explains how the client layer provides different entry points for resource providers, callers, and node operators: \passthrough{\lstinline!oan-sdk-ts!} supports programmatic integration, the community skill package supports agent-assisted resource organization, registration, discovery, and verification, and the official website provides browser-based interaction. All three reuse the OpenAgenet (OAN) protocol and do not change the semantics of DID, control proofs, registration VCs, Root proof, or governance status. Subject private keys, resource-control private keys, and Local Identity backups remain on the client side; servers receive only public materials and necessary proofs.

{\def\LTcaptype{none} 
\begin{longtable}[]{@{}>{\raggedright\arraybackslash}p{0.2400\linewidth}>{\raggedright\arraybackslash}p{0.2400\linewidth}>{\raggedright\arraybackslash}p{0.2400\linewidth}>{\raggedright\arraybackslash}p{0.2400\linewidth}@{}}
\toprule\noalign{}
Entry point & Primary users & Core responsibilities & Responsibilities it does not assume \\
\midrule\noalign{}
\endhead
\bottomrule\noalign{}
\endlastfoot
TypeScript SDK & Application developers and platform integrators & Types, request construction, registration and discovery calls, and verification assistance & Making trust decisions for users \\
Community skill package & Resource providers and agent operators & Organizing resources, assisting with registration and discovery, and explaining results & Holding governance keys or replacing node services \\
Official website & Browser users & Interactive registration, discovery, preview, and local download & Hosting user private keys or guaranteeing endpoint security \\
Operations assistance & Node operators & Building, configuration checks, deployment, and smoke testing & Automatically deciding production keys and rollback strategies \\
\end{longtable}
}

\section{TypeScript SDK Scope}\label{sec:16.1}

The TypeScript SDK consists of protocol types, an HTTP client, governance reads, and workflow assistance. It is suitable for Node.js, browser build environments, and runtimes compatible with \passthrough{\lstinline!fetch!}. Callers can configure the Registrar node, Discovery node, Root platform, and content distribution platform addresses separately. The SDK handles serialization, requests, and response parsing; it does not decide whether a resource is trustworthy on behalf of the caller.

In terms of code organization, \passthrough{\lstinline!packages/protocol-types!} provides shared types such as \passthrough{\lstinline!ResourceRegistrationSubmission!}, \passthrough{\lstinline!ResourcePackage!}, discovery queries, and governance responses; \passthrough{\lstinline!packages/client-ts!} provides \passthrough{\lstinline!OanClient!} and HTTP calls to the various nodes; \passthrough{\lstinline!packages/sdk-ts!} provides DID Document drafts, resource-package binding, hash formatting, trust summaries, and lifecycle assistance; and \passthrough{\lstinline!packages/governance-ts!} targets governance reads from the off-chain trust indexer. This layering allows an application to depend only on the HTTP client or to combine resource validation and lifecycle assistance when needed, without duplicating protocol structures.

{\def\LTcaptype{none} 
\begin{longtable}[]{@{}>{\raggedright\arraybackslash}p{0.3200\linewidth}>{\raggedright\arraybackslash}p{0.3200\linewidth}>{\raggedright\arraybackslash}p{0.3200\linewidth}@{}}
\toprule\noalign{}
Capability area & Provided by the SDK & Still the caller\textquotesingle s responsibility \\
\midrule\noalign{}
\endhead
\bottomrule\noalign{}
\endlastfoot
DID and resource model & Types, parsing, and construction assistance & Confirming resource content and control relationships \\
Registration & Request construction, submission, and status reads & Safeguarding private keys and confirming the submission target \\
Discovery & Exact DID queries, semantic queries, and result parsing & Deciding whether relevance satisfies business requirements \\
Verification & Reading proof fields and verification assistance & Defining authorization policies and making final invocation decisions \\
Lifecycle & Aggregating registration, publication, distribution, and visibility status & Handling timeouts, retries, and business compensation \\
\end{longtable}
}

The SDK supports two common endpoint-selection patterns. For the official website and community users, the caller can provide a \passthrough{\lstinline!baseUrl!}, from which the client derives service addresses under the official gateway. For multi-node or third-party-node testing, the caller can provide \passthrough{\lstinline!registrarEndpoint!}, \passthrough{\lstinline!discoveryEndpoint!}, \passthrough{\lstinline!rootEndpoint!}, and \passthrough{\lstinline!cdnEndpoint!} separately. Endpoint configuration determines where requests are sent; it does not perform node authorization, resource-control proof generation, or governance decisions for the caller.

\section{DID Identity and Key Operations in the SDK}\label{sec:16.2}

Identity and key operations take place on the client. The SDK generates subject identities, resource DIDs, and verification methods, uses private keys to sign control challenges, resource versions, or requests, and writes public keys into the DID Document. A backup may contain subject and resource-control materials, but it must not enter logs, HTTP requests, or indexes. During import, the implementation should validate the structure, the relationship between the DID and public key, and signing capability.

In a Node.js scenario, \passthrough{\lstinline!identity-store-node!} can be used to save subject, agent, and node identities in separate buckets in a local directory. In a browser scenario, the current session is held by page state or browser storage and is exported through a user operation. The common boundary of both implementations is that the DID Document may be publicly distributed, while private keys may only be used for client-side signing; the server receives signed control proofs or node request envelopes, not the original private keys.

\begin{itemize}
\tightlist
\item[$\boxtimes$]
  Private-key generation, signing, and backup export occur on the client.
\item[$\boxtimes$]
  Registration requests carry only the public DID Document, resource information, and necessary control proofs.
\item[$\boxtimes$]
  Backup format, version, DID, public key, and signing capability are validated before import.
\item[$\square$]
  Private keys must not be written to URLs, logs, telemetry, or server-side storage.
\end{itemize}

\section{Registration Flow Example}\label{sec:16.3}

{\def\LTcaptype{none} 
\begin{longtable}[]{@{}>{\raggedright\arraybackslash}p{0.3200\linewidth}>{\raggedright\arraybackslash}p{0.3200\linewidth}>{\raggedright\arraybackslash}p{0.3200\linewidth}@{}}
\toprule\noalign{}
Client scenario & Local identity input & Proof the SDK must generate \\
\midrule\noalign{}
\endhead
\bottomrule\noalign{}
\endlastfoot
Self-controlled resource & Resource identity also acting as controller & Resource document proof and subject control proof \\
Existing controller & Resource identity plus imported controller identity & Resource-material proof and controller authorization for this registration \\
New dedicated controller & Local identity of the new controller & Self-control proof for ordinary registration; usable as controller only after success \\
\end{longtable}
}

A registration flow should first prepare the Local Identity, resource DID, DID Document, metadata, ResourcePackage, hashes, and control proof, and then call the Registrar node. The following example shows only the client-call boundary; \passthrough{\lstinline!submission!} should be constructed from data that has passed local validation and signing:

\begin{lstlisting}
import { OanClient } from "@openagenet/oan-sdk-ts/client";

const client = new OanClient({
  registrarEndpoint: "https://registrar.example.org",
});
const result = await client.registerResource(submission);

console.log({
  resourceDid: result.resourceDid,
  registrationVc: result.registrationVc ?? null,
});
\end{lstlisting}

The registration VC may temporarily be absent. The client should preserve the resource DID, request ID, and status; a network timeout cannot by itself prove that registration did not occur. The client should first confirm an explicit success or failure, then query the status by request ID or resource DID, and distinguish the three stages of Registrar node acceptance, Root platform publication, and Discovery node visibility.

The minimum pre-submission checks can be summarized as ``object consistency, digest consistency, valid control, and explicit target'': \passthrough{\lstinline!didDocument.id!} must equal \passthrough{\lstinline!resourceDid!}; the resource type must match the DID subject encoding; the digests of the DID Document, metadata, and ResourcePackage must correspond to the actual materials; the control proof must be generated by the corresponding resource-control private key; and the registration endpoint must be a Registrar node explicitly selected by the user. The skill package can assist in identifying missing items, but the final submission must still be confirmed by the client or user.

\begin{lstlisting}
Registration accepted -> Root accepted -> Content distribution completed -> Visible in discovery index
          |                    |                         |                              |
       Registrar              Root                      CDN                         Discovery
\end{lstlisting}

\section{Exact DID Discovery Example}\label{sec:16.4}

Exact DID discovery uses the resource DID as the lookup key and does not depend on name similarity. After calling the discovery interface, the client should check the returned DID, version, ResourcePackage, and proof references, and then verify the DID Document, version hash, lifecycle, Root proof, and endpoint policy.

\begin{lstlisting}
const resourceDid = "did:oan:K7mQ9:7YpQm9Kx2VnRb6Ts3WfHa4Cd5Ej8LgNz";
const resource = await client.getDiscoveryResource(resourceDid);
if (resource.resourceDid !== resourceDid) {
  throw new Error("discovery_did_mismatch");
}
\end{lstlisting}

An exact hit proves only that the Discovery node returned a record for the corresponding identifier. It does not by itself prove that the version is still valid, that the node remains authorized, or that the business endpoint can be safely called.

After obtaining candidate results, an application can first perform a consistency check using \passthrough{\lstinline!resourceDid!}, then select a candidate version according to the requested version policy, and finally load the ResourcePackage or DID Document to complete evidence verification. If the detail interface returns 404, this usually means that a candidate summary exists in the discovery index but the corresponding detail is not yet readable. The client should preserve this distinction and report it to the caller instead of treating the candidate result as a complete callable object.

\section{Semantic Discovery Example}\label{sec:16.5}

Semantic discovery can combine a natural-language task, capability tags, and structured conditions. The English query used on the official website, ``I need a tool that can search code repositories and summarize the project structure.'', can serve as an example input:

\begin{lstlisting}
const discovery = await client.discoverResources({
  query: "I need a tool that can search code repositories and summarize the project structure.",
  resourceType: "tool",
  limit: 10,
});

for (const candidate of discovery.candidates) {
  console.log(candidate.resourceDid, candidate.score);
}
\end{lstlisting}

The response should retain hit fields, tags, resource type, protocol, index version, source, and matching reasons. Semantic relevance answers ``possibly applicable''; structured conditions answer ``satisfies explicit filters''; and trusted evidence answers ``whether the current version can be verified.'' None of these can replace the others.

The SDK\textquotesingle s HTTP-response wrapper should preserve protocol stages rather than compressing every interface into one Boolean success value. A registration call should at least preserve the Registrar node\textquotesingle s \passthrough{\lstinline!status!}, resource DID, registration VC, and Root response. A discovery call should preserve \passthrough{\lstinline!discoveryDid!}, the candidate array, creation time, and any optional response proof. Dynamic status or explanation responses should preserve the original JSON so that the upper layer can process actual fields. This allows the SDK to distinguish ``registration accepted,'' ``verified by the Root platform and queued,'' ``resource discovered,'' and ``evidence verification completed.''

Semantic-query result handling is best divided into two steps: first narrow candidates with query text, resource type, tags, or protocols; then check each candidate\textquotesingle s lifecycle, authorization domain, service binding, and proof status. A Discovery node may use a semantic index or fall back to keyword search when its backend is unavailable. Callers should focus on the returned candidates and explanation information, rather than hard-coding a particular search-backend name as a business dependency.

\section{Verification and Trust Evidence Handling}\label{sec:16.6}

Verification should first check the relationship between the DID Document and resource DID, the public key, and the control proof. It should then separately check the issuer, subject, version, signature, validity period, status, authorization domain, and capability scope of the registration VC, Root proof, and node authorization VC. When a critical proof is missing or expired, the client should mark the resource as unconfirmed or reject its use.

{\def\LTcaptype{none} 
\begin{longtable}[]{@{}>{\raggedright\arraybackslash}p{0.3200\linewidth}>{\raggedright\arraybackslash}p{0.3200\linewidth}>{\raggedright\arraybackslash}p{0.3200\linewidth}@{}}
\toprule\noalign{}
Evidence & Main fact it proves & Fact it cannot prove \\
\midrule\noalign{}
\endhead
\bottomrule\noalign{}
\endlastfoot
Control proof & The submitter can use the corresponding control private key & The Registrar node accepted the registration or the resource was published \\
Registration VC & The Registrar node\textquotesingle s statement about the registration fact & The Root platform published the resource or the endpoint is secure \\
Root proof & The specified resource version passed Root platform verification and entered the publication process & Every Discovery node has indexed the resource \\
Node authorization VC & The node has the corresponding qualification within the authorization domain and validity period & Every piece of business data returned by the node is correct \\
Governance status & A status fact recorded by the governance side at the current time & Resource functional quality or business suitability \\
\end{longtable}
}

A reproducible verification fixture should at least contain the resource DID, DID Document, ResourcePackage, corresponding digests, a control proof or registration VC, and the node DID Document and authorization materials used to validate node eligibility. The test should first confirm success with untampered materials, then separately modify the DID, version, digest, issuer, or lifecycle status and confirm that the SDK returns a clear verification error. Whether a failure blocks use depends on the evidence type: control-proof or digest failure should normally block registration; insufficient semantic matching affects only candidate ranking; and uncertain governance status should block high-risk invocation or require human confirmation.

\section{Community Skill Package Responsibilities}\label{sec:16.7}

The community skill package uses scripts and documentation to assist with resource organization, tag recommendation, DID generation, registration requests, discovery queries, evidence verification, and result interpretation. It is not a Registrar node, Root platform, or Discovery node, and it must not hold governance keys. User confirmation, private-key use, sensitive paths, and the final publication decision remain under the control of the user or operator.

The typical interaction follows the sequence ``read materials → form a draft → user confirmation → local signing → call the service → explain the result.'' The skill package can identify missing fields, recommend semantic tags, and call the SDK, but it must not silently replace resource endpoints, expand authorization domains, or present recommendations as statements made by the resource provider.

The public workflow entry points of \passthrough{\lstinline!OanSkill!} include \breakablecode{draftRegistrationFromResourceDescription()}, \breakablecode{registerFromResourceDescription()}, \passthrough{\lstinline!registerBatchFromResourceDescriptions()!}, \passthrough{\lstinline!discover()!}, \passthrough{\lstinline!lifecycle()!}, and read-only governance assistance methods. A resource-description workflow returns candidate information, missing inputs, quality issues, and recommended actions. This allows batch onboarding to repair items individually instead of losing the diagnostics for the entire batch because one resource failed. The skill package uses the official public gateway by default; third-party nodes participate only when explicitly configured by the user.

\section{Resource Authoring and Registration Assistance}\label{sec:16.8}

Resource-authoring assistance should uniformly handle agent services, skill packages, MCP servers, and tool APIs, and guide the user in filling in the name, description, use cases, endpoints, protocols, version, capability tags, authorization domains, and ResourcePackage information. Recommended tags are not final declarations; before generation, the implementation should check the consistency of the DID Document, metadata, and ResourcePackage hashes.

{\def\LTcaptype{none} 
\begin{longtable}[]{@{}>{\raggedright\arraybackslash}p{0.3200\linewidth}>{\raggedright\arraybackslash}p{0.3200\linewidth}>{\raggedright\arraybackslash}p{0.3200\linewidth}@{}}
\toprule\noalign{}
Input category & Minimum checks & Output purpose \\
\midrule\noalign{}
\endhead
\bottomrule\noalign{}
\endlastfoot
Identity and control & Subject DID, resource DID, and verification methods & Generate the control proof \\
Capability semantics & Name, description, use cases, and capability tags & Semantic governance and discovery \\
Invocation information & Protocol, endpoint, and authentication hints & Filtering and negotiation before invocation \\
Version materials & Version, predecessor version, and ResourcePackage digest & Updates, idempotency, and integrity checks \\
Governance boundary & Resource type, authorization domain, and lifecycle & Admission and discovery filtering \\
\end{longtable}
}

When organizing resources, \passthrough{\lstinline!authorizedDomains!} and \passthrough{\lstinline!capabilityTags!} should be processed separately. The former is a machine-readable domain used by the Registrar node for authorization-scope decisions; the latter is a semantic-discovery signal. The skill package may propose tags from the description, but it must not infer an authorization domain from the DID subject encoding, endpoint hostname, or natural-language description. When the fact cannot be determined, the missing information should be returned to the user for completion.

\section{Node Operators and Deployment Assistance}\label{sec:16.9}

Node-operations assistance can perform dependency checks, configuration checks, Rust compilation, service startup, health checks, and post-deployment registration and discovery smoke tests. The server, domain, keys, production configuration, and rollback plan require human confirmation.

\begin{itemize}
\tightlist
\item[$\boxtimes$]
  Check processes, listening ports, and health endpoints.
\item[$\boxtimes$]
  Check database connectivity and required table structures.
\item[$\boxtimes$]
  Check the off-chain trust indexer status and whether its synchronization sequence continues to advance.
\item[$\boxtimes$]
  Verify the minimum path for structured registration responses, exact DID discovery, and semantic discovery.
\item[$\square$]
  Do not treat the success of a single health endpoint as a substitute for a complete business smoke test.
\end{itemize}

The acceptance sequence for operations assistance should follow the business propagation sequence: first check processes and health endpoints, then check the database and node authorization status, next verify that a resource registration request receives a structured response, and finally confirm the discovery path using an exact DID query and a semantic query. When the Registrar node returns success but the Root platform, content distribution platform, or Discovery node is unavailable, the result should identify the corresponding stage rather than compressing the entire path into a single ``deployment successful'' conclusion.

\section{Package Distribution and Compatibility Management}\label{sec:16.10}

Published artifacts must include runtime entry points, built JavaScript, type declarations, scripts, necessary documentation, and dependencies. When the community skill package is published to npm, \passthrough{\lstinline!SKILL.md!}, scripts, or resource files must not be omitted. Before publishing the SDK, it should be installed from a clean directory and its imports, registration-request construction, and discovery-request construction should be verified.

Release acceptance should include:

\begin{enumerate}
\def\labelenumi{\arabic{enumi}.}
\tightlist
\item
  Preview the actual package file list and confirm that the presence of a file in the source repository does not mean that it is included in the published package.
\item
  Install the package to an empty directory and verify the module entry points it declares as supported.
\item
  Run minimum tests for registration, exact DID discovery, and semantic-discovery request construction.
\item
  Check that the README, type declarations, scripts, and resource files are accessible from the installed package.
\item
  Increase the major version for breaking interface or protocol-semantic changes and describe the migration boundary.
\end{enumerate}

For the community skill package, the npm artifact must also retain \passthrough{\lstinline!SKILL.md!}, \passthrough{\lstinline!references!}, \passthrough{\lstinline!agents!}, \passthrough{\lstinline!src!}, and the built \passthrough{\lstinline!dist!}; checking only that TypeScript compilation succeeds is insufficient. For the SDK, verify that the root entry point and exports such as \passthrough{\lstinline!./client!}, \passthrough{\lstinline!./governance!}, \passthrough{\lstinline!./protocol-types!}, and \passthrough{\lstinline!./identity-store-node!} all point to actual JavaScript and declaration files. Release validation should run in an empty directory without using the source directory, so that omissions in \passthrough{\lstinline!package.json!} \passthrough{\lstinline!files!}, entry points, or dependency declarations can be detected.

\section{Example Resources and Integration Templates}\label{sec:16.11}

{\def\LTcaptype{none} 
\begin{longtable}[]{@{}>{\raggedright\arraybackslash}p{0.2400\linewidth}>{\raggedright\arraybackslash}p{0.2400\linewidth}>{\raggedright\arraybackslash}p{0.2400\linewidth}>{\raggedright\arraybackslash}p{0.2400\linewidth}@{}}
\toprule\noalign{}
Component & Applicable form & Current responsibility & Key boundary \\
\midrule\noalign{}
\endhead
\bottomrule\noalign{}
\endlastfoot
Identity Runtime & Embedded application or trusted local process & Encrypted identity catalog, controlled signing, evidence verification, session and replay protection & No arbitrary \passthrough{\lstinline!sign(bytes)!} and no replacement for business authorization \\
Identity Gateway & Enterprise ingress, multi-tenant or private proxy & Routing, tenant isolation, identity envelope, Runtime or controlled signer integration & Transport authentication is not resource-identity verification \\
Identity Sidecar & Existing Agent, MCP, or HTTP service difficult to modify & Verify inbound calls, sign outbound calls, inject read-only \passthrough{\lstinline!VerifiedPrincipal!} & No separate identity system and no acceptance of client-forged identity headers \\
\end{longtable}
}

The minimum integration flow is: create or import a Local Identity; generate the resource DID, DID Document, metadata, ResourcePackage, and control proof; submit to a Registrar node; perform an exact DID query; perform natural-language, tag-based, and structured discovery; and, before invocation, verify the version, proofs, governance status, and endpoint.

\pandocbounded{\includegraphics[width=\linewidth,height=0.72\textheight,keepaspectratio]{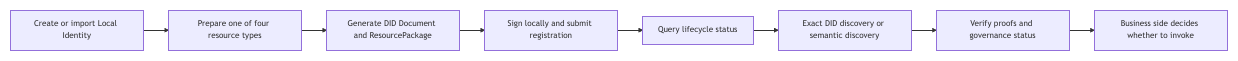}}

The SDK is the programmatic entry point, the skill package is the operational assistant, and the official website is the interactive entry point. The four resource templates can share identity, version, semantic, and proof fields, but the protocol, endpoint, and ResourcePackage contents must reflect their individual differences.

The four resource types can reuse the same onboarding skeleton, but their invocation materials differ. An agent service normally needs a callable service endpoint; a skill package needs a manifest or download entry point; an MCP server needs an MCP protocol binding; and a tool API needs an API endpoint and input/output schema. The \passthrough{\lstinline!protocolBindings!}, service endpoints, resource descriptions, and package information in a template should be filled in according to the actual resource. The fact that all four types can be discovered does not justify copying one resource type\textquotesingle s protocol fields into another.

\section{Client Security and Secret Handling}\label{sec:16.12}

Private keys and Local Identity backups are generated, used, and stored only on the client. DID Documents, registration VCs, and ResourcePackages should be handled according to their respective disclosure policies; logs should record only the DID, version, request ID, status, and error code, and must not record private keys, complete Local Identity data, or unsanitized signature material.

\begin{quote}
\textbf{Client security boundary:} A Local Identity file that can be downloaded or exported is control-rights material. It does not become public merely because it was generated by the official website. Registration VCs, DID Documents, and ResourcePackages should also be handled according to their respective disclosure policies. Their being downloaded in one archive does not give them the same disclosure level.
\end{quote}

SDK and skill-package tests must not write real private keys into fixed fixtures or commit records. Automated tests can use temporary directories, temporary keys, and placeholder DIDs. After a test, identity files should be cleaned up, and the request body, exception text, and logs should be checked to ensure that they contain no private-key fields. If an application needs to retain a Local Identity for a long period, the user should choose encrypted storage, an operating-system credential store, or an offline backup method; the SDK should not default to uploading it to a remote service.

\section{Browser Registration and Local Identity Backup Behavior}\label{sec:16.13}

The browser registration page treats the Local Identity as user-controlled material. The user can create and download a timestamped \passthrough{\lstinline!oan-registration-identity.bundle.<timestamp>.json!}, or import a backup; the browser sends only the public DID Document, resource data, and control proof to the Registrar node. The registration-result package must be generated only after submission succeeds and the resource DID has been confirmed.

{\def\LTcaptype{none} 
\begin{longtable}[]{@{}>{\raggedright\arraybackslash}p{0.3200\linewidth}>{\raggedright\arraybackslash}p{0.3200\linewidth}>{\raggedright\arraybackslash}p{0.3200\linewidth}@{}}
\toprule\noalign{}
User action & Browser-side change & Server-side effect \\
\midrule\noalign{}
\endhead
\bottomrule\noalign{}
\endlastfoot
Create and back up & Generate the identity, update the current subject, and trigger a download & None \\
Import from local & Validate and load the backup selected by the user & None \\
Preview DID Document & Generate or reuse a preview based on the current form and identity & No registration submission \\
Submit registration & Validate current input and send public materials and proofs & Creates a registration request \\
Download after success & Package the DID Document, registration VC, and existing Local Identity & Does not create a second registration \\
\end{longtable}
}

\subsection{Browser-Side Key Generation}\label{subsec:16.13.1}

The browser uses Web Crypto or the SDK to generate keys and establish the DID-to-public-key binding. The page must not place private keys in the DOM, URLs, public APIs, or error messages. Refreshing the page or clearing site data may lose an unbacked-up state, so the user should be prompted to back it up promptly.

\subsection{Local Identity Import}\label{subsec:16.13.2}

The import operation reads the JSON file explicitly selected by the user, checks the structure of the version, subject configuration, resource configuration, and private-key materials, and verifies the relationship between the DID Document and public key. A failed import must not overwrite an existing valid identity. After a successful import, the preview and submission use the imported identity and recompute the control proof.

\subsection{Local Backup Download}\label{subsec:16.13.3}

Local backup download uses the browser\textquotesingle s standard download capability. The filename contains \passthrough{\lstinline!oan-registration-identity.bundle!} and the generation time to avoid overwriting an older backup. The server does not participate in file generation; the user should keep the file as a sensitive key backup. Different browsers and mobile operating systems support automatic downloads differently. A failed download must not be marked as a valid backup, and an empty or incomplete file must not be treated as a valid backup.

\subsection{Registration-Result Package Download}\label{subsec:16.13.4}

Registration-result packaging and download are triggered only after the server clearly returns success and the page confirms the resource DID. The archive name replaces the colons in the resource DID with hyphens. It contains the registration VC, DID Document, and existing Local Identity file. An empty VC may remain as an empty item, but no archive may be downloaded after failure, timeout, or a missing DID.

\section{DID, VC, and ResourcePackage Download Formats}\label{sec:16.14}

DID Documents, registration VCs, ResourcePackages, and Local Identity backups use recognizable JSON filenames and stable encodings. Public files must not contain private keys. Importers should verify the DID, version, and hashes from the content fields rather than trusting only the extension or filename.

{\def\LTcaptype{none} 
\begin{longtable}[]{@{}>{\raggedright\arraybackslash}p{0.2400\linewidth}>{\raggedright\arraybackslash}p{0.2400\linewidth}>{\raggedright\arraybackslash}p{0.2400\linewidth}>{\raggedright\arraybackslash}p{0.2400\linewidth}@{}}
\toprule\noalign{}
File & Suggested contents & Sensitivity & Minimum validation \\
\midrule\noalign{}
\endhead
\bottomrule\noalign{}
\endlastfoot
DID Document & DID, public keys, services, and OAN metadata & Public & DID, verification methods, and structure version \\
Registration VC & Issuer, subject, claims, signature, and status references & According to issuer and disclosure policy & Issuer, subject, signature, and validity period \\
ResourcePackage & Resource description, endpoint, capabilities, and version materials & According to resource policy & Resource DID, version, and digest \\
Local Identity backup & Subject and resource-control private-key materials & Highly sensitive & Package version, DID, and public/private-key correspondence \\
\end{longtable}
}

Unknown fields may be retained within a compatible version. Missing required fields, unrecognized structure versions, inconsistent digests, or private keys mixed into a public file must cause import or publication to be rejected.

The following is an \textbf{example ResourcePackage} used to explain field relationships. Its DID, digests, timestamps, and signature values are placeholders and do not represent an online resource. The example shows how a public ResourcePackage can reference the DID Document, metadata digest, ResourcePackage digest, and Root proof at the same time. An actual implementation may also retain extension fields allowed by the protocol type.

The ResourcePackage above can serve as input for public publication and discovery, but it cannot replace the Local Identity backup. The latter also contains subject or resource-control private keys and must be exported, imported, and stored separately. Putting the Local Identity and ResourcePackage into the same registration-result archive only consolidates the download operation; it does not change their sensitivity levels or verification methods. Compatibility decisions for a download format should be based on the content structure. The timestamp and DID in a filename are only for human recognition and avoiding overwrites; they cannot prove authenticity, issuer, or control rights.

\section{References}\label{sec:16.15}

{\def\LTcaptype{none} 
\begin{longtable}[]{@{}>{\raggedright\arraybackslash}p{0.3200\linewidth}>{\raggedright\arraybackslash}p{0.3200\linewidth}>{\raggedright\arraybackslash}p{0.3200\linewidth}@{}}
\toprule\noalign{}
Source & Type & Link \\
\midrule\noalign{}
\endhead
\bottomrule\noalign{}
\endlastfoot
\passthrough{\lstinline!oan-sdk-ts!} & Code repository: DID, identity, ResourcePackage, registration, and discovery client capabilities & \url{https://github.com/OpenAgenet/oan-sdk-ts} \\
\passthrough{\lstinline!oan-community-skill!} & Code repository: community Skill, scripts, and developer workflows & \url{https://github.com/OpenAgenet/oan-community-skill} \\
\passthrough{\lstinline!oan-protocol-commons!} & Code repository: protocol types and error contracts used by the SDK & \url{https://github.com/wolfbrother/oan-protocol-commons} \\
OAN Yellow Paper & arXiv yellow paper: client, resource-publication, and verification flows & \url{https://arxiv.org/abs/2606.03163} \\
OAN White Paper & arXiv white paper: developer ecosystem and open-infrastructure positioning & \url{https://arxiv.org/abs/2606.03161} \\
\end{longtable}
}

\chapter{Security, Privacy, and Abuse Prevention}\label{chap:17}

{\def\LTcaptype{none} 
\begin{longtable}[]{@{}>{\raggedright\arraybackslash}p{0.2400\linewidth}>{\raggedright\arraybackslash}p{0.2400\linewidth}>{\raggedright\arraybackslash}p{0.2400\linewidth}>{\raggedright\arraybackslash}p{0.2400\linewidth}@{}}
\toprule\noalign{}
Security object & Primary attack surface & Primary control point & Signal that must not be mistaken for security \\
\midrule\noalign{}
\endhead
\bottomrule\noalign{}
\endlastfoot
Resource control & Copied DID, forged document, replayed signature & Control key, challenge, version, signature & Same DID string or HTTP 200 \\
Node authorization & Node impersonation, excess privilege, expiration & Node DID, authorization VC, governance state & Reachability or historical success \\
Publication/distribution & Package tampering, forged publication state & Digests, Root proof, batch validation & Registration acceptance or cache hit \\
Discovery/invocation & Index poisoning, malicious endpoint, SSRF & Provenance, lifecycle, endpoint policy, pre-invocation verification & High rank or semantic relevance \\
Availability/privacy & Flooding, exhaustion, credential leakage & Quotas, isolation, redaction, least privilege & One successful health endpoint \\
\end{longtable}
}

This chapter explains how the security boundaries of OpenAgenet (OAN) cover resource control, node authorization, root-platform publication, discovery results, governance state, invocation endpoints, and users\textquotesingle{} local identities. An attacker may send public requests, submit forged or low-quality resources, issue repeated requests, observe public DIDs and VCs, or operate a compromised node. Security controls must state the protected object, attack preconditions, control point, failure result, verification evidence, and recovery action. The trustworthiness of discovery results does not imply the security of the business endpoint.

Security requirements uniformly use the following decision chain. The tables and checklists in the subsequent sections should all be traceable to this chain:

\begin{lstlisting}
Attack preconditions -> Protected object -> Verification or control point -> Failure result -> Audit evidence -> Recovery action
\end{lstlisting}

{\def\LTcaptype{none} 
\begin{longtable}[]{@{}>{\raggedright\arraybackslash}p{0.2400\linewidth}>{\raggedright\arraybackslash}p{0.2400\linewidth}>{\raggedright\arraybackslash}p{0.2400\linewidth}>{\raggedright\arraybackslash}p{0.2400\linewidth}@{}}
\toprule\noalign{}
Security object & Primary attack surface & Primary control point & Signals that must not be misinterpreted as security \\
\midrule\noalign{}
\endhead
\bottomrule\noalign{}
\endlastfoot
Resource control & DID copying, document forgery, signature replay & Control key, challenge, version, and signature & Identical DID string, HTTP 200 \\
Node authorization & Node impersonation, privilege escalation, expired authorization & Node DID, authorization VC, and governance state & Network reachability, historical success \\
Publication and distribution & ResourcePackage tampering, forged publication status & Hash, Root proof, and batch validation & Registration acceptance, cache hit \\
Discovery and invocation & Index poisoning, malicious endpoint, SSRF & Source, lifecycle, endpoint policy, and pre-invocation validation & High ranking, semantic relevance \\
Availability and privacy & Flooding, resource exhaustion, credential leakage & Quotas, isolation, redaction, and least privilege & Success of a single health endpoint \\
\end{longtable}
}

\section{Threat Model and Trust Assumptions}\label{sec:17.1}

The threat model treats the resource provider, Registrar node, Root platform, content distribution platform, Discovery node, off-chain trust indexer, browser, and invocation endpoint separately. The resource-control private key is held by the controlling party, governance keys are used only for governance actions, and the indexer is a read-side projection of on-chain facts. The protocol cannot automatically repair a leaked key, a fully compromised host, or a vulnerability in a business endpoint.

Trust assumptions should be recorded separately by component and material:

\subsection{Resource-Control Threats}\label{subsec:17.1.1}

Resource-control threats include copying a public DID, DID Document, and VC to impersonate the controller, or using a new document to overwrite the same DID. The control points are the verification method, control challenge, signature, and version relationship. An incorrect public key, incorrect challenge, forged signature, or overwrite by an old version must be rejected.

\subsection{Node-Trust Threats}\label{subsec:17.1.2}

Node-trust threats include an unauthorized node impersonating a Registrar node, an authorized node exceeding its privileges, and continued operation after suspension or revocation. The control points are the node DID, node authorization VC, governance events, authorized domain, and capability scope. Network reachability or historical success cannot replace current authorization.

\subsection{Data-Poisoning and Abuse Threats}\label{subsec:17.1.3}

Data poisoning includes batches of fake resources, duplicate DIDs, low-quality descriptions, malicious tags, false endpoints, and forged ranking signals. Control points include control proofs, idempotency rules, lifecycle, independent root-platform verification, tag boundaries, review, and rate limiting. Rejection records should distinguish specific reasons.

\subsection{Availability Threats}\label{subsec:17.1.4}

Availability threats include registration flooding, complex semantic queries, oversized ResourcePackages, concurrent synchronization, database lock contention, and unavailable external dependencies. Control points are size limits, timeouts, concurrency quotas, pagination, query degradation, queue isolation, and backoff. Retries must not be unlimited.

\subsection{Invocation-Endpoint Security Threats}\label{subsec:17.1.5}

Invocation-endpoint threats include malicious URLs, internal-network callbacks, insecure protocols, redirect bypasses, and compromised endpoints. Control points are protocol, domain, resolved address, port, redirect, and authentication validation. When validation fails, high-risk access must be blocked and the redacted reason recorded.

\section{Resource Impersonation and DID Control Protection}\label{sec:17.2}

The control relationship of a resource DID is established by the public key and control proof in the DID Document. A public DID, DID Document, registration VC, or Root proof cannot independently grant update authority. An update to the same DID must use the original control relationship or a control-transfer process permitted by the method; it must not be accepted merely because a new document is internally self-consistent.

The resource-control relationship can be checked item by item as follows:

{\def\LTcaptype{none} 
\begin{longtable}[]{@{}>{\raggedright\arraybackslash}p{0.3200\linewidth}>{\raggedright\arraybackslash}p{0.3200\linewidth}>{\raggedright\arraybackslash}p{0.3200\linewidth}@{}}
\toprule\noalign{}
Check item & Validation & Failure handling \\
\midrule\noalign{}
\endhead
\bottomrule\noalign{}
\endlastfoot
DID and document & The DID declared by the document matches the resource DID in the request & Reject the request and do not enter the publication process \\
Public-key relationship & The verification method matches the control relationship & Mark the control relationship invalid \\
Signature challenge & The signature covers the current request, version, or challenge & Reject old signatures and signatures from another request \\
Version chain & The update declares the preceding version, digest, and control relationship & Do not allow an old version to overwrite the current record \\
State constraints & The resource has not been suspended, revoked, or restricted from updating & Reject or suspend according to governance state \\
\end{longtable}
}

``A signature with a valid format'' alone is insufficient to prove control. The verifier must confirm that the signing public key is declared by the target DID Document and that the signature covers this specific operation.

\section{Protection Against Forged Registration and Unauthorized Updates}\label{sec:17.3}

Registration and update requests separately validate the source, node authorization, control proof, document hash, metadata hash, ResourcePackage hash, version, and idempotency state. An old version, old nonce, or incorrect hash must not overwrite the current record. Tests cover missing proofs, invalid signatures, invalid versions, incorrect hashes, and duplicate submissions.

Rejection conditions for registration and updates should be distinguishable to clients:

{\def\LTcaptype{none} 
\begin{longtable}[]{@{}>{\raggedright\arraybackslash}p{0.3200\linewidth}>{\raggedright\arraybackslash}p{0.3200\linewidth}>{\raggedright\arraybackslash}p{0.3200\linewidth}@{}}
\toprule\noalign{}
Situation & Control failure point & Permitted result \\
\midrule\noalign{}
\endhead
\bottomrule\noalign{}
\endlastfoot
Missing control proof & Resource control has not been established & Reject the registration or update \\
Invalid node authorization & The request source is not eligible for access & Reject and record the authorization reason \\
Invalid version or preceding version & The update chain is discontinuous & Reject the overwrite without changing the current version \\
Inconsistent hash & The document, metadata, or ResourcePackage has been tampered with & Reject publication \\
Retry of a valid request already processed & The idempotency key or request ID already exists & Return a stable result without producing a duplicate side effect \\
\end{longtable}
}

\section{Malicious or Compromised Registrar Nodes}\label{sec:17.4}

The Root platform must not trust only the conclusion forwarded by a Registrar node. It should re-verify the node identity, authorization, control proof, DID Document, ResourcePackage, and request signature. If the node exceeds its authority or the materials have been altered, publication is rejected. A Registrar node\textquotesingle s historical record does not automatically become a root-platform publication fact.

\section{Malicious or Compromised Discovery Nodes}\label{sec:17.5}

Callers should verify the source of the discovery response, the identity of the query node, index version, resource DID, DID Document, version, Root proof, lifecycle, authorized domain, and governance state. High relevance or HTTP success cannot replace verification. If a Discovery node is compromised, its authorization can be revoked and its index can be stopped from being used.

\section{Root-Platform Compromise and Key Protection}\label{sec:17.6}

A compromised Root platform may alter materials pending publication, forge Root proof, block propagation, or expose configuration. Publication keys and databases should be isolated, with least privilege, controlled signing, hashes, auditing, backup validation, and node resynchronization. Recovery must not change resource-control private keys or on-chain facts.

\section{Spam Registration and Resource-Catalog Poisoning}\label{sec:17.7}

Catalog-poisoning governance focuses on resource authenticity, availability, and semantic quality. The registration entry retains the control proof and subject, the Root platform independently verifies and isolates suspicious resources, and Discovery nodes filter by lifecycle, governance state, and publication evidence. Cleanup must distinguish rejection, suspension, revocation, and node-authorization revocation.

Spam-registration governance should not rely on a single score. Before processing, it should combine the control proof, resource type, tag namespace, endpoint verifiability, duplication, source node, and governance state. After processing, it should record the action and propagation scope.

\begin{itemize}
\tightlist
\item[$\square$]
  Apply multidimensional rate limits by subject, node, interface, and time window to batch requests.
\item[$\square$]
  Distinguish network retries, idempotent submissions, and actual version updates for duplicate DIDs.
\item[$\square$]
  Preserve the distinctions among normative tags, aliases, and inferred matches for low-quality or malicious tags.
\item[$\square$]
  Distinguish suspension, revocation, taking offline, and node-authorization revocation for published resources.
\item[$\square$]
  After cleanup, verify convergence of the Root platform, content distribution platform, and Discovery nodes.
\end{itemize}

\subsection{Fake Resource Registration}\label{subsec:17.7.1}

Fake registrations are identified by signals such as missing control proofs, inconsistent DID Documents, unverifiable endpoints, or forged metadata, and are rejected before entering publication. Public errors must not echo private keys or complete signature materials.

\subsection{Duplicate Registration and Version Poisoning}\label{subsec:17.7.2}

Duplicate registration may be a network retry, an update to the same DID, or a malicious submission. The system distinguishes these cases using the resource DID, version, document and ResourcePackage hashes, and request ID. A duplicate valid request returns a stable result, while a real update follows control and version rules.

\subsection{Semantic-Tag Poisoning}\label{subsec:17.7.3}

Tag poisoning includes occupying popular tags, disguising non-normative tags as normative tags, cross-type mismatches, and misleading keywords. Registration validates tag IDs, vocabulary versions, type relationships, and namespaces. Discovery distinguishes normative tags, aliases, original text, and semantically inferred matches.

\subsection{Rate Limiting, Review, and Cleanup}\label{subsec:17.7.4}

Rate limiting, review, and cleanup are applied by source, subject, node DID, resource DID, interface, and risk level, while retaining the reason, operator, and impact scope. After cleanup, convergence of the Root platform, content distribution platform, and discovery index should be verified.

\section{Denial of Service and Resource Exhaustion}\label{sec:17.8}

Registration, discovery, synchronization, health, and governance queries are configured with separate resource budgets. Limits apply to request bodies, batches, decompression expansion, query depth, concurrency, database connections, and external-call time. A semantic backend failure may fall back to a constrained path, but must not relax trusted filtering. Degradation must be marked as incomplete or unconfirmed.

Resource budgets should be isolated by path so that high-cost discovery queries do not bring down registration and health checks:

{\def\LTcaptype{none} 
\begin{longtable}[]{@{}>{\raggedright\arraybackslash}p{0.2400\linewidth}>{\raggedright\arraybackslash}p{0.2400\linewidth}>{\raggedright\arraybackslash}p{0.2400\linewidth}>{\raggedright\arraybackslash}p{0.2400\linewidth}@{}}
\toprule\noalign{}
Path & Primary budget & Action on excess & Security baseline \\
\midrule\noalign{}
\endhead
\bottomrule\noalign{}
\endlastfoot
Registration & Request body, signature-verification count, concurrency, and external-dependency time & Rate-limit, queue, or reject & Do not skip control proof \\
Discovery & Query length, page size, candidate count, and execution time & Truncate, cache, or degrade & Do not bypass governance filtering \\
Node synchronization & Batch size, cursor, parallelism, and retry count & Back off, pause, or reset cursor & Do not repeatedly advance a processed sequence \\
Database/index & Connections, locks, disk, and index growth & Isolate, alert, or enter protective read-only mode & Prioritize verified data that has already been accepted \\
\end{longtable}
}

\pandocbounded{\includegraphics[width=\linewidth,height=0.72\textheight,keepaspectratio]{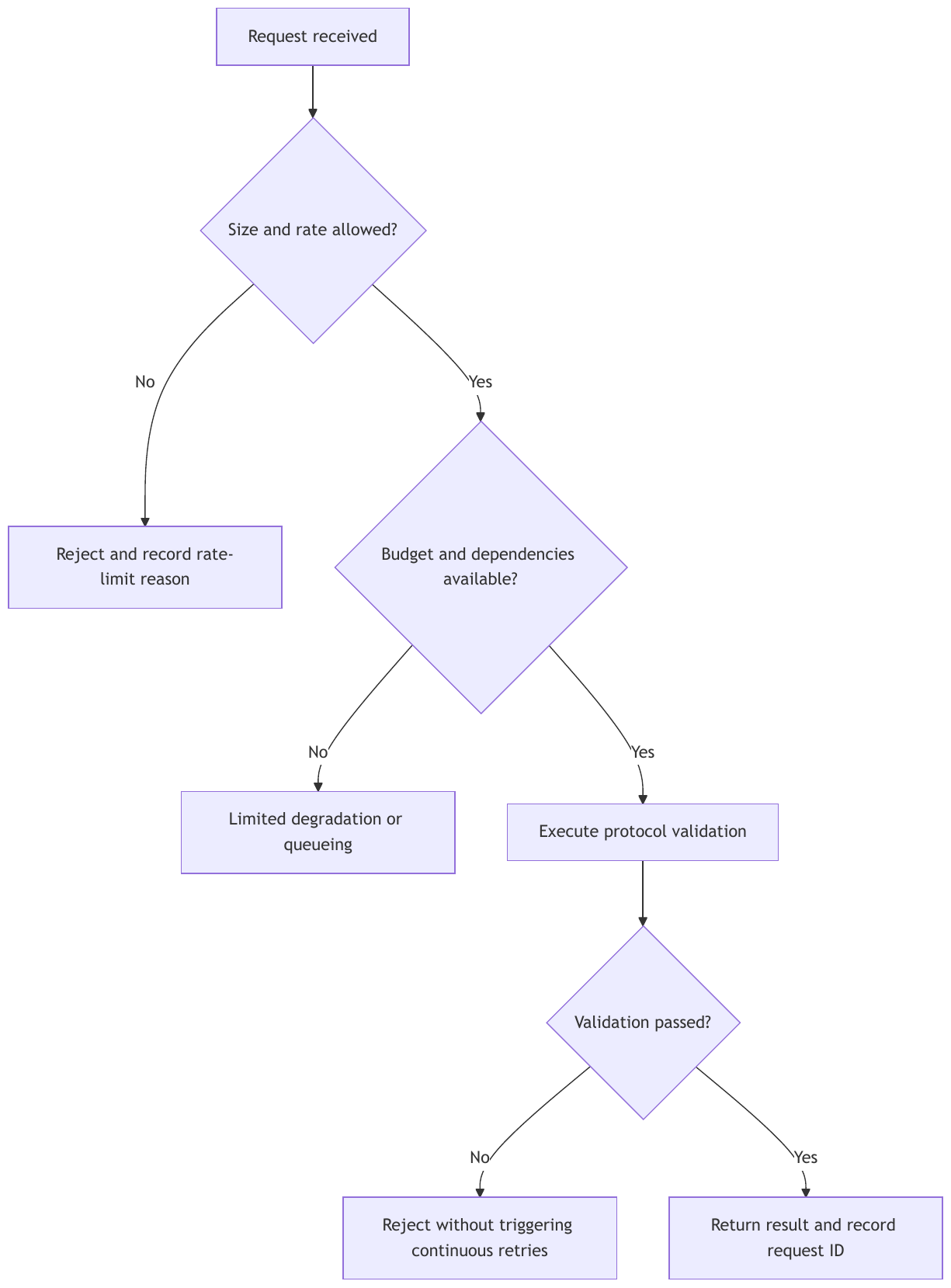}}

\subsection{Registration-Interface Attacks}\label{subsec:17.8.1}

Registration attacks include high-frequency submissions, oversized JSON, duplicate DIDs, expensive signature verification, and dependency calls. The entry point first performs size, rate, idempotency, and format checks, and then performs high-cost validation. Failed requests must not trigger continuous publication retries.

\subsection{Discovery-Query Attacks}\label{subsec:17.8.2}

Discovery attacks include unbounded broad queries, complex semantic retrieval, duplicate queries, and result amplification. Limit input length, page size, candidate count, execution time, and concurrency, and use stable cursors and caching. Degradation must not bypass governance filtering.

\subsection{Inter-Node Synchronization Pressure}\label{subsec:17.8.3}

Synchronization pressure includes large batches, cursor replay, excessive parallelism, retry storms, and unavailable distribution. The synchronizing party uses batch and concurrency limits, exponential backoff, and cursor idempotency. A duplicate batch must not be written or advance state again.

\subsection{Database and Index Resource Exhaustion}\label{subsec:17.8.4}

Database and index exhaustion includes connection-pool depletion, locks, disk exhaustion, and index bloat. Field and batch limits, index maintenance, timeouts, isolated queues, retention policies, and capacity alerts should be provided. When resources are insufficient, valid data and health queries should be protected first.

\section{Replay, Duplicate Submission, and Expired Publication}\label{sec:17.9}

Replay protection distinguishes requests, registrations, publications, synchronization batches, and old caches. Request IDs, nonces, time windows, and signatures prevent a request from being accepted again. DID, version, preceding version, and hashes determine business updates, while cursors and batch sequences ensure idempotent synchronization.

The decision order for replay and duplicate submissions should prioritize idempotency and handle state updates cautiously:

\pandocbounded{\includegraphics[width=\linewidth,height=0.72\textheight,keepaspectratio]{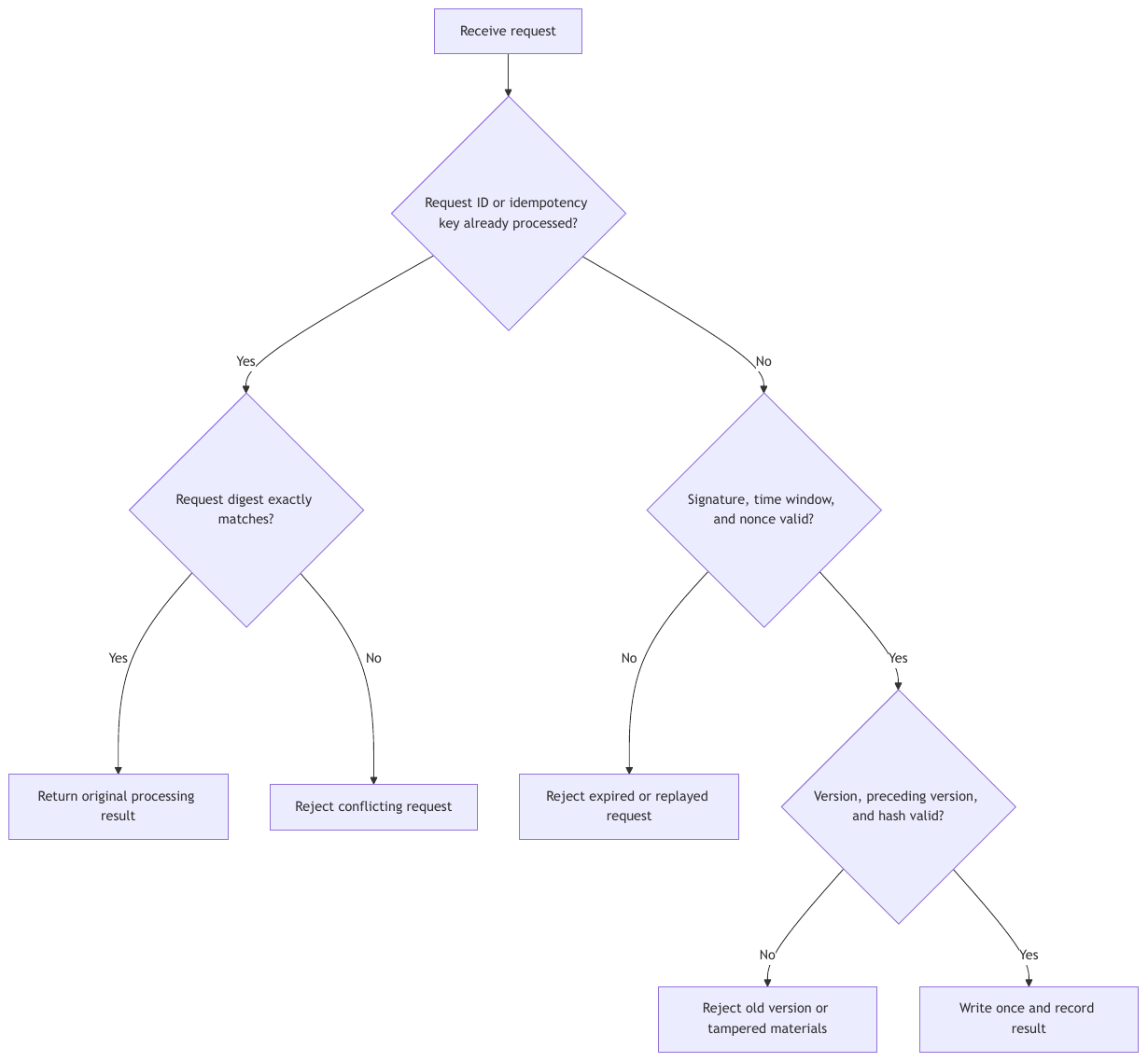}}

\section{Endpoint Abuse and Unsafe Resource Invocation}\label{sec:17.10}

Before invocation, check the endpoint protocol, certificate, domain, resolved address, authentication, version, lifecycle, governance state, and publication evidence. Where necessary, negotiate business permissions, sign requests, and validate responses. OAN does not guarantee that a business endpoint is free of vulnerabilities or malicious behavior.

Before using discovery results, the caller should complete at least the following checks:

\begin{itemize}
\tightlist
\item[$\square$]
  The resource DID, DID Document, public key, and version digest are consistent.
\item[$\square$]
  The Root proof, lifecycle, and governance state remain verifiable.
\item[$\square$]
  The endpoint protocol, domain, certificate, resolved address, port, and redirects comply with local policy.
\item[$\square$]
  The authentication, authorized domain, request signature, and response validation required for invocation are defined.
\item[$\square$]
  The freshness and source of the Discovery node\textquotesingle s result are not mistaken for a guarantee of the business service.
\end{itemize}

\section{Sensitive Metadata and Credential Exposure}\label{sec:17.11}

Private keys, Local Identity backups, internal endpoints, database paths, administrative tokens, unpublished resources, complete signature originals, and log fields that can be associated with individuals are sensitive information. Public interfaces must not return Local Identity backups. Server logs require allowlisting, redaction, and access control.

{\def\LTcaptype{none} 
\begin{longtable}[]{@{}>{\raggedright\arraybackslash}p{0.2400\linewidth}>{\raggedright\arraybackslash}p{0.2400\linewidth}>{\raggedright\arraybackslash}p{0.2400\linewidth}>{\raggedright\arraybackslash}p{0.2400\linewidth}@{}}
\toprule\noalign{}
Information category & May appear in a public response & Log handling & Action after failure \\
\midrule\noalign{}
\endhead
\bottomrule\noalign{}
\endlastfoot
DID, version, and public metadata & According to the resource disclosure policy & May be recorded with a retained correlation ID & Process discovery or publication according to state \\
Registration VC and Root proof & According to the disclosure policy & Record digest and status; the complete sensitive material need not be recorded & Reject use if validation fails \\
Private key and Local Identity backup & No & Must not be recorded & Immediately block and investigate the leakage path \\
Administrative token, database path, and internal endpoint & No & Record only after redaction or allowlisting & Rotate credentials and limit the exposure surface \\
Complete signature original and personally correlated logs & Normally should not be public & Redact, minimize, and restrict access & Delete unnecessary copies and review access \\
\end{longtable}
}

\section{Logging, Auditing, and Incident Response}\label{sec:17.12}

Security logs correlate the request ID, resource DID, node DID, version, event or cursor, source, result, error code, and time, and distinguish on-chain evidence, service logs, and manual records. Log-write permissions should be restricted. Alerts should locate isolation, revocation, rate limiting, compensation, and recovery verification actions.

Incident response should retain the minimum evidence that remains correlatable:

\begin{lstlisting}
request_id / resource_did / node_did / version / event_or_cursor
source / result / error_code / observed_at / operator_action
\end{lstlisting}

\pandocbounded{\includegraphics[width=\linewidth,height=0.72\textheight,keepaspectratio]{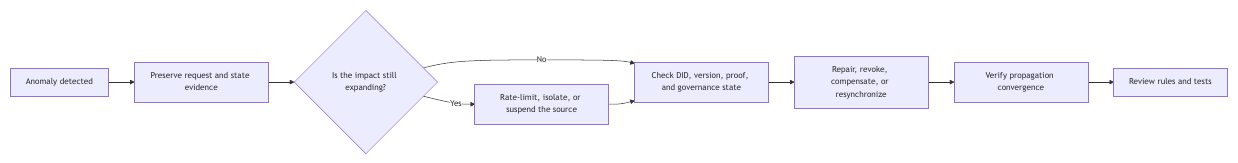}}

\section{Security Requirements for Third-Party Nodes}\label{sec:17.13}

Third-party nodes protect their own hosts, node private keys, authorization VCs, configurations, databases, dependencies, and ports, and connect through the node DID, DID Document, node authorization, and request signature. In an abnormal situation, authorization may be suspended or revoked. A resource provider must not be required to surrender a subject or resource-control private key.

Before a third-party node connects, an auditable minimum inspection record should be created containing the node DID, DID Document, node authorization VC, authorized domain, validity period, public endpoint, software version, key custody method, and revocation contact. A resource provider submits only the resource control proof and should not surrender a subject or resource-control private key.

\section{Endpoint Validation, Callback Security, and SSRF Considerations}\label{sec:17.14}

Before reading or invoking an endpoint, callback, or synchronization address, validate the protocol, host, port, resolved address, redirect, and authentication. Private, loopback, link-local, and cloud-metadata addresses are blocked according to policy. DNS rebinding must be prevented between resolution and connection. Rejection reasons must be auditable without exposing topology.

SSRF protection should cover four points: before resolution, after resolution, before connection, and after each redirect. Checking a URL only at the string level is not sufficient. When rejecting a request, record the policy-match category and request ID, but do not echo internal addresses, network topology, or resolution details.

{\def\LTcaptype{none} 
\begin{longtable}[]{@{}>{\raggedright\arraybackslash}p{0.3200\linewidth}>{\raggedright\arraybackslash}p{0.3200\linewidth}>{\raggedright\arraybackslash}p{0.3200\linewidth}@{}}
\toprule\noalign{}
Stage & Required checks & Failure result \\
\midrule\noalign{}
\endhead
\bottomrule\noalign{}
\endlastfoot
Before resolution & Scheme, host, port, credentials, and encoding & Reject invalid URL \\
After resolution & IPv4/IPv6, loopback, private, link-local, and cloud-metadata addresses & Reject restricted address \\
Before connection & Actual connection target, certificate, timeout, and policy & Fail closed \\
After redirect & Protocol, host, resolved address, and port at every hop & Reject out-of-policy redirect \\
\end{longtable}
}

\subsection{Endpoint Format and Protocol Validation}\label{subsec:17.14.1}

Allow only supported protocols and permitted port ranges. Reject endpoints with a missing host, invalid encoding, embedded credentials, or an incompatible type. HTTPS certificates and host names are validated according to risk. HTTP in a development environment must not become the default public security configuration.

\subsection{Internal and Loopback Address Restrictions}\label{subsec:17.14.2}

Before every connection, check IPv4, IPv6, loopback, private, link-local, reserved, and cloud-metadata addresses. Use standard resolution and network policies rather than relying only on a string blacklist.

\subsection{Callback Request Security}\label{subsec:17.14.3}

Bind callbacks to a one-time request ID, source context, and target. Limit methods, request headers, response size, and timeout. A callback must not change authorization or governance state. Failed retries must be limited.

\subsection{Redirect and DNS-Resolution Risks}\label{subsec:17.14.4}

DNS resolution results, connection targets, and the redirect chain are evaluated as one decision. Revalidate after a redirect to a different host, protocol, or restricted address. Timeouts, certificate changes, or inconsistent resolution should fail closed.

\section{Package and Dependency Supply-Chain Integrity}\label{sec:17.15}

Source code, Cargo/npm dependencies, lock files, build artifacts, and ResourcePackages form the supply-chain boundary. Builds use locked dependencies, controlled sources, hash or signature verification, and traceable commits. ResourcePackages are hashed again before entering the Root platform or a Discovery node.

Supply-chain verification covers the ``source-lock-build-artifact-deployment'' chain:

\begin{enumerate}
\def\labelenumi{\arabic{enumi}.}
\tightlist
\item
  Check the source commit, dependency lock files, and controlled package sources.
\item
  Build Rust, TypeScript, and community Skill artifacts in a reproducible or controlled environment.
\item
  Save a digest or signature for binaries, npm packages, ResourcePackages, and distribution materials.
\item
  Compare artifact digests before deployment, and record the version and rollback materials after deployment.
\item
  If a dependency or artifact is abnormal, stop publication, preserve evidence, and rebuild.
\end{enumerate}

\section{Availability Objectives and Graceful Degradation}\label{sec:17.16}

Security state and availability are expressed separately. An old snapshot may support a read-only result with a freshness marker, but must not approve registration, publication, authorization, or high-risk invocation. Dependency failure, index lag, unknown governance state, and resource exhaustion should have different states and recovery actions.

{\def\LTcaptype{none} 
\begin{longtable}[]{@{}>{\raggedright\arraybackslash}p{0.2400\linewidth}>{\raggedright\arraybackslash}p{0.2400\linewidth}>{\raggedright\arraybackslash}p{0.2400\linewidth}>{\raggedright\arraybackslash}p{0.2400\linewidth}@{}}
\toprule\noalign{}
State & Content that may be returned & Actions that must be prohibited & Facts that must be marked \\
\midrule\noalign{}
\endhead
\bottomrule\noalign{}
\endlastfoot
Normal and fresh & Current query, state, and proof & No additional restriction & Source and time \\
Old snapshot available & Read-only result with freshness & Registration, publication, authorization, and high-risk invocation & Snapshot time and lag \\
Index lagging & Confirmed partial facts & Treating unsynchronized data as nonexistent or current & Current sequence and target sequence \\
Governance unknown & Limited query or error state & Approving invocation through an unknown state & Reason for unknown state \\
Dependency unavailable & Health error or restricted cache & Unlimited retries or relaxed trusted filtering & Dependency and recovery state \\
\end{longtable}
}

The acceptance focus for graceful degradation is not merely that ``the page still has content.'' Users and callers must be able to see the freshness, source, and uncertainty of the content, and old data must not automatically execute registration, publication, authorization, or high-risk business invocation.

\section{References}\label{sec:17.17}

{\def\LTcaptype{none} 
\begin{longtable}[]{@{}>{\raggedright\arraybackslash}p{0.3200\linewidth}>{\raggedright\arraybackslash}p{0.3200\linewidth}>{\raggedright\arraybackslash}p{0.3200\linewidth}@{}}
\toprule\noalign{}
Source & Type & Link \\
\midrule\noalign{}
\endhead
\bottomrule\noalign{}
\endlastfoot
\passthrough{\lstinline!oan-protocol-commons!} & Code repository: signatures, hashes, credentials, and secure request structures & \url{https://github.com/wolfbrother/oan-protocol-commons} \\
\passthrough{\lstinline!oan-root-nodes!} & Code repository: root-platform verification, publication, and node authorization & \url{https://github.com/OpenAgenet/oan-root-nodes} \\
\passthrough{\lstinline!oan-registrar-nodes!} & Code repository: registration entry point, control proofs, and abuse controls & \url{https://github.com/OpenAgenet/oan-registrar-nodes} \\
\passthrough{\lstinline!oan-discovery-nodes!} & Code repository: queries, indexing, and result security boundaries & \url{https://github.com/OpenAgenet/oan-discovery-nodes} \\
\passthrough{\lstinline!oan-trust-indexer!} & Code repository: governance state and revocation-state projections & \url{https://github.com/OpenAgenet/oan-trust-indexer} \\
OAN Yellow Paper & arXiv yellow paper: security, privacy, and trust boundaries & \url{https://arxiv.org/abs/2606.03163} \\
\end{longtable}
}

\chapter{Resource Lifecycle, Governance, and Version Management}\label{chap:18}

This chapter specifies the temporal, state, and evidence relationships of resources in the OpenAgenet (OAN) network. Resources such as Agent Services, Skill packages, MCP services, and Tool APIs use DIDs to identify subjects, use DID Documents to describe control relationships, capabilities, and service endpoints, and use resource packages to carry publishable content. A registration VC proves that a Registrar node has accepted a registration, while a Root proof proves that a specific resource package has passed Root platform verification and entered the publication pipeline. ``Registered'', ``published'', ``indexed'', and ``currently callable'' are different facts and must be proved separately.

The state relationships in this chapter can be understood by using a resource version as the association key:

\begin{lstlisting}
Resource DID
  -> DID Document version
  -> Resource package version + document/metadata/package hashes
  -> Registration VC
  -> Root proof
  -> Distribution cursor and discovery index version
\end{lstlisting}

The chain above represents the association between pieces of evidence. It does not mean that all steps complete synchronously, nor does it mean that a response from a Registrar node means that the resource is already visible to a Discovery node.

\section{Resource Lifecycle State Machine}\label{sec:18.1}

{\def\LTcaptype{none} 
\begin{longtable}[]{@{}>{\raggedright\arraybackslash}p{0.3200\linewidth}>{\raggedright\arraybackslash}p{0.3200\linewidth}>{\raggedright\arraybackslash}p{0.3200\linewidth}@{}}
\toprule\noalign{}
State & Factual meaning & Default discovery behavior \\
\midrule\noalign{}
\endhead
\bottomrule\noalign{}
\endlastfoot
Draft & Formed only in the publisher\textquotesingle s local environment & Not returned \\
Registered & Accepted and recorded by a Registrar node & Does not yet represent network-wide visibility \\
Pending publication & Waiting for Root platform verification or a propagation task & Processing status may be queried \\
Published & Root platform verification has completed and publication evidence has been produced & Eligible for indexing \\
Discoverable & Verified and indexed by a Discovery node & May be returned to the querying party \\
Suspended, revoked, deactivated & Temporarily unavailable, explicitly withdrawn, or terminated by the subject & Not returned as a valid resource by default \\
\end{longtable}
}

States must be triggered by events, proofs, or persistent records. They cannot be inferred from a page button or a single successful HTTP response. Registration success indicates, at most, that the Registrar node has completed acceptance. The resource can enter the Published state only after Root platform verification, authorization checks, resource package digest verification, and completion of the publication task. It can enter the Discoverable state only after a Discovery node has completed verification and indexing. Failures must preserve the reason, associated identifiers, and retry boundaries; they must not all be described simply as completion of network-wide registration.

\section{Initial Registration}\label{sec:18.2}

The publisher creates or imports a local identity in the browser and uses a local private key to sign the DID Document and registration request; the private key should not be uploaded to the Registrar node. The Registrar node verifies the format, control relationship, resource type, required metadata, and node authorization. After successful acceptance, it records the registration and returns a registration VC according to the interface contract. The Root platform then independently verifies the resource package, Registrar node authorization, and publication conditions. The content distribution platform and Discovery nodes subsequently receive and index the content asynchronously. Missing local identity, invalid signatures, missing fields, network failures, an empty VC, and propagation delays must be reported separately and must not be summarized as completion of network-wide registration.

{\def\LTcaptype{none} 
\begin{longtable}[]{@{}>{\raggedright\arraybackslash}p{0.2400\linewidth}>{\raggedright\arraybackslash}p{0.2400\linewidth}>{\raggedright\arraybackslash}p{0.2400\linewidth}>{\raggedright\arraybackslash}p{0.2400\linewidth}@{}}
\toprule\noalign{}
Stage & Publisher-side local action & Network-side fact & Evidence that can be verified \\
\midrule\noalign{}
\endhead
\bottomrule\noalign{}
\endlastfoot
Generation & Create or import a local identity and generate a DID Document & Not yet registered & Local identity backup, DID Document, and control signature \\
Acceptance & Send the signed request & Registrar node records the registration & Response, request identifier, and registration VC (if any) \\
Publication & Wait for asynchronous processing & Root platform forms a publication fact & Root proof \\
Discoverability & Query the resource & Discovery node completes indexing & Query result, source, and indexing time \\
\end{longtable}
}

\section{Updating a Resource with the Same DID}\label{sec:18.3}

The basis for an update with the same DID is continuity of the control relationship, not equality of the DID string. An update request should include the preceding version, resource DID, version, and content digest, and should be signed by the existing control public key or an explicitly authorized update key. The Registrar node verifies control authority and the version relationship. The Root platform re-verifies the resource package and publication authorization. A Discovery node accepts only versions that have passed through the Root platform publication process. Repeated submission of identical content and the same preceding version should be handled idempotently; a content change must advance the version and digest. Requests without authorization, with a version rollback, or with a fork must be rejected. Before the new version has completed propagation, the old valid version may continue to be returned.

{\def\LTcaptype{none} 
\begin{longtable}[]{@{}>{\raggedright\arraybackslash}p{0.2400\linewidth}>{\raggedright\arraybackslash}p{0.2400\linewidth}>{\raggedright\arraybackslash}p{0.2400\linewidth}>{\raggedright\arraybackslash}p{0.2400\linewidth}@{}}
\toprule\noalign{}
Update scenario & Control proof & Version relationship & Processing result \\
\midrule\noalign{}
\endhead
\bottomrule\noalign{}
\endlastfoot
Same DID, same digest, same preceding version & Valid & Duplicate & Return idempotently without creating a new valid version \\
Same DID, new digest & Valid & Advancing & Re-verify, publish, and establish a new version \\
Same DID, new digest & Invalid or unauthorized & Any & Reject without overwriting the old version \\
Same DID, old version & Valid & Rollback & Reject or mark as a version conflict \\
\end{longtable}
}

\section{DID Document Version Management}\label{sec:18.4}

DID Document versions express changes to the identity document, control keys, service endpoints, or capability metadata. They are not equivalent to the \passthrough{\lstinline!did:oan!} method version or the resource package version. The current version should be determined jointly by the stable DID, the predecessor relationship, a sequence or equivalent version rule, and the controller\textquotesingle s signature; timestamps are for audit purposes only. A resolver must not resolve a fork at the same predecessor version solely according to arrival time. By default, a Discovery node returns the highest-sequence version that is published and has a valid state, and supports querying historical versions by version or content digest.

Resource package versioning and interface compatibility are different dimensions. A Skill package may update only its documentation or examples, while a Tool API may change its input/output schema and endpoint at the same time. An MCP service may also change its protocol binding or exposed capabilities. Therefore, a version number cannot replace a change description. The \passthrough{\lstinline!packageVersion!}, \passthrough{\lstinline!packageHash!}, \passthrough{\lstinline!metadataHash!}, and \passthrough{\lstinline!lifecycleState!} in a resource package should be consistent with the corresponding values in the DID Document metadata and the Root proof.

A minimal binding record for audit purposes can be represented as follows:

The DID, version, and digest values above are illustrative placeholders. They express the binding relationship between fields, not an official resource or fixed version.

\section{Resource Package Version Management}\label{sec:18.5}

A resource package is the publication unit verified by the Root platform, stored by the content distribution platform, and indexed by Discovery nodes. A version must point to a set of normalized metadata, entry points, protocol bindings, dependencies, and content digests. The Root proof should bind the resource DID, resource type, resource package version, and package digest. A resource DID may remain unchanged during a normal upgrade. An old resource package may be marked as superseded while its audit evidence is retained. By default, a Discovery node returns the latest valid package; when a caller requests a specific version, its publication proof and credential status must still be verified.

The update target can be selected according to the scope of the change. If only the published package content changes, at least the resource package version and package hash must be updated. If service endpoints, public keys, or resource metadata in the DID Document change, the DID Document and its digest should be updated together. If capability tags, authorization domains, or lifecycle state change, semantic governance and visibility must also be checked again. Any change that affects a field bound by the Root proof cannot be made only by modifying a local index at a Discovery node.

\section{Metadata and Endpoint Changes}\label{sec:18.6}

Changes to capability tags, descriptions, use cases, protocol bindings, endpoints, authentication requirements, and dependencies that affect discovery or invocation decisions should produce a new DID Document version, a new resource package version, or a combination of both, followed by re-signing, verification, and indexing. An endpoint change must also check TLS, authentication, protocol compatibility, and actual capabilities. Taking an old endpoint offline does not mean that the resource DID has been revoked; the caller should decide whether to invoke the resource according to the new version and state.

\section{Registration VC Lifecycle}\label{sec:18.7}

A registration VC proves that a Registrar node accepted a registration. It is not a guarantee of resource quality and does not replace the Root proof. A verifier should check the issuer\textquotesingle s authorization, subject DID, resource type, version or content digest, issuance time, validity period, and revocation status. When the VC is missing or invalid, the issuer is unauthorized, or the status is unknown, the resource must not be marked as a trusted registration. Credential revocation affects only the fact proved by that credential and does not automatically rewrite the DID Document. Deactivation of a DID does not mean that historical credentials disappear.

When processing credentials, the querying party should record ``credential exists'' and ``credential is currently valid'' separately. Even if the registration VC for resource version \passthrough{\lstinline!v1!} has a valid signature, it cannot directly prove that \passthrough{\lstinline!v2!} has been registered. Likewise, the Root proof for \passthrough{\lstinline!v2!} cannot replace the Registrar node\textquotesingle s evidence for the registration fact. After a resource update, credential revocation, or subject deactivation, a Discovery node should filter results according to the latest verifiable state while retaining historical credentials as audit material.

\section{Resource Suspension and Recovery}\label{sec:18.8}

Suspension is a reversible state. Its causes should distinguish controller action, governance restrictions, temporarily unverifiable evidence, endpoint maintenance, and node synchronization failures. A network timeout or missing metrics must not directly change the resource lifecycle state. Suspension should have a control signature, governance event, or other verifiable evidence. Before recovery, the control proof, VC, Root proof, endpoint, and governance conditions must be checked again, and the Discovery node must resynchronize according to the publication sequence rather than merely changing its local state.

\section{Resource Revocation and Deactivation}\label{sec:18.9}

Resource revocation targets publication eligibility or a version. DID deactivation targets the resource subject. Credential revocation targets a particular proof. Endpoint withdrawal targets a service entry point. Revocation and deactivation should record the initiator, basis, target DID or version, reason, effective time, and propagation scope. Historical materials may be retained for audit, but must not be returned as currently callable resources. Before propagation converges, a node must label the state and freshness instead of claiming that revocation has taken effect across the entire network immediately.

The handling of node authorization changes and resource states can be checked using the following relationships:

{\def\LTcaptype{none} 
\begin{longtable}[]{@{}>{\raggedright\arraybackslash}p{0.3200\linewidth}>{\raggedright\arraybackslash}p{0.3200\linewidth}>{\raggedright\arraybackslash}p{0.3200\linewidth}@{}}
\toprule\noalign{}
Observed fact & What it can indicate & What it cannot indicate directly \\
\midrule\noalign{}
\endhead
\bottomrule\noalign{}
\endlastfoot
Registrar node authorization revoked & The node no longer has authorization to accept new trusted publication inputs & Already published resources are automatically revoked \\
Discovery node suspended & Queries or indexes from the node require degraded handling & Resource control has become invalid \\
Resource version revoked & The specified resource or version should no longer be treated as a valid result & Other versions are necessarily revoked \\
Node recovered and synchronization completed & The node has regained the corresponding operating conditions & All historical missing data has been restored automatically \\
\end{longtable}
}

\section{Node Suspension and Resource Visibility}\label{sec:18.10}

Node authorization is a governance object, while resource control belongs to the resource DID. After a Registrar node loses authorization, the Root platform should stop accepting new trusted publication inputs from it and re-evaluate existing publications according to governance rules. After a Discovery node loses authorization, it should not continue publishing results as a normally authorized node, although its data may be retained for recovery and audit. Node recovery must first restore identity and authorization and then use the cursor to fill in missing data. Clients should distinguish node failure, synchronization in progress, and resource revocation.

Version selection is not simply ``last write wins''. For example, if a Discovery node receives \passthrough{\lstinline!1.1.0!} and \passthrough{\lstinline!1.2.0!} for the same resource at the same time, it should first confirm that both have verifiable publication sources and valid states, and then select the current result according to the publication sequence, predecessor relationship, and version rules. If the digest or proof for \passthrough{\lstinline!1.2.0!} fails verification, it cannot overwrite \passthrough{\lstinline!1.1.0!} merely because its version number is larger. The selection result should also retain the adopted version and the reasons for excluding other versions, so that propagation delays and version conflicts can be diagnosed.

\section{Version Selection by a Discovery Node}\label{sec:18.11}

A Discovery node first filters by state, signatures, registration VC, Root proof, version validity, and index freshness, and then performs semantic and structured ranking. By default, for the same DID it returns the version that is published, has a valid state, and has the highest publication sequence. When there is a version conflict, a sequence gap, or unknown proof status, it should report an anomaly instead of selecting randomly. Both exact DID queries and semantic queries should return the version, digest, source, evidence status, and indexing time.

\section{Historical Versions and Auditability}\label{sec:18.12}

Audit records should associate the DID, document version, resource package version, digest, control signature, registration VC, Root proof, publication sequence, node receipt time, and indexing time. Historical queries should distinguish current valid, superseded, revoked, archived, and not-yet-synchronized-by-the-node states. Historical evidence is used to reconstruct publication facts and must not bypass current-state checks. Node reconstruction or database migration must not silently discard evidence that the specification requires to be retained.

Historical records should be able to answer at least three questions: who controlled and signed a particular version; whether the Root platform accepted and published that version; and when and in what state a specified Discovery node observed that version. If only a resource package file can be found, without a control signature, publication proof, or index observation record, it proves only that the material was once stored. It does not prove that the material was once a valid version in the network.

\section{Resource VC and Publication Evidence Lifecycle}\label{sec:18.13}

Registration VCs and Root proofs have different issuers and factual scopes. The former proves registration acceptance, while the latter proves that a specific resource package passed Root platform verification and entered the publication pipeline. Both must bind an explicit DID, version, and digest. When an update produces a new digest, the corresponding evidence should be produced again. Revocation, deactivation, expiration, and authorization loss must be checked separately. A proof file downloaded to the publisher\textquotesingle s local environment does not change the server-side validity judgment.

The evidence association can be checked item by item by version:

\begin{lstlisting}
Same resource DID
  -> Same resource package version
  -> Corresponding DID Document/metadata/package digest
  -> Registration fact in the registration VC
  -> Root proof
  -> Receipt and indexing observation by a Discovery node
\end{lstlisting}

If any link points to a different DID, version, or digest, it should be treated as a binding inconsistency. File names or download times must not be used to infer that the evidence belongs to the same publication.

\section{Update Propagation and Stale Index Convergence}\label{sec:18.14}

\begin{lstlisting}
Registrar node -> Root platform verification and publication -> Content distribution platform -> Discovery node receipt -> Verification, indexing, and visibility
\end{lstlisting}

Each segment is associated using the DID, version, digest, publication sequence, or event cursor. Repeated delivery must be idempotent, and interrupted retries must continue from the confirmed cursor. When there is a sequence gap, digest mismatch, invalid proof, or unknown governance state, a Discovery node must not update its valid projection. Convergence of an old index is determined by completion of verification and projection of the new version, not by the fact that a request has been sent.

{\def\LTcaptype{none} 
\begin{longtable}[]{@{}>{\raggedright\arraybackslash}p{0.4800\linewidth}>{\raggedright\arraybackslash}p{0.4800\linewidth}@{}}
\toprule\noalign{}
Convergence check & Passing condition \\
\midrule\noalign{}
\endhead
\bottomrule\noalign{}
\endlastfoot
Content & The resource package can be obtained and its digest is consistent \\
Evidence & The target DID, version, and digest in the publication proof are consistent \\
Ordering & The publication sequence or synchronization cursor has no unexplained gap \\
State & The lifecycle state of the new version has been re-evaluated \\
Query & Exact and semantic queries no longer return the old version as valid \\
\end{longtable}
}

Propagation checks should record the state observed by each node rather than comparing only resource counts. Equal counts may mean that an old version is still indexed, or that a new version has replaced the old one. The resource DID, version, digest, lifecycle state, publication sequence, and observation time must be compared. When an anomaly occurs, retain the last successful cursor and the original response so that missing or out-of-order events are not concealed by resetting the index.

\section{Resource Ownership Transfer Strategy}\label{sec:18.15}

An ownership transfer must be authorized by the former controller, confirmed by the new controller, and made effective after verification by the Root platform or a governance process. Resubmitting the same DID or presenting a new registration VC is not sufficient to prove a transfer. Operational maintenance authority and publication delegation may be authorized separately, but their scope, duration, and revocation method should be recorded separately from control authority. If the original private key is lost, control must not be forcibly taken over by bypassing signatures; an auditable governance recovery process should be used. Where the current implementation does not provide a complete transfer interface, this section is a protocol design requirement.

Before the control transfer is completed, the former controller remains the key source of evidence for resource updates and revocation decisions. A statement from the new controller, operational authorization, or resource package signature cannot independently override the old control relationship. After the transfer is completed, maintenance records should retain the former controller\textquotesingle s authorization, the new controller\textquotesingle s confirmation, the effective fact established by the Root platform or governance process, and the observation that the Discovery node has completed the update. This distinguishes ``transfer proposed'', ``transfer approved'', and ``nodes converged''.

\section{References}\label{sec:18.16}

{\def\LTcaptype{none} 
\begin{longtable}[]{@{}>{\raggedright\arraybackslash}p{0.3200\linewidth}>{\raggedright\arraybackslash}p{0.3200\linewidth}>{\raggedright\arraybackslash}p{0.3200\linewidth}@{}}
\toprule\noalign{}
Source & Type & Link \\
\midrule\noalign{}
\endhead
\bottomrule\noalign{}
\endlastfoot
\passthrough{\lstinline!oan-root-nodes!} & Code repository: publication status, versions, and Root proofs & \url{https://github.com/OpenAgenet/oan-root-nodes} \\
\passthrough{\lstinline!oan-registrar-nodes!} & Code repository: registration, updates, and registration credentials & \url{https://github.com/OpenAgenet/oan-registrar-nodes} \\
\passthrough{\lstinline!oan-discovery-nodes!} & Code repository: index versions, lifecycle, and discovery status & \url{https://github.com/OpenAgenet/oan-discovery-nodes} \\
\passthrough{\lstinline!oan-trust-indexer!} & Code repository: governance, suspension, and revocation status & \url{https://github.com/OpenAgenet/oan-trust-indexer} \\
\passthrough{\lstinline!oan-protocol-commons!} & Code repository: shared structures for versions, lifecycle, and status & \url{https://github.com/wolfbrother/oan-protocol-commons} \\
OAN Yellow Paper & arXiv yellow paper: resource lifecycle, versioning, and governance mechanisms & \url{https://arxiv.org/abs/2606.03163} \\
\passthrough{\lstinline!did:oan!} DID Method Specification & Standard/method specification: resource identity and update control & \url{https://github.com/OpenAgenet/oan-public-docs/blob/main/did-oan-specs/doc/OAN\%20DID\%20Method\%20Specification.md} \\
\end{longtable}
}

\chapter{Observability, Metrics, and Operations}\label{chap:19}

This chapter explains how observability answers, at the same time, whether a service is alive, whether business functions are available, whether data is fresh, and whether an anomaly is converging. The Root platform, Registrar nodes, Discovery nodes, content distribution platform, off-chain trust indexer, database, and official website backend each produce operational and business evidence. The website may use an old snapshot to improve the user experience, but an old snapshot cannot replace the actual backend health status. Metrics must include their source, time, status, and latency; unknown values must not be treated as zero.

Observational evidence is divided into four layers: raw facts defined by service interfaces or source code, operational metrics aggregated by the backend, display data used by official website pages, and smoke tests and alert records produced by operational procedures. When recording a metric, at least \passthrough{\lstinline!source!}, \passthrough{\lstinline!observedAt!}, \passthrough{\lstinline!dataStatus!}, and \passthrough{\lstinline!lag!} should be retained together so that ``the interface has no data,'' ``the data is zero,'' and ``an old snapshot is being displayed'' can be distinguished. A display result on the official website only shows that the frontend obtained a response at a certain time; it cannot replace independent verification of a node, database, or off-chain trust indexer.

{\def\LTcaptype{none} 
\begin{longtable}[]{@{}>{\raggedright\arraybackslash}p{0.3200\linewidth}>{\raggedright\arraybackslash}p{0.3200\linewidth}>{\raggedright\arraybackslash}p{0.3200\linewidth}@{}}
\toprule\noalign{}
Observation layer & Core question & Typical evidence \\
\midrule\noalign{}
\endhead
\bottomrule\noalign{}
\endlastfoot
Liveness & Can the process respond? & liveness response, process status \\
Readiness & Are dependencies and configuration satisfied? & readiness, database and upstream status \\
Business & Are registration, publication, and discovery available? & smoke test request, response, and correlation identifiers \\
Data & Are metrics and indexes fresh? & time window, cursor, latest sequence \\
Operations & Has the failure been detected and converged? & logs, alerts, recovery records \\
\end{longtable}
}

\section{Service Health and Readiness Model}\label{sec:19.1}

The checking sequence should progress from ``can it respond?'' to ``can it complete the business operation?'' First verify the process and \passthrough{\lstinline!/health!}, then verify readiness and the database connection, and finally check the dependencies among the Root platform, Registrar nodes, Discovery nodes, content distribution platform, and off-chain trust indexer. A 200 response from an interface can only serve as evidence for that checkpoint. A service may be marked as business-available only after dependency checks, data status, and registration/discovery smoke tests have all passed. If \passthrough{\lstinline!summary!} continues to fail, the backend should remain in a failed state even if the page can still display an old snapshot.

Liveness only proves that the process can respond; readiness should also check necessary dependencies and configuration; business smoke tests verify real paths such as registration, publication, and discovery. A health check should record the target, response code, duration, dependency status, and time. A connectable database does not mean that the indexer has caught up with upstream events, and a responsive summary interface does not mean that the statistics are complete. Recovery confirmation requires at least consecutive successful health checks together with the corresponding business smoke tests.

{\def\LTcaptype{none} 
\begin{longtable}[]{@{}>{\raggedright\arraybackslash}p{0.3200\linewidth}>{\raggedright\arraybackslash}p{0.3200\linewidth}>{\raggedright\arraybackslash}p{0.3200\linewidth}@{}}
\toprule\noalign{}
Status & What it proves & What it does not prove \\
\midrule\noalign{}
\endhead
\bottomrule\noalign{}
\endlastfoot
liveness & The process can respond to the check & Dependencies are normal or business functions are available \\
readiness & Configuration and necessary dependencies are ready & Network-wide publication and indexing are complete \\
Business smoke test passed & The specified path is available under the test conditions & All resources and all loads are normal \\
Data expired & Recent data was not updated in time & The resource has been revoked \\
Metric missing & The metric cannot currently be obtained & The metric value is zero \\
\end{longtable}
}

\section{Registration Metrics}\label{sec:19.2}

Registration metrics should establish a traceable chain around each resource registration. It is recommended that \passthrough{\lstinline!requestId!} be bound to the registration request, \passthrough{\lstinline!resourceDid!} and the resource version or content digest be bound to the registration fact, and the publication sequence, distribution batch, and index version be bound to subsequent stages. Client-side repeated clicks or network retries may create multiple request records, but must not cause the same resource version to be counted as multiple successful registrations. When the VC is empty, the returned result should be recorded truthfully; an empty value must not be interpreted as a valid credential. Periodic statistics should be calculated in the aggregation layer using a unified time zone; the frontend must not accumulate counts based on page refreshes.

Registration metrics should distinguish request receipt, validation rejection, successful registration, VC return, Root platform acceptance, publication completion, and discovery visibility. Deduplication should use a request or resource-version correlation identifier rather than performing rough deduplication by IP address. Production statistics should exclude smoke tests that are explicitly marked as such. TODAY, THIS WEEK, and THIS MONTH must use a unified time zone and half-open time intervals; periodic values must not be confused with cumulative values because of caching or degradation.

{\def\LTcaptype{none} 
\begin{longtable}[]{@{}>{\raggedright\arraybackslash}p{0.2400\linewidth}>{\raggedright\arraybackslash}p{0.2400\linewidth}>{\raggedright\arraybackslash}p{0.2400\linewidth}>{\raggedright\arraybackslash}p{0.2400\linewidth}@{}}
\toprule\noalign{}
Metric & Counting unit & Correlation key & Statistical considerations \\
\midrule\noalign{}
\endhead
\bottomrule\noalign{}
\endlastfoot
Registration requests & Request or registration attempt & requestId & Distinguish retries from duplicate submissions \\
Successful registrations & Resource-version registration & resourceDid + version & Does not equal Root platform publication \\
VC returned & Returned credential result & resourceDid + credentialId & An empty VC is not counted as a valid credential \\
Root platform publication & Publication fact or publication batch & resourceDid + sequence & Asynchronous status is counted separately \\
Discoverable & Version queryable in the index & resourceDid + indexVersion & Must be combined with freshness \\
\end{longtable}
}

Registration-side metrics separately record client submissions, Registrar admission, and Root authoritative receipts. A successful client response must not be treated directly as a Root receipt. Metrics are deduplicated by resource DID and version so that retries do not inflate resource totals.

\section{Publication and Propagation Metrics}\label{sec:19.3}

The asynchronous path is sampled as three independent loops. The Root authoritative deduplicated receipt count represents completed work in the registration-request loop; the content distribution platform\textquotesingle s deduplicated persistence-confirmation count represents completed work in the publication loop; and the deduplicated validated-receipt count of each discovery target represents completed work in that target\textquotesingle s synchronization loop. Counts for multiple Discovery nodes are plotted separately and must not be summed before comparison with Root. A failed sample is recorded as a missing value with an error reason, not as zero.

{\def\LTcaptype{none} 
\begin{longtable}[]{@{}>{\raggedright\arraybackslash}p{0.2400\linewidth}>{\raggedright\arraybackslash}p{0.2400\linewidth}>{\raggedright\arraybackslash}p{0.2400\linewidth}>{\raggedright\arraybackslash}p{0.2400\linewidth}@{}}
\toprule\noalign{}
Stage & Start time & Completion time & Main evidence \\
\midrule\noalign{}
\endhead
\bottomrule\noalign{}
\endlastfoot
Registration request & Client starts submission & Root authoritative version is written & Resource DID, version, Root latest-version record \\
Root publication loop & Database publication task becomes ready & CDN confirms persistence and marks completion & Task, publication cursor, ResourcePackage digest \\
Content distribution & Distribution batch is sent & Content can be retrieved and verified & Batch, cursor, content digest \\
Per-target synchronization loop & Discovery receives an authorized digest or starts polling & That target validates the resource and advances its cursor & Target DID, synchronization cursor, indexing time \\
\end{longtable}
}

Propagation metrics carry the resource DID, version, content digest, and publication cursor through all stages, and separately record Root authoritative admission, CDN persistence confirmation, and validated receipt by each Discovery target. The three benchmark progress series count deduplicated resources at Root, at CDN, and at each Discovery node. Multiple Discovery counts must not be summed before comparison with Root. A failed sample is \passthrough{\lstinline!null!} or a gap, not zero. End-to-end latency is the time at which the final required target reaches the expected value minus the test start time; sampling can estimate that completion only to the sampling interval.

\pandocbounded{\includegraphics[width=\linewidth,height=0.72\textheight,keepaspectratio]{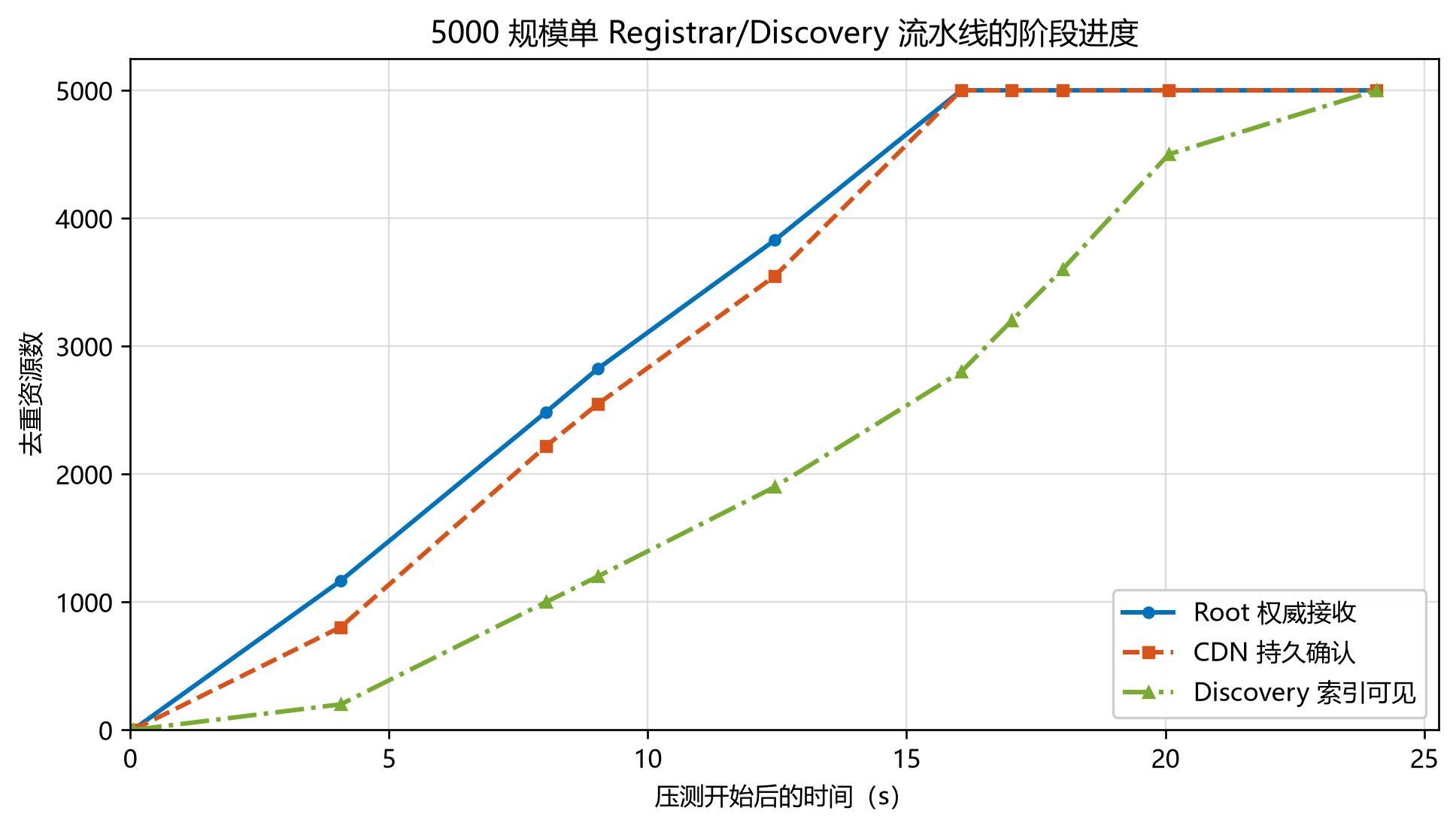}}

The data in the figure comes from repetition 1 of the 5000-resource \passthrough{\lstinline!pipeline-benchmark!} in the formal local benchmark of 2026-11-11. It used four registration processes, total concurrency 48, and a one-second target sampling interval. The curves show deduplicated cumulative counts at each stage rather than successful client responses: Root and CDN reached 5000 at approximately 16 seconds, while Discovery reached 5000 at approximately 24 seconds. In this run, the tail backlog was therefore in the single discovery target\textquotesingle s fetch, validation, and indexing path. Sampling requests themselves take time, so the horizontal axis records actual sample-completion time and completion can be estimated only at sampling precision. The editable plotting script, input CSV, PNG, and PDF are stored in \passthrough{\lstinline!figs/experiments!}.

The propagation chain should at least answer ``at which stage did the resource stop?'': the Registrar node accepted it but the Root platform has not published it; the Root platform published it but content distribution has not completed; distribution completed but the Discovery node has not indexed it; or indexing completed but the query conditions did not match. The following correlation can be used for verification, rather than treating the resource count on a page as a substitute for propagation facts:

\begin{lstlisting}
resourceDid + version
  -> publicationSequence
  -> distributionBatch / contentDigest
  -> discoveryIndexVersion / indexedAt
  -> queryRequestId / visibleAt
\end{lstlisting}

Retries of the same resource version should reuse an idempotency key or be marked as retries. If a record exists for only one stage, the operational result should explicitly state ``accepted,'' ``propagating,'' or ``not yet queryable,'' rather than simply reporting success.

\section{Discovery Query Metrics}\label{sec:19.4}

{\def\LTcaptype{none} 
\begin{longtable}[]{@{}>{\raggedright\arraybackslash}p{0.3200\linewidth}>{\raggedright\arraybackslash}p{0.3200\linewidth}>{\raggedright\arraybackslash}p{0.3200\linewidth}@{}}
\toprule\noalign{}
Query type & Must record & Interpretation boundary \\
\midrule\noalign{}
\endhead
\bottomrule\noalign{}
\endlastfoot
Exact DID query & DID, duration, result count, source & A hit does not mean that it is currently trusted \\
Structured query & Conditions, duration, hits, and filtered results & A condition match does not mean that it is callable \\
Semantic query & Original description, converted conditions, and ranking results & Relevance does not replace control and governance verification \\
Failed query & Status code, error type, requestId & An empty result cannot be equated with a service failure \\
\end{longtable}
}

Discovery metrics distinguish exact DID queries, structured-condition queries, and semantic queries, and record request duration, result count, empty results, rejections, timeouts, and server errors. Relevance is not the same as trustworthiness. Results still need status, credential, publication proof, and freshness checks. Error responses retain correlation identifiers so that a request from the official website can be traced to Discovery node logs.

Query metrics should record the query mode, a digest of the normalized input, request identifier, service node, hit count, duration, response status, and data status. The original task description of a semantic query may be used for troubleshooting and quality analysis, but logs and statistics should not retain unnecessary personal or sensitive content; a digest, length, and hash can be used instead of the full text. An empty result must be evaluated together with whether the request was valid, whether the index is fresh, and whether the resource is within the node\textquotesingle s coverage; it cannot be directly classified as a Discovery node failure. After an exact DID query hits, the resource status, publication evidence, and version must still be checked.

\section{Index Freshness and Synchronization Metrics}\label{sec:19.5}

\begin{lstlisting}
Synchronization status = upstream latest sequence - local cursor + latest successful time + latest error
Uncertain freshness -> mark as unknown; do not fill it with 0
\end{lstlisting}

Index rebuilding, incremental synchronization, synchronization pauses, and synchronization failures should use different status values. Operations personnel can then determine whether there are no new events, processing is delayed, or the synchronization path is interrupted. Synchronization observation includes the current event cursor, upstream latest sequence, lag distance, latest successful time, latest failure reason, index update time, and pending item count. A continuously increasing \passthrough{\lstinline!latest\_sequence!} only shows that new observable events exist upstream; it does not by itself prove that this node has completed indexing. When the lag distance is unknown, display unknown rather than zero. Index rebuilding and incremental synchronization should be marked separately.

Freshness assessment should observe the upstream latest sequence, the locally processed cursor, the latest successful synchronization time, and the latest error together. If the \passthrough{\lstinline!latest\_sequence!} of the off-chain trust indexer continues to increase, this only proves that observable upstream events still exist; the local cursor must also be checked for progress, along with any sequence or digest gaps and whether the Discovery node has completed the corresponding index. If there is no upstream sequence, the interface fails, or the node has not yet produced data, the status should be \passthrough{\lstinline!unknown!} or \passthrough{\lstinline!unavailable!}; it must not be disguised as a lag distance of zero.

\section{Governance and Authorization Metrics}\label{sec:19.6}

Governance metrics record the event source, on-chain location, event type, index cursor, node status changes, and latest synchronization time. Authorization eligibility, resource control, and publication results are observed separately. The indexing of a governance event does not mean that all resource statuses have been updated, and a normal node does not mean that every authorization artifact is valid. Verification failures should retain their evidence and reasons. Governance observation should also retain the original event digest and separately record the receipt, decoding, projection, and external-serving times of the off-chain trust indexer. Resource control is proven by the resource\textquotesingle s own signature and DID document; node authorization is proven by the governance path. These must remain separate in metrics and failure analysis.

\section{Network-Level and Node-Level Statistics}\label{sec:19.7}

Metrics are interpretable only together with completion conditions. A cumulative count describes completed work; a rate describes change within a window; queue depth and oldest-task age describe backlog; cursors and failure counts describe target progress; and CPU, memory, and connection-pool metrics describe only resource state. Low CPU alone does not show that a system has no bottleneck: a client may be waiting in a closed loop, a serialized critical section may be active, or downstream connections may not yet be filled.

The Network page may display both network-level aggregates and node-level details, but the aggregation source, update time, and missing nodes must be traceable. All periodic statistics use the same time zone, boundaries, and deduplication rules. When an old snapshot is used, it carries the snapshot time and degradation marker. Aggregated values must not simply add cumulative values with different definitions.

\pandocbounded{\includegraphics[width=\linewidth,height=0.72\textheight,keepaspectratio]{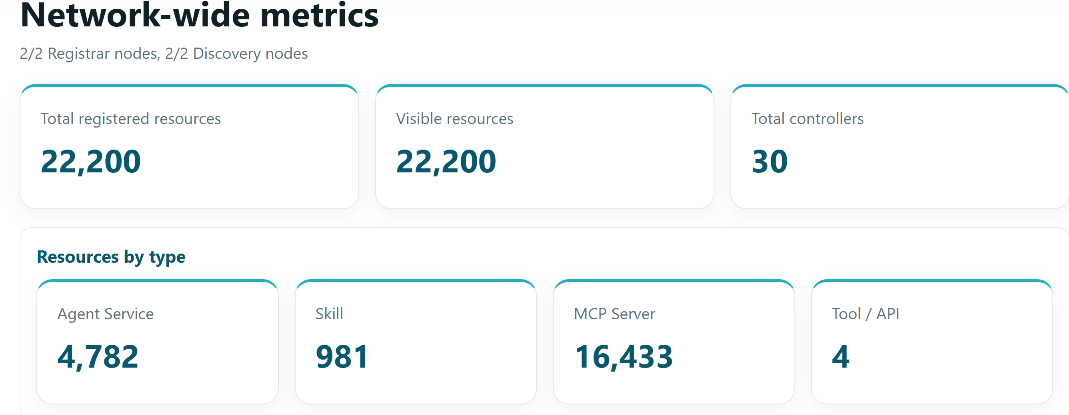}}

This figure is a historical deployment snapshot from the official Network page and illustrates how network-level status presents Registrar, Discovery, and cumulative resource counts. The values apply only to the capture time. They are not current network constants and cannot replace node-status interfaces, synchronization cursors, or business smoke tests.

{\def\LTcaptype{none} 
\begin{longtable}[]{@{}>{\raggedright\arraybackslash}p{0.2400\linewidth}>{\raggedright\arraybackslash}p{0.2400\linewidth}>{\raggedright\arraybackslash}p{0.2400\linewidth}>{\raggedright\arraybackslash}p{0.2400\linewidth}@{}}
\toprule\noalign{}
Window & Start & End & Applicable value \\
\midrule\noalign{}
\endhead
\bottomrule\noalign{}
\endlastfoot
TODAY & Midnight of the current day & Midnight of the following day & Current-day periodic value \\
THIS WEEK & Agreed start of the week & Start of the next period & Periodic weekly value \\
THIS MONTH & First day of the current month & First day of the next month & Monthly periodic value \\
ALL TIME & Defined statistical start point & Current query time & Cumulative value \\
\end{longtable}
}

All windows use the same time zone and half-open interval \passthrough{\lstinline![start, end)!}. Missing data, request failure, and an actual zero value must be encoded separately; missing values must not be filled with zero.

\subsection{Network-Level Metrics}\label{subsec:19.7.1}

Network-level metrics describe nodes and resources within the statistical scope, including node count, registration, publication, discovery, governance synchronization, and network status. Each metric should state its inclusion conditions, deduplication method, source interface, and update time. If a node lacks data, the aggregate result should state its coverage and missing data; the node must not be counted as zero merely because it did not report.

\subsection{Node-Level Metrics}\label{subsec:19.7.2}

Node-level metrics record health status, authorization status, resource processing volume, synchronization cursor, index freshness, and latest error by node DID or stable node identifier. Offline, unauthorized, not-yet-synchronized, and statistics-interface failure states use different values, avoiding the concealment of real conditions with zeros or empty strings.

\subsection{Registration, Publication, and Discovery Metrics}\label{subsec:19.7.3}

Registration, publication, and discovery metrics respectively represent acceptance by a Registrar node, trusted publication by the Root platform, and index visibility at a Discovery node. They must not be merged into one ``success count.'' The same resource version is correlated by DID, version, and publication sequence. Retries, duplicate indexing, and old snapshots must not create artificial growth.

\subsection{Metric Time Windows}\label{subsec:19.7.4}

Time windows use a unified time zone and explicit boundaries. Periodic values use fixed start and end times, while cumulative values are calculated from the defined statistical start point. Interfaces, the official website Network page, and backend reports must use the same definition to avoid abnormal periodic values caused by differences in server time zones, request times, or cache generation times.

\subsection{Missing Data and Degraded Display}\label{subsec:19.7.5}

Missing metrics, interface failures, data that has not yet been produced, and old-snapshot display are marked separately. A page may display old data with its time, but must show the degraded state. The backend continues to record the actual failure. Missing data must not be converted into zero, and a snapshot must not be treated as a real-time health conclusion.

\section{Visibility of On-Chain Governance Events}\label{sec:19.8}

Governance-event display should distinguish the on-chain event time, receipt and indexing time of the off-chain trust indexer, the time when the official website interface was generated, and the time when the page displayed the event. The event sequence or on-chain location is used for ordering and traceability; bilingual descriptions are display-layer mappings and cannot replace the original event. When the chain or indexing service is temporarily unavailable, the page should show an unavailable or delayed state.

\section{Logs, Traces, and Correlation Identifiers}\label{sec:19.9}

Cross-service records should at least correlate the request identifier, resource DID, version, publication sequence, event cursor, and node identifier, and should record the time, level, result, and error digest. Logs must not write private keys or sensitive request content. Error information exposed by public interfaces should be sufficient to locate a problem without revealing internal configuration.

Correlation fields at the interface layer should correspond to the protocol types: node-to-node requests use \passthrough{\lstinline!requestId!}, \passthrough{\lstinline!requestNonce!}, \passthrough{\lstinline!bodyHash!}, and \passthrough{\lstinline!upstreamAuth!}; resource propagation uses \passthrough{\lstinline!publicationCursor!}, batch sequence, and resource package version; discovery queries use the request identifier, Discovery node DID, candidate count, retrieval backend, and duration; the off-chain trust indexer should also record the last processed checkpoint, governance sequence, and synchronization time. These fields should be organized as structured logs rather than merely concatenated into a free-form message.

The example only shows public or low-sensitivity fields that can be used to correlate a processing flow. Production logs must still filter resource descriptions, endpoint credentials, signature material, and internal error stacks according to the deployment policy. \passthrough{\lstinline!requestId!} supports cross-service tracing, but is not an authorization credential and cannot replace resource control proof or the governance-event sequence.

\section{Snapshot, Cache, and Degraded-Mode Behavior}\label{sec:19.10}

{\def\LTcaptype{none} 
\begin{longtable}[]{@{}>{\raggedright\arraybackslash}p{0.3200\linewidth}>{\raggedright\arraybackslash}p{0.3200\linewidth}>{\raggedright\arraybackslash}p{0.3200\linewidth}@{}}
\toprule\noalign{}
Status & Frontend behavior & Backend meaning \\
\midrule\noalign{}
\endhead
\bottomrule\noalign{}
\endlastfoot
live & Display the latest response & Interface and data are normal \\
stale & Display the snapshot and mark its time & Backend refresh failed or data expired \\
unavailable & Display the reason for unavailability & Interface is unreachable or returned an error \\
unknown & Make no trust judgment & There is insufficient evidence \\
\end{longtable}
}

Snapshot replacement must satisfy four conditions: ``complete response, structural validation, source update, and time update.'' A partial response cannot replace a complete old snapshot. An old snapshot is a user-experience mechanism, not health proof. The frontend may first display a browser-side snapshot and refresh in the background. After success, it replaces the snapshot with complete new data and updates the source and time. After failure, it keeps the snapshot but displays the degraded state; continued failure still triggers a backend alert. A loading state is displayed only on a cache-free first visit; loading must not permanently cover data that is already available.

A snapshot record should contain at least the data source, generation time, schema version, and data status, for example:

Here, \passthrough{\lstinline!resourceCount: 12!} is a business value from one particular snapshot and does not represent the current real-time count. The frontend may continue to display it when a refresh fails, but the backend must still record the refresh failure and its duration. When the interface recovers, only a complete response that passes structural validation may replace the snapshot.

\subsection{Loading an Old Snapshot}\label{subsec:19.10.1}

When the page opens, it first reads an available browser-side snapshot and marks its generation time and source. If snapshot parsing fails, the schema version is incompatible, or the content integrity check fails, the snapshot should be discarded and the page should enter the first-load flow; damaged data must not be rendered as normal statistics.

\subsection{Background Data Refresh}\label{subsec:19.10.2}

After displaying an old snapshot, the frontend requests the latest interface in the background without blocking the user\textquotesingle s view of the existing content. The request should have timeout and failure handling. The refresh must not change the resource details or query results currently being viewed unless the new response passes complete structural validation.

\subsection{New Data Replacement}\label{subsec:19.10.3}

Only a complete and structurally valid new response may replace the old snapshot, while the cache time, source, and data status are updated together. If the new response is partially missing, retain the usable old data and mark the degraded state; an empty response or zero value must not replace valid content.

\subsection{Long-Term Failure and Alerts}\label{subsec:19.10.4}

Consecutive failures, long-term absence of data, and a synchronization cursor that has not advanced for an extended period should enter backend alerting. Record the interface, first failure time, latest successful time, and evidence of the error. Continuing to display a snapshot must not suppress an alert. Recovery must be confirmed by consecutive successes and a business check.

\subsection{First Load Without a Cache}\label{subsec:19.10.5}

When no snapshot exists, display the first-load state and explicitly finish it after success, failure, or timeout. A failure state should explain that the data is temporarily unavailable. It must not display loading text indefinitely, and it must not fabricate a zero value or an empty health conclusion.

\section{Backup, Recovery, and Disaster Recovery}\label{sec:19.11}

{\def\LTcaptype{none} 
\begin{longtable}[]{@{}>{\raggedright\arraybackslash}p{0.3200\linewidth}>{\raggedright\arraybackslash}p{0.3200\linewidth}>{\raggedright\arraybackslash}p{0.3200\linewidth}@{}}
\toprule\noalign{}
Backup object & Check after recovery & Handling when the check fails \\
\midrule\noalign{}
\endhead
\bottomrule\noalign{}
\endlastfoot
Node identity and authorization & DID, public key, authorization status & Stop external service \\
Database & Critical records, migrations, and indexes & Do not start write tasks \\
Publication queue & Incomplete tasks and idempotency keys & Pause duplicate delivery \\
Synchronization cursor & Cursor continuity and upstream sequence & Resume synchronization from a trusted position \\
Website artifacts & Correspondence among submission, hash, and configuration & Roll back or rebuild \\
\end{longtable}
}

Backups cover node identity and authorization material, databases, resource packages, publication queues, synchronization cursors, configuration versions, and website deployment artifacts. Private keys and credentials are encrypted separately and access to them is restricted. After recovery, first verify integrity, sequence continuity, and key correspondence; then start write tasks, and confirm recovery through health checks, registration/discovery smoke tests, and index catch-up. The recovery order should first restore configuration and node identity, then the database and its migrations, resource packages, and publication queue. Next verify the synchronization cursor, governance authorization status, and website build artifacts, and finally restore external writes. The recovery process must prevent old queue items from being published repeatedly and prevent cursor rollback from causing duplicate projections. The service may first start in read-only mode or with writes paused; open writes only after checking \passthrough{\lstinline!resourceDid!}, version, publication sequence, and authorization status. A passing \passthrough{\lstinline!/health!} only means that the process has recovered; it cannot by itself prove that historical indexing, governance projection, and incomplete publication tasks have been restored.

\section{Deployment Smoke Tests and Functional Verification}\label{sec:19.12}

Minimum smoke-test loop:

\begin{enumerate}
\def\labelenumi{\arabic{enumi}.}
\tightlist
\item
  Record the deployment commit, artifact digest, and environment configuration.
\item
  Check service health, database connection, and index cursor.
\item
  Use an isolated identity to complete registration and an expected-failure request.
\item
  Check Root platform publication, distribution, and exact/semantic discovery queries.
\item
  Save responses, logs, the timeline, and the cleanup result for test resources. Smoke tests use isolated test identities and test resources, and save the submission time, correlation identifiers, responses, logs, and cleanup results. Successful registration cannot replace discovery verification. Root platform publication, content distribution, index convergence, and database status must also be checked. If any key step fails, the response is uncertain, or production data remains uncleared, the deployment must not be declared successful. Smoke-test results should form a reviewable timeline rather than output only a ``passed'' string. Each step records the target address or service role, commit or artifact digest, request identifier, response status, duration, key response fields, and side-effect cleanup result. Registration acceptance, Root platform publication, and a discovery query hit are assessed separately. When writing a real resource to a production environment is inappropriate, use a controlled failure request to verify validation logic and an existing isolated test resource to verify the discovery path. If any key step is uncertain, the deployment status should remain pending verification.
\end{enumerate}

\subsection{Registration Function Smoke Test}\label{subsec:19.12.1}

Use an isolated identity and a controlled resource to verify request validation, control signatures, Registrar node acceptance, duplicate submission, and error responses. If the interface returns a registration VC, check its structure and its association with the resource DID, version, or digest. Clean up the test data after the test and save the evidence.

\subsection{Discovery Function Smoke Test}\label{subsec:19.12.2}

Use a published isolated test resource to execute an exact DID query and a semantic query, and check the resource DID, version, status, source, publication evidence, and index time. If the result is empty, an old version is returned, or the interface fails, use the publication record and synchronization cursor to determine whether this is propagation delay or a functional failure.

\subsection{Node Health Check}\label{subsec:19.12.3}

Check the liveness, readiness, and dependency status of the Root platform, Registrar nodes, Discovery nodes, content distribution platform, off-chain trust indexer, and official website backend one by one. Record the response code, duration, version, and check time; do not check only whether the process exists.

\subsection{Database and Indexer Check}\label{subsec:19.12.4}

Check the database connection, migration status, critical records, and indexer cursor. Verify the relationship and continuous change between the upstream latest sequence, local cursor, and \passthrough{\lstinline!latest\_sequence!}, and distinguish the absence of new events from a stopped synchronization.

\subsection{Deployment Handling after Failure}\label{subsec:19.12.5}

When a key smoke test fails, stop publication confirmation and retain the logs, commit, artifact hash, and test-resource information, then handle the issue through the repair or rollback process. Only after a retry succeeds, data side effects have been cleaned up, and evidence is complete may an acceptance result be issued again.

\section{Data Retention, Deletion, and Privacy Operations}\label{sec:19.13}

Public metrics return only necessary aggregate data. Access statistics, request correlation identifiers, and operational logs should be controlled by backend permissions. Browser-side local identity private keys, backup contents, and sensitive material submitted by users must not be written to access statistics. Deletion operations should explain the different retention rules for online data, backup copies, and immutable audit records.

\section{Operational Alerts and Long-Term Failure Detection}\label{sec:19.14}

An alert record should contain at least the service, interface, first failure time, latest successful time, consecutive failure count, error digest, impact scope, and current status.

{\def\LTcaptype{none} 
\begin{longtable}[]{@{}>{\raggedright\arraybackslash}p{0.3200\linewidth}>{\raggedright\arraybackslash}p{0.3200\linewidth}>{\raggedright\arraybackslash}p{0.3200\linewidth}@{}}
\toprule\noalign{}
Status & Trigger basis & Recovery condition \\
\midrule\noalign{}
\endhead
\bottomrule\noalign{}
\endlastfoot
Intermittent failure & A small number of requests fail within the window & Subsequent requests succeed consecutively \\
Consecutive failure & Consecutive requests fail beyond the threshold & Health and business checks succeed consecutively \\
Long-term stagnation & A metric or cursor has not updated beyond the allowed duration & Data recovers and the sequence advances again \\
Dependency failure & An upstream service or database is unavailable & Dependency recovers and regression smoke tests complete \\
\end{longtable}
}

Use a stable fingerprint to deduplicate the same failure. A recovery alert must have both consecutive successful health checks and the corresponding business smoke test; it must not rely only on restarting the process. Long-term failure detection covers \passthrough{\lstinline!summary!}, governance synchronization, publication queue, and index-cursor stagnation. Alerts record the service, interface, consecutive failure count, first failure time, latest successful time, error digest, and impact scope, and deduplicate the same failure. Recovery requires consecutive success and business verification. When the website continues to display an old snapshot, backend alerts must not be suppressed. Periodic checks such as the \passthrough{\lstinline!summary!} interface should retain two facts: ``the latest success'' and ``the latest failure,'' rather than only retaining the current status. An alert event should at least correlate the service role, interface, time window, consecutive failure count, latest successful sequence or time, error digest, and impact scope. Short-lived network jitter for the same interface may be merged into one failure fingerprint, while continued failure should be escalated. During recovery, first confirm consecutive interface success, then confirm that the data time or cursor has advanced again, and finally execute the affected registration/discovery business smoke test. Continuing to display an old snapshot in the browser must not pause or clear this alert.

\section{Public Service Status and Maintenance Communication}\label{sec:19.15}

Public status information should state the affected service, start time, current status, data delay, and next update, without exposing keys, internal addresses, or user data. During maintenance, read-only queries or old snapshots may be retained, but the time point and unavailable functions should be marked. A recovery announcement should be based on actual checks and business smoke tests.

\section{Reference Performance and Resource-Footprint Baselines}\label{sec:19.16}

The latest formal local baseline uses four registration processes, 48 total in-flight requests, a 4/8 DID verification executor, a database pool of 8, and one-second read-only sampling. The benchmark script must independently perform startup, load generation, read-only sampling, and result aggregation. A large language model may prepare parameters before the run or perform non-interfering read-only observation during the run, but it must not become a control step that the load processes wait for. Each of four pipeline workloads uses three scales and five repetitions, for 60 runs in total, with a maximum scale of approximately 5000 resources. The observer records deduplicated counts separately for Root, CDN, and every Discovery node; end-to-end completion requires every target to reach the expected value. A remote isolated test also records whether SSH compression is enabled. This profile is a reproducible experiment configuration, not a production capacity threshold.

At the largest scale, the median end-to-end times for the four workloads are approximately 21.021, 21.011, 22.021, and 24.077 seconds. These results may be cited only together with the corresponding commits, the structured path with the semantic engine disabled, and the isolated data state. One-second sampling makes each measured completion time an upper bound at sampling precision. Resource-stability experiments also retain a post-load observation window to distinguish live objects, allocator retention, caches, and continuing growth.

\subsection{Dynamic Footprint for 5000 Resources}\label{subsec:19.16.1}

The Yellow Paper separately measures the resource consumption of core services in an isolated 5000-resource run with one Registrar and one Discovery node. This run is separate from the formal throughput matrix. After structured propagation completes, observation continues for 600 seconds to distinguish the load-period peak from the stable working set. CPU in the table is the peak equivalent-core count during the observation window. PostgreSQL is aggregated across host processes and therefore uses a different scope from an individual managed node process.

{\def\LTcaptype{none} 
\begin{longtable}[]{@{}lrr@{}}
\toprule\noalign{}
Role & Final observed working set (MiB) & Peak equivalent CPU cores \\
\midrule\noalign{}
\endhead
\bottomrule\noalign{}
\endlastfoot
Root & 41.5 & 2.66 \\
Registrar & 33.9 & 1.37 \\
Discovery & 76.8 & 0.62 \\
CDN Node & 18.8 & 0.39 \\
CDN Publisher & 16.9 & 0.18 \\
NATS & 29.6 & 0.11 \\
PostgreSQL host aggregate & 371.7 & Not applicable \\
\end{longtable}
}

Root\textquotesingle s peak working set in this run was 64.7 MiB and fell to 41.5 MiB at the ten-minute observation point. The other managed roles likewise showed no continuing post-load growth. This supports only the statement that no continuing growth was observed for the specified version and configuration; it is not a general proof of absence of memory leaks. A repeat run should retain the peak, completion point, stable window, and restart baseline rather than capturing RSS at only one instant.

\subsection{Steady-State Snapshot of a Populated Deployment}\label{subsec:19.16.2}

The Yellow Paper also records a memory snapshot from a different Linux deployment with more than 22,000 registered resources, semantic inference enabled, and low traffic. It describes the steady-state composition of a populated deployment and must not be added to or directly compared with the preceding isolated Windows dynamic run. Registrar and Discovery values are the arithmetic per-instance mean across two instances. PostgreSQL uses the service-cgroup scope, including child processes and file-page cache charged to that cgroup.

{\def\LTcaptype{none} 
\begin{longtable}[]{@{}lr@{}}
\toprule\noalign{}
Component & Memory (MiB) \\
\midrule\noalign{}
\endhead
\bottomrule\noalign{}
\endlastfoot
Root & 41.5 \\
Registrar (per-instance mean) & 22.6 \\
Discovery (per-instance mean) & 104.4 \\
Trust Indexer & 8.2 \\
CDN Node & 9.5 \\
CDN Publisher & 9.7 \\
NATS & 12.4 \\
Discovery Embedding & 1320.4 \\
PostgreSQL (service cgroup) & approximately 1170 \\
\end{longtable}
}

The embedding service\textquotesingle s approximately 1.29 GiB resident memory is mainly the loaded model, inference runtime, and reusable work buffers. It does not mean that every resource duplicates a model or vector collection inside the process. Resource vectors are stored in PostgreSQL; at 768 \passthrough{\lstinline!float32!} dimensions, the raw numeric payload for 22,000 vectors is approximately 64.5 MiB, excluding record and index overhead. The PostgreSQL cgroup also contains reclaimable database file-page cache. These values are therefore neither an empty-database baseline nor mutually exclusive physical-memory values that can simply be added. Capacity planning should separately account for application processes, the shared semantic model, persistent vectors and indexes, and database cache.

\section{References}\label{sec:19.17}

{\def\LTcaptype{none} 
\begin{longtable}[]{@{}>{\raggedright\arraybackslash}p{0.3200\linewidth}>{\raggedright\arraybackslash}p{0.3200\linewidth}>{\raggedright\arraybackslash}p{0.3200\linewidth}@{}}
\toprule\noalign{}
Source & Type & Link \\
\midrule\noalign{}
\endhead
\bottomrule\noalign{}
\endlastfoot
\passthrough{\lstinline!oan-root-nodes!} & Code repository: Root platform status, publication, and distribution metrics & \url{https://github.com/OpenAgenet/oan-root-nodes} \\
\passthrough{\lstinline!oan-registrar-nodes!} & Code repository: Registrar node status, registration statistics, and health interfaces & \url{https://github.com/OpenAgenet/oan-registrar-nodes} \\
\passthrough{\lstinline!oan-discovery-nodes!} & Code repository: synchronization, indexing, querying, and freshness metrics & \url{https://github.com/OpenAgenet/oan-discovery-nodes} \\
\passthrough{\lstinline!oan-trust-indexer!} & Code repository: synchronization cursor, projection status, and governance metrics & \url{https://github.com/OpenAgenet/oan-trust-indexer} \\
OAN Yellow Paper & arXiv Yellow Paper: operational status and trust-evidence boundaries & \url{https://arxiv.org/abs/2606.03163} \\
\end{longtable}
}

\chapter{Deployment and Conformance Requirements}\label{chap:20}

This chapter explains how deployment conformance demonstrates what was deployed, how it was built, whether its dependencies are ready, whether the business functions are available, and how failure is handled through rollback. The node services, website backend, and frontend static resources of OpenAgenet (OAN) should separately record the source commit, build artifacts, configuration version, and acceptance evidence. The official deployment describes a runnable instance, while the protocol requirements apply to every implementation that claims conformance to the specification.

\section{Reference Deployment Topology}\label{sec:20.1}

The reference topology consists of the Root platform, Registrar nodes, Discovery nodes, the content distribution platform, the off-chain trust indexer, databases, and the official website. Only necessary endpoints protected by TLS and a reverse proxy are exposed externally. Nodes communicate using configured service addresses and authorization materials. Third-party nodes may be deployed independently, but they must pass conformance checks for node identity, authorization, interfaces, and synchronization before joining.

\pandocbounded{\includegraphics[width=\linewidth,height=0.72\textheight,keepaspectratio]{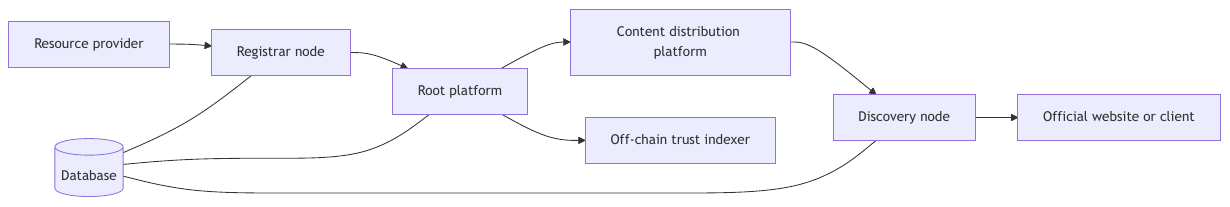}}

The deployment topology should separately identify public entry points, internal services, databases, and governance data sources. Public addresses, ports, and node counts are environment configuration and should not be fixed in the diagram as protocol requirements.

\pandocbounded{\includegraphics[width=\linewidth,height=0.72\textheight,keepaspectratio]{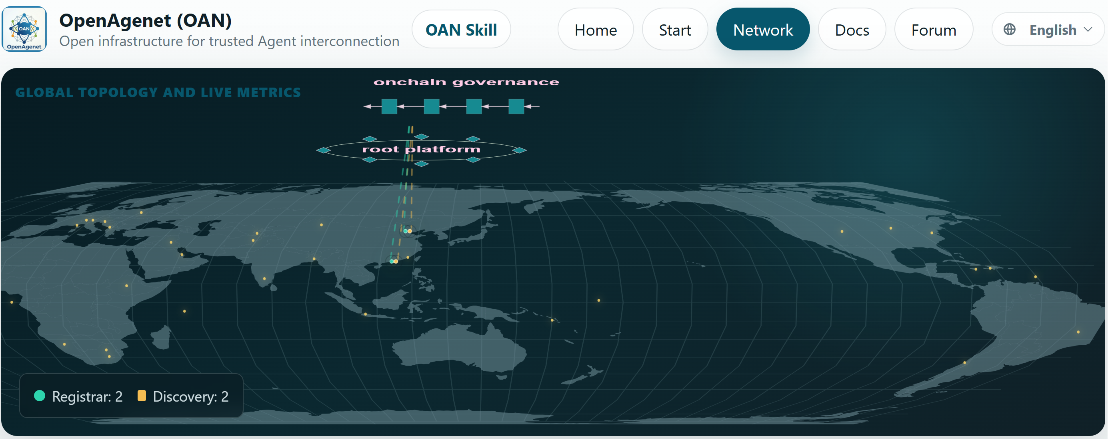}}

This deployment snapshot shows the geographic locations and connection relationships visible on the official Network page at one capture time. The map is an observability view of deployment data obtained by the page. It neither defines how many nodes the protocol requires nor proves that a displayed inter-node link remains healthy after the screenshot. Formal acceptance still correlates each displayed node with its node DID, authorization state, health endpoint, and synchronization state.

\section{Configuration and Secret Management}\label{sec:20.2}

{\def\LTcaptype{none} 
\begin{longtable}[]{@{}>{\raggedright\arraybackslash}p{0.2400\linewidth}>{\raggedright\arraybackslash}p{0.2400\linewidth}>{\raggedright\arraybackslash}p{0.2400\linewidth}>{\raggedright\arraybackslash}p{0.2400\linewidth}@{}}
\toprule\noalign{}
Configuration category & Source & Deployment check & Exposure boundary \\
\midrule\noalign{}
\endhead
\bottomrule\noalign{}
\endlastfoot
Node identity & Controlled key file or key service & DID, public key, and permissions match & Must not enter public interfaces \\
Database & Environment configuration & Connection, migration, and permissions are valid & Must not be exposed through the website proxy \\
Upstream endpoint & Environment profile & Address is reachable and protocol is compatible & Accessed only by necessary services \\
Governance configuration & Authorization configuration and version record & Package, object, and cursor are consistent & Must not return secrets \\
\end{longtable}
}

Configuration is injected per environment. The code package must not carry production private keys, database credentials, or external service tokens. The node DID private key, Root platform authorization materials, database connection information, on-chain package or object configuration, and discovery and distribution endpoints should have restricted file permissions and separate backups. Required fields should be checked before startup; during rotation, the audit record for the old and new configurations should be retained. A failed pre-start check should prevent the service from entering the ready state. Configuration rotation should record the expiration time of the old configuration, the activation time of the new configuration, and the verification result.

The following minimal record can be used as release evidence for configuration checks. The values indicate only whether a value exists and whether it passed validation; secrets must not be written into the report:

If a required endpoint, database, node identity, or governance configuration cannot be validated, the deployment should remain in the configuration stage. A successful process start or accessible page must not be used to conceal a missing item.

\section{Build and Release Artifacts}\label{sec:20.3}

A release record should include the source code version, lockfile, Rust and frontend build environments, build steps, artifact hashes, configuration version, and release time. Deployment may synchronize a versioned source archive or an equivalent package containing the confirmed version, then generate the Rust backend and frontend static resources in the target build environment. The source archive, binaries, static resources, and configuration should be mutually traceable. Before release, a traceable chain of ``source version → build → artifact → deployment'' should be established. Temporary source changes or unrecorded build directories in the target environment must not serve as the official source.

{\def\LTcaptype{none} 
\begin{longtable}[]{@{}>{\raggedright\arraybackslash}p{0.3200\linewidth}>{\raggedright\arraybackslash}p{0.3200\linewidth}>{\raggedright\arraybackslash}p{0.3200\linewidth}@{}}
\toprule\noalign{}
Artifact & Source & Verification information \\
\midrule\noalign{}
\endhead
\bottomrule\noalign{}
\endlastfoot
Source archive & Explicit Git commit & Commit, file list, and archive digest \\
Rust binary & Locked dependencies and build environment & Artifact hash, version, and target platform \\
Frontend static resources & Official website build script & Build commit, resource list, and digest \\
Configuration & Environment-specific material & Configuration version or fingerprint, without exposing secrets \\
\end{longtable}
}

A release record can describe the correspondence between artifacts with a summary that contains no secrets:

The values enclosed in angle brackets are record templates, not fixed official OAN values. During an actual release, they should be replaced with real digests, and the source archive, build log, artifact hashes, and server deployment record should be kept in the same acceptance record.

\section{Database and Storage Requirements}\label{sec:20.4}

{\def\LTcaptype{none} 
\begin{longtable}[]{@{}>{\raggedright\arraybackslash}p{0.3200\linewidth}>{\raggedright\arraybackslash}p{0.3200\linewidth}>{\raggedright\arraybackslash}p{0.3200\linewidth}@{}}
\toprule\noalign{}
State & Primary purpose & Backup/migration requirement \\
\midrule\noalign{}
\endhead
\bottomrule\noalign{}
\endlastfoot
PostgreSQL business databases & Registration, authoritative versions, publication tasks, Discovery read models & Migrate and verify constraints, versions, and task state \\
Root nonce SQLite & Replay-reservation state & Preserve on production upgrade; loss shortens replay memory \\
JetStream store & Publication notifications and consumer positions & Preserve or safely rebuild from database tasks; never treat as authoritative resource store \\
CDN package storage & Durable ResourcePackages and cursors & Preserve, or replay from Root under digest verification \\
Discovery structured/vector indexes & Query read models & Rebuildable from CDN; mark unavailable or stale during rebuild \\
Trust Indexer cursor/projection & On-chain governance read model & Preserve cursor or replay from trusted checkpoint and verify final position \\
\end{longtable}
}

Before deployment, confirm the database version, connection, migration, and index status. Migrations must be idempotent, traceable, and preceded by a backup. Storage should cover resource registration, versions, credential associations, publication tasks, synchronization cursors, governance events, and the data required for statistics. After recovery, check key records, sequence continuity, queue status, and index consistency; checking only whether the process starts is insufficient.

Database acceptance should separately record that the database can be connected to, migrations are complete, data is available, and consistency remains after recovery. Important checks include the uniqueness of the resource DID and version, idempotency keys for publication tasks, records required by the discovery index, and the continuity of governance events and synchronization cursors. If migrations are complete but a required index is missing, the service should not enter a writable ready state. Database checks must not return internal table structures through public website interfaces. Reports should retain only the necessary version, count, digest, and verification results.

\subsection{Classifying State by Authority and Recoverability}\label{subsec:20.4.1}

Storage technology is not a protocol requirement, but each state category must preserve its authority and recovery conditions. Before choosing a database, file, queue, or cache, a deployment manifest asks what a state proves and whether it can be rebuilt from other evidence if lost.

{\def\LTcaptype{none} 
\begin{longtable}[]{@{}>{\raggedright\arraybackslash}p{0.3200\linewidth}>{\raggedright\arraybackslash}p{0.3200\linewidth}>{\raggedright\arraybackslash}p{0.3200\linewidth}@{}}
\toprule\noalign{}
State category & Typical content & Persistence and recovery requirement \\
\midrule\noalign{}
\endhead
\bottomrule\noalign{}
\endlastfoot
Root authoritative state & Admitted resource versions, predecessor/controller relationships, Root proof, publication tasks, and bases for governance decisions & Must be persisted with atomic relationships; must not be inferred and rebuilt from a Discovery index or public page \\
Registrar business state & Draft flow, control proof, registration VC, submission attempts, and reconciliation records for indeterminate responses & Must support reconciliation for the same DID and credential recovery; a timeout must not trigger blind creation of a new identity \\
CDN distribution state & Persisted ResourcePackages, manifests, digests, confirmation cursors, and history & Proves content availability but does not make trust decisions; a controlled Root retransmission must still verify digest and sequence \\
Discovery derived state & Synchronization cursor, package-validation result, rejection reason, structured index, and semantic-index state & Should be persisted or rebuildable from validated packages; if structured synchronization succeeds but semantic writing fails, retain a retryable state \\
Trust Indexer derived read model & On-chain cursor, event projection, and node-authorization state & Replayable from a trusted checkpoint or on-chain events, but recovery must verify the final position and projected content \\
Replay-protection state & Request nonce, time window, and reservation result & Must remain effective across restarts for the whole request-validity window; its loss cannot be treated as ordinary cache loss \\
Runtime statistics and public metrics & Aggregated counts, latency, page snapshots, and chart caches & Recomputable views; retention policy does not change protocol admission, authorization, or publication facts \\
Temporary UI and response cache & Page-rendering state, short-lived response cache, and health-probe result & Rebuildable; cached content must not overwrite newer authoritative revocation, version, or governance state \\
\end{longtable}
}

Recovery acceptance must prove more than the ability to open a database file. At a minimum, it reconstructs the Root admission chain, Registrar submission history, CDN content integrity, visibility at every Discovery node, and semantic-index progress, and verifies that a nonce still rejects replay within its validity window. Rebuildable state also records its reconstruction source, target version, cursor, and completion criterion; ``rebuildable'' is not a reason to omit recovery evidence.

\section{TLS, Domains, Reverse Proxy, and Network Exposure}\label{sec:20.5}

The domain and TLS termination point should match the actual service entry point. The reverse proxy should forward only necessary paths while preserving the correct Host, protocol, and client association information. Public ports, cross-origin sources, firewalls, and node-to-node access rules should be configured per environment. Databases, management interfaces, and internal queues must not become exposed to ordinary Internet users because of website proxy configuration. During acceptance, check from both the public network and the internal network the allowed paths, certificate hostname, HTTP-to-HTTPS redirect, CORS scope, and rejection results for non-public paths.

\section{Deployment Order and Dependency Checks}\label{sec:20.6}

The recommended order is to record the baseline, back up and synchronize the code, build the artifacts, prepare the database, start the Root platform and its dependent services, then start the Registrar nodes, distribution, Discovery nodes, and indexer, and finally deploy the official website and perform business smoke tests. If any required dependency, migration, health check, or critical smoke test fails, stop at the current step, retain the evidence, and enter repair or rollback. The pass condition for each step should be defined by the dependency actually required by the next step; a started process alone is not a release gate.

\subsection{Code Synchronization and Build}\label{subsec:20.6.1}

Generate an archive or equivalent release package from an explicit source code version. After transferring it to the target build environment, verify the files and version information. Compile the node services using the Rust dependencies locked by the project, and generate static resources and backend artifacts using the frontend build script. Do not release from temporary workspace files or unrecorded local build directories.

\subsection{Database Preparation}\label{subsec:20.6.2}

Confirm that the backup can be restored, the database is reachable, the migration version is correct, and required indexes are present before running an idempotent migration or startup check. A failed migration must not be followed by starting a service that writes new data. If manual handling is required, retain the failure log, database version, and recovery point.

\subsection{Node Startup Order}\label{subsec:20.6.3}

Start the Root platform and its database and publication dependencies first, followed by the Registrar nodes, content distribution platform, Discovery nodes, and off-chain trust indexer. The actual order may be adjusted according to the deployment topology, but before a downstream service starts, confirm that the upstream address, identity, authorization, and interfaces are ready. The existence of a process must not be treated as a substitute for dependency availability.

\subsection{Service Health Checks}\label{subsec:20.6.4}

Check process liveness, readiness interfaces, critical dependencies, and configuration loading results one by one, and record the response code, latency, version, and time. A passing health check does not mean that the business path has passed. In particular, separately verify Root platform publication, registration acceptance, discovery queries, distribution, and index synchronization.

\subsection{Post-deployment Functional Verification}\label{subsec:20.6.5}

Post-deployment functional verification should cover at least:

\begin{enumerate}
\def\labelenumi{\arabic{enumi}.}
\tightlist
\item
  Health and readiness interfaces.
\item
  Valid requests and expected failure requests for the Registrar nodes.
\item
  Root platform publication, content distribution, and synchronization to the Discovery nodes.
\item
  Exact DID queries and semantic queries.
\item
  Database migration status, cursors, logs, and alerts.
\end{enumerate}

When any critical item fails, retain the submission, artifact, configuration, and response evidence, and stop the release completion confirmation.

Interface verification should record the request method, actual URL, HTTP status, response type, key business fields, and correlation identifiers. For dynamic JSON interfaces, checking only for a 2xx response is insufficient. Also check that the registration response contains the resource DID and registration status, that the Root platform response is \passthrough{\lstinline!resource-verified-and-queued!}, that the content distribution response contains a publication cursor, that the discovery response can be deserialized as \passthrough{\lstinline!ResourceDiscoveryResponse!}, and that the indexer status identifies the current synchronization position. When accessed through the official website proxy, additionally verify that the proxy-path mapping to the original node path has not changed the request body or error semantics.

\section{Registrar Node Functional Smoke Tests}\label{sec:20.7}

Use an isolated test identity and controlled resource to verify form or request validation, control signatures, Registrar node acceptance, duplicate submission, and error responses. If the interface returns a registration VC, check its association with the resource DID, version, or digest. Subsequent publication and cleanup of the test resource must not pollute normal resource statistics. Successful registration proves only that the Registrar node accepted the request; it does not automatically prove that the Root platform has published it or that the Discovery nodes have indexed it.

{\def\LTcaptype{none} 
\begin{longtable}[]{@{}>{\raggedright\arraybackslash}p{0.3200\linewidth}>{\raggedright\arraybackslash}p{0.3200\linewidth}>{\raggedright\arraybackslash}p{0.3200\linewidth}@{}}
\toprule\noalign{}
Acceptance step & Result & Failure handling \\
\midrule\noalign{}
\endhead
\bottomrule\noalign{}
\endlastfoot
Valid registration & Acceptance result and correlation identifier are verifiable & Stop confirmation and retain request evidence \\
Duplicate registration & Idempotent; no incorrect new registration fact is added & Check deduplication and database changes \\
Invalid signature & Clearly rejected and the original state remains unchanged & Check for side effects \\
VC result & If returned, consistent with the DID, version, or digest & Mark as missing or invalid; do not fabricate success \\
Cleanup & Test resource can be identified and cleaned up & Enter manual review \\
\end{longtable}
}

\section{Discovery Node Functional Smoke Tests}\label{sec:20.8}

Perform exact DID queries and semantic queries against an isolated test resource that has been published. Check the DID, resource type, version, status, source, Root proof status, and index time. If a query returns no result, an old version, or an interface failure, use the cursor and publication record to determine whether the condition is propagation delay or a functional fault. The pass condition for a semantic query is an explainable candidate and source, not a fixed ranking or an undefined relevance score.

\section{Root Platform, Distribution, and Indexer Verification}\label{sec:20.9}

Independently check the publication status of the Root platform, resource or cursor records of the content distribution platform, synchronization status of the off-chain trust indexer, and index results of the Discovery nodes. Each check must state its source, time, and normal criterion. Observing success at the Root platform does not prove that downstream services are already discoverable. The same isolated resource version should be used throughout the verification, and the Root platform publication sequence, distribution digest, index version, and query result should be recorded. Equality of total counts across interfaces must not substitute for a content-consistency check.

\section{Governance Synchronization Verification}\label{sec:20.10}

Select a governance event that can be confirmed, and check its on-chain position, event type, trust indexer cursor, node-state read model, and update time. Across two or more consecutive checks, the relationship between the cursor and \passthrough{\lstinline!latest\_sequence!} should match the synchronization state of the deployment environment. The absence of new events must be distinguished from a stopped synchronization process. If governance events have been read but the projection has not been updated, classify the synchronization path as incomplete rather than allowing it to pass solely because the indexer interface responded successfully.

\section{Upgrade, Rollback, and Compatibility Checks}\label{sec:20.11}

Before and after an upgrade, check database migrations, protocol versions, resource schemas, historical versions, node interoperability, and rollback paths. Continue an upgrade only when compatibility and backup conditions are satisfied. If the new artifact fails health or business verification, roll back according to the recorded version and configuration, then rerun the key acceptance checks. Upgrade verification should use representative resource DIDs, resource packages, registration requests, and discovery queries rather than checking only whether the service starts.

\subsection{Pre-upgrade Checks}\label{subsec:20.11.1}

Record the current software version, binary hash, configuration, database state, and cursor state. Confirm that the backup is readable, the target version can be built, the migration can be rolled back or recovered, and the old version remains available.

\subsection{Data and Configuration Compatibility}\label{subsec:20.11.2}

Check newly added fields, resource package and credential structures, protocol routes, environment variables, node authorization materials, and frontend interface expectations. Compatible conversions should be repeatable. Historical data that cannot be converted must stop the release before deployment rather than be silently discarded during runtime.

\subsection{Rollback Conditions}\label{subsec:20.11.3}

Evaluate rollback when a database migration fails, a critical health check fails, registration or discovery smoke tests fail, the publication queue becomes abnormal, indexing remains stalled, or a security configuration is incorrect. For irreversible data migrations, first pause writes and process the recovery plan; replacing only the binary is insufficient.

\subsection{Post-rollback Verification}\label{subsec:20.11.4}

After rollback, check the service version, configuration, database records, node identity, publication sequence, and synchronization cursor again. Then perform health, registration, discovery, Root platform, distribution, and index verification. Rollback may be closed only when the old version has returned to a demonstrable working state.

\section{Third-party Node Onboarding and Conformance Testing}\label{sec:20.12}

Third-party nodes should submit node identity, authorization materials, public endpoints, supported interfaces, and environment information, and pass tests for signature verification, authorization, protocol responses, resource publication, synchronization, querying, security boundaries, and failure recovery. Passing the tests grants onboarding eligibility only within the explicitly defined scope; it does not replace subsequent operational monitoring and governance management. The admission report should specify the node role, authorization domain, interface version, test environment, and validity period. A role, domain, or interface that was not tested must not be inferred to be compliant from another passing item.

The minimum acceptance evidence for a third-party node can be organized in the following order:

\begin{lstlisting}
node identity -> authorization -> protocol response
              -> controlled registration/publication
              -> synchronization/discovery
              -> failure recovery and evidence review
\end{lstlisting}

Testing should first be completed in an independent database and isolated endpoints. A node\textquotesingle s ability to call an interface does not mean that it has authorization, and its ability to synchronize data does not mean that it may publish resources. Every conclusion should be bound to the actual test scope and evidence.

\section{Reproducible Builds and Release Provenance}\label{sec:20.13}

Build records should trace artifacts back to the commit and lockfile, and should retain at least the build environment, commands, dependency versions, artifact hashes, and operator. The running binaries, frontend static resources, and source archive on the server should be mutually verifiable. Configuration and secrets are not included in public artifact hashes, but their version or fingerprint should be recorded. For the Rust backend and the official website frontend, build results should be retained separately, and the deployment record should explain how they are combined into the same release batch.

\section{Environment-specific Configuration Files}\label{sec:20.14}

Development, test, staging, and production environments should use separate profiles, endpoints, databases, and keys. Test authorization materials and production private keys must not be reused across environments. Configuration changes should have a commit, approval, or operation record. Differences between environments running the same code should be concentrated in configuration rather than implemented through unrecorded manual source edits. During acceptance, verify that the current profile matches the endpoints, database, node identity, and governance objects, so that a passing result in a test environment is not mistakenly recorded as a production pass.

\section{Post-deployment Functional and Security Verification}\label{sec:20.15}

The acceptance record summarizes versions, environment, configuration, health interfaces, business smoke tests, database, governance synchronization, permissions, TLS, public ports, and log checks. It should also confirm that private keys have not entered logs or public interfaces, management data is not exposed to ordinary users, and error responses do not disclose secrets. Retain reproducible command output or response summaries. The final conclusion should state the scope, evidence time, uncovered items, and whether the service is permitted to provide services externally. If any critical item is \passthrough{\lstinline!FAIL!} or uncertain, the overall result must not be written as passed.

\section{References}\label{sec:20.16}

{\def\LTcaptype{none} 
\begin{longtable}[]{@{}>{\raggedright\arraybackslash}p{0.3200\linewidth}>{\raggedright\arraybackslash}p{0.3200\linewidth}>{\raggedright\arraybackslash}p{0.3200\linewidth}@{}}
\toprule\noalign{}
Source & Type & Link \\
\midrule\noalign{}
\endhead
\bottomrule\noalign{}
\endlastfoot
\passthrough{\lstinline!oan-root-nodes!} & Code repository: Root platform services and deployment configuration & \url{https://github.com/OpenAgenet/oan-root-nodes} \\
\passthrough{\lstinline!oan-registrar-nodes!} & Code repository: Registrar node and deployment configuration & \url{https://github.com/OpenAgenet/oan-registrar-nodes} \\
\passthrough{\lstinline!oan-discovery-nodes!} & Code repository: Discovery node and deployment configuration & \url{https://github.com/OpenAgenet/oan-discovery-nodes} \\
\passthrough{\lstinline!oan-trust-indexer!} & Code repository: Trust indexer and chain endpoint configuration & \url{https://github.com/OpenAgenet/oan-trust-indexer} \\
\passthrough{\lstinline!oan-protocol-commons!} & Code repository: Cross-service protocols, configuration-related types, and error contracts & \url{https://github.com/wolfbrother/oan-protocol-commons} \\
OAN Yellow Paper & arXiv yellow paper: Deployment, trust, and verification boundaries & \url{https://arxiv.org/abs/2606.03163} \\
\end{longtable}
}

\chapter{External Protocols and Ecosystem Mapping}\label{chap:21}

This chapter treats external protocol integration as a traceable transformation process: first identify the external object and its version; then map verifiable fields to OpenAgenet (OAN) resource metadata or a resource package; next verify resource control, registration credentials, the Root platform\textquotesingle s Root proof, and governance status separately; and finally explain to the caller which information can be used for discovery and which information must still be rechecked before invocation. An adapter must preserve the external original, source, summary, and transformation version, and must not treat a transformed summary directly as an OAN trusted publication fact.

{\def\LTcaptype{none} 
\begin{longtable}[]{@{}>{\raggedright\arraybackslash}p{0.2400\linewidth}>{\raggedright\arraybackslash}p{0.2400\linewidth}>{\raggedright\arraybackslash}p{0.2400\linewidth}>{\raggedright\arraybackslash}p{0.2400\linewidth}@{}}
\toprule\noalign{}
External material & OAN carrying location & Content available for discovery & Content requiring independent verification \\
\midrule\noalign{}
\endhead
\bottomrule\noalign{}
\endlastfoot
DID Core / DID Document & Resource DID Document and verification methods & \passthrough{\lstinline!id!}, verification methods, service endpoints & Control relationship, signatures, status, and publication evidence \\
A2A Agent Card & Resource metadata, protocol binding, or associated description & Capabilities, skills, use cases, inputs and outputs, and interaction entry points & Card publisher\textquotesingle s relationship to OAN control, registration credentials, and the Root platform\textquotesingle s Root proof \\
MCP service metadata & MCP service resource package and its protocol binding & Server capabilities, tool inventory, transport, and authentication declarations & Service endpoint, version, resource package hash, and lifecycle \\
OpenAPI or tool API description & Tool API resource package and its protocol binding & Operations, parameters, schemas, errors, and endpoints & Document digest, control signature, business authorization, and actual availability \\
\end{longtable}
}

\begin{quote}
A mapping result answers ``how the external system describes and connects to this resource''; it does not directly answer ``whether this resource has been authorized by the OAN network.'' The latter determination must return to the OAN DID, signatures, registration credentials, Root proof, and governance status.
\end{quote}

External identifiers and OAN identities retain parallel evidence chains. The adapter establishes a sourced association; it does not replace either identifier or copy a successful state from one system into a trust conclusion in the other.

\pandocbounded{\includegraphics[width=\linewidth,height=0.72\textheight,keepaspectratio]{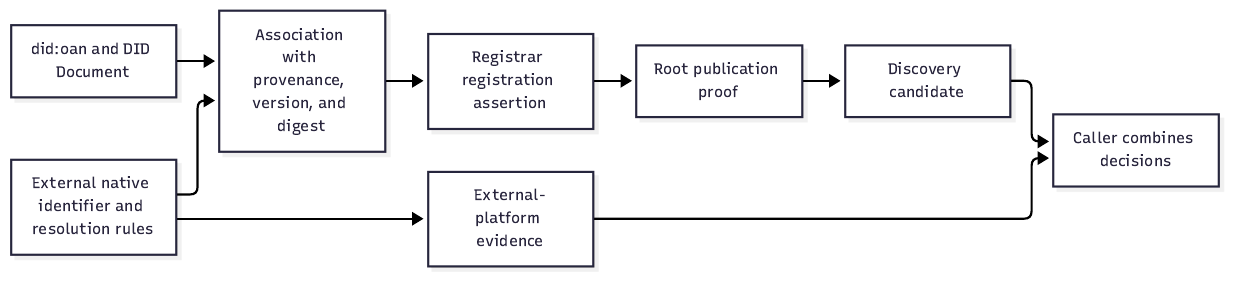}}

The association in this diagram is not identifier substitution: the external ID retains its namespace, \passthrough{\lstinline!did:oan!} remains the OAN identity, and the caller validates both evidence sets independently.

\section{\texorpdfstring{W3C DID Core and \texttt{did:oan}}{W3C DID Core and did:oan}}\label{sec:21.1}

An adapter should store common DID Core fields and \passthrough{\lstinline!did:oan!} method fields in separate layers. Common fields allow DID tools to understand the document, while OAN fields express the resource type, capabilities, version, resource package, and publication relationships. The following object is only an example of a mapping record, not a new DID Document:

This record only describes the transformation process. \passthrough{\lstinline!did:oan!} uses the common identity and service expression mechanisms of DID Core, and supplements the information required for resource discovery and publication through OAN resource metadata, resource packages, and associated proofs.

\subsection{DID Core Data Structure Correspondence}\label{subsec:21.1.1}

The \passthrough{\lstinline!id!}, \passthrough{\lstinline!verificationMethod!}, control relationships, and \passthrough{\lstinline!service!} in a DID Document correspond to the common data model of DID Core. OAN resource types, capability tags, use cases, versions, resource package references, and publication-proof associations belong to the OAN resource model or method extensions. An adapter should preserve the original fields and original paths, and must not disguise OAN extension fields as DID Core standard fields.

{\def\LTcaptype{none} 
\begin{longtable}[]{@{}>{\raggedright\arraybackslash}p{0.3200\linewidth}>{\raggedright\arraybackslash}p{0.3200\linewidth}>{\raggedright\arraybackslash}p{0.3200\linewidth}@{}}
\toprule\noalign{}
DID Core field & OAN corresponding content & Verification focus \\
\midrule\noalign{}
\endhead
\bottomrule\noalign{}
\endlastfoot
\passthrough{\lstinline!id!} & Resource or node DID & Do not rewrite it because of a name, URL, or display format \\
\passthrough{\lstinline!verificationMethod!} & Verification method of the controller or node & Public key, control relationship, and signed object are consistent \\
\passthrough{\lstinline!authentication!} / \passthrough{\lstinline!assertionMethod!} & Authentication and assertion purposes & Do not add an arbitrary public key to an authorization relationship \\
\passthrough{\lstinline!service!} & Resource endpoint and protocol entry point & Readable does not mean callable \\
\end{longtable}
}

\subsection{\texorpdfstring{\texttt{did:oan} Method Extensions}{did:oan Method Extensions}}\label{subsec:21.1.2}

The \passthrough{\lstinline!did:oan!} method extensions express the resource subject, resource type, capability set, endpoints, protocol bindings, version references, and publication-evidence associations. Method extensions must specify field types, required conditions, normalization rules, and signature coverage. Changes to extension fields should enter DID Document or resource package version management.

The method-extension checks proceed in the following order:

\begin{enumerate}
\def\labelenumi{\arabic{enumi}.}
\tightlist
\item
  Parse the method name and method-specific identifier.
\item
  Use the five-character routing code to select or verify the Registrar route.
\item
  Check metadata, services, versions, and resource package references.
\item
  Independently verify the control proof, hash, registration VC, and Root proof.
\end{enumerate}

An extension that affects signature coverage, resource type, version selection, or governance decisions should be treated as a breaking change and entered into the change record in Chapter~\ref{chap:24}.

\subsection{Resolution Behavior}\label{subsec:21.1.3}

A resolver receives the complete \passthrough{\lstinline!did:oan!}, first validates the method name and method-specific identifier, then reads the corresponding DID Document and verifies its structure, control relationship, status, and version. Not found, method unsupported, invalid document, missing proof, and unknown status should be returned separately; errors must not be concealed by an empty document or the most recent cache.

{\def\LTcaptype{none} 
\begin{longtable}[]{@{}>{\raggedright\arraybackslash}p{0.3200\linewidth}>{\raggedright\arraybackslash}p{0.3200\linewidth}>{\raggedright\arraybackslash}p{0.3200\linewidth}@{}}
\toprule\noalign{}
Resolution status & Meaning & Follow-up handling \\
\midrule\noalign{}
\endhead
\bottomrule\noalign{}
\endlastfoot
resolved & The document is readable and structurally valid & Continue verifying control, status, and publication evidence \\
not-found & No corresponding document was found & Do not substitute a file with the same name \\
method-unsupported & The resolver does not support this method & Stop high-risk invocation \\
invalid-document & The structure or binding is invalid & Reject and retain error evidence \\
status-unknown & The status source is stale or unavailable & Degrade service or require manual review \\
\end{longtable}
}

\subsection{Interoperability Boundaries}\label{subsec:21.1.4}

DID Core interoperability ensures that the common identity structure can be understood; it does not ensure that an external resolver understands OAN resource semantics, the Root platform\textquotesingle s Root proof, or governance status. An external system can read the DID Document and obtain entry points, but before invocation it must still verify the resource version, publication evidence, and endpoint according to OAN rules.

An interoperability report should record format interoperability and OAN semantic interoperability separately. The former proves that an external resolver can read common fields; the latter proves that the caller can understand the resource type, capabilities, version, resource package, and publication evidence. They must not be merged into a single ``compatible'' conclusion.

\section{A2A Agent Card Mapping}\label{sec:21.2}

The relationship between OAN and an A2A Agent Card can be distinguished as ``network-level resource discovery'' and ``service-level invocation description'': OAN provides stable identity, source, version, and governance association through the DID Document, resource metadata, and resource package; an A2A Agent Card describes the agent service provider, supported interaction entry points, capabilities, skills, input and output modes, and security declarations. The two can reference each other, but a mapping does not change the identity, signature, or authorization semantics of either side.

{\def\LTcaptype{none} 
\begin{longtable}[]{@{}>{\raggedright\arraybackslash}p{0.3200\linewidth}>{\raggedright\arraybackslash}p{0.3200\linewidth}>{\raggedright\arraybackslash}p{0.3200\linewidth}@{}}
\toprule\noalign{}
Mapping direction & Content preserved & Conclusions that cannot be derived \\
\midrule\noalign{}
\endhead
\bottomrule\noalign{}
\endlastfoot
OAN → A2A & Resource DID, name, description, capability tags, use cases, service endpoints, protocol bindings, and version & The mapping cannot automatically obtain A2A session authorization or business permissions \\
A2A → OAN & Service provider, capabilities, skills, input and output modes, security declarations, and original digest & The Card cannot automatically obtain OAN registration eligibility or the Root platform\textquotesingle s Root proof \\
\end{longtable}
}

The capability descriptions, use cases, endpoints, and protocol bindings in an OAN resource profile can provide a trusted discovery entry point for an A2A Agent Card. For reverse integration, an Agent Card can only serve as a source for a registration draft or supplementary description; the resource controller must still generate and sign the resource materials through the OAN process. Detailed invocation flows are determined by A2A, MCP, a tool API, or the resource\textquotesingle s own protocol.

\subsection{Capability Description Mapping}\label{subsec:21.2.1}

OAN\textquotesingle s \passthrough{\lstinline!capabilityTags!}, \passthrough{\lstinline!capabilityDescription!}, and use-case information can be mapped to the \passthrough{\lstinline!skills!}, skill names, descriptions, and example tasks of an A2A Agent Card. During mapping, preserve the tag-tree path, original text, resource package version, and source. The semantic matching result from an OAN Discovery node can help locate candidate resources, but it cannot prove that a resource is authorized or necessarily possesses the declared capability.

It is recommended to preserve both normalized tags and the original description:

Normalized tags serve filtering and semantic governance; the original text serves explanation and review.

\subsection{Use-Case and Skill Mapping}\label{subsec:21.2.2}

OAN use-case fields and skill package information can be used to generate the skill summary, example tasks, and input and output descriptions of an Agent Card. The generated result should retain the resource package version and field sources. If an external Card provides only a summary, reverse registration can produce only a resource description requiring verification; it cannot fill in control, version, endpoint, or governance information that OAN does not contain.

{\def\LTcaptype{none} 
\begin{longtable}[]{@{}>{\raggedright\arraybackslash}p{0.3200\linewidth}>{\raggedright\arraybackslash}p{0.3200\linewidth}>{\raggedright\arraybackslash}p{0.3200\linewidth}@{}}
\toprule\noalign{}
Use-case field & Fact it should express & Handling when missing \\
\midrule\noalign{}
\endhead
\bottomrule\noalign{}
\endlastfoot
Task & The work the user wants to complete & Mark the description as incomplete \\
Input & Type, source, and prerequisites & Do not invent content \\
Output & Result form or deliverable & Do not treat a tag as a promise \\
Constraints & Permission, cost, timeliness, and data boundaries & Recheck before invocation \\
\end{longtable}
}

\subsection{Service Endpoint Mapping}\label{subsec:21.2.3}

Service endpoints, protocol types, transport methods, and authentication requirements in an OAN DID Document or resource package can be mapped to the \passthrough{\lstinline!supportedInterfaces!}, protocol bindings, and security declarations of an Agent Card. Endpoint mapping only provides connection-preparation information and does not mean that the endpoint is available; the caller must still perform TLS, authentication, version, status, and business-authorization checks. Endpoint mappings can record ``entry point, protocol, transport, authentication, version, and status'' item by item. When any item comes from an external description, retain the original path or digest so that it can be checked again before invocation.

{\def\LTcaptype{none} 
\begin{longtable}[]{@{}>{\raggedright\arraybackslash}p{0.3200\linewidth}>{\raggedright\arraybackslash}p{0.3200\linewidth}>{\raggedright\arraybackslash}p{0.3200\linewidth}@{}}
\toprule\noalign{}
Endpoint information & Use during discovery & Recheck before invocation \\
\midrule\noalign{}
\endhead
\bottomrule\noalign{}
\endlastfoot
Entry-point URL & Locate the service & TLS, domain name, and reachability \\
Protocol/transport & Select a client & Version and negotiation result \\
Authentication requirements & Prepare credentials & User authorization and permission scope \\
Version & Select a compatible implementation & Whether it is still supported \\
Status & Indicate availability & Latest status and governance evidence \\
\end{longtable}
}

\subsection{Identity, Signature, and Trusted-Evidence Mapping}\label{subsec:21.2.4}

The \passthrough{\lstinline!provider!}, signature, or security declaration of an Agent Card can be stored as external input, but the OAN resource identity is determined by \passthrough{\lstinline!did:oan!} and its control relationship. The overall signature of an Agent Card proves the Card\textquotesingle s content and its publisher\textquotesingle s statement; a registration VC, the Root platform\textquotesingle s Root proof, and governance status prove different facts in the OAN process and must be retained separately rather than replacing one another.

Identity and evidence mappings should be stored in layers:

\begin{itemize}
\tightlist
\item
  The resource control relationship answers ``who can control this DID.''
\item
  The Agent Card or external declaration answers ``how does the external system describe this resource.''
\item
  The registration VC answers ``what registration fact did the Registrar node declare.''
\item
  The Root proof answers ``what content did the Root platform verify and publish.''
\end{itemize}

\section{MCP Resource and Server Metadata Mapping}\label{sec:21.3}

OAN treats an MCP service as a discoverable resource. Mapping should record three layers: ``server, tool, and invocation.'' The server layer describes identity, protocol version, transport, and endpoint; the tool layer describes the tool or resource inventory, names, descriptions, and input schemas; the invocation layer describes sessions, authentication, and error constraints. The resource package should preserve the inventory original or a stable reference, version, and hash; the discovery index may store only a retrieval summary. A successful MCP initialization only means that the service completed one protocol interaction; it does not mean that the Root platform has verified and published the resource.

\subsection{MCP Server Mapping}\label{subsec:21.3.1}

The name, description, protocol version, transport, endpoint, and server capabilities of an MCP Server are mapped to the metadata of an OAN MCP service resource. The server DID, control proof, resource version, resource package hash, and Root proof are stored independently to distinguish ``server self-description'' from ``network-verified publication.'' Each tool record should also retain its server and resource package version to avoid deduplicating across services solely by tool name.

\subsection{Tool Definition Mapping}\label{subsec:21.3.2}

MCP tool names, descriptions, input schemas, and available capabilities can be mapped to OAN tool entries, capability descriptions, and tags. A tool entry should include its server, resource package version, content digest, and location in the original inventory. Identical tool names do not mean that the tools belong to the same controller. A discovery index may use summary fields, but the caller should be able to return to the complete inventory to verify the tool boundary.

\subsection{Input and Output Constraint Mapping}\label{subsec:21.3.3}

MCP input schemas, result structures, and error constraints are mapped to OAN input and output descriptions, while preserving the original schema, media type, version, and hash. If mapping fails or fields are incomplete, the resource may be registered as incompletely described, but the caller must not be told that the constraints have been verified. Input and output mappings can use a schema version and digest to establish a stable reference. When field types, required relationships, enumeration values, or error structures change, treat the change as a new description version rather than merely updating display text.

\subsection{Invocation and Authentication Information Mapping}\label{subsec:21.3.4}

MCP transport, session establishment, authentication requirements, and pre-invocation conditions can form part of OAN service information and support connection preparation after discovery. OAN publication proof proves only the declared materials and their source; it cannot grant business permissions to an MCP server or guarantee session success. Invocation and authentication mappings should distinguish ``access conditions visible during discovery'' from ``permissions granted by the service at runtime.'' OAN registers and verifies declarations and their sources, but does not replace the business authentication of an MCP service or tool API.

\section{Tool and API Description Mapping}\label{sec:21.4}

OpenAPI mapping should establish stable entries at the path-and-operation level:

A normalized summary can be used for discovery and deduplication; the original document is used for pre-invocation review.

A tool API should use the path and operation as the minimum audit unit and retain at least the operationId, original document path, server endpoint, schema version, authentication requirements, and digest. The normalized summary is used for discovery and deduplication; the original document is used for pre-invocation review. A readable document does not mean that the interface is callable.

\section{VC and VP Ecosystem Interoperability}\label{sec:21.5}

VC/VP processing should follow the sequence ``parse → cryptographic verification → status check → trust-policy decision.'' Each failure must produce a clear intermediate status; ``valid JSON'' must not be displayed directly as ``trusted.''

{\def\LTcaptype{none} 
\begin{longtable}[]{@{}>{\raggedright\arraybackslash}p{0.3200\linewidth}>{\raggedright\arraybackslash}p{0.3200\linewidth}>{\raggedright\arraybackslash}p{0.3200\linewidth}@{}}
\toprule\noalign{}
Stage & Output & Is it sufficient for acceptance? \\
\midrule\noalign{}
\endhead
\bottomrule\noalign{}
\endlastfoot
Parsing & Subject, issuer, type, and claims & No \\
Signature verification & Signature matches the public key & No; the relationship must also be checked \\
Status check & Validity period and suspension/revocation status & No; the authorization scope must also be checked \\
OAN trust policy & Decision based on source and governance & Yes / No / Review required \\
\end{longtable}
}

An adapter can parse the subject, issuer, credential type, validity period, status, and proof materials of an external VC/VP, and retain the original object and verification result. Being parseable only means that the format is readable; acceptance also depends on issuer authorization, control relationship, governance status, and the OAN trust policy. For input whose status information cannot be obtained, whose signature verification fails, or whose issuer relationship is unclear, retain the failure reason and route it to rejection or manual review.

\section{External Organization and Platform Identity Mapping}\label{sec:21.6}

Organizations, domains, and platform accounts are association evidence and must not be merged into an OAN controller merely because names match. A Registrar may verify a real enterprise or individual under its operating rules and express responsibility for that verification in the registration VC it issues. The resource DID Document then uses its controller relationship to establish a verifiable chain of update responsibility. If that chain cannot be traced to a verified entity, the result should state that evidence is insufficient rather than letting an adapter guess ownership.

An external organization, platform account, domain name, or its DID can be stored as associated information for a resource provider, operator, delegate, or endpoint host, but the relationship type, evidence source, responsibility scope, and validity period should be recorded. A domain name, enterprise account, platform authentication record, and resource-control DID are not the same field. An external account name cannot replace control of the resource DID, and domain or platform authentication cannot directly create eligibility for Root platform publication.

OAN records external namespace associations in \passthrough{\lstinline!oanMetadata.externalIdentifiers!} of the DID Document. Each entry has a required \passthrough{\lstinline!id!} and an optional \passthrough{\lstinline!resolutionServiceEndpoint!}; they are not concatenated into a string that obscures the distinction between identifier and resolution entry point:

\begin{lstlisting}
{
  "externalIdentifiers": [
    {
      "id": "urn:example:agent-service:contract-review",
      "resolutionServiceEndpoint": "https://resolver.example.org/resources/contract-review"
    }
  ]
}
\end{lstlisting}

This example shows structure only and makes no claim about ownership of a real resource. External values are retained verbatim and interpreted by their own namespaces. If an external namespace, resolution protocol, or proof rule is unclear, the association can be displayed only as a declaration by the resource provider.

\section{External Credential and Trust-Evidence Mapping}\label{sec:21.7}

An external evidence record is recommended to include \passthrough{\lstinline!issuer!}, \passthrough{\lstinline!subject!}, \passthrough{\lstinline!credentialType!}, \passthrough{\lstinline!issuedAt!}, \passthrough{\lstinline!expiresAt!}, \passthrough{\lstinline!statusSource!}, \passthrough{\lstinline!sourceHash!}, \passthrough{\lstinline!verifiedAt!}, and \passthrough{\lstinline!verifier!}. When no status endpoint is available, record ``status not checked'' rather than assuming validity. OAN treats it as supplementary evidence, while node authorization, resource control, and Root platform publication are still determined by their respective OAN evidence chains. An external identifier uses a namespace, identifier, relationship type, source, and optional evidence reference rather than placing an opaque string directly into a resource record. A registration VC may carry a normalized list of external-identifier \passthrough{\lstinline!id!} values in its credential subject, while resolution remains in the DID Document\textquotesingle s \passthrough{\lstinline!oanMetadata!}, protected by the document proof and \passthrough{\lstinline!didDocumentHash!}. A verifier first checks the VC issuer, resource DID, version, and three digests, then compares the complete external-identifier entries in the corresponding DID Document. An ID appearing in a VC proves only that the Registrar included it in a registration statement; it does not prove that the Registrar verified ownership under the external platform\textquotesingle s rules.

Identity Runtime, Gateway, and Sidecar provide three adaptation forms: in-code or trusted local process, shared ingress, and workload sidecar. They can associate OAN proofs with A2A, MCP, or HTTP calls while preserving business payloads, streaming responses, and upstream error semantics. Final business authentication remains the responsibility of the resource endpoint.

{\def\LTcaptype{none} 
\begin{longtable}[]{@{}>{\raggedright\arraybackslash}p{0.2400\linewidth}>{\raggedright\arraybackslash}p{0.2400\linewidth}>{\raggedright\arraybackslash}p{0.2400\linewidth}>{\raggedright\arraybackslash}p{0.2400\linewidth}@{}}
\toprule\noalign{}
Form & Key location & Proof encapsulation point & Streaming and failure boundary \\
\midrule\noalign{}
\endhead
\bottomrule\noalign{}
\endlastfoot
Runtime & Local encrypted store or controlled Key Provider & Native client/service adapter & Does not mark a result as trusted before verification completes \\
Gateway & Runtime, KMS/HSM, or controlled remote signer & Proxy envelope at an enterprise ingress & A forwarding failure does not prove that the business action did not execute; blind retry is prohibited \\
Sidecar & Managed by Runtime; Sidecar retains no long-term key & Inbound/outbound boundary beside the business process & May forward a buffered stream; fails closed when the terminal proof is missing \\
\end{longtable}
}

These forms do not redefine A2A task/message semantics, MCP tools/resources/prompts, or HTTP state and error semantics. Existing reference combinations show that several synchronous and SSE paths work; they do not prove live interoperability for every asynchronous operation or third-party implementation.

\pandocbounded{\includegraphics[width=\linewidth,height=0.72\textheight,keepaspectratio]{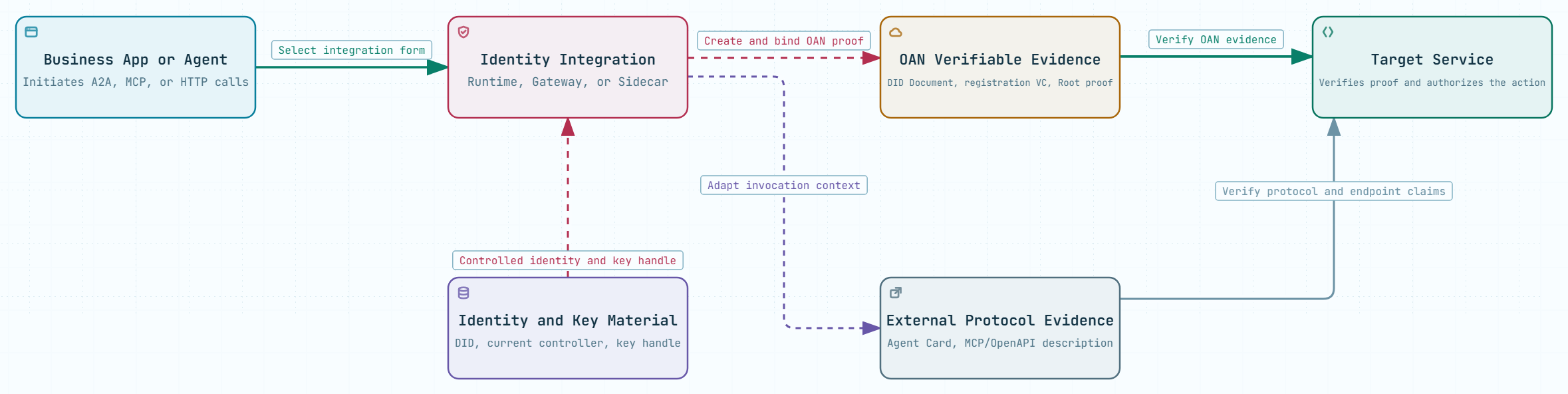}}

``Identity-integration component'' in the figure is a common abstraction for three deployment forms, not a fourth product. An application selects Runtime, Gateway, or Sidecar according to its runtime boundary. The component reads controlled identity and key handles and binds an OAN proof to the invocation context. OAN-verifiable evidence and external-protocol descriptions enter separate verification paths at the target service, which remains responsible for final business authorization. The figure uses a blueprint visual treatment without a legend; editable Archify JSON and HTML and PNG/PDF outputs are stored under \passthrough{\lstinline!figs/archify!}.

\section{Identity-Integration Depth and Integration Paths}\label{sec:21.8}

Completing field mapping for an external protocol shows only that a caller can interpret resource descriptions and protocol entry points. It does not show that actual communication is protected by OAN identity mechanisms. Integration depth depends on whether both parties support the required capabilities, whether identity checks lie on the real request path, and whether validation failure prevents business execution or prevents a result from being accepted as trusted. Installation of Community Skill alone is insufficient to establish integration depth.

{\def\LTcaptype{none} 
\begin{longtable}[]{@{}>{\raggedright\arraybackslash}p{0.2400\linewidth}>{\raggedright\arraybackslash}p{0.2400\linewidth}>{\raggedright\arraybackslash}p{0.2400\linewidth}>{\raggedright\arraybackslash}p{0.2400\linewidth}@{}}
\toprule\noalign{}
Level & Required capability & Result that can be established & Capability that must not be claimed \\
\midrule\noalign{}
\endhead
\bottomrule\noalign{}
\endlastfoot
L0: registration and discovery & Manage OAN identities; register, discover, and query resources; obtain a native-protocol entry point & Resource enters the network directory and the caller locates candidates by DID or conditions & Does not prove that a business call authenticated its peer \\
L1: single-ended identity integration & This endpoint securely holds identity, verifies obtainable OAN evidence, and applies local policy & Forms a single-ended check and audit result before or after calling an unadapted service & Cannot claim that the unadapted remote holds the corresponding private key or verified this endpoint\textquotesingle s identity \\
L2: mutual identity and message protection & Both sides prove possession of their resource private keys; requests and responses bind audience, operation, digest, freshness, and anti-replay state & Mutually authenticates identities on the actual call path and protects the source of a task, tool parameters, response, or result & Does not automatically provide end-to-end encryption, instance-level revocation, or complete business delegation \\
L3: runtime and business extensions & Adds instance keys, constrained delegation, instance revocation, end-to-end encryption, or runtime attestation to L2 as the scenario requires & Supports multiple instances, multiple hops, private data exchange, and fine-grained enterprise policy & Passing one extension does not establish that other L3 extensions are implemented \\
\end{longtable}
}

L3 is a set of extensions that can be selected, implemented, and validated independently, not one higher authentication badge. For example, an implementation may support constrained delegation without runtime attestation, or instance revocation without end-to-end encryption. A conformance report lists every extension capability and its evidence separately.

\subsection{Integration Form and Enforcement Point}\label{subsec:21.8.1}

An integration form is selected according to whether agent code can be changed, whether inbound and outbound interception points exist, who holds keys, and whether the peer is adapted. ``Reachable level'' in the table is an architectural upper bound, not a level obtained automatically after deployment.

{\def\LTcaptype{none} 
\begin{longtable}[]{@{}>{\raggedright\arraybackslash}p{0.2400\linewidth}>{\raggedright\arraybackslash}p{0.2400\linewidth}>{\raggedright\arraybackslash}p{0.2400\linewidth}>{\raggedright\arraybackslash}p{0.2400\linewidth}@{}}
\toprule\noalign{}
Integration form & Condition required of the agent or platform & Normally reachable level & Enforcement point \\
\midrule\noalign{}
\endhead
\bottomrule\noalign{}
\endlastfoot
Community Skill & Can read and execute Skill workflows and provide a controlled local directory for identity material & Currently mainly L0 & Tool boundary of registration, discovery, and query actions \\
OAN MCP tool service or CLI & Can call MCP tools or local commands & Direct L0; can use L2 capability together with protected-call tools & Tool service or command executor; ordinary business requests do not become protected automatically \\
Identity Sidecar or protocol proxy & Can configure a proxy address, service endpoint, or launch command & L1 at one endpoint; L2 when both parties use a compatible profile & Inbound and outbound boundary beside the business process \\
Protocol middleware or Identity Runtime plugin & A2A, MCP, or HTTP stack has request, response, and tool-execution interception points & L1--L2, with optional L3 extensions & Before and after native protocol processing; validation failure must fail closed \\
Native SDK integration & Agent or service code can be modified & L0--L2, with optional L3 extensions & Creation and consumption of tasks, messages, results, and business objects \\
Identity Gateway or controlled remote signer & Can configure a trusted gateway, key provider, and tenant/policy boundary & L1--L3 depending on protocol and evidence support & Enterprise ingress and centralized signing and policy boundary \\
\end{longtable}
}

The table includes current reference components and optional interface forms. ``OAN MCP tool service'' and ``controlled remote signer'' describe integration boundaries; they do not assert that the current public implementation necessarily provides independent services with those names. Implementation status is determined from the content status in Chapter~\ref{chap:00}, public repositories, and corresponding deployment evidence.

The existence of a signature-verification tool and protection of an invocation are different facts. If an agent can bypass the verification tool and access the business endpoint directly, or if the business logic still consumes a result after verification fails, the path cannot be recorded as L2. Ingress convergence, protected clients, reverse-proxy policy, or mandatory hooks in the business handler must cause validation failure to prevent request entry or prevent a response from being accepted as trusted.

\subsection{Integration Priorities for Different Protocols and Runtime Environments}\label{subsec:21.8.2}

{\def\LTcaptype{none} 
\begin{longtable}[]{@{}>{\raggedright\arraybackslash}p{0.3200\linewidth}>{\raggedright\arraybackslash}p{0.3200\linewidth}>{\raggedright\arraybackslash}p{0.3200\linewidth}@{}}
\toprule\noalign{}
Object or protocol & L0/L1 integration priority & Additional work required for L2 \\
\midrule\noalign{}
\endhead
\bottomrule\noalign{}
\endlastfoot
Skill-capable user assistant & Use Community Skill to manage resources and perform exact DID discovery; retain identity material within a controlled local boundary & Integrate the identity component into the actual A2A, MCP, or HTTP invocation entry point rather than relying on the model to decide whether to verify \\
A2A agent & Register and discover the Agent; verify Agent Card, DID, endpoint, and registration/publication evidence & Establish request-response binding, signature verification, and anti-replay checks for task requests, status updates, streaming events, callbacks, and Artifacts \\
MCP Host, Client, and Server & Register and discover the MCP Server; verify tool inventory, transport, and server evidence & After initialization, bind mutual identities, parameter and result digests, request correlation, and tool allowlists to tool calls \\
ANP or another HTTP agent with its own identity mechanism & Verify external and OAN identity materials separately and retain both evidence chains & Define and enforce a verifiable cross-system binding; a string mapping between external DIDs is not identity mutual recognition \\
Custom HTTP or framework agent & Use SDK, middleware, or proxy for registration, discovery, and single-ended checks & Validate audience, operation, digest, freshness, and response proof on the service\textquotesingle s real inbound and outbound paths \\
Closed-platform agent & Integrate through MCP, HTTP, tool, or webhook interfaces permitted by the platform & L1/L2 is possible only when the complete communication boundary can be controlled and validation enforced; a wholly non-configurable platform cannot obtain these capabilities transparently \\
\end{longtable}
}

MCP \passthrough{\lstinline!stdio!} requires separate treatment. It mainly relies on local process and IPC permissions. Adding a signature outside an ordinary \passthrough{\lstinline!stdio!} tool can prove only that a trusted proxy forwarded or signed a result on the tool\textquotesingle s behalf; it does not prove that the tool process itself holds an independent OAN private key. Process-level identity requires the tool process to use a controlled identity handle directly or independent material proving the proxy\textquotesingle s authorization relationship.

\subsection{Multiple-Identity Selection and User Authorization}\label{subsec:21.8.3}

A user assistant may communicate under its own DID while managing multiple resource identities, but each operation selects an identity according to purpose. ``Current logged-in user,'' ``assistant identity,'' and ``target resource identity'' must not be merged into one default signer:

\begin{itemize}
\tightlist
\item
  Registration confirmation, DID Document management, and controller changes use the controller identity authorized by the current control rules.
\item
  Tasks, tool results, and business-data signatures use the corresponding resource identity or an instance identity with an explicit delegation scope.
\item
  User approval is completed in an independent, trusted, auditable interface; model output stating that ``the user approved'' does not replace actual authorization.
\item
  An identity service returns a constrained handle, public DID, and necessary state to an assistant; it does not expose the long-term private keys of multiple resources to model context.
\end{itemize}

Selecting the wrong identity changes proof semantics as well as key use. A controller\textquotesingle s authorization of a resource update cannot replace the resource\textquotesingle s signature over a business result, and a resource signature cannot replace user approval of a high-risk action. Adapters and audit records retain the DID, verification method, operation purpose, resource version, and authorization source used, so these facts remain distinguishable.

\subsection{Participation Stage and Integration Depth Are Separate Dimensions}\label{subsec:21.8.4}

The White Paper summarizes ecosystem participation as observing, publishing, discovering, integrating, and operating. It describes what work a participant performs in the network; L0--L3 describes the identity-protection depth of one agent-interaction path. The dimensions are orthogonal. Operating infrastructure does not automatically mean the operator\textquotesingle s business calls are L2, and an L2 business application does not acquire qualification to operate a Registrar or Discovery node.

{\def\LTcaptype{none} 
\begin{longtable}[]{@{}>{\raggedright\arraybackslash}p{0.3200\linewidth}>{\raggedright\arraybackslash}p{0.3200\linewidth}>{\raggedright\arraybackslash}p{0.3200\linewidth}@{}}
\toprule\noalign{}
Participation stage & Typical action & Relationship to L0--L3 \\
\midrule\noalign{}
\endhead
\bottomrule\noalign{}
\endlastfoot
Observe & View the public portal, Network page, documentation, and redacted metrics & Produces understanding and observation evidence only; does not pass any integration level \\
Publish & Create identities for a small number of resources and complete governed registration & Usually begins at L0 and may prepare identity and evidence for later L1/L2 use \\
Discover & Obtain trusted candidates by DID or condition and verify their origin & A core L0 capability; whether to continue to L1/L2 depends on the invocation path \\
Integrate & Consume identity capability in an SDK, Runtime, Sidecar, Gateway, or protocol adapter & May remain at L0/L1 or reach L2 through mutual mandatory enforcement \\
Operate & Run a Registrar, Discovery, or other infrastructure role under governance admission & Acquires node qualification only within a limited role and authorization domain; grants neither resource control nor business invocation permission \\
\end{longtable}
}

Adoption can begin wherever appropriate; partners do not have to follow one fixed sequence. An existing MCP, A2A, HTTP, or industry system may first consume Discovery, first operate a constrained Discovery node, or first add L1 verification at an existing ingress. Each step separately records the participant role, authorization domain, actual enforcement point, and level passed so that ``integrated with OAN'' does not conceal the scope of capability.

\section{Field Source, Version, and Hash Requirements}\label{sec:21.9}

Each mapped field records the source object, source path, original version, normalized version, mapping-rule version, signing subject, content digest, and verification time. Normalization is replayable. When ordering, case folding, aliases, or field trimming changes the digest, the digest and mapping version are updated and the old result retained for audit. A crawled description proves only the data source, not that the crawler controls the corresponding resource. When no chain through the controller and Registrar responsibility can reach a verified entity, the evidence gap must be displayed explicitly.

\section{Compatibility Profiles and Migration Paths}\label{sec:21.10}

An interoperability profile should specify the supported DID, A2A Agent Card, MCP, and OpenAPI versions; required and optional fields; signature algorithms; transport; error policy; and migration method. During migration, first retain the original evidence, read the old and new profiles in parallel, check field digests and signature boundaries, and then switch to single-write operation. If migration fails, return to the old profile without overwriting the sole source. The profile itself should have a name and version so that callers know which set of mapping rules is being used.

\section{External Protocol Capability Negotiation}\label{sec:21.11}

Capability results may be \passthrough{\lstinline!compatible!}, \passthrough{\lstinline!partially-compatible!}, or \passthrough{\lstinline!incompatible!}, while reporting the common protocol, version, transport, and security capabilities and the missing items. OAN currently has no unified runtime-negotiation endpoint or standard response object. An adapter can determine a profile only from an actual Agent Card, MCP initialization result, OpenAPI document, or local configuration. Successful negotiation means only that the format and interaction capability are available; resource identity, publication evidence, governance status, and business permission still require separate verification.

OAN currently places governance facts such as node authorization and revocation on chain, while keeping DID Documents, ResourcePackages, discovery indexes, and business invocation data off chain. An external blockchain identity scheme is integrated by mapping verifiable facts and references, not by copying all resource content on chain.

\section{External Signature and Evidence Retention}\label{sec:21.12}

An adapter\textquotesingle s output should store the original text, normalized digest, external signature, and OAN signature separately, and assign a source identifier to each item of material. When repackaging or forwarding, do not retain only the transformed object; it should at least be possible to answer ``what was the original object, who signed it, what did they sign, did the transformation change the content, and which OAN party signed again.''

The verification order for a resource with external identifiers is shown below. When any evidence is unknown, the unknown state is retained rather than being converted into success by another evidence chain.

\begin{lstlisting}
Resolve did:oan and the DID Document
  -> verify resource/controller control proof
  -> recompute DID Document, metadata, and ResourcePackage digests
  -> verify Registrar registration VC
  -> verify Root publication proof and governance state
  -> read externalIdentifiers and resolution entry points
  -> verify external evidence under the external namespace's rules
  -> combine both results into an invocation or manual-review decision
\end{lstlisting}

The adaptation record also retains credential references, object version, digest, verification time, verifier, and failure reason. When re-encoding or generating an OAN ResourcePackage, the external signature, resource-control signature, registration VC, and Root trusted-publication proof remain separate. An audit result can answer what each proof covers, whether transformation changed content, and when verification occurred.

OAN Identity Runtime, Gateway, and Sidecar respectively provide embedded identity services, a centralized ingress proxy, and a workload sidecar. They map OAN DIDs, request/response proofs, and verification results into A2A, MCP, or ordinary HTTP sessions without redefining those protocols\textquotesingle{} task, tool, streaming, cancellation, or error models. A deployment selects the form matching its key-custody and network boundaries; all three are not required simultaneously.

An external identity or credential may be an input to Registrar real-world identity verification, relationship binding, or invocation policy, but it does not automatically become an OAN controller. A mapping states the external issuer, validation and revocation policy, OAN authorizer, relationship validity period, and update path after invalidation. An external system name or runtime description does not itself prove control.

\section{References}\label{sec:21.13}

{\def\LTcaptype{none} 
\begin{longtable}[]{@{}>{\raggedright\arraybackslash}p{0.3200\linewidth}>{\raggedright\arraybackslash}p{0.3200\linewidth}>{\raggedright\arraybackslash}p{0.3200\linewidth}@{}}
\toprule\noalign{}
Source & Type & Link \\
\midrule\noalign{}
\endhead
\bottomrule\noalign{}
\endlastfoot
\passthrough{\lstinline!oan-protocol-commons!} & Code repository: resource metadata, protocol types, and external-description bindings & \url{https://github.com/wolfbrother/oan-protocol-commons} \\
\passthrough{\lstinline!oan-sdk-ts!} & Code repository: client-side resource materials and protocol-mapping helpers & \url{https://github.com/OpenAgenet/oan-sdk-ts} \\
\passthrough{\lstinline!oan-community-skill!} & Code repository: community integration scripts and external-resource registration flows & \url{https://github.com/OpenAgenet/oan-community-skill} \\
OAN White Paper & arXiv white paper: ecosystem interoperability and open-infrastructure positioning & \url{https://arxiv.org/abs/2606.03161} \\
OAN Yellow Paper & arXiv yellow paper: resources, trust, and verification mechanisms & \url{https://arxiv.org/abs/2606.03163} \\
Agentic Overlay Network Architecture & IETF draft: agent network architecture mapping & \url{https://datatracker.ietf.org/doc/draft-xu-agentic-overlay-network-architecture/} \\
Efficient Agent Discovery Profile & IETF draft: discovery metadata and query profile & \url{https://datatracker.ietf.org/doc/draft-xu-efficient-agent-discovery-profile/} \\
\end{longtable}
}

\chapter{End-to-End Reference Flows}\label{chap:22}

The flows in this chapter consistently use the state chain of ``preparation, submission, verification, propagation, discovery, invocation, update or recovery.'' Interface names, website buttons, and deployment addresses in the reference implementation belong to the example environment; the specification does not require every implementation to use the same interface, but it does require key facts to be independently verifiable. An end-to-end record must be associated with at least the resource DID, resource type, version, request identifier, source node, observation time, and evidence summary. ``Completion'' in a flow must be determined by the corresponding facts and evidence: the Registrar node returning an acceptance result, the Root platform forming a publication fact, the content distribution platform completing material transfer, and the Discovery node establishing an index are different stages and must not be merged into one success state.

{\def\LTcaptype{none} 
\begin{longtable}[]{@{}>{\raggedright\arraybackslash}p{0.3200\linewidth}>{\raggedright\arraybackslash}p{0.3200\linewidth}>{\raggedright\arraybackslash}p{0.3200\linewidth}@{}}
\toprule\noalign{}
Stage & Primary fact & Qualified output \\
\midrule\noalign{}
\endhead
\bottomrule\noalign{}
\endlastfoot
Preparation & Local Identity, DID Document, resource metadata, and ResourcePackage & Verifiable control materials \\
Registration & Registrar node acceptance and upstream submission & Registration result, VC (if any), and associated ID \\
Publication & Root platform verification and formation of a publication fact & Root proof \\
Propagation & content distribution platform transfer and cursor recording & Verifiable ResourcePackage and synchronization result \\
Discovery & Discovery node verification and index establishment & Candidate results with source and freshness \\
Invocation & Caller re-verification and execution of local policy & Decision to allow or reject invocation \\
\end{longtable}
}

\pandocbounded{\includegraphics[width=\linewidth,height=0.72\textheight,keepaspectratio]{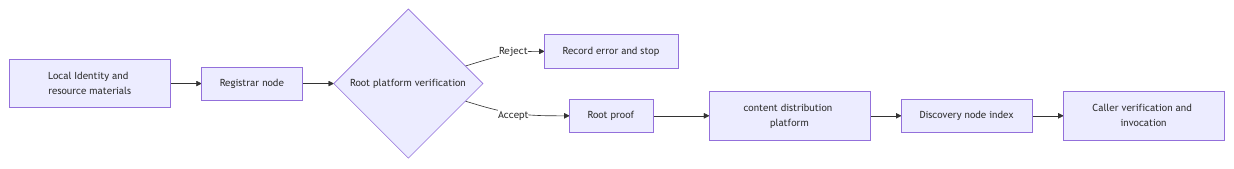}} This chapter assumes that the Local Identity and test resources are ready, node endpoints are accessible, and the test and production environments are isolated. It provides a reference path from resource creation, registration, trusted publication, discovery, and verification to update and fault handling. The examples show reproducible interaction relationships; they do not elevate a particular website button, current deployment address, or one successful response into a mandatory interface for all implementations. During execution, test resources, test subjects, and request identifiers should be isolated from production resources, and the associations among request materials, response summaries, DID Documents, registration credentials, Root proofs, synchronization cursors, and query results should be preserved.

\pandocbounded{\includegraphics[width=\linewidth,height=0.72\textheight,keepaspectratio]{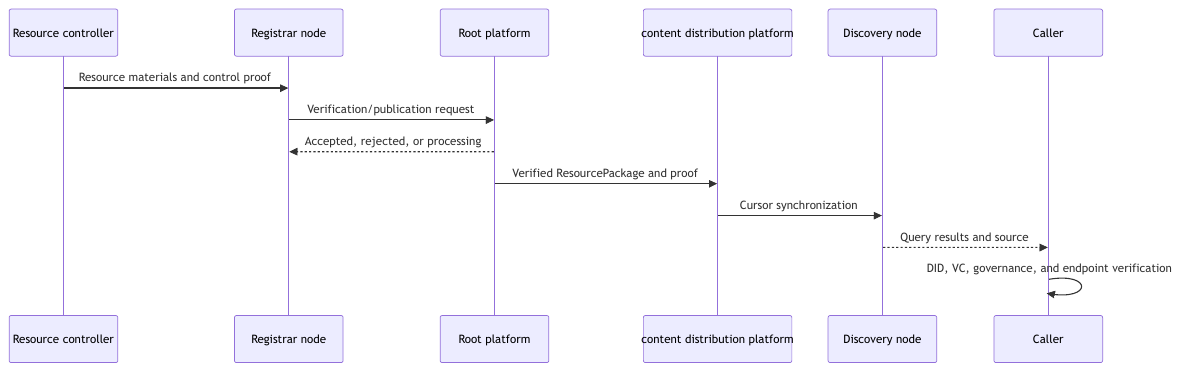}}

Every step should record the input source, fact producer, verifier, state change, and evidence; a page prompt or a single successful HTTP response cannot represent end-to-end completion.

\section{Create a Local Identity}\label{sec:22.1}

The client generates the subject keys and subject DID locally, establishes a Local Identity record, and exports an identity backup file with a time identifier. When registering an agent service, skill package, MCP service, or tool API, the subject then generates or associates the resource identity and resource control proof locally. The private key remains within the publisher's control and is not sent with an ordinary registration request; during recovery, the file structure, correspondence between the public and private keys, consistency of the subject DID, and the ability to sign again must be re-verified. Acceptance evidence includes the DID, public key, backup summary, and equivalence after recovery, rather than treating successful private-key upload as the result.

{\def\LTcaptype{none} 
\begin{longtable}[]{@{}>{\raggedright\arraybackslash}p{0.3200\linewidth}>{\raggedright\arraybackslash}p{0.3200\linewidth}>{\raggedright\arraybackslash}p{0.3200\linewidth}@{}}
\toprule\noalign{}
Stage & Local action & Acceptance evidence \\
\midrule\noalign{}
\endhead
\bottomrule\noalign{}
\endlastfoot
Generate & Generate the subject DID, public key, and private key & The public key is consistent with the subject DID Document \\
Establish record & Save the subject and the resource identity records under its control & The record structure and version are valid, and the control relationship is clear \\
Back up & Export the identity package from the browser or client & The file is parseable, and the private key was not sent to a node \\
Recover & Import and verify the key relationship & The DID, public key, and signing capability are equivalent before and after recovery \\
\end{longtable}
}

\section{Register an Agent Service}\label{sec:22.2}

Take as an example an agent service that declares ``code repository search and project structure summarization.'' The publisher first prepares the resource DID, capability tags, capability description, use cases, endpoint, protocol declaration, authentication conditions, and version, and then signs the DID Document or ResourcePackage materials with the local control key. The client submits the public registration materials and control proof to the Registrar node; the Registrar node validates the request and submits it to the Root platform, returning a registration result and a registration VC if one is available. The publisher must then wait for Root platform publication, content distribution, and Discovery node indexing, and obtain the corresponding states and evidence separately.

Registration materials, request summaries, the registration VC (if any), Root platform status, ResourcePackage summary, and discovery results should be retained separately. ``Submission succeeded'' must not be written as ``discovered across the network.'' When submitting the same resource again, first determine whether the resource DID, version, or materials have changed; repeated requests for the same version should be judged by the actual response of the Registrar node as to whether they are idempotent, and a new version must not be inferred from a page prompt.

\section{Register a Skill Package}\label{sec:22.3}

A skill package example should provide its name, capabilities, inputs and outputs, use cases, dependencies, version, installation or invocation entry point, tags, subject DID, and resource DID, and calculate the content digest of the ResourcePackage. The publisher first normalizes and signs the materials locally, then submits them to the Registrar node; after submission, it checks the acceptance result, resource type, version, digest, and registration credential if any, and then checks Root platform publication and the Discovery node index in the order specified in 22.6. A skill package being installable or runnable locally does not mean that it has received Root platform publication or a Discovery node index.

Skill package acceptance checks two lines at the same time: the \textbf{product line} checks the entry point, dependencies, inputs and outputs, and use cases; the \textbf{network line} checks the DID, ResourcePackage version, digest, control signature, registration, publication, and index status. The results of the two lines are recorded separately. Failure of the product line means that the resource itself cannot be used as declared; failure of the network line means that the resource has not formed a verifiable network publication chain. The pass result of one line cannot replace the other.

\section{Register an MCP Service}\label{sec:22.4}

An MCP service example should prepare the server identity, tool or resource inventory, transport method, endpoint, authentication method, protocol version, and compatibility declaration. The publisher includes the inventory digest, service endpoint, and resource version in the resource materials and signs them locally. The Registrar node verifies the fields and control proof; after discovery, the caller must still establish a session according to the MCP flow and check the DID, ResourcePackage, publication evidence, and governance status according to OpenAgenet (OAN) rules.

{\def\LTcaptype{none} 
\begin{longtable}[]{@{}>{\raggedright\arraybackslash}p{0.3200\linewidth}>{\raggedright\arraybackslash}p{0.3200\linewidth}>{\raggedright\arraybackslash}p{0.3200\linewidth}@{}}
\toprule\noalign{}
Check result & Fact represented & Fact not represented \\
\midrule\noalign{}
\endhead
\bottomrule\noalign{}
\endlastfoot
Field validation passed & The Registrar node can understand the materials & The Root platform has published them \\
Registration VC returned & The registration fact has a credential & An MCP session will necessarily succeed \\
Root platform publication & A network publication fact has formed for a specific resource version & Business permission is automatically granted \\
MCP session succeeded & One connection interaction succeeded & The resource will remain available or its governance status is valid \\
\end{longtable}
}

\section{Register a Tool API}\label{sec:22.5}

A tool API takes an OpenAPI or equivalent description as input and specifies the operations, parameters, request body, responses, errors, authentication, server endpoint, and version. The publisher first determines the source and version of the description file, then binds the normalized description digest to the ResourcePackage and signs it; the Registrar node verifies the materials, and the caller can confirm through the digest that the obtained API contract has not been replaced. Digest verification cannot prove that the interface is currently available; actual business permissions are still determined by the API server.

The original description, normalized digest, and OAN control signature should be saved separately; being readable as a document does not mean that the interface is callable.

\section{Publish and Distribute Resources}\label{sec:22.6}

After a resource is accepted by the Registrar node, the Root platform verifies the control proof, node authorization, resource type, version, content digest, and publication conditions, and forms a Root proof. The content distribution platform saves and transfers the verified materials according to the publication list or cursor. The Discovery node receives, verifies signatures, validates, and establishes structured and semantic indexes according to the cursor. Each step records its result independently: successful registration must not substitute for discoverability, and material distribution must not substitute for completed indexing.

\pandocbounded{\includegraphics[width=\linewidth,height=0.72\textheight,keepaspectratio]{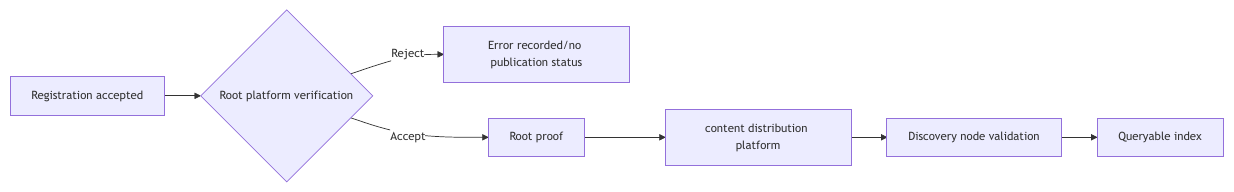}}

\subsection{Registration Result}\label{subsec:22.6.1}

The registration result must contain at least the resource DID, resource type, version or digest, acceptance status, registration VC if the interface returns one, and an associated identifier. When the VC is empty, it can only be truthfully recorded as empty; a credential must not be fabricated. When the Registrar node rejects acceptance, the error reason should be retained, and the client may retry according to the request identifier or an idempotent method allowed by the protocol. The client should also record which node returned the registration result and the observation time, so that later asynchronous publication status is not mixed into the registration result.

\passthrough{\lstinline!submitted!} only represents the registration-stage result; it cannot be interpreted as Root platform publication or discovery visibility.

\subsection{Root Platform Publication}\label{subsec:22.6.2}

The Root platform checks the Registrar node authorization, resource control signature, ResourcePackage integrity, version relationship, and governance conditions, then records the publication sequence and forms a Root proof after acceptance. Rejection, duplication, and asynchronous processing should each have an explicit status. When the Root platform has not accepted the resource, content distribution or local caching must not be marked as trusted publication. Publication evidence must be bound at least to the resource DID, version, and each digest; repeated processing must not produce mutually conflicting current versions.

The Root platform publication record must be associated at least with the resource DID, version, DID Document hash, metadata hash, package hash, Registrar node DID, publication sequence, and processing status; repeated processing should remain idempotent.

\subsection{Content Distribution Platform}\label{subsec:22.6.3}

The content distribution platform provides the DID Document, ResourcePackage, digests, and proof materials according to the Root platform publication list or cursor. The recipient validates integrity and then records the receipt cursor. Resumable transfer and repeated delivery should be idempotent; successful distribution only means that the materials can be obtained, not that the Discovery node has established a queryable index. If the recipient finds that the digest, version, or proof is inconsistent, it should reject writing to the visible index and retain diagnostic information for the failed materials.

The distribution validation order is ``list, content, digest, proof, cursor.'' The sender records the batch and cursor; the recipient first verifies each version and digest and then commits the local state. Successful transfer does not mean that indexing is complete.

\subsection{Discovery Node Indexing}\label{subsec:22.6.4}

After receiving the materials, the Discovery node verifies the Root proof, digest, version, governance status, and authorized domain, and then writes to the structured and semantic indexes. Completed indexing should have the resource DID, version, source, indexing time, and cursor evidence. A semantic index hit cannot override invalid proof or a revoked resource. The querying party should judge whether the result is fresh based on the indexing time and synchronization cursor, and return to the verifiable source for re-checking when the latest status is required.

The index record must contain at least \passthrough{\lstinline!resourceDid!}, version, source, authorized domain, lifecycle, \passthrough{\lstinline!indexedAt!}, and the synchronization cursor. Only verified materials enter the visible index; a semantic hit only provides a retrieval entry point.

\section{Discover by DID}\label{sec:22.7}

The caller submits the complete resource DID to the Discovery node as an exact query condition, for example by placing the DID in the query field when requesting \passthrough{\lstinline!/discovery/resources/query!}. The Discovery node returns the resource version, digest, source, lifecycle status, indexing time, and Root proof status in its current index. Pre-invocation verification must still be performed after obtaining the result. Not found, node not synchronized, stale index, and revoked resource should be handled separately; an empty result must not be directly interpreted as proof that the resource does not exist.

Acceptance should distinguish an exact hit, not found, not yet synchronized, stale index, suspended/revoked, and query error, and retain the querying node, request identifier, observation time, and response summary.

\section{Discover by Structured Conditions}\label{sec:22.8}

The caller queries by combining conditions such as resource type, subject type, tags, protocol, version, status, and authorized domain, and specifies field encoding, intersection/union relationships, filter order, pagination, and sorting. The Discovery node first applies hard conditions such as authorized domain and lifecycle, and then performs tag, protocol, or text-relevance processing; the actual filter order for each interface is subject to its protocol description. A tag match in the result only means that the retrieval condition was satisfied; the caller must still check the DID, signature, credential, publication evidence, version, and endpoint.

{\def\LTcaptype{none} 
\begin{longtable}[]{@{}>{\raggedright\arraybackslash}p{0.3200\linewidth}>{\raggedright\arraybackslash}p{0.3200\linewidth}>{\raggedright\arraybackslash}p{0.3200\linewidth}@{}}
\toprule\noalign{}
Condition & Function & Example \\
\midrule\noalign{}
\endhead
\bottomrule\noalign{}
\endlastfoot
resourceType & Restrict the resource type & \passthrough{\lstinline!skill!} \\
protocol & Restrict the protocol & \passthrough{\lstinline!mcp!} \\
capabilityTags & Capability matching & \passthrough{\lstinline!code.repository!} \\
authorizedDomains & Query visibility scope & \passthrough{\lstinline!technology.software\_engineering!} \\
limit & Quantity boundary & Range allowed by the implementation \\
\end{longtable}
}

\section{Discover by Natural-Language Task Description}\label{sec:22.9}

The caller submits a task description, for example, ``I need a tool that can search code repositories and summarize the project structure.'' The Discovery node or its semantic component extracts capabilities, use cases, resource types, protocols, tags, and status conditions from the description, and then performs hard-condition filtering and semantic ranking. The current implementation may use a lexical, structured, vector, or hybrid path, depending on the deployment configuration. When returning results, it should retain the matched resource DID, resource version, source, and explanation fields when provided by the interface. Semantic relevance is not control or security authorization.

\begin{lstlisting}
Task description -> capability/use-case extraction -> tags and structured conditions -> discovery query -> match rationale
\end{lstlisting}

The explanation result should distinguish the sources of text matching, tag matching, structured filtering, and semantic relevance.

\section{Verify a Discovered Resource Before Invocation}\label{sec:22.10}

The caller obtains the DID Document, ResourcePackage, Root proof, registration VC, and status information from the discovery result, verifies DID resolution, control relationship, credential, publication evidence, governance authorization, version, and endpoint in sequence, and then decides whether to invoke according to local policy. A discovery result is a candidate resource and an evidence entry point, not invocation authorization. When verification fails, status is unknown, evidence is stale, or endpoint authentication conditions are not met, high-risk invocation should be stopped and the reason recorded.

\begin{itemize}
\tightlist
\item[$\square$]
  DID and DID Document verification passed.
\item[$\square$]
  ResourcePackage, version, and digest are consistent.
\item[$\square$]
  The registration VC and Root proof are verifiable.
\item[$\square$]
  Governance status and freshness satisfy the policy.
\item[$\square$]
  Endpoint, authentication, and business permissions satisfy the invocation conditions.
\end{itemize}

The caller should not use the candidate object returned by the discovery interface directly as the final invocation object. A candidate is usually an indexed representation of \passthrough{\lstinline!ResourceDiscoveryCandidate!}; the complete DID Document and ResourcePackage must be obtained through the corresponding read interfaces. The content distribution platform's resource package read returns \passthrough{\lstinline!ResourcePackage!}, the DID Document read returns \passthrough{\lstinline!DidDocument!}, and the metadata read returns \passthrough{\lstinline!ResourceMetadata!}. After successful reads, the caller should verify the resource DID, version, three types of hashes, registration VC, Root proof, and governance status in the same verification context.

\subsection{DID Resolution}\label{subsec:22.10.1}

Resolve the complete resource DID, confirm that the method is \passthrough{\lstinline!did:oan!}, read the DID Document, and validate consistency among \passthrough{\lstinline!id!}, verification methods, control relationship, service endpoints, and resource metadata. Resolution not found, unsupported method, invalid document structure, and abnormal status should be handled separately; arbitrary same-name JSON must not be used as a substitute.

The resolution result should provide at least the following fields for continued verification in subsequent steps:

{\def\LTcaptype{none} 
\begin{longtable}[]{@{}>{\raggedright\arraybackslash}p{0.3200\linewidth}>{\raggedright\arraybackslash}p{0.3200\linewidth}>{\raggedright\arraybackslash}p{0.3200\linewidth}@{}}
\toprule\noalign{}
Field & Function & Handling when missing \\
\midrule\noalign{}
\endhead
\bottomrule\noalign{}
\endlastfoot
\passthrough{\lstinline!did!} / \passthrough{\lstinline!id!} & Confirm that the request identifier and document identifier are consistent & Stop \\
\passthrough{\lstinline!verificationMethod!} & Find the control relationship and public-key materials & Stop control-rights verification \\
\passthrough{\lstinline!service!} & Find the resource access or invocation entry point & Mark endpoint as incomplete \\
\passthrough{\lstinline!resourceMetadata!} & Compare the resource type, version, and capabilities & Mark description as incomplete \\
Resolution status & Distinguish success, not found, and method error & Must not default to success \\
\end{longtable}
}

\subsection{Control-Rights Verification}\label{subsec:22.10.2}

Verify the resource signature according to the verification methods and control relationships in the DID Document, confirming the normalized content covered by the signature, the resource DID, version, and predecessor relationship. Mathematical consistency between the public key and signature does not mean that the publisher is authorized to control another DID; verification must also bind the target DID and the existing control relationship. Verification inputs should come from the currently obtained DID Document and ResourcePackage, rather than only using truncated display fields in the discovery result.

Control-rights verification should be completed in the following order:

\begin{enumerate}
\def\labelenumi{\arabic{enumi}.}
\tightlist
\item
  Obtain the verification method and public-key reference from the DID Document.
\item
  Confirm that the public key belongs to the control relationship of the target DID.
\item
  Reconstruct the signature input according to the declared normalization rules.
\item
  Verify that the signature covers the DID, version, ResourcePackage digest, and predecessor relationship.
\item
  Associate the verification result with the source information returned by the Registrar node, Root platform, and Discovery node.
\end{enumerate}

\subsection{Registration VC Verification}\label{subsec:22.10.3}

Verify the issuer, subject DID, resource type, version or digest, signature, validity period, and revocation/suspension status of the registration VC. When the VC is missing or the interface does not return one, handle it as unknown or incomplete as defined by the protocol; the page text ``submission succeeded'' must not be treated as a credential. A registration VC only proves the registration fact issued by the Registrar node; it does not automatically prove that Root platform publication and Discovery node indexing are complete.

{\def\LTcaptype{none} 
\begin{longtable}[]{@{}>{\raggedright\arraybackslash}p{0.3200\linewidth}>{\raggedright\arraybackslash}p{0.3200\linewidth}>{\raggedright\arraybackslash}p{0.3200\linewidth}@{}}
\toprule\noalign{}
Check object & Facts to verify & Meaning of failure \\
\midrule\noalign{}
\endhead
\bottomrule\noalign{}
\endlastfoot
Issuer & Whether it is an allowed Registrar node or an issuer specified by the protocol & The credential source is untrusted \\
Subject & Whether it points to the current resource DID & The credential does not match the resource \\
Claims & Type, version, digest, and registration fact & The content cannot support the current version \\
Proof & Signature, public key, and signature coverage & The credential cannot be verified \\
Status & Validity period, suspension, or revocation information & Can only be marked unknown or rejected \\
\end{longtable}
}

\subsection{Root Proof Verification}\label{subsec:22.10.4}

Verify the issuer, target resource DID, ResourcePackage version, package digest, metadata digest, publication sequence, and signature of the Root proof, and confirm that the Root platform's own authorization status is valid. The proof only covers the facts it declares; it does not guarantee endpoint business quality.

The Root proof verification result is recorded separately from the registration VC. Each evidence state distinguishes at least \passthrough{\lstinline!verified!}, \passthrough{\lstinline!missing!}, \passthrough{\lstinline!invalid!}, and \passthrough{\lstinline!not\_checked!}, while governance state separately distinguishes \passthrough{\lstinline!active!}, \passthrough{\lstinline!suspended!}, \passthrough{\lstinline!revoked!}, and \passthrough{\lstinline!unknown!}. This is a classification requirement for a verification report, not a unified JSON response currently returned by every node. Only when the evidence refers to the same DID, version, and digests may the caller treat registration and Root acceptance as consecutive facts; neither may replace a missing other.

\subsection{Governance Status Check}\label{subsec:22.10.5}

Check node authorization, authorized domain, suspension or revocation events, and their freshness from a trusted governance-status source, and record the on-chain position, index cursor, and check time. When governance data is stale or the source is unavailable, sensitive invocations should be rejected or sent for manual review rather than being allowed by default.

The governance check must output at least the following decisions rather than only a Boolean:

{\def\LTcaptype{none} 
\begin{longtable}[]{@{}>{\raggedright\arraybackslash}p{0.4800\linewidth}>{\raggedright\arraybackslash}p{0.4800\linewidth}@{}}
\toprule\noalign{}
Decision & Meaning \\
\midrule\noalign{}
\endhead
\bottomrule\noalign{}
\endlastfoot
Authorization eligibility & Whether the current node has the corresponding role and authorization scope \\
Resource status & Whether the target resource is suspended, revoked, or still valid \\
Data freshness & Whether the event sequence, index cursor, and most recent successful time are acceptable \\
Evidence location & The event, transaction, or index record corresponding to the status \\
Invocation decision & Allow, reject, downgrade to display-only, or send for manual review \\
\end{longtable}
}

Registration responses, Root acceptance state, durable CDN confirmation, and the target Discovery node\textquotesingle s indexing result are retained separately. Only completion of the last state establishes that this target Discovery node can return the new version; earlier states remain useful recovery and diagnostic evidence.

For a resource that also has an external identifier, the complete decision path is:

\begin{lstlisting}
OAN identity/control validation + Registrar/Root/governance validation
  AND the external identifier has an acceptable result under its namespace rules
  AND the business endpoint authenticates and authorizes the current caller
  -> only then proceed to business invocation
\end{lstlisting}

External evidence is not mandatory for every resource. It is required only when invocation policy depends on that external fact. If it cannot be obtained, return ``not checked'' or ``status unknown'' rather than pretending that it is inapplicable or passed.

\section{Update a Resource While Retaining Its DID}\label{sec:22.11}

The original controller generates a new DID Document or ResourcePackage version with the same DID, carries the predecessor version and content digest, and signs it using the existing control relationship. The Registrar node and Root platform re-verify it and produce new registration or publication evidence; the distribution and discovery nodes converge according to the publication sequence. Repeated submission of identical content should be idempotent, and an unauthorized new document must not overwrite the old version. Completion of an update must be determined from the new publication proof, downstream synchronization status, and query result together, rather than only from another successful registration-interface response.

{\def\LTcaptype{none} 
\begin{longtable}[]{@{}>{\raggedright\arraybackslash}p{0.4800\linewidth}>{\raggedright\arraybackslash}p{0.4800\linewidth}@{}}
\toprule\noalign{}
Check item & Requirement before and after update \\
\midrule\noalign{}
\endhead
\bottomrule\noalign{}
\endlastfoot
resourceDid & Remains the same \\
Resource version & Increases according to the rules and never rolls back \\
Control proof & Covers the new normalized materials \\
Publication proof & Binds the new version and digests \\
Discovery result & Prioritizes the current version while keeping history auditable \\
\end{longtable}
}

\section{Suspend, Revoke, and Recover a Resource}\label{sec:22.12}

Resource suspension is triggered by the controller, governance, or a valid evidence status; revocation targets a resource or version; recovery requires rechecking the control proof, credential, publication proof, and endpoint. Each node should record the triggering fact, processing time, and propagation status according to its responsibility. The status in a discovery result is the status to which the node has synchronized and does not necessarily equal the time when the operation just occurred. A network timeout or temporarily offline node does not mean that the resource has been revoked. When propagation to the Discovery node is incomplete, the status and freshness should be displayed and historical evidence retained.

\pandocbounded{\includegraphics[width=\linewidth,height=0.72\textheight,keepaspectratio]{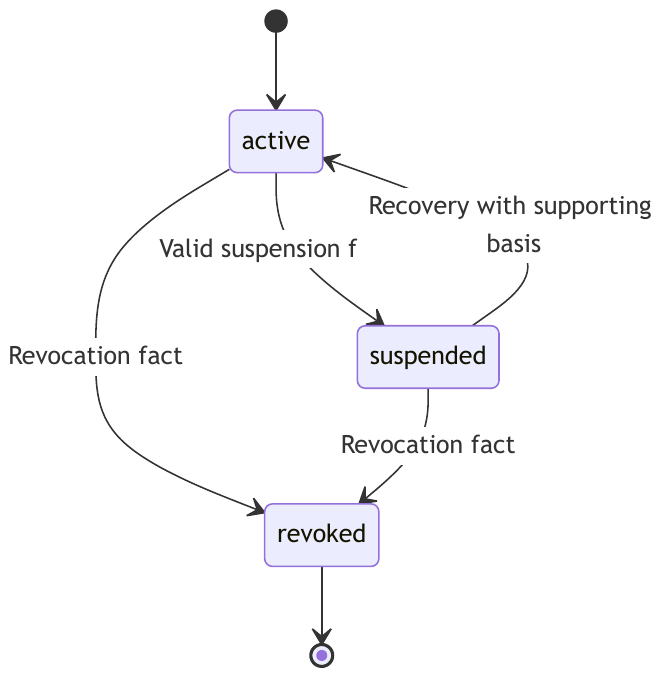}}

\section{Connect a Third-Party Registrar Node}\label{sec:22.13}

The third party submits the node DID Document, authorization request, authorized domain, public endpoint, and runtime configuration, and completes identity, signature, interface, resource-validation, security-boundary, and upstream-submission tests. After passing, it obtains registration eligibility within the explicitly defined scope only; while the node is running, its authorization status, versions, and failed-request retries must still be checked, and one acceptance test cannot permanently replace governance status. Connection tests should use isolated resources and retain test requests, responses, error branches, authorization scope, and validity period, so that test resources are not mistaken for formal network resources.

\begin{itemize}
\tightlist
\item[$\square$]
  The node DID Document, role, endpoint, and protocol version are consistent.
\item[$\square$]
  The authorized domain is consistent with the governance authorization materials.
\item[$\square$]
  Evidence exists for positive, negative, retry, and recovery cases.
\item[$\square$]
  The private key and governance secrets are not included in public materials.
\end{itemize}

\section{Connect a Third-Party Discovery Node}\label{sec:22.14}

The third-party Discovery node obtains the node identity, authorization materials, and publication-source configuration, first completes a full synchronization to establish a baseline, and then performs incremental synchronization, digest validation, version selection, DID/structured/semantic queries, deduplication, freshness, and degradation tests. When providing results externally, it retains the source, version, indexing time, and authorized domain; local sorting must not be rewritten as a Root platform publication fact. A query interface being available does not mean that synchronized data covers the latest publication sequence.

The connection order is ``establish a full baseline → incremental synchronization → inject duplicates/out-of-order/missing data → query and deduplicate → freshness and recovery verification.'' Local sorting only affects display order.

\section{Observe Governance Events Through the Network}\label{sec:22.15}

After a governance event is produced on-chain, the off-chain trust indexer reads it in order and forms a node-authorization status read model. The Root platform and Discovery node then execute policies according to their synchronization status. An observer should record the event type, on-chain position, index cursor, projection time, and page display time, distinguishing on-chain facts from website display. Continued growth of \passthrough{\lstinline!latest\_sequence!} only proves that the indexer is reading new events; whether the authorization status of a specific node has changed must also be checked against whether the corresponding event has entered the read model.

\begin{lstlisting}
On-chain event -> indexer cursor -> authorization-status read model -> Root platform/Discovery node decision -> page observation
\end{lstlisting}

\section{Resolve and Verify a Resource DID}\label{sec:22.16}

The client first resolves the resource DID and DID Document, and then verifies the control relationship, resource metadata, service endpoint, version, and related publication evidence. The resolution result may be encapsulated by an SDK or local verifier, but the caller should still be able to obtain the original document, digest, and error status for auditing and review. Successful resolution only proves that the document can be read and passes structural checks; it cannot bypass checks of the ResourcePackage, credential, Root proof, and governance status.

It is recommended to output distinguishable statuses such as \passthrough{\lstinline!resolved!}, \passthrough{\lstinline!invalid!}, \passthrough{\lstinline!not-found!}, and \passthrough{\lstinline!status-unknown!}, rather than compressing every failure into an empty object.

\section{Retrieve Historical Resource Versions}\label{sec:22.17}

The caller requests historical ResourcePackages and their DID Documents, digests, and proofs by DID, version, or publication sequence, and explicitly identifies whether each has been superseded, revoked, or is for audit only. Historical content must not automatically become the currently discoverable or callable version merely because it can be read. When a historical version lacks the corresponding publication proof or status evidence, it should be marked as incompletely verifiable rather than supplemented with the current version's proof.

Historical records must be associated at least with the DID, version/sequence, digest, publication proof, current status, and query time. ``Readable'' and ``currently callable'' must be two separate fields.

\section{Handle an Unavailable or Stale Discovery Node}\label{sec:22.18}

The caller determines whether discovery results are usable based on response errors, node health, indexing time, cursor, and declared freshness. An existing old snapshot may be displayed first to improve the experience, but its time and degraded status must be marked. For actions requiring the latest authorization or version, the caller should re-verify from a trusted source and must not rely only on an old cache.

{\def\LTcaptype{none} 
\begin{longtable}[]{@{}>{\raggedright\arraybackslash}p{0.3200\linewidth}>{\raggedright\arraybackslash}p{0.3200\linewidth}>{\raggedright\arraybackslash}p{0.3200\linewidth}@{}}
\toprule\noalign{}
Situation & Read-only display & Authorization/invocation action \\
\midrule\noalign{}
\endhead
\bottomrule\noalign{}
\endlastfoot
Node timeout & An expired snapshot may be displayed with an explicit marker & Stop or switch and re-verify \\
Stale index & Display the expiration time & Retrieve the current evidence again \\
Unknown source & Do not display a trusted indicator & Reject \\
Recovery completed & Update the snapshot & Perform complete verification again \\
\end{longtable}
}

\subsection{Discovery Node Unavailable}\label{subsec:22.18.1}

When a discovery request times out, the connection fails, or an explicit service error is returned, record the node and associated identifier, and retry or switch to another authorized Discovery node according to policy. Unavailability represents infrastructure status; it does not automatically mean resource revocation, and an empty result must not be treated as proof that the resource does not exist on the network.

A retry record must contain at least the endpoint, request identifier, error type, count, backoff time, and last known indexing time. A connection failure must not be confused with a valid empty result.

\subsection{Return an Old Snapshot}\label{subsec:22.18.2}

An old snapshot may be used for limited read-only display, but it must carry the generation time, source, and degraded-status marker. It may display the resource name, type, historical match results, and last indexing time, but it cannot by itself be used to determine current authorization, revocation status, endpoint availability, or invocation eligibility. If a resource in the snapshot concerns invocation, authorization, or a sensitive decision, the current DID Document, credential, publication proof, and governance status must be obtained and verified again before the action. After online evidence is obtained again, the new result should supersede the snapshot while retaining the temporal relationship between the old and new sources.

\subsection{Determine Data Freshness}\label{subsec:22.18.3}

Determine freshness from the indexing time, event cursor, latest upstream sequence, and allowed delay. A \passthrough{\lstinline!latest\_sequence!} that does not increase may mean that there are no new events, or that synchronization has stopped; the most recent successful time and error evidence must also be considered. Unknown freshness must not default to latest.

A freshness report should provide \passthrough{\lstinline!indexedAt!}, \passthrough{\lstinline!lastSuccessfulSyncAt!}, \passthrough{\lstinline!latestSequence!}, \passthrough{\lstinline!cursor!}, and error status together; it must not display only one sequence number.

\subsection{Switch to Another Discovery Node}\label{subsec:22.18.4}

Before switching, confirm the candidate node's authorization status, endpoint, and supported query profiles, and retain the source node in the result. Results from different nodes need to be deduplicated and cross-validated by DID, version, digest, and publication proof; they must not simply be concatenated as equally trusted results.

Switching checklist:

\begin{itemize}
\tightlist
\item[$\square$]
  The candidate node DID, role, and authorization status are valid.
\item[$\square$]
  The query profile and authorized domain meet the caller's requirements.
\item[$\square$]
  Results are deduplicated by DID, version, and digest.
\item[$\square$]
  Conflicting results are checked against Root platform publication evidence.
\end{itemize}

\subsection{Re-synchronize After Recovery}\label{subsec:22.18.5}

After the node recovers, fill gaps from the persisted cursor or trusted publication sequence, re-verify affected versions and revocation status, and then update the index projection. Recovery is complete when the cursor catches up within the allowed range, key resource queries are consistent, and error alerts have recovered; process restart alone is not sufficient.

Recovery verification should sample at least one newly added resource, one updated resource, one revoked or suspended resource, and one resource that did not change during the outage, confirming that recovery did not only restore the count while missing lifecycle states.

The recovery completion criterion can be written as:

\begin{lstlisting}
Cursor caught up + key resource consistency + lifecycle re-filtering + error alerts recovered
\end{lstlisting}

\section{References}\label{sec:22.19}

{\def\LTcaptype{none} 
\begin{longtable}[]{@{}>{\raggedright\arraybackslash}p{0.3200\linewidth}>{\raggedright\arraybackslash}p{0.3200\linewidth}>{\raggedright\arraybackslash}p{0.3200\linewidth}@{}}
\toprule\noalign{}
Source & Type & Link \\
\midrule\noalign{}
\endhead
\bottomrule\noalign{}
\endlastfoot
\passthrough{\lstinline!oan-protocol-commons!} & Code repository: shared objects and protocol types for end-to-end flows & \url{https://github.com/wolfbrother/oan-protocol-commons} \\
\passthrough{\lstinline!oan-root-nodes!} & Code repository: Root platform verification, publication, and governance interfaces & \url{https://github.com/OpenAgenet/oan-root-nodes} \\
\passthrough{\lstinline!oan-registrar-nodes!} & Code repository: Registrar node flows & \url{https://github.com/OpenAgenet/oan-registrar-nodes} \\
\passthrough{\lstinline!oan-discovery-nodes!} & Code repository: Discovery synchronization, indexing, and query flows & \url{https://github.com/OpenAgenet/oan-discovery-nodes} \\
\passthrough{\lstinline!oan-trust-indexer!} & Code repository: governance-status synchronization and query flows & \url{https://github.com/OpenAgenet/oan-trust-indexer} \\
\passthrough{\lstinline!oan-sdk-ts!} & Code repository: client-side end-to-end invocation helpers & \url{https://github.com/OpenAgenet/oan-sdk-ts} \\
\passthrough{\lstinline!oan-community-skill!} & Code repository: community end-to-end operation examples & \url{https://github.com/OpenAgenet/oan-community-skill} \\
OAN White Paper & arXiv white paper: overall vision and ecosystem flows & \url{https://arxiv.org/abs/2606.03161} \\
OAN Yellow Paper & arXiv yellow paper: protocol flows and verification mechanisms & \url{https://arxiv.org/abs/2606.03163} \\
GRAIL Semantic Discovery Paper & arXiv paper: semantic discovery flows & \url{https://arxiv.org/abs/2605.02489} \\
\passthrough{\lstinline!did:oan!} DID Method Specification & Standard/method specification: resource identity in end-to-end flows & \url{https://github.com/OpenAgenet/oan-public-docs/blob/main/did-oan-specs/doc/OAN\%20DID\%20Method\%20Specification.md} \\
OAN Resource Identity and Discovery & IETF draft: identity, registration, and discovery flows & \url{https://datatracker.ietf.org/doc/draft-xu-oan-resource-identity-discovery/} \\
Agentic Overlay Network Architecture & IETF draft: node collaboration flows & \url{https://datatracker.ietf.org/doc/draft-xu-agentic-overlay-network-architecture/} \\
Efficient Agent Discovery Profile & IETF draft: discovery query flows & \url{https://datatracker.ietf.org/doc/draft-xu-efficient-agent-discovery-profile/} \\
\end{longtable}
}

\chapter{Conformance Profiles and Acceptance Criteria}\label{chap:23}

This chapter explains how conformance profiles turn each requirement into a ``test object, input, steps, criteria, evidence, and conclusion,'' without labeling an untested capability as conformant. A publishable offline evaluation package contains a manifest, case matrix, fixed inputs, expected results, requirement checklist, and evidence index, forming an auditable baseline. This manual lists such a package as externally verifiable evidence only when those materials are public and reproducible. A real Root platform, content distribution platform, on-chain governance system, and production recovery are identified separately.

{\def\LTcaptype{none} 
\begin{longtable}[]{@{}>{\raggedright\arraybackslash}p{0.3200\linewidth}>{\raggedright\arraybackslash}p{0.3200\linewidth}>{\raggedright\arraybackslash}p{0.3200\linewidth}@{}}
\toprule\noalign{}
Conclusion & Meaning & Follow-up action \\
\midrule\noalign{}
\endhead
\bottomrule\noalign{}
\endlastfoot
PASS & All criteria are satisfied within the declared scope & Archive the evidence \\
FAIL & At least one mandatory criterion is not satisfied & Block release or admission \\
NOT TESTED & No reproducible evidence is available & Must not claim conformance \\
CONDITIONAL & Valid only under explicitly stated limitations & Record the limitations and deadline \\
\end{longtable}
}

Conformance review presupposes an isolated test environment, fixed test vectors, a clearly specified specification version, and results that can be retained as evidence. A profile\textquotesingle s "pass" means only that the corresponding role has completed the prescribed capability and rejection-behavior tests; it does not automatically prove that the entire OpenAgenet (OAN) network or the quality of a resource business has passed. Test records should reproduce the inputs, steps, responses, logs, versions, and determinations.

The applicable scope of a profile should be stated clearly at the beginning of the report, for example, "resource provider - ResourcePackage generation," "Registrar node - registration interface," and "Discovery node - query and indexing." The same test resource may run through multiple profiles, but each role is responsible only for the facts it produces or verifies; the state of a test resource during registration, publication, distribution, and discovery should be correlated through the DID, version, digest, and request identifier, rather than by name alone.

\pandocbounded{\includegraphics[width=\linewidth,height=0.72\textheight,keepaspectratio]{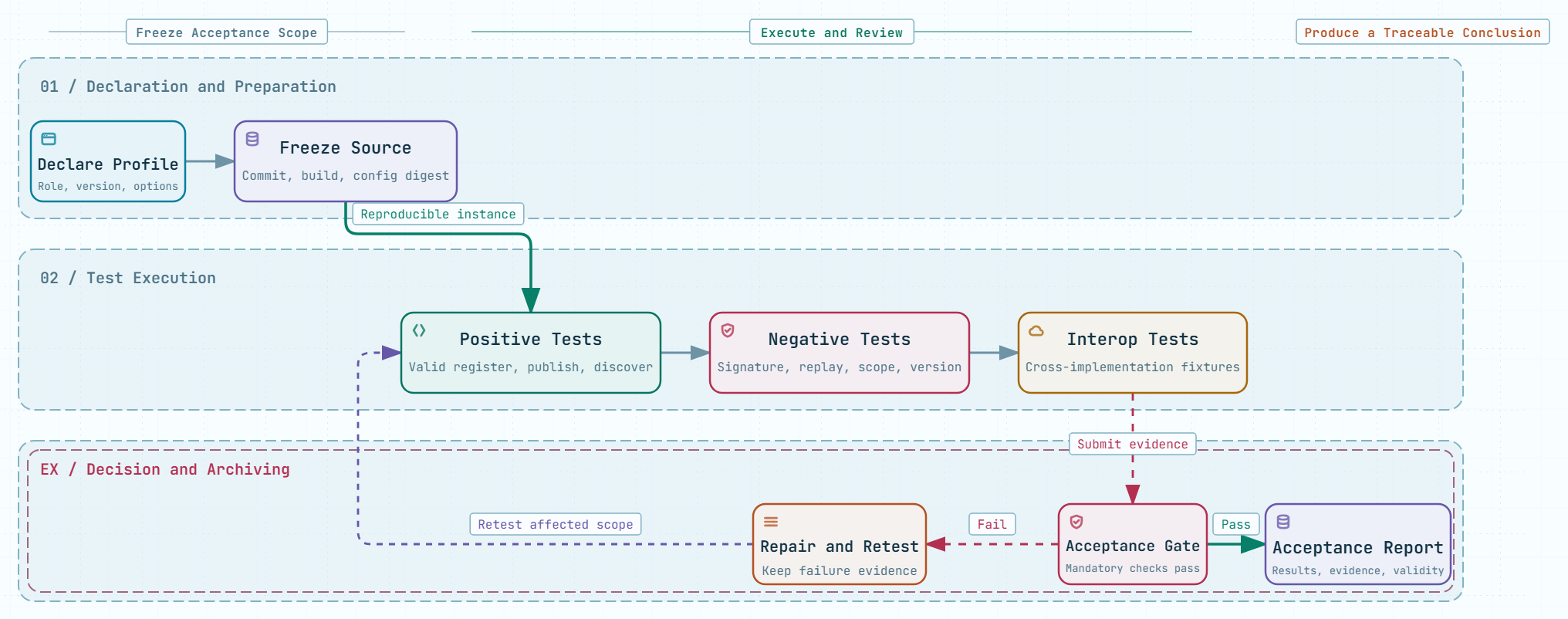}}

Acceptance begins by declaring a profile and freezing the implementation source. Positive tests show that valid materials complete the prescribed flow; negative tests show that invalid or unauthorized materials are rejected without side effects; and interoperability tests then verify cross-implementation fixtures and evidence. Failure of any mandatory gate returns the affected scope for retesting and cannot be offset by success in another test. Editable Archify JSON and HTML and PNG/PDF outputs are stored under \passthrough{\lstinline!figs/archify!}.

\section{Resource Provider Conformance Profile}\label{sec:23.1}

A resource provider should be able to generate or import a Local Identity, create a resource DID, form a valid DID Document, resource metadata, and ResourcePackage, and sign them with the local control key. Acceptance tests cover the required fields, capability tags, use cases, endpoints, versions, and digests of four resource types; qualifying evidence consists of verifiable signature materials, structured inputs, and rejection results for invalid inputs. Private keys must not appear in upload requests or logs.

{\def\LTcaptype{none} 
\begin{longtable}[]{@{}>{\raggedright\arraybackslash}p{0.2400\linewidth}>{\raggedright\arraybackslash}p{0.2400\linewidth}>{\raggedright\arraybackslash}p{0.2400\linewidth}>{\raggedright\arraybackslash}p{0.2400\linewidth}@{}}
\toprule\noalign{}
Acceptance object & Test input & Pass criterion & Evidence to retain \\
\midrule\noalign{}
\endhead
\bottomrule\noalign{}
\endlastfoot
Local Identity & Create a new identity, import a valid backup, import a corrupted backup & A subject control relationship can be generated or restored, and the corrupted file is rejected & DID, public key, import result; do not retain the private key \\
Resource DID & One instance of each of the four resource types & The DID is consistent with the control relationship in the DID Document & DID Document, normalized digest, signature verification result \\
Resource metadata & Required fields, missing fields, and fields with incorrect types & Valid data can generate a ResourcePackage, and invalid data is explicitly rejected & Input samples, error responses, validation logs \\
ResourcePackage & Content, version, hash, and endpoint & The digest can be recomputed, and verification fails after tampering & ResourcePackage digest, recomputation result, version information \\
\end{longtable}
}

The acceptance order should be fixed as "identity available -\textgreater{} DID Document generation -\textgreater{} metadata validation -\textgreater{} ResourcePackage digest and signature -\textgreater{} submission-material check." If any step fails, the materials must not continue to be marked as publishable; test records should be associated with the resource DID, test vector, and specification version.

The resource provider\textquotesingle s pass condition is not "a webpage can download a file," but the ability to reconstruct the same materials locally and verify the control relationship. Test personnel should retain at least one valid set of materials, one set missing a required field, and one set with a modified signature or digest, and check the generation result, explicit error, and no-side-effect result respectively.

\section{Registrar Node Conformance Profile}\label{sec:23.2}

Using an isolated test identity, submit valid materials, materials with missing fields, materials with invalid signatures, materials from an unauthorized node, duplicate materials, and expired materials, and check the Registrar node\textquotesingle s status codes, error structures, registration records, idempotent results, VC (if any), and upstream submission. The qualifying condition is that only valid requests within the authorized scope are accepted, rejected requests do not change the state of valid resources, and the processing path can be reviewed using logs and correlation identifiers.

Registrar node acceptance should cover at least the following request sequence:

\begin{enumerate}
\def\labelenumi{\arabic{enumi}.}
\tightlist
\item
  Submit valid materials and confirm the response structure, registration state, and correlation request identifier.
\item
  Resubmit the same DID, version, and digest and confirm an idempotent result that does not produce a second valid registration fact.
\item
  Replace the control signature, node authorization, or resource digest and confirm that the request is rejected and the original registration remains unchanged.
\item
  Submit expired materials or materials outside the authorized scope and confirm that the rejection reasons are distinguishable.
\item
  Check the failure state, retry boundary, and pending record when the upstream is temporarily unavailable.
\end{enumerate}

For every request, retain the normalized request digest from the sending side and the correlation identifier returned by the Registrar node; for asynchronous upstream calls, record "accepted by the Registrar node" and "processed by the Root platform" as two independent records. If no registration VC is present in a successful response, record that the field is missing or was not returned rather than fabricating a credential in the acceptance materials.

{\def\LTcaptype{none} 
\begin{longtable}[]{@{}>{\raggedright\arraybackslash}p{0.3200\linewidth}>{\raggedright\arraybackslash}p{0.3200\linewidth}>{\raggedright\arraybackslash}p{0.3200\linewidth}@{}}
\toprule\noalign{}
Checkpoint & What to observe & Qualifying result \\
\midrule\noalign{}
\endhead
\bottomrule\noalign{}
\endlastfoot
Request validation & Required fields, version, digest, and signature & Invalid requests are rejected before a registration is created \\
Control & Relationship between the resource DID and the control signature & A signature from another subject cannot overwrite the resource \\
Authorization scope & Registrar node identity and permitted resource scope & An out-of-scope submission does not change the valid state \\
VC result & Issuer, subject, type, and status & The VC, if returned, is verifiable and corresponds to the registration fact \\
Upstream submission & Root platform or subsequent publication task & Failure has a clear state and does not impersonate network-wide visibility \\
\end{longtable}
}

\section{Root Platform Conformance Profile}\label{sec:23.3}

Validate valid and tampered ResourcePackages, control proofs, Registrar node authorization, version sequences, and digests. The Root platform forms a Root proof for a trusted publication only when the control relationship, node authorization, content integrity, and governance conditions are all satisfied; repeated publication should be idempotent, and rejection or asynchronous failure must not produce a publication fact that a Discovery node could treat as a valid version.

Root platform acceptance uses a four-step check: "validate inputs, determine eligibility, form the publication fact, and read the result again":

\begin{itemize}
\tightlist
\item[$\square$]
  The control signature binds the resource DID, version, and content to be published.
\item[$\square$]
  Registrar node authorization is valid, and the submission scope matches the authorization scope.
\item[$\square$]
  The ResourcePackage, DID Document, and digest validate against one another.
\item[$\square$]
  The version sequence has no rollback, unexplained gap, or unexplainable fork.
\item[$\square$]
  The issuer, subject, covered content, and status of the Root proof are verifiable.
\item[$\square$]
  Rejection, duplication, and asynchronous failure do not generate a valid publication record.
\end{itemize}

Acceptance evidence should retain both the initial processing result and the publication status obtained by a subsequent query, so that "request received," "validation passed," "publication fact formed," and "indexed by a Discovery node" can be distinguished as separate stages.

The key criterion for the Root platform profile is that "rejected input must not form a valid publication fact that can propagate." Therefore, in addition to checking interface responses, test personnel should reread the publication record, content distribution manifest, or publication sequence, and confirm that an incorrect digest, incorrect control signature, or out-of-scope Registrar node has not entered a subsequent queue; these checks should use independent test resources and request identifiers.

\section{Content Distribution Service Conformance Profile}\label{sec:23.4}

Deliver complete packages, incomplete packages, incorrect digests, duplicate batches, checkpoint cursors, and old versions to the content distribution platform, and check the manifest, response, receipt cursor, integrity, and audit records. The qualifying condition is that only materials permitted by the Root platform are delivered, duplication and checkpointing do not lose content or repeatedly change facts, and the receiver can verify the content using the digest and proof.

{\def\LTcaptype{none} 
\begin{longtable}[]{@{}>{\raggedright\arraybackslash}p{0.3200\linewidth}>{\raggedright\arraybackslash}p{0.3200\linewidth}>{\raggedright\arraybackslash}p{0.3200\linewidth}@{}}
\toprule\noalign{}
Test step & Constructed condition & Pass criterion \\
\midrule\noalign{}
\endhead
\bottomrule\noalign{}
\endlastfoot
Complete delivery & Manifest, content, digest, and publication proof are all present & The platform accepts the content and returns a traceable batch \\
Integrity failure & Delete content or modify the digest & Reject the content and do not update the version available for discovery \\
Duplicate delivery & Submit the same batch and cursor again & Idempotent; do not count the publication fact twice \\
Checkpoint recovery & Continue from an intermediate cursor & Fill only the gap, with a consistent final digest \\
Old-version delivery & Below the currently valid version & Do not overwrite newer valid content \\
\end{longtable}
}

At the end of testing, recheck the receiver\textquotesingle s content against the manifest and digest, and record the delivery batch, cursor, response time, retry count, and final state. Successful distribution means only that the content can be retrieved and verified; it does not replace the Root proof.

Checkpoint and duplicate tests should be run again after restarting the receiver or interrupting the network, to check whether the persisted cursor still points to the last confirmed batch. If the receiver has saved the content but not the corresponding proof or cursor, treat it as an incomplete distribution state and do not directly hand it to a Discovery node for establishing a trusted index.

\section{Discovery Node Conformance Profile}\label{sec:23.5}

First perform full and incremental synchronization, then use published test resources for DID, structured, and semantic queries, while adding duplicate materials, old versions, revoked states, sequence gaps, and expired data. The qualifying condition is that indexing occurs only after validation, the latest valid version is selected by default for the same DID, results are not duplicated, provenance and freshness are returned, and relevance does not override control or governance checks.

The Discovery node test sequence is as follows:

\begin{enumerate}
\def\labelenumi{\arabic{enumi}.}
\tightlist
\item
  Check the resource count, versions, and digests after full synchronization.
\item
  Inject an incremental publication and confirm that only the corresponding index record is added or updated.
\item
  Perform an exact DID query, a structured-condition query, and a semantic query separately.
\item
  Add duplicate materials, an old version, a revoked resource, and an expired snapshot, and check filtering and deduplication.
\item
  Verify the source node, version, freshness, and publication-evidence reference in the results.
\end{enumerate}

{\def\LTcaptype{none} 
\begin{longtable}[]{@{}>{\raggedright\arraybackslash}p{0.3200\linewidth}>{\raggedright\arraybackslash}p{0.3200\linewidth}>{\raggedright\arraybackslash}p{0.3200\linewidth}@{}}
\toprule\noalign{}
Query type & Must verify & Check that it cannot replace \\
\midrule\noalign{}
\endhead
\bottomrule\noalign{}
\endlastfoot
Exact DID query & DID, current version, and source & Control and governance status \\
Structured query & Type, tags, protocol, and status conditions & Actual resource availability \\
Semantic query & Conversion from the input description to tag/text conditions & Authorization and signature verification \\
\end{longtable}
}

Multiple records for the same DID should be selected deterministically according to version and publication status; when freshness, governance status, or publication evidence cannot be confirmed, mark the result as uncertain and hand it to the caller for review.

Query acceptance for the Discovery node profile should use the same set of published resources to cover six result categories: "exact hit, structured hit, semantic hit, duplicate record, old version, and invisible resource." The report should record the query input, filter conditions, returned resource DID, version, source, index time, and duplicate count, rather than recording only the total number of results.

\section{Off-Chain Trust Indexer Conformance Profile}\label{sec:23.6}

Provide sequential governance events, a checkpoint, duplicate events, missing events, and temporarily unavailable upstream data, and check the authorization-state read model, event cursor, latency, and error state. The qualifying condition is correct ordering and idempotency, recovery from the cursor after restart, and no treatment of unread events or manual displays as on-chain facts; evidence includes event positions, cursors, and projection records.

{\def\LTcaptype{none} 
\begin{longtable}[]{@{}>{\raggedright\arraybackslash}p{0.3200\linewidth}>{\raggedright\arraybackslash}p{0.3200\linewidth}>{\raggedright\arraybackslash}p{0.3200\linewidth}@{}}
\toprule\noalign{}
Input scenario & Expected projection & Evidence fields \\
\midrule\noalign{}
\endhead
\bottomrule\noalign{}
\endlastfoot
Continuous events & Authorization state updates in order & Event position, cursor, projection time \\
Duplicate events & State does not change repeatedly & Event unique identifier, idempotent result \\
Sequence gap & Pause or mark as incomplete & Gap range, error state, alert \\
Restart recovery & Continue from the persisted position & Cursors before and after restart, completion records \\
Upstream unavailable & Retain the last trusted state and mark it as stale & Most recent successful time, failure reason \\
\end{longtable}
}

The "state updated" status of the off-chain trust indexer must point back to a governance event or an explicit upstream proof; the existence of a projection record in the database alone cannot serve as a governance fact.

When test events arrive continuously, check event order, duplicate events, and sequence gaps separately; when \passthrough{\lstinline!latest\_sequence!} temporarily stops increasing, use the most recent successful read time and error state to determine whether there are "no new events" or synchronization has "stalled." A test can be recorded as passed only when the cursor, event position, and authorization-state read model correspond to one another.

\section{Parser and Client Conformance Profile}\label{sec:23.7}

Use inputs consisting of a valid DID, an unknown DID, an unsupported method, an invalid structure, an incorrect control signature, an abnormal status, a historical version, and a missing proof, and check the result types and stop behavior. The qualifying condition is distinguishable errors, no silent acceptance of invalid materials, no presentation of a historical version as the current version, and availability of the original document, digest, and error information for review.

{\def\LTcaptype{none} 
\begin{longtable}[]{@{}>{\raggedright\arraybackslash}p{0.3200\linewidth}>{\raggedright\arraybackslash}p{0.3200\linewidth}>{\raggedright\arraybackslash}p{0.3200\linewidth}@{}}
\toprule\noalign{}
Input state & What the client should display or return & Follow-up action \\
\midrule\noalign{}
\endhead
\bottomrule\noalign{}
\endlastfoot
Valid and resolved & Current DID Document and resolution status & Continue control, publication, and governance verification \\
Not found & Explicit not-found & Do not substitute a same-name cache \\
Method unsupported & Explicit method-unsupported & Stop high-risk calls \\
Document or signature invalid & invalid-document or verification failure & Reject and retain the original evidence \\
Status unknown or expired & status-unknown & Degrade the display or request re-verification \\
Historical version & Explicit version and time & Do not use it as the current version \\
\end{longtable}
}

Acceptance should also check that error paths do not mislabel partial resolution results as trusted resources, and do not bypass control, credential-status, or governance-status checks through automatic retries.

\section{SDK Conformance Profile}\label{sec:23.8}

The SDK should cover Local Identity, resource DID, metadata, control signatures, registration, exact DID discovery, semantic discovery, VC verification, and Root proof verification. Use the test inputs for \passthrough{\lstinline!oan-sdk-ts!} to check serialization, error mapping, and cross-environment endpoint configuration; the qualifying condition is that the SDK does not upload private keys, does not lose version or evidence fields, and enables callers to distinguish registration success from discoverability.

SDK acceptance checklist:

\begin{itemize}
\tightlist
\item[$\square$]
  The results of creating and importing a Local Identity can be reviewed, and the private key does not enter network requests.
\item[$\square$]
  When generating a resource DID, DID Document, and signature, the fields are consistent with the Rust protocol types.
\item[$\square$]
  Registration success, registration failure, and "not yet discoverable" map to different results.
\item[$\square$]
  Exact DID queries and semantic queries retain source, version, and freshness information.
\item[$\square$]
  Verification results for the VC, Root proof, and governance status are not merged into a single Boolean.
\item[$\square$]
  Network errors, server errors, and format errors can be distinguished by the caller.
\end{itemize}

Cross-environment tests should use fixed node endpoint configuration and test vectors; one online address or one successful request cannot serve as proof of the SDK\textquotesingle s permanent compatibility. The test report should also record the SDK unit tests, node integration tests, and actual requests to the official website separately, so that local object-test results are not mistakenly written as end-to-end interoperability results. The acceptance conclusion should be associated with at least the SDK commit, protocol public type version, test vector, target endpoint, and response digest.

\section{Security and Key Custody Acceptance Criteria}\label{sec:23.9}

Test the generation, storage, backup, recovery, rotation, and permission boundaries of subject, resource, and node keys, and check the verification of signatures, VCs, and Root proofs. The qualifying condition is that private keys are used only within the controller\textquotesingle s boundary, backups are recoverable and protected, and incorrect keys and unauthorized signatures are rejected; evidence should be redacted and must not submit the private key in plaintext.

Security acceptance task checklist:

\begin{itemize}
\tightlist
\item[$\square$]
  Establish the usage boundaries of the subject private key, resource control private key, and node runtime credentials.
\item[$\square$]
  Check the generation, import, permissions, and recovery result of backup files, without writing private keys to logs or request bodies.
\item[$\square$]
  Perform rejection tests using an incorrect public key, an incorrect signature, and a signature from another subject.
\item[$\square$]
  Verify the issuer, subject, signature coverage, and status of the VC and Root proof.
\item[$\square$]
  Check whether sensitive fields in exceptions, debug logs, and evidence packages have been redacted.
\end{itemize}

The test report should make clear that "the controller can sign" and "the service node can verify" are two different facts; the latter does not mean that the node has obtained the former\textquotesingle s private key. Security acceptance should treat the client, Registrar node, Root platform, Discovery node, and evidence storage as separate observation boundaries, and record item by item whether private keys appear in request bodies, logs, temporary files, error stacks, or test archives. Control signatures, registration VCs, and Root proofs should each be separately checked for issuer, subject, signature coverage, and status.

\section{Interoperability and Compatibility Acceptance Criteria}\label{sec:23.10}

Use the same set of vectors for Rust protocol types, the TypeScript SDK, the official website API, the Registrar node, the Discovery node, and A2A/MCP mappings, and compare DID Documents, ResourcePackages, digests, signatures, errors, and version results. The qualifying condition is consistent normalization and field semantics across languages, with mappings retaining provenance and without weakening OAN control and governance verification through adaptation.

{\def\LTcaptype{none} 
\begin{longtable}[]{@{}>{\raggedright\arraybackslash}p{0.2400\linewidth}>{\raggedright\arraybackslash}p{0.2400\linewidth}>{\raggedright\arraybackslash}p{0.2400\linewidth}>{\raggedright\arraybackslash}p{0.2400\linewidth}@{}}
\toprule\noalign{}
Comparison dimension & Rust implementation & TypeScript/HTTP implementation & Qualifying condition \\
\midrule\noalign{}
\endhead
\bottomrule\noalign{}
\endlastfoot
Serialization & Public protocol types & SDK and official website request bodies & Fields, types, and defaulting rules are consistent \\
Digest & Digest of normalized content & Digest generated/read by the frontend or SDK & The same input produces the same digest \\
Signature & Control signature and proof verification & Verification result returned by the SDK/node & Signature coverage is consistent \\
Error & Node error structure & SDK error mapping & The failure reason is not lost \\
Mapping & A2A Agent Card, MCP metadata & Adaptation or display result & Preserve the original text and provenance \\
\end{longtable}
}

Cross-language comparison must use the same version, input, and normalization rules; comparing only the final page text cannot prove protocol interoperability. The interoperability conclusion should also distinguish three levels: "the same input produces the same object," "different implementations can parse the object," and "the cross-node end-to-end flow succeeds." Record the corresponding evidence separately; do not determine interoperability passed merely because the final page displays the same text.

Cross-language acceptance may use the method of "one fixture, three comparisons": first compare the normalized results of the Rust protocol types and TypeScript objects, then compare the HTTP request sent by the SDK or official website with the result actually parsed by the node, and finally compare the resource DID, version, and digest throughout the registration, publication, and discovery paths. Any difference at any level should record the differing field and applicable version, and must not be attributed to a network problem alone.

\section{Performance and Availability Acceptance Criteria}\label{sec:23.11}

Measure the success rate, latency, backlog, recovery time, and resource consumption of registration, publication, synchronization, query, and governance reads in a test environment with clearly defined concurrency, resource scale, network conditions, and statistical windows. Thresholds should be defined in advance in the profile or deployment acceptance record; without measurement data, the result can only be marked as not tested, and the performance of one demonstration must not be written as a general performance commitment.

{\def\LTcaptype{none} 
\begin{longtable}[]{@{}>{\raggedright\arraybackslash}p{0.3200\linewidth}>{\raggedright\arraybackslash}p{0.3200\linewidth}>{\raggedright\arraybackslash}p{0.3200\linewidth}@{}}
\toprule\noalign{}
Path & Fields recommended for recording & Boundary of interpretation \\
\midrule\noalign{}
\endhead
\bottomrule\noalign{}
\endlastfoot
Registration & Request count, success rate, P50/P95 latency, error classification & Not equivalent to network-wide discovery time \\
Publication & Publication batches, queue time, failed retries & Not equivalent to completed distribution \\
Synchronization & Cursor, backlog, time per batch, recovery time & Not equivalent to query quality \\
Query & Query type, hit rate, P50/P95, duplicate rate & Semantic relevance must be explained separately \\
Governance read & Latest sequence, latency, abnormal duration & Do not interpret a stalled sequence as no events \\
\end{longtable}
}

The report should distinguish at least the cold-start, steady-state, and failure-recovery windows, and should provide the environment, data scale, concurrency, and statistical method at the same time.

Performance results should state whether they include network round trips, database writes, semantic retrieval, and waiting for asynchronous propagation; registration interface latency cannot directly represent network-wide resource visibility latency, and query hit rate cannot directly represent semantic quality. When timeouts, retries, or partial failures occur, also report the request count, failure classification, and final state after recovery. The minimum directory for an evidence package may be fixed as:

\begin{lstlisting}
evidence/
  manifest.json
  spec-and-profile/
  fixtures/
  requests-and-responses/
  logs-and-metrics/
  deployment/
  review-and-result/
\end{lstlisting}

\passthrough{\lstinline!manifest.json!} should list the digest, source, generation time, test environment, and corresponding clauses for each material; the directory structure specifies only how evidence is organized and does not require production configuration or sensitive data to be made public.

The evidence manifest may use the following record format that contains no secrets:

\begin{lstlisting}
{
  "profile": "discovery-node",
  "profileVersion": "v1-alpha",
  "implementationCommit": "<git-commit>",
  "fixtureId": "valid-resource-v1",
  "environment": "isolated-test",
  "result": "PASS",
  "artifacts": [
    {"path": "requests-and-responses/query.json", "sha256": "<digest>"},
    {"path": "logs-and-metrics/discovery.log", "sha256": "<digest>"}
  ],
  "reviewedAt": "<timestamp>"
}
\end{lstlisting}

The angle-bracketed content consists of placeholder values in the evidence-manifest template; a formal report should replace them with the actual commit, digest, and time, but must not write private keys, database credentials, or unredacted internal endpoints.

An automated evaluator closes evidence item by item rather than reading only a process exit code:

\begin{lstlisting}
for each requirement in manifest.requirements:
    assert implementation revision and artifact digest are fixed
    load every fixture and verify its digest
    run positive, negative and required interoperability cases
    compare HTTP result, persisted state, cursor movement and logs
    if environment prerequisite is absent:
        record ENVIRONMENT_UNMET, never PASS
    else if any mandatory assertion fails or evidence is missing:
        record FAIL with the failing evidence paths
    else:
        record PASS with request, response and state evidence paths

overall PASS only if every mandatory requirement is PASS
\end{lstlisting}

\section{Evidence Required for Conformance Implementation}\label{sec:23.12}

Review materials should include at least the specification and protocol versions, code commit, configuration digest, test vectors, interface requests and responses, logs, metrics, deployment records, and failure handling. Submit sensitive content as digests, redacted excerpts, or through controlled viewing; evidence must identify the test time, environment, operator, tool version, and covered clauses.

\subsection{Protocol and Data-Structure Evidence}\label{subsec:23.12.1}

Provide examples of DIDs, DID Documents, resource metadata, ResourcePackages, VCs, Root proofs, and API responses, identify the schema, normalization, version, and content digest, and explain the expected results of valid, missing, and invalid examples. Each data sample should identify at least \passthrough{\lstinline!fixtureId!}, object type, schema/profile version, normalization method, content digest, expected conclusion, and covered clause; valid samples and negative samples should be retained in pairs.

Protocol evidence should also cover HTTP methods and paths, the deserialization type of request JSON, response types or dynamic fields, error status, and key validation results. For registration, Root platform validation and publication, content distribution, and discovery queries, retain at least one valid request and one key failed request with its response for each. Shared types should be traceable to \passthrough{\lstinline!oan-protocol-commons!}, and node-local types and dynamic JSON should be traceable to the corresponding handler. Retaining only an interface address or one successful screenshot cannot prove that the input/output contract is complete.

\subsection{Cryptographic and Identity Evidence}\label{subsec:23.12.2}

Provide public keys, signatures, verification methods, control relationships, credential issuers, and verification results; provide private keys only as evidence of existence or a digest in a controlled test environment. The materials should prove the object covered by the signature and the target DID, rather than merely displaying a signature string. Cryptographic evidence should allow recomputation of the normalized object covered by the signature and the source of the public key used for verification, and should record the verification results for the control signature, registration VC, and Root proof separately.

\subsection{Node-Behavior Evidence}\label{subsec:23.12.3}

Provide requests and responses, events, cursors, logs, and timelines for registration, Root platform publication, distribution, discovery, governance synchronization, and error handling, associated with the resource DID, version, and request identifier. For asynchronous flows, also record stages such as "received, verified, published, and indexed"; do not retain only the last successful response.

\subsection{Deployment and Smoke-Test Evidence}\label{subsec:23.12.4}

Provide the source commit, build artifacts, configuration digest, service health, database status, registration and discovery smoke-test results, and failure-handling records. Public materials should be redacted, and production keys and internal management endpoints must not be included in the evidence. Deployment evidence should at least link the source commit, build artifacts, runtime configuration, and smoke-test results; the smoke test should include one success path and one expected failure path.

\subsection{Version and Provenance Evidence}\label{subsec:23.12.5}

Provide the code commit, lock files, schema/protocol version, ResourcePackage and proof digests, artifact hashes, and publication records, so that reviewers can confirm that the test object and deployment object came from the same source. Provenance checking should at least compare the code commit, dependency lock files, build-artifact digest, and runtime version identifier; when the objects do not match, they must not be considered the same test conclusion.

\section{Test Vectors and Interoperability Fixtures}\label{sec:23.13}

Test fixtures should fix valid DIDs, DID Documents, resource metadata, ResourcePackages, hashes, control signatures, registration VCs, Root proofs, governance events, and invalid inputs, and provide forms readable by Rust, TypeScript, or JSON. Each vector should be marked with the expected success/failure, version, digest, and covered clause, avoiding dependence on live public-network data.

The recommended fixed-fixture record form is as follows:

The same \passthrough{\lstinline!fixtureId!} should point to the same original input in Rust, TypeScript, and HTTP tests; if the adaptation layer must convert it, retain the digests before and after conversion and the conversion rules at the same time.

\section{Negative Tests and Rejection Behavior}\label{sec:23.14}

Negative tests cover invalid DIDs, missing fields, incorrect signatures, unauthorized nodes, expired or revoked credentials, old versions, duplicate requests, incorrect digests, malicious endpoints, and sequence gaps. Acceptance checks not only the returned error, but also confirms that there are no side effects such as writing an invalid index, generating an incorrect publication proof, advancing a cursor, or leaking secrets.

{\def\LTcaptype{none} 
\begin{longtable}[]{@{}>{\raggedright\arraybackslash}p{0.3200\linewidth}>{\raggedright\arraybackslash}p{0.3200\linewidth}>{\raggedright\arraybackslash}p{0.3200\linewidth}@{}}
\toprule\noalign{}
Negative scenario & Rejection result that must be observed & Side effect that must not occur \\
\midrule\noalign{}
\endhead
\bottomrule\noalign{}
\endlastfoot
Invalid DID or missing structure & Distinguishable format/field error & No registration or index write \\
Incorrect or unauthorized signature & Control/authorization failure & Do not overwrite the valid resource \\
Expired or revoked credential & Invalid status & Do not generate a valid publication proof \\
Incorrect digest or malicious endpoint & Integrity/policy failure & Do not advance the publication or synchronization cursor \\
Duplicate or old version & Idempotency or version conflict & Do not create an erroneous new version \\
\end{longtable}
}

Each negative test must check the response, database or index changes, event cursor, sensitive information in logs, and subsequent query results at the same time.

\subsection{Security-Mechanism Degradation Ablation}\label{subsec:23.14.1}

In copied reference-service workspaces, the Yellow Paper separately removes authorization-domain filtering, registration-credential verification, and freshness checks, then runs the complete and degraded paths against the same invalid cases designed for the missing mechanism. The complete path showed no erroneous acceptance or exposure in these cases; each degraded variant exposed its targeted case. This demonstrates that these checks participate in actual trust decisions rather than existing only as descriptive fields.

This result is mechanism-isolation evidence, not a general security proof. A conformance rerun fixes the implementation commit, generated cases, exact removed checkpoint, and observation location, and compares at least the following results.

{\def\LTcaptype{none} 
\begin{longtable}[]{@{}>{\raggedright\arraybackslash}p{0.3200\linewidth}>{\raggedright\arraybackslash}p{0.3200\linewidth}>{\raggedright\arraybackslash}p{0.3200\linewidth}@{}}
\toprule\noalign{}
Isolated mechanism & Focus of degraded case & Rejection behavior required of the complete implementation \\
\midrule\noalign{}
\endhead
\bottomrule\noalign{}
\endlastfoot
Authorization-domain filtering & Whether an out-of-domain resource is indexed or returned by Discovery & Does not expose the out-of-domain resource; semantic relevance cannot expand visibility \\
Registration-credential verification & Whether a credential with missing material, the wrong subject, or an invalid proof is accepted & Rejects the mismatched credential and creates no trusted-publication or indexing side effect \\
Freshness check & Whether stale state, an old version, or stale authorization continues to pass & Adopts a restrictive state and stops sensitive writes or exposure of current results \\
\end{longtable}
}

A degraded variant runs only in an isolated copy with controlled test data; it must not modify a normal service or production binary. The report retains inputs, responses, persistent side effects, and subsequent query results for both the complete and degraded paths. When the case set, implementation commit, or measurement definition differs, erroneous-acceptance counts must not be concatenated into a trend.

\section{Conformance Evidence Retention and Review}\label{sec:23.15}

Archive evidence by specification version, code commit, environment, time, operator, and test fixture. When a change affects the protocol, schema, signature, governance, node behavior, or deployment path, the relevant profile must be retested. The review conclusion should state pass, fail, not tested, or conditional pass, and must not conceal a key failure behind a single overall score.

{\def\LTcaptype{none} 
\begin{longtable}[]{@{}>{\raggedright\arraybackslash}p{0.3200\linewidth}>{\raggedright\arraybackslash}p{0.3200\linewidth}>{\raggedright\arraybackslash}p{0.3200\linewidth}@{}}
\toprule\noalign{}
Conclusion & Review requirement & Retest condition \\
\midrule\noalign{}
\endhead
\bottomrule\noalign{}
\endlastfoot
PASS & All mandatory criteria and rejection behaviors have evidence & Retest when the specification, code, or deployment object changes \\
FAIL & State the failed input, impact scope, and blocking action & Rerun the original and regression vectors after the fix \\
NOT TESTED & List the missing environment or evidence & Do not reuse the old conclusion before the prerequisites are completed \\
CONDITIONAL & State the limitation, deadline, and responsible person & Review before the limitation is removed or the deadline expires \\
\end{longtable}
}

The review conclusion should also distinguish "node profile passed," "end-to-end path passed," and "production deployment passed," so that a local test result is not overgeneralized as conformance of the entire network.

\section{Cross-Role Protocol Invariants}\label{sec:23.16}

In addition to role-profile acceptance, conformance checks cover protocol invariants that span multiple components. The table converts Yellow Paper invariants I1--I9 into acceptance assertions for this manual. They constrain observable results rather than prescribing a database, queue, or programming language. If any invariant fails, end-to-end conformance fails even when every service health endpoint and local positive case passes.

{\def\LTcaptype{none} 
\begin{longtable}[]{@{}>{\raggedright\arraybackslash}p{0.3200\linewidth}>{\raggedright\arraybackslash}p{0.3200\linewidth}>{\raggedright\arraybackslash}p{0.3200\linewidth}@{}}
\toprule\noalign{}
Invariant & Relationship that must always hold & Minimum negative or recovery validation \\
\midrule\noalign{}
\endhead
\bottomrule\noalign{}
\endlastfoot
I1 Root admission integrity & A Root admission record binds the resource DID, version, representation digest, state, and Root proof; a later package claiming that admission fact matches the bound digest & After package content, version, or digest is altered, CDN/Discovery validation rejects it and does not advance a trusted cursor \\
I2 Registrar authorization & When Root accepts a Registrar submission, that Registrar is within valid authorization scope at submission time and its node identity verifies the upstream signed envelope & A submission from an unauthorized, revoked, or incorrectly signing Registrar is rejected without an admission record \\
I3 Discovery scope & Every resource indexed and exposed by Discovery satisfies the node authorization domain, resource state, and query visibility scope; capability tags may narrow matching or affect ranking but cannot expand authorization & Inject an out-of-domain but highly relevant resource; an ordinary query must not expose its existence \\
I4 Current-version visibility & After Discovery processes a replacement or revocation fact, the old version is no longer returned as the current acceptable version & Publish v2 or revoke v1; after the corresponding cursor is processed, a default query no longer returns v1 as current \\
I5 Invocation binding & Before a Service Agent accepts a trusted invocation, signature, timestamp, nonce, target DID, request-body digest, and caller credential all bind correctly & Replace any one of target, body, nonce, or caller credential; the business handler must not execute the protected action \\
I6 DID proof and controller authorization & A DID Document proof verifies under its declared verification method; an external-controller proof binds the resource, controller, Registrar, document digest, purpose, validity period, and nonce; for an existing resource, the protocol validates authorization against the controller declared by Root\textquotesingle s current admitted version & Self-takeover by a candidate new controller, update by an old controller after transfer, and replay across Registrars are all rejected \\
I7 Root identity directory is non-authoritative & After obtaining a public key from the Root identity directory, an application still verifies a fresh challenge signature; a directory hit grants neither resource control nor a substitute for DID Document and VC verification & Login fails with only a directory record and no valid challenge signature; a successful login still cannot directly authorize a controller operation \\
I8 Semantic recovery & If structured synchronization succeeds and semantic indexing fails, a durable retry state remains; semantic work must not be silently discarded & Inject an embedding/upsert failure; confirm that the structured record remains, semantic state is pending/failed, and recovery completes after the dependency returns \\
I9 Public interface is non-authoritative & When a portal, public-metrics cache, or curated governance-event table is stale or unavailable, protocol decisions still derive from Root admission, governance projections, package verification, Discovery authorization, and response signatures & Make a public page return an old snapshot; nodes must not restore a revoked resource or permit a sensitive write based on it \\
\end{longtable}
}

Each invariant maps one-to-many to stable Requirement IDs. For example, I6 covers at least first registration under an external controller, update of an existing DID, controller transfer, rejection of the former controller, and atomic concurrent update. One positive signature test cannot represent all of these semantics. Evidence for I8 spans fault injection, persistent state, and recovery retry. Evidence for I9 simultaneously shows that the public surface and protocol fact can diverge while the system still makes a restrictive decision.

An invariant evidence package records all involved roles, implementation commits, configuration digest, input fixture, requests and responses, persistent side effects, cursor changes, and final query results. An asynchronous invariant waits for an explicit cursor or state condition rather than reading once after a fixed sleep. A timeout is reported as non-convergence or failure; it cannot be converted into an overall pass from the last local success.

\section{References}\label{sec:23.17}

{\def\LTcaptype{none} 
\begin{longtable}[]{@{}>{\raggedright\arraybackslash}p{0.3200\linewidth}>{\raggedright\arraybackslash}p{0.3200\linewidth}>{\raggedright\arraybackslash}p{0.3200\linewidth}@{}}
\toprule\noalign{}
Source & Type & Link \\
\midrule\noalign{}
\endhead
\bottomrule\noalign{}
\endlastfoot
\passthrough{\lstinline!oan-protocol-commons!} & Code repository: protocol types, schemas, and error contracts & \url{https://github.com/wolfbrother/oan-protocol-commons} \\
\passthrough{\lstinline!oan-root-nodes!} & Code repository: Root platform conformance and publication test objects & \url{https://github.com/OpenAgenet/oan-root-nodes} \\
\passthrough{\lstinline!oan-registrar-nodes!} & Code repository: registration flow and server-side validation & \url{https://github.com/OpenAgenet/oan-registrar-nodes} \\
\passthrough{\lstinline!oan-discovery-nodes!} & Code repository: discovery queries, indexing, and semantic test objects & \url{https://github.com/OpenAgenet/oan-discovery-nodes} \\
\passthrough{\lstinline!oan-trust-indexer!} & Code repository: governance state and synchronization test objects & \url{https://github.com/OpenAgenet/oan-trust-indexer} \\
\passthrough{\lstinline!oan-sdk-ts!} & Code repository: client conformance and cross-language invocation verification & \url{https://github.com/OpenAgenet/oan-sdk-ts} \\
\passthrough{\lstinline!oan-community-skill!} & Code repository: community Skill smoke tests and operational verification & \url{https://github.com/OpenAgenet/oan-community-skill} \\
OAN Yellow Paper & arXiv yellow paper: protocol security properties and conformance boundaries & \url{https://arxiv.org/abs/2606.03163} \\
OAN Resource Identity and Discovery & IETF draft: resource identity and discovery test directions & \url{https://datatracker.ietf.org/doc/draft-xu-oan-resource-identity-discovery/} \\
Efficient Agent Discovery Profile & IETF draft: discovery profile test directions & \url{https://datatracker.ietf.org/doc/draft-xu-efficient-agent-discovery-profile/} \\
\end{longtable}
}

\chapter{Versioning, Compatibility, and Change Management}\label{chap:24}

This chapter explains that the version object of OpenAgenet (OAN) is not a single number, but a set of coordinates across multiple layers:

{\def\LTcaptype{none} 
\begin{longtable}[]{@{}>{\raggedright\arraybackslash}p{0.3200\linewidth}>{\raggedright\arraybackslash}p{0.3200\linewidth}>{\raggedright\arraybackslash}p{0.3200\linewidth}@{}}
\toprule\noalign{}
Layer & Example & Purpose \\
\midrule\noalign{}
\endhead
\bottomrule\noalign{}
\endlastfoot
Specification & Unified technical specification major/minor/patch versions & Defines semantics and requirements \\
Protocol & Node API, error, and event versions & Defines interaction contracts \\
Data schema & DID Document, ResourcePackage, VC, and governance event schema/profile & Defines parseable structures \\
Method & \passthrough{\lstinline!did:oan!} method rules & Defines identifier and resolution semantics \\
Implementation & Rust crate, npm package, and website build commit & Provides concrete code \\
Deployment & Configuration, binary, and service commit & Defines the runtime environment \\
\end{longtable}
}

Release records must associate the layers above and must not use Cargo or npm version numbers as substitutes for protocol, schema, or deployment versions. One implementation version may support multiple protocol or schema versions, and one resource DID may correspond to multiple historical DID Documents and ResourcePackage versions; therefore, releases, compatibility, migration, and rollback must all record the actual mappings. Version records must also distinguish "code committed," "artifact built," "service deployed," and "change effective," so that repository state is not mistaken for production state.

The relationships among version layers can be checked as follows:

{\def\LTcaptype{none} 
\begin{longtable}[]{@{}>{\raggedright\arraybackslash}p{0.3200\linewidth}>{\raggedright\arraybackslash}p{0.3200\linewidth}>{\raggedright\arraybackslash}p{0.3200\linewidth}@{}}
\toprule\noalign{}
Change object & Direct impact & Must not be inferred directly \\
\midrule\noalign{}
\endhead
\bottomrule\noalign{}
\endlastfoot
Rust crate or npm package & A specific implementation or client capability & The protocol, DID method, or production deployment has been upgraded \\
Protocol/API & Requests, responses, and error handling between nodes & Historical ResourcePackages have already been migrated \\
Resource schema/profile & Parsing of DID Documents, ResourcePackages, VCs, or events & Control rights or the resource DID have changed \\
\passthrough{\lstinline!did:oan!} method & Identifier resolution and control semantics & Ordinary resource versions must use a new DID \\
Deployment artifacts and configuration & Runtime behavior in a particular environment & All nodes or third-party implementations have been upgraded \\
\end{longtable}
}

\section{Specification Version Management}\label{sec:24.1}

The unified technical specification uses major, minor, and patch versions to record structural, semantic, and errata changes. A major version change means that existing implementations may no longer interoperate directly; a minor version is used for backward-compatible extensions; and a patch version is used for clarifications or corrections that do not change specification semantics. Each release retains the affected sections, release date, status, and implementation mapping.

{\def\LTcaptype{none} 
\begin{longtable}[]{@{}>{\raggedright\arraybackslash}p{0.3200\linewidth}>{\raggedright\arraybackslash}p{0.3200\linewidth}>{\raggedright\arraybackslash}p{0.3200\linewidth}@{}}
\toprule\noalign{}
Level & Applicable change & Must retain \\
\midrule\noalign{}
\endhead
\bottomrule\noalign{}
\endlastfoot
Major & Breaks existing interoperability or semantics & Impact scope, migration, and approval \\
Minor & Backward-compatible extension & New fields, compatibility tests, and implementation mapping \\
Patch & Clarification or erratum & Clause, original intent, and release date \\
\end{longtable}
}

\section{Protocol Version Management}\label{sec:24.2}

Node protocols and public APIs should explicitly define version identifiers, supported ranges, request/response schemas, error semantics, and negotiation methods. During the transition period, old clients may use only profiles that the server still declares as supported. The server must not silently change signatures, versions, publication sequences, or governance decisions merely to accommodate legacy requests; when compatibility is not possible, it must return an identifiable incompatibility error.

\begin{lstlisting}
client-supported profiles intersection server-supported profiles
empty intersection -> explicit incompatible response
\end{lstlisting}

Protocol compatibility therefore requires more than checking whether a request can be deserialized. It must also verify whether the same input still produces consistent results for signature verification, publication sequencing, governance decisions, error codes, and retry behavior. The server must not silently change these semantics to accommodate legacy requests. If an old version can still be read but cannot be safely written, read, write, and query capabilities must be declared separately.

{\def\LTcaptype{none} 
\begin{longtable}[]{@{}>{\raggedright\arraybackslash}p{0.3200\linewidth}>{\raggedright\arraybackslash}p{0.3200\linewidth}>{\raggedright\arraybackslash}p{0.3200\linewidth}@{}}
\toprule\noalign{}
Capability & Compatibility issue & Recording method \\
\midrule\noalign{}
\endhead
\bottomrule\noalign{}
\endlastfoot
Read & Whether an old client can parse the response and unknown fields & Supported response profiles and examples \\
Write & Whether the request will generate a different signature, version, or publication event & Supported write versions and rejection conditions \\
Query & Whether pagination cursors, sorting, and semantic filters retain their meaning & Query version, errors, and boundary examples \\
Retry & Whether a retry causes duplicate registration or publication & Idempotency keys, sequence numbers, and duplicate-request tests \\
\end{longtable}
}

\section{Resource Schema Version Management}\label{sec:24.3}

Resource metadata, ResourcePackages, VCs, governance events, and discovery responses record schema or profile versions separately. They are not one shared version: \passthrough{\lstinline!packageVersion!} in a ResourcePackage describes the resource content version; the schema version of a DID Document describes the document structure; the schema version of a VC describes the credential claims and proof structure; and the event version of a governance event describes its payload and event semantics. A reader should identify each version first, then select the parser and validation rules. It must not use the ResourcePackage version as a substitute for the credential or event schema version.

The handling boundaries for field changes are as follows:

{\def\LTcaptype{none} 
\begin{longtable}[]{@{}>{\raggedright\arraybackslash}p{0.3200\linewidth}>{\raggedright\arraybackslash}p{0.3200\linewidth}>{\raggedright\arraybackslash}p{0.3200\linewidth}@{}}
\toprule\noalign{}
Change & Default handling & Additional requirements \\
\midrule\noalign{}
\endhead
\bottomrule\noalign{}
\endlastfoot
Add an optional field & Compatibility candidate & Old readers should ignore or retain it; new readers provide fallback behavior \\
Add a required field & Potentially breaking & Provide default semantics, upgrade the profile, or migrate old data \\
Rename a field, change its type, or change its unit & Breaking & New version, conversion rules, and negative examples \\
Change the digest algorithm or signature coverage & Breaking & New validation rules, test vectors, and republishing \\
Add an unknown status or governance event & Requires explicit handling & Old implementations must not treat an unknown status as valid or completed \\
\end{longtable}
}

The example above illustrates field relationships only; the angle-bracketed values are not fixed OAN values. For historical materials, readers should retain the original version, digest, and verification result. Required fields, digest algorithms, signature structures, or status values that cannot be understood must be rejected or placed into an explicitly isolated state; they must not be silently treated as currently valid data.

Readers should identify the schema/profile first, then validate required fields, the digest algorithm, the signature structure, and the status value. When deleting or renaming a field, changing its type, or changing signature coverage, a new profile, migration notes, and negative examples should be produced.

\section{DID Method Compatibility}\label{sec:24.4}

Changes to the \passthrough{\lstinline!did:oan!} method must assess the DID Core data structures, method-specific identifiers, resolver, DID Documents, historical resources, and Root proofs together. A normal resource version upgrade usually keeps the resource DID unchanged and expresses the version change through a new DID Document, ResourcePackage, content digest, control proof, or Root proof. This is a different layer of change from the resolver implementation version or node software version. If method resolution rules, identifier normalization, or control semantics change incompatibly, a strategy must be defined for continued resolution, migration, or deactivation of old DIDs; updating only the implementation version is insufficient.

{\def\LTcaptype{none} 
\begin{longtable}[]{@{}>{\raggedright\arraybackslash}p{0.4800\linewidth}>{\raggedright\arraybackslash}p{0.4800\linewidth}@{}}
\toprule\noalign{}
Object & Must check \\
\midrule\noalign{}
\endhead
\bottomrule\noalign{}
\endlastfoot
Identifier & Syntax, routing-code case, suffix character set, and parsing behavior \\
DID Document & id, verification methods, services, and extension fields \\
Control semantics & Public keys, signature coverage, and update authorization \\
Historical resources & Old DIDs, versions, VCs, and Root proofs \\
Resolver & Success, error, unknown method, and migration behavior \\
\end{longtable}
}

An unchanged resource DID does not mean that the document content is immutable. The updater must still prove control over the DID and ensure that the new document, ResourcePackage version, VC, and Root proof can validate one another. A Discovery node should select the currently available version according to its supported versions and freshness rules, while retaining traceable relationships to historical versions. Conversely, simply changing a server, SDK, or resolver does not automatically create a new resource version.

\section{Backward-Compatible and Breaking Changes}\label{sec:24.5}

Compatibility is determined by verifiable results for old clients, old nodes, historical data, and signature verification behavior, not merely by whether fields can still be deserialized. The assessment should cover at least registration submission, Root platform publication, content distribution platform processing, discovery indexing, governance validation, SDK/Skill invocation, and failure recovery. If any stage interprets the same input as a different resource, version, status, or trust result, it must not be deemed compatible merely because it is parseable.

{\def\LTcaptype{none} 
\begin{longtable}[]{@{}>{\raggedright\arraybackslash}p{0.3200\linewidth}>{\raggedright\arraybackslash}p{0.3200\linewidth}>{\raggedright\arraybackslash}p{0.3200\linewidth}@{}}
\toprule\noalign{}
Change type & Default determination & Required action \\
\midrule\noalign{}
\endhead
\bottomrule\noalign{}
\endlastfoot
Add an optional field & Compatibility candidate & Regression tests with old clients \\
Clarify an error & Requires assessment & Check whether rejection behavior changes \\
Remove a required field & Breaking & Increase the version and migrate \\
Change signature or governance semantics & Breaking & Governance approval, republishing, and retesting \\
\end{longtable}
}

\subsection{Compatible Changes}\label{subsec:24.5.1}

Compatible changes preserve existing required fields, signature semantics, status meanings, and interface success/failure rules. New content should be safely ignorable by old implementations or identifiable through negotiation. Even if the code compiles, results must be verified with old clients, old nodes, and historical data.

\begin{itemize}
\tightlist
\item[$\square$]
  Existing required fields and error semantics remain valid.
\item[$\square$]
  Existing signature, status, and version-selection results remain unchanged.
\item[$\square$]
  Old clients can read or safely ignore the new content.
\item[$\square$]
  Historical resources, DID Documents, and credentials remain verifiable.
\end{itemize}

\subsection{Breaking Changes}\label{subsec:24.5.2}

Breaking changes include removing or renaming required fields, changing DID or signature semantics, changing resource status, protocol paths, version selection, or governance authorization decisions. Such changes must increase the version, publish the impact, schedule a migration window, and prevent nodes that have not completed migration from mistakenly mixing new and old materials. The following checks should be completed before releasing a breaking change:

\begin{itemize}
\tightlist
\item[$\square$]
  List affected nodes, clients, SDKs/Skills, DID Documents, ResourcePackages, VCs, Root proofs, and indexes.
\item[$\square$]
  Specify the read, write, and discovery behavior of the old version during the transition period.
\item[$\square$]
  Provide before-and-after migration examples, negative examples, rollback data, and retest records.
\item[$\square$]
  Prevent unmigrated components from interpreting new and old materials as the same valid version.
\end{itemize}

Before release, freeze the impact scope and identify affected nodes, clients, DID Documents, ResourcePackages, VCs, Root proofs, governance status, and databases. Unmigrated objects must not mix new and old materials. For historical objects that cannot be migrated in one step, explicitly state whether they remain readable, verifiable only, prohibited from updates, or placed in an isolation area, and ensure that discovery results do not display expired or unverified objects as the currently valid version.

\subsection{Version Identifiers}\label{subsec:24.5.3}

A version identifier should state its layer, format, source, and comparison rules. For example, specification, protocol, ResourcePackage, schema, Rust crate, npm package, and deployment commit versions cannot substitute for one another. Record the ResourcePackage version separately from resource DID stability, and store both the version and content digest in the evidence.

\subsection{Migration and Transition Period}\label{subsec:24.5.4}

The transition period should specify the parallel support range for old and new versions, start and end times, the dual-read/single-write data strategy, node upgrade order, failure rollback, and the conditions for stopping the old version. During the transition period, different versions must not create unexplainable forks of valid versions for the same DID. At the end, retest the registration, Root platform publication, content distribution platform, discovery query, and governance paths. If the old version is read-only, the server must state this clearly in its protocol capability declaration and error responses; callers must not have to infer migration status through trial and error. The migration plan should record at least:

{\def\LTcaptype{none} 
\begin{longtable}[]{@{}>{\raggedright\arraybackslash}p{0.3200\linewidth}>{\raggedright\arraybackslash}p{0.3200\linewidth}>{\raggedright\arraybackslash}p{0.3200\linewidth}@{}}
\toprule\noalign{}
Phase & Permitted operations & Must observe \\
\midrule\noalign{}
\endhead
\bottomrule\noalign{}
\endlastfoot
Dual read & Read old and new formats & Parsing and digest consistency \\
Single-write switch & Generate only the new format & Whether old-client errors are identifiable \\
Convergence & Complete historical data and indexes & Consistency of DIDs, versions, and Root proofs \\
Completion & Close the old path & Rollback entry point and residual data \\
\end{longtable}
}

It is recommended to extend read capabilities first, then switch writers, and finally close the old read path. Throughout the process, retain evidence of database migration, rollback, and discovery/publication retesting.

\section{Specification Changes for Governance Controls}\label{sec:24.6}

Changes involving node authorization, governance events, resource status, Root proofs, trust decisions, or node protocols must go through a formal proposal, impact assessment, governance approval, release, and effectiveness record. The proposal, approval, code implementation, build artifact, deployment status, and production effectiveness are different types of evidence and must be recorded separately. Signers, verifiers, and affected nodes must be explicit; code being merged or a website being updated cannot substitute for governance effectiveness.

\pandocbounded{\includegraphics[width=\linewidth,height=0.72\textheight,keepaspectratio]{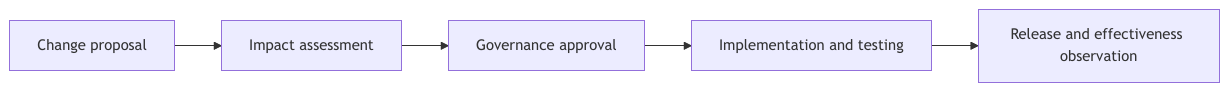}}

\subsection{Change Proposal}\label{subsec:24.6.1}

The proposal should describe the motivation, scope, affected sections and objects, compatibility level, migration plan, test evidence, risks, rollback method, and intended effective time. Proposals involving signatures, authorization, or DID resolution should also list the impact on historical data and third-party nodes.

{\def\LTcaptype{none} 
\begin{longtable}[]{@{}>{\raggedright\arraybackslash}p{0.4800\linewidth}>{\raggedright\arraybackslash}p{0.4800\linewidth}@{}}
\toprule\noalign{}
Item & Content to provide \\
\midrule\noalign{}
\endhead
\bottomrule\noalign{}
\endlastfoot
Motivation and scope & Reasons, roles, and objects \\
Compatibility & Major/minor/patch level and basis \\
Migration & Data, nodes, clients, and evidence \\
Verification & Test vectors, integration tests, and rollback drills \\
Effectiveness & Approving authority, conditions, and time \\
\end{longtable}
}

\subsection{Governance Approval}\label{subsec:24.6.2}

Approval should produce a verifiable governance record describing the decision-making authority, approved content, version, scope, conditions, and basis for effectiveness. An ordinary implementation commit, an operational verbal confirmation, or a local configuration on a single node cannot replace a fact that requires governance authorization. The governance approval record should answer at least "who approved it, what was approved, when it takes effect, who verifies it, and which objects are affected":

\subsection{Release and Effectiveness}\label{subsec:24.6.3}

The release record associates the specification version, protocol/schema version, code commit, artifact hash, and configuration. Effectiveness should be executed according to the declared time, governance event, or node upgrade condition. Backups, compatibility checks, and conformance retesting must be completed before effectiveness; updating a page does not mean that the protocol has taken effect.

After effectiveness, observe cross-node status, client errors, publication cursors, indexing latency, version distribution, and rollback feasibility to confirm that implementation status matches the specification declaration. If deployment is complete but a Discovery node is still using an old protocol, old schema, or old governance configuration, record it as "deployed but not fully effective" rather than marking it complete directly.

\subsection{Rollback and Emergency Changes}\label{subsec:24.6.4}

Emergency changes are limited to fixing security, data integrity, or major availability risks, and must record the reason, authorization, impact scope, and temporary validity period. Rollback must consider the database, publication sequence, governance status, and historical evidence; it is not enough to replace the service binary with an older version. Complete the formal review record afterward. Rollback checks for an emergency change should be performed by object:

{\def\LTcaptype{none} 
\begin{longtable}[]{@{}>{\raggedright\arraybackslash}p{0.4800\linewidth}>{\raggedright\arraybackslash}p{0.4800\linewidth}@{}}
\toprule\noalign{}
Object & Rollback requirement \\
\midrule\noalign{}
\endhead
\bottomrule\noalign{}
\endlastfoot
Binary and configuration & Restore verifiable build artifacts and configuration digests \\
Database & Retain the pre-migration backup and recovery validation \\
Synchronization cursor & Prevent duplicate processing or skipped events \\
Publication status & Do not leave a failed publication recorded as a valid fact \\
Cache and index & Define the cleanup and resynchronization scope \\
\end{longtable}
}

Rollback checks should also cover synchronization cursors, caches, and unfinished publication tasks. After an emergency change ends, complete the formal approval and retesting.

\section{Deprecation and Migration Strategy}\label{sec:24.7}

Deprecated objects may include interface paths, fields, digest algorithms, protocol versions, node binaries, SDK/Skill packages, or configuration items. A notice should state the object, replacement, support period, affected versions, migration tools, verification method, and service-stop conditions. It should separately specify whether the old object is readable, writable, discoverable, or callable during the transition period. Old interfaces or fields may return explicit deprecation information during the transition period, but must not silently change results. After nodes and clients complete migration, stop the old version according to the recorded conditions and retain audit evidence.

The notice should provide at least before-and-after migration examples, verification commands, the support end time, and a failure rollback entry point.

\section{Change Records and Release Notes}\label{sec:24.8}

Each change must record the affected sections, data structures, interfaces, code repositories, database migrations, tests, deployment, rollback, operators, and known impacts. It must also identify the responsible person and deadline when specification clauses and implementation status are inconsistent. Release notes should tell maintainers "what changed, why it changed, who needs to act, and how to verify it," and should allow them to trace the change identifier to approval, commit, artifact, deployment, and effectiveness checks.

The YAML above is an example of a release record, not a fixed OAN configuration. Status checks can proceed in the following order: after code is committed, record the commit identifier; after the build is complete, record the artifact digest; after deployment is complete, record the target environment and service status; after the smoke test passes, record the test result; and only after the protocol, data, and governance effectiveness conditions are satisfied, record \passthrough{\lstinline!effective!}. If any stage fails, retain the failure reason and rollback action instead of overwriting the status as successful.

\subsection{Specification Change Records}\label{subsec:24.8.1}

Record the sections, clauses, version, change type, original text and new intent, compatibility conclusion, approval record, and effective time. Mark errata, clarifications, and specification semantic changes separately so that substantive changes are not presented as textual revisions. Specification change records can be checked using the following fields:

{\def\LTcaptype{none} 
\begin{longtable}[]{@{}>{\raggedright\arraybackslash}p{0.4800\linewidth}>{\raggedright\arraybackslash}p{0.4800\linewidth}@{}}
\toprule\noalign{}
Field & Content \\
\midrule\noalign{}
\endhead
\bottomrule\noalign{}
\endlastfoot
affectedChapters & Affected sections and clauses \\
changeType & Erratum, clarification, compatible extension, or semantic change \\
approval & Approval record \\
effectiveAt & Effectiveness condition or time \\
implementationMap & Corresponding code, schema, and deployment versions \\
\end{longtable}
}

\subsection{Protocol Change Records}\label{subsec:24.8.2}

Record changes to routes, requests/responses, status codes, errors, signatures, cursors, version negotiation, and node interoperability, and associate them with protocol tests and the support period for old versions. At minimum, state whether old requests can still register, publish, query, and retry, and what stable error semantics are returned when they are incompatible. Protocol change records should list routes, requests/responses, status codes, errors, signatures, cursors, and negotiation behavior separately, and should include at least one successful old-version example, one successful new-version example, and one incompatible example.

\subsection{Implementation Change Records}\label{subsec:24.8.3}

Record the affected repositories, commits, dependencies, Cargo/package versions, database migrations, build artifacts, and deployment environments. If an implementation fix has not yet been synchronized with the specification, clearly state its temporary status and who is responsible for the subsequent specification update. Rust crates, npm packages, website builds, and node binaries must be recorded separately; none of their version numbers may be treated as the version number of the entire OAN implementation. Implementation change records should distinguish the four statuses "code merged," "build completed," "deployed," and "smoke test passed," so that the existence of a commit is not interpreted as production effectiveness.

\subsection{Migration Notices and Known Impacts}\label{subsec:24.8.4}

List the nodes, clients, SDKs/Skills, configurations, and databases that need upgrades. Explain the impact on old resources, historical DIDs, VCs, Root proofs, indexes, and caches, as well as known limitations after migration failure and rollback. Migration notices should list actions and impacts by role and distinguish "must upgrade" from "can continue reading":

{\def\LTcaptype{none} 
\begin{longtable}[]{@{}>{\raggedright\arraybackslash}p{0.4800\linewidth}>{\raggedright\arraybackslash}p{0.4800\linewidth}@{}}
\toprule\noalign{}
Role & What to watch \\
\midrule\noalign{}
\endhead
\bottomrule\noalign{}
\endlastfoot
Resource provider & ResourcePackage, DID Document, and signature-generation methods \\
Node operator & Configuration, database, cursor, and service upgrade order \\
Client and SDK & Version negotiation, error handling, and cache invalidation \\
Reviewer & Old resources, historical evidence, and retesting scope \\
\end{longtable}
}

\section{References}\label{sec:24.9}

{\def\LTcaptype{none} 
\begin{longtable}[]{@{}>{\raggedright\arraybackslash}p{0.3200\linewidth}>{\raggedright\arraybackslash}p{0.3200\linewidth}>{\raggedright\arraybackslash}p{0.3200\linewidth}@{}}
\toprule\noalign{}
Source & Type & Link \\
\midrule\noalign{}
\endhead
\bottomrule\noalign{}
\endlastfoot
\passthrough{\lstinline!oan-protocol-commons!} & Code repository: protocol, schema, error codes, and version-related structures & \url{https://github.com/wolfbrother/oan-protocol-commons} \\
\passthrough{\lstinline!oan-root-nodes!} & Code repository: Root platform publication and compatibility changes & \url{https://github.com/OpenAgenet/oan-root-nodes} \\
\passthrough{\lstinline!oan-registrar-nodes!} & Code repository: registration protocol and version migration & \url{https://github.com/OpenAgenet/oan-registrar-nodes} \\
\passthrough{\lstinline!oan-discovery-nodes!} & Code repository: indexing, querying, and synchronization compatibility & \url{https://github.com/OpenAgenet/oan-discovery-nodes} \\
\passthrough{\lstinline!oan-trust-indexer!} & Code repository: governance events and projection version migration & \url{https://github.com/OpenAgenet/oan-trust-indexer} \\
\passthrough{\lstinline!oan-sdk-ts!} & Code repository: client versions and compatibility & \url{https://github.com/OpenAgenet/oan-sdk-ts} \\
\passthrough{\lstinline!oan-community-skill!} & Code repository: scripts, examples, and release compatibility & \url{https://github.com/OpenAgenet/oan-community-skill} \\
OAN White Paper & arXiv white paper: ecosystem evolution and the positioning of open infrastructure & \url{https://arxiv.org/abs/2606.03161} \\
OAN Yellow Paper & arXiv yellow paper: protocol versions, trust, and change boundaries & \url{https://arxiv.org/abs/2606.03163} \\
\passthrough{\lstinline!did:oan!} DID Method Specification & Standard/method specification: DID method compatibility & \url{https://github.com/OpenAgenet/oan-public-docs/blob/main/did-oan-specs/doc/OAN\%20DID\%20Method\%20Specification.md} \\
OAN Resource Identity and Discovery & IETF draft: resource identity and discovery version direction & \url{https://datatracker.ietf.org/doc/draft-xu-oan-resource-identity-discovery/} \\
Agentic Overlay Network Architecture & IETF draft: network architecture evolution direction & \url{https://datatracker.ietf.org/doc/draft-xu-agentic-overlay-network-architecture/} \\
Efficient Agent Discovery Profile & IETF draft: discovery profile evolution direction & \url{https://datatracker.ietf.org/doc/draft-xu-efficient-agent-discovery-profile/} \\
\end{longtable}
}

\appendix
\chapter{Normative Data Schemas}\label{chap:appendix-a}

The schemas in this appendix are the common basis for protocol implementations, test vectors, and interface validation. Field names are governed by the actual JSON serialization result. Rust types, TypeScript types, and HTTP examples must not independently invent synonymous fields. If a field exists only in a reference implementation or a specific resource type, it should be explicitly marked as optional or as a profile extension. Schema versions, content digests, signature coverage, and unknown-field handling rules must remain consistent with Chapters~\ref{chap:04}~through~\ref{chap:07}, Chapter~\ref{chap:15}, and Chapter~\ref{chap:24} of the main text.

\section{DID Document Schema}\label{sec:appendix-a.1}

The public \passthrough{\lstinline!DidDocument!} fields are \passthrough{\lstinline!@context!}, \passthrough{\lstinline!id!}, optional \passthrough{\lstinline!controller!}, \passthrough{\lstinline!verificationMethod!}, \passthrough{\lstinline!authentication!}, \passthrough{\lstinline!assertionMethod!}, optional \passthrough{\lstinline!capabilityInvocation!}, \passthrough{\lstinline!service!}, optional \passthrough{\lstinline!proof!}, and optional \passthrough{\lstinline!oanMetadata!}. \passthrough{\lstinline!id!} is a valid \passthrough{\lstinline!did:oan!}; its five-character case-sensitive Base58 routing code selects a Registrar route, while its 32-character Base58 suffix identifies the subject. Neither encodes resource type. The top-level \passthrough{\lstinline!controller!} and \passthrough{\lstinline!oanMetadata.controllerDid!} express the same single controller. An update to an existing DID is authorized against the current document held by the Root, not merely the controller claimed by the candidate document.

\section{Resource Metadata Schema}\label{sec:appendix-a.2}

The four resource types share information such as the resource DID, resource type, name, description, version, capability tags, use cases, endpoints, protocol bindings, and implementation links. However, the current public types carry this information at different levels: \passthrough{\lstinline!ResourceMetadata!} uses \passthrough{\lstinline!resourceDid!}, \passthrough{\lstinline!resourceType!}, \passthrough{\lstinline!subjectType!}, \passthrough{\lstinline!name!}, \passthrough{\lstinline!description!}, \passthrough{\lstinline!capabilityTags!}, \passthrough{\lstinline!authorizedDomains!}, \passthrough{\lstinline!protocolBindings!}, \passthrough{\lstinline!services!}, \passthrough{\lstinline!lifecycleState!}, \passthrough{\lstinline!packageVersion!}, hashes, and \passthrough{\lstinline!updatedAt!}; the DID Document expresses the corresponding information through \passthrough{\lstinline!oanMetadata.resourceDescription!}, \passthrough{\lstinline!capabilityTags!}, \passthrough{\lstinline!protocolBindings!}, and \passthrough{\lstinline!packageInfo!}. The enumeration values for the four resource types are \passthrough{\lstinline!agent\_service!}, \passthrough{\lstinline!skill!}, \passthrough{\lstinline!mcp\_server!}, and \passthrough{\lstinline!tool\_api!}; the display name \passthrough{\lstinline!agent!} must not be used as a protocol value. Length, duplicate-item, and format restrictions for strings, arrays, and URIs should be supplemented by the specific profile or node validation rules. Unknown resource types, an empty DID, missing digests, or unresolvable endpoints must not be treated as complete metadata. New optional fields may be added compatibly; breaking changes should be handled through version management.

\passthrough{\lstinline!externalIdentifiers!} is an array in DID Document \passthrough{\lstinline!oanMetadata!}, not in the strongly typed \passthrough{\lstinline!ResourceMetadata!} projection. Each item requires a non-empty \passthrough{\lstinline!id!}; \passthrough{\lstinline!resolutionServiceEndpoint!} is optional and is a supported absolute URI. The current public validation limits the array to eight items, rejects control characters and \passthrough{\lstinline!did:oan:!} in an external ID, rejects dangerous URI schemes and embedded credentials, and limits the total serialized size. The following JSON Schema captures the structural portion; UTF-8 byte limits and compound deduplication remain implementation checks.

\begin{lstlisting}
{
  "$schema": "https://json-schema.org/draft/2020-12/schema",
  "$id": "urn:oan:schema:external-identifiers:2026-11",
  "title": "OAN externalIdentifiers",
  "type": "array",
  "maxItems": 8,
  "items": {
    "type": "object",
    "additionalProperties": false,
    "required": ["id"],
    "properties": {
      "id": {
        "type": "string",
        "minLength": 1,
        "pattern": "^(?!did:oan:)[^\\u0000-\\u001F\\u007F]+$"
      },
      "resolutionServiceEndpoint": {
        "type": "string",
        "format": "uri",
        "pattern": "^(?!(?:file|data|javascript):)(?![^:/?#]+://[^/?#]*@).+$"
      }
    }
  }
}
\end{lstlisting}

A complete value embedded under \passthrough{\lstinline!oanMetadata!} may be:

\begin{lstlisting}
{
  "subjectType": "agent_service",
  "resourceType": "agent_service",
  "externalIdentifiers": [
    {
      "id": "urn:example:agent-service:contract-review",
      "resolutionServiceEndpoint": "https://resolver.example.org/resources/contract-review"
    }
  ],
  "capabilityTags": ["document.review"],
  "authorizedDomains": ["professional_services.legal"]
}
\end{lstlisting}

\section{ResourcePackage Schema}\label{sec:appendix-a.3}

A ResourcePackage is a verifiable, distributable versioned object defined by the current public \passthrough{\lstinline!oan-package!} types. Its fields include \passthrough{\lstinline!packageVersion!}, \passthrough{\lstinline!resourceDid!}, \passthrough{\lstinline!resourceType!}, the complete \passthrough{\lstinline!didDocument!}, \passthrough{\lstinline!didDocumentHash!}, \passthrough{\lstinline!metadataHash!}, \passthrough{\lstinline!packageHash!}, \passthrough{\lstinline!hashAlgorithm!}, \passthrough{\lstinline!metadata!}, \passthrough{\lstinline!rootProof!}, and \passthrough{\lstinline!createdAt!}. A resource package has no generic top-level \passthrough{\lstinline!manifest!}, \passthrough{\lstinline!contentDigest!}, or \passthrough{\lstinline!resourceMetadata!} fields. If a specific resource requires a file manifest, it should be placed in an agreed extension field or external resource material, with the profile made explicit. Package digests currently cover the DID Document, resource metadata, and package declaration separately. The ResourcePackage version cannot replace the DID Document or resource version. Validation must be rejected when the resource DID, version, digest, metadata, or Root proof from the Root platform is missing; when digests do not match; or when resource types do not match.

Publication validation is performed in the following order:

\begin{enumerate}
\def\labelenumi{\arabic{enumi}.}
\tightlist
\item
  Parse the package version, resource type, and required structure.
\item
  Validate consistency among the resource DID, DID Document, metadata, and package declaration.
\item
  Recalculate content digests according to the normalization rules.
\item
  Verify the Root proof before entering the distribution or discovery process.
\end{enumerate}

\section{Root proof Schema}\label{sec:appendix-a.4}

The Root platform trusted publication proof contains at least \passthrough{\lstinline!rootDid!} in the current public \passthrough{\lstinline!RootProof!} type and may contain \passthrough{\lstinline!bulletinEventHash!}, \passthrough{\lstinline!signature!}, \passthrough{\lstinline!packageClaims!}, \passthrough{\lstinline!proof!}, \passthrough{\lstinline!cryptoSuite!}, and \passthrough{\lstinline!hashAlgorithm!}. The resource DID, resource type, version, DID Document digest, metadata digest, package digest, lifecycle state, and authorized domains are primarily located in \passthrough{\lstinline!packageClaims!}, rather than being required as flat fields at the top level of the proof object. The issuer must be a Root platform identity currently authorized to perform Root platform publication. Verifiers should check the Root platform signature, claim scope, hash bindings, version relationships, bulletin event reference, and governance status, rather than checking only that JSON fields exist. When the proof is invalid, digests do not match, the issuer is unauthorized, or the target version has been revoked, the Discovery node must not treat the corresponding package as a currently valid resource.

\section{Registration VC Schema}\label{sec:appendix-a.5}

The public response for a registration VC uses the W3C-style \passthrough{\lstinline!type!}, \passthrough{\lstinline!issuer!}, \passthrough{\lstinline!issuanceDate!}, \passthrough{\lstinline!credentialSubject!}, \passthrough{\lstinline!credentialStatus!}, and \passthrough{\lstinline!proof!} to express the registration fact issued by the Registrar node. The public Rust credential model also provides internal structures in the forms of \passthrough{\lstinline!subject!}, \passthrough{\lstinline!status!}, \passthrough{\lstinline!issuedAt!}, \passthrough{\lstinline!expiresAt!}, and \passthrough{\lstinline!claims!}; the two forms must not be mixed within the same profile. The registration VC currently returned by the official Registrar node uses claims such as \passthrough{\lstinline!credentialSubject.resourceDid!}, \passthrough{\lstinline!resourceType!}, \passthrough{\lstinline!didDocumentHash!}, \passthrough{\lstinline!metadataHash!}, \passthrough{\lstinline!packageHash!}, \passthrough{\lstinline!packageVersion!}, \passthrough{\lstinline!hashAlgorithm!}, \passthrough{\lstinline!authorizedDomains!}, and \passthrough{\lstinline!lifecycleState!}. It proves that the Registrar node accepted the registration request and issued a registration credential; it does not automatically prove that the Root platform has published the resource or that the resource endpoint is available. Verification fails when the issuer, target DID, claim bindings, or proof is missing. When a VC is empty, expired, revoked, or has an unknown status, callers should distinguish among incomplete, invalid, and temporarily indeterminate states.

\begin{quote}
A registration VC proves that the Registrar node accepted the registration and issued a registration credential. A resource can be used as a network-wide discoverable resource only when both the Root platform publication proof and the Discovery node index status meet the requirements.
\end{quote}

\section{Node Authorization VC Schema}\label{sec:appendix-a.6}

In the public credential model, a node authorization VC expresses node eligibility through \passthrough{\lstinline!credential\_type!}, \passthrough{\lstinline!issuer!}, \passthrough{\lstinline!subject!}, \passthrough{\lstinline!role!}, \passthrough{\lstinline!status!}, \passthrough{\lstinline!issuedAt!}, optional \passthrough{\lstinline!expiresAt!}, \passthrough{\lstinline!claims!}, and \passthrough{\lstinline!proof!}. After JSON serialization, the field names are respectively \passthrough{\lstinline!type!}, \passthrough{\lstinline!issuer!}, \passthrough{\lstinline!subject!}, \passthrough{\lstinline!role!}, \passthrough{\lstinline!status!}, \passthrough{\lstinline!issuedAt!}, \passthrough{\lstinline!expiresAt!}, \passthrough{\lstinline!claims!}, and \passthrough{\lstinline!proof!}. Authorized domains and capability scope are normally located in \passthrough{\lstinline!claims!}, rather than being fixed flat fields. Role values should be consistent with the node-role enumeration and deployment profile. Authorized domains and capability scope should use comparable structured values. Verifiers must check the credential signature, governance status, and current validity period together; the existence of a historical VC must not cause suspension or revocation events to be ignored. The \passthrough{\lstinline!credentialSubject!}, \passthrough{\lstinline!validFrom!}, and other fields below are provided only as display-layer compatibility examples; when using the public credential model, the fields above should be used.

\section{Governance Event Schema}\label{sec:appendix-a.7}

The public bulletin model for governance events consists of \passthrough{\lstinline!Bulletin!}, \passthrough{\lstinline!BulletinEvent!}, and \passthrough{\lstinline!BulletinEventCore!}. The event core fields are \passthrough{\lstinline!sequence!}, \passthrough{\lstinline!previousHash!}, \passthrough{\lstinline!eventType!}, \passthrough{\lstinline!subjectDid!}, \passthrough{\lstinline!actorDid!}, \passthrough{\lstinline!payload!}, and \passthrough{\lstinline!createdAt!}. The event body additionally contains \passthrough{\lstinline!eventHash!} and \passthrough{\lstinline!signature!}, as well as the optional \passthrough{\lstinline!proof!}, \passthrough{\lstinline!cryptoSuite!}, and \passthrough{\lstinline!hashAlgorithm!}. Current event types include Root platform initialization, content distribution platform information updates, authorization/revocation of Registrar nodes, authorization/domain updates/revocation of Discovery nodes, node revocation, and capability-tag tree updates. An off-chain projection may additionally record receipt time, projection time, a cursor, and errors, but these are not public core fields of \passthrough{\lstinline!BulletinEvent!}. The original on-chain or bulletin-chain event is the source of truth; page display text and projection status are derived information. When an event lacks a sequence, previous hash, signature, or target subject, or when projection order is discontinuous, authorization status must not be silently updated. The anomaly should be retained until the missing events are filled.

\section{Registration Request and Response Schema}\label{sec:appendix-a.8}

The current official Registrar node browser/client entry point is \passthrough{\lstinline!POST /resources/register!}. The request is deserialized as \passthrough{\lstinline!ResourceRegistrationSubmission!} and contains \passthrough{\lstinline!resourceDid!}, \passthrough{\lstinline!resourceType!}, \passthrough{\lstinline!didDocument!}, \passthrough{\lstinline!didDocumentHash!}, \passthrough{\lstinline!metadata!}, \passthrough{\lstinline!packageVersion!}, \passthrough{\lstinline!packageHash!}, \passthrough{\lstinline!metadataHash!}, \passthrough{\lstinline!hashAlgorithm!}, nullable \passthrough{\lstinline!registrationCredential!}, and \passthrough{\lstinline!subjectControlProof!}. Only when forwarding to the Root platform internally does the node wrap the request with \passthrough{\lstinline!registrarDid!} and \passthrough{\lstinline!upstreamAuth!}, using \passthrough{\lstinline!POST /root/resources/verify-and-publish!}. Therefore, the internal forwarding package must not be treated as a public registration request. The response usually contains \passthrough{\lstinline!status!}, \passthrough{\lstinline!resourceDid!}, \passthrough{\lstinline!resourceType!}, \passthrough{\lstinline!registrationCredential!}, and \passthrough{\lstinline!rootResponse!}; different deployments may add status or correlation identifiers. Whether retries with the same DID, version, and digests are fully idempotent depends on the node storage and Root platform implementation. A valid registration must not be written when the control proof is missing, fields are incomplete, or the previous version does not match.

\section{Discovery Request and Response Schema}\label{sec:appendix-a.9}

The fields of the public \passthrough{\lstinline!ResourceDiscoveryQuery!} are optional \passthrough{\lstinline!query!}, \passthrough{\lstinline!resourceType!}, \passthrough{\lstinline!capabilityTags!}, \passthrough{\lstinline!protocol!}, and \passthrough{\lstinline!version!}, together with \passthrough{\lstinline!versionMode!} and \passthrough{\lstinline!limit!}, which have default values. Exact queries by complete DID, structured-condition queries, tag queries, and natural-language queries are ways for the node to use these fields; they are not separate core \passthrough{\lstinline!queryType!} fields. Pagination and freshness fields may appear in a node read model or an extended profile. The public \passthrough{\lstinline!ResourceDiscoveryResponse!} contains the Discovery node DID, a candidate array, a creation time, and an optional response proof. A candidate contains the resource DID, resource type, score, version, lifecycle, capability tags, authorized domains, services, protocol bindings, resource package information, and a reference to the Root proof. Index time, source descriptions, and query explanations are node- or official-website-derived display information. Semantic scores cannot replace trusted evidence. Unknown conditions, invalid DIDs, expired cursors, or unavailable query services should return identifiable errors rather than hiding failure behind an empty result.

\section{Error Response Schema}\label{sec:appendix-a.10}

The current public error structure of nodes is based on \passthrough{\lstinline!\{ "error": "machine\_readable\_error\_code" \}!}; a trusted-call rejection may also include \passthrough{\lstinline!reason!}. Future or specific gateways may extend this to an object containing a message, a correlation request identifier, and retry guidance, but these are not fields currently guaranteed by all nodes. Error codes should distinguish input errors, signature/credential errors, unauthorized access, not found, duplicates, version conflicts, unavailable dependencies, synchronization lag, and internal errors. Public-facing messages must not disclose private keys, databases, or internal configuration. Clients receiving an unknown error code should handle it as a failure rather than defaulting to success.

If a gateway uses an extended error object, the client should first determine whether \passthrough{\lstinline!error!} is a string or an object, and then read optional fields. It must not assume an object shape or treat the HTTP status code alone as the business error code. The Registrar node currently maps some upstream errors to its own HTTP error responses; callers should retain both the HTTP status and the response body.

\section{Node Authorization and Governance Status Schema}\label{sec:appendix-a.11}

Node authorization status and governance status are usually formed as read-side projections by an off-chain trust indexer or a node status interface. They are not equivalent to the \passthrough{\lstinline!NodeAuthorizationCredential!} or the bulletin event body. A projection may contain the node DID, node role, authorized domains, current status, status reason, validity period, most recent governance event, event sequence, synchronization cursor, data time, and freshness. The specific fields are governed by the corresponding interface. Status values should at minimum distinguish active, suspended, revoked, expired, unknown, and syncing. When the governance source is missing or the indexer is behind, an active status must not be fabricated. For sensitive write actions, use a reject or manual-confirmation policy when the status is unknown or exceeds the freshness permitted by the deployment. A display interface may separately mark the delay.

\section{Distribution Manifest and Synchronization Cursor Schema}\label{sec:appendix-a.12}

The current public structures on the distribution side fall into two categories. A content distribution platform index response uses \passthrough{\lstinline!items!}, \passthrough{\lstinline!count!}, \passthrough{\lstinline!afterCursor!}, \passthrough{\lstinline!nextCursor!}, and \passthrough{\lstinline!hasMore!}, with each item containing \passthrough{\lstinline!cursor!} and a complete ResourcePackage. A batch notification sent by the Root platform to a Discovery node uses a batch identifier, Root platform DID, target Discovery node DID, authorized domains, a sequence range, items, a content distribution platform manifest address, an update address, and proof. An item additionally carries the resource DID, operation, resource package version, DID Document digest, metadata digest, package digest, hash algorithm, lifecycle state, bulletin sequence, and bulletin event hash. Manifest version, generation time, error, or confirmation fields may be added by a specific interface; fields from one interface must not be applied directly to another. Cursors must advance monotonically, repeated batches must be safely replayable, and processing of an item must be paused when there is a sequence gap or digest mismatch. A cursor is transmission progress; it does not mean that the resource has already been indexed by the Discovery node.

The minimum-shape example for a Root platform notification item is:

\section{Search Explanation and Evidence Schema}\label{sec:appendix-a.13}

Search explanations should state the query type, parsed conditions, matched capability tags or use cases, structured filters, relevance score, result source, version, resource status, index time, Root proof, and freshness. An explanation is derived information that helps callers understand the result; it cannot serve as signature or authorization evidence. When fields are missing, the semantic-model version is incompatible, or the source cannot be verified, the result should be explicitly marked as unknown or incomplete rather than supplemented with speculative reasons.

\section{References}\label{sec:appendix-a.14}

{\def\LTcaptype{none} 
\begin{longtable}[]{@{}>{\raggedright\arraybackslash}p{0.3200\linewidth}>{\raggedright\arraybackslash}p{0.3200\linewidth}>{\raggedright\arraybackslash}p{0.3200\linewidth}@{}}
\toprule\noalign{}
Source & Type & Link \\
\midrule\noalign{}
\endhead
\bottomrule\noalign{}
\endlastfoot
\passthrough{\lstinline!oan-protocol-commons!} & Code repository: normative data structures and schemas & \url{https://github.com/wolfbrother/oan-protocol-commons} \\
\passthrough{\lstinline!oan-sdk-ts!} & Code repository: TypeScript data types and client validation & \url{https://github.com/OpenAgenet/oan-sdk-ts} \\
\passthrough{\lstinline!did:oan!} DID Method Specification & Standard/method specification: DID Documents and identifier data structures & \url{https://github.com/OpenAgenet/oan-public-docs/blob/main/did-oan-specs/doc/OAN\%20DID\%20Method\%20Specification.md} \\
OAN Resource Identity and Discovery & IETF draft: resource identity data structures & \url{https://datatracker.ietf.org/doc/draft-xu-oan-resource-identity-discovery/} \\
\end{longtable}
}

\chapter{Protocol and API Reference}\label{chap:appendix-b}

This appendix provides rules for reading and implementing the interfaces. Paths, fields, and statuses are cross-checked against each node\textquotesingle s service entry points, \passthrough{\lstinline!oan-protocol-commons/schemas/agent-contract/openapi.yaml!}, the frontend and backend API wrappers on the official website, and actual deployment configuration; capabilities declared only in the common contract must not be written as if they were exposed by all nodes. Interface changes must be associated with the protocol version, error pattern, authentication method, idempotency rules, and the change record in Chapter~\ref{chap:24}.

\section{Endpoint Catalog}\label{sec:appendix-b.1}

The endpoint catalog lists health, DID, resource registration, Root platform publication, content distribution, discovery, governance status, and statistics interfaces by service role, and distinguishes public entry points, node-to-node entry points, internal-only entry points, and deployment prefixes. The official website currently forwards some registration and discovery requests through a same-origin backend; the nodes themselves also expose independent service endpoints, so website paths cannot automatically be inferred as protocol paths for all nodes. Example base URLs use placeholders and do not hard-code the website\textquotesingle s current domain or a particular server address as a protocol requirement; each row records at least the call direction, HTTP method, authentication, request/response pattern, failure behavior, and version. The table below lists only representative entry points that can be verified in the current code; management, statistics, and debugging interfaces not listed here remain subject to the routes of their corresponding nodes.

{\def\LTcaptype{none} 
\begin{longtable}[]{@{}>{\raggedright\arraybackslash}p{0.2400\linewidth}>{\raggedright\arraybackslash}p{0.2400\linewidth}>{\raggedright\arraybackslash}p{0.2400\linewidth}>{\raggedright\arraybackslash}p{0.2400\linewidth}@{}}
\toprule\noalign{}
Service role & Call direction, method, and path & Authentication boundary & Main returns \\
\midrule\noalign{}
\endhead
\bottomrule\noalign{}
\endlastfoot
Registrar node & Client -\textgreater{} Registrar node: \passthrough{\lstinline!POST /resources/register!}; Client -\textgreater{} Registrar node: \passthrough{\lstinline!GET /registrar/status!}, \passthrough{\lstinline!GET /registrar/did!} & The registration endpoint receives a resource control proof; node-to-node forwarding to the Root platform uses node-to-node signatures & Registration VC, acceptance status, Root platform response \\
Root platform & Registrar node -\textgreater{} Root platform: \passthrough{\lstinline!POST /root/resources/verify-and-publish!}; Client or node -\textgreater{} Root platform: \passthrough{\lstinline!GET /root/resources/\{did\}!}, \passthrough{\lstinline!GET /root/status!} & Publication writes are limited to authorized nodes or administrative calls; public reads depend on deployment & Root proof, resource version, status, and cursor \\
content distribution platform & Root platform or publisher -\textgreater{} content distribution platform: \passthrough{\lstinline!POST /cdn/resources!}, \passthrough{\lstinline!POST /cdn/resources/batch!}; Client or node -\textgreater{} content distribution platform: \passthrough{\lstinline!GET /cdn/resources/\{did\}!}, \passthrough{\lstinline!GET /cdn/resources/index!} & Publication uses node-to-node authentication; resource reads and index reads depend on deployment & ResourcePackage, index entries, synchronization status \\
Discovery node & Client -\textgreater{} Discovery node: \passthrough{\lstinline!POST /discovery/resources/query!}; Node or operations personnel -\textgreater{} Discovery node: \passthrough{\lstinline!GET /discovery/index/resources/\{did\}!}, \passthrough{\lstinline!GET /discovery/status!} & Queries may be public; management and synchronization interfaces are restricted & Resource results, source, and proof references \\
Operations API & Operations personnel -\textgreater{} node: \passthrough{\lstinline!GET /health!} and node status, metrics, and audit interfaces & Operations authentication and network boundary & Status summary, without secrets \\
\end{longtable}
}

The correspondence between core payload types and endpoints is as follows. When shared types come from \passthrough{\lstinline!oan-protocol!} or \passthrough{\lstinline!oan-package!}, field names follow the JSON mapping of \passthrough{\lstinline!serde!}; \passthrough{\lstinline!Value!} represents a dynamic response in the current implementation and cannot be used to infer a fixed schema that does not exist.

{\def\LTcaptype{none} 
\begin{longtable}[]{@{}>{\raggedright\arraybackslash}p{0.2400\linewidth}>{\raggedright\arraybackslash}p{0.2400\linewidth}>{\raggedright\arraybackslash}p{0.2400\linewidth}>{\raggedright\arraybackslash}p{0.2400\linewidth}@{}}
\toprule\noalign{}
Endpoint & Request JSON type & Response type or main fields & Notes \\
\midrule\noalign{}
\endhead
\bottomrule\noalign{}
\endlastfoot
\passthrough{\lstinline!POST /resources/register!} & \passthrough{\lstinline!ResourceRegistrationSubmission!} & Dynamic JSON: registration status, resource DID, registration VC, Root platform response & \passthrough{\lstinline!/resources/submit!} is bound to the same handler \\
\passthrough{\lstinline!POST /root/resources/verify-and-publish!} & \passthrough{\lstinline!ResourceVerifyAndPublishRequest!} & Dynamic JSON: \passthrough{\lstinline!resource-verified-and-queued!}, version, three categories of hashes, and CDN queue status & Node-to-node call limited to authorized Registrar nodes \\
\passthrough{\lstinline!POST /cdn/resources!} & \passthrough{\lstinline!ResourceCdnPublishRequest!} & Dynamic JSON: \passthrough{\lstinline!published!}, version, and publication cursor & ResourcePackage must pass Root platform source and hash validation \\
\passthrough{\lstinline!POST /discovery/resources/query!} & \passthrough{\lstinline!ResourceDiscoveryQuery!} & \passthrough{\lstinline!ResourceDiscoveryResponse!} & \passthrough{\lstinline!limit!} defaults to 10, \passthrough{\lstinline!versionMode!} defaults to \passthrough{\lstinline!latest!} \\
\passthrough{\lstinline!POST /discovery/query/explain!} & \passthrough{\lstinline!ResourceDiscoveryQuery!} & Dynamic explanation JSON & Not equivalent to a normal discovery response \\
\passthrough{\lstinline!GET /cdn/resources/\{did\}!} & Path parameter \passthrough{\lstinline!did!} & \passthrough{\lstinline!ResourcePackage!} & Reads a ResourcePackage directly \\
\passthrough{\lstinline!POST /cdn/resources/batch-get!} & Local \passthrough{\lstinline!ResourceBatchGetRequest!} & \passthrough{\lstinline!requestedCount!}, \passthrough{\lstinline!foundCount!}, \passthrough{\lstinline!items!} & The request body contains a \passthrough{\lstinline!resourceDids!} array \\
\end{longtable}
}

\section{Request Header Fields and Signatures}\label{sec:appendix-b.2}

Request headers and signature materials are defined by the interface profile. The current node-to-node \passthrough{\lstinline!SignedRequestEnvelope!} uses JSON fields such as \passthrough{\lstinline!requestId!}, \passthrough{\lstinline!protocolVersion!}, \passthrough{\lstinline!purpose!}, \passthrough{\lstinline!method!}, \passthrough{\lstinline!path!}, \passthrough{\lstinline!aud!}, \passthrough{\lstinline!requestTimestamp!}, \passthrough{\lstinline!requestNonce!}, \passthrough{\lstinline!bodyHash!}, and \passthrough{\lstinline!proof!} to carry signature semantics; the browser-facing public registration and discovery entry points on the official website must not arbitrarily add a set of required \passthrough{\lstinline!X-OAN-*!} headers that are not defined at the service entry point. If the interface uses a protocol envelope to carry this information, the JSON fields take precedence; if the deployment layer additionally uses HTTP headers, they may only be documented as deployment-profile details. The server must validate the normalized request covered by the signature, the target DID, the time window, and the nonce; it must reject missing, expired, replayed, or misdirected requests.

The roles of node-to-node protocol envelope fields can be checked against the table below. \passthrough{\lstinline!method!} and \passthrough{\lstinline!path!} bind the protected HTTP request, \passthrough{\lstinline!aud!} binds the recipient, and \passthrough{\lstinline!bodyHash!} binds the request body; together, these fields prevent a valid proof issued by one node from being moved to another request. When a browser submits a resource, the resource control proof is located in the request body\textquotesingle s \passthrough{\lstinline!subjectControlProof!}; it must not be mistakenly written as a node-to-node HTTP header.

{\def\LTcaptype{none} 
\begin{longtable}[]{@{}>{\raggedright\arraybackslash}p{0.3200\linewidth}>{\raggedright\arraybackslash}p{0.3200\linewidth}>{\raggedright\arraybackslash}p{0.3200\linewidth}@{}}
\toprule\noalign{}
Field & Purpose & Validation focus \\
\midrule\noalign{}
\endhead
\bottomrule\noalign{}
\endlastfoot
\passthrough{\lstinline!requestId!} & Identifies one node-to-node request & Used for log correlation; cannot replace replay protection \\
\passthrough{\lstinline!protocolVersion!}, \passthrough{\lstinline!purpose!} & Identifies the protocol version and call purpose & Must match the profile allowed by the receiving interface \\
\passthrough{\lstinline!method!}, \passthrough{\lstinline!path!}, \passthrough{\lstinline!aud!} & Binds the HTTP method, path, and recipient & Must not differ from the actual request or target node \\
\passthrough{\lstinline!requestTimestamp!}, \passthrough{\lstinline!requestNonce!} & Restricts the time range and prevents reuse & Check the clock window and nonce replay cache \\
\passthrough{\lstinline!bodyHash!}, \passthrough{\lstinline!proof!} & Binds the request body and provides an integrity proof & Recalculate the request-body digest and validate the proof \\
\end{longtable}
}

\begin{lstlisting}
POST /resources/register
Host: registrar.example
Content-Type: application/json
\end{lstlisting}

The \passthrough{\lstinline!upstreamAuth.proof!} of a node-to-node request should cover the method, normalized path, recipient, timestamp/nonce, content digest, and actual request body; the server should perform the write only after completing the time-window, replay-cache, signer-authorization, and resource-control checks. In a browser registration request, the resource control proof is located in the request body\textquotesingle s \passthrough{\lstinline!subjectControlProof!}; it is not the same type of credential as node-to-node \passthrough{\lstinline!upstreamAuth!}.

\section{Response Header Fields and Cache Directives}\label{sec:appendix-b.3}

A response may provide the source, protocol/mode version, data time, freshness, ETag, cache control, correlation identifier, and necessary security headers when applicable; these fields are not guaranteed to be uniform public response headers across all current node interfaces, and the \passthrough{\lstinline!X-*!} items below may only serve as deployment-layer or interface-profile extension conventions. An ETag or cache hit only indicates that the content version has not changed; it does not prove that the resource status or governance authorization remains valid. When returning an old snapshot, the response or body should include identifiable time and degradation indicators; caching must not be used to conceal an upstream failure.

{\def\LTcaptype{none} 
\begin{longtable}[]{@{}>{\raggedright\arraybackslash}p{0.3200\linewidth}>{\raggedright\arraybackslash}p{0.3200\linewidth}>{\raggedright\arraybackslash}p{0.3200\linewidth}@{}}
\toprule\noalign{}
Field or directive (when applicable) & Purpose & Usage notes \\
\midrule\noalign{}
\endhead
\bottomrule\noalign{}
\endlastfoot
\passthrough{\lstinline!ETag!} & Identifies the response content version & Does not mean that the governance status is valid \\
\passthrough{\lstinline!Cache-Control!} & Specifies caching and revalidation policy & An old snapshot must also identify its freshness \\
\passthrough{\lstinline!X-Request-Id!} & Correlates the client, nodes, and logs & Must not contain private information \\
\passthrough{\lstinline!X-OAN-Data-Time!} & Indicates the data generation or collection time & Distinguish it from the server response time \\
\passthrough{\lstinline!X-OAN-Freshness!} & Indicates statuses such as \passthrough{\lstinline!current!} and \passthrough{\lstinline!stale!} & Must not conceal an upstream failure \\
\end{longtable}
}

\section{Status Codes and Error Codes}\label{sec:appendix-b.4}

HTTP status codes express transport-layer and generic processing results, while OpenAgenet (OAN) error codes express specific semantics such as field errors, invalid signatures, unauthorized access, not found, duplication, version conflicts, dependency failures, synchronization lag, and internal errors. The current public error-code vocabulary is defined by \passthrough{\lstinline!oan-protocol-commons/docs/error-codes.md!}, including \passthrough{\lstinline!invalid\_did\_document\_structure!}, \passthrough{\lstinline!invalid\_registration\_credential!}, \passthrough{\lstinline!invalid\_request\_signature!}, \passthrough{\lstinline!unauthorized\_domains!}, \passthrough{\lstinline!invalid\_root\_proof!}, and \passthrough{\lstinline!package\_decode\_failed!}; that file defines error semantics but does not bind every error code to a unique HTTP status. Different nodes may map these errors through an HTTP status or response body. Clients should determine whether to retry from the actual error structure; they must not treat every 2xx response as publication success, nor conflate 404, empty results, and synchronization incomplete.

{\def\LTcaptype{none} 
\begin{longtable}[]{@{}>{\raggedright\arraybackslash}p{0.3200\linewidth}>{\raggedright\arraybackslash}p{0.3200\linewidth}>{\raggedright\arraybackslash}p{0.3200\linewidth}@{}}
\toprule\noalign{}
HTTP status & OAN error code examples & Client action \\
\midrule\noalign{}
\endhead
\bottomrule\noalign{}
\endlastfoot
\passthrough{\lstinline!400!} & \passthrough{\lstinline!invalid\_did\_document\_structure!}, \passthrough{\lstinline!invalid\_authorized\_domains!}, \passthrough{\lstinline!package\_decode\_failed!} & Correct the input or ResourcePackage and retry; the specific status depends on the interface implementation \\
\passthrough{\lstinline!401!}/\passthrough{\lstinline!403!} & \passthrough{\lstinline!invalid\_request\_signature!}, \passthrough{\lstinline!unauthorized\_domains!} & Re-authenticate, use an authorized node, or stop the request \\
\passthrough{\lstinline!404!} & Resource-not-found error returned by the node & Distinguish not found from an unsynchronized index \\
\passthrough{\lstinline!409!} & Duplicate or version-conflict error returned by the node & Obtain the current version and process it according to the update rules \\
\passthrough{\lstinline!429!} & Deployment-layer rate-limit error & Follow the retry indication and back off \\
\passthrough{\lstinline!5xx!} & Dependency unavailable or internal error & Back off and retry only when the interface marks the request as retryable \\
\end{longtable}
}

\section{Pagination and Cursor Rules}\label{sec:appendix-b.5}

List interfaces use explicit \passthrough{\lstinline!limit!}, cursor, and sorting rules. Cursors may only be generated and interpreted by the server; clients must not assume that they are readable or modifiable. The current public \passthrough{\lstinline!ResourceDiscoveryQuery!} type always provides \passthrough{\lstinline!limit!}, with a default value of 10, but the request model of \passthrough{\lstinline!POST /discovery/resources/query!} has no cursor field; the Discovery node truncates candidate results according to \passthrough{\lstinline!limit!}. The interface that actually provides \passthrough{\lstinline!afterCursor!}, \passthrough{\lstinline!nextCursor!}, and \passthrough{\lstinline!hasMore!} is the content distribution platform\textquotesingle s \passthrough{\lstinline!GET /cdn/resources/index!}; its cursor is not the same type as the discovery query\textquotesingle s \passthrough{\lstinline!limit!}. Page size should have a deployment limit, and the sort key must remain stable during pagination; expired, invalid, or filter-mismatched cursors should return an error. Publication cursors, synchronization cursors, event cursors, and pagination cursors must not be used interchangeably.

\begin{lstlisting}
GET /cdn/resources/index?limit=20&afterCursor=1800
Accept: application/json

{
  "items": [],
  "count": 0,
  "afterCursor": 1800,
  "nextCursor": 1800,
  "hasMore": false
}
\end{lstlisting}

A pagination cursor only represents the reading position in the content distribution platform\textquotesingle s list; the \passthrough{\lstinline!limit!} in a discovery query only controls the number of candidate results for that request and does not produce a resumable discovery pagination cursor. Publication sequences, distribution synchronization cursors, and governance event cursors must use the fields and validation rules defined by their respective interfaces.

\section{Version and Compatibility Header Fields}\label{sec:appendix-b.6}

Protocol versions, resource schema versions, capability declarations, and compatibility policies are expressed through defined protocol envelope fields, request-body fields, or deployment profiles, and remain traceable in requests and responses. Current public protocol constants include \passthrough{\lstinline!oan-resource-2026!}, while the default value of \passthrough{\lstinline!ResourceDiscoveryQuery.versionMode!} is \passthrough{\lstinline!latest!}; such protocol fields must not be directly rewritten as uniform HTTP headers without code support. When the server encounters an unsupported version, it should return an explicit incompatibility result; compatible extensions must not change the semantics of DID, signatures, status, or publication sequences. See Chapter~\ref{chap:24} for version selection rules.

\begin{lstlisting}
{
  "query": "a tool that can search code repositories and summarize project structure",
  "capabilityTags": ["code-search", "project-structure-summary"],
  "versionMode": "latest",
  "limit": 10
}
\end{lstlisting}

The example above is a discovery query request for \passthrough{\lstinline!POST /discovery/resources/query!}; all fields correspond to the current \passthrough{\lstinline!ResourceDiscoveryQuery!}. \passthrough{\lstinline!protocolVersion!} belongs to the node-to-node \passthrough{\lstinline!SignedRequestEnvelope!} and must not be mixed into this query object. The current query model has no unified \passthrough{\lstinline!capabilities!} field; capability conditions should be expressed through \passthrough{\lstinline!capabilityTags!} or an extension explicitly supported by the target interface. When the server does not support the requested version, it should return an explicit version-incompatibility error and the range of acceptable versions; clients must not ignore the response-mode version merely because they received an HTTP 2xx response.

\section{Health, Readiness, and Liveness Response Formats}\label{sec:appendix-b.7}

The common minimum structure of the public \passthrough{\lstinline!/health!} response across current nodes is \passthrough{\lstinline!HealthResponse!}: \passthrough{\lstinline!status!}, \passthrough{\lstinline!nodeType!}, and an optional node DID. A node status interface may additionally provide database, upstream, synchronization cursor, or error summaries. Operations systems can use this information to distinguish process liveness, dependency readiness, business availability, and data synchronization status, but must not treat every field as a uniform schema present on every node. Passing liveness does not imply readiness, and passing readiness does not imply that registration or discovery smoke tests have passed; sensitive configuration and tokens must not appear in the response.

\begin{lstlisting}
{
  "status": "ok",
  "nodeType": "discovery",
  "did": "did:oan:INDS:7YpQm9Kx2VnRb6Ts3WfHa4Cd5Ej8LgNz"
}
\end{lstlisting}

If the deployment provides an extended status interface, it may add check time, dependency status, and synchronization cursors beyond the minimum response above; extension fields should identify their source and version, and the extended response must not be treated in reverse as the mandatory response structure of \passthrough{\lstinline!/health!} on every node.

{\def\LTcaptype{none} 
\begin{longtable}[]{@{}>{\raggedright\arraybackslash}p{0.3200\linewidth}>{\raggedright\arraybackslash}p{0.3200\linewidth}>{\raggedright\arraybackslash}p{0.3200\linewidth}@{}}
\toprule\noalign{}
Check type & What it proves & What it does not represent \\
\midrule\noalign{}
\endhead
\bottomrule\noalign{}
\endlastfoot
liveness & The process can still respond & Dependency and business availability \\
readiness & The service is ready to receive requests & Successful registration or discovery smoke tests \\
dependency & Database, upstream, or off-chain trust indexer status & All business paths are functioning normally \\
smoke test & A critical business flow can be completed & Long-term performance and full correctness \\
\end{longtable}
}

\section{Public, Authenticated, and Operations-Only API Matrix}\label{sec:appendix-b.8}

The interface matrix must at least distinguish public read-only interfaces, authenticated publication/registration interfaces, node-to-node interfaces, and operations-only interfaces. Public discovery, DID, status, and necessary statistics may be accessible from the Internet; write, management, database, full-log, detailed access-statistics, and key-related interfaces must be protected by authentication, network boundaries, and auditing. The relay interface of the official website\textquotesingle s backend must not expand downstream service permissions, and ordinary users must not obtain backend statistics or secrets through public interfaces.

{\def\LTcaptype{none} 
\begin{longtable}[]{@{}>{\raggedright\arraybackslash}p{0.2400\linewidth}>{\raggedright\arraybackslash}p{0.2400\linewidth}>{\raggedright\arraybackslash}p{0.2400\linewidth}>{\raggedright\arraybackslash}p{0.2400\linewidth}@{}}
\toprule\noalign{}
Access level & Examples & Permitted data & Protection requirements \\
\midrule\noalign{}
\endhead
\bottomrule\noalign{}
\endlastfoot
Public read-only & DID, discovery, necessary status & Published ResourcePackages and public proofs & Rate limiting, input validation, auditing \\
Authenticated write & Registration, updates, node-to-node synchronization & Objects within the authorized scope & Signatures, nonce, idempotency, and authorization checks \\
Operations only & Metrics, full logs, administrative operations & Internal operational data & Strong authentication, network isolation, and least privilege \\
Prohibited from public access & Database, private keys, detailed access statistics & Not open to Internet users & Must not be leaked through the official website\textquotesingle s relay \\
\end{longtable}
}

\section{Reference Sources}\label{sec:appendix-b.9}

{\def\LTcaptype{none} 
\begin{longtable}[]{@{}>{\raggedright\arraybackslash}p{0.3200\linewidth}>{\raggedright\arraybackslash}p{0.3200\linewidth}>{\raggedright\arraybackslash}p{0.3200\linewidth}@{}}
\toprule\noalign{}
Source & Type & Link \\
\midrule\noalign{}
\endhead
\bottomrule\noalign{}
\endlastfoot
\passthrough{\lstinline!oan-protocol-commons!} & Code repository: protocol types, OpenAPI contract, and error codes & \url{https://github.com/wolfbrother/oan-protocol-commons} \\
\passthrough{\lstinline!oan-root-nodes!} & Code repository: Root platform API & \url{https://github.com/OpenAgenet/oan-root-nodes} \\
\passthrough{\lstinline!oan-registrar-nodes!} & Code repository: registration API & \url{https://github.com/OpenAgenet/oan-registrar-nodes} \\
\passthrough{\lstinline!oan-discovery-nodes!} & Code repository: discovery API & \url{https://github.com/OpenAgenet/oan-discovery-nodes} \\
\passthrough{\lstinline!oan-sdk-ts!} & Code repository: client API wrappers & \url{https://github.com/OpenAgenet/oan-sdk-ts} \\
OAN Resource Identity and Discovery & IETF Internet-Draft: resource identity and discovery API & \url{https://datatracker.ietf.org/doc/draft-xu-oan-resource-identity-discovery/} \\
Efficient Agent Discovery Profile & IETF Internet-Draft: discovery API profile & \url{https://datatracker.ietf.org/doc/draft-xu-efficient-agent-discovery-profile/} \\
\end{longtable}
}

\chapter{Cryptography and Encoding}\label{chap:appendix-c}

This appendix defines the cryptographic and encoding boundaries used by OpenAgenet (OAN) for identity, credentials, resource publication, and cross-node verification. The current shared Rust implementation is primarily based on \passthrough{\lstinline!oan-crypto!}, \passthrough{\lstinline!oan-credentials!}, and \passthrough{\lstinline!oan-did-oan!}; the browser-side TypeScript SDK uses the Web Cryptography API to generate Ed25519 Local Identity. Any cross-language implementation must recompute the same digest from the same canonicalized input and must handle algorithms, encodings, and error states explicitly.

\section{Supported Algorithms}\label{sec:appendix-c.1}

The current \passthrough{\lstinline!CryptoSuite!} includes \passthrough{\lstinline!Ed25519Sha256Legacy!}, \passthrough{\lstinline!Ed25519Sha256!}, and \passthrough{\lstinline!Sm2Sm3!}: the first two use Ed25519 signatures and SHA-256 hashing, while the latter uses SM2 signatures and SM3 hashing. The corresponding verification method types are \passthrough{\lstinline!Ed25519VerificationKey2020!} or \passthrough{\lstinline!SM2VerificationKey2020!}, and the proof types are \passthrough{\lstinline!Ed25519Signature2020!} or \passthrough{\lstinline!SM2Signature2020!}. The algorithm suite must be consistent with the public key, proof, and hash algorithm; unknown suites are rejected, and legacy suites may only be retained under a compatibility policy and must not be silently downgraded.

{\def\LTcaptype{none} 
\begin{longtable}[]{@{}>{\raggedright\arraybackslash}p{0.2400\linewidth}>{\raggedright\arraybackslash}p{0.2400\linewidth}>{\raggedright\arraybackslash}p{0.2400\linewidth}>{\raggedright\arraybackslash}p{0.2400\linewidth}@{}}
\toprule\noalign{}
Algorithm Suite & Signature Algorithm & Hash Algorithm & Typical Use \\
\midrule\noalign{}
\endhead
\bottomrule\noalign{}
\endlastfoot
\passthrough{\lstinline!Ed25519Sha256Legacy!} & Ed25519 & SHA-256 & Compatibility with existing credential or digest flows \\
\passthrough{\lstinline!Ed25519Sha256!} & Ed25519 & SHA-256 & Current Ed25519 proof flow \\
\passthrough{\lstinline!Sm2Sm3!} & SM2 & SM3 & Identity and proof flows supporting SM2 \\
\end{longtable}
}

The algorithm suite, verification method, public key encoding, and \passthrough{\lstinline!proof!} type must match as a set; unknown or deprecated suites must not silently switch to another algorithm suite.

In implementation terms, \passthrough{\lstinline!Ed25519Sha256Legacy!} and \passthrough{\lstinline!Ed25519Sha256!} use the same Ed25519 key and SHA-256 hash, but their signing inputs differ: the former first computes SHA-256 over the canonicalized JSON, then signs the UTF-8 bytes of the resulting 64 lowercase hexadecimal characters; the latter directly signs the UTF-8 bytes of the canonicalized JSON. \passthrough{\lstinline!Sm2Sm3!} directly uses SM2 to sign the UTF-8 bytes of the canonicalized JSON, and its digest function uses SM3. A cross-language implementation must not determine the input solely from the signature algorithm name; it must read both \passthrough{\lstinline!cryptoSuite!} and \passthrough{\lstinline!hashAlgorithm!}.

{\def\LTcaptype{none} 
\begin{longtable}[]{@{}>{\raggedright\arraybackslash}p{0.3200\linewidth}>{\raggedright\arraybackslash}p{0.3200\linewidth}>{\raggedright\arraybackslash}p{0.3200\linewidth}@{}}
\toprule\noalign{}
Selection & Decision in the Current Implementation & Failure Handling \\
\midrule\noalign{}
\endhead
\bottomrule\noalign{}
\endlastfoot
\passthrough{\lstinline!cryptoSuite!} & Determines the signature algorithm, hash algorithm, and signature input form & Reject when unrecognized \\
\passthrough{\lstinline!proof.type!} & The \passthrough{\lstinline!Ed25519Signature2020!} or \passthrough{\lstinline!SM2Signature2020!} corresponding to the suite & Reject when the type does not match \\
Verification method type & Derives the suite from \passthrough{\lstinline!Ed25519VerificationKey2020!} or \passthrough{\lstinline!SM2VerificationKey2020!} & Reject when it cannot be derived \\
Public key carrier & Use either \passthrough{\lstinline!publicKeyMultibase!} or \passthrough{\lstinline!publicKeyJwk!} & Reject when both are missing or cannot be parsed \\
\end{longtable}
}

\subsection{Current Status and Boundary of Chinese Commercial Cryptography Support}\label{subsec:appendix-c.1.1}

In this manual, ``Chinese commercial cryptography support'' primarily means SM2 signature and SM3 digest capability. SM2 and SM3 correspond to the relevant Chinese cryptographic algorithm standards. The ability of code to execute an algorithm does not mean that a particular deployment has completed commercial cryptography product selection, key management, testing and certification, or other compliance work. OAN\textquotesingle s current core registration and publication path primarily uses digital signatures, digests, and control proofs. The core modules do not yet define a protocol use or end-to-end implementation of SM4 symmetric encryption. The current state therefore must not be described as full support for an SM2, SM3, and SM4 suite.

The current implementation should be understood at the following levels.

{\def\LTcaptype{none} 
\begin{longtable}[]{@{}>{\raggedright\arraybackslash}p{0.2400\linewidth}>{\raggedright\arraybackslash}p{0.2400\linewidth}>{\raggedright\arraybackslash}p{0.2400\linewidth}>{\raggedright\arraybackslash}p{0.2400\linewidth}@{}}
\toprule\noalign{}
Level & Current status & Supported conclusion & Unsupported conclusion \\
\midrule\noalign{}
\endhead
\bottomrule\noalign{}
\endlastfoot
Public \passthrough{\lstinline!did:oan!} method baseline & The current baseline specifies Ed25519Signature2020; other cryptographic algorithms require a separate profile & The method design leaves room for deployment extensions & The baseline \passthrough{\lstinline!did:oan!} document directly specifies SM2/SM3 \\
Shared Rust cryptographic layer & Implements SM2 key generation and parsing, JWK/multibase public-key handling, SM2 signing and verification, and SM3 digests, with cryptographic-primitive tests & OAN code has executable foundational SM2/SM3 capability & Network ingress, VCs, SDKs, and production services all accept SM2/SM3 end to end \\
General proofs, request envelopes, and ResourcePackages & Some structures can select a cryptographic suite from a proof, verification method, or signing key; a ResourcePackage digest can be calculated according to the suite & Common protocol structures provide an implementation foundation for algorithm agility & Every object already has cross-language fixed vectors and production gates for Chinese commercial cryptography \\
DID Document registration ingress & The formal Registrar and Root network paths first perform current standard DID-proof verification; compatibility fallback remains constrained by the current Ed25519 DID Document rules & The public registration path has a definite Ed25519 baseline & Constructing a \passthrough{\lstinline!SM2Signature2020!} DID Document is sufficient for successful registration \\
Core-node runtime identity & Registrar, Root, and content-distribution-related nodes currently load and use node keys under Ed25519/SHA-256 & Existing inter-node signing has a common deployment baseline & A node can switch to SM2/SM3 merely by changing an identity file \\
VCs and Root proofs & Shared credential issuance, registration VCs, node authorization, and the main publication-proof paths remain primarily Ed25519/SHA-256 & Existing credentials can be mutually verified under the current baseline & SM2/SM3 VCs have completed an end-to-end issuance, status, propagation, and Discovery-side verification cycle \\
TypeScript SDK and web client & Identity generation, DID Data Integrity signing, and cross-language baselines currently focus on Ed25519 & A browser can generate and submit the materials required by the current production path & The browser can already generate, import, and sign complete SM2 identity materials \\
Production deployment & SM2 node identities and a complete Chinese-commercial-cryptography registration---publication---discovery gate are not the current production baseline & The verified Ed25519/SHA-256 path can continue to be used & The live network can be claimed to have completed end-to-end migration to Chinese commercial cryptography or compliance certification \\
\end{longtable}
}

The current status should therefore be stated as follows: \textbf{the shared Rust cryptographic layer contains substantive SM2/SM3 implementations, and some general protocol objects provide a foundation for algorithm extension; the public \passthrough{\lstinline!did:oan!} baseline, core-node identities, credential issuance, TypeScript SDK, and live primary path do not yet form an end-to-end SM2/SM3 profile.} References to \passthrough{\lstinline!sm2-sm3!}, \passthrough{\lstinline!SM2VerificationKey2020!}, or \passthrough{\lstinline!SM2Signature2020!} in older design or method materials are sources for extension design and implementation; they do not override the current public baseline or runtime facts.

To elevate this capability into externally claimable system-level support for Chinese commercial cryptography, at least the following work must be completed together:

\begin{enumerate}
\def\labelenumi{\arabic{enumi}.}
\tightlist
\item
  Publish an SM2/SM3 profile aligned with the current \passthrough{\lstinline!did:oan!} baseline, fixing verification methods, proofs, signature inputs, public-key encoding, the SM2 user identifier, and error handling.
\item
  Carry DID Documents, controller-authorization proofs, node request envelopes, registration VCs, node-authorization VCs, Root publication proofs, and ResourcePackage digests through one profile without silently downgrading to Ed25519 on failure.
\item
  Make the runtime cryptographic suites, key materials, and trust anchors of Registrar, Root, content distribution, and Discovery explicit configuration, and verify admission and rejection rules in mixed-suite environments.
\item
  Add SM2 key generation, import, signing, backup, and recovery to the TypeScript SDK or a controlled independent signing provider, and define the security boundary when browsers lack native support.
\item
  Establish fixed test vectors shared by Rust and client implementations, and pass positive, tampering, wrong-suite, and downgrade-attack tests across identity generation, registration, Root admission, content distribution, Discovery verification, and client reverification.
\item
  If the target also includes commercial cryptography compliance, separately implement requirements for cryptographic products, key custody, runtime environments, testing and certification, and applicable standards. Protocols and open-source algorithm implementations do not replace that work.
\end{enumerate}

\section{Canonicalized JSON and Hashing}\label{sec:appendix-c.2}

The shared implementation first converts the serializable object to JSON, then recursively orders object keys lexicographically; arrays retain their original order, strings use JSON escaping, numbers use their serialized representation, and the rule for omitting empty fields is determined by the Rust/TypeScript data model. \passthrough{\lstinline!Ed25519Sha256Legacy!} signs the SHA-256 hexadecimal result of the canonicalized JSON, while \passthrough{\lstinline!Ed25519Sha256!} and \passthrough{\lstinline!Sm2Sm3!} sign the UTF-8 bytes of the canonicalized JSON. Cross-language implementations must fix the rules for UTF-8, key ordering, numbers, and empty fields; any difference will cause digest or signature verification to fail.

\begin{lstlisting}
{
  "b": 2,
  "a": {
    "y": 1,
    "x": "OAN"
  },
  "items": ["one", "two"]
}
\end{lstlisting}

During canonicalization, keys are recursively ordered lexicographically, arrays retain their original order, and the input uses UTF-8. The corresponding canonicalized result is:

\begin{lstlisting}
{
  "a": {
    "x": "OAN",
    "y": 1
  },
  "b": 2,
  "items": ["one", "two"]
}
\end{lstlisting}

Cross-language tests should fix the rules for null values, number representations, string escaping, and omission of object fields at the same time, rather than relying solely on the default serialization behavior of a particular language runtime.

For example, the SHA-256 result of the canonicalized string above is \passthrough{\lstinline!03332fb018e2f0376e263dabf4b2231e3bd70767e19fb7e263060efb68476e02!}. This value corresponds only to the example object, example key order, and example UTF-8 encoding; it is not a protocol constant. On the Rust side, \passthrough{\lstinline!oan-crypto::canonical\_json!} and \passthrough{\lstinline!hash\_json\_with\_suite!} can be used to produce a baseline; on the TypeScript side, the object should first be recursively sorted according to the same rules, and \passthrough{\lstinline!crypto.subtle.digest("SHA-256", ...)!} should then be used to compute the digest over the UTF-8 bytes.

Implementation must distinguish three changes in particular: changing object member order should not change the result after canonicalization, changing array element order will change the result, and whether changing a number from \passthrough{\lstinline!1!} to \passthrough{\lstinline!1.0!} is considered equivalent depends on the JSON number representation before the input enters the canonicalization function. Production code must not first convert JSON to indented text and then sign it directly, nor may it rely on the order of fields returned by a database.

\section{JWS and Detached Signatures}\label{sec:appendix-c.3}

The current OAN shared implementation uses \passthrough{\lstinline!DataIntegrityProof!} to express detached proofs, containing \passthrough{\lstinline!type!}, \passthrough{\lstinline!creator!}, \passthrough{\lstinline!created!}, \passthrough{\lstinline!proofPurpose!}, \passthrough{\lstinline!proofValue!}, \passthrough{\lstinline!cryptoSuite!}, \passthrough{\lstinline!hashAlgorithm!}, and \passthrough{\lstinline!verificationMethod!}. \passthrough{\lstinline!proofValue!} is an unpadded Base64url-encoded signature value; the signature input is determined by the canonicalized result of the object to be signed and the algorithm suite. The verifier must first determine the suite, reconstruct the signature input, and verify the public key and signature, then check the creator, verification method, target DID, validity period, and status; it must not merely compare \passthrough{\lstinline!proofValue!} strings.

\begin{lstlisting}
{
  "type": "DataIntegrityProof",
  "creator": "did:oan:K7mQ9:7YpQm9Kx2VnRb6Ts3WfHa4Cd5Ej8LgNz#key-1",
  "created": "2026-09-07T01:02:03Z",
  "proofPurpose": "assertionMethod",
  "verificationMethod": "did:oan:K7mQ9:7YpQm9Kx2VnRb6Ts3WfHa4Cd5Ej8LgNz#key-1",
  "cryptoSuite": "Ed25519Sha256",
  "hashAlgorithm": "SHA-256",
  "proofValue": "base64url-without-padding"
}
\end{lstlisting}

Signature verification failures should distinguish inconsistent canonicalized input, a mismatched public key, an invalid signature, an incorrect proof purpose, and an unacceptable status, so that the caller can decide whether to correct the request, retrieve the data again, or stop processing.

The minimum signature verification sequence is as follows:

\begin{enumerate}
\def\labelenumi{\arabic{enumi}.}
\tightlist
\item
  Determine the suite from \passthrough{\lstinline!proof.cryptoSuite!} or a compatible \passthrough{\lstinline!proof.type!}, and confirm that it is consistent with the verification method and public key format.
\item
  Locate the public key in the DID Document by \passthrough{\lstinline!verificationMethod!}, and check the relationship among the method identifier, controller, and target DID.
\item
  Remove \passthrough{\lstinline!proof!} from the object being verified and reconstruct the canonicalized signature input according to the suite; legacy suites must also first compute the SHA-256 hexadecimal digest.
\item
  Decode \passthrough{\lstinline!proofValue!} as unpadded Base64url and perform signature verification, then check \passthrough{\lstinline!created!}, purpose, issuer authorization, and credential status.
\end{enumerate}

The fact that \passthrough{\lstinline!proofValue!} can be decoded only shows that its encoding is readable; it does not show that the signature is correct. A correct signature also does not automatically show that the issuer has authorization from the Root platform or that the resource remains available.

\section{DID and Key Encoding}\label{sec:appendix-c.4}

The current syntax is \passthrough{\lstinline!did:oan:<5-character Base58 routing-code>:<32-character Base58 suffix-code>!}. Both components are case-sensitive and exclude \passthrough{\lstinline!0!}, \passthrough{\lstinline!O!}, \passthrough{\lstinline!I!}, and \passthrough{\lstinline!l!}. The routing code selects a Registrar route and does not encode resource type. Depending on the algorithm, public keys may use a \passthrough{\lstinline!z!}-prefixed Base58 multibase or a JWK: Ed25519 uses \passthrough{\lstinline!kty=OKP!}, \passthrough{\lstinline!crv=Ed25519!}, and an unpadded Base64url \passthrough{\lstinline!x!}; SM2 uses \passthrough{\lstinline!kty=EC!}, \passthrough{\lstinline!crv=SM2!}, and \passthrough{\lstinline!x!} and \passthrough{\lstinline!y!}. Private keys must not enter a DID Document, ordinary protocol request, or service log.

\begin{lstlisting}
did:oan:<5-character Base58 routing-code>:<32-character Base58 suffix-code>
\end{lstlisting}

{\def\LTcaptype{none} 
\begin{longtable}[]{@{}>{\raggedright\arraybackslash}p{0.3200\linewidth}>{\raggedright\arraybackslash}p{0.3200\linewidth}>{\raggedright\arraybackslash}p{0.3200\linewidth}@{}}
\toprule\noalign{}
Material & Recommended Representation & Must Not Appear \\
\midrule\noalign{}
\endhead
\bottomrule\noalign{}
\endlastfoot
Resource DID & \passthrough{\lstinline!did:oan!} string & Incorrect routing-code length, illegal Base58 character \\
Ed25519 public key & multibase or Ed25519 JWK & Private key, raw string with no declared algorithm \\
SM2 public key & SM2 JWK or a public key representation supported by the implementation & Incomplete \passthrough{\lstinline!x!}/\passthrough{\lstinline!y!} coordinates \\
Private key & Client-protected storage & DID Document, ordinary request, service log \\
\end{longtable}
}

\passthrough{\lstinline!publicKeyMultibase!} currently consists of a \passthrough{\lstinline!z!} prefix followed by the Base58 encoding of the public key bytes; Ed25519 uses a 32-byte public key, while SM2 uses uncompressed SEC1 public key bytes. JWK uses unpadded Base64url-encoded fields; after decoding, Ed25519 \passthrough{\lstinline!x!} should be 32 bytes, and SM2 \passthrough{\lstinline!x!} and \passthrough{\lstinline!y!} should each be 32 bytes. \passthrough{\lstinline!oan-crypto!} parses these carriers according to the verification method or explicit \passthrough{\lstinline!cryptoSuite!}, and performs signature verification only after parsing succeeds.

The routing code of a DID selects a Registrar route; neither resource type nor the public key is encoded in the identifier. A randomly generated DID and its key pair are generated in two independent steps. If repeatable derivation is required, \passthrough{\lstinline!DidOan::derive!} control material and a nonce must be supplied explicitly. That derivation is an implementation capability, and a random DID must not be treated as reversible from a public key.

\section{VC and VP Encoding}\label{sec:appendix-c.5}

Registration VCs and node authorization VCs use JSON objects to express the issuer, subject, status, issuance/expiration times, claims, and proof; the shared credential implementation is based on \passthrough{\lstinline!AgentRegistrationCredential!}, \passthrough{\lstinline!NodeAuthorizationCredential!}, and \passthrough{\lstinline!DataIntegrityProof!}. At issuance, a signature is generated over the object without \passthrough{\lstinline!proof!}; at verification, \passthrough{\lstinline!proof!} is removed to reconstruct the original payload, which is then verified according to the suite specified by the proof. If a VP is presented by an external system, its original structure, credential references, and verification results should be retained; parseability does not mean acceptance under the OAN trust policy, and expired, revoked, subject-mismatched, or unauthorized issuers must be rejected or marked unacceptable.

Successful VC parsing only means that the structure is readable; the caller must still verify the signature, issuer authorization, subject, validity period, and credential status. When a VP serves as an external presentation container, the credential source and verification result should be retained; the act of presentation itself must not be treated as authorization.

The verification object for a registration VC is the \passthrough{\lstinline!AgentRegistrationCredential!} with \passthrough{\lstinline!proof!} removed; the verification object for a node authorization VC is the \passthrough{\lstinline!NodeAuthorizationCredential!} with \passthrough{\lstinline!proof!} removed. The following table can be used as the implementation order for the verification chain:

{\def\LTcaptype{none} 
\begin{longtable}[]{@{}>{\raggedright\arraybackslash}p{0.3200\linewidth}>{\raggedright\arraybackslash}p{0.3200\linewidth}>{\raggedright\arraybackslash}p{0.3200\linewidth}@{}}
\toprule\noalign{}
Check & Object of Attention & Meaning When Not Passed \\
\midrule\noalign{}
\endhead
\bottomrule\noalign{}
\endlastfoot
Structure & \passthrough{\lstinline!type!}, \passthrough{\lstinline!issuer!}, \passthrough{\lstinline!subject!}, time, \passthrough{\lstinline!claims!}, \passthrough{\lstinline!proof!} & The credential cannot be parsed or proceed to subsequent verification \\
Signature & \passthrough{\lstinline!proof!}, public key in the issuer\textquotesingle s DID Document & The credential content is not proven by the corresponding private key \\
Binding & \passthrough{\lstinline!issuer!}, \passthrough{\lstinline!subject!}, resource DID, DID Document hash, and ResourcePackage hash & The credential does not correspond to the current resource \\
Authorization & Issuer identity, node role, authorization domain, and governance status & The issuer cannot make this statement within the current boundary \\
Status & \passthrough{\lstinline!status!}, \passthrough{\lstinline!issuedAt!}, \passthrough{\lstinline!expiresAt!}, and available revocation facts & The credential is invalid or should no longer be used \\
\end{longtable}
}

The verifier should retain enough of the original credential, DID Document, verification time, and error code to support re-verification; it should not store only a boolean value of ``verification succeeded''.

\section{Key Backup and Recovery Format}\label{sec:appendix-c.6}

Local Identity backups are sensitive client files and may contain the version, update time, default subject/agent identifiers, subject/agent/node records, DID Documents, public and private key JWKs, and resource profiles. The TypeScript SDK currently uses an identity storage snapshot with version \passthrough{\lstinline!1!}; identity backup files downloaded through the website registration flow include a time identifier in the filename, and after a successful resource submission, the registration VC, DID Document, and Local Identity file may also be packaged for storage. Backup files should be kept privately by the user. On import, the JSON structure, version, public/private key correspondence, and DID consistency must be checked; unsupported versions, missing keys, or parse failures must not result in a partial load.

The import flow should complete the integrity and version checks first, then load subject, agent, and node records as one operation. If a public and private key do not match, the DID is inconsistent, or the version is unsupported, the entire backup should be rejected; a partially usable identity must not be generated.

The private key JWK in a backup file is highly sensitive material. Browser download, import, and re-export should all be actively initiated by the user; the page, Registrar node, Root platform, and Discovery node do not need to receive this private key. After import, the client may sign a test message with the private key and verify it with the public key from the backup to confirm that the key pair matches; the private key or test signature must not be written to logs. The time identifier in the backup filename is used only to distinguish downloaded files and does not participate in identity or signature computation.

The key values in the example above are unusable placeholders intended only to show field positions; a real backup must not use this example content, and production private keys must not be placed in public documentation or test output.

\section{DID and Resource Identifier Test Vectors}\label{sec:appendix-c.7}

Test vectors cover valid resource and node DIDs, routing-code case changes, omissions, incorrect separators, invalid lengths, prohibited Base58 characters, and extra segments. The DID string does not encode \passthrough{\lstinline!agent\_service!}, \passthrough{\lstinline!skill!}, \passthrough{\lstinline!mcp\_server!}, or \passthrough{\lstinline!tool\_api!}; resource type comes from DID Document metadata and registration material, so DID syntax and resource-type consistency are tested separately. Each vector fixes the input and parse result or error; generated DIDs validate format only, and random suffixes need not repeat across runs.

{\def\LTcaptype{none} 
\begin{longtable}[]{@{}>{\raggedright\arraybackslash}p{0.3200\linewidth}>{\raggedright\arraybackslash}p{0.3200\linewidth}>{\raggedright\arraybackslash}p{0.3200\linewidth}@{}}
\toprule\noalign{}
Vector Category & Input & Passing Criterion \\
\midrule\noalign{}
\endhead
\bottomrule\noalign{}
\endlastfoot
Valid DID & Four resource types and infrastructure DIDs & Parsing succeeds and the type relationship is correct \\
Format error & Incorrect case, separator, or length & A parse error is returned consistently \\
Prohibited character & \passthrough{\lstinline!0!}, \passthrough{\lstinline!O!}, \passthrough{\lstinline!I!}, \passthrough{\lstinline!l!}, etc. & Reject without silently correcting \\
Type mismatch & Semantic code does not match the resource type & Reject or return an explicit error \\
\end{longtable}
}

Test vectors also record the input, expected error, and side-effect criteria. \passthrough{\lstinline!did:oan:K7mQ9:7YpQm9Kx2VnRb6Ts3WfHa4Cd5Ej8LgNz!} passes syntax parsing; resource type is checked separately across the registration submission, DID Document metadata, and ResourcePackage. Replacing a routing-code or suffix character with \passthrough{\lstinline!0!}, \passthrough{\lstinline!O!}, \passthrough{\lstinline!I!}, or \passthrough{\lstinline!l!} fails parsing, and the client must not rewrite it before submission.

{\def\LTcaptype{none} 
\begin{longtable}[]{@{}>{\raggedright\arraybackslash}p{0.3200\linewidth}>{\raggedright\arraybackslash}p{0.3200\linewidth}>{\raggedright\arraybackslash}p{0.3200\linewidth}@{}}
\toprule\noalign{}
Test Field & Example Value & Expected Result \\
\midrule\noalign{}
\endhead
\bottomrule\noalign{}
\endlastfoot
Valid routing code & \passthrough{\lstinline!K7mQ9!} & Parsed as a five-character case-sensitive Base58 Registrar routing code \\
Valid suffix & \passthrough{\lstinline!7YpQm9Kx2VnRb6Ts3WfHa4Cd5Ej8LgNz!} & Length is 32 and contains only permitted Base58 characters \\
Resource type match & \passthrough{\lstinline!SKFI!} + \passthrough{\lstinline!skill!} & Type check passes \\
Resource type conflict & \passthrough{\lstinline!SKFI!} + \passthrough{\lstinline!mcp\_server!} & Returns \passthrough{\lstinline!ResourceTypeMismatch!} or the corresponding error code \\
\end{longtable}
}

\section{Hash and Signature Verification Fixtures}\label{sec:appendix-c.8}

Fixtures should include at least the original canonicalization input, canonicalized JSON, algorithm suite, hash algorithm, digest, signature input, public key, proof, and expected verification result, and should separately cover the correct payload, field-order changes, number/string changes, public key changes, truncated signatures, and proof-suite mismatches. The Rust implementation may use \passthrough{\lstinline!oan-crypto!} tests to produce the baseline, and the TypeScript implementation should read the same fixture to recompute it. Fixtures must not contain production private keys; dedicated test keys should be used when generating fixtures, with the public keys and signature results made available.

A fixture is recommended to use one JSON file that can be read by both Rust and TypeScript, with the input object and expected result stored separately. An example structure is:

The public key, signature, and digest in a fixture must be generated from the same dedicated test key set; the \passthrough{\lstinline!x!} and \passthrough{\lstinline!proofValue!} in the example remain structural placeholders and cannot be used directly as inputs for a passing test. At least the following negative variants should be run: change only JSON member order, change array order, change one string, replace the public key, truncate the signature, and cross-replace \passthrough{\lstinline!cryptoSuite!} and \passthrough{\lstinline!proof.type!}. A correct implementation should accept only the original payload paired with the matching public key, and should produce stable failures for the other variants.

\section{References}\label{sec:appendix-c.9}

{\def\LTcaptype{none} 
\begin{longtable}[]{@{}>{\raggedright\arraybackslash}p{0.3200\linewidth}>{\raggedright\arraybackslash}p{0.3200\linewidth}>{\raggedright\arraybackslash}p{0.3200\linewidth}@{}}
\toprule\noalign{}
Source & Type & Link \\
\midrule\noalign{}
\endhead
\bottomrule\noalign{}
\endlastfoot
\passthrough{\lstinline!oan-protocol-commons!} & Code repository: cryptographic suites, keys, signatures, and hash implementations & \url{https://github.com/wolfbrother/oan-protocol-commons} \\
\passthrough{\lstinline!oan-sdk-ts!} & Code repository: client cryptographic materials and DID operations & \url{https://github.com/OpenAgenet/oan-sdk-ts} \\
\passthrough{\lstinline!did:oan!} DID Method Specification & Standard/method specification: DID identifiers and verification methods & \url{https://github.com/OpenAgenet/oan-public-docs/blob/main/did-oan-specs/doc/OAN\%20DID\%20Method\%20Specification.md} \\
OAN Yellow Paper & arXiv yellow paper: cryptographic proofs and trust model & \url{https://arxiv.org/abs/2606.03163} \\
\end{longtable}
}

\chapter{Conformance Test Matrix}\label{chap:appendix-d}

This appendix defines ``conformance'' as reproducible input, rule execution, and evidence judgment, rather than treating successful access to an online service on one occasion as complete proof. The test executor should record the test suite version, fixture version, version of the implementation under test, configuration summary, execution time, environment type, and redacted output; every case must produce a stable case ID, a \passthrough{\lstinline!PASS!} or \passthrough{\lstinline!FAIL!} conclusion, the actual error cause, and locatable evidence files. For rejection cases, the pass condition is not merely that the request fails. It must also show that the rejection occurred at the correct boundary, that no publication or index state that should not exist was written, and that the error code or reason conforms to the agreed contract. The currently directly reproducible offline evaluation is centered on \passthrough{\lstinline!oan-third-party-node-admission-test-kit/fixtures/v1-alpha!}; its \passthrough{\lstinline!case-matrix.md!} and \passthrough{\lstinline!requirements-traceability-matrix.md!} define the fixed fixtures, expected outputs, and checks for 49 requirements. This suite verifies the evaluation package and protocol judgment materials; it does not automatically amount to end-to-end proof against the official online Root, chain, content distribution platform, or off-chain trust indexer.

Every test record should contain at least the following fields:

{\def\LTcaptype{none} 
\begin{longtable}[]{@{}>{\raggedright\arraybackslash}p{0.4800\linewidth}>{\raggedright\arraybackslash}p{0.4800\linewidth}@{}}
\toprule\noalign{}
Field & Recording requirement \\
\midrule\noalign{}
\endhead
\bottomrule\noalign{}
\endlastfoot
Case identifier & Use the stable case ID in the manifest; do not use a temporary line number in a test file as the unique identifier. \\
Preconditions & Record the fixture, protocol version, authorization-domain set, governance snapshot, simulated responses, and necessary service status. \\
Operation and input & Record the called interface, HTTP method, request-body summary, query conditions, and signature or hash inputs; private keys, tokens, and database connection strings must be redacted. \\
Actual result & Record the status code, normalized result, error code, lifecycle, version, cursor, and necessary response summary. \\
Pass criterion & Compare the actual result item by item with the \passthrough{\lstinline!outcome!}, \passthrough{\lstinline!reasonCode!}, and normative clauses in the expected fixture. \\
Evidence artifacts & Save the JSON report, Markdown report, references to the input and expected fixtures, version, and redaction rules. \\
\end{longtable}
}

Test failures fall into three categories: protocol or security failures that block publication, identity or authorization failures that block node admission, and operational or freshness failures that do not block publication but must trigger alerts and leave an audit trail. The test suite must not silently retry, automatically correct input, or ignore unknown fields merely to obtain a ``pass.'' If the implementation has an explicit compatibility mode, it should be marked separately in the report. On-chain events, real content distribution, and cross-node network timing that are not covered by the current V1-alpha fixture should be marked as uncovered, rather than treating offline pass results as evidence that they are covered.

Test execution results should distinguish ``the test result is a pass'' from ``the object under test accepts the input.'' For an input that is expected to be rejected, the service returning a rejection is the correct behavior of the object under test, so the test case itself should be recorded as \passthrough{\lstinline!PASS!} rather than counting the business rejection directly as a failure. The minimum record in a report may use the following structure:

\begin{lstlisting}
{
  "caseId": "fixture.registrar.rejects-did-mismatch",
  "expectedOutcome": "invalid",
  "actualOutcome": "invalid",
  "verdict": "PASS",
  "reasonCode": "DID_ID_MISMATCH",
  "sideEffect": "no_resource_record_written",
  "evidence": ["evidence/fixture.registrar.rejects-did-mismatch.json"]
}
\end{lstlisting}

\passthrough{\lstinline!expectedOutcome!} describes the fixture expectation, \passthrough{\lstinline!actualOutcome!} describes the implementation\textquotesingle s actual judgment, and \passthrough{\lstinline!verdict!} describes whether the test passed; these three meanings must not be merged into an ambiguous \passthrough{\lstinline!status!} field. Write tests should also compare the number of resource records, publication-queue entries, and index entries before and after the request to prove that the rejection produced no subsequent side effects.

Different test levels have different conclusion scopes: offline fixtures check rules and report generation, node integration tests check HTTP routes, persistence, and calls between nodes, and end-to-end tests check the propagation of publication, distribution, discovery, and governance state. The report should record the level and must not merge offline rule checks, service health checks, and complete business acceptance into a single ``system passed'' conclusion.

\section{Resource Registration Tests}\label{sec:appendix-d.1}

Registration tests should use \passthrough{\lstinline!agent\_service!}, \passthrough{\lstinline!skill!}, \passthrough{\lstinline!mcp\_server!}, and \passthrough{\lstinline!tool\_api!} as the primary four resource categories. Positive samples should simultaneously satisfy consistency among \passthrough{\lstinline!resourceDid!}, the DID Document \passthrough{\lstinline!id!}, DID Document metadata, ResourcePackage metadata, the VC subject, and hash fields, and should contain a verifiable subject-control proof, resource type, and authorization domain. Registrar interface-layer tests should cover \passthrough{\lstinline!/resources/register!} and its compatibility entry point \passthrough{\lstinline!/resources/submit!}, verifying shape validation, authorization-domain validation, submission to the Root platform, and the boundary at which the registration credential and Root platform response are returned. Negative samples should replace one fact at a time, such as a DID mismatch, VC subject mismatch, package-hash mismatch, missing \passthrough{\lstinline!packageProof!}, missing credential proof, inconsistent control challenge, incomplete metadata, or a suspended resource, so that one fixture does not contain multiple indistinguishable errors.

{\def\LTcaptype{none} 
\begin{longtable}[]{@{}>{\raggedright\arraybackslash}p{0.2400\linewidth}>{\raggedright\arraybackslash}p{0.2400\linewidth}>{\raggedright\arraybackslash}p{0.2400\linewidth}>{\raggedright\arraybackslash}p{0.2400\linewidth}@{}}
\toprule\noalign{}
Test item & Prerequisite data and steps & Pass and rejection criteria & Evidence and impact \\
\midrule\noalign{}
\endhead
\bottomrule\noalign{}
\endlastfoot
Minimal positive registration & Load \passthrough{\lstinline!registrar-resource.valid.json!} and submit a fixed ResourcePackage to the test Registrar. & Accept; resource type, DID, VC, authorization domain, and hashes are all bound; the returned result must not claim a publication stage beyond the one actually reached. & Request summary, response summary, resource record, and expected fixture; after passing, allow entry into subsequent publication tests. \\
Field and type consistency & Use \passthrough{\lstinline!did-mismatch!}, \passthrough{\lstinline!credential-subject-mismatch!}, and \passthrough{\lstinline!metadata-incomplete!} separately. & Reject with a stable reason; no valid resource record readable by a Discovery node may be written. & Error response and before/after storage comparison; failure blocks registration publication. \\
Proof and hash binding & Use \passthrough{\lstinline!tampered-proof!}, \passthrough{\lstinline!credential-hash-mismatch!}, \passthrough{\lstinline!package-hash-mismatch!}, and missing-proof fixtures. & Reject; validation must cover the proof subject, issuer, DID Document hash, metadata hash, and package hash. & Validation path and redacted proof summary; failure blocks publication. \\
Duplicate submission and update & Submit the same DID and version twice, then submit an incremented version with a controller signature. & Duplicate submission must not create two conflicting current versions; an update must pass control and version rules; historical and current versions must have distinguishable meanings. & Two responses, resource records, version list, and retry log; an abnormal duplicate write blocks publication. \\
\end{longtable}
}

The current fixed fixtures directly cover positive registration and multiple rejection conditions. Real Registrar database idempotency, concurrent writes, and propagation of updates for the same DID still need to be executed in a running-service scenario and marked in the report as live integration coverage; offline fixtures cannot substitute for them.

The four resource categories use the same binding checks with their respective valid \passthrough{\lstinline!resourceType!} values. DID routing codes are tested independently and do not encode resource category. Test data may be organized as follows so that assertions can be reused without applying fields from one category incorrectly to another:

{\def\LTcaptype{none} 
\begin{longtable}[]{@{}>{\raggedright\arraybackslash}p{0.2400\linewidth}>{\raggedright\arraybackslash}p{0.2400\linewidth}>{\raggedright\arraybackslash}p{0.2400\linewidth}>{\raggedright\arraybackslash}p{0.2400\linewidth}@{}}
\toprule\noalign{}
Resource type & Example routing requirement & Minimal business materials & Key assertion \\
\midrule\noalign{}
\endhead
\bottomrule\noalign{}
\endlastfoot
Agent service & Valid five-character routing code and 32-character suffix & DID Document, service endpoint, capability tags & Endpoint and \passthrough{\lstinline!agent\_service!} metadata agree \\
Skill package & Same common DID requirement & DID Document, package information, download or manifest address & Package digest, version, and \passthrough{\lstinline!skill!} metadata agree \\
MCP service & Same common DID requirement & DID Document, MCP endpoint, protocol binding & Protocol, endpoint, and \passthrough{\lstinline!mcp\_server!} metadata agree \\
Tool API & Same common DID requirement & DID Document, API endpoint, input/output description & Interface description and \passthrough{\lstinline!tool\_api!} metadata agree \\
\end{longtable}
}

The same negative case should change only one fact under test, such as replacing only the DID Document \passthrough{\lstinline!id!} while keeping the VC, hashes, authorization domain, and lifecycle unchanged. This allows the failure reason to point accurately to the corresponding clause and prevents the service from masking the target defect with an earlier validation failure.

\section{Root Platform Publication Tests}\label{sec:appendix-d.2}

Root platform publication tests should separate Registrar acceptance from Root platform acceptance, validation, and publication. The input contains the registration service node\textquotesingle s identity materials, the resource DID Document, the ResourcePackage, registration credential, subject-control proof, version, and hashes. The Root platform first validates node authorization and the request signature, then validates resource binding; only after all checks pass does it create a trusted Root platform publication proof and a publication fact available for distribution. Tests should check that a failure leaves no intermediate ``publish first, validate later'' state and that there is no path by which a Discovery node accepts a result merely because the Registrar returned success. For duplicate delivery, reprocessing must remain idempotent: the same publication fact must not generate a conflicting version, while changed content or proof must be judged again under version and control rules.

The current third-party admission test suite does not include the Root, chain, or real publication queue within the V1-alpha evaluation scope. Therefore, the offline part of this section can validate only publication input materials and expected rules; it cannot claim complete coverage of real Root platform publication. When running services, record the Root platform interface response, publication job ID, resource version, trusted Root platform publication proof summary, rejection reason, and reprocessing result. Logs containing private governance materials must not enter the public report.

The key assertion for Root platform publication tests is ``validate first, then create a publishable fact.'' They may be executed in the following order:

\begin{enumerate}
\def\labelenumi{\arabic{enumi}.}
\tightlist
\item
  Prepare the identity materials for an authorized Registrar node and a resource registration submission.
\item
  Separately tamper with the node-to-node signature, resource control proof, DID Document hash, ResourcePackage hash, and authorization domain, and call \passthrough{\lstinline!POST /root/resources/verify-and-publish!} for each case.
\item
  Check the HTTP result, machine-readable error, resource version, publication queue, and trusted Root platform publication proof for each rejection.
\item
  Send the original valid request again and confirm that duplicate publication does not create a conflicting current version.
\end{enumerate}

The storage and queue comparison in step 3 is necessary evidence; saving only the HTTP response cannot prove that the Root platform left no partial publication state for a failed request.

\section{Distribution and Synchronization Tests}\label{sec:appendix-d.3}

Distribution tests center on ``the publication fact reaches an authorized node in a verifiable manner.'' First establish a baseline with full synchronization, then perform incremental synchronization using a cursor, injecting duplicate events, out-of-order events, missing events, checkpoint recovery, and content-hash errors separately. A receiving node must not overwrite data based only on arrival time; it should decide whether to accept it according to the resource DID, version, publication proof, and cursor rules. Materials that are missing or have an incorrect hash should remain pending or failed and must not enter a visible index. Recovery tests should record the last successful cursor, retry count, backoff result, and final convergence state, distinguishing ``transfer complete,'' ``local validation complete,'' and ``visible to discovery.'' A content distribution platform cache hit or HTTP success can prove only a transport-layer result; it cannot replace the receiver\textquotesingle s validation of the trusted Root platform publication proof.

The boundary between this section and the current V1-alpha fixture should be written explicitly into the report: the existing test suite validates authorization domains, resource visibility, provenance, and fixed evidence rules, but its manifest explicitly excludes a real CDN, Root queue, and on-chain synchronization. After the implementation is connected to complete distribution tests, use repeatable resource versions and manually constructed cursors, and save four categories of evidence from the sender, receiver, validator, and indexer. Any failure that could cause an old version to become visible again, a revocation state to be lost, or an unvalidated resource to enter the index should be treated as a publication-blocking issue.

Distribution tests must at least verify that cursor advancement and resource acceptance are related but distinct results: a receiving node may successfully read one page of data while rejecting one resource in that page because of a proof or hash error. Whether the cursor advances and whether the failed item enters a retry queue must be recorded according to the established interface and implementation rules. Re-sending the same publication fact should be safely replayable; sending different content while reusing the same resource DID and version should enter a conflict-handling path and must not silently overwrite accepted content.

\begin{lstlisting}
{
  "afterCursor": 1800,
  "nextCursor": 1812,
  "hasMore": true,
  "accepted": 11,
  "rejected": 1,
  "rejectedReasons": ["invalid_root_proof"]
}
\end{lstlisting}

The example above is an illustrative result-summary structure for a test report; the fields should follow the actual content distribution interface response. Advancement of \passthrough{\lstinline!nextCursor!} does not mean that every item has been validated and accepted. The report must also save the DID and reason for each rejected item.

\section{Discovery and Semantic Search Tests}\label{sec:appendix-d.4}

Discovery tests should cover exact DID queries for \passthrough{\lstinline!/discovery/resources/query!}, resource type, protocol, capability tags, authorization domain, and lifecycle filtering, as well as natural-language queries and the query-explanation interface, if enabled by the implementation. Test input first passes syntax, scope, and authorization-domain checks, then is converted into structured conditions, tag conditions, or semantic-retrieval conditions. Results must retain the resource DID, version, source, governance status, and necessary proof summary. Relevance scores, keyword hits, tag matches, and semantic similarity may be used only for ranking or explanation; they must not bypass authorization-domain checks, suspension/revocation filtering, DID Document validation, or trusted Root platform publication proof checks.

{\def\LTcaptype{none} 
\begin{longtable}[]{@{}>{\raggedright\arraybackslash}p{0.3200\linewidth}>{\raggedright\arraybackslash}p{0.3200\linewidth}>{\raggedright\arraybackslash}p{0.3200\linewidth}@{}}
\toprule\noalign{}
Query category & Representative input & Behavior that must be validated \\
\midrule\noalign{}
\endhead
\bottomrule\noalign{}
\endlastfoot
Exact DID query & A fixed resource DID or a nonexistent DID. & Hit the unique resource or return a stable empty result; do not return another resource because of textual similarity. \\
Structured compound query & A combination of \passthrough{\lstinline!resourceType!}, \passthrough{\lstinline!protocol!}, \passthrough{\lstinline!capabilityTags!}, \passthrough{\lstinline!limit!}, and authorization domain. & Parse fields according to protocol rules; reject invalid enumerations, negative values, or values over the limit; pagination must neither duplicate nor omit accepted results. \\
Semantic query & For example, \passthrough{\lstinline!Find an MCP server for security audit!}. & Convert query text into explainable candidate conditions; returned results must identify semantic, tag, text, or other sources, and must not treat similarity as a trust proof. \\
Visibility boundary & In-domain, out-of-domain, suspended, revoked, protocol-mismatched, and type-mismatched resources. & Return only resources within the authorization domain and with an available status; an empty result is valid and must not be filled with recommendations. \\
Deduplication and explanation & Multiple versions of the same DID, duplicate index records, and an explain request. & Deduplicate according to DID and version rules, and explain candidate sources and filtering reasons. \\
\end{longtable}
}

\passthrough{\lstinline!fixtures/v1-alpha/discovery!} already provides fixed cases for visibility, empty results, arbitrary tag matching, invalid filters, invalid tags, invalid resource types, invalid protocols, \passthrough{\lstinline!limit!} boundaries, and overly short queries; these cases should be used as the regression baseline. The specific quality of semantic models, vector stores, and natural-language ranking must not be inferred from fixed-rule fixtures; record the model version, index build time, and explanation result separately.

Discovery-result tests should record candidate generation, filtering, ranking, and proof checking separately. Even if a resource is highly matched by text or capability tags, it must not appear in callable results if the authorization domain does not cover it, its lifecycle is unavailable, its DID Document is inconsistent with the ResourcePackage, or its trusted Root platform publication proof is invalid. Conversely, an empty result does not mean that the discovery service is faulty. It constitutes a recall or indexing defect only when a valid resource satisfying all filter conditions exists but the response is still empty.

{\def\LTcaptype{none} 
\begin{longtable}[]{@{}>{\raggedright\arraybackslash}p{0.3200\linewidth}>{\raggedright\arraybackslash}p{0.3200\linewidth}>{\raggedright\arraybackslash}p{0.3200\linewidth}@{}}
\toprule\noalign{}
Test stage & Observed value & Pass condition \\
\midrule\noalign{}
\endhead
\bottomrule\noalign{}
\endlastfoot
Input parsing & Query text, DID, resource type, tags, protocol, and \passthrough{\lstinline!limit!} & Valid input enters the corresponding retrieval path; invalid input fails in a controlled manner \\
Candidate generation & Candidate DIDs, match sources, and original scores & Candidates come from locally indexed data; resources are not invented \\
Governance filtering & Authorization domain, lifecycle, and publication-proof status & Candidates outside the boundary are excluded \\
Result output & DID, version, source, status, and proof summary & Ranking does not change the trust conclusion; results are reviewable \\
\end{longtable}
}

\section{Governance Authorization Tests}\label{sec:appendix-d.5}

Governance tests must treat node authorization, resource control, and resource publication results as three independent judgments. A node may perform actions allowed by its role only when its node DID Document, authorization credential, authorization domain, protocol version, and lifecycle status are consistent. The resource owner\textquotesingle s control proof establishes only the right to operate the resource and cannot replace node authorization. Use \passthrough{\lstinline!active!}, \passthrough{\lstinline!suspended!}, \passthrough{\lstinline!revoked!}, unknown, and non-canonical case variants separately, and verify that sensitive operations reject or fail closed when the status cannot be confirmed or has become invalid. For authorization domains, cover the empty set, wildcard, parent and child domains, sibling domains, duplicates, case variations, and unknown domains, confirming that a tag tree or resource description does not enlarge the node\textquotesingle s authorization scope.

The \passthrough{\lstinline!governance!}, \passthrough{\lstinline!node!}, \passthrough{\lstinline!domains!}, and \passthrough{\lstinline!lifecycle!} fixtures provide stable inputs for offline judgment. The source, event sequence, or projection time in the governance snapshot must be saved together with the test report. Projection latency of the off-chain trust indexer, on-chain finality, and real revocation propagation belong to the running integration scope and cannot be claimed complete based only on snapshot tests.

Governance authorization tests should preferably use a closed-loop assertion of ``authorization fact -\textgreater{} node action -\textgreater{} resource visibility.'' For a revoked Discovery node, do not merely verify that the authorization snapshot status is \passthrough{\lstinline!revoked!}; also verify that its synchronization or query action is rejected, that its existing authorization scope does not continue to expand, and that already indexed resources do not become valid results again because of an old cache. For an authorization-domain change, check both the new and old domains to avoid validating only the added domain while missing whether the old permission was correctly withdrawn.

\section{Version Update and Lifecycle Tests}\label{sec:appendix-d.6}

Lifecycle tests center on state transitions and version relationships rather than using timestamp order as the sole basis. Establish an active resource, publish a new version with a valid proof from the original controller, and then verify the relationship among the current version, historical versions, and same-DID query results. Next apply suspension, revocation, and restoration facts separately, and check whether the statuses seen by the Registrar, Root platform, content distribution platform, Discovery node, and caller change within their allowed propagation boundaries. Updates without controller authorization, version rollback, same-version content conflicts, and restoration without a basis should all be rejected. Discovery queries should return only the available current version by default. If the interface supports historical queries, it must state the status of historical records clearly and must not allow an old version to bypass current governance status and become callable again.

The existing \passthrough{\lstinline!lifecycle!} fixture can validate the normative form of status values and active/inactive semantics. Same-DID updates, real historical persistence, and cross-node convergence should use service integration tests, retaining the DID Document hash, ResourcePackage hash, version, control proof, publication proof, and index time for every submission.

Records for a same-DID update should contain at least the following relationships:

\begin{lstlisting}
same resource DID
  -> controller proof for the update
  -> strictly selected package version
  -> new DID document/package hashes
  -> Root publication evidence
  -> Discovery current-version result
\end{lstlisting}

Even when only the description or endpoint is updated, recompute the affected DID Document hash, metadata hash, or ResourcePackage hash and validate the corresponding proof; incrementing the version number cannot replace content binding. Historical-query tests should also confirm that historical versions are traceable but do not bypass the current suspension, revocation, or authorization status.

\section{Security and Abuse-Prevention Tests}\label{sec:appendix-d.7}

Security tests should validate ``rejection with no side effects.'' Fixed signature samples confirm that a valid Ed25519 signature can be verified and that the same payload after tampering cannot pass. At the service level, also replace the request nonce, request ID, audience, path, protocol version, and signer, and check replay, cross-interface reuse, and unauthorized calls. Request size, query \passthrough{\lstinline!limit!}, text length, invalid JSON, incorrect Content-Type, malicious endpoints, and abnormal authorization domains should be limited before expensive semantic retrieval, network requests, or persistence are entered. Rate-limit and timeout tests should record response status, wait time, resource consumption, and recovery, and must not use unlimited retries to hide rejection.

Sensitive-evidence tests must ensure that private keys, access tokens, database connection strings, internal file paths, and unredacted control materials do not enter reports, public website interfaces, or ordinary node responses. \passthrough{\lstinline!security/sensitive-evidence.json!} and report-redaction checks are the currently reproducible baseline. Real services should additionally sample application logs, reverse-proxy logs, and error responses to confirm that exceptional branches do not disclose internal stacks or secrets.

Side-effect checks in security tests should cover three locations: input buffers or temporary files before request processing, server-side persistent records, and downstream publication or index queues. For oversized requests, malicious endpoints, and invalid signatures, a passing result should include request limiting, no resource-record write, no node-to-node publication call, and no public error stack. For rate-limit tests, distinguish ``the service rejects the request'' from ``the service process is unavailable'': the former is a policy result, while the latter is a stability defect.

\section{Third-Party Node Admission Tests}\label{sec:appendix-d.8}

Third-party node tests begin with node identity consistency, requiring no contradiction among configuration, DID Document, status response, authorization credential, role, endpoint, protocol version, lifecycle, and authorization domain. After admission, verify separately whether a Registrar node can accept and submit within its authorization domain, and whether a Discovery node can synchronize and query only Root platform publication content within its authorized scope. Also verify that invalid nodes, out-of-domain resources, protocol-version mismatches, and revoked status do not continue to receive permission because of configuration caches or historical credentials. Logging and failure-recovery tests should check that the request ID, cursor, rejection reason, and retry record are sufficient to reconstruct the event without containing private keys or internal secrets.

The current suite\textquotesingle s \passthrough{\lstinline!node-identity.*!}, authorization-domain, governance-snapshot, and discovery-visibility fixtures are suitable as offline admission gates for third-party nodes. Real node admission must additionally use isolated endpoints and an independent database; test nodes must not be connected directly to the official production data plane. The report should state the node under test, authorization domain, configuration version, and revocation/restoration times.

The admission conclusion for a third-party node should be split into five observation areas: identity, protocol, authorization, business, and operations. Consistent identity materials establish ``who the node is''; successful protocol calls establish ``whether the interface can communicate''; only a passing authorization and business loop establishes ``whether the node can complete registration, synchronization, or discovery within its authorization scope''; logging, rate-limiting, and recovery evidence are used to judge whether the node can operate continuously. Any failed observation area should retain its specific failure reason in the report and must not be offset by another area passing.

\section{DID Resolution and Method Conformance Tests}\label{sec:appendix-d.9}

Resolution tests validate the method prefix, five-character routing code, 32-character suffix, Base58 character set, and case-sensitive parsing of \passthrough{\lstinline!did:oan!}; parseability or a server returning JSON is not proof of control. DID Document tests check \passthrough{\lstinline!id!}, controller continuity, public keys, authentication/assertion references, service endpoints, resource type, and metadata, with deterministic failures for an incorrect method, segment count, routing length, invalid character, or service reference. Historical resolution additionally records version, digest, and status.

Test evidence should contain identifiers before and after normalization, parsed fields, the verification method, and the error reason. Private keys may be used for signing at test runtime but must not be written to fixture reports.

Method conformance tests should also distinguish the syntax, structure, and control layers: the syntax layer confirms that the string has the \passthrough{\lstinline!did:oan!} form; the structure layer confirms that the DID Document\textquotesingle s \passthrough{\lstinline!id!}, verification methods, and service references are mutually consistent; the control layer requires generating a proof with the corresponding private key and verifying it with the public key. Passing only the first two layers cannot establish that a particular subject controls the resource. Fixed suffixes used in test vectors can reproduce resolution results but should not be interpreted as key-derivation results in a production environment.

\section{API Compatibility and Error-Contract Tests}\label{sec:appendix-d.10}

API tests should check successful and failed contracts at the same time. For each interface, record at least the method, path, Content-Type, required fields, handling of unknown fields, status code, error-body structure, error reason, pagination fields, and retry guidance. A client must not treat any 2xx response as business success, nor automatically interpret a network timeout as a resource rejection. For compatibility versions, distinguish ignorable extension fields, protocol versions that must be rejected, and differences in field order or JSON representation that must not change semantics. Retries apply only to explicitly retryable transport or temporary dependency failures. Requests carrying signatures and write side effects must use request ID, nonce, or server-side idempotency rules to prevent duplicate submission.

\passthrough{\lstinline!oan-protocol-commons/docs/error-codes.md!}, protocol types, and the \passthrough{\lstinline!requests!} fixtures in the test suite are the main sources for checking the error contract. The current fixtures cover an empty request body, invalid method, malformed JSON, missing fields, oversized request, incorrect Content-Type, and incorrect query type. When a service adds an error code, it should update the common error documentation, client mapping, and expected fixture together.

Compatibility tests should assert request-delivery results and business-semantic results separately. An unknown ignorable field should not change the meaning of existing fields; an unknown protocol version should produce a clear incompatibility result; a missing required field should fail before business write; and a network timeout can be recorded only as an unknown transport result, not directly judged as registration failure. For interfaces with write side effects, retry tests must use a fixed \passthrough{\lstinline!requestId!}, nonce, or equivalent idempotency condition, and check whether the server generated duplicate VCs, duplicate versions, or duplicate publication tasks.

{\def\LTcaptype{none} 
\begin{longtable}[]{@{}>{\raggedright\arraybackslash}p{0.3200\linewidth}>{\raggedright\arraybackslash}p{0.3200\linewidth}>{\raggedright\arraybackslash}p{0.3200\linewidth}@{}}
\toprule\noalign{}
Scenario & Expected test conclusion & Evidence to save \\
\midrule\noalign{}
\endhead
\bottomrule\noalign{}
\endlastfoot
Valid request and known version & Business processing succeeds & Status code, response summary, and resource status \\
Add an unknown extension field & Process according to interface compatibility rules & Original request and field-processing result \\
Missing required field & Controlled rejection with no write side effect & Error code and before/after storage counts \\
Retry after network timeout & No duplicate business fact is created & Two request identifiers, responses, and final record \\
\end{longtable}
}

\section{Freshness, Convergence, and Recovery Tests}\label{sec:appendix-d.11}

Freshness tests should record observation time, fact-production time, synchronization cursor, index-update time, and permitted maximum delay separately. The test flow may first publish a resource with a unique version and hash, confirm Root platform acceptance, observe the states of the content distribution platform, Discovery node, and off-chain trust indexer, and finally compare whether all layers converge on the same DID, version, status, and proof. After deliberately interrupting the network, restarting a service, deleting temporary transfer state, or injecting duplicate events, recovery must continue from a verifiable cursor or a full rebuild; it must not infer missing facts from ``the latest response.'' An old snapshot may be used for non-authoritative display or limited degradation, but authorization, revocation, and new-publication decisions must follow explicit freshness boundaries.

The report should list each layer\textquotesingle s \passthrough{\lstinline!observedAt!}, \passthrough{\lstinline!source!}, \passthrough{\lstinline!cursor!}, \passthrough{\lstinline!version!}, \passthrough{\lstinline!lifecycleState!}, validation result, and latency at a minimum. For results exceeding the threshold, mark whether the outcome blocks sensitive operations, degrades to read-only, or only produces an operational alert. The current V1-alpha suite does not provide a complete real-network convergence evaluation, so this part should be published separately from integration reports for the Root, CDN, Discovery node, and off-chain trust indexer.

Convergence tests should use a traceable resource version as the correlation key rather than comparing only the number of resources returned by each interface. The following minimum evidence chain may be retained in the report:

The times, DID, and digests in the example are for illustrating report fields only. Actual acceptance should judge \passthrough{\lstinline!converged!} according to the threshold configured for the test environment, while retaining raw responses, log correlation identifiers, and cursors from every layer. Only then can propagation delay, index failure, lagging governance state, and version-binding errors be distinguished.

\section{References}\label{sec:appendix-d.12}

{\def\LTcaptype{none} 
\begin{longtable}[]{@{}>{\raggedright\arraybackslash}p{0.3200\linewidth}>{\raggedright\arraybackslash}p{0.3200\linewidth}>{\raggedright\arraybackslash}p{0.3200\linewidth}@{}}
\toprule\noalign{}
Source & Type & Link \\
\midrule\noalign{}
\endhead
\bottomrule\noalign{}
\endlastfoot
\passthrough{\lstinline!oan-protocol-commons!} & Code repository: test data structures, schemas, and error contract & \url{https://github.com/wolfbrother/oan-protocol-commons} \\
\passthrough{\lstinline!oan-root-nodes!} & Code repository: Root platform validation and publication test objects & \url{https://github.com/OpenAgenet/oan-root-nodes} \\
\passthrough{\lstinline!oan-registrar-nodes!} & Code repository: registration protocol and validation test objects & \url{https://github.com/OpenAgenet/oan-registrar-nodes} \\
\passthrough{\lstinline!oan-discovery-nodes!} & Code repository: discovery, indexing, and semantic-query test objects & \url{https://github.com/OpenAgenet/oan-discovery-nodes} \\
\passthrough{\lstinline!oan-trust-indexer!} & Code repository: governance-state and synchronization test objects & \url{https://github.com/OpenAgenet/oan-trust-indexer} \\
OAN Yellow Paper & arXiv yellow paper: security properties and verification requirements & \url{https://arxiv.org/abs/2606.03163} \\
Efficient Agent Discovery Profile & IETF draft: testing directions for the discovery profile & \url{https://datatracker.ietf.org/doc/draft-xu-efficient-agent-discovery-profile/} \\
\end{longtable}
}

\chapter{Reference Deployments and Examples}\label{chap:appendix-e}

This appendix provides reference paths that can be used for development, verification, and operations handover. The specification requires attention to role boundaries, evidence binding, state propagation, and interface semantics; the reference implementation uses Rust nodes, a TypeScript SDK, community skill packages, PostgreSQL/NATS, and a website gateway; specific domain names, ports, keys, database URLs, model directories, and system service names are all environment configuration. Deployment personnel should first confirm the trust root, authorized domains, node DIDs, and protocol version of the target network before writing configuration, and should not copy private keys, tokens, or online addresses from the examples. Server-side build artifacts, logs, caches, and database data are kept separate from the source archive, and secrets are injected only through protected runtime configuration.

When synchronizing versioned source archives, the archived content should come from a confirmed commit or release version, rather than unreviewed files in the working tree. The following command represents one general method for creating a source archive:

\begin{lstlisting}[language=bash]
git archive --format=tar --prefix=release/ <COMMIT_OR_TAG> | gzip -c > <RELEASE_ARCHIVE>.tar.gz
\end{lstlisting}

The archive represents source deliverables only; build directories, dependency caches, database files, runtime keys, and logs should not be mixed into the source package. Source code, build artifacts, configuration, runtime data, and website static files should be stored separately in controlled directories defined by the deployment environment, with their versions and digests associated through release records. Directory names, filesystem layout, and transfer methods are environment implementation details, not protocol requirements.

The recommended general startup chain is as follows:

\begin{lstlisting}
Governance state/node authorization
        |
        v
Root platform verification and publication -----> content distribution platform -----> Discovery node -----> Caller
        ^                        ^                 |
        |                        |                 v
Registrar node -----------------+           Structured/semantic query
        ^
        |
Resource controller, SDK, or community skill package
\end{lstlisting}

The startup order should first satisfy dependencies such as the database, message system, and off-chain trust indexer, then start the Root platform, Registrar node, content distribution components, and Discovery node, and finally start the website backend, static frontend, and reverse proxy. During shutdown or upgrade, new writes should be stopped first, and the publication and distribution queues should be allowed to reach an acceptable state before the program is replaced, while the database, cursors, version records, and redacted logs are retained. If a dependency is temporarily unavailable, the service should perform limited retries and backoff according to its own configuration, and report \passthrough{\lstinline!unknown!}, unavailable, or pending synchronization externally, rather than generating false publication or authorization facts.

\section{Official Network Deployment}\label{sec:appendix-e.1}

The reference layering of the official network is as follows: the Registrar node handles resource onboarding, the Root platform performs network-level verification, publication, and coordination, the content distribution platform transfers Root platform-verified ResourcePackages and proofs, the Discovery node establishes structured and semantic indexes within the authorized domain, the off-chain trust indexer projects governance events into queryable state, and the website backend provides the frontend with display and user-operation entry points through a public gateway. The website is not a trust authority; a website failure should not change the protocol state of a resource or node.

Production deployment should verify at least the following chain:

\begin{enumerate}
\def\labelenumi{\arabic{enumi}.}
\tightlist
\item
  The database, message system, and listening addresses of all nodes are ready, and the public Internet exposes only the public HTTPS entry protected by the reverse proxy.
\item
  The Root platform, Registrar node, Discovery node, content distribution platform, and off-chain trust indexer each return an interpretable status through \passthrough{\lstinline!/health!} or their role-status interface.
\item
  The Registrar node can complete controlled registration-request validation and submission to the Root platform; registration smoke tests should use dedicated test resources or controlled failure inputs and must not pollute the production resource directory.
\item
  Root platform publication results can be consumed by the content distribution platform, validated and indexed by the Discovery node, and then observed through exact and semantic queries.
\item
  The website frontend\textquotesingle s \passthrough{\lstinline!VITE\_*!} endpoints point to the current public gateway, and build artifacts contain no residual \passthrough{\lstinline!localhost!} or \passthrough{\lstinline!127.0.0.1!}; after website deployment, perform smoke tests for registration, discovery, Network, routing, and the browser console.
\end{enumerate}

Deployment records should retain the build version, configuration summary, service-switch time, health-check results, registration and discovery smoke-test results, index cursor, and failed-retry information. Public domains, node addresses, and certificates belong to the target environment configuration; when migrating to another environment, endpoints and certificates authorized by that environment must be used.

Each service should be released through a reversible replacement order: first verify the new binary and static artifacts, then switch the corresponding service through the target environment\textquotesingle s service manager, and finally check routing through the public entry point. Website static files may first be uploaded to a controlled temporary release directory; after checking the file count, entry file, and endpoint references, switch to the current version. For node services, retain the old binary, configuration, and runtime state before replacing the new version. During rollback, restore only a verified old version; do not reset the database, publication cursor, or governance state.

{\def\LTcaptype{none} 
\begin{longtable}[]{@{}>{\raggedright\arraybackslash}p{0.3200\linewidth}>{\raggedright\arraybackslash}p{0.3200\linewidth}>{\raggedright\arraybackslash}p{0.3200\linewidth}@{}}
\toprule\noalign{}
Layer & Typical observations & Handling boundary on failure \\
\midrule\noalign{}
\endhead
\bottomrule\noalign{}
\endlastfoot
Process layer & Service-manager status, listening ports, and startup logs & Fix the configuration or binary and restart; process liveness is not business availability \\
Interface layer & \passthrough{\lstinline!/health!}, node status, and public-route status codes & Distinguish reverse-proxy, upstream-service, and request-format problems \\
Business layer & Registration acceptance, Root platform publication, and discovery queries & Use controlled test data; block publication or roll back on failure \\
Data layer & Database connection, cursor, index statistics, and version records & Retain the scene and backups first; do not directly delete or rebuild business data \\
\end{longtable}
}

\section{Private and Alliance Network Deployment}\label{sec:appendix-e.2}

{\def\LTcaptype{none} 
\begin{longtable}[]{@{}>{\raggedright\arraybackslash}p{0.3200\linewidth}>{\raggedright\arraybackslash}p{0.3200\linewidth}>{\raggedright\arraybackslash}p{0.3200\linewidth}@{}}
\toprule\noalign{}
Configuration item & Private/consortium handling & Fact that must not change \\
\midrule\noalign{}
\endhead
\bottomrule\noalign{}
\endlastfoot
Trust root & Inject the organization\textquotesingle s or consortium\textquotesingle s Root DID, keys, and governance process securely & Node authorization remains verifiable and is not reduced to an IP allowlist \\
Resource visibility & Limit scope through authorization domains, Discovery partitioning, and reverse proxies & Visibility filtering must not relabel out-of-domain resources \\
Data distribution & Use an internal CDN, object store, or node synchronization & Receiver still verifies digest, version, and Root proof \\
Operations boundary & Keep databases, Trust Indexer, and management interfaces on controlled networks & Reachability gives no caller private keys or governance write permission \\
\end{longtable}
}

Private or alliance networks may deploy the Root platform, Registrar node, Discovery node, and content distribution platform within an organization\textquotesingle s internal or isolated network, but they should retain the same protocol responsibilities: the Root platform is the hub for network-level verification, trust authorization, data distribution, and semantic governance; the Registrar node is responsible for resource onboarding; and the Discovery node establishes indexes only for published materials. Network isolation changes reachability, not the rules for judging DIDs, control proofs, Root proofs, authorized domains, or lifecycle states.

Typical configuration boundaries are as follows:

When an alliance member joins, it should first submit its node identity, role, endpoint, authorized domain, and protocol-version materials, and then obtain authorization from the governance authority; revocation or suspension should be able to propagate to the Root platform and Discovery node. During integration, use an independent database and test domain to avoid accidentally synchronizing private-network test resources to the official public network.

The boundary of an alliance network should also appear in operations accounts and data backups: node operators may manage processes, configuration, and logs, but should not read the resource controller\textquotesingle s private key from application configuration; resource publishers submit only public DID Documents, proofs, and ResourcePackages and do not gain governance write access because the network is internal. During cross-member integration, record each member\textquotesingle s node DID, authorized domain, configuration version, and revocation status; after testing, clean up temporary credentials and test-resource indexes while retaining necessary audit summaries.

\section{Minimal Registrar Node Deployment}\label{sec:appendix-e.3}

The minimal Registrar node needs only its own DID Document and runtime key, resource-record storage, Root platform address, Root platform DID or audience, listening address, CORS/network policy, and necessary database configuration. The reference Rust implementation provides \passthrough{\lstinline!/health!}, \passthrough{\lstinline!/registrar/did!}, \passthrough{\lstinline!/resources/register!}, a compatible submission entry point, registration status, resource records, the capability-tag tree, tag suggestions, and authorization queries; the specific available interfaces should be determined by the source and configuration of the target version. The registration flow is: receive a ResourcePackage, check the DID and resource type, metadata, authorized domain, control proof, and hash, generate a signed request for the Root platform, wait for Root platform verification and publication, then save the registration record and return the registration credential and Root platform response.

Minimum verification should include:

\begin{lstlisting}[language=bash]
curl -fsS "${REGISTRAR_BASE_URL}/health"
curl -fsS "${REGISTRAR_BASE_URL}/registrar/did"
curl -X POST "${REGISTRAR_BASE_URL}/resources/register" \
  -H 'content-type: application/json' \
  --data-binary @registration-request.json
\end{lstlisting}

\passthrough{\lstinline!$\{REGISTRAR\_BASE\_URL\}!} is an example variable injected by the deployment environment, not a fixed address or port requirement. The Registrar node must not receive the resource controller\textquotesingle s private key; the private key is used in the resource controller\textquotesingle s SDK, browser, or local Skill, while the node receives only the public key, signature, and public proof. When the Root platform is unreachable, the node must return a distinguishable upstream failure and must not display local storage or acceptance as Root platform publication.

\section{Minimal Discovery Node Deployment}\label{sec:appendix-e.4}

The minimal Discovery node needs its own identity and authorization materials, a synchronization endpoint for the Root platform or content distribution platform, authorized domains, a database, index configuration, and a listening address. It first obtains or receives Root platform publication materials, verifies the Root proof, ResourcePackage hash, DID Document, version, and lifecycle, and then writes them into the local structured index; semantic retrieval may use the tag tree, text fields, and vector index on this basis, but must not skip publication and authorization verification. The reference implementation provides role interfaces such as \passthrough{\lstinline!/discovery/did!}, \passthrough{\lstinline!/discovery/resources/query!}, query explanations, query suggestions, the capability-tag tree, synchronization history, and index statistics.

The criterion for completed initial synchronization is not ``the HTTP request succeeded,'' but rather: the synchronization cursor has been recorded, the ResourcePackage\textquotesingle s source and hash checks have passed, resources within the authorized domain have entered the index, out-of-domain, suspended, revoked, or protocol-mismatched resources have been filtered, and index statistics are consistent with the synchronization record. It is recommended to use a fixed resource set to perform the following query:

\begin{lstlisting}
{
  "query": "Find an MCP server for security audit",
  "resourceType": "mcp_server",
  "limit": 10
}
\end{lstlisting}

Returned results should distinguish matching sources such as text, tags, structured conditions, and semantic relevance, and retain the resource DID, version, status, source, and freshness fields. If the Root platform or content distribution platform is temporarily unavailable, a known old snapshot may be displayed, but it must not be used to produce a new trusted publication or authorization conclusion.

The synchronization status of the Discovery node should be distinguished according to the following table for shared use by the Network page, operations interfaces, and alerting rules:

{\def\LTcaptype{none} 
\begin{longtable}[]{@{}>{\raggedright\arraybackslash}p{0.3200\linewidth}>{\raggedright\arraybackslash}p{0.3200\linewidth}>{\raggedright\arraybackslash}p{0.3200\linewidth}@{}}
\toprule\noalign{}
Status & Meaning & Permitted action \\
\midrule\noalign{}
\endhead
\bottomrule\noalign{}
\endlastfoot
\passthrough{\lstinline!ready!} & The cursor is advancing, the index is queryable, and the most recent verification succeeded & Allow normal queries \\
\passthrough{\lstinline!syncing!} & Full or incremental synchronization is in progress & Query confirmed data and mark the synchronization status \\
\passthrough{\lstinline!stale!} & An old index remains, but it has exceeded the freshness boundary configured for the environment & Degrade the display or block sensitive invocations \\
\passthrough{\lstinline!unknown!} & The upstream, cursor, or governance status cannot be confirmed & Must not be used to confirm that a resource is trusted or callable \\
\end{longtable}
}

These statuses are observations at the operations and presentation layers, not substitutes for resource lifecycle states. The specific freshness threshold should be provided by the deployment environment and operations policy; it cannot be inferred as a fixed protocol delay from one query response.

The minimal Registrar node configuration can be abstracted as the following environment-configuration example; all values are placeholders and cannot be used directly in production:

\begin{lstlisting}
REGISTRAR_LISTEN_ADDR=<LISTEN_ADDRESS>
ROOT_ENDPOINT=<AUTHORIZED_ROOT_ENDPOINT>
DATABASE_URL=<DATABASE_CONNECTION_REFERENCE>
NODE_DID=<NODE_DID>
AUTHORIZED_DOMAINS=<AUTHORIZED_DOMAIN_SET>
\end{lstlisting}

After startup, at least three results should be recorded separately: ``node reachable,'' ``request validated locally,'' and ``request submitted to the Root platform.'' The latter two must not be ambiguously represented by the same HTTP 200 status code; if the Root platform times out or returns an unknown result, the Registrar node should retain retryable request-correlation information and mark the result as unknown or pending.

\section{SDK Registration Example}\label{sec:appendix-e.5}

The SDK example should show that the private key is used only locally by the caller, that the DID Document is bound to the ResourcePackage, that the registration request is accepted by the designated Registrar node, and that the registration result requires continued observation of publication and discovery status. The following code shows the interface relationship; endpoints, domain names, and resource content must be replaced according to the environment:

\pandocbounded{\includegraphics[width=\linewidth,height=0.72\textheight,keepaspectratio]{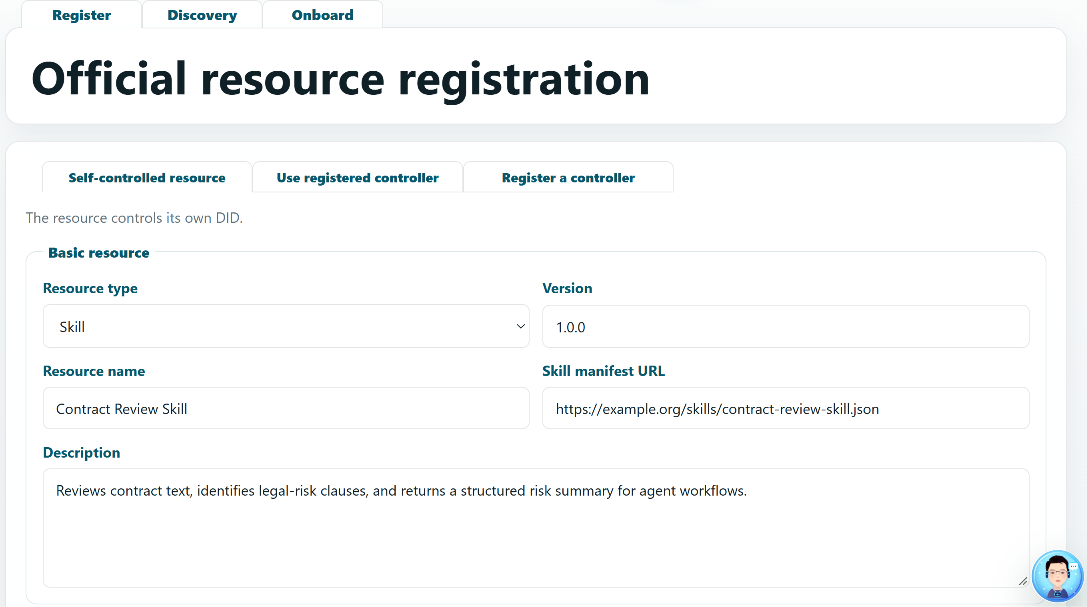}}

This snapshot shows how the official registration page collects resource type, description, capabilities, and endpoint input. The form is not itself a protocol object. Before submission, the local SDK still generates the DID Document, digests, and control proof; the private key must not be uploaded to the web backend.

\begin{lstlisting}
import {
  createAgentIdentity,
  createRegistrationSubmissionFromIdentity,
  finalizeRegistrationSubmissionWithProof,
} from "@openagenet/oan-sdk-ts";
import { OanClient } from "@openagenet/oan-sdk-ts/client";

const registrarDid = process.env.OAN_REGISTRAR_DID!;
const identity = await createAgentIdentity(
  "Repository structure summarizer",
  "skill",
  undefined,
  {
    registrarDid,
    authorizedDomains: ["technology.software_engineering"],
    description: "Searches code repositories and summarizes project structure.",
    capabilityTags: ["code.repository", "analysis.summary"],
  },
);

const draft = createRegistrationSubmissionFromIdentity(identity, {
  endpoint: process.env.OAN_RESOURCE_ENDPOINT!,
  packageVersion: "1.0.0",
});
const submission = await finalizeRegistrationSubmissionWithProof(draft, {
  resourceIdentity: identity,
  controllerIdentity: identity,
  registrarDid,
});

const client = new OanClient({
  registrarEndpoint: process.env.OAN_REGISTRAR_ENDPOINT!,
});

const result = await client.registerResource(submission);
console.log(result.status, result.resourceDid, result.registrationCredential);
\end{lstlisting}

\passthrough{\lstinline!finalizeRegistrationSubmissionWithProof()!} prepares canonical material, computes the bound digests, and generates the subject control proof under the current model. The caller must not enter fictitious hashes to simulate success. The caller retains the registration VC, Root response, and resource DID and continues checking their states; the SDK neither holds private keys for the caller nor grants business permission at the resource endpoint.

When processing registration results, network errors, Registrar rejection, Registrar acceptance with an unknown Root platform result, and Root platform publication should be recorded separately. The example code shows only result branches and does not treat a particular field of \passthrough{\lstinline!result!} as a fixed field present in every version:

\begin{lstlisting}
try {
const result = await client.registerResource(submission);
console.log("registrar response received", result);
  // The caller continues to read and verify the VC, Root platform publication status, and discovery visibility.
} catch (error) {
  console.error("registration transport or client error", error);
  // Do not automatically generate a new DID or resubmit unconditionally when the result cannot be confirmed.
}
\end{lstlisting}

Real integration should securely retain the request-correlation identifier, resource DID, and Local Identity backup; error logs may contain only redacted summaries and must not output signing private keys, the complete Local Identity backup, or database connection strings.

\section{DID and Semantic Discovery Examples}\label{sec:appendix-e.6}

DID exact queries are used to confirm whether a known resource is visible in the authorized view of a specified Discovery node; semantic queries are used to find candidate resources from a task description. Both should return the candidate resource\textquotesingle s DID, resource type, version, status, capability tags, authorized domain, endpoint or endpoint reference, source, and verification summary. A caller must not skip DID Document, Root proof, VC, lifecycle, or endpoint verification because of a high semantic-match score, a tag hit, or a recommendation displayed by the website.

\pandocbounded{\includegraphics[width=\linewidth,height=0.72\textheight,keepaspectratio]{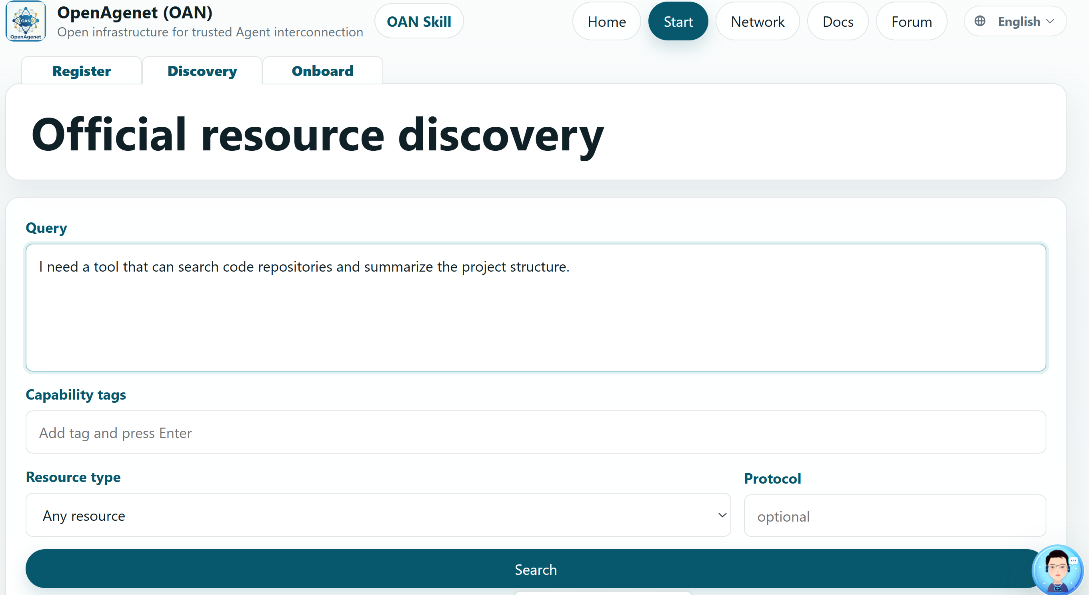}}

This snapshot shows the structured and semantic discovery entry points and candidate presentation. A page hit means only that the Discovery service returned a candidate at that time; the caller still verifies the candidate version, lifecycle, Root proof, endpoint, and business authorization before invocation.

\begin{lstlisting}
const exact = await client.getDiscoveryResource(
  "did:oan:K7mQ9:7YpQm9Kx2VnRb6Ts3WfHa4Cd5Ej8LgNz",
);

const semantic = await client.suggestDiscoveryQuery({
  query: "Find an MCP server for security audit",
});
const candidates = await client.discoverResources({
  query: "Find an MCP server for security audit",
  resourceType: "mcp_server",
  capabilityTags: semantic.capabilityTags,
  limit: 10,
});
\end{lstlisting}

If the implementation supports query explanations, the caller should further read the explain result to describe how the query was converted into structured conditions, tags, or semantic conditions, and why candidates were retained or filtered. When there is no match, return a stable empty result; overly short queries, out-of-domain conditions, or invalid filter conditions should return controlled errors. Pagination and deduplication should follow DID and version rules; multiple indexed copies of the same resource must not be displayed as multiple independent resources.

Exact and semantic queries may use the same discovery entry point, but their acceptance assertions differ: exact queries focus on uniqueness, complete DID matching, and version selection; semantic queries focus on candidate sources, filtering reasons, and ranking explanations. A minimal manual acceptance record may be written as follows:

Here, \passthrough{\lstinline!example-identifier!} is only a placeholder. If the result contains only text similarity and no verifiable source, governance state, or proof summary, it should be treated as a candidate recommendation and not directly as a trusted callable object.

\section{Resource Verification Example}\label{sec:appendix-e.7}

After receiving a discovery result, the caller should verify it in the order of ``identity, binding, source, governance, endpoint'': first parse \passthrough{\lstinline!did:oan!} and check the public keys, controllers, and service references in the DID Document; then check the resource DID, type, version, DID Document hash, metadata hash, and package hash in the ResourcePackage; next verify the Root proof and its lifecycle status; then check the issuer, subject, authorized domain, hash bindings, and proof of the registration VC; finally verify the endpoint, protocol, certificate, and invocation permissions according to business policy. Any failure should block high-risk invocation and retain a redacted failure reason.

\begin{lstlisting}
import { verifyResourcePackageShape, assertUsableLifecycle } from "@openagenet/oan-sdk-ts";

verifyResourcePackageShape(resourcePackage);
assertUsableLifecycle(resourcePackage);
// Continue with the caller's own VC, node-authorization, and endpoint-verification policies.
\end{lstlisting}

The SDK\textquotesingle s structural validation functions cannot replace governance-state reads, network-freshness judgments, or business authorization. Verification results should record the version, source URL, hash, issuer, status time, and verification-component version on which they rely; private keys, tokens, and complete sensitive requests must not be recorded.

\section{Operations Smoke-Test Example}\label{sec:appendix-e.8}

Post-deployment smoke tests should run in dependency order and record ``service reachable,'' ``protocol action succeeded,'' and ``state converged'' separately. The recommended check sequence is: health interfaces and ports; database connection and migration status; Root platform status and governance/authorization reads; Registrar node health and controlled registration request; content distribution queue and publication status; Discovery node synchronization cursor, index statistics, exact query, and semantic query; latest cursor of the off-chain trust indexer; website backend health, frontend routes, registration/discovery transport, and browser console. The deployment process should be considered complete only after the public endpoint, registration transport, discovery query, Network summary, and browser-route checks pass.

A smoke-test record may use the following structure:

\begin{lstlisting}
{
  "deploymentId": "2026-09-07T12:00:00Z-<commit>",
  "environment": "<environment>",
  "checks": [
    {"name": "root-health", "status": "pass", "observedAt": "..."},
    {"name": "registration-transport", "status": "pass", "resourceDid": "..."},
    {"name": "discovery-query", "status": "pass", "candidateCount": 1},
    {"name": "trust-indexer-freshness", "status": "unknown", "reason": "..."}
  ]
}
\end{lstlisting}

The DIDs, times, and statuses in the example illustrate a record format. When a smoke test fails, first determine whether the cause is network transport, build artifacts, configuration, database, governance state, synchronization delay, or a functional error, and then decide whether to retry, roll back, or block public release; do not obtain superficial success by deleting logs, resetting cursors, or skipping registration/discovery checks.

\section{Same-DID Update Example}\label{sec:appendix-e.9}

A same-DID update must be generated by the original resource controller or an explicitly authorized controller using a new control proof; the DID itself remains unchanged, while updated content in the ResourcePackage or DID Document forms a new version and hash. The process is: read the Local Identity and current version, modify metadata or the ResourcePackage, increment the version according to the applicable versioning scheme, recalculate the metadata and package hashes, sign the update materials, submit them to the Registrar node, have the Root platform re-verify them and form a new Root proof, and then synchronize through the content distribution platform and Discovery node. The Discovery node responds with the latest available version by default, but must retain auditable version and status information.

Update tests should cover a missing control proof, a control proof corresponding to old materials, version rollback, same-version content conflicts, hash inconsistency after an update, and republication of an old version. A failed update must not overwrite the current valid version, and a Discovery node must not treat a VC returned by the Registrar node as a new version; the registration credential, Root proof, and discovery status should be recorded separately.

The comparison record after an update is complete should contain at least the old and new resource DIDs, DID Document hashes, ResourcePackage hashes, control-proof summary, Root proof summary, discovery-index version, and query time. If the content has not changed, the client may reuse an existing preview or idempotent result; if any field affecting a binding relationship changes, the hashes must be recalculated and re-verified, and the display-only version string must not be changed by itself.

\section{Resource Suspension and Recovery Example}\label{sec:appendix-e.10}

Suspension, revocation, and recovery are lifecycle facts; they are not equivalent to deleting a DID or erasing history. The governance authority or resource controller generates a state change through the permitted governance process, the Root platform and off-chain trust indexer read and project the fact, the Discovery node filters unavailable resources after synchronization, and the caller rechecks the status during verification. Suspension usually means temporarily unavailable, while revocation means that the current credential or resource status is no longer available; recovery must have a clear basis and the Discovery node must not independently change the resource back to \passthrough{\lstinline!active!}.

\begin{lstlisting}
State-change fact -> governance event/authorization materials -> Root platform and indexer read
                  -> Discovery node updates filtering state -> caller re-verifies before deciding whether to invoke
\end{lstlisting}

Audit records should include the resource DID, original status, new status, source of the fact, event or request identifier, observation time, propagation time, and execution result. When governance status has not synchronized or has exceeded the permitted freshness, sensitive invocation should fail closed; the website or an old snapshot may display ``pending synchronization,'' but must not display the resource as currently callable.

\section{Cross-Protocol A2A and MCP Mapping Example}\label{sec:appendix-e.11}

The goal of cross-protocol mapping is to reduce the cost for existing Agent platforms, MCP clients, and tool callers to integrate with OpenAgenet (OAN), not to force objects from different protocols to be treated as the same object. During mapping, the names, capability tags, use cases, inputs, outputs, service endpoints, protocol bindings, and versions in the OAN DID Document and resource description may be mapped to an A2A Agent Card, MCP service metadata, or tool API description; during reverse conversion, the original DID, original fields, version, source, and hash must be retained. An A2A Agent Card or MCP metadata describes capabilities and connection methods, but does not replace OAN control proofs, VCs, Root proofs, governance status, or authorized domains.

\begin{lstlisting}
OAN DID Document/ResourcePackage
  - id, resourceType, capabilityTags, useCases
  - protocolBindings, service endpoint, version, hashes
        |
        +--> A2A Agent Card: identity reference, skills, interfaces
        +--> MCP metadata: server capabilities, tools, transport
        +--> Tool/API description: operation, schema, endpoint
\end{lstlisting}

The mapper should make the source and conversion direction of each field explicit: capability names may be used for display and retrieval, DIDs and hashes are used for identity and integrity traceability, signatures and VCs are retained as independent verification materials, and endpoint mapping represents only a connection entry point and does not automatically authorize invocation. If the target protocol has no equivalent field, use an extension field or an additional reference, and indicate lost information in the result; semantic similarity, protocol compatibility, or successful metadata conversion must not be interpreted as proof that the resource is trusted or usable for business purposes.

The mapping relationship below only describes how information may be carried; it does not require A2A Agent Cards, MCP metadata, or tool API descriptions to use OAN field names:

{\def\LTcaptype{none} 
\begin{longtable}[]{@{}>{\raggedright\arraybackslash}p{0.3200\linewidth}>{\raggedright\arraybackslash}p{0.3200\linewidth}>{\raggedright\arraybackslash}p{0.3200\linewidth}@{}}
\toprule\noalign{}
OAN information & Possible location in the external description & Retention requirement \\
\midrule\noalign{}
\endhead
\bottomrule\noalign{}
\endlastfoot
Resource DID, DID Document source & Identity reference or extension field & Retain the original DID and source; do not rewrite it as an ordinary name \\
Capability tags, use cases, and description & Agent Card skills, MCP capabilities, or tool description & May be used for display and retrieval, but cannot replace proofs \\
Protocol bindings, endpoints, and inputs/outputs & Interface, transport, and schema descriptions & Retain the version and source; endpoint availability requires separate verification \\
VC, hashes, and Root proof & Verification attachment or reference in the external object & Do not delete or overwrite the original proofs because conversion succeeded \\
\end{longtable}
}

If a mapped object re-enters the OAN registration or discovery flow, DID, hash, control-proof, authorized-domain, and lifecycle checks must be performed again; an external protocol object may serve as onboarding material, but cannot independently become an OAN publication fact.

\section{References}\label{sec:appendix-e.12}

{\def\LTcaptype{none} 
\begin{longtable}[]{@{}>{\raggedright\arraybackslash}p{0.3200\linewidth}>{\raggedright\arraybackslash}p{0.3200\linewidth}>{\raggedright\arraybackslash}p{0.3200\linewidth}@{}}
\toprule\noalign{}
Source & Type & Link \\
\midrule\noalign{}
\endhead
\bottomrule\noalign{}
\endlastfoot
\passthrough{\lstinline!oan-root-nodes!} & Code repository: Root platform deployment and publication examples & \url{https://github.com/OpenAgenet/oan-root-nodes} \\
\passthrough{\lstinline!oan-registrar-nodes!} & Code repository: Registrar node deployment and registration examples & \url{https://github.com/OpenAgenet/oan-registrar-nodes} \\
\passthrough{\lstinline!oan-discovery-nodes!} & Code repository: Discovery node deployment, synchronization, and query examples & \url{https://github.com/OpenAgenet/oan-discovery-nodes} \\
\passthrough{\lstinline!oan-trust-indexer!} & Code repository: trust indexer deployment and governance synchronization examples & \url{https://github.com/OpenAgenet/oan-trust-indexer} \\
\passthrough{\lstinline!oan-sdk-ts!} & Code repository: client integration examples & \url{https://github.com/OpenAgenet/oan-sdk-ts} \\
\passthrough{\lstinline!oan-community-skill!} & Code repository: community registration and discovery operation examples & \url{https://github.com/OpenAgenet/oan-community-skill} \\
OAN Yellow Paper & arXiv yellow paper: reference deployment and end-to-end mechanisms & \url{https://arxiv.org/abs/2606.03163} \\
OAN Resource Identity and Discovery & IETF draft: boundaries of resource identity and discovery examples & \url{https://datatracker.ietf.org/doc/draft-xu-oan-resource-identity-discovery/} \\
\end{longtable}
}

\chapter{Repository and Implementation Mapping}\label{chap:appendix-f}

This appendix describes implementation through the chain ``specification clause -\textgreater{} data structure/interface -\textgreater{} implementation file -\textgreater{} test evidence -\textgreater{} deployment configuration -\textgreater{} maintenance responsibility'', rather than treating the repository name itself as proof of functionality. The current core runtime implementation consists of the shared protocol library, Root platform, Registrar node, Discovery node, content distribution platform, and off-chain trust indexer; the SDK, community skill package, and official website are client and public entry points; the third-party node admission test kit provides independent evaluation support and is not a production runtime node. \passthrough{\lstinline!oan-onboard-tests!} and the root-level \passthrough{\lstinline!genesis!} are not independently verifiable directories in the current workspace. Therefore, historical references involving them can only be treated as source-material conventions and cannot serve as evidence of the current implementation.

Mapping statuses have the following meanings:

{\def\LTcaptype{none} 
\begin{longtable}[]{@{}>{\raggedright\arraybackslash}p{0.4800\linewidth}>{\raggedright\arraybackslash}p{0.4800\linewidth}@{}}
\toprule\noalign{}
Status & Meaning \\
\midrule\noalign{}
\endhead
\bottomrule\noalign{}
\endlastfoot
Implemented & The corresponding logic can be located in the current code, tests, or configuration, and can be reproduced or checked through the specified method. \\
Partially implemented & Data structures, interfaces, or partial workflows exist, but end-to-end, production-grade, or complete failure-scenario coverage is still missing. \\
Evaluation support & Provided by fixed fixtures, report generators, or verification scripts; does not mean that the runtime node has implemented the capability. \\
Reference / to be verified & Required by the specification or proposed by the deployment design, but the current workspace lacks sufficient code or test evidence for confirmation. \\
\end{longtable}
}

Each implementation fact should record both the source version and the verification time. For inputs, explain how they are converted into validation conditions, authorization-domain filters, or query conditions; for outputs, explain whether they are raw facts, verification results, cached projections, search rankings, or display summaries; for Root proof, VCs, signatures, and governance status, the issuer, verifier, bound content, and status source must be retained. Capability labels, semantic relevance, A2A Agent Card, or MCP metadata only describe or assist capability discovery and cannot replace control and governance verification. Mapping records must distinguish at least three types of evidence: source-code evidence explains where the logic and fields are defined, test evidence explains what result the logic produces for fixed inputs, and runtime evidence explains what the service, configuration, and dependencies can actually complete in the specified environment. They answer ``what was written'', ``what was tested'', and ``what was run'', respectively, and cannot replace one another with a README, a single successful compilation, or one health endpoint. For dynamic data, also record the commit, configuration summary, database/index state, and request correlation identifier at the time the evidence was generated.

\section{\texorpdfstring{\texttt{oan-protocol-commons}}{oan-protocol-commons}}\label{sec:appendix-f.1}

This repository provides shared Rust workspace-level foundations and primarily maps to Chapters~\ref{chap:04}~through~\ref{chap:08}, Chapter~\ref{chap:15}, and Appendices~\ref{chap:appendix-a}~through~\ref{chap:appendix-c}. \passthrough{\lstinline!crates/oan-core!} carries DID Document, resource types, and core domain objects; \passthrough{\lstinline!oan-crypto!} handles cryptographic suites, keys, and hash-related logic; \passthrough{\lstinline!oan-credentials!} handles credential construction or verification; \passthrough{\lstinline!oan-package!} handles ResourcePackage and its bindings; \passthrough{\lstinline!oan-protocol!} handles cross-node requests, responses, and protocol constants; \passthrough{\lstinline!oan-publication-events!} handles publication events; \passthrough{\lstinline!oan-storage!} provides JSON, SQLite, or PostgreSQL storage abstractions; \passthrough{\lstinline!oan-service-security!} provides shared logic for request envelopes, nonce, request ID, and signatures; \passthrough{\lstinline!oan-semantic-recommender!} and its \passthrough{\lstinline!data!} directory provide capability-label trees, authorization-domain classification, and recommendation/normalization assistance.

{\def\LTcaptype{none} 
\begin{longtable}[]{@{}>{\raggedright\arraybackslash}p{0.2400\linewidth}>{\raggedright\arraybackslash}p{0.2400\linewidth}>{\raggedright\arraybackslash}p{0.2400\linewidth}>{\raggedright\arraybackslash}p{0.2400\linewidth}@{}}
\toprule\noalign{}
Mapping object & Main location & Corresponding specification facts & Current verification method and boundary \\
\midrule\noalign{}
\endhead
\bottomrule\noalign{}
\endlastfoot
DID Document and resource types & \passthrough{\lstinline!crates/oan-core/src/lib.rs!} & \passthrough{\lstinline!did:oan!} identifiers, subject types, and core fields & Rust unit tests and calls from each node; complete coverage still requires confirmation through specification-pattern and method tests. \\
Hashes, signatures, and cryptographic suites & \passthrough{\lstinline!crates/oan-crypto/src/lib.rs!} & Control proofs, request signatures, package/metadata hashes & Valid/tampered-signature fixtures in \passthrough{\lstinline!oan-third-party-node-admission-test-kit!}; not equivalent to key custody or a production HSM guarantee. \\
Credentials and ResourcePackage & \passthrough{\lstinline!crates/oan-credentials!}, \passthrough{\lstinline!crates/oan-package!} & Structural bindings for registration VC, ResourcePackage, and Root proof & Subject, hash, and proof consistency checks in registration/admission fixtures; the real path requires service integration verification. \\
Protocol types and errors & \passthrough{\lstinline!crates/oan-protocol!}, \passthrough{\lstinline!docs/error-codes.md!}, \passthrough{\lstinline!schemas/agent-contract!} & HTTP payloads, protocol versions, error contracts, and external descriptions & Node compilation, client tests, and fixed request fixtures; version compatibility must be retested by release version. \\
Storage and events & \passthrough{\lstinline!crates/oan-storage!}, \passthrough{\lstinline!crates/oan-publication-events!} & Resource records, cursors, publication events, and database boundaries & Tests for each service and local integration; transaction, backup, and recovery strategies cannot be proved by the shared library alone. \\
Semantic and authorization-domain assistance & \passthrough{\lstinline!crates/oan-semantic-recommender/data/*!} & Capability-label trees, semantic governance vocabulary, and authorization-domain classification & Recommender tests and Discovery/Registrar consumer tests; recommendation output is not an authorization decision. \\
\end{longtable}
}

The shared library's maintenance responsibility is to keep data structures, serialization, error codes, and cryptographic bindings consistent across nodes. Any field change should trigger synchronized checks of the Root platform, Registrar node, Discovery node, SDK, community skill package, and fixed fixtures. The shared library does not have authority to decide node authorization, resource publication, or business calls on behalf of upper-layer services.

When tracing downward from an HTTP interface, the contract can be located through the implementation chain: \passthrough{\lstinline!services/registrar-node/src/main.rs!} in the Registrar node binds \passthrough{\lstinline!/resources/register!} and \passthrough{\lstinline!/resources/submit!} to \passthrough{\lstinline!register\_resource!}; \passthrough{\lstinline!services/root-node/src/main.rs!} in the Root platform binds \passthrough{\lstinline!/root/resources/verify-and-publish!} to \passthrough{\lstinline!verify\_resource\_and\_publish!}; the Discovery node's service directory of the same name binds \passthrough{\lstinline!/discovery/resources/query!} to \passthrough{\lstinline!resource\_query!}; \passthrough{\lstinline!services/cdn-node/src/main.rs!} in the content distribution platform handles \passthrough{\lstinline!/cdn/resources!}, batch publication, and resource reads; and \passthrough{\lstinline!src/api.rs!} in the off-chain trust indexer handles read interfaces such as \passthrough{\lstinline!/v1/status!} and \passthrough{\lstinline!/v1/events!}. A handler signature is direct evidence for confirming input and output types, while a route table is direct evidence for confirming HTTP methods and paths.

When maintaining interfaces, record ``shared DTO changes'' separately from ``dynamic response-field changes'': the former requires checking \passthrough{\lstinline!oan-protocol!}, \passthrough{\lstinline!oan-package!}, all deserialization callers, and cross-language SDKs; the latter requires checking the handler that generates the JSON, error-code tests, public proxies, and the endpoint table in Appendix~\ref{chap:appendix-b}. Successful Rust compilation only proves that the type layer compiles; it does not prove that JSON fields, HTTP paths, authentication boundaries, and asynchronous state semantics still match the external documentation. When locating a shared field, maintainers should verify it in the order ``domain structure -\textgreater{} serialization implementation -\textgreater{} callers -\textgreater{} fixture''. For example, first confirm the ResourcePackage structure and validation methods in \passthrough{\lstinline!oan-package!}, then check the request/response types in \passthrough{\lstinline!oan-protocol!} and the deserialization paths in the Registrar node and Discovery node, and finally use resource fixtures to verify behavior for missing, tampered, and version-conflicting inputs. The fact that a single node compiles after modification does not show that the cross-repository protocol remains compatible; at minimum, the corresponding SDK exports and the community skill package's inputs and outputs must also be checked. When a field in the shared library changes, maintainers should first determine whether the change is a serialization-compatibility change, a validation-semantics change, or a purely internal implementation change. Serialized fields, enum values, error codes, and hash calculation methods belong to the cross-repository impact surface; even if the Rust workspace compiles, node HTTP payloads, TypeScript types, community skill package scripts, and admission-test fixtures must still be checked. If only an internal helper function changes and external payloads and error behavior remain unchanged, the relevant unit tests should be retained as boundary evidence, without expanding the protocol-version scope.

\section{\texorpdfstring{\texttt{oan-root-nodes}}{oan-root-nodes}}\label{sec:appendix-f.2}

This repository corresponds to Chapter~\ref{chap:03}, Chapters~\ref{chap:09}~through~\ref{chap:10}, Chapters~\ref{chap:13}~through~\ref{chap:15}, and the related lifecycle, distribution, and operations chapters. The main Root platform service is located at \passthrough{\lstinline!services/root-node/src/main.rs!}. Its routes include the root DID Document, status, Registrar node and Discovery node directories, resource details and versions, capability-label trees, announcements/events, authorization, revocation, and internal interfaces for the content distribution platform and Discovery node. \passthrough{\lstinline!services/cdn-publisher!} consumes publication tasks, prepares ResourcePackage from the Root platform, publishes them to the CDN in batches, and writes back publication results; \passthrough{\lstinline!services/cdn-node!} provides the data-plane service for the content distribution platform. Configuration examples are located in each service directory's \passthrough{\lstinline!config.example.toml!}.

The Root platform implementation boundary can be checked as follows:

{\def\LTcaptype{none} 
\begin{longtable}[]{@{}>{\raggedright\arraybackslash}p{0.2400\linewidth}>{\raggedright\arraybackslash}p{0.2400\linewidth}>{\raggedright\arraybackslash}p{0.2400\linewidth}>{\raggedright\arraybackslash}p{0.2400\linewidth}@{}}
\toprule\noalign{}
Root platform action & Input & Output or persisted fact & Evidence and deviation \\
\midrule\noalign{}
\endhead
\bottomrule\noalign{}
\endlastfoot
Node authorization and status reads & Node DID Document, authorization materials, governance/status data & Authorized-node directory, status, and authorization domain & Root service routes and authorization-handling logic; on-chain finality depends on the off-chain trust indexer and runtime configuration. \\
Resource verification and publication & Registrar node submission, resource DID Document, ResourcePackage, control proof, and hash & Accept/reject result, resource version, Root proof, or publication task & Root service tests and shared types; service integration tests should confirm that failures do not produce an erroneous published state. \\
Content distribution coordination & Verified resource version, distribution task, and cursor & CDN publication queue, batch response, and published marker & Queue, batching, and write-back logic in \passthrough{\lstinline!cdn-publisher/src/main.rs!}; not equivalent to external reachability of the CDN. \\
Semantic governance & Capability-label tree, aliases, authorization-domain classification & Semantic materials that Registrar and Discovery nodes can synchronize & \passthrough{\lstinline!oan-semantic-recommender/data!} and Root platform interfaces; vocabulary evolution requires versioning. \\
\end{longtable}
}

Root platform maintainers must also monitor operational boundaries involving idempotency, version conflicts, publication retries, queue backlogs, database backups, and loss of governance-status connectivity. The Root platform issues Root proof, which proves only facts within its verification and publication scope; it does not prove the business quality of resource endpoints and does not hold resource control private keys. Root platform code review should record public query routes separately from internal write routes. \passthrough{\lstinline!/root/resources/\{did\}!}, version queries, and capability-label queries are read paths; Registrar node submissions, node authorization, revocation, content distribution queues, and Discovery node notifications are protected write or coordination paths. When modifying \passthrough{\lstinline!services/root-node/src/main.rs!}, maintainers should also check publication-task consumption in \passthrough{\lstinline!services/cdn-publisher/src/main.rs!}, ResourcePackage reads in \passthrough{\lstinline!services/cdn-node/src/main.rs!}, and the queues, database, and upstream endpoints in Root platform configuration; a Root platform HTTP response alone cannot establish that content distribution has completed. The key verification point for Root platform changes is when the publication fact is created. Maintainers should record the Registrar node submission time, Root platform verification result, resource version and hash, Root proof, content distribution task status, and Discovery node notification status separately. Only after Root platform verification has produced a publication fact do the subsequent distribution and discovery paths have a citable root source. Root platform query interfaces, content distribution platform caches, and Discovery node index entries should be retained as separate evidence.

\section{\texorpdfstring{\texttt{oan-registrar-nodes}}{oan-registrar-nodes}}\label{sec:appendix-f.3}

This repository primarily corresponds to Chapters~\ref{chap:05}~through~\ref{chap:08}, Chapter~\ref{chap:15}, and the resource lifecycle section. Its implementation entry point is \passthrough{\lstinline!services/registrar-node/src/main.rs!}. Current routes include \passthrough{\lstinline!/health!}, \passthrough{\lstinline!/registrar/did!}, \passthrough{\lstinline!/resources/register!}, the compatible \passthrough{\lstinline!/resources/submit!}, resource lists and details, status, statistics, Root platform authorization, capability-label trees, label suggestions, label normalization, registration-domain directories, and registration recommendations. The configuration example is located at \passthrough{\lstinline!services/registrar-node/config.example.toml!}; data directories, key directories, and optional SQLite/PostgreSQL databases are determined after configuration parsing.

The Registrar node processing chain is:

\begin{lstlisting}
Resource controller submits public materials and control proof
  -> validate shape, DID, resource type, metadata, and authorization domain
  -> construct a signed request for the Root platform
  -> receive the Root platform verification/publication response
  -> generate and store the registration VC and registration record
  -> return the registration result and Root platform response to the caller
\end{lstlisting}

The registration VC here is issued by the Registrar node for the registration fact and cannot replace Root proof. When the Root platform rejects a request, the Registrar must not treat local acceptance as network-wide publication. The Registrar node can see public resource materials, public keys, signatures, and credentials, but the normal flow must not obtain the resource control private key. \passthrough{\lstinline!oan-third-party-node-admission-test-kit/fixtures/v1-alpha/resources!} provides fixed verification examples for DID, resource type, authorization domain, VC subject, hash, control challenge, proof integrity, and lifecycle; SDK and official website tests cover client-call shapes but do not replace server-side validation by the Registrar.

Maintenance priorities are to keep the fields of \passthrough{\lstinline!/resources/register!} consistent with the protocol types, make error codes and HTTP statuses explicit, ensure that upstream failures do not create false success, prevent duplicate submissions from producing conflicting records, and update configuration examples, public error documentation, the SDK, and the community skill package together. Registration issue diagnosis should follow request correlation: first confirm whether the request was rejected locally at the input-validation and error-conversion points of \passthrough{\lstinline!register\_resource!}; then check whether \passthrough{\lstinline!ResourceVerifyAndPublishRequest!} for the Root platform was constructed correctly; finally compare the registration record, VC, and Root platform response to determine whether the result was only local acceptance. A reusable diagnostic record should retain at least the resource DID, request identifier, version, error code, upstream status, and database-write result; ``the page displayed success'' must not replace the actual result between nodes. The Registrar node's implementation mapping should also state which layer each result belongs to: local request validation, registration-record write, submission to the Root platform, registration VC return, and Root platform publication result must not be collapsed into one ``registration succeeded'' outcome. For retries after network timeouts or incomplete responses, use the original request's correlation information to check whether the server has already produced a registration fact; only after confirming idempotency conditions and final state can the client decide whether to submit again.

\section{\texorpdfstring{\texttt{oan-discovery-nodes}}{oan-discovery-nodes}}\label{sec:appendix-f.4}

This repository corresponds to Chapter~\ref{chap:03}, Chapter~\ref{chap:06}, Chapters~\ref{chap:10}~through~\ref{chap:12}, Chapters~\ref{chap:14}~through~\ref{chap:15}, and freshness and lifecycle requirements. \passthrough{\lstinline!services/discovery-node/src/main.rs!} provides interfaces for node DID, status, authorization domain, synchronization history, index statistics, query statistics, resource index, resource query, query suggestions, query explanations, rejection packages, and capability-label trees; \passthrough{\lstinline!semantic\_aliases.v1.json!} contains Discovery-side semantic alias materials. \passthrough{\lstinline!services/embedding-service!} provides an optional text-vector service. Its configuration, tests, and real-model smoke files are located in that service directory; enabling a semantic model must not be written as a mandatory dependency for every deployment.

The Discovery node data path should be distinguished into four layers:

\begin{enumerate}
\def\labelenumi{\arabic{enumi}.}
\tightlist
\item
  The synchronization layer receives materials from the Root platform or content distribution platform and records the source, cursor, and synchronization result.
\item
  The verification layer checks Root proof, ResourcePackage hashes, DID Document, version, authorization domain, and lifecycle.
\item
  The indexing layer writes verified materials into a structured index and builds retrieval indexes by capability labels, text fields, or optional vectors.
\item
  The query layer converts DID, structured conditions, and natural-language input into conditions and returns candidates with source, version, status, and explanation information.
\end{enumerate}

{\def\LTcaptype{none} 
\begin{longtable}[]{@{}>{\raggedright\arraybackslash}p{0.3200\linewidth}>{\raggedright\arraybackslash}p{0.3200\linewidth}>{\raggedright\arraybackslash}p{0.3200\linewidth}@{}}
\toprule\noalign{}
Implementation point & Code/material & Boundary to preserve \\
\midrule\noalign{}
\endhead
\bottomrule\noalign{}
\endlastfoot
Synchronization and rejection & \passthrough{\lstinline!src/main.rs!}, configuration, and synchronization-related tests & Unverified, out-of-domain, paused, revoked, or hash-inconsistent materials must not enter the visible index. \\
Exact and structured queries & \passthrough{\lstinline!/discovery/resources/query!} and query models & DID exact queries, resource type, protocol, labels, domain, and limit should have stable validation and error contracts. \\
Semantic retrieval & \passthrough{\lstinline!services/embedding-service!}, semantic aliases, and recommendation data & Relevance affects only ranking/explanation and cannot replace control, Root platform publication, or governance status. \\
Deduplication and versions & Resource index and version fields & Multiple copies of the same version of one DID must not be displayed as independent resources; current and historical versions must be distinguishable. \\
Freshness and degradation & Synchronization status, index statistics, and query responses & Old snapshots may support limited display, but must not create new publication, authorization, or call permissions. \\
\end{longtable}
}

Current fixed fixtures cover in-domain visibility, out-of-domain invisibility, empty results, arbitrary-label matching, and invalid queries. Real synchronization, vector-model quality, concurrent indexing, and cross-node convergence still require confirmation through service integration or deployment tests. Discovery node maintainers are responsible for index consistency, query error contracts, source retention, governance-status filtering, and semantic-configuration versioning. Maintenance diagnosis for discovery queries can trace backward from the entry point: \passthrough{\lstinline!/discovery/resources/query!} handles request parsing and query dispatch; index tables provide the read model for verified resources; semantic aliases and capability-label materials provide normalization and candidate assistance; \passthrough{\lstinline!services/embedding-service/src/server.ts!} and \passthrough{\lstinline!src/embedding.ts!} participate in text-vector generation only when vector mode is enabled. When investigating ``duplicate results'', separately check same-DID version deduplication, index-copy deduplication, pagination cursors, and frontend display mapping; do not only adjust ranking or the frontend array. The Discovery node's code mapping must also retain the conversion relationship ``raw materials -\textgreater{} verification results -\textgreater{} index records -\textgreater{} query candidates''. An original ResourcePackage or Root platform publication notification must not be treated directly as a queryable result; it may enter the index only after source, hash, version, authorization-domain, and lifecycle checks are complete. Relevance scores, label hits, and text snippets in query responses are ranking or explanation evidence; the resource DID, version, source, and governance status are the factual entry points for callers to continue verification.

\section{\texorpdfstring{\texttt{oan-trust-indexer}}{oan-trust-indexer}}\label{sec:appendix-f.5}

This repository corresponds to Chapter~\ref{chap:13}, Chapters~\ref{chap:19}~through~\ref{chap:20}, and authorization-status and recovery requirements. \passthrough{\lstinline!src/event\_source.rs!} handles event sources, \passthrough{\lstinline!src/sync.rs!} handles cursor advancement and synchronization, \passthrough{\lstinline!src/store.rs!} handles projection storage, \passthrough{\lstinline!src/domain.rs!} handles domain state, \passthrough{\lstinline!src/api.rs!} and \passthrough{\lstinline!src/grpc.rs!} provide query interfaces, and \passthrough{\lstinline!src/config.rs!} handles runtime configuration. \passthrough{\lstinline!configs/mainnet.example.toml!}, \passthrough{\lstinline!mainnet.current-events.toml!}, and the single-checkpoint smoke configuration distinguish continuous synchronization, current-event observation, and diagnostic runtime modes.

The indexer's trust path is ``read on-chain events -\textgreater{} process them in order or by cursor -\textgreater{} validate event structure and state transitions -\textgreater{} write the off-chain read model -\textgreater{} provide state queries externally''. The read model must retain the event source, cursor, last successful processing time, projection time, errors, and necessary version information. The off-chain projection is a queryable cache of governance facts and does not create on-chain authorization; when using it, the Root platform, Registrar node, and Discovery node should account for delay, staleness, and fail-closed boundaries. The latest sequence returned by \passthrough{\lstinline!trust/v1/status!} or an equivalent interface can only serve as an observation of index progress and cannot alone prove that a particular on-chain transaction has reached finality.

Existing admission fixtures can verify consistency for active, suspended, and revoked states and governance snapshots, but they are not equivalent to proof of a real chain RPC, GraphQL endpoint, finality, checkpoint recovery, or continuously growing data. Maintenance responsibilities include event-source compatibility, non-regressing cursors, idempotent duplicate-event handling, exception retries, database backups, and state-freshness monitoring; when changing the chain endpoint, do not only change the address while overlooking the linked checks for package, objects, event types, and parsing logic. The interface response of the off-chain trust indexer can only describe the state of its read model. Maintainers should place the event source, last processed cursor, projection time, errors, and chain-endpoint configuration in the same runtime record, and record ``on-chain event read'', ``off-chain state projected'', and ``Root platform adopted this state'' separately; a successful health endpoint must not merge the three into one conclusion. An off-chain trust indexer version upgrade should perform at least one small-scope replay or checkpoint-recovery verification: use a known event sequence to check parsing results, cursors, duplicate-event handling, and state projections, then compare the subject, node, credential issuer, and event sequence returned by the API. If the package, object, or event type changes, determine separately whether the old read model can continue to be used; a restarted process with a successful \passthrough{\lstinline!/health!} response does not establish that historical governance state has been correctly recovered.

\section{\texorpdfstring{\texttt{oan-sdk-ts}}{oan-sdk-ts}}\label{sec:appendix-f.6}

This repository corresponds to Chapters~\ref{chap:04}~through~\ref{chap:08}, Chapter~\ref{chap:11}, Chapters~\ref{chap:15}~through~\ref{chap:16}, and Appendices~\ref{chap:appendix-a}~through~\ref{chap:appendix-c}. \passthrough{\lstinline!packages/sdk-ts/src/index.ts!} provides DID Document drafts, ResourcePackage shape validation, lifecycle checks, registration, discovery, query explanations, capability-label assistance, and governance reads; \passthrough{\lstinline!identity.ts!} handles Local Identity and resource records, key-pair generation, identity-package import/export, and registration materials; \passthrough{\lstinline!did-oan.ts!} handles \passthrough{\lstinline!did:oan!} parsing and routing-code behavior. \passthrough{\lstinline!packages/protocol-types/src/index.ts!} provides cross-package protocol types. Tests are located in \passthrough{\lstinline!tests/sdk-core-tests.ts!}, \passthrough{\lstinline!identity-store-node-tests.ts!}, and \passthrough{\lstinline!client-workflow-tests.ts!}.

The SDK is a caller tool, not the authoritative implementation of the Registrar, Root, or Discovery. It can generate keys locally, construct DID Documents and registration materials, call public interfaces, and structurally validate ResourcePackage, but it cannot confirm on-chain finality for the caller and must not treat recommendation results as authorization conclusions. Endpoint stubs, fixed DIDs, and placeholder hashes in tests are used to verify interface behavior; publication examples must replace them with real endpoints, package hashes, authorization domains, and resource descriptions. Private keys and Local Identity packages are stored by the caller and are not sent to OpenAgenet (OAN) nodes.

{\def\LTcaptype{none} 
\begin{longtable}[]{@{}>{\raggedright\arraybackslash}p{0.2400\linewidth}>{\raggedright\arraybackslash}p{0.2400\linewidth}>{\raggedright\arraybackslash}p{0.2400\linewidth}>{\raggedright\arraybackslash}p{0.2400\linewidth}@{}}
\toprule\noalign{}
SDK capability & Corresponding implementation & Verification evidence & Maintenance boundary \\
\midrule\noalign{}
\endhead
\bottomrule\noalign{}
\endlastfoot
DID and identity & \passthrough{\lstinline!identity.ts!}, \passthrough{\lstinline!did-oan.ts!} & SDK core and identity-store tests & Preserve binding among DID, subject type, public key, and local private key; does not implement network governance. \\
Registration materials & \passthrough{\lstinline!index.ts!}, identity-record conversion functions & Client workflow tests & Materials are still revalidated by the Registrar and Root platform after generation. \\
Discovery and explanation & Discovery client in \passthrough{\lstinline!index.ts!} & Stub workflow and query-assistance tests & Result ranking/explanation cannot replace the caller's final verification. \\
Resource validation & ResourcePackage shape and lifecycle functions & SDK core tests & Structural validation is not equivalent to verification of real VC, governance, or endpoint availability. \\
\end{longtable}
}

SDK release maintenance must also check build output, export entry points, Node/browser runtime boundaries, and protocol-version compatibility, to avoid inconsistencies among the README, npm package contents, and source API. The SDK's public entry points are governed by \passthrough{\lstinline!package.json!} \passthrough{\lstinline!exports!}: the root entry, \passthrough{\lstinline!./client!}, \passthrough{\lstinline!./governance!}, \passthrough{\lstinline!./protocol-types!}, and \passthrough{\lstinline!./identity-store-node!} correspond to modules in the build output. After changing \passthrough{\lstinline!packages/*/src!}, check whether \passthrough{\lstinline!npm run build!} generates the corresponding JavaScript and declaration files, then run \passthrough{\lstinline!tests/sdk-core-tests.ts!}, \passthrough{\lstinline!tests/client-workflow-tests.ts!}, and \passthrough{\lstinline!tests/identity-store-node-tests.ts!}; even if only internal implementation changes but export paths, request fields, or error types change, it should still be treated as a client-compatibility change. SDK mapping should record the boundary between ``local generation'' and ``remote confirmation'': identity, DID Document drafts, ResourcePackage shape, and signature materials can be generated or verified locally by the caller, whereas registration VC, Root proof, Discovery index status, and governance status come from nodes or subsequent caller queries. The existence of an SDK method name does not mean that remote business processing has completed; examples and tests should make clear which responses are supplied by stubs and which require a real Registrar node or Discovery node.

\section{\texorpdfstring{\texttt{oan-community-skill}}{oan-community-skill}}\label{sec:appendix-f.7}

This repository corresponds to Chapter~\ref{chap:16} and the community-caller operation paths in Chapters~\ref{chap:04}~through~\ref{chap:08}, 11, 15, and 18. Its \passthrough{\lstinline!SKILL.md!} makes clear that the skill package is used for organizing resource descriptions, managing Local Identity, generating registration materials, submitting to the Registrar node, performing Discovery queries, explaining semantic queries, observing lifecycle, and assisting with read-only governance; it does not start or maintain the Root platform, Registrar node, Discovery node, content distribution platform, NATS, PostgreSQL, or off-chain trust indexer, and it does not perform on-chain governance writes. The implementation description requires continued checking of \passthrough{\lstinline!src/index.ts!}, \passthrough{\lstinline!validation.ts!}, \passthrough{\lstinline!registration.ts!}, \passthrough{\lstinline!resource-description.ts!}, \passthrough{\lstinline!discovery.ts!}, \passthrough{\lstinline!lifecycle.ts!}, \passthrough{\lstinline!capability-assist.ts!}, and \passthrough{\lstinline!tests/oan-community-skill-tests.ts!}.

Typical inputs to the community skill package are a resource README, skill-package description, or product description supplied by the user. It first generates candidate registration materials, then reports missing fields and quality issues, requires the user to supplement facts such as name, description, public access address, authorization domain, capability labels, use cases, inputs and outputs, license, and maintainer, and finally calls the SDK's normal registration path. It must not invent endpoints, maintainers, licenses, or authorization domains, and must not treat capability labels as an authorization scope. For batch resources, report success, skipped items, failures, missing inputs, and next steps item by item; do not silently discard other results because one item failed.

The skill's maintenance responsibility is to keep user-operation semantics, SDK calls, public endpoint defaults, error classification, and privacy boundaries consistent; a published package must include the scripts, documentation, and resources required for runtime, rather than only TypeScript compilation output. It may call public or explicitly user-provided Registrar, Discovery, and optional trust-query endpoints, but must not expose private node governance materials or read the Root platform's private indexer. Implementation tracing for the community skill package should start at the public call entry in \passthrough{\lstinline!src/index.ts!}, follow the inputs and outputs of \passthrough{\lstinline!registration.ts!}, \passthrough{\lstinline!resource-description.ts!}, \passthrough{\lstinline!validation.ts!}, \passthrough{\lstinline!discovery.ts!}, and \passthrough{\lstinline!lifecycle.ts!} separately, and then compare them with \passthrough{\lstinline!tests/oan-community-skill-tests.ts!} and the published-package file list. The skill package organizes and calls; it does not generate unconfirmed facts for the user. Therefore, when adding automatic completion, label suggestions, or retry logic, confirm that inferred content is not written directly into registration materials and that submissions are not repeated unconditionally when the result is unknown. Changes to the community skill package and SDK should also distinguish ``helping generate materials'' from ``writing network facts''. The skill package can organize descriptions, prompt for missing fields, and call public interfaces, but it cannot treat inferred endpoints, labels, authorization domains, or licenses as user-confirmed registration facts; when the request result is unknown, it also cannot create duplicate VCs, duplicate versions, or duplicate publication tasks through unconditional retries.

\section{Official Website and Public Developer Entry Points}\label{sec:appendix-f.8}

The official website is a public display and operation entry point, corresponding to parts of Chapters 15, 16, 19, 20, 22, and 24. The frontend uses public Registrar, Discovery, Root platform, content distribution platform, and off-chain trust indexer endpoints; the backend provides the aggregation, statistics, access analytics, or proxy capabilities required by the website through \passthrough{\lstinline!/api/*!} and \passthrough{\lstinline!/health!}. Production deployment instructions, build and smoke-test scripts, and Nginx and systemd configurations describe build, configuration, reverse-proxy, and verification boundaries. The website's Home, Register, Discover, Network, and Docs pages can call or display protocol data, but page copy, maps, statistics, cached snapshots, and recommendation ranking belong to the presentation layer.

The website's registration page can generate or import Local Identity in the browser, construct a DID Document and registration materials, and submit them to a specified Registrar node; private keys must not be sent to the website backend. The discovery page can submit a DID or natural-language task description and display candidate resources and DID Document previews returned by the Discovery node; users must still perform final verification based on the resource DID, VC, Root proof, status, and endpoints. When the Network page displays nodes and metrics, it should distinguish current real-time data, old snapshots, unknown status, and statistical aggregation, and must not treat UI state as governance authority.

Hard checks for website deployment maintenance include: production frontend build variables must not retain local-machine addresses; the backend should be exposed only through a reverse proxy; static assets and backend versions must match; public HTTPS, language switching, registration transmission, discovery queries, the Network summary, and browser routes must all pass smoke tests. SEO, access statistics, and page caching can only be loosely coupled enhancements and must not change the protocol behavior of registration, discovery, governance, or node services. Website issue diagnosis should first distinguish the browser layer, website backend aggregation layer, and node-service layer: frontend pages and \passthrough{\lstinline!frontend/src/api.ts!} initiate requests and present results; backend \passthrough{\lstinline!backend/src/http/router.rs!} and the corresponding handlers provide public aggregation interfaces; the Registrar node and Discovery node are the direct handlers of registration/discovery business logic. When access-statistics writes, SEO metadata, browser caching, or old-snapshot reads fail, the page may degrade or omit the record, but must not block registration submission, discovery queries, or node-status judgment; any cross-layer change must run page-route, public-interface, and node-business smoke tests separately. The change boundary between website components and protocol implementation can be checked by interface direction: \passthrough{\lstinline!/api/*!} from the browser to the website backend is a public aggregation or proxy path, requests from the website backend to nodes are service-to-service calls, and the node's own \passthrough{\lstinline!/resources/register!} and \passthrough{\lstinline!/discovery/resources/query!} are the protocol business entry points. When changing website caching, statistics, SEO, or page copy, verify that failures in these features do not change node request bodies, response semantics, registration/discovery results, or governance status; when changing node interfaces, do not perform only web regression testing while skipping protocol and client tests.

\section{Specification Chapter and Implementation Module Mapping}\label{sec:appendix-f.9}

The following table provides the primary index for maintenance and release work. Chapter numbers are the entry points for specification responsibility; specific implementations still need to be checked against the preceding repository sections, the current commit, and the target deployment configuration.

{\def\LTcaptype{none} 
\begin{longtable}[]{@{}>{\raggedright\arraybackslash}p{0.1920\linewidth}>{\raggedright\arraybackslash}p{0.1920\linewidth}>{\raggedright\arraybackslash}p{0.1920\linewidth}>{\raggedright\arraybackslash}p{0.1920\linewidth}>{\raggedright\arraybackslash}p{0.1920\linewidth}@{}}
\toprule\noalign{}
Specification scope & Main implementation repository/module & Main interfaces or data & Main tests/evidence & Implementation status and maintenance responsibility \\
\midrule\noalign{}
\endhead
\bottomrule\noalign{}
\endlastfoot
Resource model, DID Document, ResourcePackage & \passthrough{\lstinline!oan-protocol-commons!}, \passthrough{\lstinline!oan-sdk-ts!} & \passthrough{\lstinline!DidDocument!}, resource types, ResourcePackage, hash fields & Rust/TS unit tests, resource fixtures & Basic structure implemented; shared-library maintainers lead protocol-type changes. \\
Registration and control proofs & \passthrough{\lstinline!oan-registrar-nodes!}, \passthrough{\lstinline!oan-sdk-ts!}, \passthrough{\lstinline!oan-community-skill!} & \passthrough{\lstinline!/resources/register!}, registration materials, subject control proof, registration VC & \passthrough{\lstinline!registrar-resource.*!}, client workflow tests & Partially implemented; real idempotency, concurrency, and upstream failures require continuous-integration verification. \\
Root platform verification and trusted publication & \passthrough{\lstinline!oan-root-nodes!}, \passthrough{\lstinline!oan-protocol-commons!} & Root resource verification/publication interface, Root proof, version & Root platform unit tests, Root/CDN harness, publication logs & Partially implemented; on-chain governance and production finality cannot be inferred from offline tests. \\
Content distribution and synchronization & \passthrough{\lstinline!oan-root-nodes/services/cdn-publisher!}, \passthrough{\lstinline!cdn-node!}, Discovery node synchronization logic & Publication queue, ResourcePackage, cursor, hash & Root/CDN workflow tests, synchronization logs & Partially implemented; checkpointing, out-of-order processing, and real fault recovery require deployment-level verification. \\
Discovery and semantic governance & \passthrough{\lstinline!oan-discovery-nodes!}, \passthrough{\lstinline!oan-protocol-commons/oan-semantic-recommender!} & Query requests, capability-label tree, semantic aliases, index records & Discovery fixtures, query-explanation and semantic-service tests & Structured/semantic paths exist; model quality and cross-node convergence require separate records. \\
Node authorization and governance status & \passthrough{\lstinline!oan-trust-indexer!}, Root platform, admission test kit & Governance events, authorization domain, lifecycle, cursor & governance/node/domain/lifecycle fixtures & Partially implemented; indexer delay, chain endpoint, and finality are runtime-verification concerns. \\
Public APIs and error contracts & \passthrough{\lstinline!oan-protocol-commons!}, each node's \passthrough{\lstinline!main.rs!}, \passthrough{\lstinline!oan-sdk-ts!} & HTTP paths, status codes, error codes, pagination & \passthrough{\lstinline!docs/error-codes.md!}, request fixtures, SDK tests & Main interfaces implemented; new fields/error codes must be synchronized across all clients. \\
Community developer experience & \passthrough{\lstinline!oan-community-skill!}, \passthrough{\lstinline!oan-sdk-ts!} & Local Identity, resource descriptions, registration/discovery assistance & Skill tests, SDK tests, npm build output & User path implemented; package completeness and version compatibility are the package maintainer's responsibility. \\
Official website and operations entry point & Website frontend, website backend, and deployment workflow & \passthrough{\lstinline!/api/*!}, static frontend, health and smoke tests & Production deployment instructions, build/smoke-test scripts, browser checks & Presentation/aggregation layer; not protocol authority, deployment maintained by operations maintainers. \\
Third-party node conformance & \passthrough{\lstinline!oan-third-party-node-admission-test-kit!} & Fixed manifest, fixtures, expected results, reports & 49 requirement checks, case matrix, traceability matrix & Evaluation support; current V1-alpha does not cover a real Root platform, chain, CDN, or indexer. \\
\end{longtable}
}

Cross-index maintenance follows four rules: first, when adding or modifying a shared field, update the specification and \passthrough{\lstinline!oan-protocol-commons!} first, then check all production services and clients; second, when adding an interface, update the error contract, SDK, community skill package, and fixed request examples at the same time; third, governance status, Root proof, and semantic matching must retain their respective sources of fact and must not be merged in an aggregation layer into an unexplained ``trusted'' boolean; fourth, passing deployment scripts and smoke tests only proves that the target environment met the conditions checked at that time and must not turn environment variables, domain names, models, node counts, or database selections into fixed protocol requirements.

When implementation and specification diverge, maintainers should state in the change record the affected chapters, input/output differences, compatibility strategy, test evidence, whether data migration is created, and rollback method. Only when code, tests, configuration, and runtime verification jointly support it may ``Reference / to be verified'' be promoted to ``Implemented''; any aggregate result that lacks a source of fact, verification subject, or status boundary can only be observation information and cannot serve as cross-node trust proof.

The minimum tracking record for a cross-repository change can be established in the following order:

\begin{lstlisting}
Chapter or field change
  -> oan-protocol-commons types/validation/errors
  -> Root, Registrar, Discovery, or indexer call paths
  -> SDK and community Skill inputs and outputs
  -> fixtures, unit tests, and integration tests
  -> configuration, systemd/reverse proxy, and deployment smoke tests
  -> compatibility conclusion, release commit, and rollback basis
\end{lstlisting}

If a change concerns only website copy, SEO, or access statistics, explicitly record it as a presentation-layer website change and do not list it as a protocol-version change; if a change concerns protocol fields, DID Document, ResourcePackage, authorization domain, error codes, or lifecycle status, it must return to the shared types and all consumers for item-by-item verification. This tracking chain is for maintenance and review and does not change the original responsibility boundaries of the repositories.

``Implemented'' in the cross-index should be based on the minimum evidence set: the corresponding code entry point is locatable, there is at least one targeted test or repeatable check, the runtime configuration can resolve the required dependencies, and there is no conflict with current protocol fields or error contracts. If a directory only contains a file but has no call path, test result, or runtime configuration, it should remain ``Reference / to be verified''; if only test fixtures pass without service integration, it should be marked ``Evaluation support'' or ``Partially implemented'' and must not be promoted directly to production capability.

\section{References}\label{sec:appendix-f.10}

{\def\LTcaptype{none} 
\begin{longtable}[]{@{}>{\raggedright\arraybackslash}p{0.3200\linewidth}>{\raggedright\arraybackslash}p{0.3200\linewidth}>{\raggedright\arraybackslash}p{0.3200\linewidth}@{}}
\toprule\noalign{}
Source & Type & Link \\
\midrule\noalign{}
\endhead
\bottomrule\noalign{}
\endlastfoot
\passthrough{\lstinline!oan-protocol-commons!} & Code repository: shared protocol foundations & \url{https://github.com/wolfbrother/oan-protocol-commons} \\
\passthrough{\lstinline!oan-root-nodes!} & Code repository: Root platform and content distribution coordination & \url{https://github.com/OpenAgenet/oan-root-nodes} \\
\passthrough{\lstinline!oan-registrar-nodes!} & Code repository: Registrar node & \url{https://github.com/OpenAgenet/oan-registrar-nodes} \\
\passthrough{\lstinline!oan-discovery-nodes!} & Code repository: Discovery node & \url{https://github.com/OpenAgenet/oan-discovery-nodes} \\
\passthrough{\lstinline!oan-trust-indexer!} & Code repository: off-chain trust indexer & \url{https://github.com/OpenAgenet/oan-trust-indexer} \\
\passthrough{\lstinline!oan-sdk-ts!} & Code repository: TypeScript SDK & \url{https://github.com/OpenAgenet/oan-sdk-ts} \\
\passthrough{\lstinline!oan-community-skill!} & Code repository: community Skill & \url{https://github.com/OpenAgenet/oan-community-skill} \\
\end{longtable}
}

\end{document}